\documentclass[a4]{report}

\usepackage{etoolbox}
\usepackage[T1]{fontenc}
\patchcmd{\thebibliography}{\section*{\refname}}{}{}{}
\usepackage{todonotes}
\usepackage{amssymb}
\usepackage{amsmath}
\usepackage{pgfplots}
\usepackage{amsthm}
\usepackage[utf8]{inputenc}
\usepackage{mathrsfs}
\usepackage{dsfont}
\usepackage{mathtools}
\usepackage{physics}
\usepackage{bm}
\usepackage{thmtools}
\usepackage{verbatim}
\usepackage[margin=3.8cm,top=3.4cm,bottom=3.6cm]{geometry}
\usepackage{hyperref}

\usepackage[capitalize]{cleveref}
\hypersetup{
	colorlinks=true,
	linkcolor=blue,      
	urlcolor=cyan,
}

\usepackage{xcolor}
\usepackage{qcircuit}
\newcommand{\Ngate}[2]{*+<.6em>{#1} \POS ="i","i"+UR;"i"+UL **\dir{-};"i"+DL **\dir{-};"i"+DR **\dir{-};"i"+UR **\dir{-},"i"  #2}
\newcommand{\Nmultigate}[3]{*+<1em,.9em>{\hphantom{#2}} \POS [0,0]="i",[0,0].[#1,0]="e",!C *{#2},"e"+UR;"e"+UL **\dir{-};"e"+DL **\dir{-};"e"+DR **\dir{-};"e"+UR **\dir{-},"i" #3}
\newcommand{\Nghost}[2]{*+<1em,.9em>{\hphantom{#1}} #2}
\newcommand{\ww}[1][-1]{\ar@{~} [0,#1]}
\newcommand{\dw}[1][-1]{\ar@{.} [0,#1]}
\newcommand{\dbw}[1][-1]{\ar@{=} [0,#1]}
\newcommand{\dbww}[1][-1]{\ar@{$\approx$} [0,#1]}
\usepackage{tikz}
\usetikzlibrary{positioning}
\usetikzlibrary{calc}
\usepackage{contour}
\usepackage{ulem}

\contourlength{0.8pt}
\newcommand{\myuline}[1]{%
	\uline{\phantom{#1}}%
	\llap{\contour{white}{#1}}%
}
\makeatletter
\newcommand\footnoteref[1]{\protected@xdef\@thefnmark{\ref{#1}}\@footnotemark}
\makeatother
\makeatletter
\def\Ddots{\mathinner{\mkern1mu\raise\p@
		\vbox{\kern7\p@\hbox{.}}\mkern2mu
		\raise4\p@\hbox{.}\mkern2mu\raise7\p@\hbox{.}\mkern1mu}}
\makeatother

\theoremstyle{plain}
\newtheorem{Thm}{Theorem}[section]
\newtheorem{Prop}[Thm]{Proposition}
\newtheorem{Cor}[Thm]{Corollary} 
\newtheorem{Lem}[Thm]{Lemma}
\newtheorem{Conj}[Thm]{Conjecture}
\newtheorem{OP}[Thm]{Open Problem}
\declaretheorem[style=definition,qed=$\blacksquare$, sibling=Thm]{Definition}
\declaretheorem[style=definition,qed=$\blacklozenge$,sibling=Thm]{Example}
\declaretheorem[style=definition,qed=$\maltese$,sibling=Thm]{Remark}
\declaretheorem[style=definition,qed=$\maltese$,sibling=Thm]{Warning}
\newcommand{\N}{\mathbb{N}}

\newcommand{\R}{\mathbb{R}}
\newcommand{\C}{\mathbb{C}}

\newcommand{\bbT}{\mathbb{T}}
\newcommand{\bbE}{\mathbb{E}}
\newcommand{\bbI}{\mathbb{I}}

\newcommand{\bbD}{\mathbb{D}}
\newcommand{\bbH}{\mathbb{H}}
\newcommand{\bbP}{\mathbb{P}}

\newcommand{\cG}{\mathcal{G}}
\newcommand{\cF}{\mathcal{F}}
\newcommand{\cL}{\mathcal{L}}

\newcommand{\cS}{\mathcal{S}}
\newcommand{\cR}{\mathcal{R}}

\newcommand{\cV}{\mathcal{V}}

\newcommand{\cB}{\mathcal{B}}

\newcommand{\cH}{\mathcal{H}}
\newcommand{\cP}{\mathcal{P}}
\newcommand{\cW}{\mathcal{W}}
\newcommand{\cU}{\mathcal{U}}

\newcommand{\cE}{\mathcal{E}}
\newcommand{\cK}{\mathcal{K}}

\newcommand{\cX}{\mathcal{X}}
\newcommand{\cY}{\mathcal{Y}}

\newcommand{\cZ}{\mathcal{Z}}

\newcommand{\fF}{\mathfrak{F}}

\newcommand{\fC}{\mathfrak{C}}

\newcommand{\scrA}{\mathscr{A}}
\newcommand{\scrB}{\mathscr{B}}
\newcommand{\scrT}{\mathscr{T}}

\newcommand{\scrS}{\mathscr{S}}
\newcommand{\scrD}{\mathscr{D}}

\newcommand{\scrF}{\mathscr{F}}

\newcommand{\scrE}{\mathscr{E}}

\newcommand{\scrC}{\mathscr{C}}
\newcommand{\scrI}{\mathscr{I}}

\newcommand{\bfC}{\mathbf{C}} 
\newcommand{\bfE}{\mathbf{E}}

\newcommand{\bfD}{\mathbf{D}}
\newcommand{\bfP}{\mathbf{P}}

\newcommand{\bbX}{\mathbb{X}}
\newcommand{\bbY}{\mathbb{Y}}
\newcommand{\bbZ}{\mathbb{Z}}
\newcommand{\bbG}{\mathbb{G}}

\newcommand{\sfP}{\mathsf{P}}

\newcommand{\sfJ}{\mathsf{J}}

\newcommand{\sfx}{\mathsf{x}}
\newcommand{\sfy}{\mathsf{y}}
\newcommand{\sfz}{\mathsf{z}}
\newcommand{\sfe}{\mathsf{e}}
\newcommand{\sfa}{\mathsf{a}}

\newcommand{\sfc}{\mathsf{c}}
\newcommand{\sfk}{\mathsf{k}}
\newcommand{\sfm}{\mathsf{m}}
\newcommand{\sfh}{\mathsf{h}}
\newcommand{\sfone}{\mathsf{1}}
\newcommand{\sftwo}{\mathsf{2}}

\newcommand{\pow}[1]{\mathcal{P}(#1)}
\newcommand{\into}{\rightarrow}

\newcommand{\bone}{\mathds{1}}
\newcommand{\id}{\textup{id}}

\newcommand{\up}[1]{\textup{#1}}

\newcommand{\Span}[1]{\textup{span}\{#1\}}
\newcommand{\End}[1]{\textup{End}(#1)}
\newcommand{\supp}[1]{\textup{supp}(#1)}

\newcommand{\sfA}{\mathsf{A}}
\newcommand{\sfB}{\mathsf{B}}
\newcommand{\sfp}{\mathsf{p}}
\newcommand{\sfq}{\mathsf{q}}

\newcommand{\Theory}{\mathbf{\Theta}}
\newcommand{\CIT}{\mathbf{CIT}}
\newcommand{\QIT}{\mathbf{QIT}}
\newcommand{\IC}[1]{\textup{IC}(#1)}

\newcommand{\channel}[3]{\scalemyQ{.8}{0.7}{0.5}{ & \push{#1}  \qw & \gate{#2} & \push{#3} \qw & \qw}}

\newcommand{\state}[2]{\scalemyQ{.8}{0.7}{0.5}{ &  \Ngate{#1} & \push{#2} \qw & \qw}}

\newcommand{\bistate}[3]{\scalemyQ{.8}{0.7}{0.5}{ &  \Nmultigate{1}{#1} & \push{#2} \qw & \qw \\
& \Nghost{#1} & \push{#3} \qw & \qw } }

\newcommand{\onetwochannel}[4]{\scalemyQ{.8}{0.7}{0.5}{  & \push{#1}  \qw & \multigate{1}{#2} & \push{#3} \qw & \qw \\
& & \Nghost{#2} & \push{#4} \qw & \qw} }

\newcommand{\twoonechannel}[4]{\scalemyQ{.8}{0.7}{0.5}{ & \push{#1}  \qw & \multigate{1}{#3} &  \\
		& \push{#2}  \qw & \ghost{#3} & \push{#4} \qw & \qw} }

\newcommand{\oneext}[4]{\scalemyQ{.8}{0.7}{0.5}{ & & \Nmultigate{1}{#2} & \push{#4} \ww & \ww\\ 
	& \push{#1}  \qw & \ghost{#2} & \push{#3} \qw & \qw}}

\newcommand{\twoext}[5]{\scalemyQ{.8}{0.7}{0.5}{ & \push{#2}  \ww& \Nmultigate{1}{#3}{\ww} & \push{#5} \ww & \ww\\ 
			& \push{#1}  \qw & \ghost{#3} & \push{#4} \qw & \qw} }

		\newcommand{\Sys}[1]{\up{Sys}_{#1}}
		\newcommand{\triv}{\mathbf{1}}
		\newcommand{\AllTrans}[1]{\up{Trans}_{#1}}
		\newcommand{\Trans}[3]{\up{Trans}_{#1}(#2, #3)}
				\newcommand{\Chan}[3]{\up{Chan}_{#1}(#2, #3)}
						
		\newcommand{\Sets}{\mathbf{Sets}}
		\newcommand{\St}[1]{\up{St}(#1)}
		\newcommand{\bbK}{\mathbb{K}}
	\newcommand{\OblQIT}{\mathbf{OblQIT}}
		\newcommand{\OblCIT}{\mathbf{OblCIT}}
		\newcommand{\NCIT}{\mathbf{NCIT}}
		
		\newcommand{\Top}{\mathbf{Top}}
		\newcommand{\Groups}{\mathbf{Groups}}
		\newcommand{\FinSets}{\mathbf{FinSets}}
		\newcommand{\Vect}[1]{\mathbf{Vect}_{#1}}

		\usepackage{stmaryrd}
\newcommand{\og}{\talloblong}
\newcommand{\after}{\, \vcenter{\hbox{\scalebox{0.4}{$\bigodot$}}} \, }
\newcommand{\swap}[2]{\sigma_{#1, #2}}

\newcommand{\bbW}{\mathbb{W}}
\newcommand{\bbA}{\mathbb{A}}

\usepackage{caption}
\graphicspath{{"/Users/nicholashoughton-larsen/Dropbox/Uni/PhD/PhD Thesis/Pictures/"}}
\usepackage{float}
\newcommand{\Graphs}{\mathbf{Graphs}}
\usepackage{tikz-cd}
\usepackage{adjustbox}
\usepackage[toc,page]{appendix}

\newcommand{\der}{{\trianglerighteq}}
\newcommand{\red}{{\trianglelefteq}}
\usepackage{epsdice}

\newcommand{\ports}[1]{\mathsf{ports}(#1)}
\newcommand{\ICC}[1]{\up{ICC}(#1)}

\newcommand{\bbB}{\mathbb{B}}

\usepackage{stackengine,scalerel}
\newcommand{\contr}[2]{{\fC_{#2}( #1 )}}
\newcommand{\cder}{\trianglerighteq}

\newcommand{\bbQ}{\mathbb{Q}}
\newcommand{\sfb}{\mathsf{b}}

\newcommand{\Nctrlo}[2]{\controlo \qwx[#1] #2}

\newcommand{\myQ}[3]{\begin{array}{c}\Qcircuit@C=#1em@R=#2em{#3} \end{array}}
\newcommand{\scalemyQ}[4]{\scalebox{#1}{$\myQ{#2}{#3}{#4}$}}
\newcommand{\Dil}[1]{\up{Dil}(#1)}
\newcommand{\CDil}[2]{\up{CausDil}^{#2}(#1)}
\newcommand{\OneDil}[1]{\up{Dil}_0(#1)}

\newcommand{\TREX}{\mathbf{T}-\mathbf{REX}}

\newcommand{\Logic}{\mathbf{Logic}}
\newcommand{\res}{\up{res}}
\newcommand{\scrU}{\mathscr{U}}
\newcommand{\Int}[1]{\up{Int}(#1)}

\newcommand{\Cns}[1]{\up{Cons}(#1)}

\usepackage{afterpage}

\begin{document}
				\pagenumbering{gobble}

\thispagestyle{empty}
{
	\ \\[-1cm]
	\vfill
	{\raggedleft\large\bfseries Nicholas Gauguin Houghton-Larsen \par}
	\vfill
	\linethickness{0.45mm}
	\begin{center}
		\line(1,0){414}
	\end{center}
	\ \\[0.1cm]
	{\centering\LARGE\bfseries A Mathematical Framework for \\ Causally Structured Dilations  and its \\ Relation to Quantum Self-Testing \par}
	\ \\[0.1cm]
		\linethickness{0.45mm}
			\begin{center}
			\line(1,0){414}
		\end{center}
	
\vfill
	
	\emph{This PhD thesis has been submitted for assessment to the PhD School of the Faculty of Science,
	University of Copenhagen, on the 1st of December 2020.}\vfill\noindent

	\vfill
	{PhD thesis\hfill{}$\bullet$\hfill{}Department of Mathematical Sciences\hfill{}$\bullet$\hfill{}University of 
		Copenhagen\\}
	\newpage

	\noindent	\textsc{PhD Thesis by:}\\[0.1cm]
Nicholas Gauguin Houghton-Larsen\\
	Department of Mathematical Sciences,
	University of Copenhagen\\
	Universitetsparken 5,
	2100 K{\o}benhavn {\O},
	Denmark\\[0.1cm]
	\texttt{nicholas.gauguin@gmail.com}
	
		\vspace{1cm}
	
	\noindent \textsc{Date of Submission:} \\[0.1cm]
	1st of December 2020
	
	\vspace{1cm}
	
		\noindent \textsc{Date of Defence:} \\[0.1cm]
	11th of February 2021
	
	\vspace{1cm}

	\noindent 	\textsc{Supervisor:}\\[0.1cm]
	Matthias Christandl (professor), University of Copenhagen, Denmark
	\vspace{1cm}
	
	\noindent \textsc{Assessment Committee:} \\[0.1cm]
	Roger Colbeck (professor), University of York, UK  \\
			Tobias Fritz (assistant professor), University of Innsbruck, Austria\\
			Nathalie Wahl (professor), University of Copenhagen, Denmark



\vfill
	
%
%
%
%


\noindent This is the first version of my thesis uploaded to \texttt{www.arxiv.org}. The University of Copenhagen holds a version with ISBN 978-87-7125-039-8 (at the time of writing it is stored at \url{https://www.math.ku.dk/english/research/phd-theses/}). In the present version, a few typographical errors have been rectified and some details have been added to the proof of \cref{thm:AppInfoDist}. \\ [0.3cm]
Comments and further corrections are very welcome and may be sent to my private email address found on the top of this page.\\[0.3cm]

\begin{flushright}	\emph{Nicholas Gauguin Houghton-Larsen} \\
	\emph{Copenhagen, March 2021} \end{flushright}

	\newpage

{\centering
	\subsubsection*{Abstract}}
	This is a PhD thesis within the sub-field of mathematical physics that pertains to \emph{quantum information theory}. Most of its results can be interpreted in the mathematical language of category theory, and may as such be of interest also outside of quantum information theory.

	In high-level terms, I present a framework in which one can argue mathematically about aspects of the following fundamental question: \emph{How do two given implementations of the same physical process compare to each other?} Though of independent interest, the main motivation for this question comes from the area of \emph{quantum self-testing} (\cite{MY98, MY04}), where one desires to understand all the different ways in which a given set of measurement statistics can be produced by an implementation of local measurements on a multipartite quantum state. The problem which motivated the thesis is that although the traditional envision of quantum self-testing is mathematically precise, the language in which it is cast has no clear operational interpretation.
	
	According to the framework proposed in the thesis, a collection of measurement statistics is regarded as the input-output behaviour of an information channel, and the various implementations of this channel correspond to causally structured computations which may be secretly executed in the environment of the channel during our interaction with it. The main contribution of the thesis is to introduce a formalism which makes the previous sentence precise, and to provide its relation to the usual definition of quantum self-testing. The relation  is essentially that quantum self-testing corresponds to the existence of an  implementation from which all others can be derived, and which moreover holds no pre-existing information about the outputs of the channel. This constitutes a first step towards recasting quantum self-testing in purely operational (theory-independent) terms.

\cref{chap:Theories} reviews a variation on a category-theoretic model for physical theories. This model includes quantum information theory and classical information theory, but also more mathematical examples such as any category with finite products (e.g. the categories of sets or groups), and any partially ordered commutative monoid, when suitably interpreted. The key feature of the model is that it facilitates the notion of \emph{marginals} (as known from e.g. classical probability theory), and the dual notion of \emph{dilations}. 
	
Dilations are the topic of \cref{chap:Dilations}. The results presented there are conceptually independent of quantum self-testing, but rather initiate a systematic study of dilations and constitute an original proof of concept, by demonstrating that several features of information theories can be derived from a handful of principles which reference only the structure of dilations.  

 \cref{chap:Metric} contains some initial thoughts as to how to make an approximate (metric) version of the theory of dilations, and a new metric for quantum channels, the \emph{purified diamond-distance} is introduced. It generalises the purified distance of Refs. \cite{Toma10,Toma12}. 

\cref{chap:Causal} lays out a formalism for arguing about information channels whose outputs are causally contingent on their inputs. This can be seen as a generalised alternative to the framework of quantum combs (\cite{Chir09combs}), but can also be viewed as generalising the abstract notion of traces in symmetric monoidal categories (\cite{JSV96}). The formalism allows us to make precise the notion of a \emph{causal dilation}, which captures the above-mentioned causally structured side-computations.

Finally, in \cref{chap:Selftesting}, the connection to quantum self-testing is established. This chapter also contains simple proofs of a few general results about self-testing, and a novel recharacterisation of the set of quantum behaviours in terms of non-signalling properties of their Stinespring dilations.

\newpage

{\centering
	\subsubsection*{Resumé}}
Dette er en ph.d.-afhandling inden for den gren af matematisk fysik der vedrører \emph{kvanteinformationsteori}. De fleste af dens resultater kan fortolkes i et matematisk kategori-teoretisk sprog og kan som sådan være af interesse også uden for kvanteinformationsteorien.

I overordnede træk præsenteres en teoretisk ramme, i hvilken man matematisk kan tale om aspekter ved følgende grundlæggende spørgsmål: \emph{Hvad er forholdet mellem to givne implementeringer af den samme fysiske proces?} Spørgsmålet er af uafhængig interesse, men dets vigtigste motivation kommer fra feltet \emph{`quantum self-testing'} (\cite{MY98, MY04}), hvor man ønsker at forstå alle de forskellige måder, hvorpå et givent sæt af fordelinger for måleresultater kan fremkomme ved lokale målinger på en kvantetilstand delt mellem flere parter. Det problem der motiverede afhandlingen er, at omend den traditionelle opfattelse af `quantum self-testing' er matematisk præcis, så har det sprog i hvilket fænomenet er defineret ikke nogen klar operational fortolkning. 

Ifølge den teoretiske ramme der udlægges i afhandlingen betragtes et sæt af fordelinger for måleudfald som input-output-opførslen for en informationskanal, og de mulige implementeringer af denne kanal svarer til kausalt strukturerede processer som hemmeligt udføres i kanalens omgivelser i løbet af vores interaktion med den. Afhandlingens hovedbidrag er at indføre en formalisme der gør forudgående sætning præcis, samt at bestemme formalismens relation til den sædvanlige definition af `quantum self-testing'. Relationen er essentielt set, at `quantum self-testing' svarer til eksistensen af en implementering, hvorfra alle andre kan udledes, og som desuden ikke indeholder forhånds-eksisterende information om kanalens outputs. Dette udgør et første skridt i retning af en omarbejdning af `quantum self-testing' til rent operationelle (teori-uafhængige) termer.

Kapitel \ref{chap:Theories} gennemgår en variation af en kategori-teoretisk model for fysiske teorier. Modellen inkluderer kvanteinformationsteori og klassisk informationsteori, men også mere matematiske eksempler, såsom enhver kategori med endelige produkter (f.eks. kategorierne bestående af mængder eller grupper), og ethvert partielt ordnet kommutativt monoid, passende fortolket. Nøgleegenskaben ved modellen er, at den tillader begrebet \emph{marginalisering} (som det kendes eksempelvis fra sandsynlighedsteorien) og det duale begreb \emph{udvidelse} (eng. \emph{`dilations'}).

Udvidelser er emnet for Kapitel \ref{chap:Dilations}. Resultaterne, der præsenteres dér, er konceptuelt uafhængige af `quantum self-testing', men indleder snarere en systematisk undersøgelse af udvidelser og udgør et originalt `proof of concept' ved at demonstrere, at flere informations-teoretiske egenskaber kan udledes fra kun en håndfuld af principper, der alene refererer til strukturen af udvidelser.

Kapitel \ref{chap:Metric} indeholder nogle indlende tanker om, hvordan man kan lave en approksimativ (metrisk) teori for udvidelser, og en ny metrik for kvantekanaler, \emph{`purified diamond-distance'}, introduceres. Denne generaliserer `purified distance' fra Ref. \cite{Toma10,Toma12}.

Kapitel \ref{chap:Causal} udlægger en formalisme, hvori man kan tale om informationskanaler hvis outputs er kausalt betingede af deres inputs. Denne kan ses som et generaliseret alternativ til `quantum combs' (\cite{Chir09combs}), men kan også anskues som generalisering af abstrakte spor (eng. `traces') i symmetriske monoidiale kategorier (\cite{JSV96}). Formalismen giver os mulighed for at præcisere forestillingen om en \emph{kausal udvidelse} (eng. \emph{`causal dilation'}), der netop indfanger de ovennævnte kausalt strukturede sideprocesser.

Afslutningsvist etableres forbindelsen til `quantum self-testing' i Kapitel \ref {chap:Selftesting}. Dette kapitel indeholder også simple beviser for et par generelle resultater om `quantum self-testing', samt en ny karakterisering af mængden af `quantum behaviours' i termer af `non-signalling'-egenskaber ved deres Stinespring-udvidelser.

\newpage

{\centering
	\section*{Acknowledgements}} 
	\vspace{.8cm}

Though a PhD study appears most of the time to be a very lonely endeavour, it goes without saying that it rests upon the support and engagement of many actors. I owe my gratitude to several people and institutions, each of whom and which have played an instrumental role. \\

First and foremost, I would like to thank my supervisor, professor Matthias Christandl. He dreams big, is profoundly adaptive to new ideas, and firmly believes in the resolution of problems. When he suggested to me to investigate the meaning of quantum self-testing, he likely did not expect the present dissertation as outcome. I am deeply grateful towards him for always keeping open boundaries and giving me space to pursue ways and methods I believed to be interesting and illuminating. I hope that he finds my final results and presentation to duly honour his confidence in me.
When I think back at the time we spent together  in addition to discussing the project, I am reminded in particular of many entertaining lunch discussions we had when I visited him at MIT in August 2018, and of our exhaustive but exciting period of planning and executing the first year mathematics course Analyse 1 in April--June 2019. With Matthias, I learned a fundamentally different view on science and mathematics, and his vivid personality has made a lasting impression on me.  \\

Secondly, I am highly indebted to associate professor Laura Man\v{c}inska. Despite of having no contractual obligations to my studies, she has effectively co-supervised this thesis project, taken part in the vast majority of supervision meetings with Matthias, and even filled his role enthusiastically in the 6 months of his absence staying at MIT. I am thankful for her commitment and for sharing with me her expert knowledge of the field of quantum self-testing, contributing with many suggestions and points that Matthias and I would have probably overlooked on our own. \\

As a PhD student, I have been employed at the Villum Centre of Excellence for the Mathematics of Quantum Theory (QMATH) at the Department of Mathematical Sciences at the University of Copenhagen. I would like in this regard to acknowledge the funding by Sapere Aude, the European Research Council (ERC Grant Agreement no. 337603) and VILLUM FONDEN via the QMATH Centre (Grant no. 10059). 

At QMATH, I have enjoyed the help and company of many colleagues, and have been astounded by the engagement and ambition with which the centre is run by its scientific founders, professors Jan Philip Solovej, Matthias Christandl and Bergfinnur Durhuus, and by centre administrator Suzanne Andersen. I wish their enterprise the best of luck in the future.

The Department of Mathematical Sciences is a unique work place, and as I was also a bachelor and master student there, it is hard not to think of it with nostalgic warmth. Many of its employees have influenced and inspired me through time, and I would like to express my gratitude in particular towards Niels Grønbæk, Ernst Hansen, Niels Richard Hansen, Magdalena Musat, Jan Philip Solovej and (my peer) Asbjørn C. Nordentoft. I would also like to acknowledge the support from people with whom I have engaged in their administrative roles, in particular PhD secretary Nina Weisse, administrative officer Mette Fulling, PhD coordinator Morten S. Risager, and head of section Henrik Laurberg Pedersen.\\

During my PhD studies, I have had the pleasure of spending three weeks at MIT in Boston in August 2018 and two months at the ICMAT and Universidad Complutense in Madrid in October--September 2019. Both of these experiences were very enjoyable, and I would like to thank professor Aram Harrow at MIT and professor David P\'{e}rez-Garc\'{i}a at Complutense for their hospitality.  \\

The PhD committee comprises professor Roger Colbeck from the University of York, assistant professor Tobias Fritz from the University of Innsbruck, and (chair of committee) professor Nathalie Wahl from the University of Copenhagen. I am both honoured and happy about the engagement of these three, and would like to thank them for their time. I hope they will all find sentences of value within the thesis.\\ %

Last but not least, I am infinitely thankful towards my family and my friends for always supporting and encouraging me in my endeavours. They are too many to list, and I would not risk leaving any of them out. \\

\vfill

In the past, I often smiled at dedications of highly technical academic works to people without prerequisites for understanding their content. Now, I see that such dedications serve to recognise that those people were indispensable in shaping and sustaining the person who ultimately grew capable of materialising a product of such ridiculously demanding scope.  \\

With that in mind, I dedicate this work to my family. To my mother, who was one of the most extraordinary, ambitious and giving persons I have ever known; to my father, who ignited my interest in science and whose insights and advice continue to guide me; and to my brother, who always knows how to challenge me and whom I admire for his kindness and intellect more than he could possibly imagine.

\newpage

\begin{figure}[H]	\label{fig:Mondrian}
	\begin{center}
		\includegraphics[scale=0.3]{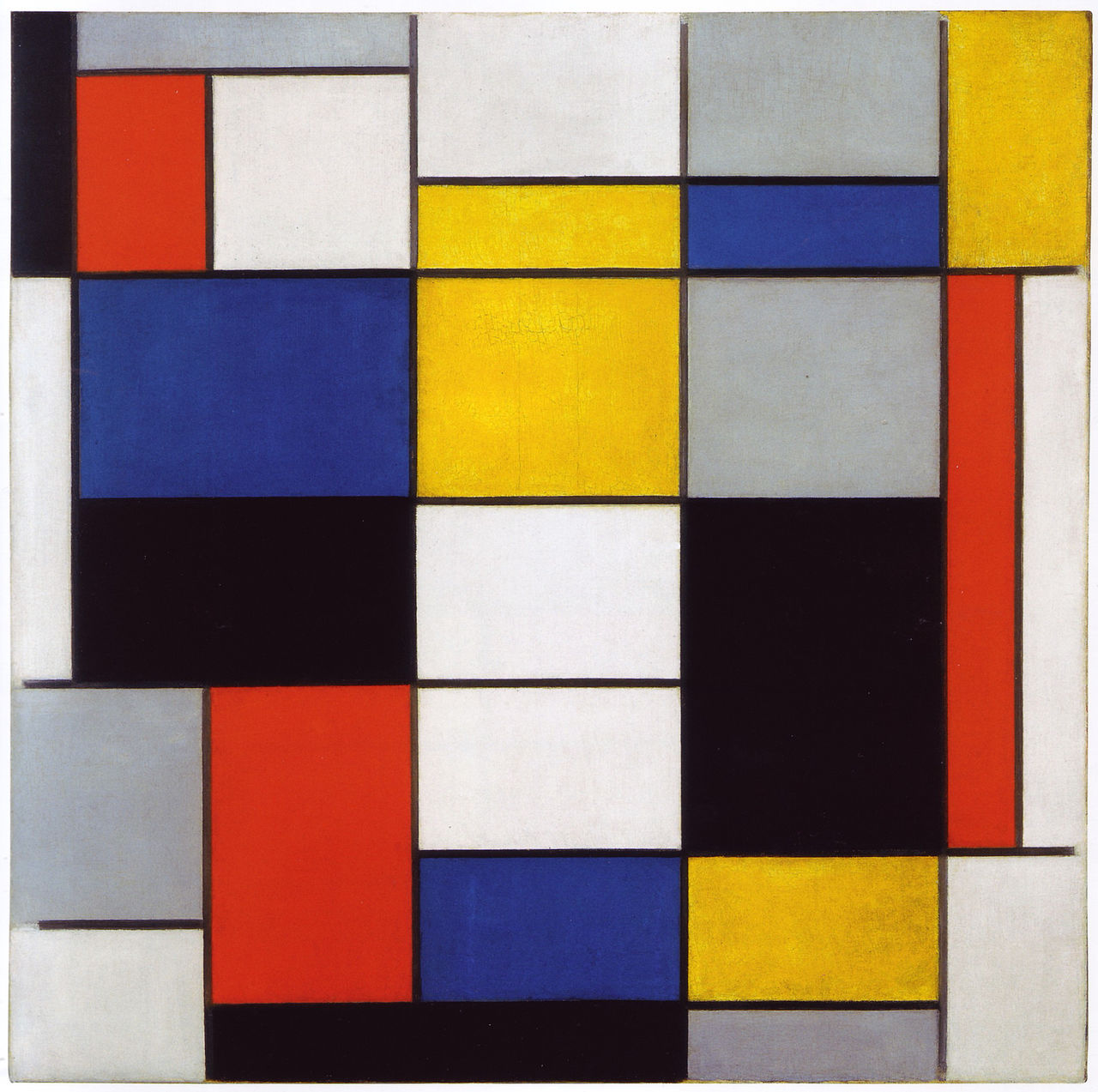}
		\caption*{\footnotesize 
			 \emph{Composition A} by Piet Mondrian (1923)} \caption*{\footnotesize  Galleria Nazionale d'Arte Moderna e Contemporanea}
	\end{center}
\end{figure}

{\centering
	\section*{A Note to the Reader}}

\vspace{0.8cm}

 Everyone who has written down something for anyone to read is familiar with the trivial but crucial condition that statements must be structured \emph{sequentially}, one sentence following the other, paragraph by paragraph, chapter after chapter. 
 
 A novelist can use this to advantage, by introducing characters and revealing plot twists according to a carefully crafted schedule. The author of an academic dissertation essentially has to do the same, but as a general rule this circumstance is hindering rather than advantageous. The reason is, of course, that abstract ideas are not connected in a linearly ordered fashion. 
 
An additional dare is posed for academic writing because a reader cannot be counted upon to read every single sentence from the beginning to the end. Few (if any) readers of a fictional novel start by reading the first pages, then read the last, and then sporadically glance through the chapters -- in contrast, the order of things which a PhD student envisions for a thesis might in the end not be the one most suitable to any given reader.

 In writing this document, I have strived for the storyline to emerge clearly from the general introduction and the individual introductions to the five chapters, so as to guide you as much as possible. This is, however, my first PhD thesis, and so I hope for forgiveness in cases where I have not succeeded in coping sublimely with structural challenges.

\newpage

$\phantom{x}$
\newpage

\tableofcontents

\newpage

\pagenumbering{roman}

\chapter*{Introduction}
\addcontentsline{toc}{chapter}{Introduction}

\section*{For Everyone}
\addcontentsline{toc}{section}{For Everyone}

A friend once told me that if you ask people whether they would rather be born in 100 years than live today, the vast majority say no. Ironically, if you then ask that majority whether they would prefer having lived in the world 100 years \myuline{ago}, they shake their heads again. Some of them probably realise their risk aversion. 

Most people assess living standards in terms of health and wealth, freedom of choice, security to education, and the like; as such, the state of humanity has indeed been on a steady rise during the last centuries, if not millennia (\cite{Rosl18, Pink18}). When gauging their lives, few might think of humankind's enterprises within mathematics and the natural sciences. Nonetheless, these too have undergone tremendous improvements during the same time.\footnote{Pondering the relationship between these two developments is left as an exercise to the reader.} \\

Popular consensus has it that \emph{modern natural science} began less than 500 years ago, owing to the impact of significant figures like the notorious stone-dropping Galileo Galilei (1564–-1642), a main proponent  and pioneer of the paradigmatic conviction that knowledge about the physical world should be acquired by experimental observation and formulated in mathematical terms  (\cite{Galileo}).\footnote{It goes without saying that these paragraphs represent gross simplifications of the history; proper accounts could easily fill hundreds of pages. The entry \cite{ScientificMethod} in the  Stanford Encyclopedia of Philosophy gives a decent overview of the history of \emph{`the scientific method'}.} As this program unfolded over the centuries, it instilled in its practitioners the aspiration to identify a small set of valid principles, \emph{laws of Nature}, which were not to be further explained themselves, but from which all other observed phenomena could be logically derived. (For example, Newton's laws of motion and gravitation are simple and universal, yet allow us to derive information sufficient to safely send members of our species away from our planet and land them 380.000 km away on the Moon.) 

The idea of compressing all truth to a small set of postulates is an imprint from the \emph{mathematical science}, which itself dates back more than 2000 years as the very institutionalisation of logical inference, similarly personified by the iconic geometry-obsessed Euclid of Alexandria (ca. 300 BC). In contrast to the natural sciences, mathematics refuses external physical inputs for certification of its initial axioms and for justification of its desired conclusions; this shifts the emphasis from the actual content of statements to the logical interdependencies among statements themselves. (For example, it is known (\cite{BanachTarski}) that the so-called \emph{axiom of choice}\footnote{A formal version of the seemingly obvious statement that given any non-zero number of bags each of which contains at least one marble, it is possible to form a collection containing precisely one marble from each bag.} formally implies the absurd statement that a solid ball can be dissected into finitely many pieces which can be reassembled into two solid balls each identical to the original.) \\

It is difficult to find a word befitting of the scale of advancement that physics and mathematics have experienced since their conceptions -- the study of their evolution is a science in itself. All scientific activities are bound to progress in a trivial sense simply because knowledge is accumulative over time, at least insofar as it is recorded; as such, advancement would seem only a matter of speed. However, as articulated by the science philosopher Thomas Kuhn (1922-1996) (\cite{Kuhn12}), transitions of a much more disruptive character occasionally occur in the sciences, and they cause profoundly new mentalities to ascend.\\

Roughly 100 years ago, \myuline{both} physics and mathematics found themselves at such bewildering points of disruption, after many years of marching steadily and obliviously towards them.

In mathematics, the continued process of rigorously formalising its concepts in the language of set theory had approached a landscape inhibited by more and more intriguing entities; objects such as Peano's space-filling curve (\cite{Peano1890}), Weierstra\ss' nowhere differentiable but everywhere continuous function (\cite{Weier1872}), and Cantor's uncountable infinities (\cite{Cantor1884}) were proved to formally exist by abstract arguments, though their interpretation stretched the intuition of contemporaries. This growing balloon of peculiarities was building up tensions that forced mathematicians to question the very foundations of mathematical thinking, and exhibits such as Russell's paradox around 1903 (\cite{Russell}) eventually became so incriminating that the balloon cracked wide open. It was exposed that mathematics ultimately did not rest on solid formal grounds, and the so-called \emph{Foundational Crisis of Mathematics} was burning at its fullest.

In physics, a revolution of remarkably similar significance was playing out. The physicist Albert A. Michelson\footnote{A similar quote is often falsely attributed to William Thomson (Lord Kelvin).} had barely uttered the words (\cite{MichelsonThomson}) \emph{"[...] it seems probable that most of the grand underlying principles [in Physics] have been firmly established"} in 1894, before, as if orchestrated by the Goddess of Irony, chaos began to sprout -- among other things, Maxwell's equations for the successful theory of electromagnetism seemed to display a conflict with the principle of Galilean relativity, the so-called `ultraviolet catastrophe' plagued statistical mechanics, and Nature appeared to exhibit a weird discretized behaviour with respect to the emission of light from atoms. The tendency of these beauty flaws to resist elimination and rather conspire to unite in opposition was stressing and aggravating the physical community.  \\

Eventually, thanks to exceptional thinkers in both disciplines, these tensions were unravelled and new paradigms arose in mathematics and physics alike.\\

Physicists had understood that we needed to profoundly revise some of our dearest conceptions about how the world works. Albert Einstein realised that the notions of \emph{space} and \emph{time} behaved in surprising and malleable ways defying thousands of years of human intuition, resolving not only in 1905 the problem from Maxwell's equations (\cite{Einstein05}), but also providing over the years 1907--1915 a new and radically different theory of gravitation (\cite{Einstein16}). Today, his \emph{theory of general relativity} remains a landmark in physics. Similarly, a list of people too long to reproduce -- but including (Einstein and) Max Planck, Niels Bohr, Werner Heisenberg, Louis de Broglie and Erwin Schr\"{o}dinger -- progressively and collectively grasped through the period 1900--1930 that the discrete, quantised behaviour of Nature was covering over an underlying reality inherently different from the one we experience in our daily lives. This theory, which became known as \emph{quantum physics}, was not only puzzling because it seemed  best phrased in unexpectedly sophisticated mathematical realms of \emph{complex linear algebra} and \emph{Hilbert spaces} -- it also challenged the very idea that questions about the properties of a physical object are meaningful.

Mathematicians, meanwhile, came to terms with their own crisis. They managed to repair the axioms of set theory and to make precise what formal reasoning in general \emph{is}, thus effectively making the analysis of reasoning \myuline{part} \myuline{of} \myuline{mathematics} \myuline{itself}. Two of the most striking insights were due to Kurt G\"{o}del around 1930, who demonstrated that a formal statement can be given a finite, checkable proof provided that it is true under every possible interpretation of its content (the \emph{Completeness Theorem}, \cite{GodelComplete}), but also that any potent system of reasoning will spawn formal statements which are true under some interpretations and false under others, and thus cannot be settled by checkable mathematical proofs (the \emph{Incompleteness Theorem}, \cite{GodelIncomplete}).\footnote{For example, even some statements about the natural numbers $1,2,3, \ldots$ and the arithmetic operations $+$ and $\cdot$ cannot be decided-- they simply have different truth values under different interpretations of what these entities mean.  No matter how well we try to contain them by specifying how they interact with one another (by axioms such as \emph{for all $a,b,c$, it holds that  $(a+b) \cdot c = a \cdot c + b \cdot c$}), we will not succeed in eliminating undecidable statements.} In the primeval soup of these ideas about `checkable' procedures -- as contemplated also by contemporaries such as Alonzo Church, Alan Turing and Emil Post -- eventually emerged the formal notions of \emph{algorithms} and \emph{computability}, which previously had only intuitive meaning. Not long after this, Claude Shannon in 1948 (\cite{Shannon48}) conceived of a mathematical theory of \emph{information},  and on these two pillars -- the theories of computation and information -- was built the field of \emph{computer science}. Amusingly, the desire to rigorously treat an abstract mathematical universe of infinite sets had led us to create the finitistic framework of computation, to which we now owe the existence of every digital computer on Earth (and in space).\footnote{If nothing else, let this be a testament to the fact that basic research in mathematics should always be supported.} \\

It may very well have been accidental that the two crises of the sciences raged at the same time. There is, however,  a poetic glow to the fact that quantum theory and computer science were conceived and born simultaneously, and, as it turns out, destined to meet again later in life. \emph{Information is physical}, said the physicist Rolf Landauer in 1961 (\cite{Landauer61}), and he thereby ushered an era devoted to the thesis that the theory of computation and information processing cannot be separated from physics, since the processing is ultimately executed by physical entities. The specific cocktail of \emph{\myuline{quantum} information theory} was given shape in the early 1980s, when various people apprehended that quantum physics may affect the efficiency of computation (Richard Feynman \cite{Feynman82} and David Deutsch \cite{Deutsch85}), that it fundamentally prohibits certain standard information-theoretic tasks such as duplicating information (William K. Wootters and Wojciech H. Zurek \cite{Woot82}), and that it provides the means for cryptographic schemes not conceivable in classical information theory (Stephen Wiesner \cite{Wiesner83}, Charles Bennett and Gilles Brassard \cite{Bennett84}). \\

Over the years, the field of quantum information theory grew larger, and though it is today still relatively young, it is a well-established area of research, tri-disciplinary between physics, computer science and mathematics. Whether we will ever be able to build an operational \emph{quantum computer} which outperforms the most powerful digital computers is a question of intense dispute, but regardless of this a vast number of insights has been gained in information theory from the influence of quantum theory, and in quantum theory from the influence of information theory (\cite{NC02}). \\

Now, one of the subfields of quantum information theory, known as \emph{quantum foundations}, seeks to better understand what are the core principles of quantum information theory, and how can they be phrased in general, abstract terms. Research within this subfield attempts to define a mathematical universe of \emph{physical theories} and to understand what makes quantum theory special among them. \\

This PhD thesis confines to that line of thought, and aims to recast a specific phenomenon in quantum information theory, \emph{quantum self-testing}, in general, abstract terms. In doing so, it presents a new theory of so-called \emph{dilations}, a concept which is well known in the field but has not been studied systematically before. Intuitively, a dilation of an information channel (an information channel being for example a device which accepts an input, computes the value of a function, and then returns an output) can be thought of as encoding `secret computations' which take place in the course of our interaction with the information channel. The main conclusions of this thesis are that quantum self-testing can be understood in the language of such dilations, and that in fact many features of quantum information theory itself can be derived from principles phrased exclusively in terms of dilations. In developing the formalism necessary for these conclusions, it uses the mathematical language of \emph{category theory}, a field which arose in the 1940s (\cite{MacLane}) and today has wide applicability. As such, it is my hope that some of the ideas and results presented here may find application also in pure mathematics, or other fields outside of quantum information theory. \\

No one can say with certainty what the future of science is like. Physics and mathematics -- and computer science, the newcomer -- will probably again face critical and disruptive periods. I am thankful for having lived 100 years after the groundbreaking work that led to the exciting scientific landscape of today, and I hope that this landscape will be even more exciting to those who gaze upon it 100 years from now.

\section*{For Someone}
\addcontentsline{toc}{section}{For Someone}

In order to understand what quantum self-testing is, and how it came to be, we must first return to the turbulent early years of quantum physics. \\

Though Einstein had played a major role in establishing quantum theory,\footnote{In 1921, he was rewarded the Nobel prize in for his discovery of the \emph{photoelectric effect}, which posited the quantised nature of light.} he was famously non-pleased with the philosophical inclinations it seemed to require. One of the strange features of quantum theory is that it is probabilistic: When we measure the same property in two physical systems prepared identically, we might get different results. Quantum theory predicts the \emph{probability distributions} which the measurement results follow, but generally cannot predict the exact values obtained. When quantum theory was still young, there were (at least) two different opinions about how to interpret this circumstance.\footnote{See e.g. the witty descriptions in Ref. \cite{Grif05}, from which I have borrowed the terms `realist position' and `orthodox position'.} 

According to the \emph{realist position}, as held by Einstein, a measurement of a physical system reveals a property which the system already possessed in advance; though we may not know e.g. what the velocity (momentum) of a particle is before we measure it, the particle surely \myuline{had} a velocity prior to our measurement. As such, if quantum theory predicts randomness in measurement outcomes, it must be because the theory itself falls short of giving a complete description of reality.

 On the other hand, according to the \emph{orthodox position}, as defended by others,  the randomness of quantum theory is \myuline{fundamental} and exempt from ordinary intuition. It simply makes no sense to speak of a physical system having a particular property before we measure it; this was the message of quantum theory, and it needed no fix. That idea was absurd to the realists, and in 1935, Einstein and colleagues Boris Podolsky and Nathan Rosen presented a thought experiment (\cite{EPR35}) meant to expose that it was flawed. \\

In high-level terms, Einstein, Podolsky and Rosen argued that in certain experimental scenarios, the outcome of one measurement seemed to be definite (i.e. non-random), yet quantum theory failed to predict its value.

More precisely, they imagined a source emitting pairs of particles going off to two different sites, $\sfA$ and $\sfB$. At each site, an experimenter is waiting for the respective particle and can choose to measure one of two properties\footnote{In their paper \cite{EPR35}, these two properties were the \emph{momentum} or the \emph{position} of the particle, but this is not essential.} of it, corresponding to measurements $M^0_\sfA$ or $M^1_\sfA$ at site $\sfA$, and $M^0_\sfB$ or $M^1_\sfB$ at site $\sfB$. Like any other physical theory, quantum theory has a notion of \emph{state} of a physical system. A `physical system' is a somewhat abstract concept, but for example the two emitted particles considered together form a physical system; as such, quantum theory mathematically associates to this system a set of possible states, $\psi$.\footnote{In the case of two particles, the states correspond more or less to functions called \emph{wave functions}, but again this is inessential.} 

Einstein, Podolsky and Rosen (a trio which became known as `EPR') now pointed out that according to the mathematical formalism of quantum theory, there exists a state $\psi$, and measurements  $M^0_\sfA$, $M^1_\sfA$,  $M^0_\sfB$ and $M^1_\sfB$, for which the theory predicts the following: If the two particles are in the state $\psi$ and the measurement $M^x_\sfA$ ($x = 0,1$) is performed at site $\sfA$ and yields outcome\footnote{For example, in the case of momentum and position, $y^x_\sfA$ is some real number.}  $y^x_\sfA$, then the outcome $y^x_\sfB$ of the measurement $M^x_\sfB$ (same $x$) at site $\sfB$ can be inferred \myuline{with} \myuline{certainty} from $y^x_\sfA$, i.e. there are pre-determined functions $f_0$ and $f_1$ such that $y^0_\sfB = f_0(y^0_\sfA)$ and $y^1_\sfB = f_1(y^1_\sfA)$. 

Now, if the sites $\sfA$ and $\sfB$ are sufficiently separated, and if the measurements are performed within suitable time spans, then the principle of special relativity (that no signal can travel faster than light) ensures that the measurement at site $\sfA$ cannot affect the measurement at site $\sfB$, and vice versa. Consequently, they argued, it must be the case that the two measurement outcomes $y^0_\sfB$ and $y^1_\sfB$ \emph{were really determined all along}. Nevertheless, quantum theory \myuline{also} says that the state $\psi$ does not yield definite (non-random) values for both of measurements $M^0_\sfB$ and $M^1_\sfB$ -- in fact, the measurements $M^0_\sfB$ and $M^1_\sfB$ have the property that every quantum state whatsoever will give random outcomes for at least one of them. They drew from this the conclusion that the quantum states $\psi$ simply did not model all information about the particles, and they expressed the belief that it was possible to find another theory which resolved this problem.  \\

However, their criticism backfired spectacularly. Three decades later,  in 1964, the physicist John S. Bell (\cite{Bell64}), inspired by their paper, astounded the scientific community by demonstrating that \myuline{nothing} could be done to repair the alleged incompleteness of quantum theory. His insight was striking, because it ultimately meant that the `realist' and `orthodox' positions towards quantum theory were not a matter of philosophical taste -- quantum theory was plainly \myuline{incompatible} with the former, and this incompatibility could moreover be subjected to an \emph{experimental test}.

Bell considered a version of the EPR-scenario in which the relevant quantum state $\psi$ of the two particles is the so-called \emph{singlet state}, and for which the quantum measurements were measurements of so-called \emph{spins} of the particles, meaning in particular that the possible measurement outcomes were $+1$ or $- 1$. More specifically, there exist according to the formalism of quantum theory, for any unit vector $v \in \R^3$, a `spin measurement in direction $v$', $M(v)$, and for unit vectors $v_\sfA, v_\sfB \in \R^3$ the spin measurements $M_\sfA(v_\sfA)$ at site $\sfA$ and $M_\sfB(v_\sfB)$ at site $\sfB$ are such that when measuring two particles in the singlet state, the probability of obtaining measurement outcomes $y_\sfA, y_\sfB \in \{+1,-1\}$ is given by $\frac{1}{4}    - \frac{y_\sfA  y_\sfB}{4} v_\sfA \cdot v_\sfB$, where $v_\sfA \cdot v_\sfB$ is the scalar product of $v_\sfA$ and $v_\sfB$. Thus, if the four measurements $M^0_\sfA$, $M^1_\sfA$, $M^0_\sfB$ and $M^1_\sfB$ in the EPR-scenario are chosen as spin measurements, with $M^{x_\sfA}_\sfA = M_\sfA(v^{x_\sfA}_\sfA)$ and $M^{x_\sfB}_\sfB = M_\sfB(v^{x_\sfB}_\sfB)$ for $x_\sfA, x_\sfB\in \{0,1\}$ and some unit vectors $v^0_\sfA, v^1_\sfA, v^0_\sfB, v^1_\sfB \in \R^3$, then the probability distributions predicted by quantum theory are

\begin{align} \label{eq:Quantum}
P^{x_\sfA, x_\sfB}_{\up{quant.}}(y_\sfA, y_\sfB) = \frac{1}{4}    - \frac{y_\sfA  y_\sfB}{4} v^{x_\sfA}_\sfA \cdot v^{x_\sfB}_\sfB
\end{align}

Now, \myuline{if} there is, as Einstein, Podolsky and Rosen hoped, a complete theory meeting their standards of \emph{realism}, then the measurement outcomes merely reveal pre-existing properties which can be described by $\pm 1$-valued random variables $Y^{x_\sfA, x_\sfB}_\sfA$ (the measurement outcome at site $\sfA$) and $Y^{x_\sfA, x_\sfB}_\sfB$ (the measurement outcome at site $\sfB$). If moreover this assumed theory is \emph{local}, meaning that it complies to the non-signalling principle from special relativity, then, when $\sfA$ and $\sfB$ are suitably separated, $Y_\sfA$ cannot depend on  $x_\sfB$ and $Y_\sfB$ not on $x_\sfA$. As such, what we have is really \myuline{four} random variables, $Y^{x_\sfA}_\sfA$ for $x_\sfA \in \{0,1\}$, and $Y^{x_\sfB}_\sfB$ for $x_\sfB \in \{0,1\}$, and their probability distributions are simply 

	\begin{align} \label{eq:LocHid}
P^{x_\sfA, x_\sfB}_{\up{loc. real.}}(y_\sfA, y_\sfB) = \up{Pr} \left(Y^{x_\sfA}_\sfA = y_\sfA,Y^{x_\sfB}_\sfB = y_\sfB \right).
\end{align}

	What Bell then did was to derive an inequality that the probabilities \eqref{eq:LocHid} are bound to obey due to the mere fact that they arise as distributions of random variables as indicated, but which the  probabilities \eqref{eq:Quantum} as predicted by the formalism of quantum theory do \myuline{not} obey (for suitable choices of the vectors $v^{x_i}_i$). As such, \emph{Bell's inequality} by itself is not a result about quantum theory; it is about any theory which meets the requirements of \emph{realism} (so as to infer the existence of random variables) and \emph{locality} (so as to conclude the independence of the outcomes at site $\sfA$ from the measurement chosen at site $\sfB$, and vice versa). The result about quantum theory is that it \myuline{violates} Bell's inequality, and hence cannot be both local and realistic.\footnote{It is known, incidentally, that quantum theory \myuline{can} be given a realistic interpretation (i.e. one in which measurable properties are described by random variables) known as \emph{de Broglie-Bohm theory}, or simply \emph{Bohmian mechanics} (\cite{Bohm52}), but it is, of course, non-local.} \\

From an abstract vantage point, the collections $P= (P^{x_\sfA, x_\sfB}_{\up{loc. real.}})_{x_\sfA, x_\sfB \in \{0,1\}}$ of probability distributions which arise from local realism (i.e. which are of the form \eqref{eq:LocHid}) form a convex set,\footnote{In the sense that if  $P_1 = (P^{x_\sfA, x_\sfB}_1)_{x_\sfA, x_\sfB \in \{0,1\}}$ and  $P_2= ({P}^{x_\sfA, x_\sfB}_2)_{x_\sfA, x_\sfB \in \{0,1\}}$ are two such collections, and if $\alpha \in [0,1]$, then $P= (P^{x_\sfA, x_\sfB})_{x_\sfA, x_\sfB \in \{0,1\}}$ is also such a collection, with $P^{x_\sfA, x_\sfB} := \alpha P^{x_\sfA, x_\sfB}_1 + (1-\alpha)P^{x_\sfA, x_\sfB}_2$. This is because the weight $\alpha$ can be encoded as the success probability of a $\{0,1\}$-valued random variable $Z$, which we may include into the random variables giving rise to $P_1$ and $P_2$.} and Bell's inequality corresponds to a \emph{half-space} which confines this convex set. (This is similar to the way in which a pyramid is confined by half-spaces, four half-spaces corresponding to its tilted sides, and one to its horizontal bottom.) In honour of Bell, we generally refer to such half-space inequalities as \emph{Bell-inequalities}. One of the simplest derivations of a Bell-inequality is not Bell's original, but was given a few years later (\cite{CHSH69}),  by J. Clauser, M. A. Horne, A. Shimony and R. A. Holt. They first observed that the random variables $Y^{x_i}_i$ must with unit probability satisfy the inequality

\begin{align} \label{eq:CHSHineq0}
(Y^0_\sfA + Y^1_\sfA ) \cdot Y^0_\sfB + (Y^0_\sfA - Y^1_\sfA) \cdot Y^1_\sfB \leq 2 .
\end{align}

Indeed, since $Y^{x_\sfA}_\sfA$ has values $\pm 1$, either the sum $Y^0_\sfA + Y^1_\sfA $ or the difference $Y^0_\sfA - Y^1_\sfA$ is $\pm 2$ while the  other is $0$, and in each case the above expression then takes one of the values $\pm 2$ (since also $Y^{x_\sfB}_\sfB$ is $\pm 1$). But now, the inequality \eqref{eq:CHSHineq0} must also hold for the expectation values, that is, 

\begin{align} \label{eq:CHSHineq}
\up{E}(Y^0_\sfA  \cdot Y^0_\sfB )  + \up{E} (Y^1_\sfA \cdot Y^0_\sfB )+ \up{E}(Y^0_\sfA  \cdot Y^1_\sfB ) - \up{E}(Y^1_\sfA \cdot Y^1_\sfB ) \leq 2.
\end{align}

Each of these four expectation values can be re-expressed using the probabilities \eqref{eq:LocHid}, since 

\begin{align}
\begin{split}
\up{E}(Y^{x_\sfA}_\sfA  \cdot Y^{x_\sfB}_\sfB )  & = \up{Pr} (Y^{x_\sfA}_\sfA  = Y^{x_\sfB}_\sfB  ) - \up{Pr} (Y^{x_\sfA}_\sfA \neq   Y^{x_\sfB}_\sfB  ) \\ & = P^{x_\sfA, x_\sfB}_{\up{loc. real.}}(1,1)+ P^{x_\sfA, x_\sfB}_{\up{loc. real.}}(-1,-1)-P^{x_\sfA, x_\sfB}_{\up{loc. real.}}(1,-1)-P^{x_\sfA, x_\sfB}_{\up{loc. real.}}(-1,1),
\end{split}
\end{align}

but we may equivalently keep the inequality in the form \eqref{eq:CHSHineq}. This is the so-called \emph{CHSH-inequality}. To see that it can be violated in quantum theory, note that, by \cref{eq:Quantum},  

\begin{align}
P^{x_\sfA, x_\sfB}_{\up{quant.}}(1,1)+ P^{x_\sfA, x_\sfB}_{\up{quant.}}(-1,-1)-P^{x_\sfA, x_\sfB}_{\up{quant.}}(1,-1)-P^{x_\sfA, x_\sfB}_{\up{quant.}}(-1,1)= - v^{x_\sfA}_\sfA\cdot v^{x_\sfB}_\sfB ,
\end{align}

so if quantum theory were locally realistic, the CHSH-inequality would read 

\begin{align} \label{eq:quantumCHSH}
- v^{0}_\sfA\cdot v^{0}_\sfB - v^{1}_\sfA\cdot v^{0}_\sfB - v^{0}_\sfA\cdot v^{1}_\sfB  +  v^{1}_\sfA\cdot v^{1}_\sfB \leq 2.
\end{align}

However, by choosing $v^0_\sfA = (-1,0,0)$, $v^1_\sfA = (0,-1,0)$, $v^0_\sfB = (1/\sqrt{2}, 1/\sqrt{2},0)$ and $v^1_\sfB = (1/\sqrt{2}, -1/\sqrt{2},0)$, we easily compute that each of the four terms attain the value $1/\sqrt{2}$, so that the entire expression equals $2 \sqrt{2}$ which is evidently larger than $2$.\\

And \myuline{now} we come to quantum self-testing. \\

Though Bell's theorem was a shock, it was not a shock that extended to comatose paralysis. On the contrary, the result stimulated a renewed interest in the set-ups from the thought experiment envisioned by Einstein, Podolsky and Rosen. An obvious question was the following: \emph{By how much can quantum theory violate the principles of local realism?} 

A precise incarnation of this question was by how much quantum theory can violate the CHSH-inequality. This problem was solved in 1980 by the mathematician Boris Cirelson (\cite{Cir80}), who showed that the violation is at most $2 \sqrt{2}$, and also coined the term \emph{behaviour} (\cite{Cir93}) about the collections $P=(P^{x_\sfA, x_\sfB})_{x_\sfA, x_\sfB \in \{0,1\}}$ of probability distributions producible within a given theory. To prove that there was no \emph{quantum} behaviour which exceeded the value $2 \sqrt{2}$ was not simply a matter of optimising the expression \eqref{eq:quantumCHSH} over unit vectors, as the formula \eqref{eq:Quantum} applies only to give those quantum behaviours which result from spin measurements on particles in the singlet state. Rather, Cirelson's argument was rooted in the general formalism of quantum theory, in terms of linear operators on Hilbert spaces.\footnote{Though I will not reproduce it here, Cirelson's proof was not particularly technical; what he did was basically to establish an operator inequality.} The set of quantum behaviours can be shown to be convex like the smaller collection of locally realistic behaviours, and \emph{Cirelson's inequality} (or \emph{Cirelson's bound}) is thus a quantum analogue of Bell's inequality, namely an inequality corresponding to a half-space which confines the set of possible behaviours.

One of the questions raised by his work was the following: \emph{What are the configurations of quantum states and quantum measurements whose behaviour reach the Cirelson bound $2 \sqrt{2}$?} 

A number of results (\cite{SW87,PR92,BMR92,Cir93}) soon demonstrated that the value $2 \sqrt{2}$ could in fact, in a certain sense, only be obtained by measuring the singlet state using the above spin measurements. While this was curious, it was mainly considered interesting for foundational reasons. \\

Probably the first person to acknowledge that the scenarios considered by Bell and Cirelson could have applications in the newly emerging field of quantum information theory was the physicist Artur Ekert. In 1991, he pointed out (\cite{Ek91}) that because the values of the CHSH-expression which exceed $2$ signify the lack of local realism, such values must certify \emph{genuine} randomness in measurement outcomes, randomness which may be used for cryptographic purposes,\footnote{For example, it is often of interest to generate \emph{shared randomness} so that one may use this to establish a secret key for encryption. However, it is of course important that this randomness is the genuine randomness that comes from quantum measurements, and not randomness which was known to the potentially adversarial manufacturer of the devices in advance.} since by Bell's argument not even a potentially untrusted manufacturer of the measurement devices could have known it in advance. Using the fact that the particular value $2 \sqrt{2}$ more or less \myuline{uniquely} determines the configuration of state and measurements, this idea was made even more explicit at the turn of the millennium, in the papers \cite{MY98} and \cite{MY04} by Dominic Mayers and Andrew Yao, who gave the name \emph{self-testing} to this phenomenon, that devices could be used to `test themselves'. 
It is important to appreciate that the idea of exploiting quantum self-testing for applications constituted an almost paradigmatic change in mindset relative to the perspective of Bell and Cirelson. Whereas they had been thinking about trustworthy experimenters who wished to establish the supremacy of quantum theory over local realism, the new ideas took the point of view that the whole experimental set-up was like a \emph{game}, a potentially vicious scheme in which untrustworthy agents had prepared an experiment whose purpose was to fool us to believe that a certain state was being subjected to certain measurements.  \\

The mathematical definition of self-testing (which took its modern standard form in Ref. \cite{MYS12}) is as follows: 

In quantum information theory, the \emph{physical systems} at sites $\sfA$ and $\sfB$ are modelled by (finite-dimensional) Hilbert spaces $\cH_\sfA$ and $\cH_\sfB$ over the complex field $\C$.  When considering the two systems as one (as we did above), the associated Hilbert space is the tensor product, $\cH_\sfA \otimes \cH_\sfB$. A \emph{state} on this system is modelled\footnote{Two comments are in place here. First of all, only the so-called \emph{pure} states are modelled as such (we will return to this shortly). Secondly, it is more correct to say that pure states are modelled by \emph{rank-one projections} (or, what is equivalent, one-dimensional subspaces of the Hilbert space), since for any $\alpha \in \C$ of unit modulus, the vectors $\psi$ and $\alpha \psi$ correspond to the same state.} by a unit vector $\psi \in \cH_\sfA \otimes \cH_\sfB$. Finally, the \emph{measurement} $M^{x_i}_i$ ($x_i = 0,1$) is modelled\footnote{Again, there are more general kinds of measurements than PVMs, and we shall return to this point.} by a so-called \emph{projection-valued measure (PVM)} on $\cH_i$, that is, by orthogonal projections $\Pi^{x_i}_i(1)$, $\Pi^{x_i}_i(-1)$ on $\cH_i$ (one for each possible measurement outcome $y_i = 1,-1$), which sum to the identity operator on $\cH_i$, $\Pi^{x_i}_i(1)+ \Pi^{x_i}_i(-1) = \bone_{\cH_i}$. In summary, a configuration of states and measurements is defined by a triple $(\psi, \Pi_\sfA, \Pi_\sfB)$, where $\psi \in \cH_\sfA \otimes \cH_\sfB$ is a unit vector, and where $\Pi_\sfA = (\Pi^{x_\sfA}_\sfA)_{x_\sfA \in \{0,1\}}$ and $\Pi_\sfB = (\Pi^{x_\sfB}_\sfB)_{x_\sfB \in \{0,1\}}$ are collections of PVMs on $\cH_\sfA$ and $\cH_\sfB$, respectively. Such a triple is called a \emph{(tensor-product) quantum strategy}, and the formalism of quantum theory stipulates that it gives rise to the behaviour $P= (P^{x_\sfA, x_\sfB})_{x_\sfA, x_\sfB \in \{0,1\}}$ given by the inner products 

\begin{align} \label{eq:Born}
P^{x_\sfA, x_\sfB}(y_\sfA, y_\sfB) = \langle \psi, [\Pi^{x_\sfA}_\sfA(y_\sfA) \otimes \Pi^{x_\sfB}_\sfB(y_\sfB)] \psi \rangle, \quad x_i \in \{0,1\}, \, y_i \in \{1,-1\}.
\end{align}

(The quantum behaviour \eqref{eq:Quantum} then arises from a suitable choice of such a quantum strategy. In particular, the \emph{singlet state} corresponds to the vector $\psi = \frac{e_0 \otimes e_1 + e_1\otimes e_0}{\sqrt{2}} \in \C^2 \otimes \C^2$, where $(e_0,e_1)$ is the standard basis in $\C^2$, and the \emph{spin measurements} correspond to projections $\Pi^{x_i}_i(\pm 1)$ which project onto various $1$-dimensional subspaces of $\C^2$.) 

Now, in the case of the CHSH-inequality, quantum self-testing formally means that if $(\psi, \Pi_\sfA, \Pi_\sfB)$ is any quantum strategy for which the associated behaviour reaches the Cirelson bound $2 \sqrt{2}$, then this strategy is `reducible', or `equivalent', in a certain sense, to a fixed, \emph{canonical} strategy $(\tilde{\psi}, \tilde{\Pi}_\sfA, \tilde{\Pi}_\sfB)$, namely the one described by the singlet state and the spin measurements from above. Precisely, this reducibility criterion is expressed by the existence of so-called \emph{residual} Hilbert spaces $\cH^\res_\sfA$ and $\cH^\res_\sfB$, a \emph{residual} state $\psi^\res \in \cH^\res_\sfA \otimes \cH^\res_\sfB$, and isometries $W_i : \cH_i \to \tilde{\cH}_i \otimes \cH^\res_i$ such that 

\begin{align} \label{eq:Reducible}
[W_\sfA \otimes W_\sfB ] [\Pi^{x_\sfA}_\sfA(y_\sfA) \otimes \Pi^{x_\sfB}_\sfB(y_\sfB) ]\psi = [\tilde{\Pi}^{x_\sfA}_\sfA(y_\sfA) \otimes \tilde{\Pi}^{x_\sfB}_\sfB(y_\sfB) ]\tilde{\psi}  \otimes \psi^\res, \quad x_i \in \{0,1\}, \, y_i \in \{1,-1\}.
\end{align}  

In quantum theory, the local application of an isometry $W_i$ is like a change of coordinates, so \cref{eq:Reducible} is supposed to express that, up to such local changes of coordinates, the strategy $(\psi, \Pi_\sfA, \Pi_\sfB)$ is really just the canonical strategy $(\tilde{\psi}, \tilde{\Pi}_\sfA, \tilde{\Pi}_\sfB)$, except possibly augmented by a state $\psi^\res$ which is shared between the two sites $\sfA$ and $\sfB$, but which is not acted upon by the measurements. \\

Of course, the above definition generalises significantly beyond the CHSH-scenario. (In fact, the scenario considered by Mayers and Yao was a different one.) In general, we use the term \emph{(bipartite)\footnote{There is also a more or less obvious generalisation from two sites $\sfA$ and $\sfB$ to more sites, but we mostly consider the bipartite scenario.} Bell-scenario} about a quadruple of finite non-empty sets $(X_\sfA, X_\sfB, Y_\sfA, Y_\sfB)$, with $X_i$ corresponding to a set of possible measurement settings (`inputs') at site $i$, and $Y_i$ a set of possible measurement results (`outputs') at site $i$.\footnote{Unfortunately, the symbol $Y_i$ is now used for a set, whereas we previously used it for a random variable; hopefully this causes no confusion.} The definition of a quantum strategy for this Bell-scenario generalises in the obvious way, as a triple $(\psi, \Pi_\sfA, \Pi_\sfB)$, where $\Pi_i = (\Pi^{x_i}_i)_{x_i \in X_i}$ is a collection of PVMs $(\Pi^{x_i}_i(y_i))_{y_i \in Y_i}$ on $\cH_i$, i.e. orthogonal projections on $\cH_i$ summing to $\bone_{\cH_i}$. The behaviour of such a strategy is given as in \cref{eq:Born}. Moreover, we no longer talk of a specific \myuline{inequality} being saturated, we will simply say that the quantum \myuline{behaviour} $P$ \emph{self-tests the quantum strategy $(\tilde{\psi}, \tilde{\Pi}_\sfA, \tilde{\Pi}_\sfB)$}, if any quantum strategy $(\psi, \Pi_\sfA, \Pi_\sfB)$ with behaviour $P$ is reducible to $(\tilde{\psi}, \tilde{\Pi}_\sfA, \tilde{\Pi}_\sfB)$, by means of a residual state $\psi^\res$ and isometries $W_\sfA$ and $W_\sfB$ as in \cref{eq:Reducible}.\\

The traditional definition of quantum self-testing as laid out above is mathematically unambiguous. The circumstance that motivated this PhD thesis is that its \emph{operational} significance is unclear. The most convincing argument for this is by observing that the definition is intimately intertwined with the very formalism of quantum information theory: It is carved in the stones of Hilbert spaces, linear operators and vectors, and it is not at all obvious how one would formulate it independently of this, despite the fact that the narrative of self-testing -- namely, `there is essentially only one way of realising the behaviour $P$' - suggests that a general formulation should be possible. \\

Not only is a reformulation desirable in order to understand the significance of the phenomenon in \myuline{other} theories than quantum information theory. It is also desirable in order to better understand its significance \myuline{within} quantum information theory. 

First of all, there is consensus among many that the Hilbert space formulation of quantum information theory is mysteriously obscure. Grounded in this opinion, a number of works (see e.g. Refs. \cite{Hard01, Chir11}, and the book \cite{Foils}) have demonstrated that remarkable reformulations of the theory are possible, namely formulations which do not refer to Hilbert spaces or linear algebra, but are cast instead in a universal language pertaining to general theories of information processing. It is conceivable that quantum information will eventually be best understood and studied from such an abstract point of view, and as such it is highly relevant to have a definition of self-testing which is compatible with that mode of abstraction. 

Secondly, even within the usual formalism, the significance of a `quantum strategy' is somewhat unclear. For example, the most general kind of quantum states are not represented by unit vectors, but so-called \emph{density matrices}. Similarly, the most general kinds of quantum measurements are not represented by PVMs, but \emph{POVMs (positive operator-valued measures)}. Whereas a number of mathematical results imply that general states and POVMs can be seen as `arising' in a precise way from pure states, respectively PVMs, the meaning of these results as they apply to quantum strategies is obfuscated, at best (this point is detailed in \cref{chap:Selftesting}). In fact, if we really take literally the assumption that the experimental set-up in a self-testing scenario is crafted by untrusted agents, then it seems presumptuous to believe in the first place that the two devices establish their outputs by the simple process of sharing a quantum state and performing measurements on it.\footnote{For example, there could be an intricate procedure by which a sequence of local operations is first executed to decide which of several shared states to use in a subsequent protocol, etc.} Though this worry might seem ludicrous to those who find it intuitively clear that we can always standardise the form of more general `strategies' to triple-form $(\psi, \Pi_\sfA, \Pi_\sfB)$, it is not clear how to give a formal argument for this without having an accepted notion of `general strategy', and at any rate the meaning of the components $\psi$, $\Pi_\sfA$ and $\Pi_\sfB$ certainly does not crystallise in the process of this standardisation.

 Lastly, in order for quantum self-testing to be a practical significance, it is important that self-testing results be \emph{robust}, such that if a strategy $(\psi, \Pi_\sfA, \Pi_\sfB)$ gives rise to a behaviour which is merely \myuline{close} to $P$, then it is \myuline{close} to being reducible to the canonical strategy $(\tilde{\psi}, \tilde{\Pi}_\sfA, \tilde{\Pi}_\sfB)$, in suitable senses of the word `close'. (The reason for this is not only that real experiments are prone to measurement errors, but also that the probabilities $P^{x_\sfA, x_\sfB}(y_\sfA, y_\sfB)$ can never be determined precisely, but only estimated based on finitely many observations.) It seems obvious that a sensible notion of `closeness' should be operational (the standard choice from Ref. \cite{MYS12} of using the Hilbert space norm of the difference between left and right hand sides in \cref{eq:Reducible} is not); the problem of defining such a distance measure is left open by this thesis, but it is certainly necessary that there first exist an operational definition in the \myuline{exact} case.

\section*{For Anyone}
\addcontentsline{toc}{section}{For Anyone} 

In this thesis, I present a framework which offers a fundamentally different way of looking at quantum self-testing. \\

Let us consider the behaviour $P= (P^{x_\sfA, x_\sfA})_{x_\sfA \in X_\sfA, x_\sfB \in X_\sfB}$ observed in a Bell-scenario not simply as a collection of probability distributions on the outcome set $Y_\sfA \times Y_\sfB$, but as a dynamic \emph{information channel} which receives local inputs $x_\sfA, x_\sfB$ and produces local outputs $y_\sfA, y_\sfB$. We make no assumptions about the constituents of this channel, but merely assume that it can indeed be `constructed' from basic constituents, and that it adheres to the locality assumption so as to produce at site $i$ the output $y_i$ given only the input $x_i$. 

Now, instead of asking what the individual components of the channel might be, we ask a different and purely operational question: \\

\emph{What are the possible side-computations that may secretly be executed in the environment during our interaction with the channel?} \\

To intuitively understand the idea of `side-computations in the environment', three simple examples of information channels are helpful. They can be pictorially displayed as

\begin{align}
\myQ{0.7}{0.5}{& \push{A} \qw & \gate{f} & \push{B} \qw & \qw },    \quad \myQ{0.7}{0.5}{& \push{\{0,1\}} \qw & \gate{T} & \push{\{0,1\}} \qw & \qw }, \quad \myQ{0.7}{0.5}{& \push{\C^2} \qw & \gate{\id} & \push{\C^2} \qw & \qw },
\end{align}

of which the first represents the computation of an ordinary function $f: A \to B$, the third represents the identity channel on the system $\C^2$ in quantum information theory, and the one in the middle represents the `\emph{bit refreshment}' channel in classical information theory, which accepts as input any bit and outputs a uniformly random bit, regardless of the input. 

(In each case, we tacitly assume that we can interact an arbitrary number of times with independent copies of the channel, so as to establish that the input-output behaviour of the channel is really as declared.) \\

Suppose we interact with the first channel, $f$. We do so by providing an input $a \in A$ to the \emph{input interface} of the channel, and receiving the output $b= f(a)$ at the \emph{output interface}. (For example, this is the kind of interaction we have with an ordinary digital computer.) Now, we imagine an \emph{environment}, consisting of additional interfaces which we do not see, but which other agents -- be they untrustworthy, or simply `Nature' itself -- can access. (For example, when interacting with a digital computer, there might be hidden interfaces within the computer, to and from which another party can send and receive information.) Provided that we really see the behaviour $f$ at our interfaces, what computations might be going on simultaneously between these hidden interfaces? 

It is quite easy to analyse this question. Of course, the environment may, simultaneously with our use of the channel, perform a computation which is completely independent, given by some function $g: C \to D$. In this case, the \myuline{total} channel describing the situation is the parallel composition $f \times g: A \times C \to B \times D$. More interestingly, the environment might \emph{copy our input} $a \in A$, and use it to compute some function $g: A \to D$, so that the total channel is given by the function $(f,g ): A \to B \times D $, $a \mapsto (f(a),    g(a))$; the value $f(a)$ is returned to us, but the value $g(a)$ is kept secret in the environment, possibly to be used in other computations. Even more generally, the environment can copy our input $a \in A$ in order to decide which of several functions $g_a : C \to D$ to apply on the side. In a sense, we are describing the obvious fact that if we want someone to compute a function value $f(a)$ for us, we cannot do this without sharing with them the value of $a$, and thereby allowing them to keep it in memory. On the other hand, it is intuitively clear that the value of the input $a$ is the `strongest' possible information the environment can extract from our use of the channel; every other side-computation can be `derived' from the one that corresponds to copying the input. \\

The various channels that formalise side-computations in the presence of $f$ will be called \emph{dilations} of $f$. The notion of dilation is dual to that of a \emph{marginal}, in the sense that a dilation of $f$ is precisely a channel whose marginal is $f$. \\

Suppose instead we interact with the third channel, $\id_{\C^2}$. This channel is in a sense the quantum analogue of the identity function from $\{0,1\}$ to $\{0,1\}$; it accepts as input a quantum state on the $2$-dimensional system $\C^2$, and does nothing to it. Again, we may ask about the  various possible side-computations, or, more precisely, the various possible dilations of $\id_{\C^2}$. 

Readers unfamiliar with quantum information theory might think that, once again, the environment can keep a copy of our input in memory. This, however, is \myuline{not} the case, due to the so-called \emph{No-Cloning Theorem} of quantum information theory (\cite{Woot82}). According to this result, quantum information, in contrast to classical information, cannot be copied; in fact, every dilation of $\id_{\C^2}$ must factor in the same way as the independent side-computations for $f$ above,\footnote{We will establish this result by an abstract argument  in \cref{prop:IsoSelfUniversal} in \cref{chap:Dilations}.} with the exception that the environment may \emph{stall} its secret computations until we feed an input to our accessible interface. Hence, there is again a `strongest possible' dilation of the channel $\id_{\C^2}$, namely the one which simply registers in the environment that an input has been provided. \\

Finally, suppose we interact with the `bit refreshment' channel, $T$. As it turns out, every dilation will be derivable from one of two possible dilations, but those two should be considered genuinely different. They intuitively correspond to two different \emph{implementations} of $T$, which can easily be described in words. (Here, I use the word `implementation' in an intuitive sense, but a fundamental point of the work in this thesis is that this intuitive notion can be formalised by the precise notion of dilation.) 

The first such implementation of $T$ is the obvious one; our input to the channel is discarded, and as output we are given a completely fresh random bit.  This seems to be merely the description of the input-output behaviour of the channel, so it may come as a surprise that it could be implemented in other ways. Indeed it can, however: 

In the second implementation, our input is not discarded, but instead the environment generates a random bit and uses it to decide whether to give us back as output our original input, or to give instead the \myuline{opposite} of our original input. From our point of view, the input-output behaviour of the channel is still a bit refreshment.

The two corresponding dilations are given as follows. The first one can be pictorially represented as 

\begin{align}  \label{eq:Introdil1}
\myQ{0.7}{0.5}{& \qw& \push{\{0,1\}} \qw & \multigate{1}{\id} & \ww &  \push{\{0,1\}} \ww & \ww \\ && & \Nghost{\id} & \push{\triv} \ww & \Nmultigate{1}{\id}{\ww} &\push{\{0,1\}} \qw  & \qw \\&
	& & \Nmultigate{1}{\up{Cop}} & \push{\{0,1\}}\ww & \Nghost{\id}{\ww} \\ & \Ngate{r}   & \push{\{0,1\}} \ww &  \Nghost{\up{Cop}}{\ww}  &\ww &  \push{\{0,1\}} \ww & \ww }\quad,
\end{align}

where $r$ denotes a uniformly random bit, where `Cop' is the copy channel, where `$\triv$' denotes a trivial system which is used to stall computation,  and where the wiggly lines correspond to inaccessible interfaces belonging to the environment. As such, the diagram should be read as follows: A random bit $r$ is generated and copied. One copy is stored in the memory of the environment, while the other is saved to be eventually revealed as output to us. When we provide an input to the accessible input interface, this input is recorded in the memory of the environment, and the release of $r$ as output at the accessible interface is triggered.

The second implementation of the channel $T$ corresponds to the dilation represented as

\begin{align}  \label{eq:Introdil2}
\myQ{0.7}{0.5}{& &  & \Nmultigate{1}{\up{Cop}} & \ww &  \push{\{0,1\}} \ww & \ww \\ & \qw& \push{\{0,1\}} \qw & \ghost{\up{Cop}} & \push{\{0,1\}} \ww & \Nmultigate{1}{\up{XOR}}{\ww} &\push{\{0,1\}} \qw  & \qw \\&
	& &  \Nmultigate{1}{\up{Cop}}& \push{\{0,1\}} \ww & \Nghost{\up{XOR}}{\ww} \\ & \Ngate{r}   & \push{\{0,1\}} \ww &  \Nghost{\up{Cop}}{\ww}  & \ww & \push{\{0,1\}} \ww & \ww} \quad,
\end{align}

where `XOR' denotes the \emph{exclusive OR}, namely the function which output $0$ if its two inputs bits are identical, and $1$ if they are distinct. This time, a random bit $r$ is generated and copied, one copy stored in memory, and the other used to decide whether, when our input bit comes it (and is copied to the memory of the environment), it should be given back to us as output as it is, or first flipped.\\

We will ultimately see (in \cref{chap:Causal}) that the two dilations  \eqref{eq:Introdil1} and \eqref{eq:Introdil2} correspond to formally distinct situations, but to appreciate the significance of this it is important to first realise a sense in which the two dilations are \emph{equivalent}: In equations, the channel \eqref{eq:Introdil1} can be written as 

\begin{align} \label{eq:EqDil1}
\delta_b \mapsto  \frac{1}{2} \tilde{\delta}_b  \otimes \delta_0 \otimes \tilde{\delta}_0 +   \frac{1}{2} \tilde{\delta}_b \otimes \delta_1 \otimes \tilde{\delta}_1 
\end{align}

where $\delta_z$ denotes the classical state which is $z$ with certainty (i.e. the degenerate probability distribution in $z$), and where, somewhat intermittently, we have used the symbol $\tilde{\phantom{x}}$ to indicate information belonging to the environment. In words, on input $b \in \{0,1\}$, the total output of the channel is the uniform mixture of the states $\tilde{\delta}_b  \otimes \delta_0 \otimes \tilde{\delta}_0$ (corresponding to the random bit being $0$) and $\tilde{\delta}_b \otimes \delta_1 \otimes \tilde{\delta}_1 $  (corresponding to the random bit being $1$). Likewise, the channel \eqref{eq:Introdil2} is given equationally by 

\begin{align} \label{eq:EqDil2}
\delta_b \mapsto  \frac{1}{2} \tilde{\delta}_b  \otimes \delta_b \otimes \tilde{\delta}_0  +   \frac{1}{2} \tilde{\delta}_b  \otimes \delta_{b \oplus 1} \otimes \tilde{\delta}_1 ,
\end{align}

where $\oplus$ denotes addition modulo $2$. Now, if \emph{in the environment} of the channel \eqref{eq:EqDil1} one applies the channel $\tilde{\delta}_b \otimes \tilde{\delta}_k \mapsto \tilde{\delta}_b \otimes \tilde{\delta}_{b \oplus k}$, then one effectuates the change $\tilde{\delta}_b  \otimes \delta_k \otimes \tilde{\delta}_k \mapsto \tilde{\delta}_b  \otimes \delta_k \otimes \tilde{\delta}_{b \oplus k}$ and thereby obtains altogether the channel \eqref{eq:EqDil2}, as can be verified by comparing the outputs for $b=0$ and $b=1$. Conversely, if the channel  $\tilde{\delta}_b \otimes \tilde{\delta}_k \mapsto \tilde{\delta}_b \otimes \tilde{\delta}_{b \oplus k}$ is applied in the environment of \eqref{eq:EqDil2}, the channel \eqref{eq:EqDil1} is obtained.

\myuline{However}, this apparent equivalence of the two dilations is deceiving, because the demonstrated  `equivalence' ignores \emph{causality}: The side-information encoded by the copies of the bit $r$ is available in the environment \myuline{before} we feed our input to the accessible interface -- the channel needed to go from e.g. \eqref{eq:Introdil1} to \eqref{eq:Introdil2} needs the copy of the random bit \myuline{as} \myuline{well} \myuline{as} a copy of our input, and therefore does not reproduce the correct causal structure in \eqref{eq:Introdil2}, according to which the copy of $r$ exists before our input was presented. As it turns out, \myuline{no} channel which preserves the causal structure will lead us between the two dilations \eqref{eq:Introdil1} and \eqref{eq:Introdil2}. They should be considered different, formalising the intuition that  the side-information in one dilation (pre-existing knowledge of which bit will be given as output) is information about something entirely different than the side-information in the second (pre-existing knowledge of whether or not the input will be flipped). \\
	
What I will do in the thesis is to demonstrate that quantum self-testing can be viewed on the same footing as the above examples. The various `implementations' of the observed quantum behaviour (i.e. the various quantum strategies) appear as \emph{causally structured dilations} (or, as we will say, simply \emph{causal dilations}) of the behaviour channel, formalising various possible side-computations. The self-testing phenomenon is then more or less\footnote{There are two caveats to this equivalence, but at this point it only makes sense to describe them in high-level terms: First of all, some dilations of the behaviour channel will be very strange and not be derivable from any dilation corresponding to a quantum strategy. The root of this problem is that quantum measurements turn out to have causal dilations which go against the intuition about what a measurement is (\cref{ex:MeasDil}). We will exclude the strange dilations by introducing the notion of a \emph{classically bound} dilation. Secondly, quantum self-testing actually also implies the existence of a certain simple representative in the equivalence class of the strongest possible dilation (this representative essentially corresponds to the canonical quantum strategy), but I conjecture that such a representative can always be found (\cref{conj:Selftesting}).} the existence of a strongest possible causal dilation, which moreover has the property that it holds no pre-existing side-information about the outputs at the accessible interface (in line with Ekert's early observation). \\

Even though this connection to self-testing is one of the main contributions of thesis --  and certainly the unique problem which motivated the project -- it is important for me to stress that the emphasis in the thesis is first and foremost on initiating an abstract and general study of \emph{dilations}. This is not only because the structure of dilations in a given theory turns out to be very interesting in its own right, but also because the fact that quantum self-testing can be interpreted as a dilational phenomenon implies, in my opinion, that the general study of dilations is necessary and valuable by extension. A systematic study of dilations has, to the best of my knowledge, not been attempted before; I hope that the results presented in this thesis will find interest, and that the strands left open will be even as interesting as to attract the curiosity and contemplation of others. \\

\newpage

\chapter*{Structure of the Thesis}
\addcontentsline{toc}{chapter}{Structure of the Thesis}
It is assumed that the reader of this section has already been through the general introduction. From this point onwards, the thesis contains the following elements: 

\begin{itemize}
	\item Preliminaries 
	\item \cref{chap:Theories} -- Theories
	
	\item \cref{chap:Dilations} -- Dilations
	
	\item  \cref{chap:Metric} -- Metric Theories 
	
	\item \cref{chap:Causal} -- Contractible Theories and Causal Dilations
	
	\item \cref{chap:Selftesting} -- Rigidity and Quantum Self-Testing
	
	\item Conclusion
	
\end{itemize}

The section \textbf{Preliminaries} collects a few non-standard mathematical facts, mostly pertaining to the formalism of quantum information theory. Some of them are used quite extensively, and it is advisable for the reader to skim them in advance. \\

Each of the five chapters begins with a prelude, divided into two--four subsections, more or less following the self-explanatory pattern \emph{§1. Introduction and Motivation} -- \emph{§2. Comparison to Existing Literature} -- \emph{§3. Contributions}. Each of them moreover concludes with a summary and the mentioning of several open ends and ideas for future work.  \\

Below, I will briefly sketch the role of each chapter -- it might afterwards be beneficial for the reader to read in series the preludes to the individual chapters. This not only gives a more precise idea of their content (under \emph{§1. Introduction and Motivation}), but also details the relations to existing literature (under \emph{§2. Comparison to Existing Literature}) and provides overviews of the technical contributions (under \emph{§3. Contributions}), which it would not make much sense to reproduce here before the relevant concepts have been introduced.\\

First, we must in \textbf{\cref{chap:Theories}} agree on a mathematical framework in which to even discuss physical theories, channels and dilations. This chapter reviews a variation on the \emph{categorical framework} for discussing operational aspects of theories (\cite{Foils}). More precisely, a \emph{theory} will be modelled by a \emph{symmetric monoidal category} in which the monoidal unit is \emph{terminal} (these concepts will be explained and heavily exemplified). This framework constitutes a natural and minimal language in which to eventually make sense of the key ingredients required for the definitions we desire.

The role of \cref{chap:Theories} is mostly that of introductory review, and it contains only few original observations. My advice for readers who believe themselves familiar with the content of \cref{chap:Theories}, would be to start by skimming the introductory section and the summary (\cref{sec:SummaryTheories}). \\

In \textbf{\cref{chap:Dilations}}, \emph{dilations} are introduced formally, but completely disregarding causal structure. As mentioned in the general introduction, ignoring causality may effectively change the relationship among dilations -- for example, the two dilations of the bit refreshment channel will be equivalent in the  \emph{dilational ordering} of \cref{chap:Dilations}, but not in the \emph{\myuline{causal}-dilational ordering} of \cref{chap:Causal}, which will eventually be the correct formalisation of `derivability' among causal dilations. 

However, the causality-free setting of \cref{chap:Dilations} turns out to be enlightening for other reasons, namely that it allows us speak of \emph{dilational principles} which a given \myuline{theory} might comply to. The power of these principles will be demonstrated in \cref{chap:Dilations} by deriving from them a number of features, which previously relied on specifics of the formalism of quantum information theory, or on probabilistic concepts. 

The results of \cref{chap:Dilations} for the most part play no role whatsoever in establishing the connection of the framework to quantum self-testing. Rather, they are included because they are interesting in their own right, and because I believe a thorough study of dilations has to begin in the special case where causality is trivial. \\

\textbf{\cref{chap:Metric}} contains a rather general definition of \emph{metrics} on a theory, and discusses some properties which are natural to require of such metrics, with special emphasis on compatibility with dilations. This idea  leads us to introduce the \emph{purified diamond-distance}, which is a particularly well-behaved metric in quantum information theory, generalising the purified distance of Refs. \cite{Toma10,Toma12}.  

The most important thing to say about this chapter is probably that I was not sure whether to include it in the thesis or not -- the observations in \cref{chap:Metric} should be considered introductory and somewhat detached from the remainder of the thesis. Nevertheless, I believe that it adds a further perspective to the theory of dilations in \cref{chap:Dilations}, and that the open problem of extending the metric theory to the causal setting of the two later chapters might be one of the most interesting left from the thesis.\\

\textbf{\cref{chap:Causal}} is the longest chapter of the thesis. Here, we introduce the formal apparatus which we will use to speak about \emph{causality}, in particular the notions of \emph{causal dilations} (which formalise causally structured side-computations) and the \emph{causal-dilational ordering} (which formalises the idea that some causal dilations are derivable from others). In theory, \cref{chap:Causal} is a `causal version' of \cref{chap:Dilations}, but in practice things are more subtle.

 First of all, owing to the causal structure, a new operation among channels arises, namely that of \emph{contraction}. For example, in the channel \eqref{eq:Introdil2} which we saw a few pages ago, the wiggly output wire at the bottom can be `contracted' with the straight input wire, thus creating a new circuit;\footnote{There is no reason why one would want to do so in the particular channel \eqref{eq:Introdil2}, I am merely using it as example since we have not yet seen other causal channels than \eqref{eq:Introdil1} and \eqref{eq:Introdil2}.} it is not clear that this operation can be defined solely in terms of the total input-output behaviour of the channel \eqref{eq:Introdil2} without reference to a particular circuit-representation, but as demonstrated in \cref{chap:Causal} it often can. This is important because we have to allow such contractions to occur in the environment when defining the causal-dilational ordering (`derivability').\footnote{The example just given is \myuline{not} a contraction within the environment, as the input interface involved in the contraction does not belong to the environment, but we will see plenty of such examples.} As detailed later, we can view the contraction operation as an instance of abstract \emph{notions of contraction}, which are related to so-called \emph{traces} in symmetric monoidal categories (\cite{JSV96}). 
 
 Secondly, it is relevant to prove a number of \emph{stability results} to consolidate the concept of a causal dilation. For example, we will see the non-obvious fact that causal dilations are actually stable under contractions in the environment as described above, and we will see (less surprisingly) that the derivability relation is `composable', e.g. in the sense that derivability is preserved under serial and parallel composition of channels. 
 
 Finally, since the causal-dilational ordering is more complicated than the dilational ordering, it will not be possible to replicate the precision of \cref{chap:Dilations} in its analysis. This however gives rise to the idea of \emph{rigidity} of a causal channel, which asserts the existence of a strongest possible causal dilation. This is the concept which we will ultimately link with quantum self-testing. \\

That link is established finally in \textbf{\cref{chap:Selftesting}}. Here, we essentially identify the traditional quantum strategies as causal dilations from which all other (sensible) dilations are derivable. We then establish that self-testing as ordinarily conceived implies the equivalence in the causal-dilational ordering of all causal dilations corresponding to quantum strategies, and thus in particular the existence of a causal dilation from which all others can be derived and which has no pre-existing side-information about the outputs at the accessible interface.  

This chapter also contains a surprising recharacterisation of quantum behaviours as those causal channels which admit a causally structured Stinespring dilation which is non-signalling. \\

The thesis ends with a common \textbf{Conclusion} which is kept rather short in light of the individual chapter conclusions.

\chapter*{Preliminaries}
\addcontentsline{toc}{chapter}{Preliminaries}

The thesis can in principle be read by someone with little knowledge about quantum theory, whereas it requires exposition to a wide range of various elementary mathematical constructs and ideas (graphs, metric spaces, mathematical standards of formalisation and proof, etc.). The thesis can be read without previous acquaintance with category theory, though superficial or intuitive understanding of the subject is beneficial.

A few notions which are needed in the thesis, but may not be covered by standard mathematical experience, are listed below. The reader with further interest in quantum information theory may consult the standard reference \cite{NC02}, or one of many excellent lecture notes available online, e.g. \cite{Wat11}.

{\centering
	\subsection*{§1. Dirac Notation.}}

Many practitioners of quantum physics fancy the so-called `Dirac notation' (\cite{Dirac81}) for vectors, whereas mathematicians tend to dislike it, perhaps in lack of a rigorous presentation. We will not need this notation overwhelmingly, but it is used on occasion. It can easily be introduced in a precise fashion.

Let $\cH$ be a Hilbert space over $\C$ with inner product $\langle \cdot, \cdot \rangle$, which we take to be linear in its \myuline{second} argument (and thus anti-linear in the first). Given a vector $\psi \in \cH$, let us denote by $\ket{\psi}$  (`ket $\psi$') the linear map $\C \to \cH$ given by $z \mapsto z \psi$ and by $\bra{\psi}$ (`bra $\psi$') the linear map $\cH \to \C$ given by $\phi \mapsto \langle \psi, \phi \rangle$. One easily checks by definition of adjoints that $\ket{\psi}^* = \bra{\psi}$ and $\bra{\psi}^* = \ket{\psi}$. By virtue of the Riesz representation theorem every linear functional $\cH\to \C$ is of the form $\bra{\psi}$ for some $\psi \in \cH$. The merits of these bizarre-looking conventions are now threefold: \\

\begin{itemize}
	\item In equations, we can replace vectors $\psi \in \cH$ by their kets $\ket{\psi}$ without disturbing the content. For example, it is easy to check that $z_1 \ket{\psi_1} + z_2 \ket{\psi_2} = \ket{z_1 \psi_1 + z_2 \psi_2}$ for $z_1, z_2 \in \C$, and that if $A: \cH\to \cK$ is a linear operator with $A \psi = \phi$ then  $A \ket{\psi} = \ket{\phi}$. As a result, we can ultimately forget about the vectors $\psi$ and think of the kets $\ket{\psi}$ as fundamental and `belonging' to $\cH$. The corresponding bras $\bra{\psi}$ can be thought of as simply alternative representations of the same underlying objects, `belonging' to the dual space $\cH^*$.
	\item The operator $\braket{\phi}{\psi} := \bra{\phi} \circ \ket{\psi}$ is the linear map $\C \to \C$ given by $z \mapsto \langle \phi, \psi \rangle z$, naturally identified with the number $\langle \phi, \psi \rangle $ itself. This justifies the suggestive identity $\braket{\phi}{\psi} = \langle \phi, \psi \rangle $ and makes explicit mentioning of an inner product on $\cH$ unnecessary; it has effectively been merged with the notation for vectors. 
	\item We have a succinct way of writing the operator $\ketbra{\phi}{\psi} := \ket{\phi} \circ \bra{\psi}$ given by $ \chi \mapsto \langle \psi, \chi \rangle \phi$; in particular, for $\psi \in \cH$ a unit vector, we have a succinct notation for the projection onto the subspace spanned by $\psi$, namely $\ketbra{\psi}{\psi}$. 
\end{itemize}

Now, once the bra-ket notation gains a life of its own, it is tempting to forget so much about the initial vectors that we insert into the symbol $\ket{\phantom{\psi}}$ an arbitrary name for the ket rather than an actual vector; in particular, the kets in the standard basis of $\C^n$ are customarily named $\ket{0}, \ket{1}, \ldots, \ket{n-1}$. (As such, $\ket{0}$ denotes not, as the previous convention would dictate, the zero operator $\C \to \cH$.) 

In a similar spirit of inconsistency, we will actually from now on use letters $\psi, \phi, \ldots$ from the end of the Greek alphabet to denote \emph{rank-1 projections} (i.e. orthogonal projections onto $1$-dimensional subspaces), and then write $\ket{\psi}, \ket{\phi}, \ldots$ for \emph{vector representatives}, i.e. unit vectors in the corresponding subspaces. This convention not only overwrites the above, but also abuses notation, since `$\ket{\psi}$' is only determined from `$\psi$' up to multiplication by a complex number $\alpha$ of unit modulus; however, whenever we use in an equation the `vector representative' $\ket{\psi}$ of the projection $\psi$, it will be the case that the content of the equation is insensitive to the choice of the scalar $\alpha$.

{\centering
	\subsection*{§2. General CPTP Maps and Their Representations.}}

As we will see, systems in quantum information theory are modelled by (separable) Hilbert spaces, and the processing of quantum information between such systems by \emph{completely positive trace-preserving (CPTP)} maps on associated operator algebras. We will mostly be interested in the case where the Hilbert spaces are finite-dimensional, but the definitions are presented generally below.  \\

\textbf{Complete Positivity (CP).} Given a Hilbert space $\cH$, recall that an operator $A$ on $\cH$ is said to be \myuline{positive}, denoted $A \geq 0$, if it can be written in the form $B^* B$ for some operator $B$ on $\cH$, with $B^*$ denoting the adjoint (Hermitian conjugate) of $B$. Given Hilbert spaces $\cH$ and $\cK$, a linear map $\Lambda: B(\cH) \to B(\cK)$ from (bounded) operators on $\cH$ to (bounded) operators on $\cK$ is called \emph{positive} if $\Lambda(A) \geq 0$ for all $A \geq 0$. 

The map $\Lambda$ is called \emph{completely positive} if for any Hilbert space $\cR$, the linear map $\Lambda  \otimes \id_\cR : B(\cH) \otimes B(\cR)  \to B(\cK) \otimes B(\cR)$ is positive. (Observe the isomorphisms $ B(\cH) \otimes B(\cR) \cong B(\cH \otimes \cR)$ and $B(\cK) \otimes B(\cR) \cong B(\cK \otimes \cR)$.) 

Clearly any completely positive map is positive, but there are positive maps which are not completely positive, for example the map $B(\C^2) \to B(\C^2)$ which maps a $2 \times 2$ matrix to its transpose.

For any linear operator $ S: \cH \to \cK$, the map $B(\cH) \ni A \mapsto SAS^* \in B(\cK)$ is an example of a completely positive map; it is called \emph{conjugation by $S$}. We will be mostly interested in the case where $S$ is an \myuline{isometry} (i.e. satisfies $S^*S = \bone_\cH$).  \\

\textbf{Trace-Preservation (TP).} Let $B_1(\cH) \subseteq B(\cH)$ denote the subspace of \myuline{trace class} operators on $\cH$. (When $\cH$ is finite-dimensional, $B_1(\cH)=B(\cH)=\End{\cH}$, the space of all linear operators on $\cH$.) Let us call a linear map $\Lambda: B_1(\cH) \to B_1(\cK)$ \emph{trace-preserving} if $\tr(\Lambda(A))= \tr(A)$ for all $A \in B_1(\cH)$, \myuline{and} if $\Lambda$ is continuous w.r.t. the trace norm $\norm{\cdot}_1$, given by $\norm{A}_1 = \tr(\abs{A})= \tr(\sqrt{A^*A})$. (When $\cH$ is finite-dimensional, the continuity requirement is void.) 

Every isometric conjugation $A \mapsto SAS^*$ (restricted to $B_1(\cH)$) is an example of a trace-preserving map, since $\tr(SAS^*)= \tr(S^*SA)$ by cyclicity of the trace. Another example of a trace-preserving map is the trace itself, that is, the map $\tr : B_1(\cH) \to B_1(\C) \cong \C$. \\

\textbf{CPTP Maps.} A linear map $\Lambda : B_1(\cH) \to B_1(\cK)$ is called \emph{CPTP} if is it completely positive and trace-preserving. Both isometric conjugations and traces are examples of CPTP maps. Moreover, the serial composition of any two CPTP maps is CPTP, and the tensor product of any two CPTP maps is also CPTP (observing again isomorphisms of the sort $B_1(\cH_1) \otimes B_1(\cH_2) \cong B_1(\cH_1 \otimes \cH_2)$). 

A CPTP map from $B_1(\C) \cong \C$ to $B_1(\cK)$ is called a \emph{state (on $\cK$)}, and it is easily seen that the states on $\cK$ are precisely the maps of the form $ \C \ni a \mapsto a \varrho \in B_1(\cK)$ where $\varrho$ is a positive trace class operator on $\cK$ with $\tr(\varrho)=1$ (a so-called \emph{density operator}).  \\

\textbf{Kraus Representations.} It can be shown that $\Lambda : B_1(\cH) \to B_1(\cK)$ is CPTP if and only if there exists a (countable) family $(K_i)_{i \in I}$ of linear operators $K_i : \cH\to \cK$ such that $\sum_{i \in I} K^*_i K_i = \bone_\cH$ and 

\begin{align} \label{eq:Kraus}
\Lambda(A) = \sum_{i \in I} K_i A K^*_i  \quad \text{for all $A \in B_1(\cH)$.} 
\end{align}

A representation such as \eqref{eq:Kraus} is called a \emph{Kraus representation of $\Lambda$}. \\ 

\textbf{Stinespring Representations.} It can be shown that $\Lambda : B_1(\cH) \to B_1(\cK)$ is CPTP if and only if there exists a Hilbert space $\cE$ and an isometry $S: \cH \to \cK\otimes \cE$ such that 

\begin{align} \label{eq:Stinespring}
\Lambda(A) =  [\id_{B_1(\cK)} \otimes \tr_\cE](SAS^*) \quad \text{for all $A \in B_1(\cH)$}, 
\end{align}

where $\tr_\cE : B_1(\cE) \to \C$ denotes the trace on $B_1(\cE)$. This statement is known as \emph{Stinespring's Dilation Theorem} (\cite{Stine55}) and the isometric conjugation $A \mapsto SAS^*$ in the representation \eqref{eq:Stinespring} is known as a \emph{Stinespring dilation of $\Lambda$} (sometimes, the term `Stinespring dilation' is used to refer to the isometry $S$ itself).

 Stinespring's theorem also contains a clause of uniqueness up to isometries, that is, if $S: \cH\to \cK \otimes \cE$ and $S' : \cH \to \cK \otimes \cE'$ are two isometries which both define a Stinespring dilation of $\Lambda$, and if $\dim \cE \leq \dim \cE'$, then there exists an isometry $W: \cE \to \cE'$ such that $(\bone_\cK \otimes W)S = S'$. 

In the special case where $\cH = \C$ so that $\Lambda$ is a state, identifiable with a density operator $\varrho$ on $\cK$, the Stinespring dilations are defined by isometries $\C \to \cK \otimes \cE$, i.e. unit vectors $\ket{\psi} \in  \cK \otimes \cE$, and they are typically called \emph{purifications of $\varrho$}. \\

{\centering
	\subsection*{§3. Special CPTP Maps and Their Representations.}}

\textbf{Classical Systems.} Given a countable (often finite) set $X$, the associated Hilbert space of square-summable sequences $\ell^2(X)$ (which coincides with $\C^X$ when $X$ is finite) is the quantum analogue of the set $X$. We will call Hilbert spaces of the form  $\ell^2(X)$ \emph{classical}. Any (separable) Hilbert space $\cH$ is \myuline{isomorphic} to a classical one, but for classicality we require strict equality. In effect, this is a matter of there being chosen a preferred orthonormal basis in $\cH$, indeed $\ell^2(X)$ ($\C^X$) has the canonical orthonormal basis $(\ket{e_x})_{x \in X}$, where $e_x$ is the sequence given by $e_x(x)=1$ and $ e_x(x')=0$ for $x' \neq x$. By abuse of notation, we write the basis elements $\ket{e_x}$ as $\ket{x}$. In quantum information theory, the basis $(\ket{x})_{x \in X}$ is often called the \emph{computational basis}. \\

\textbf{Decoherence and Classical States.} Given a function\footnote{Some mathematicians use the term `map' in place of `function', reserving the term `function' for maps which take values in $\R$ or $\C$. We do not employ this convention.} $f:X \to Y$ between countable sets, it can be naturally represented as a CPTP map $\hat{f} : B_1(\ell^2(X)) \to B_1(\ell^2(Y))$, namely the one defined by $\hat{f}(A)= \sum_{x \in X} \bra{x} A \ket{x}  \ketbra{f(x)}$, which in particular satisfies $\hat{f}(\ketbra{x}) = \ketbra{f(x)}$. 

As such, the representation of the identity function  $x \mapsto x$ on $X$ is the CPTP map $\Delta_X$ given by $\Delta_X(A) = \sum_{x \in X} \bra{x} A \ket{x}  \ketbra{x}$. We will call $\Delta_X$ the \emph{decoherence map associated to $X$}, and a state $\varrho$ on $\ell^2(X)$ is called \emph{classical} if $\Delta_X(\varrho)= \varrho$. It is a simple exercise to verify that $\varrho$ is classical precisely if $\varrho = \sum_{x \in X} p(x) \ketbra{x}$ for some probability density $p: X \to [0,1]$, so classical states on $B_1(\ell^2(X))$ can be identified with probability distributions on $X$. \\

\textbf{Quantum Measurements.} If $M: B_1(\cH) \to B_1(\ell^2(Y))$ is a CPTP map whose domain is represented by a classical system, we say that $M$ has \emph{classical outcomes} if $\Delta_Y \circ M= M$. More commonly, such a CPTP map is called a \emph{measurement on $\cH$ with outcomes in $Y$}. Using the Kraus representation of $M$, it is easy to verify that if $M$ is classical then there exists a \emph{Positive Operator-Valued Measure (POVM) on $\cH$}, i.e. a family $(E_y)_{y \in Y}$ of positive operators $E_y$ on $\cH$ with $\sum_{y \in Y} E_y = \bone_\cH$, such that

\begin{align} \label{eq:POVM}
\Lambda(A) = \sum_{y \in Y} \tr(E_y A)  \ketbra{y} \quad \text{for all $A \in B_1(\cH)$}; 
\end{align}

conversely, any POVM $(E_y)_{y \in Y}$ defines a measurement. Thus, we can identity measurements with POVMs. 

A measurement $M: B_1(\cH) \to B_1(\ell^2(Y))$ is said to be \emph{projective} if the associated POVM $(E_y)_{y \in Y}$ is a PVM (Projection-Valued Measure), i.e. if $E_y$ is a projection on $\cH$ for all $y \in Y$.\\

\textbf{Naimark's Theorem} For any measurement $M: B_1(\cH) \to B_1(\ell^2(Y))$, there exists a Hilbert space $\cK^\up{Nai}$, a \myuline{projective} measurement $M^\up{Nai} : B_1(\cH \otimes \cK^\up{Nai}) \to B_1(\ell^2(Y))$ and a pure state $\phi^\up{Nai}$ on $\cK^\up{Nai}$, such that 

\begin{align} \label{eq:Naimark}
M = M^\up{Nai} \circ ( \id_{B_1(\cH)} \otimes \phi^\up{Nai}).
\end{align}

This statement is known as \emph{Naimark's (Dilation) Theorem} (\cite{Neum40}), and the representation \eqref{eq:Naimark} is called a \emph{Naimark representation} of $M$. (Sometimes, Naimark's name is transcribed as `Neumark'.)\\

\textbf{Ensembles of CPTP Maps.} The decoherence maps $\Delta_X$ facilitate more refined notions of classicality too. In particular, if $\Lambda : B_1(\cH) \otimes B_1(\ell^2(X)) \to B_1(\cK)$ is a CPTP map for which a factor of the domain is a classical system, we may say that $\Lambda$ is classical on this factor if $\Lambda \circ (\id_{B_1(\cH)} \otimes \Delta_X) = \Lambda$. It is easy to verify that this is the case precisely if there exists a family $(\Lambda^x)_{x \in X}$ of CPTP maps $\Lambda^x : B_1(\cH) \to B_1(\cK)$ such that 

\begin{align}
\Lambda (A \otimes B) = \sum_{x \in X} \Lambda^x(A) \bra{x} B \ket{x}   \quad \text{for all $A \in B_1(\cH)$, $B \in B_1(\ell^2(X))$} .
\end{align}

Thus, to specify a CPTP map $\Lambda : B_1(\cH) \otimes B_1(\ell^2(X)) \to B_1(\cK)$ which is classical on $B_1(\ell^2(X))$ is precisely to specify an \myuline{ensemble} of CPTP maps $\Lambda^x : B_1(\cH) \to B_1(\cK)$, indexed by $x \in X$. In this case we will often use the terminology that $\Lambda$ `measures' of `reads off' the classical value $x$ and applies the according map $\Lambda^x$.\\

{\centering
	\subsection*{§4. Miscellaneous.}}

\textbf{Pre-Orders.} Let $P$ be a class of objects (e.g. a set). Recall that a \myuline{relation} on $P$ is a subclass $R$ of $P \times P$, and that we tend to write $pRq$ rather than $(p,q) \in R$. Recall that a relation $R$ is \myuline{reflexive} if $pRp$ for all $p \in P$, and \myuline{transitive} if for all $p,q,r \in P$ the conditions  $p Rq$ and $q R r$ imply the condition $p R r$. A relation $R$ is called a  \emph{pre-order} if it is reflexive and transitive.\footnote{It is worth observing that an \myuline{equivalence relation} is thus a pre-order which is additionally \myuline{symmetric}, meaning that $p Rq$ implies $q R p$.} Pre-orders are typically denoted with directional symbols, such as $\geq$, $\succeq$, $\cder$ etc., with the implicit convention that mirroring the symbol inverts the order (e.g. `$p \leq q$' means $q \geq p$). If the conditions $p R q$ and $q R p$ imply $p=q$, the pre-order is commonly called a \emph{partial order}. Most pre-orders of interest to us will not have this property, but it in general the relation $\sim_R$ defined by $p \sim_R q \Leftrightarrow p R q \land q R p$ is an equivalence relation on $P$.

\textbf{Special Elements of Pre-Orders.} Let $\geq$ be a pre-order on $P$. An element $u \in P$ is called a \emph{largest (greatest) element} if $u \geq p$ for all $p \in P$. An element $m \in P$ is called a \emph{maximal element} if for all $p \in P$ with $p \geq m$ it also holds that $m \geq p$ (i.e. $m \sim_{\geq} p$). 

Any largest element is a maximal element, but not necessarily conversely. For instance, in the pre-order on $\{0,1,2\}$ defined precisely by the reflexive conditions and the two conditions $1 \geq 0$ and $2 \geq 0$, both $1$ and $2$ are maximal though neither is largest. \emph{Smallest (least) elements} and \emph{minimal elements} are defined dually, by inverting the order.

 Given a subclass $P_0 \subseteq P$ we can naturally restrict the pre-order to that subclass, and we may consequently speak of largest and maximal (respectively smallest and minimal) elements \emph{in $P_0$} by minding this restriction. For instance, in the previous example, the element $2$ is a largest element \myuline{in $\{0,2\}$}.

\textbf{Dense Subclasses of Pre-Orders.} A subclass $D \subseteq P$ is called \emph{dense in $P$}, if for any $p \in P$ some $d \in D$ satisfies $d \geq p$. (As such, `dense' means `dense at the top'.) By extension, a class $D$ is called dense in $P_0 \subseteq P$, if $D \subseteq P_0$ and $D$ is dense in $P_0$ considered as a pre-order on its own. 

This terminology has been imported from the subject of forcing in axiomatic set theory, cf. Ref. \cite{Kunen80}.  \\

\textbf{The Schmidt Decomposition.} If $\ket{\psi} \in \cH_1 \otimes \cH_2$ is any vector in a tensor-product of Hilbert spaces, then there exist a family $(p(j))_{j \in J}$ of strictly positive numbers, and orthonormal systems $(\ket{\psi_1(j)})_{j \in J}$ in $\cH_1$ and $(\ket{\psi_2(j)})_{j \in J}$ in $\cH_2$, such that 

\begin{align} \label{eq:Schmidt}
\ket{\psi} = \sum_{j \in J} \sqrt{p(j)} \ket{\psi_1(j)} \otimes \ket{\psi_2(j)},
\end{align}

 and $\sum_{j \in J} p(j) = \norm{\ket{\psi}}^2$. An expression of the form \eqref{eq:Schmidt} is called a \emph{Schmidt decomposition} of $\ket{\psi}$. In fact, the cardinality $\abs{J}$ is unique, as is the family $(p(j))_{j \in J}$ (up to permutation). They are referred to as the \emph{Schmidt rank} and \emph{Schmidt coefficients} of $\ket{\psi}$, respectively.

\chapter{Theories}
\label{chap:Theories}
\pagenumbering{arabic}

{\centering
\subsection*{§1. Introduction and Outline.}}

In this chapter we set up a mathematical framework for investigating general physical theories. The chapter has three sections, all of which serve mainly as review. It contains no original observations, except for a few examples in \cref{sec:Examples}, the comment about functors in \cref{rem:QITCIT}, the failure of the Cantor-Schr\"{o}der-Bernstein property as described in \cref{ex:Graphs}, and the definition of `normal' theories (\cref{def:Normal}). \\

\textbf{General Theories.} The first item on the agenda is to define mathematically what is meant by a \emph{(physical) theory}. We will define a theory as a certain type of mathematical structure (like a group, or a measurable space), and as usual the concept is abstracted from a selection of prominent examples. One example with which every reader will be familiar is the theory of \emph{sets and functions}: 

We may think of a set $X$ as a \emph{(physical) system}, and think of a function $f:X \to Y$ as a \emph{(physical) transformation} from the system $X$ to the system $Y$. Functions can be composed \emph{serially}, one following the other, by forming from $f: X \to Y$ and $g: Y \to Z$ the composite $g \circ f : X \to Z$. But they can also be composed \emph{parallelly}, one next to the other; given functions $f_1 : X_1 \to X_2$ and $f_2: X_2 \to Y_2$, we have a function $f_1 \times f_2 : X_1 \times X_2 \to Y_1 \times Y_2$ defined by $(f_1 \times f_2)(x_1, x_2)=(f_1(x_1), f_2(x_2))$. The parallel composition of functions involves a composition of  the underlying systems (sets), namely the formation of the product set $Z_1 \times Z_2$ from the individual sets $Z_1$ and $Z_2$. 

In general, a \emph{theory} will be a structure encompassing systems, transformations, and notions of composing transformations serially and parallelly. Whereas the theory of sets and functions is undoubtedly the example known to most readers, the two most \myuline{important} examples for us is \emph{Classical Information Theory}, $\CIT$, and \emph{Quantum Information Theory}, $\QIT$. The systems of $\CIT$ are (finite) sets and its transformations are so-called \emph{classical channels} (Markov kernels) between them, which can be thought of as probabilistic functions. The systems of $\QIT$ are (finite-dimensional) Hilbert spaces and its transformations are so-called \emph{quantum channels} (CPTP maps) between them. These two theories are described in \cref{sec:ModelforTheories} (\cref{ex:QIT} and \cref{ex:QIT}), where also the general definition of a theory (\cref{def:Theory}) and some surrounding terminology is provided.\\

\textbf{Specific Theories.} \cref{sec:Examples} comprises a large collection of further examples of theories. Some of these will be merely curious, helping to paint a landscape, but most will serve to illustrate points later. I have categorised the examples into classes, and included among them many mathematical ones (though none of them very technical), which admittedly stretch the boundaries of what one might call a `physical' theory. In particular, the example classes include all categories with finite products (\cref{subsec:Cart}), and monoid-like structures related resource theories in the sense of Ref. \cite{CFS16} (\cref{subsec:Thin}).

It is not necessary for the reading of the thesis to be intimately acquainted with all of the examples presented in \cref{sec:Examples}, but it likely yields an elevated reading experience to familiarise oneself with one example from each class.  \\

\textbf{Pictorial Syntax.} The mathematical structure that defines a theory is an algebraic entity equipped with two binary operations, serial and parallel composition of transformations. This complexity can make equations difficult to interpret and consequently obscure intuition. In the last section of the chapter, \cref{sec:Pictorial}, we review a widely used \emph{pictorial syntax} (\cite{Sel10survey}) for arguing about transformations in a theory. This replaces algebraic expressions by pictures, and may thus tremendously clarify algebraic manipulations. We shall use the pictorial syntax in many instances throughout the thesis, and have already seen it exemplified in the general introduction when discussing dilations of the `bit refreshment' channel.

In \cref{subsec:Inter}, we formally introduce the concepts of \emph{interfaces} and \emph{channels}, as opposed to \emph{systems} and \emph{transformations}. The distinction between the two (which arise from the finer points of the pictorial syntax, but which is often ignored in other presentations) might seem at this point overly formal, but it will be important later on, in particular in \cref{chap:Causal}.      \\

{\centering
	\subsection*{§2. Comparison to Existing Literature.}}

\textbf{On the Definition of a Theory.} Our model of `theories' does not aim to capture every single construct that a physicist might call a theory (for example, Einstein's theory of special relativity \cite{Einstein05} is not a theory in that sense). Rather, it aims to capture \emph{operational} aspects of theories, in line with a `pragmatist' tradition of physics (cf. Ref. \cite{Foils}): The interest is not in the ultimate explanation about what or why Nature is, but instead in what intelligent beings can and cannot do with the physical systems and physical transformations handed to them. 

Roughly speaking, there are two pillars of mathematical frameworks which intend to capture operational aspects of theories. One is the \emph{categorical} pillar (pioneered by Refs. \cite{AbCo04,Sel04, Baez06}), according to which the fundamental objects of interest are systems and transformations which can be serially and parallelly combined, as outlined above. It uses \emph{symmetric monoidal categories} (\cite{MacLane}) as a model for theories. The second pillar is the \emph{convex} or \emph{probabilistic} framework, often in the incarnation of \emph{generalised probabilistic theories} (\cite{Barn16}). In this framework, an underlying categorical structure  is often implicitly present (\cite{Barr07, Barn11}), but the emphasis is on probabilities and convexity, and the study of how \emph{state spaces} morph under the composition of systems. There has been work which quite explicitly merges the categorical and probabilistic pillars (e.g. Refs. \cite{Chir10, Hardy10}), and the book \cite{Foils} gives a fairly recent overview of various tendencies within the field. 
In developing the theory presented in this thesis, I have made an effort to stay within a purely categorical framework. This is not (only) because it is more general than merged frameworks, but also because almost all defined concepts are completely independent of probabilistic structure. Precisely, the definition chosen here for a theory (\cref{def:Theory}) is that of a \emph{symmetric monoidal category in which the unit object is terminal.} As such, theories in our sense are more restricted than those modelled by arbitrary symmetric monoidal categories (\cite{Coecke16Gen}), but on the other hand do not assume additional structures like \emph{dagger compactness} or \emph{$*$-autonomy}, which were and still are instrumental ingredients in some works (e.g. Refs. \cite{AbCo04, Kiss17}).\footnote{Somewhat confusingly, treatments employing dagger compactness tend to define the transformations in quantum theory as linear operators between Hilbert spaces, rather than as CPTP maps between operator algebras (see also Ref. \cite{Coecke10Guises}). We shall use the symmetric monoidal categories exclusively as they pertain to the latter depiction.} 	

It is well-established that symmetric monoidal categories with terminal unit object can be interpreted as theories in which future events cannot signal to the past, and as such these are often termed \emph{causal theories} (\cite{Chir10, CoLal13, Coecke14}). In fact, our notion of theories exactly coincides with that of a \emph{causal deterministic theory} in the words of Ref. \cite{Chir10}. However, in other treatments the terminality assumption is mostly accompanied by further standing assumptions, and in practice this renders the scope of those treatments smaller than the one presented here. Accordingly, many of the examples in \cref{sec:Examples} would be ruled out in other works (for example in Ref. \cite{Chir10} the assumption of `non-determinism' rules out our cartesian theories, and the assumption that transformations are determined by their action on states rules out our thin theories). 

In the mathematical literature our notion of theories are commonly referred to as \emph{(symmetric) semi-cartesian categories}, or \emph{monoidal categories with projections} (see \cite{SemicartesianWebsite}, and the comments between Remarks 2.3 and 2.4 in Ref. \cite{Fritz20Synthetic}), but here a systematic study of the class also seems to be absent. \\

\textbf{On the Use of the Pictorial Syntax.}  Ref. \cite{Sel10survey} reviews a large class of pictorial syntaxes for monoidal categories, including the one for symmetric monoidal categories, attributed to \cite{Pen71}. Nowadays, its use and interpretation are fairly standardised, with minor differences in the choice of layout (e.g. some prefer that diagrams be read from top to bottom rather than left to right). As observed pedantically in \cref{subsec:Inter}, however, the ambiguity in its representation of composite systems means that the pictures do not strictly correspond to transformations between systems, but rather to transformations between \emph{interfaces}, that is, tuples of systems labelled by port names. We shall use the term \emph{channels} about such transformations. The distinction is minute and for most purposes insignificant (which is probably why it has not been pointed out before), but we need it for the precise definition of marginalisation and dilations (\cref{chap:Dilations}), and it will become even more pressing in \cref{chap:Causal}. \\

\newpage
\section{A Mathematical Model for Physical Theories} \label{sec:ModelforTheories}

The precise definition we choose for a theory is a \emph{symmetric monoidal category} in which \emph{the monoidal unit} is \emph{terminal}. These words might intimidate certain readers, but I should like to emphasise that the concept is intuitively simple and ubiquitous, in fact intelligible to anyone who has interacted with the real world. Readers who prefer concrete rather than abstract mind-sets will not lose much by fixing `theory' to mean either $\CIT$ (classical information theory) or $\QIT$ (quantum information theory). \\

Category theory was created in the 1940s by Samuel Eilenberg and Saunders Mac Lane (\cite{Eil45}). It was developed for applications in algebraic topology, but it soon grew wildly beyond this scope and is nowadays considered a universal language for many mathematical ideas and constructions (the original go-to reference is \cite{MacLane}; Ref. \cite{Awo10} offers a modern and less overwhelming treatment).

In recent times, category theory has been successfully implemented also in areas outside of pure mathematics, of which Ref. \cite{Rosetta} provides a very readable overview. One of these areas is the study of foundational physics, where it was realised that (symmetric monoidal) categories can be used to model physical theories.\\

What is a category?  Formally, it is a type of mathematical structure. Poetically, it is the incarnation of the abstract idea of `serial composition'. More precisely, a \emph{category} $\bfC$ comprises a collection of \emph{objects}, $\cX, \cY, \cZ, \ldots$, and a collection of \emph{morphisms} between these objects, $T, S, R, \ldots$. For example, the objects could be \myuline{sets} and the morphisms from the set $\cX$ to the set $\cY$ could be \myuline{functions} from $\cX$ to $\cY$. Alternatively, the objects could be \myuline{groups} and the morphisms from the group $\cX$ to the group $\cY$ could be \myuline{group homomorphisms} from $\cX$ to $\cY$. We write $T: \cX \to \cY$ to signify that $T$ is a morphism from $\cX$ to $\cY$. A category $\bfC$ is defined by its collection\footnote{Readers who are used to defining a mathematical structure as a \emph{set} equipped with certain operations or additional material, and who know something about the axioms of set theory, might worry that it is dangerous to define a mathematical structure whose underlying universe it too big to be a set (e.g. the collection of all sets, which is a proper class). There are however at least two formal escape routes: One is to use a different frame of axioms in which the notion of a \emph{(proper) class} has formal meaning, for example the set theory of von Neumann-Bernays-G\"{o}del. Another is to consider proper classes as entities which exist in the metalanguage, namely as predicates in first-order logic which intuitively define the class. See e.g. Ref. \cite{MacLane} for further discussions.}  of objects and morphisms, and by a notion of composition of morphisms: Given $T: \cX \to \cY$ and $S: \cY \to \cZ$, there is a morphism $S \after T : \cX\to \cZ$, called the (serial) composition of $T$ with $S$. In the cases of functions or group homomorphisms this composition is ordinary functional composition, but in general $\after$ is just an abstract binary operation. It is subject to the associativity requirement $(R \after S) \after T = R \after (S \after T)$, and it is moreover required that to every object $\cZ$ is associated a morphism $\id_\cZ : \cZ \to \cZ$, called identity, such that $T \after \id_\cX = T = \id_\cY \after T$ for any morphism $T: \cX \to \cY$. And that is it. \\

Like other mathematical structures, a category may be equipped with additional architecture, making it a more refined object. One such additional architecture is that of a \emph{symmetric monoidal} structure, which adds one further mode of composition. Whereas the composition inherent in every category is \emph{serial}, one morphism following another, a (symmetric) monoidal structure facilitates a notion of \emph{parallel} composition. The category of sets and functions is an example of a category which allows such a structure -- as discussed in the introduction, the parallel composition of $f_1 :X_1 \to Y_1$ and $f_2: X_2 \to Y_2$ is the function $f_1 \times f_2 : X_1 \times X_2 \to Y_1 \times Y_2$, given by $(f_1 \times f_2)(x_1, x_2)= (f_1(x_1), f_2(x_2))$. Clearly, the category of groups sustains a similar construction. An example of a category with no obvious monoidal structure is that of \myuline{Boolean algebras} and homomorphisms between Boolean algebras. 

In general, a \emph{monoidal structure} on a category $\bfC$ is a binary operation, $\og$, additional to the existing serial composition. This operation has two components: For any two objects $\cX$ and $\cY$ in $\bfC$, it defines an object $\cX \og \cY$ in $\bfC$, the composite of $\cX$ and $\cY$; and for any two morphisms $T_1: \cX_1 \to \cY_1$ and $T_2: \cX_2 \to \cY_2$ in $\bfC$ it defines a parallel composite $T_1 \og T_2 : \cX_1 \og \cX_2 \to \cY_1 \og \cY_2$ in $\bfC$. Just as the serial composition in a bare category, the parallel composition is subject to an associativity requirement,\footnote{Although the associativity requirement is cumbersome to state precisely. The reason is that in most cases of interest, the composition is only `almost' associative; for example, given sets $X, Y$ and $Z$, the two sets $(X \times Y) \times Z$ and $X \times (Y \times Z)$ are easily identifiable but not formally identical.} and also required to interplay sensibly with the serial composition (for example, one requires $(S_1 \after T_1) \og (S_2 \after T_2)= (S_1 \og S_2) \after (T_1 \og T_2)$). Moreover, one requires the existence of a special system, $\triv$, which acts as a unit for the $\og$-operation on systems: $\cX \og \triv = \cX = \triv \og \cX$. In the example of sets and functions, we can declare as unit object any set $\{*\}$ which contains a single element.\footnote{Again, we do not strictly have the equalities $X \times \{*\} = X = \{*\} \times X$, but the three sets are naturally identifiable.} In the example of groups and group homomorphisms we can take for the object $\triv$ any trivial group. %

Finally, the word \emph{symmetric} in `symmetric monoidal' refers to the fact that the parallel composition is required to be in a certain sense symmetric. This is again well illustrated in the example of sets and functions: Whereas for functions $f:X \to X$ and $g: X \to X$ the serial compositions $g \circ f$ and $f \circ g$ are generally very different, there is a sense in which, for $f_1: X_1 \to Y_1$ and $f_2: X_2 \to Y_2$, the functions $f_1 \times f_2$ and $f_2 \times f_1$  are just two different ways of looking at the same function. Similarly, the sets $X \times Y$ and $Y \times X$ can be easily identified. \\

Readers who are interested in an accessible and more detailed introduction to symmetric monoidal categories may consult one of many well-written expositions, e.g. Refs. \cite{Rosetta, Coecke10Guises}. Readers in want of more knowledge about mere categories may consult Ref. \cite{Awo10}. \\

For the sake of completeness -- and out of respect for mathematically minded readers -- I find it appropriate to reproduce below a precise definition of symmetric monoidal categories. On the other hand, any reader who feels comfortable with an intuitive impression of symmetric monoidal categories (or is creative enough to assemble a definition based on the many examples in \cref{sec:Examples}), is invited to save eye power by skipping \cref{def:SMC} and going now directly to \cref{subsec:CITQIT}.\\

\begin{Remark} (For those Intending to Read the Definition.) \label{rem:Strict}\\
To avoid as much formalism as possible, only the definition of an especially simple kind of symmetric monoidal category is stated, namely a so-called \emph{strict} one. This is essentially means doing away with the issues surrounding the precise relation between $(\cX \og \cY) \og \cZ$ and $\cX \og (\cY \og \cZ)$, and between $\cX \og \triv$, $\cX$ and $\triv \og \cX$. This approach is standard, and it is justified by Mac Lane's `Strictification Theorem' (\cite{MacLane}, Chapter XI, Section 3), according to which any monoidal category is equivalent to a strict monoidal category (via a pair of `strong monoidal functors'). 

In practice, this means that we need never formally consider non-strict categories. Thus, we adopt the commonly held attitude that for \textbf{Theorems} and \textbf{Definitions} we assume categories to be strict, whereas for \textbf{Examples} we have no hesitations about exposing non-strict categories. 

The strictification theorem does not go as far as to drown the similar problem of the relationship between $\cX \og \cY$ and $\cY \og \cX$. Rather, this relationship must be formalised in terms of \emph{swapping morphisms} $\sigma_{\cX, \cY} : \cX \og \cY \to \cY\og \cX$. Unfortunately, the conditions imposed on these morphisms take up a part of \cref{def:SMC} which in size is disproportional to their significance.
\end{Remark}

	\begin{Definition} (Symmetric (Strict) Monoidal Categories (\cite{MacLane}).)\label{def:SMC} \\
	A \emph{symmetric (strict) monoidal category} is a quadruple $(\bfC, \triv,\og, \sigma)$ comprised as follows: 
	
	\begin{itemize}
		\item[1.] $\bfC$ is a category.
		\item[2.] $\triv$ is an object in $\bfC$.
		\item[3a.] $\og$ is a map which maps pairs of objects $(\cX, \cY)$ to objects $\cX \og \cY$, and pairs of morphisms $(T_1:\cX_1 \to \cY_1, T_2: \cX_2 \to \cY_2)$ to morphisms $T_1 \og T_2: \cX_1 \og \cX_2 \to \cY_1 \og \cY_2$.
		
		\item[3b.] $\og$ is associative on objects and morphisms, with $\triv$ and $\id_\triv$ as units, in the sense that 
		\begin{itemize}
			\item[$\bullet$]  for any objects $\cX, \cY$ and $\cZ$ in $\bfC$, 
			\begin{align} 
			(\cX \og \cY) \og \cZ &= \cX \og (\cY \og \cZ), \\ 
			\cX \og \triv = & \cX = \triv \og \cX;
			\end{align}
				\item[$\bullet$] for any morphisms $T: \cX_1 \to \cX_2$, $S : \cY_1 \to \cY_2$ and $R: \cZ_1 \to \cZ_2$,
			\begin{align}
			(T \og S) \og R &= T \og (S \og R), \\ 
			T \og \id_\triv = & T = \id_\triv \og T.
			\end{align}
		\end{itemize}
		
		\item[3c.] $\og$ is \emph{functorial}, meaning that
		\begin{itemize}
			\item[$\bullet$]  for any objects $\cX$ and $\cY$, 
			\begin{align}
			\id_{\cX \og \cY} =	\id_\cX \og \id_\cY ;
			\end{align}
			\item[$\bullet$]  for any morphisms $T_1: \cX_1 \to \cY_1$, $T_2: \cX_2 \to \cY_2$  and $S_1: \cY_1 \to \cZ_1$, $S_2: \cY_2 \to \cZ_2$,
			\begin{align}
			(S_1 \after T_1) \og (S_2 \after T_2)= (S_1 \og S_2) \after (T_1 \og T_2) .
			\end{align}
		\end{itemize}
		\item[4.] $\sigma$ is a collection of morphisms in $\bfC$ called \emph{swappings}, one morphism $\sigma_{\cX, \cY} : \cX \og \cY \to \cY\og \cX$ for each pair $(\cX, \cY)$ of objects in $\bfC$. They are subject to the conditions 
		\begin{align}
		\swap{\cX}{\triv}= \swap{\triv}{\cX}= \id_\cX, \\
		\swap{\cY} {\cX}\after \swap{\cX} {\cY} = \id_{\cX \og \cY}, \\
	(	\swap{\cZ} {\cX} \og \id_\cY) \; \after \; \swap{(\cX \og \cY)}{\cZ}  = \id_\cX \og \swap{\cY} {\cZ}
		\end{align}
		
		for all objects $\cX, \cY$ and $\cZ$, the latter of which is to say that if in $\cX \og \cY \og \cZ$ we swap $\cX \og \cY$ for $\cZ$ and then $\cZ$ for $\cX$, this altogether amounts to swapping $\cY$ for $\cZ$. 
		
		Moreover, it must hold for any morphisms $T_1: \cX_1 \to \cY_1$ and $T_2: \cX_2 \to \cY_2$, that
		\begin{align}
		\sigma_{\cY_1, \cY_2} \after (T_1 \og T_2)=(T_2 \og T_1)\after \sigma_{\cX_1, \cX_2}.
		\end{align}
	\end{itemize}
	
\end{Definition}

As is customary in all mathematical disciplines, we shall often abbreviate the quadruple $(\bfC, \triv,\og, \sigma)$  simply by `$\bfC$', letting its family members be implicit as they are usually clear from the context.

\subsection{Definition of Theories -- $\CIT$ and $\QIT$}
\label{subsec:CITQIT}
Having defined symmetric monoidal categories, the scariest part of the section, if not the  entire chapter, is over. 

We have already touched on two examples of symmetric monoidal categories, namely sets with functions and groups with group homomorphisms. Now that formalities are in order, let us baptise them properly: 

\begin{Example} ($\Sets^*$.) \label{ex:Sets}\\
The category $\Sets^*$ has non-empty\footnote{The reason for restricting to non-empty sets will become clear later (\cref{subsec:Normal}); the problem is essentially that the empty set is very destructive in its parallel composition with other sets.} sets $X, Y, Z, \ldots$ as objects and functions $f: X \to Y$ as morphisms from $X$ to $Y$. Its serial composition is given by ordinary functional composition. The symmetric monoidal structure on $\Sets^*$ is given by $X \og Y := X \times Y$, the cartesian product of sets, and $f_1 \og f_2 := f_1 \times f_2$ for functions $f_1 : X_1 \to Y_1$, $f_2: X_2 \to Y_2$, where $(f_1 \times f_2)(x_1,x_2)=(f_1(x_1), f_2(x_2))$ for $(x_1, x_2) \in X_1 \times X_2$. As unit object $\triv$ we may take any set with a single element, say $\triv := \{\emptyset\}$ for concreteness.	It is tedious but straightforward to verify the conditions of \cref{def:SMC} (ignoring the formal difference between $(X \times Y) \times Z$ and $X \times (Y \times Z)$, and between $X \times \triv$, $X$ and $\triv \times X$). The swapping functions $\sigma_{X, Y}: X \times Y \to Y \times X$ are given by $\sigma_{X, Y}(x,y) = (y,x)$. %
\end{Example}

	\begin{Example} ($\Groups$.) \label{ex:Groups} \\
In $\Groups$ the objects are groups $G,H, K,\ldots$, and the morphisms from $G$ to $H$ are group homomorphisms $\varphi:G \to H$. Group homomorphisms are functions after all, and so we can define the compositions exactly as in $\Sets^*$: Serial composition of morphisms $\varphi: G \to H$ and $\psi:H \to K$ is given by the ordinary functional composition $\psi \circ \varphi$, and parallel composition of $\varphi_1 : G_1 \to H_1$ with $\varphi: G_2 \to H_2$ by $\varphi_1 \times \varphi_2: G_1 \times G_2 \to H_1 \times H_2$, where $K_1 \times K_2$ denotes the  product group of $K_1$ and $K_2$. As unit object $\triv$ we take some fixed trivial group. 
	\end{Example}

Here is another example of a symmetric monoidal category, indeed historically  one of the main inspirations for the very definition of the concept: 

\begin{Example} ($\mathbf{Vect}_k$.)\label{ex:VectTensor} \\
Let $k$ be a field (e.g. $k=\R$ or $k= \C$), and let $\mathbf{Vect}_k$ denote the category whose objects are (finite-dimensional) vector spaces over $k$, and whose morphisms are $k$-linear maps between these spaces, with functional composition as composition. The \emph{tensor product} $\otimes$ defines a notion of parallel composition, making $\mathbf{Vect}_k$ a symmetric monoidal category: The composition of objects $V$ and $W$ is the tensor product $V \otimes W$, and the parallel composition of the linear maps $A_1:V_1 \to W_1$ and $A_2: V_2 \to W_2$ is the linear map $A_1 \otimes A_2 : V_1 \otimes V_2 \to W_1 \otimes W_2$ determined by $(A_1 \otimes A_2)(v_1 \otimes v_2) = A_1(v_1) \otimes A_2(v_2)$. For unit object $\triv$ we take the $1$-dimensional vector space $k$. 
\end{Example}

Whereas we want to include $\Sets^*$ and $\Groups$ in our club of theories, $\mathbf{Vect}_k$ is for our purposes an imposter. (Some authors are more accommodating; see \cref{rem:CausTheory}.) The reason is that $\Sets^*$ and $\Groups$ admit a well-defined notion of \emph{marginalisation}, whereas $\textbf{Vect}_k$ does not:\\

We will come to think of objects as `systems' in a theory, and morphisms as `transformations' between those systems. The composite $\cX \og \cY$ will represent the junction of two systems into one, and the system $\triv$ will represent the `trivial system', i.e. the system corresponding to `nothing'. As such, transformations $\tr_\cX : \cX \to \triv$ correspond to various ways of \emph{discarding}, or \emph{trashing} the system $\cX$, and by extension the transformations $\tr_\cX \og \id_{\cY}  : \cX \og \cY \to \triv \og \cY = \cY$  correspond to ways of discarding only the system $\cX$ from the composite system $\cX \og \cY$. This is precisely the process known as marginalisation, and for it to exist and be unique, we need $\tr_\cX$ to exist and be unique. 

Both $\Sets^*$ and $\Groups$ have this property; there is a unique function from $X$ to $\triv = \{\emptyset\}$ for any set $X$, and there is a unique homomorphism from $G$ to $\triv$ for any group $G$. In contrast, there are many $k$-linear maps from a vector space $V$ to the vector space $k$ (these are precisely the functionals on $V$).\\

In general, an object in a category is called \emph{terminal}	if every object admits a unique morphism to it. We thus arrive at the following definition of a theory: 
	
	\begin{Definition} (Theories.)  \label{def:Theory} \\
		A \emph{theory} is a symmetric (strict) monoidal category $\Theory$, such that the monoidal unit object $\triv$ is terminal.
		
		The following terminology is employed:
		\begin{itemize}
			\item Objects in $\Theory$ are called  \emph{systems}, and we denote the class of all systems in $\Theory$ by $\Sys{\Theory}$.

			\item Given $\cX, \cY \in \Sys{\Theory}$, the system $\cX\og \cY$ is called the \emph{composite of $\cX$ and $\cY$}.

			\item Given $\cX, \cY\in \Sys{\Theory}$, the morphisms $T: \cX \to \cY$ in $\Theory$ are called \emph{transformations from $\cX$ to $\cY$}, and the class of all such transformations is denoted by $\Trans{\Theory}{\cX}{\cY}$. The class of \myuline{all} transformations in $\Theory$ is denoted by $\AllTrans{\Theory}$.

			\item Given transformations $T: \cX \to \cY$ and $S: \cY \to \cZ$ in $\Theory$, the transformation $S \after T : \cX \to \cZ$ is called the \emph{serial composition of $T$ and $S$}. 
			
			\item Given transformations $T_1: \cX_1 \to \cY_1$ and $T_2: \cX_2 \to \cY_2$ in $\Theory$, the transformation $T_1 \og T_2 : \cX_1 \og \cX_2 \to \cY_1 \og \cY_2$ is called the \emph{parallel composition of $T_1$ and $T_2$}. 
			
			\item The system $\triv$ is called \emph{the trivial system}. Given a system $\cX \in \Sys{\Theory}$, the unique transformation from $\cX$ to $\triv$ is denoted $\tr_\cX$ and called \emph{the trash of $\cX$}.
			
		\end{itemize}
	\end{Definition}
	
	\begin{Remark} (All Theories are Causal.) \label{rem:CausTheory}\\
	In some line of work, theories are simply identified with symmetric monoidal categories, and	the stricter concept defined by \cref{def:Theory} is then referred to as \emph{causal} theory, since the terminality assumption on $\triv$ can be interpreted as an impossibility of signalling from the future to the past (\cite{Chir10, Coecke14}). Deviating from this terminology is justified on the grounds that we shall have no interest in `theories' which are not causal, and that we will already use the  work `causal' to a near-excessive degree in other connections.

	\end{Remark}

\begin{Remark} (Typesetting.) \\
	Generically, we typeset theories with boldface letters ($\Theory, \Sets, \ldots$), systems of a theory with calligraphic Latin letters ($\cX, \cY, \cZ, \ldots$), and transformations of a theory with ordinary capital Latin letters ($T, S, R, \ldots$).

	In specific theories (such as $\Sets^*$, $\Groups$, and $\CIT$ and $\QIT$ defined below) we may deviate from these conventions if tradition prescribes. More systematic deviations will be mentioned as introduced (for example, the special transformations to be called `states' will be generically typeset with lower-case Latin letters $s,t, \ldots$). 

\end{Remark}

The next two examples of theories will be our most important:

\begin{Example} (Classical Information Theory, $\CIT$.) \label{ex:CIT}\\ 
	The systems of $\CIT$ are finite, non-empty sets $X, Y, Z, \ldots$. They compose under the cartesian product $\times$, as in the theory $\Sets^*$, and the trivial system is some distinguished one-element set, say $\triv := \{0\}$. A transformation $T: X \to Y$ can be thought of as a `probabilistic function'. Formally, it is a \emph{Markov kernel}, i.e. a collection $T= (t_x)_{x \in X}$ of probability distributions on $Y$; a genuine (`deterministic') function $f: X \to Y$ corresponds indeed to the collection $(\delta_{f(x)})_{x \in X}$, where $\delta_y$ denotes the degenerate distribution in the point $y \in Y$. (Observe in particular that a transformation from $\triv$ to $Y$ is simply a probability distribution on $Y$.) The serial and parallel composition of transformations is best described by appealing to intuition: If we think of a transformation $T : X \to Y$ as encoding a process by which on input $x \in X$ a random $y \in Y$ is produced according to the distribution $t_x$, then the serial composition of $T: X \to Y$ with $S: Y \to Z$ corresponds -- unsurprisingly -- to the process resulting from applying $S$ after $T$, assuming independence of the randomness in $S$ and $T$. Likewise, the parallel composition $T_1 \og T_2$ corresponds to the process of drawing simultaneously and independently outputs $y_1$ and $y_2$ based on the inputs $x_1$ and $x_2$, by means of $T_1$ and $T_2$ respectively. Formally,
	
	\begin{align}
	(t^1_{x_1})_{x_1 \in X_1} \og (t^2_{x_2})_{x_2 \in X_2} = (t^1_{x_1} \otimes t^2_{x_2})_{(x_1, x_2) \in X_1 \times X_2}\end{align}
	
(with $t^1_{x_1} \otimes t^2_{x_2}$	denoting the product distribution on $Y_1 \times Y_2$ of distributions $t^1_{x_1}$ on $Y_1$ and $t^2_{x_2}$ on $Y_2$), and 
	
	\begin{align}
	(s_y)_{y \in Y} \after (t_x)_{x \in X} = (u_x)_{x \in X}, \quad  u_x = \sum_{y \in Y} t_x(y) s_y.
	\end{align}
	
As usual, it is tedious but easy to verify that $\CIT$ satisfies the formal conditions of \cref{def:Theory}. Note that the identity transformation on $X$ is $(\delta_x)_{x \in X}$, and that the trash $\tr_X: X \to \triv$ is the $X$-indexed collection of degenerate distributions on the one-element set $\triv$.

Given a channel $T=(t_x)_{x \in X}$ we will often write $T(x)$ or $T(\delta_x)$ for the probability distribution $t_x$, when there is no risk of confusion. 	\end{Example}

\begin{Example} (Quantum Information Theory, $\QIT$.) \label{ex:QIT}\\ The systems of $\QIT$ are finite-dimensional, non-zero Hilbert spaces $\cH, \cK,\cL, \ldots$ over $\C$. They compose parallelly under the tensor product $\otimes$, and the unit object is some distinguished $1$-dimensional space, say $\triv := \C$. A morphism $\Lambda: \cH \to \cK$ is a so-called \emph{quantum channel} from $\cH$ to $\cK$, meaning a CPTP (completely positive trace-preserving) linear map from $\up{End}(\cH)$ to $\up{End}(\cK)$, where $\up{End}(\cL)$ denotes the space of linear operators on $\cL$. The fact that a \myuline{morphism} $\Lambda: \cH \to \cK$ is a \myuline{map} $\Lambda: \up{End}(\cH) \to \up{End}(\cK)$ is notationally odd-looking, but should not imply confusion in relevant instances. The serial composition of morphisms $\Lambda: \cH\to \cK$ and $\Phi: \cK \to \cL$ is given by the  functional composition $\Phi \circ \Lambda$, and the parallel composition of morphisms $\Lambda_1: \cH_1 \to \cK_1$ and $\Lambda_2 : \cH_2 \to \cK_2$ is given by the tensor product map $\Lambda_1 \otimes \Lambda_2$, determined by $(\Lambda_1 \otimes \Lambda_2)(A_1 \otimes A_2) = \Lambda_1(A_1) \otimes \Lambda_2(A_2)$ for $A_1 \in \End{\cH_1}$, $A_2 \in \End{\cH_2}$. (Observe here the isomorphisms $\End{\cH_1 \otimes \cH_2} \cong \End{\cH_1} \otimes \End{\cH_2}$ and $\End{\cK_1 \otimes \cK_2} \cong \End{\cK_1} \otimes \End{\cK_2}$.) The identity $\id_\cH$ is the identity map on $\End{\cH}$ and the trash $\tr_\cH$ is the trace $\End{\cH} \to \End{\C} \cong \C$ (this is the only trace-preserving linear map to $\End{\C}$).
		\end{Example}
	
	We will typeset, as is customary, the transformations in $\QIT$ with Greek letters, but often typeset linear operators which are used to define the transformations (e.g. isometries or Kraus operators) with Latin letters. Though this clashes somewhat unfortunately with the general convention of using Latin letters for the transformations themselves, this should not cause confusion.

\begin{Remark} (On Terminology and Notation.) \\
	The category $\CIT$ is often denoted in the literature by `$\mathbf{FinStoch}$' (see e.g. Ref. \cite{Fritz20Synthetic}; the terminology seems to have originated in Ref. \cite{Baez14}), and referred to as the \emph{category of finite  sets and stochastic maps between them}. The name and notation chosen in this thesis is meant to reflect the emphasis on the category as a \myuline{theory} of \myuline{classical} \myuline{information}, and to reinforce the physical and formal relationship with the theory $\QIT$. (The category $\QIT$ is rarely named in the literature.)
\end{Remark}

At this point, we shall not entertain any physical interpretations whatsoever of the theory $\QIT$. (Realistically, most readers of these sentences will know of such an interpretation anyway.) Suffice it to say that in the same way that $\CIT$ models the  probabilistic processing of classical information with which most of us are at least intuitively familiar, it has been determined, ultimately empirically, that $\QIT$ is the correct model for the processing of \myuline{quantum} information (\cite{NC02}), and hence for information processing as it really is in our world (to the best of our understanding). \\

That said, three points about $\QIT$ deserve mentioning before we continue with the investigation of general theories:

\begin{Remark} (A Formal Relationship between $\CIT$ and $\QIT$.) \label{rem:QITCIT}\\
First, it should be pointed out that the theory $\CIT$ is naturally contained in the theory $\QIT$, by means of the following construction: To a system $X$ in $\CIT$ we associate the Hilbert space $\Gamma(X) := \C^X$ (with its canonical inner product) in $\QIT$, and to a morphism $T=(t_x)_{x \in X}:X \to Y$ in $\CIT$ we associate the CPTP map $\Gamma(T): \Gamma(X) \to \Gamma(Y)$ given by $\Gamma(T)(A) = \sum_{x \in X, y \in Y} t_x(y) \bra{x} A \ket{x} \ketbra{y}$ for $A \in \End{\C^X}$, such that in particular $\Gamma(T)(\ketbra{x}) = \sum_{ y \in Y} t_x(y) \ketbra{y}$. The precise sense in which this construction gives a representation of $\CIT$ in $\QIT$ can be summarised by the observation that $\Gamma$ has all the properties of a (strong) monoidal functor (\cite{MacLane}) from $\CIT$ to $\QIT$, \myuline{except} that $\Gamma$ does not preserve identities, i.e. $\Gamma(\id_X) \neq \id_{\Gamma(X)}$. Explicitly, $\Gamma(S \circ T )= \Gamma(S) \circ \Gamma(T)$, $\Gamma(X \times Y) \cong \Gamma(X) \otimes \Gamma(Y)$ and $\Gamma(T_1 \times T_2) \cong \Gamma(T_1) \otimes \Gamma(T_2)$.\footnote{An earlier version of this chapter contained an entire section proposing this notion of \emph{homomorphism} between theories, defined as maps satisfying all properties of (strong) monoidal functors, except preservation of identities. An injective such homomorphism, like $\Gamma$, can be interpreted as a \emph{generalisation} from one theory to another.} Moreover, $\Gamma$ is injective. It is interesting to observe (and seems to have been not noted before), that the failure of $\Gamma$ to preserve identities cannot be fixed by a redefinition -- there simply does not exist a strong monoidal functor from $\CIT$ to $\QIT$ which is injective. In succinct terms, the reason for this, which we will come to appreciate in \cref{chap:Dilations}, is that the classical channel $\up{Cop}: X \to X \times X$ which deterministically copies the input (i.e. corresponds to the function $x \mapsto (x,x)$), would under such a functor $\Phi$ have to map to a quantum channel $\Phi(\up{Cop}) : \Phi(X) \to \Phi(X) \otimes \Phi(X)$, both of whose marginals are $\id_{\Phi(X)}$, which is by the No Broadcasting Theorem (\cite{Barn96} -- see also \cref{cor:NoBroadCast}) impossible, except when $X \cong \triv$. \end{Remark}

\begin{Remark} (Notions of Classicality.)\\
Secondly, the above-mentioned embedding $\Gamma$ of $\CIT$ in $\QIT$ allows us to define notions of `classicality' in quantum information theory. This was already reviewed in the preliminary section of the thesis, but can now be rephrased in terms of the embedding $\Gamma$. 

Specifically, we call a system in $\QIT$ \emph{classical} if it is of the form $\Gamma(X)=\C^X$ for some set $X$ (any system in $\QIT$ is \myuline{isomorphic} to $\C^X$ for some set $X$; being classical is thus a matter of being equipped with a preferred basis). As already observed, the embedded identities $\Gamma(\id_X)$ are distinct from the actual identities $\id_{\Gamma(X)}$ on $\C^X$, indeed $\Gamma(\id_X)$ is the \emph{decoherence channel on $X$}, namely the quantum channel $\Delta_X : \C^X \to \C^X$ given by $\Delta_X(A)= \sum_{x \in X} \bra{x}A \ket{x} \ketbra{x}$. The fact that $\Delta_X = \Gamma(\id_X)$ is distinct from $\id_{\Gamma(X)}$ means that the e.g. the condition $\Gamma(\id_Y) \circ M= M$ for a quantum channel $M : \cH \to \C^Y$ is non-trivial, and it makes sense in this case to say that $M$ \emph{has classical outcomes}. The more well-known term for this concept is that $M$ is a \emph{measurement}. There is similarly a notion of a channel $\Lambda : \C^X \to \cK$ having \emph{classical inputs}, which corresponds to being an ensemble $(\varrho_x)_{x \in X}$ of channels $\triv \to \cK$ (so-called \emph{states}). As mentioned in the preliminary section, if $\Lambda$ is a quantum channel between composite systems it makes sense to speak of $\Lambda$ being classical on \myuline{some} of the input or output systems, thus giving rise most generally to ensembles of so-called \emph{quantum instruments}. 

It is easy to very that the embeddings of classical channels, $\Gamma(T)$, are precisely those channels $\Lambda : \C^X \to \C^Y$ which satisfy $\Delta_Y \circ \Lambda \circ \Delta_X = \Lambda$. However, certain \myuline{compositions} of transformations in $\QIT$ could result in a transformation interpretable in $\CIT$ even though its constituents are not. For example, for serially composable transformations $\Lambda_1$ and $\Lambda_2$ in $\QIT$, the transformation $\Lambda_2 \circ \Lambda_1$ might be classical (i.e. of the form $\Gamma(T)$) even though neither $\Lambda_1$ nor $\Lambda_2$ is classical. Abstractly, this is what allows us in the first place to make statements about quantum information which are classically intelligible. By considering a more intricate combination of transformations, one can exhibit a total transformation which is classical, although \myuline{no} \myuline{classical} \myuline{choice} \myuline{of} \myuline{the} \myuline{constituents} will reproduce this transformation. This statement is essentially the famous observation of John Bell (\cite{Bell64}) described in the introduction of the thesis, and the consequences are profound: \emph{$\QIT$ is a larger theory than $\CIT$, and this can be classically observed.} 

\end{Remark}
 
 \begin{Remark} (On the Definition of $\QIT$.) \\
Finally, it is very important to appreciate the fact that though $\QIT$ was defined in \cref{ex:QIT} in terms of Hilbert spaces and linear operators, there could very well be ways of defining $\QIT$ (up to a suitable notion isomorphism) \emph{without making any reference to such entities}. As demonstrated by ground-breaking works such as \cite{Hard01} and later \cite{Chir11}, there \myuline{are} indeed completely different definitions of $\QIT$, which in their formulation are much less obscure, cast in an operational language. In fact, the quest for simple and natural definitions of $\QIT$ is an ongoing area of research (see Ref. \cite{Foils} for a review), and in many ways the question that motivated the present thesis -- that of finding an operational definition of quantum self-testing -- is very much inspired by this line of thought. 

\end{Remark}

 \subsection{States, Isomorphisms and Reversibles}
 \label{subsec:Basic}
 
 We now proceed to discuss special kinds of transformations in a given theory: \emph{States}, \emph{isomorphisms} and \emph{reversibles}. The naming of states is uncontroversial, whereas there is no consensus on the naming of the latter two (see also \cref{rem:Rev}). \\
 
 First, however, let us prove the following helpful result about trashes, which is used over and over throughout the thesis: %

\begin{Lem} (Properties of Trashes.) \label{lem:Trashes}\\
	The following holds of the trashes in a theory $\Theory$:
	
	\begin{enumerate}
	
		\item For any transformation $T: \cX \to \cY$ in $\Theory$, $\tr_\cY \after T = \tr_\cX$.
		\item For any systems $\cX, \cY$ in $\Theory$, $\tr_{\cX \og \cY} = \tr_\cX \og \tr_\cY$.
			\item $\tr_\triv = \id_\triv$.
	
		\end{enumerate}
	\end{Lem}

\begin{proof} Given $T: \cX \to \cY$, the transformation $\tr_\cY \after T$ is \emph{some} transformation from $\cX$ to $\triv$. Since there is only one, namely $\tr_\cX$, we must have $\tr_\cY \after T = \tr_\cX$, proving the first property. The second and third are proved similarly.  \end{proof}

A theory $\Theory$ has a single system which is distinguished, namely the trivial system $\triv$ -- in fact, $\triv$ may be the only system in $\Theory$. The system $\triv$ represents `nothing', and whereas we have imposed that transformations to $\triv$ are not very diverse (there is only one from each system), transformations \myuline{from} the system $\triv$ are very colourful. They physically correspond to producing something from nothing: 

\begin{Definition} (States in $\Theory$.) \label{def:state}\\
	Given a system $\cX$ in $\Theory$, the transformations from $\triv$ to $\cX$ are called \emph{states on $\cX$}. The class of all states on $\cX$ is denoted by $\St{\cX}$.
	\end{Definition}

We generically denote states by lowercase Latin letters $s, t,  \ldots$.

\begin{Example} 	The states in a theory are usually rather easy to understand:
\begin{itemize}
	\item In $\Sets^*$, a state on $X$ is a map $s: \triv \to X$, or, what is equivalent, an element $x \in X$. As such, some systems have many states and some have few. (If we consider instead of $\Sets^*$ the theory $\Sets$ in which also the empty set is included, then some systems -- namely $\emptyset$ -- has \emph{no} states.)
	
	\item In $\Groups$, a state on $G$ is a homomorphism $\sigma: \triv \to G$. There is only one such, since it must map to the identity element in $G$.
	
	\item In $\CIT$, a state on $X$ is a classical channel $\triv \to X$, or, what is equivalent, a probability distribution $p$ on $X$.
	
	\item In $\QIT$, a state on $\cH$ is a completely positive trace-preserving linear map from $\End{\C}$ to $\End{\cH}$. Since $\End{\C}\cong \C$ as vector spaces, such a map is characterised by a unique element $\varrho \in \End{\cH}$, namely the image of $1 \in \C$, and by the CPTP property this element must be positive and of unit trace. Conversely, for any positive $\varrho \in \End{\cH}$ with $\tr(\varrho)=1$ the map $\C \ni a \mapsto a \varrho \in \End{\cH}$ is CPTP and hence defines a state on $\cH$. In conclusion, we may identify the set of states on $\cH$ with the set of \emph{density matrices on $\cH$}, namely 
	
	\begin{align}
	\scrD(\cH) := \{\varrho \in \End{\cH} \mid \varrho \geq 0, \tr(\varrho)=1\}.
	\end{align}

		\item In any theory $\Theory$, there is by assumption a unique map from $\triv$ to $\triv$, namely $\tr_\triv (=\id_\triv) $, and this is a state on $\triv$. It is possible that no other systems in $\Theory$ have states.
	\end{itemize}
	\end{Example}

\begin{Example} (States from States.) \label{ex:StatesfromStates}\\
The following hold in any theory:

\begin{itemize}
	\item Since $\triv \og \triv = \triv$, the parallel composition of states $s\in \St{\cX}$ and $t\in \St{\cY}$ is a new state $s\og t : \triv \to \cX \og \cY$, on the system $\cX \og \cY$.  
	
	Succinctly, we have a map $\St{\cX} \times \St{\cY} \to \St{\cX \og \cY}$ given by $(s,t) \mapsto s \og t$.  
	
	\item If $s \in \St{\cX}$ and $T: \cX \to \cY$ is a transformation, then the serial composition $T \after s : \triv \to \cY$ is a state on $\cY$.
	
	In other words, any transformation $T: \cX \to \cY$ induces a map $\St{\cX} \to \St{\cY}$ given by $s \mapsto T \after s$. 
	\end{itemize}
	\end{Example}

\begin{Warning} (Transforming States.) \label{rem:StateDet}\\
From common use of the words, it is tempting to think that a transformation is determined by its action on states, i.e. that the abstract morphism $T: \cX \to \cY$ in $\Theory$ can be identified with the set-theoretic function $s \mapsto T \after s$ from $\St{\cX}$ to $\St{\cY}$. This identification can be done without harm in $\CIT$ and $\QIT$ (though in $\QIT$ it relies on the non-trivial fact that any operator on $\cH$ is a linear combination of density matrices). It is also unproblematic in the theory $\Sets^*$, where in fact it is subtle to even distinguish the original from the impersonator, since $\St{X} \cong X$. The principle that transformations be determined by their action on states is physically sound (what sense is there in two transformations being distinct if this cannot be observed on states?), and it is enforced in much existing literature, essentially by identifying transformations which act identically on states (e.g. as in Ref. \cite{Chir10}). Nevertheless, it is not a principle  we shall commit to. In fact, it may very well fail in more mathematical examples of theories, such as $\Groups$, where each system has only one state, whence any two transformations from $G$ to $H$ act identically on states. 
\end{Warning}

Leaving states for now, we proceed to two other important types of transformations: 

\begin{Definition} (Reversibles and Isomorphisms in $\Theory$.) \label{def:RevIso}
	
	\begin{itemize}
		\item A transformation $R: \cX\to \cY$ is called \emph{reversible} if it has a left-inverse, i.e. if there exists a transformation $R^{-} : \cY\to \cX$ such that $R^{-} \after R = \id_\cX$. 
		\item A transformation $\alpha: \cX \to \cY$ is called an \emph{isomorphism} if it has a two-sided inverse, i.e. if there exists a transformation $\beta: \cY \to \cX$ such that $\beta \after \alpha = \id_\cX$ and $\alpha \after \beta = \id_\cY$. 
		
	\end{itemize}
	
\end{Definition}

	\begin{Remark} (Terminology.) \label{rem:Rev}\\
Refs. \cite{Chir10, Chir11} use the term `reversible' differently than we -- in fact, it is used even within these references in two different ways, cf. Definitions 13 and 46 in \cite{Chir10}. One of the uses (Def. 13) is for what we call `isomorphisms'. The terminology chosen in the \cref{def:RevIso} is based on the grounds that (1) the term `isomorphism' has been established in the mathematical literature on categories for more than half a century; (2) we will need \myuline{some} word for transformations with left-inverses;\footnote{In category theory proper, there is in fact a term for morphisms with left-inverses, namely \emph{split monomorphisms}; however, I render this type of vocabulary slightly too esoteric for our purposes.} (3) the Latin verb \emph{revertere} means to \emph{turn back}.
\end{Remark}

\begin{Example} (Isomorphisms and Reversibles.)\label{ex:IsRev} 
	\begin{itemize}
		
		\item In $\Sets^*$, a map $f: X \to Y$ is an isomorphism if and only if it is bijective. It is reversible if and only if it is injective.
		
		\item In $\Groups$, the isomorphisms are precisely the group isomorphisms. Any reversible transformation is an injective homomorphism (though some of these are not reversible).

		\item In $\QIT$, every transformation $\Lambda: \cH \to \cK$ which is a unitary conjugation, $A \mapsto UAU^*$, is an isomorphism, with two-sided inverse given by $B \mapsto U^*BU$. It is not obvious that any isomorphism in $\QIT$ must take this form, but we shall later see a swift argument for this (\cref{cor:IsosinQIT}) using the machinery of \cref{chap:Dilations}. 
		
		We shall similarly be able to characterise all reversible transformations in $\QIT$ (\cref{cor:RevinQIT}). For now, let us observe that if $\Sigma : \cH \to \cK$ is an isometric conjugation, $A \mapsto SAS^*$, then it is reversible. It is tempting to provide as left-inverse the completely positive map $B \mapsto S^* B S$, but it is not trace-preserving since $\tr(S^*AS)= \tr(SS^*A)$ and $SS^* \lneq \bone$ (unless $S$ is unitary). Instead, pick an isometry $S': \cK \to \cH \otimes \cE$ such that $S' S = \bone_\cH \otimes \ket{\psi}$, where $\ket{\psi}$ is a unit vector in some space $\cE$; then the map $B \mapsto [\id_{\End{\cH}} \otimes \tr_\cE] (S' B {S'}^*)$ is a completely positive trace-preserving left-inverse. 
	\end{itemize}	\end{Example}

 	It is well-known (and easy to show) that if $\alpha: \cX \to \cY$ is an isomorphism, its two-sided inverse $\beta$ is unique. We denote it, as is customary, by $\alpha^{-1}$. Clearly, $\alpha^{-1}: \cY \to \cX$ is an isomorphism with inverse $(\alpha^{-1})^{-1}= \alpha$. If $\alpha: \cX \to \cY$ and $\beta : \cY \to \cZ$ are isomorphisms, then their serial composition is an isomorphism too, with $(\alpha \after \beta)^{-1}= \beta^{-1} \after \alpha^{-1}$. Similarly, if $\alpha_1 : \cX_1 \to \cY_1$ and $\alpha_2: \cX_2 \to \cY_2$ are isomorphisms, their parallel composition is an isomorphism with $(\alpha_1 \og \alpha_2)^{-1}  = \alpha^{-1}_1 \og \alpha^{-1}_2$. 
 	
 	It is customary to call systems $\cX, \cY$ \emph{isomorphic} if there exists an isomorphism between them. For example, systems $\cH$ and $\cK$ are isomorphic in $\QIT$ precisely when they have the same dimension, and systems $X$ and $Y$ are isomorphic in $\CIT$ precisely when they have the same number of elements. By the previous comments, isomorphism of systems is always an equivalence relation, well-behaved under parallel composition of systems. \\
 	
	Evidently, every isomorphism in a theory $\Theory$ is reversible. As \cref{ex:IsRev} shows, there may however easily be reversible transformations in $\Theory$ which are not isomorphisms. It should also be observed that, contrary to isomorphisms, reversibles need not have unique inverses:

\begin{Remark} (Non-Uniqueness of Left-Inverses.) \\
	In $\Sets^*$, the injective inclusion map $\iota: \{0,1\} \to \{0,1, \ldots, 8,9\}$ given by $\iota(x)=x$ has as left-inverse any function $g: \{0,1, \ldots, 8,9\} \to \{0,1\}$ for which $g(0)=0$ and $g(1)=1$. As such, the values $g(y)$ for $y \in \{2,3, \ldots, 8,9\}$ can be set arbitrarily as $0$ or $1$, so there are as many such maps $g$ as there are subsets of the set $\{2,3, \ldots, 8,9\}$, namely $2^8=256$.
	\end{Remark}

Let me end this section by posing a curious problem. \\

The reversible transformations in a theory $\Theory$ facilitate a rudimentary notion of dimension.  More precisely, let us define the \emph{dimensional ordering}, $\preceq$, on $\Sys{\Theory}$ by declaring that $\cX \preceq \cY$ exactly if there exists a reversible transformation $R: \cX \to \cY$. It is easy to see that $\preceq$ is a pre-order, i.e. a reflexive transitive relation.  (We also have $\cX_1 \otimes \cX_2 \preceq \cY_1 \otimes \cY_2$ if $\cX_1 \preceq \cY_1$ and $\cX_2 \preceq \cY_2$, and we have $\triv \preceq \cX$ if $\St{\cX} \neq \emptyset$.)  

In $\Sets^*$ and $\CIT$, the dimensional ordering reproduces the ordering in terms of cardinality of sets ($X \preceq Y \Leftrightarrow \abs{X} \leq \abs{Y}$), and in $\QIT$ it yields the usual ordering according to dimension ($\cH\preceq \cK \Leftrightarrow \dim \cH \leq \dim \cK$).  In general, however, the ordering $\preceq$ is simply an abstract relation, not necessarily related to cardinal numbers or enjoying properties we usually expect.

 One such property would be the  \emph{Cantor-Schr\"{o}der-Bernstein property} (named by analogy with the Cantor-Schr\"{o}der-Bernstein theorem for sets \cite{SchroderBernstein}), i.e. the principle that if $\cX \preceq \cY$ and $\cY \preceq \cX$ then $\cX$ and $\cY$ are isomorphic.  By \cref{ex:Graphs}, however, it is demonstrated that this property is not always satisfied. 
 
 Another expectable property would be 
 
 \begin{itemize}
 		\item \emph{Linearity:} For all systems $\cX, \cY$, either $\cX \preceq \cY$ or $\cY \preceq \cX$,
 	\end{itemize}
 
 or the stronger property 
 
  \begin{itemize}
 	\item\emph{Well-Foundedness:} For any set of systems $(\cX_i)_{i \in I}$, there is some $i_0 \in I$ such that $\cX_{i_0} \preceq \cX_i$ for all $i \in I$.
 \end{itemize}

Readers acquainted with the theory of ordinals and cardinals may appreciate that well-foundedness is more or less equivalent to the possibility of representing the levels of $\preceq$ using cardinals, as we can for $\Sets^*$ (where the equivalence class of $X$ is represented by the cardinality $\abs{X}$) and for $\QIT$ (where the equivalence class of $\cH$ is represented by the cardinal $\dim \cH\in \N$).\footnote{If such a representation is possible, well-foundedness of $\preceq$ follows from well-foundedness of cardinals. If on the other hand $\preceq$ is well-founded (and the collection of its equivalence classes is small enough to be a set) then by a standard result (\cite{Kunen80}) that set is order-isomorphic to an ordinal $\gamma$. As such, the levels of $\preceq$ are representable as an ordering among ordinals $\beta \in \gamma$, and from this we can obtain an ordering in terms of cardinals by means of the cardinal counting map $\beta \mapsto \kappa_\beta$ from ordinals to cardinals.}

I do not know if it is possible to construct an example in which linearity or well-foundedness fail. In fact, I do not even know the answer to the following question:

\begin{OP} 
Is every pre-order the dimensional ordering of some theory $\Theory$?
	\end{OP}

\section{A Reservoir of Examples}
\label{sec:Examples}

Examples in mathematics serve roughly two purposes, one soft and one hard. 

The soft purpose is that \emph{examples help humans fix ideas}. For instance, a person seeing the definition of a topological space for the first time may not immediately grasp what this concept is about. Exhibiting concrete examples will help that person form a view of what a topological space is; some of these examples will fit smoothly in line with those that motivated the definition in the first place, whereas others may be surprising. %

The hard purpose is that \emph{examples uncover formal interdependencies of properties}. Some examples of topological spaces will show that certain properties cannot be derived -- or are undecidable -- from the axioms defining a topological space. For instance, one cannot prove that a topological space has infinitely many open sets (for this is not always true, as e.g. the trivial topology exemplifies). In a similar vein, that a set is closed does not imply that its image under a continuous map is closed (as exemplified by the map $\R \ni x \mapsto \frac{1}{1+x^2} \in \R$, which maps $\R$ to $(0,1]$).\\

 In this section we go through a lot of examples of theories. In fact, the presented catalogue might be one the largest list of theories (in the sense of \cref{def:Theory}) existing at one place in the literature. Some of these examples serve the soft purpose, but most will the hard as well. Some of the examples will be so mathematical that a physicist would not call them `theories' (like $\Groups$, cf. \cref{rem:StateDet}). Still, such examples may easily serve the hard purpose, demonstrating that defined concepts do not always behave as expected.

\subsection{Variations of $\CIT$ and $\QIT$}

We have already seen our two most important examples of theories, $\CIT$ (\cref{ex:CIT}) and $\QIT$ (\cref{ex:QIT}). Historically, much of quantum theory was conceived in a setting of infinite-dimensional Hilbert spaces, and emphasis on the finite-dimensional setting was only recently articulated (\cite{NC02,Foils}). 

There is indeed version of quantum information theory which allows (separable) infinite-dimensional Hilbert spaces as systems (\cite{Attal14, Wolf19}):

\begin{Example} ($\QIT^\infty$.) \\
In the infinitary version of quantum information theory, $\QIT^\infty$, systems are separable Hilbert spaces $\cH, \cK, \cL, \ldots$, and a transformation from $\cH$ to $\cK$ is a CPTP map $\Lambda: B_1(\cH) \to B_1(\cK)$, as outlined in the preliminary section of the thesis. (Recall that $B_1(\cL)$ denotes the Banach space of trace-class operators on the Hilbert space $\cL$.) When $\cH$ and $\cK$ are finite-dimensional, this notion of transformation restricts to that from $\QIT$. The composite of systems in $\QIT^\infty$ is again given by the tensor product (whose construction now requires a metric completion of the algebraic tensor product), and the trivial system is $\triv = \C$. Serial and parallel compositions of transformations are given, as for $\QIT$, by the functional composition and tensor product of linear maps, respectively. States on a system $\cX$ are (by the same argument used for $\QIT$) in natural bijective correspondence with linear operators $\varrho \in B_1(\cX)$, which are positive and of unit trace. 
\end{Example}

The theory $\QIT^\infty$ does display features which $\QIT$ does not, but they also have a lot in common and for of our purposes their differences are not profound. (One of the differences is that in $\QIT^\infty$ there exists a system $\cH$, namely any space of infinite dimension, into which all systems admit a reversible transformation; in $\QIT$ there exists no such system.) \\

There is also a version of $\CIT$ which goes beyond finite sets:

\begin{Example} ($\mathbf{Stoch}$.) \\
The theory $\mathbf{Stoch}$ (following the notation of Ref. \cite{Fritz20Synthetic}) has measurable spaces $\cX = (X, \bbE)$ for systems, and  transformations from $\cX = (X, \bbE)$ to $\cY = (Y, \bbK)$ are \emph{Markov kernels from $\cX$ to $\cY$}, that is, $X$-indexed collections of probability measures on $\cY$, $(\lambda_x)_{x \in X}$, for which the function $x \mapsto \lambda_x(B)$ is measurable for any fixed $B \in \bbE$. The composite of systems $\cX= (X, \bbE)$ and $\cY=(Y, \bbK)$ is the measurable  spaces $(X \times Y, \bbE \otimes \bbK)$, where $\bbE \otimes \bbK$ is the product $\sigma$-algebra, and the parallel composition of transformations is defined in the obvious way by forming product measures (see any introductory book on measure theory, e.g. \cite{Hansen06, Schill17, Foll99}). The serial composition is also defined in a rather obvious fashion, by \emph{integrating} the Markov kernels, though the construction is somewhat shrouded in measurability technicalities. (For more details, see Ref. \cite{Fritz20Synthetic}, or the curiously historical lecture notes \cite{Law62} which apparently constitute the first categorical presentation of Markov kernels.) 

Every measurable injection $(X, \bbE) \to (Y, \bbK)$ is a reversible transformation in $\mathbf{Stoch}$, its isomorphisms are precisely the Borel-isomorphisms, and states on $(X, \bbE)$ correspond to probability measures on $(X, \bbE)$. 
\end{Example}

It is intuitively clear that the theory $\QIT^\infty$ extends $\QIT$, and that $\mathbf{Stoch}$ extends $\CIT$ (we will be more precise about this in \cref{subsec:Subtheories}). However, it would seem that $\mathbf{Stoch}$ is in a sense `too big' an extension of $\CIT$ when compared to the extension $\QIT^\infty$ of $\QIT$. Indeed, as mentioned earlier, the theory $\CIT$ embeds into $\QIT$, but there is no obvious sense in which $\mathbf{Stoch}$ embeds into $\QIT^\infty$, since there is no canonical way of associating a Hilbert space to a measurable space $(X, \bbE)$. One could speculate that by taking \myuline{measure} spaces $(X, \bbE, \mu)$ in place of measurable spaces $(X, \bbE)$ as systems (and by requiring a sufficient compatibility of the transformations, e.g. having appropriate densities w.r.t. the ground measures), one could massage $\mathbf{Stoch}$ into a theory for which such an embedding would be possible, but I do not know of any such construction. \\

So far, all of our theories have had plenty of states, with the exception of $\Groups$. It is possible to device an `information' theory which also does not have many states:

\begin{Example} (Oblivious Information Theory.) \label{ex:Oblivious}\\
	In $\QIT$, every system $\cH$ has a unique \emph{invariant state}, $\tau_\cH$, described by the density matrix $\frac{1}{\dim\cH} \bone_\cH$, and defined by the property that $\alpha \circ \tau_\cH = \tau_\cH$ for all isomorphisms (unitary conjugations) $\alpha : \cH \to \cH$. The states $\tau_\cH$ are also called \emph{fully mixed}, and they are can be interpreted as representing complete obliviousness about the system $\cH$. Let us define \emph{oblivious quantum information theory}, $\OblQIT$, as the theory in which this obliviousness is preserved:   The systems of $\OblQIT$ are finite-dimensional non-zero Hilbert spaces, and the transformations from $\cH$ to $\cK$ are the CPTP maps $\Lambda: \End{\cH} \to \End{\cK}$ for which $\Lambda(\tau_\cH)= \tau_\cK$. The composition of systems and transformations is given as in $\QIT$, and the trivial system is again $\triv := \C$. Importantly, the unique state $\tr_\triv = \id_\triv$ on the system $\triv$ is invariant, and therefore all trashes $\tr_\cH$ in $\QIT$ are valid transformations in $\OblQIT$, so $\triv$ is indeed terminal. Note also that any state $\sigma: \triv \to \cH$ in $\OblQIT$ must be invariant, i.e. the system $\cH$ admits a unique state, namely $\tau_\cH$. 

Of course, there is nothing quantum about this idea: Classical information theory, $\CIT$, also has unique invariant states (namely the uniform distributions) and by restricting to transformations that map uniform distributions to uniform distributions we similarly obtain an oblivious classical information theory, $\OblCIT$. An interesting exercise for the reader is to verify that, contrary to the what is the case in $\CIT$, every reversible transformation in $\OblCIT$ is an isomorphism.  
\end{Example}

Rather than restricting the class of transformations, we can enlarge it:

\begin{Example} (Negative Information Theory.) \\
Let us define $\NCIT$  (`negative classical information theory') as the theory whose systems and composition of systems is the same as in $\CIT$, but whose transformations from $X$ to $Y$ are collections $t= (t_x)_{x \in X}$ of `not necessarily positive probability distributions on $Y$' , that is, of functions $t_x: Y \to \R$ such that 

\begin{align} \label{eq:unity}
\sum_{y \in Y} t_x(y)=1.
\end{align}

 The potential usefulness of negative probabilities has been discussed in e.g. Ref. \cite{Feyn87}. We define serial and parallel composition in $\NCIT$ by the same equations as for $\CIT$. 
  Importantly, the system $\triv = \{0\}$ remains terminal in $\NCIT$ because $\delta_0$ is the only $\R$-valued function on $\{0\}$ satisfying the normalisation \eqref{eq:unity}.

Certainly, the theory $\NCIT$ seems quite distinct from $\CIT$, e.g. in admitting on the system $\{0,1\}$ the `unbounded' set of states $\{t \delta_0 + (1-t) \delta_1 \mid t \in \R\}$, and in admitting `convex combinations' such as $\delta_0 = \frac{1}{2}(3 \delta_0 -2 \delta_1)+ \frac{1}{2}(- \delta_0 + 2 \delta_1)$. But to prove a categorical distinction from the theory $\CIT$ we cannot refer to notions of convexity or boundedness; we have to point out a distinction visible in terms of the serial and parallel composition. 
To this end, consider in $\NCIT$ the transformations $\alpha^w  = (\alpha^w_j)_{j \in \{0,1\}}: \{0,1\}\to \{0,1\}$ given for $w \in \R$ by 
 
 \begin{align}
 \alpha^w_0 = (1-w) \delta_0 + w \delta_1,  \quad \alpha^w_1 = -w \delta_0 + (1+w) \delta_1.
 \end{align}

It is easily verified that $\alpha^0 = \id_{\{0,1\}}$ and that $\alpha^u \circ \alpha^v = \alpha^{u+v}$ for  all $u,v \in \R$, so the map $w \mapsto \alpha_w$ is an injective group homomorphism from the additive group of reals to the group of automorphisms of $\{0,1\}$ (i.e. isomorphisms $\{0,1\}\to \{0,1\}$) in $\NCIT$. In particular, any automorphism $\alpha^w$ with $w \neq 0$ has infinite multiplicative group order; in $\CIT$, on the other hand, every automorphism of every system is a bijection on a finite set and hence has finite multiplicative order.
\end{Example} 

\subsection{Cartesian Theories}
\label{subsec:Cart}

The theory $\Sets^*$ (\cref{ex:Sets}) has the following feature: Any transformation into a composite system, say $f:X \to Y_1 \times Y_2$, is given by two components, $f= (f_1, f_2)$ with $f_1: X \to Y_1$ and $f_2: X \to Y_2$. These two component functions are the \emph{marginals} of $f$, and they completely determine $f$. Similarly, a transformation $\varphi: G \to H_1 \times H_2$ in $\Groups$ is determined by its marginals. 

On the other hand, transformations in $\CIT$ (or $\QIT$) are \myuline{not} determined by marginals; for examples, the states $k:= \frac{1}{2} \delta_0 \otimes \delta_0 +\frac{1}{2} \delta_1 \otimes \delta_1 $ and $p:= \left(\frac{1}{2} \delta_0 + \frac{1}{2} \delta_1 \right) \otimes \left(\frac{1}{2} \delta_0 + \frac{1}{2} \delta_1 \right)$ on the system $\{0,1\} \times \{0,1\}$ have identical marginals, but they are not the same; the state $k$ represents two copies of a uniformly random bit, whereas $p$ represents two independent uniformly random bits.

Let us be slightly more precise:

\begin{Definition} (Marginal-Determined.)\\
	Let $\cY_1$ and $\cY_2$ be systems in a theory $\Theory$. Let $\pi_1 : \cY_1 \og \cY_2 \to \cY_1$ and $\pi_2:  \cY_1 \og \cY_2 \to \cY_2$ denote the \emph{factor projections}, $\id_{\cY_1} \og \tr_{\cY_2}$ and $ \tr_{\cY_1} \og \id_{\cY_2}$, respectively. We say that the system pair $(\cY_1, \cY_2)$ is \emph{marginal-determined}, if for any two transformations $T_1: \cX \to \cY_1$ and $T_2: \cX \to \cY_2$, there exists a unique transformation $T : \cX \to \cY_1 \og \cY_2$ such that $\pi_1 \after T= T_1$ and $\pi_2 \after T= T_2$.

	\end{Definition}

\begin{Definition}(Cartesian Theories.) \\
		A theory $\Theory$ is called \emph{cartesian} if any pair of systems in $\Theory$ is marginal-determined.
		\end{Definition}

The introductory lines serve to illustrate that $\Sets^*$ and $\Groups$ are cartesian theories, whereas $\CIT$ and $\QIT$ are not. Readers acquainted with category theory will realise that cartesian theories are precisely \emph{categories with finite products}  (\cite{MacLane,Awo10}), sometimes referred to as \emph{cartesian} categories (hence the name). This realisation immediately gives a true bombardment of theory examples, including all sorts of categories whose morphisms are functions on `structured sets' which admit a notion of product:

\begin{itemize}
	\item $\Top^*$, in which the systems are non-empty topological spaces and the transformations continuous maps;
	
	\item $\mathbf{Rings}$, in which the systems are algebraic rings and the transformations are ring homomorphisms;
	
	\item $\mathbf{Man}^\infty$, in which the systems are differentiable manifolds and the transformations are smooth maps;  
	
	\item $\vdots$
	
\end{itemize}

In all these cases, the serial composition of transformations is given by ordinary functional composition and the parallel composition is given by means of the products that these categories facilitate (products of topological spaces, of rings, of manifolds, ...). 

Another example of this is the theory $\Graphs$ of graphs and homomorphisms. However, in the same way that the empty set in $\Sets$ introduces some pathological features making the theory $\Sets^*$ nicer in the end, so do graphs with un-looped vertices cause problems in $\Graphs$ (see \cref{subsec:Normal}). Thus, we shall consider instead the theory $\Graphs^*$, whose objects are (non-empty) graphs in which every vertex has a loop: 
\begin{Example} ($\Graphs^*$.) \label{ex:Graphs} \\
	The category $\Graphs^*$ has non-empty looped graphs for objects and graph homomorphisms for morphisms, with functional composition as serial composition. Any graph with one vertex and its loop is terminal; we fix one and call it $\triv$. One can consider several `products' of graphs $G$ and $H$ (see \cite{GraphProducts}), but only one kind will make $\Graphs^*$ a \myuline{cartesian} theory, and this is the so-called \emph{direct product}, $G \times H$. It has as vertices pairs of vertices in $G$ and $H$, and it has an edge between $(u_1,v_1)$ and $(u_2, v_2)$ precisely if $u_1$ and $u_2$ are adjacent in $G$ and $v_1$ and $v_2$ in $H$.\footnote{There is a distinct product, $G \square H$, which quite confusingly is called the \emph{cartesian product} of $G$ and $H$; it can also be seen as a \emph{funny tensor product} of categories (\cite{GraphProducts}). The operation $\square$ makes $\Graphs^*$ a symmetric monoidal category in a different way. Thus, $\Graphs^*$ can be considered a theory in two distinct ways: Either by equipping it with the direct product $\times$, or by equipping it with the product $\square$.} Transformations compose parallelly are one would expect. The isomorphisms in $\Graphs$ are precisely the graph isomorphisms, and a state on the graph $G$ is simply a vertex in $G$ (for this it matters that every vertex has a loop).

	The collection of reversible transformations allows us to exhibit a  feature which we have not encountered earlier. It can be phrased as the failure in $\Graphs^*$ of the Cantor-Schr\"{o}der-Bernstein property discussed in \cref{subsec:Basic}, and explicitly it is the following: There are systems $G,H$ with the property that we can find reversible transformations $g :G  \to H$ and $h:H \to G$, though there is no isomorphism between $G$ and $H$. It is an easy exercise to see that this can only be the case if $G$ and $H$ are infinite, but we may actually take them rather simple. Indeed, let $G$ be the graph\\

	\adjustbox{scale=0.6,center}{
	\begin{tikzcd}
	\bullet \arrow[out=150,in=90,"l_1",loop, dash] \arrow[rr, "c_0", dash, bend left] \arrow[rdd, "a_0", dash] & 	& 	\bullet  \arrow[out=90,in=30,loop, dash] \arrow[rr, "d_1", dash, bend right] & &  \bullet  \arrow[out=150,in=90,loop, dash] \arrow[rr, "c_1", dash, bend left] \arrow[rdd, "a_1", dash] & 	& 	\bullet  \arrow[out=90,in=30,loop, dash] \arrow[rr, dash, bend right] & &  \bullet \arrow[out=150,in=90,loop, dash] \arrow[rr, dash, bend left] \arrow[rdd, dash] & 	& 	\bullet  \arrow[out=90,in=30,loop, dash] \arrow[rr, dash, bend right] & &  \bullet \arrow[out=150,in=90,loop, dash] \arrow[rr, dash, bend left] \arrow[rdd, dash] & & \phantom{\bullet} & \cdots    \\ & & & & & & & & & & & & & & & \cdots\\
	 & 	\bullet \arrow[ruu, "b_0", dash] \arrow[out=220,in=180,loop, dash]& 	&&  & \bullet \arrow[ruu, "b_1", dash] \arrow[out=220,in=180,loop, dash]& 	& & & \bullet \arrow[ruu, , dash] \arrow[out=220,in=180,loop, dash]& 	&&  & \phantom{\bullet}  & &\cdots
	\end{tikzcd}  
}

and $H$ the graph\\

\adjustbox{scale=0.6,center}{
	\begin{tikzcd}
\bullet \arrow[rr, "d'_1", dash, bend right]\arrow[out=90,in=30,"r'_0", loop, dash] 	& & \bullet \arrow[out=150,in=90,loop, dash] \arrow[rr, "c'_1", dash, , bend left] \arrow[rdd, "a'_1", dash] & 	& 	\bullet  \arrow[out=90,in=30,loop, dash] \arrow[rr, dash, bend right] & &  \bullet  \arrow[out=150,in=90,loop, dash] \arrow[rr, dash, bend left] \arrow[rdd, dash] & 	& 	\bullet  \arrow[out=90,in=30,loop, dash] \arrow[rr, dash, bend right] & &  \bullet \arrow[out=150,in=90,loop, dash] \arrow[rr, dash, bend left] \arrow[rdd, dash] & 	& 	\bullet  \arrow[out=90,in=30,loop, dash] \arrow[rr, dash, bend right] & &  \bullet \arrow[out=150,in=90,loop, dash] \arrow[rr, dash, bend left] \arrow[rdd, dash] & & \phantom{\bullet} & \cdots  \\ & & & & & & & & & & & & & & &  & &\cdots\\
& & 	& 	\bullet \arrow[ruu, dash, "b'_1"]\arrow[out=220,in=180,loop, dash] & 	&&  & \bullet  \arrow[ruu,  dash] \arrow[out=220,in=180,loop, dash]& 	& & & \bullet \arrow[ruu,  dash] \arrow[out=220,in=180,loop, dash]& 	&&  & \phantom{\bullet}  & &\cdots
	\end{tikzcd}  ,
}

both extending infinitely to the right in a periodic fashion. For reversible transformation $g: G \to H$ we take the unique injective homomorphism that maps the leftmost `head', $a_0b_0c_0$, in $G$ to the leftmost head, $a'_1b'_1c'_1$, in $H$. (The image of $G$ under $g$ is all of $H$ except for the edges $r'_0$ and $d'_1$ and their common vertex.) Now, $g$ has as left-inverse the map $g^- : H \to G$ which acts as the inverse of $g$ on its image, and by collapsing $r'_0, d'_1$ and their common vertex to the left `ear' of the leftmost head in $G$. A reversible $h: H \to G$ is constructed completely analogously, by mapping the structure $d'_1a'_1b'_1c'_1$ to $d_1a_1b_1c_1$. The graphs $G$ and $H$ are not isomorphic, however, since $H$ has a loop (namely $l_1$) whose vertex only has one other edge.  	\end{Example}%

Another example of a cartesian theory whose transformations are functions on `structured sets' can be obtained from linear algebra. We saw in \cref{ex:VectTensor} that $\Vect{k}$, the category of vector spaces over $k$ and $k$-linear maps between them, can be augmented to a symmetric monoidal category by means of the tensor product, $\otimes$. We also discussed, however, that this does not constitute a theory in the sense of \cref{def:Theory}, since the monoidal unit $k$ fails to be terminal. It turns out that we can augment $\Vect{k}$ with another notion of parallel composition that does make it into a theory:

 \begin{Example} ($\Vect{k}$.) \label{ex:VectCart}\\
	Let $k$ be a field and consider on the category $\Vect{k}$ the symmetric monoidal structure defined by the \emph{direct sum}, $\oplus$; the composite of systems $V$ and $W$ is $V \oplus W$, and the parallel composition of $A_1 : V_1 \to W_1$ with $A_2: V_2 \to W_2$ is $A_1 \oplus A_2 : V_1 \oplus V_2 \to W_1 \oplus W_2$. For terminal object and $\oplus$-unit we fix a zero-dimensional space, $\triv := 0$. The trashes $\tr_V: V \to 0$ must then be the zero-maps, and the factor projections $\pi_j : V_1 \oplus V_2 \to V_j$ are consequently the ordinary projections $P_j$ onto the subspaces $V_j$. The theory is cartesian because any linear map $A: W \to V_1 \oplus V_2$ is specified by the projected maps $A_j := P_j A$ and because any such pair of maps $A_1, A_2$ defines a map to $V_1 \oplus V_2$ by $x \mapsto A_1 x \oplus A_2 x$.

 \end{Example} %

We end with an example demonstrating that the systems in a cartesian theory need not be `sets with structure':

 \begin{Example} (The Interval Theory.) \label{ex:Interval}\\
Consider the real unit interval $[0,1]$, and define a theory $\Theory$ as follows: 

Systems of $\Theory$ are numbers $x, y, z, \ldots \in [0,1]$. For any $x,y \in [0,1]$ there is at most one transformation from $x$ to $y$, and there is one precisely if $x \geq y$. (It does not matter what the transformation actually \emph{is}, but for concreteness we may choose is to be the pair $(x,y)$.) Serial composition of transformations can be defined uniquely, since $x \geq y$ and $y \geq z$ implies $x \geq z$, and each system has an identity transformation since $x \geq x$. The associative and symmetric composition of systems in $\Theory$ is given by $x \og y := \max\{x,y\}$, and the system $0 \in [0,1]$ is a unit for this operation which is terminal in $\Theory$ since $x \geq 0$ always. Parallel composition of transformations is also uniquely defined, by the observation that $x_1\geq y_1, x_2 \geq y_2$ imply $\max\{x_1, x_2 \}\geq \max\{y_1, y_2\}$. Finally, the theory is cartesian since $z \geq x_1$ and $z \geq x_2$ if and only if $z \geq \max\{x_1, x_2\}$.\end{Example}

\subsection{Thin Theories}
\label{subsec:Thin}

We just saw in \cref{ex:Interval} that there exist theories with at most one transformation from one system to another. We can make an entire example class out of such theories, and they turn out to have a fairly graspable characterisation. As actual physical theories tend to have many transformations, this class of theories is mostly interesting for purely mathematical purposes, or for finding counterexamples. 

In category theory, categories with at most one morphism from one object to another are called \emph{thin} (\cite{ThinCat}), so we adopt the same terminology: 

\begin{Definition} (Thin Theories.) \\
	A theory $\Theory$ is called \emph{thin} if for any systems $\cX, \cY \in \Sys{\Theory}$ there is at most one transformation from $\cX$ to $\cY$. 
\end{Definition}

Obviously, the composition of transformations -- serial and parallel -- in a thin theory is unexciting. Really, it is the composition of its \myuline{systems} which is interesting. 

In any theory $\Theory$, the composition $\og$ on systems gives $\Sys{\Theory}$ the structure of a \emph{monoid}, $(\Sys{\Theory}, \og, \triv)$, with unit object $\triv$. And in a thin theory, the transformation structure can be compactly summarised as follows: Let us define on $\Sys{\Theory}$ a relation $\succeq$ by $\cX \succeq \cY$ if and only if there is a transformation from $\cX$ to $\cY$. By the axioms of identities and serial composition, this relation is reflexive and transitive, i.e. it is a pre-order on $\Sys{\Theory}$. Terminality of $\triv$ means $\cX \succeq \triv$ for all $\cX\in \Sys{\Theory}$, and the parallel composition of transformations means that $\cX_1 \og \cX_2 \succeq \cY_1 \og \cY_2$ when $\cX_1 \succeq \cY_1$ and $\cX_2 \succeq \cY_2$. Finally, the symmetry condition of the theory implies that $\cX \og \cY$ and $\cY \og \cX$ are equivalent under the pre-order, i.e. $\cX \og \cY \succeq \cY \og \cX$ and $\cY \og \cX \succeq \cX \og \cY$.

Conversely, it is easy to see that any monoid equipped with a pre-order subject to these conditions defines a thin theory.\\

In summary, we have proved the following:

\begin{center}
\fbox{
\textit{A thin (strict) theory is the same thing as a \emph{pre-ordered quasi-commutative monoid}, }
}
	\end{center}

where by this horrifying sequence of words I mean a quadruple $(M, \star, 1, \succeq)$, such that 
	
	\begin{itemize}
		\item $(M, \star, 1)$ is a monoid;
		\item $\succeq$ is a pre-order on $M$;
		\item $x\succeq 1$ for all $x \in M$;
		\item $x_1 \succeq y_1, x_2 \succeq y_2 \Rightarrow x_1 \star x_2 \succeq y_1 \star y_2$ for all $x_1, x_2, y_1, y_2 \in M$;
		\item $x \star y \simeq y \star x$ for all $x,y \in M$.\footnote{Here, $z \simeq w$ means $z \succeq w$ and $w \succeq z$.}
		\end{itemize}
	
The identity transformations in $\Theory$ are the relationships $x \succeq x$, and the trashes are the relationships $x \succeq 1$. 
	
	\begin{Remark} (Strictness.) \\
If we drop the strictness assumption for the theory, the strict associativity of the operation $\star$ and strict unitality of the element $1$ are replaced by $\simeq$-equivalences in the monoid $(M, \star, 1)$ (e.g. $x \star 1 \simeq x \simeq 1 \star x$ rather than $x \star 1 = x = 1 \star x$). As such, the characterisation of thin theories by means of pre-ordered monoid-like structures is not contingent on the strictness.

\end{Remark}

Some of the simplest examples of thin theories are partially ordered commutative monoids: 

\begin{Example} (One Ordering, two Compositions.)\label{ex:TwoCompositions} \\
	The non-negative integers $\N_0 = \{0,1,2, \ldots\}$ form a commutative monoid with unit $0$, both when equipped with addition, $(n,m) \mapsto n+m$, and when equipped with the max-function, $(n,m) \mapsto \max\{n,m\}$. The usual ordering $\geq$ on $\N_0$ satisfies the required compatibility conditions with these binary operations, so we have two thin theories $(\N_0, +, 0, \geq)$ and $(\N_0, \max, 0, \geq)$. 
	
	\end{Example}

\begin{Example} (One Composition, two Orderings.) \label{ex:ThinNat}\\
The natural numbers, $\N= \{1,2,3, \ldots\}$ form a commutative monoid with unit $1$ when equipped with multiplication, $(n,m) \mapsto n \cdot m$. The usual ordering $\geq$ on $\N_0$ satisfies the required compatibility conditions, and so does the \emph{divisibility ordering}, $\geq_\up{div}$, according to which $n\geq_\up{div} m$ precisely if $m$ divides $n$. Thus we have two thin theories, $(\N, \cdot, 1, \geq)$ and $(\N, \cdot, 1, \geq_\up{div})$.
	\end{Example}

\begin{Example} (Powersets.) \\
	For any set $A$, the powerset $\pow{A}$ is a commutative monoid with unit $\emptyset$ when equipped with the union-operation $\cup$. Set-theoretic inclusion $\supseteq$ is a compatible partial order, so we have a thin theory $(\pow{A}, \cup, \emptyset, \supseteq)$.
	\end{Example}

There are also examples of thin theories in which the pre-order does not meet the condition to be a partial order. The condition $x \simeq y \Rightarrow x = y$ can break down violently, or subtly:

\begin{Example} (Any Monoid is a Thin Theory.) \\
	Let $(M, \star, 1)$ be any monoid. By putting the trivial relation $\succeq$ on $M$, which renders $x \succeq y$ for \myuline{all} $x,y \in M$, we see that $(M, \star, 1)$ is augmented to a thin theory. (This is even independent of whether or not the monoid is commutative.)
	\end{Example}

\begin{Example} ($\Logic$.) \label{ex:Logic}\\
	Let $P_1, P_2, P_3,  \ldots$ be infinitely many symbols, and consider the set of all \emph{well-formed formulas} that can be generated from these symbols along with the logical connectives $\land, \lor, \lnot, \top, \bot, \to, \leftrightarrow$ and the parentheses $)$ and $($. For example, $\lnot(P_1 \land P_3) $ is a well-formed formula, whereas $) \to P_2 \lnot$ is not. The set of well-formed formulas is pre-ordered by the relation $\psi \succeq \phi$ asserting that formula $\phi$ is provable from formula $\psi$ (using some standard inference system, see e.g. Ref. \cite{Enderton01}), and by equipping it with the binary operation that maps the pair $\psi_1, \psi_2$ to $(\psi_1 \land \psi_2)$, it becomes a thin theory $\Logic$, with unit $\top$ (though it is neither strictly commutative nor strictly associative). 
\end{Example}

\begin{Remark} (Relation to Resource Theories \cite{CFS16}.)\label{ex:Resource} \\
	In Ref. \cite{CFS16} the authors propose ordered monoids $(M, \star, 1, \succeq)$ as a model of \emph{resource convertibility};  more precisely, a thin theory in our language is in their language a \emph{waste-free theory of resource convertibility}. The elements $x,y,z, \ldots \in M$ correspond to resources in some universe, and the pre-order assignment $x \succeq y$ reflects that resource $x$ can be converted into resource $y$ at no cost. The binary operation $\star$ simply represents the junction of two resources, and $1 \in M$ represents a void resource. (The adjective `waste-free' refers to the relations $x \succeq 1$.)  The scope of this interpretation is large, ranging from economy to chemistry. It also includes the theory of resource convertibility associated to \emph{(bipartite) quantum entanglement}.

\end{Remark}

States, isomorphisms and reversible transformations are strange notions in thin theories:

\begin{Remark} (States, Isomorphisms and Reversibles in Thin Theories.) \\
An isomorphism from $x$ to $y$ in a thin theory $\Theory$ is a pair of relationships $x\succeq y$ and $y\succeq x$, whose two compositions yield the identities on $x$ and $y$, respectively. However, since the serial composition of the transformation $x \succeq y$ with $y \succeq x$ must yield \myuline{some} transformation from $x$ to $x$ and since there is by assumption only one, namely the identity $x \succeq x$, \myuline{any} pair of relationships between two systems $x,y$ witnesses an isomorphism. In short, for $x$ and $y$ to be isomorphic is precisely the condition $x \simeq y$. The same argument implies that any reversible transformation is necessarily an isomorphism. 

Not all systems in a thin theory have states. Actually, for $x$ to have a state means precisely that $1 \succeq x$, which is to say that $x \simeq 1$. 
	\end{Remark}

In \cref{ex:Interval} we saw a \myuline{cartesian} thin theory. We end this subsection by classifying those thin theories which are also cartesian:

\begin{Prop} (Thin Cartesian Theories.) \\
A thin theory described by the pre-ordered quasi-commutative monoid $(M, \star, 1, \succeq)$ is cartesian if and only if for every $x, y \in M$, the element $x \star y \in M$ is a least upper bound for $x$ and $y$ in the pre-order $(M, \succeq)$.

\end{Prop} 

\begin{proof}
	The thin theory described by $(M, \star, 1, \succeq)$ is cartesian if and only if for any $x,y,z \in M$ it holds that $(z \succeq x) \land (z \succeq y) \Leftrightarrow z \succeq  x \star y $. This is precisely to say that $x \star y$ is an upper bound for $x$ and $y$ which is least among all upper bounds. 
	\end{proof}

It follows that for instance the theory $\Logic$ from \cref{ex:Logic} is also cartesian.

 \subsection{Sub-Theories}
 \label{subsec:Subtheories}
 Some of the above examples of theories were \emph{nested}, one inside the other. For example, the oblivious version of quantum information, $\OblQIT$, was a \emph{sub-theory} of $\QIT$, in the sense that all of its systems and transformations, along with their serial and parallel composition, came from $\QIT$. In a similar way, $\QIT$ itself was a sub-theory of $\QIT^\infty$, as was $\CIT$ of $\mathbf{Stoch}$. 
 
 In fact, $\CIT$ is also a sub-theory of $\QIT$, or, more precisely, $\QIT$ has a sub-theory which is `isomorphic' to $\CIT$, by the construction mentioned earlier, according to which we associate the Hilbert space $\hat{X} := \C^X$ to the finite set $X$ and the CPTP map $\hat{T}: \End{\C^X} \to \End{\C^Y}$ given by $\hat{T}(A) = \sum_{x \in X, y \in Y} t_x(y) \bra{x} A \ket{x} \ketbra{y}$ to the classical channel $T=(t_x)_{x \in X}:X \to Y$. This last example, however, is different than the others in a significant way: The identity transformations $\hat{\id}_X$ in the smaller theory do \myuline{not} coincide with the identity transformations $\id_{\hat{X}}$ in the larger theory. We have already noticed in \cref{rem:QITCIT} that this is not an artefact of this specific embedding of $\CIT$ in $\QIT$, but a living condition of any such embedding. Similarly, the correct definition of `sub-theory' should not require identities to agree.
  
 \begin{Definition} (Sub-Theories.) \\
 		Let $\Theory$ be a theory. A \emph{sub-theory of $\Theory$} is a theory $\Theory_0$ for which 	\begin{itemize}
 		\item $\Sys{\Theory_0} \subseteq \Sys{\Theory}$, and the trivial system and composition of systems in $\Theory_0$ are the same as in $\Theory$;
 		\item for any systems $\cX, \cY \in \Sys{\Theory_0}$, $\Trans{\Theory_0}{\cX}{\cY} \subseteq \Trans{\Theory}{\cX}{\cY}$ and the serial and parallel compositions in $\Theory_0$ are the same as in $\Theory$.\end{itemize}\end{Definition}
 
 \begin{Remark}
Since identity transformations (and swapping transformations) in $\Theory_0$ are not required to be the same as in $\Theory$,  a sub-theory is a weaker notion than that of a (symmetric monoidal) sub-\myuline{category}.
 \end{Remark}

The concept of sub-theory is relevant in the context of example appropriation, since it paves the way for an explosion: Whenever we have a theory $\Theory$, we can choose from it any collection of systems and transformations and consider as new example theory the sub-theory $\Theory_0$ of $\Theory$ that this collection generates.\footnote{To the extent that the collections are not so bizarre that this procedure cannot be formalised in the formal language which we use, cf. the earlier comments on sets versus proper classes.}

 \begin{Example} ($\FinSets^*$.)\\
 	$\FinSets^*$ is the the sub-theory 
 of $\Sets^*$ generated by the finite sets and all the functions between them. That is, $\FinSets^*$ has as systems finite sets $X, Y, Z, \ldots$, composing under the cartesian product, and the transformations from $X$ to $Y$ are functions $f: X \to Y$, composing serially and parallelly as in $\Sets^*$. Observe that $\FinSets^*$ can also be regarded as a sub-theory of $\CIT$, generated this time by all the systems, but only the transformations $(t_x)_{x \in X}: X \to Y$ corresponding to deterministic functions.
 \end{Example}

 \begin{Example} ($\TREX$.) \label{ex:TREX} \\
$\TREX$ is the sub-theory of $\mathbf{FinSets^*}$ with all the same systems, but consisting only of the \myuline{surjective} functions. Note that the serial and parallel compositions of surjective functions are surjective, and that the trashes and identities are surjective. Though it has the `cartesian product' for parallel composition, the theory $\TREX$ is \myuline{not} a cartesian theory (e.g. there is no transformation $D:X  \to X \times X$ with marginals equal to $\id_X$ when $\abs{X}\geq 2$). In general, the theory $\TREX$ is extremely strange, and it will provide us with many counterexamples throughout. One bizarre feature is that, though it is much more intricate than a thin theory, it retains the property of having states only on those systems which are isomorphic to $\triv$.
 \end{Example}

 Readers who know about the theory of computation and algorithms may also define a sub-theory of $\Sets^*$ whose systems are collections of strings and whose transformations are algorithms (computable functions).

\section{Pictorial Syntax }
\label{sec:Pictorial}

So far, we have used an algebraic syntax in terms of the symbols `$\og$' and `$\after$' to represent composite transformations in a theory. Whereas this is in principle unproblematic, it is more or less undebatable that, to the human eye, the nature of already rather simple compositions can be obfuscated by the algebraic notation. For example, if $s$ is a state on $\cZ_1 \og \cZ_2$ and if $T_1: \cX_1 \og \cZ_1 \to \cY_1$ and $T_2: \cX_2 \og \cZ_2 \to \cY_2$ are transformations, then what is the appropriate intuition about the transformation  $(T_1 \og T_2) \after (\id_{\cX_1} \og s \og \id_{\cX_2})$?  

\subsection{Pictures for Algebra}

A viable and very effective solution is to introduce a \myuline{pictorial} syntax for systems, transformations and the two modes of composition. The basic idea is to pictorially denote a transformation $T: \cX \to \cY$ as a box with incoming and outgoing wires, as such: 

\begin{align}
\myQ{1}{0.7}{ & \push{\cX}  \qw & \gate{T} & \push{\cY} \qw & \qw } \quad 
\end{align}

This reinforces the interpretation of $T$ as a `process' which transforms input from the system $\cX$ to outputs on the system $\cY$. If $\cX$ and $\cY$ are composite systems, say $\cX= \cX_1 \og \cX_2$ and $\cY = \cY_1 \og \cY_2 \og \cY_3$, we may detail the representation by drawing one wire for each factor: 

\begin{align} \label{eq:Stack}
\myQ{1}{0.7}{ & \push{\cX_1}  \qw & \multigate{2}{T} & \push{\cY_1} \qw & \qw \\
	& & \Nghost{T}& \push{\cY_2} \qw & \qw \\
	& \push{\cX_2}  \qw & \ghost{T} & \push{\cY_3} \qw & \qw} \quad 
\end{align}

Note that the associativity $(\cY_1 \og \cY_2) \og \cY_3 = \cY_1 \og (\cY_2 \og \cY_3)$ is built into this notation, and if we moreover agree that the trivial system $\triv$ may be represented by empty space (no wire at all), then relations as $\cZ \og \triv = \cZ = \cZ \og \triv$ are also automatic. As such, we may represent a state $s: \triv \to \cX$ by a box with no incoming wires, and a trash $\tr_\cZ: \cZ \to \triv$ by a box with no outgoing wires, as

\begin{align} 
\myQ{1}{0.7}{ &  \Ngate{s}& \push{\cX} \qw & \qw} \quad, \quad  \text{respectively} \quad  \myQ{1}{0.7}{ & \push{\cZ} \qw& \gate{\tr} } \quad . 
\end{align}

The serial composition of transformations $\channel{\cX}{T}{\cY}$ and $\channel{\cY}{S}{\cZ}$ is represented by indeed connecting them serially, as 

\begin{align}
\myQ{1}{0.7}{ & \push{\cX}  \qw & \gate{T} & \push{\cY} \qw & \gate{S} & \push{\cZ} \qw & \qw} 
\end{align}

(with suitable modifications when the systems are represented as composites with several wires). This visual representation agrees with the Western reading direction from left to right, but disagrees with the unfortunate direction of functional composition mimicked in the notation `$S \after T$'.

The parallel composition of $\channel{\cX_1}{T_1}{\cY_1}$ and $\channel{\cX_2}{T_2}{\cY_2}$ is represented by vertical juxtaposition, as

\begin{align}
\myQ{1}{0.7}{ & \push{\cX_1}  \qw & \gate{T_1} & \push{\cY_1} \qw & \qw  \\
	& \push{\cX_2}  \qw & \gate{T_2} & \push{\cY_2} \qw & \qw } 
\quad 
\end{align}

(again modified if there are more incoming and outgoing wires to each box). Importantly, this notation is consistent with the convention of representing composite systems as a stack of wires as in \cref{eq:Stack}: The parallel composition $T_1 \og T_2$ indeed has domain $\cX_1 \og \cX_2$ and codomain $\cY_1 \og \cY_2$. 

An identity transformation $\id_\cZ : \cZ \to \cZ$ can be represented simply as the wire

\begin{align}
\myQ{1}{0.7}{  & \push{\cZ} \qw & \qw } \quad ,
\end{align}

and when combined with the convention on representing serial composition by serial connection, this consistently suggests the facts that $\id_\cZ \after T = T$ and $S \after \id_\cZ = S$ for transformations $T: \cX \to \cZ$ and $S: \cZ \to \cY$. \\

Within this pictorial syntax, the transformation $T := (T_1 \og T_2) \after (\id_{\cX_1} \og s \og \id_{\cX_2})$ from above is now drawn as 

\begin{align}  \label{eq:Bellrep}
\myQ{1}{0.7}{
	& \qw	& \push{\cX_1}  \qw   & \multigate{1}{T_1} & \push{\cY_1}  \qw & \qw   \\
	& \Nmultigate{1}{s}  & \push{\cZ_1} \qw & \ghost{T_1} &  \\
	& \Nghost{s} & \push{\cZ_1} \qw  & \multigate{1}{T_2}  \\
	& \qw 	&   \push{\cX_2} \qw  & \ghost{T_2} & \push{\cY_2}  \qw & \qw \\
} 
\quad.
\end{align}

This picture aids the intuition about the transformation $T$, by providing the interpretation that a state $s$ is shared across two different sites, at each of which a transformation is then applied to form locally at site $i$ a connection from $\cX_i$ to $\cY_i$, using the part of the state $s$ which is present at site $i$. 

We can use the pictorial syntax not only to better display the nature of composite transformations, but also to manipulate them more transparently. For instance, we can apply to $T$ the trash $\tr_{\cY_1} : \cY_1 \to \triv$ to the upper wire in \eqref{eq:Bellrep}, and in parallel apply $\id_{\cY_2}$ to the lower wire, thus computing that

\begin{align} \label{eq:Bellmani}
\myQ{1}{0.7}{
	& \push{\cX_1}  \qw   & \multigate{1}{T_1} & \push{\cY_1}  \qw & \gate{\tr}   \\
	& \Nmultigate{1}{s}   & \ghost{T_1} &  \\
	& \Nghost{s}   & \multigate{1}{T_2}  \\
	&   \push{\cX_2} \qw  & \ghost{T_2} & \push{\cY_2}  \qw & \qw \\
} 
\quad = \quad 
\myQ{1}{0.7}{
	& \push{\cX_1}  \qw   & \multigate{1}{\tr}    \\
	& \Nmultigate{1}{s}  & \ghost{\tr} &  \\
	& \Nghost{s}   & \multigate{1}{T_2}  \\
	&   \push{\cX_2} \qw  & \ghost{T_2}& \push{\cY_2}  \qw & \qw  \\
} 
\quad = \quad 
\myQ{1}{0.7}{
	& \push{\cX_1}  \qw   & \gate{\tr}    \\
	& \Nmultigate{1}{s}  & \gate{\tr} &  \\
	& \Nghost{s} & \multigate{1}{T_2}  \\
	&   \push{\cX_2} \qw  & \ghost{T_2} & \push{\cY_2}  \qw & \qw \\
} 
\quad ,
\end{align}

resting for the equalities on \cref{lem:Trashes}. (For clarity, the labels `$\cX_1$' and `$\cX_2$' on the internal wires have been omitted, and we shall often omit wire labels when they are irrelevant or clear form context.) 

The transformation $
\scalemyQ{.7}{0.7}{0.5}{
	& \Nmultigate{1}{s}  & \gate{\tr} &  \\
	& \Nghost{s} & \multigate{1}{T_2}  \\
	&   \push{\cX_2} \qw  & \ghost{T_2} & \push{\cY_2}  \qw & \qw \\
} 
$ can now be renamed as $\channel{\cX_2}{T'_2}{\cY_2}$, and the above computation then altogether suggests that by trashing the system $\cY_1$ from $T$  we obtain something of the form $\tr_{\cX_1} \og T'_2$ for some transformation $T'_2 : \cX_2 \to \cY_2$. (Physically this means that the output on $\cY_2$ alone is unaffected by the input to $\cX_1$; we shall consider such \emph{non-signalling} properties systematically in the next section.)\\

The pictorial syntax laid out above is ubiquitous in the literature, and has been since the introduction of symmetric monoidal categories. Ref. \cite{Sel10survey} gives a detailed and formal survey of general \emph{graphical calculi} for monoidal categories, and therein the author essentially attributes the boxes-and-wires representation to Roger Penrose, dating back almost 50 years (\cite{Pen71}). \\

\begin{Remark} (On the Validity of Pictorial Reasoning.)\\
It is only fair for the reader to question the exact relationship between the algebraic and pictorial syntaxes. Is it really the case that one can deduce the algebraic identity 

\begin{align}
(\tr_{\cY_1} \og \id_{\cY_2}) \after \big( (T_1 \og T_2) \after (\id_{\cX_1} \og s \og \id_{\cX_2}) \big)= \tr_{\cX_1} \og T'_2
\end{align}

for some $T'_2: \cX_2 \to \cY_2$ just by reference to the graphical manipulation in \cref{eq:Bellmani}? It is instructive to consider for each step of the manipulation the translation from pictures to algebraic symbolism, and to verify that this is indeed the case.

One might worry that in general care must be taken when translating between the conclusions of pictorial and algebraic manipulations. It would be dangerous if graphical manipulations suggested algebraically invalid derivations, and conversely regrettable if some algebraic derivations had no graphical counterpart.
Fortunately, it is a mathematical fact that this does not happen (see Thm. 2.1 in \cite{Sel10survey} for a formal statement, and Ref. \cite{JS91} for an even more precise treatment). This fact is the ultimate power -- and justification -- of the pictorial syntax.

 I have only defined the pictorial syntax by examples, and I shall take the attitude of not being uptight about the formal correspondence between derivations in the two syntaxes. Rather, pictures will be used where they enlighten, and with the implicit understanding that they really represent underlying algebraic arguments which can be distilled upon desire. 
\end{Remark}

\subsection{Interfaces and Channels}
\label{subsec:Inter}

Though the pictorial approach to notation is intuitively superior, and equivalent to the algebraic with regards to deductive power, there is a sense in which it is distinct from the algebra it intends to represent. \\

The problem has to do with multiplicity of wires. According to the pictorial syntax, a composite system $\cX = \cX_1 \og \cX_2$ can, when it appears as domain or codomain of a transformation, be represented equally well as 
$
\myQ{0.7}{0.5}{ & \push{\cX} \qw & \qw } $
and as
$
\myQ{0.7}{0.5}{ & \push{\cX_1} \qw & \qw  \\ & \push{\cX_2} \qw & \qw }$. Similarly, though we made the convention that we can choose to represent the system $\triv$ by empty space, we made no convention that we \myuline{must} do so. Thus, in the thin theory $(\N, \cdot, 1, \geq)$ (\cref{ex:ThinNat}), for instance, the three pictures

\begin{align}
\myQ{1}{0.7}{  & \push{2} \qw & \qw  \\ & \push{3} \qw & \qw \\ & \push{5} \qw & \qw } \quad, \quad  
\myQ{1}{0.7}{  & \push{6} \qw & \qw  \\ & \push{5} \qw & \qw } \quad, \quad 
\myQ{1}{0.7}{  & \push{1} \qw & \qw  \\ & \push{1} \qw & \qw  \\ & \push{30} \qw & \qw  \\ & \push{1} \qw & \qw } \quad
\end{align}

are all valid pictorial representations of the system $30$, and in the theory $\QIT$ the pictures 

\begin{align} \label{eq:idqubit}
\myQ{0.7}{0.5}{  & \push{\C^2} \qw & \multigate{1}{\id_{\C^2 \otimes \C^2}} & \push{\C^2 \otimes \C^2} \qw & \qw \\ & \push{\C^2} \qw & \ghost{\id_{\C^2 \otimes \C^2}} } 
, \quad
 \myQ{0.7}{0.5}{  & \push{\C^2} \qw & \multigate{1}{\id_{\C^2 \otimes \C^2}} & \push{\C^2} \qw & \qw \\ & \push{\C^2} \qw & \ghost{\id_{\C^2 \otimes \C^2}} & \push{\C^2} \qw & \qw } , 
 \quad \myQ{0.7}{0.5}{  & \push{\C^2 \otimes \C^2} \qw & \gate{\id_{\C^2 \otimes \C^2}} & \push{\C^2 \otimes \C^2} \qw & \qw }
\end{align}

are all valid representations of the transformation $\id_{\C^2 \otimes \C^2}$ (there is one more). Now, this ambiguity is \myuline{not} in conflict with the equivalence of derivations in the algebraic and pictorial syntaxes mentioned earlier,\footnote{This is essentially because the various representations of the identity transformations are isomorphisms, and so can act as regrouping devices which  collect a bunch of wires into one, much like cable collectors used in offices.} but is does raise the following question: \\

 \emph{If it is not transformations between systems, then \myuline{what} \myuline{exactly} are we drawing when we draw a picture in the pictorial syntax? }\\

We will need the viewpoint that the various pictorial representations of the same formal object correspond to various ways of thinking about that object. For example, in $\CIT$, the difference between drawing a state $p : \triv \to X$ as 

\begin{align} \label{eq:button}
\myQ{1}{0.7}{  & \Ngate{p}  & \push{X} \qw & \qw   } \quad \text{and as} \quad 
\myQ{1}{0.7}{&\push{\triv}  \qw  & \gate{p}  & \push{X} \qw & \qw   }
\end{align}

will be that the latter represents the viewpoint that the process under consideration accepts an input to the trivial system, whereas the former does not. When we define \emph{causal specifications} in \cref{chap:Causal}, the utility of this is amplified, since it will allow the interpretation that the trivial system can act as a \emph{button} that has to be pushed before the state $p$ materialises, whereas absence of the trivial system as input signifies that the state $p$ was always present.

 Similarly, if $X$ is a composite system, say $X= X_1 \times X_2$, the two drawings 

\begin{align} \label{eq:bipart}
\myQ{1}{0.7}{& \Nmultigate{1}{p}  & \push{X_1} \qw & \qw    \\ & \Nghost{p}  & \push{X_2} \qw & \qw }
\quad \text{and} \quad 
\myQ{1}{0.7}{  & \Ngate{p}  & \push{X} \qw & \qw   }
\end{align}

will signify that the state $p$ is presented across two different ports (first drawing), or that all of $p$ is given through one single port (second drawing). Generally, we want wires to represent \emph{ports} which can be labelled by distinct \myuline{names} and associated with (possibly different) \myuline{systems}. None of this intuition is reflected a priori in the formal algebraic structure of symmetric monoidal categories, so if we want we an algebraic counterpart to the pictures we have to encode the additional structure explicitly. \\

We do this by introducing the concept of an \emph{interface}. Intuitively, an interface in a theory $\Theory$ is specified by a finite collection of ports distinguishable from each other by unique names, and each of which able to transmit information of a certain type as defined by an associated system in $\Theory$. Formally, we can pack this information compactly by defining an interface as a \emph{map} from its set of port names to the associated systems:

\begin{Definition} (Interfaces in $\Theory$.) \label{def:Inter}\\
		Let $\Theory$ be a theory. An \emph{interface in $\Theory$} is a map, $\bbX$, whose domain, denoted $\ports{\bbX}$, is a finite set of so-called \emph{ports} (or \emph{port names}), and which assigns to each port $\sfp \in \ports{\bbX}$ a system $\bbX(\sfp) \in \Sys{\Theory}$. 
\end{Definition}

\begin{Example} (Simple Interfaces.) \\
	Let us call an interface \emph{simple} if it has just a single port. A simple interface $\bbX$ is completely specified by the name $\sfp \in \ports{\bbX}$ and the associated system $\bbX(\sfp) \in \Sys{\Theory}$.
\end{Example}

\begin{Example} (Pairs of Qubits.) \label{ex:C2} \\
	Let $\bbX$ be an interface in $\mathbf{QIT}$ with a single port, say $\ports{\bbX} = \{\sfone\}$, and with $\bbX(\sfone)= \C^2 \otimes \C^2$. Let $\bbX'$ be an interface with two ports, say $\ports{\bbX'}= \{\sfa,\sfb\}$ and with $\bbX'(\sfa)=\bbX'(\sfb) = \C^2$. The interfaces $\bbX$ are $\bbX'$ are distinct, though they both represent the same total system  $\C^2 \otimes \C^2$ (i.e. a pair of qubits). The interface $\bbX$ represents a scenario in which this system is thought of as a single entity, while $\bbX'$ represents a bipartite situation where each factor is separate and clearly labelled. The pictorial depictions of $\bbX$ and $\bbX'$ are illustrated by \cref{eq:idqubit}.
\end{Example}

\begin{Example} (Interfaces in Thin Theories.)\\
	In a thin theory $\Theory$, identified with the pre-ordered quasi-commutative monoid $(M, \star, \succeq, 1)$, the distinction between a system and an interface is not foggy at all; indeed, a system in $M$ is an element of $M$ whereas an interface $\bbX$ corresponds to a \myuline{tuple} of elements in $M$, indexed by the port names $\ports{\bbX}$.
	\end{Example}

Defining interfaces as maps allows for smooth notation in many regards:

Since maps are set-theoretically identified with their graphs, the notation $\bbX_0 \subseteq \bbX$ makes sense for interfaces $\bbX_0$ and $\bbX$, and it means that $\ports{\bbX_0} \subseteq \ports{\bbX}$ with $\bbX_0(\sfp)=\bbX(\sfp)$ for all $\sfp \in \ports{\bbX_0}$. We shall say in this case that $\bbX_0$ is a \emph{sub-interface} of $\bbX$.  If $\bbX_0$ is a sub-interface of $\bbX$, we can form the \emph{complementary interface} $\bbX \setminus \bbX_0$, namely the unique sub-interface of $\bbX$ with $\ports{\bbX \setminus \bbX_0}= \ports{\bbX} \setminus \ports{\bbX_0}$.

Two interfaces $\bbX_1$ and $\bbX_2$ are called \emph{parallelly composable} if $\ports{\bbX_1} \cap \ports{\bbX_2} = \emptyset$, i.e. if they share no port names. In this case the union $\bbX_1 \cup \bbX_2$ is an interface, namely the one given by $\ports{\bbX_1 \cup \bbX_2} = \ports{\bbX_1} \cup \ports{\bbX_2}$ and $(\bbX_1 \cup \bbX_2)(\sfp_j) = \bbX_j(\sfp_j)$ for $\sfp_j \in \ports{\bbX_j}$. The interface $\bbX_1 \cup \bbX_2$ is called the \emph{composite} of $\bbX_1$ and $\bbX_2$. \textbf{Importantly, we do not allow the parallel composition of interfaces which have port names in common.} This has the positive consequence that all input wires in a drawing are identified by a unique port name, as are all output systems. (We do allow, however, that an input and output wire are labelled by the same name, as long as these correspond to the same system, cf. the definition of a channel below.) 

We define \emph{the trivial interface}, denoted $\bbI$, as the empty map, i.e. as the unique interface with no port names. The difference between the two drawings in \cref{eq:button} is exactly that the input interface of the left one is the trivial interface, whereas the input interface of the right one has a single port with associated system $\triv$. The trivial interface $\bbI$ is a sub-interface of every interface and it is parallelly composable with every interface $\bbX$, with $\bbI\cup \bbX = \bbX$. 

Finally, it is obvious that every non-trivial interface $\bbX$ factors into simple interfaces (i.e. interfaces with a single port) as $\bbX_1 \cup \ldots \cup \bbX_{\abs{\bbX}}$ where $\abs{\bbX}:= \abs{\ports{\bbX}}$ is the \emph{size of $\bbX$}, and where the interfaces $\bbX_1, \ldots, \bbX_{\abs{\bbX}}$ are the simple sub-interfaces of $\bbX$. \\

Having defined interfaces, we must replace transformations with entities whose domains and codomains are interfaces rather than systems. We will call them \emph{channels}.\footnote{Somewhat confusingly, this conflicts with the use of the term `quantum [classical] channel' for the \myuline{transformations} in $\QIT$ [$\CIT$], but the confusion should not cause any serious harm.} A channel $T$ from an interface $\bbX$ to an interface $\bbY$ is more or less a transformation from $\cX$, the total system corresponding to the interface $\bbX$, to $\cY$, the total system corresponding to the interface $\bbY$. However, channels need to be defined in such a way that they retain the information about how the various ports of the interfaces relate to these total systems. For example, if $\ports{\bbX}=\{\sfone, \sftwo\}$ and $\ports{\bbY} = \{\sfa, \sfb\}$, and if $T: \bbX(\sfone) \og \bbX(\sftwo) \to \bbY(\sfa)\og \bbY(\sfb) $ is a transformation, then  we want to think of this transformation as identical to the transformation $T': \bbX(\sfone) \og \bbX(\sftwo) \to \bbY(\sfb) \og \bbY(\sfa)$ given by $T' = \sigma_{\bbY(\sfa), \bbY(\sfb)} \after T$, provided that we know the port sequence in each case ($\sfa$-$\sfb$ versus $\sfb$-$\sfa$). Pictorially, 

\begin{align}
 \myQ{0.7}{0.5}{ & \sfone \quad  & \push{\bbX(\sfone)} \qw & \multigate{1}{T} & \push{\bbY(\sfa)} \qw & \qw  & \sfa \\ &\sftwo \quad&  \push{\bbX(\sftwo)} \qw & \ghost{T} & \push{\bbY(\sfb)} \qw & \qw  & \sfb }  \quad \text {and} \quad   \myQ{0.7}{0.5}{ & \sfone \quad  & \push{\bbX(\sfone)} \qw & \multigate{1}{T} & \push{\bbY(\sfb)} \qw & \qw  & \sfb\\ &\sftwo \quad&  \push{\bbX(\sftwo)} \qw & \ghost{T} & \push{\bbY(\sfa)} \qw & \qw  & \sfa } 
\end{align}

should be merely different drawings of the same thing. In general, we want to define a channel as an equivalence class of transformations related by swappings on their input and output systems.

 Given an interface $\bbZ$, a bijection $\ell: \{1, \ldots, \abs{\bbZ}\} \to \ports{\bbZ}$ is a choice of enumeration of the ports in $\bbZ$.  Given such an enumeration $\ell$, we define \emph{the system (in $\bbZ$) corresponding to $\ell$} as the composite 

\begin{align}
\bbZ[\ell] :=  \bbZ(\ell(1)) \og \bbZ(\ell(2)) \og  \ldots \og  \bbZ(\ell(\abs{\bbZ})),
\end{align}

with the natural convention that if $\bbZ = \bbI$ and $\ell$ is the empty enumeration, then $\bbZ[\ell]=\triv$. We formalise channels as follows:

\begin{Definition} (Channels in $\Theory$.) \label{def:Channel}\\
	Let $\bbX$ and $\bbY$ be interfaces in a theory $\Theory$, subject to the condition that if $\sfp \in \ports{\bbX} \cap \ports{\bbY}$, then $\bbX(\sfp)=\bbY(\sfp)$. A \emph{channel from $\bbX$ to $\bbY$} is a triple $(T, \bbX, \bbY)$, where $T$ is family of transformations in $\Theory$, 
	
	\begin{align}
	T= (_{\ell_\bbY}T_{\ell_\bbX}: \bbX[\ell_\bbX] \to \bbY[\ell_\bbY])_{\ell_\bbX, \ell_\bbY}, 
	\end{align}
	
	indexed by the collection of enumerations $\ell_\bbX$ of $\ports{\bbX}$ and $\ell_\bbY$ of $\ports{\bbY}$, and subject to the condition that for any permutations $\pi_\bbX$ of $\ports{\bbX}$ and $\pi_\bbY$ of $\ports{\bbY}$, we have  
	
	\begin{align}
	_{\pi_\bbY \circ \ell_\bbY}	T_{\pi_\bbX \circ \ell_\bbX} = \sigma_{\pi_\bbY} \after _{ \ell_\bbY}	T_{ \ell_\bbX}  \after \sigma^{-1}_{\pi_\bbX},
	\end{align}
	
	where $\sigma_{\pi_\bbX} : \bbX[\ell_\bbX] \to \bbX[\pi_\bbX \circ \ell_\bbX]$ and $\sigma_{\pi_\bbY}: \bbY[\ell_\bbY] \to \bbY[\pi_\bbY \circ \ell_\bbY]$ denote the transformations which swap the individual factors in accordance with the permutations $\pi_\bbX$ and $\pi_\bbY$, respectively. \end{Definition}

We will write $T: \bbX \to \bbY$ to indicate that $T$ is a channel from $\bbX$ to $\bbY$, and by abuse of notation we often write simply `$T$' for all of the individual components $_{ \ell_\bbY}	T_{ \ell_\bbX} $.\\

What we have obtained in summary is the following: 

\begin{itemize}
\item In a diagram, every \emph{wire} corresponds to a simple interface; all incoming wires are formally given distinct port names (whereas their systems may coincide), and likewise so are all outgoing wires.

 \item Every \emph{box} corresponds to a channel between the interfaces that connect to it; as such, the order in which these are drawn from top to bottom is insignificant, as the port names formally keep track of the matching. 

\item A port name may occur both as input and output, but in that case the associated systems are identical (and we will really think of it as the same physical port). 
\end{itemize}

In effect, we have replaced the theory $\Theory$ by another category, $\IC{\Theory}$, namely the \emph{category of interfaces and channels in $\Theory$}, where the objects are interfaces and the morphisms are channels, which compose serially in an obvious (but tedious) way. This category is moreover `partially' symmetric monoidal, in the sense that \myuline{some} interfaces (namely those with no overlapping port names) are deemed parallelly composable, and \myuline{some} channels are deemed parallelly composable (namely those for which the interfaces are parallelly composable, and for which the parallel composition does not result in a violation of the condition that a repeated port name can correspond to different systems). Though this might all seem very formal, we will almost never explicitly flesh out the formalities and so an intuitive understanding of these concepts suffices. \\

As a last convention, it is only natural to introduce an abbreviation which confuses the reader by overwriting the above correspondence -- namely, that in order to keep diagrams visually simple we will often write 

\begin{align}
\myQ{0.7}{0.5}{& \push{\bbX} \qw & \gate{T} & \push{\bbY} \qw & \qw} \quad \text{as abbreviation for} \quad 	\myQ{0.7}{0.5}{& & \qw & \multigate{2}{T}   & \qw & \qw  \\
	& \bbX \quad & \vdots	&\Nghost{T}	  & \vdots & \bbY \\
	& & \qw 	& \ghost{T} &  \qw  & \qw  },
\end{align}

with $T$ a channel from $\bbX$ to $\bbY$. In this way, we do not need to draw a specific number of wires (or to draw dotted lines) when arguing about general channels.

\subsection{Normal Theories}
\label{subsec:Normal}

The pictorial syntax raises another problem which can be viewed as a discrepancy between pictures and algebra, though in a more subtle way than before. Consider the equation

\begin{align}
\myQ{1}{0.7}{ & \push{\cX}  \qw & \gate{T} & \push{\cY} \qw & \qw  \\
	& \\
	& \push{\cZ}  \qw & \gate{S} & \push{\cW} \qw & \qw } 
\quad = \myQ{1}{0.7}{ & \push{\cX}  \qw & \gate{T'} & \push{\cY} \qw & \qw  \\
	& \\
	& \push{\cZ}  \qw & \gate{S'} & \push{\cW} \qw & \qw } \quad.
\end{align}

Is it necessarily that case that $T=T'$ and $S=S'$? The graphical language somehow invites the presumption that this holds, whereas the algebraic equation $T \og S = T' \og S'$ does not enforce the same suspicion. To understand whether the conclusion is legal, it clearly suffices (by symmetry) to know whether the identity $T=T'$ is implied by $T \og S = T' \og S'$. Moreover, since the identity $T \og S = T' \og S'$ evidently implies $T \og \tr_\cZ = T' \og \tr_\cZ$, it actually suffices to know the answer in the case $S=S'=\tr_\cZ$. This motivates the following definition:

\begin{Definition} (Normality.)\label{def:Normal} \\
	A theory $\Theory$ is called \emph{normal} if for any systems $\cX, \cY, \cZ \in \Sys{\Theory}$, and for any transformations $T, T': \cX \to \cY$, it holds that 
	
	\begin{align}
	T \og \tr_\cZ = T' \og \tr_\cZ \Rightarrow T=T'.
	\end{align}
	
\end{Definition}

Our reasons for regarding $\Sets$ and $\Graphs$ as poorer theories than $\Sets^*$ and $\Graphs^*$ is precisely that the latter are not normal: 

\begin{Example} ($\Sets$ is Not Normal.)\\
	For any sets $X$ and $Y$, and for any functions $f,g: X \to Y$, we have $f \times \tr_\emptyset = g \times \tr_\emptyset$, as both functions are empty. (Here, $\tr_\emptyset$ is the unique empty map $\emptyset \to \triv$.)
\end{Example}

\begin{Example} ($\Graphs$ is Not Normal.) \\
	The (direct) product of graphs $G$ and $H$, $G \times H$, renders $(u_1,v_1)$ adjacent to $(u_2,v_2)$ if and only if both $u_1$ is adjacent to $u_2$ in $G$ and $v_1$ is adjacent to $v_2$ in $H$. In particular, if $\bullet$ denotes a graph with precisely one vertex and no edges, then $G \times \bullet$ is the graph with the same vertex set as $G$, but with no edges at all. 
	Now, if $f: G \to H$ is any graph homomorphism from $G$ to some graph $H$, then $f \times \tr_\bullet : G \times \bullet \to H$ is the homomorphism which acts as $f$ on the vertex set. Since $G \times \bullet$ has no edges, however, we cannot tell from $f \times \tr_\bullet$ how $f$ acts on edges, so in particular if we choose $H$ to be a graph for which some pair of vertices has two distinct edges between them, we can find $f,f': G \to H$ distinct with $f \times \tr_\bullet = f' \times \tr_\bullet$.
\end{Example}

Luckily, we have the following:

\begin{Prop} (Virtually All Theories are Normal.) \\
	The following theories are normal:
	
	\begin{enumerate}
		\item All theories in which every system has at least one state. 
		\item All thin theories.
		\item All sub-theories of a normal theory.

	\end{enumerate}
	
	In fact, every single theory presented as an example in \cref{sec:Examples} is normal. 
\end{Prop}

\newpage

\begin{proof}
1.	If $s$ is a state on $\cZ$, then for any transformation $T : \cX \to \cY$ we have $T=  T \og \id_\triv = (T \og \tr_\cZ) \circ (\id_\cX \og s)$, so $T$ can be recovered uniquely from $T \og \tr_\cZ$. Pictorially, 

\begin{align}
\myQ{.7}{.5}{ & \qw & \gate{T} & \qw & \qw } = \myQ{.7}{.5}{ & \qw & \gate{T} & \qw & \qw  \\& \Ngate{s} &  \gate{\tr}} .
\end{align}  
	
2.	A thin theory is normal for the simple reason that \myuline{any} two transformations from $\cX$ to $\cY$ are identical, regardless of whether they satisfy additional constraints or not.
	
3.	If $\Theory$ is a sub-theory of a theory $\tilde{\Theory}$, then the normality condition for $\Theory$ concerns a smaller class of systems and transformations than that corresponding to a normality condition for $\tilde{\Theory}$. 
	
	The final statement is left for the reader to ponder. 
\end{proof}

We make the following convention: \\

\framebox{\textbf{From now on, we always use the term `theory' to mean \myuline{normal theory}.}}

\section{Summary and Outlook}

\label{sec:SummaryTheories}

In this chapter, we have seen a mathematical definition of `physical' theories (\cref{def:Theory}), and many examples of this concept (\cref{sec:Examples}). We have also defined the special kinds of transformations called \emph{states} (\cref{def:state}), \emph{reversibles} and \emph{isomorphisms} (\cref{def:RevIso}). Finally, we have discussed a pictorial syntax for representing the elements of a theory, and we have defined in this context the notions of \emph{interfaces} (\cref{def:Inter}) and \emph{channels} (\cref{def:Channel}). We also observed the notion of a theory being \emph{normal} (\cref{def:Normal}), and we will employ that assumption implicitly from now on. \\

The reign of category theory in mathematics has hatched a practice which is nowadays standard in most of its disciplines: When defining a class of mathematical structures (for examples: physical theories), one should define along with them the appropriate notion of \emph{structure-preserving maps} (or, \emph{homomorphisms}) between them -- mathematical extremists would say that one has not even understood what is integral to a structure before one has defined what are the structure-preserving maps. 

As such, it would be natural to develop a theory of homomorphisms between theories, as alluded to in \cref{rem:QITCIT}. It is not a priori clear that anything substantial can be gained by this exercise, but from initial thoughts it seems to me that the notion of a theory $\Theory$ which allows an embedding $\Gamma : \CIT \to \Theory$ is highly interesting: By virtue of such an embedding, we can basically talk about \emph{probabilities} and \emph{classicality} and construct in $\Theory$ a \emph{convex structure}, in the same way as we do in $\QIT$.\footnote{Though one apparently has to assume by axiom that there exist for any finite tuple $(T_1, \ldots, T_n)$ of transformations $T_k : \cX \to \cY$ in $\Theory$ a unique transformation $T: \cX \og \Gamma(\{1, \ldots, n\}) \to \cY$ which reads off the value $k$ in the classical system $\Gamma(\{1, \ldots, n\})$ and chooses the respective transformation $T_k$.} In existing treatments (such as Ref. \cite{Chir10}), probabilistic and convex structure have to be separately planted on top of the compositional structure in the theory. \\

\chapter{Dilations}
\label{chap:Dilations}

{\centering
	\subsection*{§1. Introduction and Outline.}}

Consider a channel $\scalemyQ{.8}{0.7}{0.5}{ & \push{\cX_1}  \qw & \multigate{2}{L} & \push{\cY_1} \qw & \qw \\
	& & \Nghost{L}& \push{\cY_2} \qw & \qw \\
	& \push{\cX_2}  \qw & \ghost{L} & \push{\cY_3} \qw & \qw}$ and suppose that, for some reason or another, we only have access to the third output port, corresponding to the system $\cY_3$. By the interpretation of trashes as discarding, the channel we really `see' is the \emph{marginal} 

\begin{align}
\myQ{0.7}{0.5}{ & \push{\cX_1}  \qw & \multigate{1}{T} &  \\
	& \push{\cX_2}  \qw & \ghost{T} & \push{\cY_3} \qw & \qw} \quad  := \quad 
\myQ{0.7}{0.5}{ & \push{\cX_1}  \qw & \multigate{2}{L} & \push{\cY_1} \qw & \gate{\tr} \\
	& & \Nghost{L}& \push{\cY_2} \qw & \gate{\tr} \\
	& \push{\cX_2}  \qw & \ghost{L} & \push{\cY_3} \qw & \qw} \quad .
\end{align}

For example, in the theory $\Sets^*$, $L$ would be a function, and $T$ would be just the third component of that function, with the two other components discarded. (In $\CIT$ or $\QIT$, the act of marginalisation may affect correlations between outputs as well.) We shall now develop the first symptoms of an obsession, namely the quest for answering the following question:
\begin{center}
\emph{Knowing $T$, what are the possible $L$ that $T$ could have come from?}\\
\end{center} 

In fact, we will include among the possible $L$ also some which have \myuline{inputs} additional to those of $\cX_1$ and $\cX_2$, as long as these inputs do not \emph{signal} to the interface accessible to us. Precisely, given a channel $\channel{\bbX}{T}{\bbY}$ we define a \emph{dilation of $T$} to be a channel $\scalemyQ{.8}{0.7}{0.5}{& \push{\bbX} \qw &  \Nmultigate{1}{L}{\qw}   & \push{\bbY} \qw  & \qw  \\
	&\push{\bbD} \ww & \Nghost{L}{\ww} & \push{\bbE} \ww  & \ww
}$ such that 

\begin{align}
\scalemyQ{1}{0.7}{0.5}{& \push{\bbX} \qw &  \Nmultigate{1}{L}{\qw}   & \push{\bbY} \qw  & \qw  \\
	&\push{\bbD} \ww & \Nghost{L}{\ww} & \push{\bbE} \ww  & \Ngate{\tr}{\ww}
}		
   = 
\myQ{0.7}{0.5}{& \push{\bbX} \qw &   \Ngate{T}{\qw}    & \push{\bbY} \qw  & \qw  \\
	&\push{\bbD} \ww  & \Ngate{\tr}{\ww}
}, 
\end{align}

and we thus desire to understand all possible dilations of $T$. In some theories, this question will be a purely mathematical curiosity (for example in thin theories), but mostly it has a clear physical interpretation, as in the information theories $\CIT$ and $\QIT$: What we will imagine is a dichotomy between interfaces $\bbX$ and $\bbY$ which are \emph{accessible} or \emph{open} to us, and interfaces $\bbD$ and $\bbE$ which are \emph{inaccessible} or \emph{hidden}. (This dichotomy is visually represented using wiggly wires above.) The hidden interfaces may be controlled by untrustworthy agents, or may represent simply parts of Nature which are not within our reach; we will often use the word \emph{the environment} as umbrella term. By interacting with the open interfaces, we may establish that we are interacting with the channel $T$,\footnote{An operationally minded reader might inquire as to \myuline{how} we would determine that the channel we have access to is $T$; even in the case where $T$ is a function between finite sets, to confirm its identity we need to check its value on every possible input. Whereas this problem is very real, we shall completely ignore it. One can either regard it as an assumption of mathematical character, or imagine iterated use of identical and independent copies of the channel.} but the point is that $T$ could be `implemented' in many different ways across the hidden interfaces, as formalised by its various dilations. \\ 

In this chapter, I introduce a very general theory of dilations, applicable to all of the theories we have seen in \cref{chap:Theories}. Indeed, the singular reason for demanding in \cref{def:Theory} that the trivial system $\triv$ be terminal, is that this requirement minimally facilitates the notion of dilations. Whereas this is well-known (\cite{Chir10, Chir14dilation}), the structure of dilations has, to the best of my knowledge, never been studied systematically in the literature before, neither for specific theories nor for theories in general. As such, all the theory and results laid out in this chapter are new.  \\

\textbf{Dilations, Non-Signalling and DiVincenzo's Property.} Naturally, we begin in \cref{sec:Trinity} by defining \emph{marginalisation} and \emph{dilations}. Due to the unconventional choice of defining dilations as two-sided (i.e. with the interface $\bbD$ possibly non-trivial), the concept of dilations will be closely related to that of \emph{non-signalling}, which is also defined and exemplified. 

Some 20 years ago, D. DiVincenzo proposed a later confirmed conjecture (\cite{Beck01,Egg02}) about the structure of non-signalling channels in $\QIT$, which is here promoted to a general property that a theory might have. In the context of dilations, the DiVincenzo property entails that any dilation is of the form 	$\scalemyQ{.8}{0.7}{0.5}{ & \push{\bbX}  \qw & \multigate{1}{L_0} & \qw & \push{\bbY} \qw & \qw &  \qw \\ & & \Nghost{L_0} &  \push{\bbE_0} \ww & \Nmultigate{1}{G}{\ww} & \push{\bbE} \ww & \ww\\
	& 	&  \push{\bbD} \ww &\ww & \Nghost{G}{\ww}}$ for some channel $G$ and some \myuline{one-sided} dilation $L_0$, thus effectively reducing the study of dilations to that of one-sided dilations. In later sections it is proved that many theories enjoy this property (but it fails for example in some thin theories and in the theory $\TREX$). \\

\textbf{A Hierarchy of Dilations.} The most important observation about the dilations of a given channel $\channel{\bbX}{T}{\bbY}$ is that they do not constitute just a messy class of channels; they are naturally \emph{ordered}. For example, consider in the theory $\CIT$ the state $\state{p}{X}$, with $p$ the uniform distribution on the set $X= \{\epsdice{1}, \epsdice{2}, \epsdice{3}, \epsdice{4}, \epsdice{5}, \epsdice{6}\}$, representing the throw of a fair die. Among its one-sided dilations $
\scalemyQ{0.8}{0.7}{0.5}{& \Nmultigate{1}{\ell}  & \push{E} \ww & \ww    \\ & \Nghost{\ell}  & \push{X} \qw & \qw  }$ are the one with $E=X$ and $\ell(x,x')= p(x) \delta_{x,x'}$ (diagonal distribution), the one with $X=\{\mathsf{even}, \mathsf{odd}\}$ and $\ell(x, k)= p(x) \delta_{\mathsf{par}(x), k}$ (where $\mathsf{par}(x)$ denotes the parity of $x$), and the ones with $E$ arbitrary and $\ell= p \otimes r$ for some state $r$ on $E$. These dilations have clear meanings in terms of \emph{side-information}: The first corresponds to the environment keeping record of the precise outcome of die throw, the second corresponds to the environment knowing only the parity, and the third kinds correspond to the environment possessing information completely independent of the die throw. 

The \emph{dilational ordering} introduced in \cref{sec:DilOrd} formalises the intuition that the second dilation may be derived from the first, and the third in turn from the second. In general, the dilational ordering (\cref{def:DilOrd}) will describe how one dilation $ \scalemyQ{0.8}{0.7}{0.5}{& \push{\bbX} \qw &  \Nmultigate{1}{L'}{\qw}   & \push{\bbY} \qw  & \qw  \\
	& & \Nghost{L'} & \push{\bbE'} \ww  & \ww 
}$ might be \emph{derivable} from another dilation $ \scalemyQ{0.8}{0.7}{0.5}{& \push{\bbX} \qw &  \Nmultigate{1}{L}{\qw}   & \push{\bbY} \qw  & \qw  \\
& & \Nghost{L} & \push{\bbE} \ww  & \ww 
}$ by constructions taking place in the environment, specifically, if there exists a channel $G$ such that

 \begin{align} \label{eq:Gderive}
\scalemyQ{1}{0.7}{0.5}{& \push{\bbX} \qw &  \Nmultigate{1}{L'}{\qw}   & \push{\bbY} \qw  & \qw  \\
	& & \Nghost{L'} & \push{\bbE'} \ww  & \ww 
} \quad = \quad  \scalemyQ{1}{0.7}{0.5}{ & \push{\bbY}  \qw & \multigate{1}{L} & \qw & \push{\bbY} \qw & \qw &  \qw \\ & & \Nghost{L} &  \push{\bbE} \ww & \Ngate{G}{\ww} & \push{\bbE'} \ww & \ww } \quad .
\end{align}

At this early point in our story, we have to require $L$ to be one-sided for the definition of the relation to be well-tempered (though $L'$ can be two-sided, cf. \cref{def:DilOrd}); in fact, the generalisation to arbitrary dilations cannot be executed in a satisfactory manner without treating causality, which we postpone to \cref{chap:Causal}.\footnote{\label{FootnoteDilCont}The essential problem can be appreciated in a concrete example: If $\scalemyQ{.8}{0.7}{0.5}{& \push{\bbX} \qw &  \Nmultigate{1}{L}{\qw}   & \push{\bbY} \qw  & \qw  \\
		&\push{\bbD} \ww & \Nghost{L}{\ww} & \push{\bbE} \ww  & \ww
	}$ is a dilation of $\channel{\bbX}{T}{\bbY}$, and $\scalemyQ{.8}{0.7}{0.5}{& \push{\bbY} \qw &  \Nmultigate{1}{M}{\qw}   & \push{\bbZ} \qw  & \qw  \\
		&\push{\bbE} \ww & \Nghost{M}{\ww} & \push{\bbK} \ww  & \ww
	}$ a dilation of $\channel{\bbY}{S}{\bbZ}$ whose hidden input interface matches the hidden output interface of $L$, then the channel $
	\scalemyQ{.8}{0.7}{0.5}{& \push{\bbX} \qw &  \Nmultigate{1}{L}{\qw}   &  \qw & \push{\bbY} \qw  & \qw &  \qw  & \multigate{1}{M} & \push{\bbZ} \qw & \qw  \\
		&\push{\bbD} \ww & \Nghost{L}{\ww} & \push{\bbE} \ww  & \ww & & \push{\bbE} \ww& \Nghost{M}{\ww} & \push{\bbK} \ww & \ww
	}$ is a dilation of $\scalemyQ{.8}{0.7}{0.5}{& \push{\bbX} \qw & \gate{T} & \gate{S} & \push{\bbZ} \qw & \qw}$; it would seem reasonable that in the dilational ordering, this dilation should be greater than $\scalemyQ{.8}{0.7}{0.5}{& \push{\bbX} \qw &  \Nmultigate{1}{L}{\qw}   &   \push{\bbY} \qw   & \multigate{1}{M} & \push{\bbZ} \qw & \qw  \\
		&\push{\bbD} \ww & \Nghost{L}{\ww}  & \push{\bbE} \ww& \Nghost{M}{\ww} & \push{\bbK} \ww & \ww
	}$ (which is also a dilation), but the operation required to realise this is that of \emph{contracting} the interface $\bbE$, an operation which at this point is not feasible.} As such, there is a sense in which the dilational ordering introduced here is provisional. Nevertheless, its study is by no means futile -- it uncovers over-arching principles which govern the structure of dilations in many theories, and whose distillation and ramifications are the topic of the remainder of the chapter. \\

\textbf{Dilational Axioms.}  In the extreme case, the dilational ordering collapses to a single level, namely when the channel $\channel{\bbX}{T}{\bbY}$ has only the \emph{trivial} dilations $\scalemyQ{.8}{0.7}{0.5}{& \push{\bbX} \qw &  \gate{T}   & \push{\bbY} \qw  & \qw  \\
	& \push{\bbD }\ww & \Ngate{S}{\ww} & \push{\bbE} \ww & \ww	
}$ obtained by parallel composition. This phenomenon will be called \emph{dilational purity} of $T$ (\cref{subsec:Purity}), and though it is significant we cannot (and want not) expect such astounding simplicity in the structure of dilations of general channels. 

However, in \cref{sec:CompLoc} we shall see two milder forms of collapse. They will be presented as possible \emph{dilational axioms} which a theory might or might not comply to, but as we will see they reign in virtually all theories of physical interest: \\

\textbf{Completeness.} The first such axiom is the existence of \emph{complete dilations}, which are simply largest (greatest) elements in the dilational ordering, i.e. dilations from which any other dilation can be derived. In the above example with the fair die, the dilation which represented a copy of the outcome was in fact complete. In general, completeness in $\CIT$ is obtained by copying (\cref{thm:CITComp}), and likewise in cartesian theories completeness is obtained by copying (\cref{thm:CartisComp}). In $\QIT$, on the other hand, Stinespring dilations serve as complete (\cref{thm:QITComp}). 

The existence of complete dilations  in a theory is ultimately an information-theoretic principle: \emph{There is a largest amount of side-information to be had.} It provides conceptual clarity in the dilational ordering, in addition to being of technical importance in many later considerations.  \\

\textbf{Localisability of Side-Information.} The other such axiom governs not the structure of dilations of a single channel, but how the dilational structure behaves under the two modes of composition in the theory. As such, it actually comprises \myuline{two} separate axioms:

Two channels $\channel{\bbX}{T}{\bbY}$ and $\channel{\bbY}{S}{\bbZ}$ in succession define a third, serially composed channel $\scalemyQ{.8}{0.7}{0.5}{& \push{\bbX} \qw & \gate{T} & \gate{S} & \push{\bbZ} \qw & \qw}$. \emph{Temporal localisability of side-information} will refer to the principle that any (one-sided) dilation of the serial composition can be derived from the composition of (one-sided) dilations of the individual channels $T$ and $S$, that is, can be `temporally localised' in the composition. Specific examples will demonstrate that this axiom is not automatic, but its information-theoretic interpretation certainly suggests that it should hold in physically sensible theories: It expresses that any side-information about a composite channel must stem from side-information about the first and second channels. Ref. \cite{Chir11} considers a related principle (`Atomicity of Composition' -- see details below), and in the words of one of the authors (\cite{Foils}, p. 194), \emph{``Although [the failure of this principle] is logically conceivable, it raises a puzzling questions: What is the extra information about?''}  

The obvious sibling to temporal localisability is \emph{spatial localisability of side-information}, which asserts that any (one-sided) dilation of a \myuline{parallel} composition $\scalemyQ{.8}{0.7}{0.5}{& \push{\bbX_1} \qw &  \gate{T_1}   & \push{\bbY_1} \qw  & \qw  \\
	& \push{\bbX_2}\qw & \gate{T_2} & \push{\bbY_2} \qw & \qw	
}$ must derive from a parallel composition of (one-sided) dilations. Its interpretation and motivation is analogous to that of temporal localisability: Side-information about the independent execution of two channels should be `spatially localisable', as side-information about the individual channels. Again, this principle is not automatic, as demonstrated by examples.

The two localisability principles hold in many theories, in particular in all cartesian theories (\cref{thm:CartLoc}), and in the two information theories $\CIT$ and $\QIT$ (\cref{thm:CITLoc} and \cref{thm:QITLoc}). The principles have interesting consequences, most of which factor through two specific consequences: Firstly, temporal localisability implies that any channel has a \myuline{reversible} dilation, a fact that will be used in later sections over and over. Secondly, spatial localisability implies the DiVincenzo property, demonstrating that this property is not in any way a quantum feature but owes its validity to a different information-theoretic principle entirely, a circumstance which has apparently not been noted before. \\

 \textbf{Universal Dilations.} For some applications, it will be important that the channel $G$ which derives in the environment a given dilation from a complete dilation (cf. Eq. \eqref{eq:Gderive}) is unique. This phenomenon is embodied in the notion of a \emph{universal} dilation (\cref{def:Univ}) presented in \cref{sec:Universal}. In the present chapter, the significance of universal dilations over complete dilations is mainly expressed by \cref{prop:Blackwell}. In \cref{chap:Causal}, however, it will be instrumental in the form of \cref{lem:UnivCont}, which ultimately implies that we can introduce a notion of \emph{contraction} for channels (\cref{thm:StdCont}). 
 
 Cartesian theories are easily proved to have universal dilations (\cref{thm:CartUniv}), and it is also not hard to prove the existence of universal dilations for $\CIT$, by cutting down the hidden system of a complete dilation (\cref{thm:CITUniv}). The theory $\QIT$ has universal dilations too, namely minimal Stinespring dilations (\cref{thm:QITUniv}), but this fact is non-trivial and can be seen as a generalisation of the injectivity of the Choi-Jamio\l{}kowski isomorphism (\cref{lem:choigeneralised}), a result which might be of independent interest. \\

\textbf{Purification.}  The two information theories $\CIT$ and $\QIT$ share the properties of completeness (even universality), and of spatial and temporal localisability. They are distinguished, however, by \emph{purifiability} (\cref{def:SelfUniv}), which most elegantly can be described by saying that every quantum channel has a dilation which is \emph{a complete dilation of itself}, or, equivalently, which is dilationally pure. This phenomenon very effectively separates the natures of $\QIT$ and $\CIT$, and \cref{sec:Selfuniv} serves to demonstrate this separation in a sequence of results about purifiable theories. Purifiable theories include other creatures than $\QIT$, though, for example we shall see that a thin theory $(M, \star, 1, \succeq)$ is purifiable precisely if it satisfies the cancellation law $x \star y \succeq x \star z \Rightarrow y \succeq z$.

\cref{subsec:IsoPure} is about isomorphisms in purifiable theories, and proves among other things an extremely general version (\cref{cor:NoBroadCast}) of the \emph{No Broadcasting Theorem} (\cite{Barn96}).

In \cref{subsec:RevPure}, we prove a structure-theorem for reversible channels (\cref{thm:StructureOfReversibles}), which in the case of $\QIT$ specialises to the result that a quantum channel is reversible precisely if it is the tensoring with a state followed by an isometric conjugation.  

Finally, in \cref{subsec:Complementarity}, I introduce a general notion of \emph{complementarity} between channels, generalising the concept of complementary for quantum channels (\cite{Devetak05}), and made possible by a combination of purifiability with the other dilational principles. In particular, this entails an abstract version of the complementarity between reversible and `completely forgetful' channels (\cref{thm:InfoDist}). This complementarity comprises a long list of impossibility (`no go') theorems, since it implies not only a strong version of the No Broadcasting Theorem (and hence the \emph{No Cloning Theorem} (\cite{Woot82}) and the \emph{No Deletion Theorem} (\cite{Pati00})), but also the \emph{No Hiding} (\cite{Braun07}) and \emph{No Masking} (\cite{Modi18}) \emph{Theorems}.

{\centering
	\subsection*{§2. Comparison to Existing Literature.}}

\textbf{Purity from Dilations.} The most well-known notion of `purity' in information theory is arguably that of \emph{probabilistic purity}, which refers simply to convex extremality. This purity notion in general differs from that of dilational purity (\cref{def:DilPure}), cf. \cref{rem:PurityNotions}. The difference is qualitative too, in the sense that convex purity requires an ambient convex structure to make sense, whereas dilational purity is well-defined in general. 

The idea of defining purity categorically in terms of dilations is not new, but has been considered also in Ref. \cite{Chir14pure}; in lack of the DiVincenzo property, the definition given there is weaker than the one here (the latter producing the intended notion), but the ideas behind the two definitions are identical. Other authors (\cite{Cunn17}) have considered a different categorically definable purity notion (in terms of so-called \emph{factorisation systems}), but that notion is distinct from dilational purity as it reproduces convex purity in the case of $\CIT$. See \cref{rem:PurityNotions} for further details.\\

\textbf{Purification Principles.} `Purifiability' in the abstract has been considered in the literature before, in the form of the \emph{Purification Postulate} of Refs. \cite{Chir10,Chir11}. The seminal result of that work is that, within a large ground class of theories, five axioms  along with the Purification Postulate characterise the theory $\QIT$. Whereas the Purification Postulate is intimately related with \emph{purifiability} in the sense proposed here (\cref{def:SelfUniv}), there are differences, as detailed in  \cref{rem:PurePrinc}. The most important difference is that the purifiability notion proposed in this chapter is not only less restrictive than that of Refs. \cite{Chir10, Chir11} within their ground class of theories -- its effective scope is also larger outside of this ground class. Indeed, in Refs. \cite{Chir10,Chir11} the Purification Postulate is often used in combination with a number of `standing assumptions' (some of which pertain to probabilistic structure), and this has the subtle side-effect that some theories which accidentally satisfy the Purification Postulate, but violate the standing assumptions, are nothing like quantum theory.\footnote{The authors are of course aware of this circumstance, cf. the remark following Cor. 6 on p. 16 in Ref. \cite{Chir10}.} For example, every cartesian theory (e.g. the theory $\Sets^*$) would comply the the Purification Principle of Refs. \cite{Chir10,Chir11}, which pertains only to purification of states. In contrast, this does not happen for the notion of purifiability presented here, essentially because it concerns purification of all kinds of channels.

It should also be mentioned that the results derived in this chapter based on the purifiability principle (most importantly those of \cref{thm:StructureOfReversibles}, \cref{thm:Complementarity} and \cref{thm:InfoDist}) do not have counterparts in the work of Refs. \cite{Chir10,Chir11}. Furthermore, they are all derived in the simplest possible framework, from few principles, which are principles exclusively about dilations. Finally, it is interesting to note that (as explained in \cref{rem:PurePrinc}),  the Purification Postulate of Refs. \cite{Chir10,Chir11} implies the existence of complete dilations, whereas the framework here separates these two phenomena.\\

\textbf{Dilations in General and Axioms about Them.} It is well-known in the literature that general theories in the sense of \cref{def:Theory} facilitate the concepts of marginalisation and of (one-sided) dilations (\cite{Chir10,Chir14dilation}). Be that as it may, the structure of dilations has to the best of my knowledge never been studied systematically before. (Refs. \cite{Chir10,Chir11} has a dilational axiom in focus -- the `Purification Principle' mentioned just above -- but this axiom is easily stated without referring to any dilational ordering.) Whereas a principle vaguely related to that of temporal localisability is mentioned in Ref. \cite{Chir11} (see details below), the dilational ordering in general, and the completeness, (spatial) localisability and universality axioms in particular have no counterparts in the existing literature. \\

\textbf{Temporal Localisability and `Atomicity of Composition'.} Ref. \cite{Chir11} considers a principle under the name \emph{Atomicity of Composition}, according to which the serial composition of two \emph{atomic} transformations is atomic; the notion of an `atomic' transformation has, as far as I can see, no exact counterpart in the framework of this thesis, because its definition ultimately relies on convex structure (it can be seen as a broadening of convex extremality), but it is meant to capture the impossibility of extracting further information from a transformation. As such, atomicity is in spirit similar to dilational purity (formally it subsumes it), and a thoughtless identification of the two notions would thus translate to the serial composition of pure transformations being pure. That statement follows rigorously from temporal localisability (\cref{def:TempLoc}), and though it is evidently less general it would seem that the two principles are really cut from the same stone. \\

\textbf{The DiVincenzo Property and Spatial Localisability.} The conjecture of DiVincenzo about the structure of non-signalling channels in $\QIT$ has been studied extensively. It was first mentioned publicly in Ref. \cite{Beck01}, where it was shown to be true for a particular class of measurement channels. In Ref. \cite{Egg02}, its validity was extended to all channels, with a proof based on the uniqueness of Stinespring dilations; the proof was then shortened in Ref. \cite{Piani06}.\footnote{When viewed correctly, however, the proof in Ref. \cite{Piani06} is conceptually equivalent to that in Ref. \cite{Egg02}, as it too uses uniqueness of Stinespring dilations albeit in the form of purification of states as mediated by the Choi-Jamio\l{}kowski isomorphism.}  Finally, in Ref. \cite{Chir10}, it was shown that the structure-theorem holds not only in $\QIT$, but more generally in the presence of the Purification Postulate, using a uniqueness of dilations similar to that of Stinespring's. 

\cref{prop:StructureNonSignalling} in this chapter shows that, contrary to what the preceding works suggest (see also p. 201 in \cite{Foils}), the structure-theorem has nothing to do with purification. Rather, its root is the principle of spatial localisability (\cref{def:SpatLoc}), from which the DiVincenzo property can be derived in one sentence. It is true that in $\QIT$ the validity of spatial localisability is best proved by an argument involving Stinespring dilations (this provides the link to the existing proofs), but the principle itself is conceptually independent of purification -- it governs many non-quantum theories, including classical information theory.  \\

{\centering
	\subsection*{§3. Contributions.}}

The original contributions of this chapter are the following:\\

\begin{enumerate}
	\item Defining the \emph{dilational ordering} among dilations of a channel $T$ in a general theory $\Theory$ (as modelled by a symmetric monoidal category with terminal unit). 
	
	\item Identifying \emph{completeness} (\cref{def:Complete}) and \emph{localisability} (\cref{def:LocTheory}) as axioms about the dilational ordering which are simple to state, easy to interpret, true in the majority of example theories, and potent in deriving consequences. Among these consequences are the existence of reversible dilations (\cref{prop:Reversibledilations}), and the DiVincenzo property (\cref{prop:StructureNonSignalling}), which has often before been presented as a quantum feature resting on Stinespring's dilation theorem.
	
	\item Identifying the concept of \emph{universal dilations}, which reveal the precise structure of the dilational ordering (\cref{prop:Blackwell}), have occasional technical significance (e.g. in \cref{thm:StructureOfReversibles}), and will be instrumental in defining in \cref{chap:Causal} a new operation (\emph{contraction}) generalising and connecting the works of Refs. \cite{JSV96} and \cite{Chir09combs}. In proving the existence of universal dilations in $\QIT$, the injectivity of the Choi-Jamio\l{}kowski isomorphism (\cite{Jiang13,Pill67,Choi75,Jam72}) is generalised in a non-trivial way from pure states of full rank to isometric channels of full rank (\cref{lem:choigeneralised}). This is the most technical result of the chapter.

		\item Presenting a notion of \emph{purifiability} (\cref{def:SelfUniv}) which broadens the scope of the `Purification Postulate' of Refs. \cite{Chir10,Chir11}, and in conjunction with other dilational principles allows the derivation of new general results, including a structure-theorem for reversible transformations (\cref{thm:StructureOfReversibles}), and an abstract notion of \emph{complementarity} between channels (\cref{thm:Complementarity}), which places the notion of complementary quantum channels (\cite{Devetak05}) in an general setting. 
	
	\item Proving for the first time a general version of the complementarity between reversible and completely forgetful\footnote{A \emph{completely forgetful} channel is a channel which is the serial composition of a trash and a state.} channels (\cref{thm:InfoDist}) known from quantum information theory. This theorem moreover comprises many of the impossibility theorems known in the case of $\QIT$, including the fairly recent \emph{No Hiding} and \emph{No Masking Theorems} (\cite{Braun07, Modi18}), which to my best knowledge have not been proved in general frameworks before.

	\end{enumerate}

Most importantly, however, all of the principles which we entertain in this chapter are \emph{principles about dilations and nothing else}. As such, the results of the chapter constitute a proof of concept, demonstrating that a remarkable amount of features, including many features of quantum information theory, can be derived from a small set of principles which pertain exclusively to the nature of dilations. In particular, neither the statement of these principles nor the proofs of their consequences depend in any way on probabilistic structure or on `dagger-compact structure' (\cite{AbCo04}). In short, the framework is the most general possible in which it even makes sense to speak of dilations.

\newpage

\section{A Terminal Trinity: Marginals, Dilations and Non-Signalling}
\label{sec:Trinity}

In \cref{def:Theory}  we imposed on a theory the requirement that the trivial system $\triv$ be terminal. This requirement supports the concepts of \emph{marginalisation}, \emph{dilation} and \emph{non-signalling}. These three concepts are reviewed and exemplified in this section.\\

Marginalisation we already saw in the form of (somewhat intermittently named) \emph{factor projections} in connection with cartesian theories in \cref{subsec:Cart}. To treat general marginalisation succinctly and precisely, however, we need to work in the realm of channels and interfaces (cf. \cref{def:Inter} and \cref{def:Channel}):

\begin{Definition} (Projections and Marginals.) \\
	Let $\bbY$ be an interface in a theory $\Theory$, and let $\bbY_0 \subseteq \bbY$ be a sub-interface. The \emph{projection from $\bbY$ to $\bbY_0$} is the channel $\pi_{\bbY \to \bbY_0}: \bbY \to \bbY_0$ given by $\id_{\bbY_0} \og \tr_{\bbY \setminus \bbY_0}$. Pictorially, $\pi_{\bbY \to \bbY_0}$ is  
	
	\begin{align} 
	\myQ{1}{1}{&  \push{\bbY_0} \qw &  \qw   & \qw  \\
		& \push{\bbY \setminus \bbY_0} \qw  	& \Ngate{\tr}{\qw}} \quad  
	\left( \text{abbreviating} \quad 
	\scalemyQ{.7}{1}{1}{& &  \qw  & \qw & \qw  \\
		&  \bbY_0 \qquad &	\vdots \\
		&	&  \qw  & \qw  & \qw \\
		& &  \qw  & \gate{\tr} \\
		& \bbY \setminus \bbY_0 \qquad &   \vdots 	& & \\
		& &  \qw  & \gate{\tr}  } \quad \right).
	\end{align}

	For a channel $T: \bbX\to \bbY$, the \emph{$\bbY_0$-marginal of $T$} is the serial composition $\pi_{\bbY\to \bbY_0}\after T : \bbX \to \bbY_0$. Pictorially, the $\bbY_0$-marginal of $T$ is  
	
		\begin{align}
	\myQ{1}{0.7}{& \push{\bbX} \qw &  \Nmultigate{1}{T}{\qw}   & \push{\bbY_0} \qw  & \qw  \\
		& & \Nghost{T} & \push{\bbY \setminus \bbY_0} \qw  & \Ngate{\tr}{\qw}
	}		
	\quad  
	\left( \text{abbreviating} \quad 	
	\scalemyQ{.7}{1}{1}{& & \qw & \multigate{5}{T}   & \qw & \qw  \\
		&& 	&  \Nghost{T} &  	\vdots & \bbY_0 \\
		& \bbX \quad & \vdots	&\Nghost{T}	  & \qw  & \qw \\
		& &  	& \Nghost{T} &  \qw  & \gate{\tr} \\
		& & & \Nghost{T} &  	\vdots & \qquad \bbY \setminus \bbY_0 \\
		& & \qw 	& \ghost{T} &  \qw  & \gate{\tr} }	\quad \right).	
	\end{align}

	\end{Definition}

A given channel $T: \bbX \to \bbY$ has $2^\abs{\bbY}$ marginals, one for each sub-interface of $\bbY$. The $\bbY$-marginal is the full channel $T$, whereas the $\bbI$-marginal is the trash $\tr_\bbX$. The intermediate marginals are typically more interesting.

\begin{Example} (Marginals in $\Sets^*$.) \\
	Let $\onetwochannel{X}{f}{Y}{Z}$ be a channel in $\Sets^*$. Since $\Sets^*$ is cartesian, $f$ must take the form $x \mapsto (f_1(x), f_2(x))$ for some functions $f_1: X \to Y$ and $f_2: X \to Z$, and these uniquely determine $f$. These two functions are precisely the marginals of $f$ corresponding to the two simple interfaces for $Y$ and $Z$. (This example generalises to any cartesian theory.) 
\end{Example}

\begin{Example} (Marginals in $\CIT$.)\\
		In $\CIT$, marginalisation specialises to the usual notion in terms of summing over elements of the discarded systems. In particular, the marginals of a state $\bistate{p}{X}{Y}$ are the states $\state{p_{\sfx}}{X}$ and $\state{p_{\sfy}}{Y}$ with densities given by $p_\sfx(x) = \sum_{y \in Y} p(x,y)$ and $p_\sfy(y) = \sum_{x \in X} p(x,y)$, respectively. Marginals of a general channel in $\CIT$ is given by marginalising each component. 
\end{Example}

\begin{Example} (Marginals in $\QIT$.)\\
	Marginalisation in $\QIT$ also specialises to the usual notion, since projections are given by the \emph{partial trace}. For example, both non-trivial marginals of the maximally entangled qubit $\bistate{\phi}{\C^2}{\C^2}$ with vector representative $\ket{\phi} = \frac{\ket{0}\otimes \ket{0}+ \ket{1} \otimes \ket{1}}{\sqrt{2}}$ are given by the fully mixed qubit state $\state{\tau_{\C^2}}{\C^2}$. 
	\end{Example}

\begin{Example} (Marginals in a Thin Theory.) \\
	Let $\Theory$ be a thin theory, identified with the pre-ordered quasi-commutative monoid $(M, \star, 1, \succeq)$. A channel $\scalemyQ{.7}{0.7}{0.5}{& & \Nmultigate{2}{\phantom{X}} & \push{y_1} \qw & \qw \\ 
	& \push{x} \qw & \ghost{\phantom{X}} & \push{y_2} \qw & \qw \\
 & & \Nghost{\phantom{X}} & \push{y_2} \qw & \qw }$ is simply the relationship  $x \succeq y_1 \star y_2 \star y_3$. In addition the two trivial marginals $x \succeq y_1 \star y_2 \star y_3$ and $x \succeq 1$, there are six marginal relationships: $x \succeq y_2 \star y_3$, $x \succeq y_1 \star y_3$, $x \succeq y_1 \star y_2$, $x \succeq y_1$, $x \succeq y_2$ and $x \succeq y_3$. 	\end{Example}

\vspace{.5cm}

Now, \emph{dilations} are usually defined as dual to marginals, but we shall employ a slightly more general definition, according to which dilations of a channel $T: \bbX\to \bbY$ may not only have outputs additional to those of $\bbY$, but also \myuline{inputs} additional to those of $\bbX$. The reason for this is both one of more robust interpretation (two-sided dilations model general \emph{side-computations}, whereas one-sided dilations model only \emph{escaping side-information}), and one of mathematical robustness.

\begin{Definition} (Dilations.)\label{def:dilation} \\
	Let $T: \bbX \to \bbY$ be a channel in $\Theory$. A \emph{dilation of $T$} is a channel $L: \bbX \cup \bbD \to \bbY \cup \bbE$ whose $\bbY$-marginal is $T \og \tr_{\bbD}$, i.e. for which
	
		\begin{align}
	\myQ{0.7}{0.5}{& \push{\bbX} \qw &  \Nmultigate{1}{L}{\qw}   & \push{\bbY} \qw  & \qw  \\
		&\push{\bbD} \ww & \Nghost{L}{\ww} & \push{\bbE} \ww  & \Ngate{\tr}{\ww}
	}		
	\quad   = \quad 
	\myQ{0.7}{0.5}{& \push{\bbX} \qw &   \Ngate{T}{\qw}    & \push{\bbY} \qw  & \qw  \\
		&\push{\bbD} \ww  & \Ngate{\tr}{\ww}
	}	\quad .
	\end{align}

The interfaces $\bbD$ and $\bbE$ are called \emph{hidden} or \emph{inaccessible (relative to $T$)}, and by convention we use wiggly lines to pictorially represent the wires corresponding to the hidden interfaces. A dilation a called \emph{one-sided} if $\bbD = \bbI$.

	\end{Definition}

\begin{Remark} (On the Importance of Normality for Dilations.) \\
The normality condition on $\Theory$, cf. \cref{def:Normal} ($T \og \tr_\cZ = T' \og \tr_\cZ \Rightarrow T=T'$), is crucial for the concept of dilations not to be pathological. Indeed, if $\Theory$ is not normal and if $T, T': \bbX \to \bbY$ are distinct channels with $T \og \tr_\bbZ = T' \og \tr_\bbZ$ for some interface $\bbZ$, then the channel $L:= T \og \tr_\bbZ = T' \og \tr_\bbZ$ is a dilation of both $T$ and $T'$. In other words, one cannot tell from the dilation $L$ which is the channel that it dilates, although we know which are the hidden interfaces.  \end{Remark}

\begin{Remark} (On the Importance of Interfaces for Dilations.)\\
	Consider in the thin theory $(\N_0, \max, 0, \geq)$ from \cref{ex:TwoCompositions} a channel $\channel{x}{\phantom{X}}{y}$ with $10 \geq x \geq y$. The channel $\scalemyQ{.8}{0.7}{0.5}{& \push{x} \qw &  \gate{\phantom{X}}   & \push{y} \qw  & \qw  \\
		& \push{10}\ww & \Ngate{\phantom{X}}{\ww} & \push{10} \ww & \ww	
	}$ is a dilation. As a \myuline{channel}, it determines the channel $\channel{x}{\phantom{X}}{y}$ that it dilates, because it formally determines its input and output interfaces which determine $x$ and $y$. As a \myuline{transformation}, however, it is simply the relationship $\max\{10,x\} \geq \max\{10,y\}$, i.e. $10 \geq 10$, with no recognition of the values of $x$ and $y$, that is, of the channel that it dilates. 
	\end{Remark}

To characterise the dilations of a given channel, it obviously suffices to characterise those whose hidden interfaces are simple. Nevertheless, it is generally a challenging task to do this (indeed, this is why it is possible to write an entire chapter about it). At this point, we have only a few such characterisations within reach:

\begin{Example} (Dilations in Cartesian Theories.) \label{ex:dilSets}\\
 Let $\channel{X}{T}{Y}$ be a channel in $\Sets^*$ (or indeed in any other cartesian theory). We can determine all of its dilations. Indeed, given a dilation $ \scalemyQ{.8}{0.7}{0.5}{ &\push{D} \ww & \Nmultigate{1}{L}{\ww} & \push{E} \ww & \ww \\ &\push{X} \qw &  \ghost{L} &\push{Y} \qw & \qw}$, consider the channel

 	\begin{align} \label{eq:DilGuess}
 \scalemyQ{1}{0.7}{0.5}{& &\push{D} \ww & \Nmultigate{1}{L}{\ww} & \push{E} \ww & \ww \\ & &  & \ghost{L} & \gate{\tr}  \\ & \push{X} \qw & \ctrlo{-1} & \gate{T} & \push{Y} \qw & \qw}, 
 \end{align}
 
 where (following Ref. \cite{Fritz20Synthetic}) the channel $\scalemyQ{.8}{0.7}{0.5}{& & & \qw \\ &\push{X} \qw & \ctrlo{-1} & \qw}$ denotes the diagonal function $x \mapsto (x,x)$ (or, in a general cartesian theory, the unique channel with both marginals equal to $\id_X$). The channel \eqref{eq:DilGuess} is easily seen to have precisely  the same marginals as $L$, so by the cartesian property it must in fact be \myuline{equal} to $L$. (In the absence of a hidden input, this is easy to understand intuitively in the theory $\Sets^*$: A one-sided dilation $L$ is determined by its marginals; one of those is $T$, and the other is $\scalemyQ{.8}{0.7}{0.5}{ & & \Nmultigate{1}{L} & \push{E} \ww & \ww \\ & \push{X} \qw  & \ghost{L} & \gate{\tr}}$.)  Now, this shows that any dilation of $T$ is of the form 
 
 	\begin{align} \label{eq:dilTall}
 \scalemyQ{1}{0.7}{0.5}{& &\push{D} \ww & \Nmultigate{1}{G}{\ww} & \push{E} \ww & \ww \\ & &  & \ghost{G} & \\ & \push{X} \qw & \ctrlo{-1} & \gate{T} & \push{Y} \qw & \qw},
 \end{align}
 
 for some channel $G$. Conversely, it is obvious that the channel \eqref{eq:dilTall} is a dilation for any choice of $G$, so these are precisely the dilations of $T$. 

\end{Example}

\begin{Example} (Dilations in Thin Theories.) \\
	Let $\Theory$ be a thin theory, described the pre-ordered quasi-commutative monoid $(M, \star, 1, \succeq)$. Interfaces $\bbX$ and $\bbY$ in $\Theory$ are simply tuples of elements $(x_1,  \ldots, x_n) \in M^n$ and $(y_1, \ldots, y_m) \in M^m$, and a channel $\bbX \succeq \bbY$ is nothing but the relationship $x_1 \star \ldots \star x_n \succeq y_1 \star \ldots \star y_m$. A dilation of $\bbX \succeq \bbY$ is determined by a pair of interfaces $\bbD = (d_1, \ldots, d_k)$ and $\bbE=(e_1, \ldots, e_\ell)$ such that $\bbX \cup \bbD \succeq \bbY \cup \bbE$, i.e. $x_1 \star \ldots \star x_n \star d_1 \star \ldots \star d_k \succeq y_1 \star \ldots \star y_m \star e_1 \star \ldots \star e_\ell$. Of course, these can be understood by understanding the dilations with simple hidden interfaces, namely the pairs of elements $d, e \in M$ which satisfy $(x_1 \star \ldots \star x_n) \star d \succeq (y_1 \star \ldots \star y_m) \star e$. (The general dilations then arise by $\star$-factorisations of $d$ and $e$.)

 	\end{Example}

In other theories, even determining all the \myuline{one-sided} dilations of \myuline{states} is non-trivial:

\begin{Example} (One-Sided Dilations of States in $\CIT$.) \label{ex:dilCIT} \\
	Let $\state{p}{X}$ be a state in $\CIT$. What are its one-sided dilations? Taking the hidden interface $\bbE$ to be simple, we are looking for states $
	\scalemyQ{0.8}{0.7}{0.5}{& \Nmultigate{1}{\ell}  & \push{E} \ww & \ww    \\ & \Nghost{\ell}  & \push{X} \qw & \qw  }$ with $
	\scalemyQ{0.8}{0.7}{0.5}{& \Nmultigate{1}{\ell}  & \push{E} \ww & \Ngate{\tr}{\ww}   \\ & \Nghost{\ell}  & \push{X} \qw & \qw  } = \state{p}{X}$. Some obvious dilations are the \emph{trivial} dilations of the form $\scalemyQ{.8}{0.7}{0.5}{& \Ngate{r}  & \push{E} \ww & \ww    \\ & \Ngate{p}  & \push{X} \qw & \qw  }$, corresponding to independent side-information; but we also have e.g. the dilation $\scalemyQ{.8}{0.7}{0.5}{& \Nmultigate{1}{\hat{p}}  & \push{X} \ww & \ww    \\ & \Nghost{\hat{p}}  & \push{X} \qw & \qw  } := \scalemyQ{.8}{0.7}{0.5}{&   & & \push{X} \ww & \ww    \\ & \Ngate{p}  & \ctrlo{-1} & \push{X} \qw & \qw  } $ given by \myuline{copying}, i.e. with density given by $\hat{p}(x,x')=  \delta_{x,x'}p(x)$. Intuitively, a copy seems like the strongest sort of side-information we can have, and this intuition can be formalised: Every one-sided dilation $\ell$ of $p$ is in fact of the form

	\begin{align}  \label{eq:CITstateext}
	\myQ{0.7}{0.5}{& \Nmultigate{1}{\hat{p}}  & \push{X} \ww & \Ngate{G}{\ww} & \push{E} \ww & \ww    \\ & \Nghost{\hat{p}}  & \push{X} \qw & \qw  & \qw & \qw}
	\end{align}
	
	for some channel $G$. It is easy to see that any $G$ yields a dilation. The converse owes to the existence of conditional distributions: Because the $X$-marginal of a dilation $\ell$ is $p$, the density $\ell: X \times E \to [0,1]$ can be expressed as $\ell(x,e) =  g_x(e) p(x)$ for probability densities $g_x : E \to [0,1]$, which we may interpret as the \emph{conditional distribution} (w.r.t. $\ell$) of the $E$-output given the $X$-output. However, this precisely means that if we take $G=(g_x)_{x \in X}$, then \eqref{eq:CITstateext} is satisfied. (Observe that $g_x$ is uniquely determined if $p(x)> 0$, whereas it can be chosen arbitrarily if $p(x)=0$.)\end{Example}

\begin{Example} (One-Sided Dilations of States in $\QIT$.) \label{ex:dilQIT}\\
Let $\state{\varrho}{\cH}$ be a state in $\QIT$, and let us determine its one-sided dilations $\scalemyQ{.8}{0.7}{0.5}{& \Nmultigate{1}{\xi}  & \push{\cE} \ww & \ww    \\ & \Nghost{\xi}  & \push{\cH} \qw & \qw  }$. By a well-known result, the state $\varrho$ has a \emph{purification}, that is, a dilation $\scalemyQ{.8}{0.7}{0.5}{& \Nmultigate{1}{\pi}  & \push{\cH} \ww & \ww    \\ & \Nghost{\pi}  & \push{\cH} \qw & \qw }$ with $\pi$ (probabilistically) pure, and this purification is unique up to isometric conjugations on the hidden system, cf. the paragraph on Stinespring Representations in the preliminary section on the thesis. Like the classical copy in \cref{ex:dilCIT}, the fixed purification $\pi$ can help us determine all possible dilations of $\varrho$: Indeed, if $\scalemyQ{.8}{0.7}{0.5}{& \Nmultigate{1}{\xi}  & \push{\cE} \ww & \ww    \\ & \Nghost{\xi}  & \push{\cH} \qw & \qw  }$ is any dilation, let $\scalemyQ{.8}{0.7}{0.5}{& \Nmultigate{2}{\pi'} & \push{\cE'} \ww & \ww \\& \Nghost{\pi'}  & \push{\cE} \qw & \qw    \\ & \Nghost{\pi'}  & \push{\cH} \qw & \qw  }$ be a purification of $\xi$; then $\pi$ and $\pi'$ are both purifications of $\varrho$, and by the uniqueness clause we find an isometric channel $\Sigma$ such that  $\scalemyQ{.8}{0.7}{0.5}{&  &    & \Nmultigate{1}{\Sigma} & \push{\cE'} \ww & \ww\\ & \Nmultigate{1}{\pi}  & \push{\cH} \ww & \Nghost{\Sigma}{\ww} & \push{\cE} \qw & \qw    \\ & \Nghost{\pi}  & \push{\cH} \qw & \qw & \qw } =\scalemyQ{.8}{0.7}{0.5}{& \Nmultigate{2}{\pi'} & \push{\cE'} \ww & \ww \\& \Nghost{\pi'}  & \push{\cE} \qw & \qw    \\ & \Nghost{\pi'}  & \push{\cH} \qw & \qw  }$. By trashing $\cE'$, we find that $\scalemyQ{.8}{0.7}{0.5}{& \Nmultigate{1}{\xi}  & \push{\cE} \ww & \ww    \\ & \Nghost{\xi}  & \push{\cH} \qw & \qw  }$ equals

	\begin{align}   \label{eq:stateQIT}
\myQ{0.7}{0.5}{& \Nmultigate{1}{\pi}  & \push{\cH} \ww & \Ngate{\Gamma}{\ww} & \push{\cE} \ww & \ww    \\ & \Nghost{\pi}  & \push{\cH} \qw & \qw  & \qw & \qw} \quad,
\end{align}

with $\Gamma$ denoting the marginal of $\Sigma$. Hence, the one-sided dilations of $\state{\varrho}{\cH}$ are precisely those states of the form \eqref{eq:stateQIT} for some channel $\Gamma$. \end{Example}

\vspace{.5cm}

Based on \cref{ex:dilSets}, \cref{ex:dilCIT} and \cref{ex:dilQIT}, it is very natural to suspect that the `one-dilation-to-rule-them all'-phenomenon is more than a mere coincidence. This will be the topic of \cref{subsec:Complete} when we introduce and study \emph{complete} dilations.\\

For now, however, we shall raise a different question, namely how we might hope to systematically understand the two-sided dilations in $\CIT$ and $\QIT$ (and for that matter in other theories) based on the one-sided dilations. To have a better perspective on this question, it is beneficial to introduce an additional concept, namely that of \emph{non-signalling} (\cite{Foils}):  

\begin{Definition} (Non-Signalling.) \label{def:NonSignalling} \\
	Let $T: \bbX \to \bbY$ be a channel in $\Theory$, and let $\bbX_0$ and $\bbY_0$ be sub-interfaces of $\bbX$ and $\bbY$, respectively. We say that \emph{$T$ is non-signalling from $\bbX_0$ to $\bbY_0$} if there exists a channel $T':\bbX \setminus  \bbX_0 \to \bbY_0$ such that
	
		\begin{align}
	\myQ{1}{0.7}{& \push{\bbX \setminus \bbX_0} \qw &  \Nmultigate{1}{T}{\qw}   & \push{\bbY_0} \qw  & \qw  \\
		&\push{\bbX_0} \qw & \Nghost{T}{\qw} & \push{\bbY \setminus \bbY_0} \qw  & \Ngate{\tr}{\qw}
	}		
	\quad   = \quad 
	\myQ{1}{0.7}{& \push{\bbX \setminus \bbX_0} \qw &   \Ngate{T'}{\qw}    & \push{\bbY_0} \qw  & \qw  \\
		&\push{\bbX_0} \qw  & \Ngate{\tr}{\qw}
	}	\quad .
	\end{align}

\end{Definition}

 The interpretation of non-signalling is disclosed by its very name: Non-signalling from $\bbX_0$ to $\bbY_0$ means that no inputs in the interface $\bbX_0$ affect the outputs in the interface $\bbY_0$.

\begin{Example} (Bell-Channels and Non-Signalling.) \label{ex:Bellch} \\
	Consider in a theory $\Theory$ a channel of form
	
	\begin{align} \label{eq:BellCh}
	\myQ{1}{0.7}{
		& \qw	& \push{\cX_1}  \qw   & \multigate{1}{T_1} & \push{\cY_1}  \qw & \qw   \\
		& \Nmultigate{1}{s}  & \push{\cZ_1} \qw & \ghost{T_1} &  \\
		& \Nghost{s} & \push{\cZ_1} \qw  & \multigate{1}{T_2}  \\
		& \qw 	&   \push{\cX_2} \qw  & \ghost{T_2} & \push{\cY_2}  \qw & \qw \\
	} \quad,
\end{align}

as in our earlier discussion of the pictorial syntax. Let us call such a channel a \emph{(bipartite) Bell-channel}, as it corresponds to the experimental set-up considered by J. Bell (\cite{Bell64}): A state $s$ is shared across two sites, and locally at each site a channel $T_i$ can be applied to part of the state and a local input. The ground-breaking demonstration of \cite{Bell64} (mentioned in the general introduction, and alluded to in connection with \cref{ex:QIT}) was that the components $T_1, T_2$ and $s$ can be chosen from $\QIT$ in such a way that the total channel \eqref{eq:BellCh} is interpretable in $\CIT$, though no choice of $T_1, T_2$ and $s$ from $\CIT$ will reproduce it. A key point in this argument is that it is quite easy to understand the Bell-channels in $\CIT$, because any randomness from the channels $T_i$ can be extracted and moved to $s$, thus effectively realising the channel \eqref{eq:BellCh} as a convex combination of products of deterministic channels (functions).

It is not clear that the Bell-channels in $\QIT$ are similarly easy to characterise, but one thing can be said, in fact about Bell-channels in any theory: They must be \myuline{non-signalling} from $\cX_1$ to $\cY_2$ and from $\cX_2$ to $\cY_1$. Indeed, as demonstrated by our earlier computation,

\begin{align} 
\scalebox{0.7}{$\myQ{1}{0.7}{
	& \push{\cX_1}  \qw   & \multigate{1}{T_1} & \push{\cY_1}  \qw & \gate{\tr}   \\
	& \Nmultigate{1}{s}   & \ghost{T_1} &  \\
	& \Nghost{s}   & \multigate{1}{T_2}  \\
	&   \push{\cX_2} \qw  & \ghost{T_2} & \push{\cY_2}  \qw & \qw \\
} 
\quad = \quad 
\myQ{1}{0.7}{
	& \push{\cX_1}  \qw   & \multigate{1}{\tr}    \\
	& \Nmultigate{1}{s}  & \ghost{\tr} &  \\
	& \Nghost{s}   & \multigate{1}{T_2}  \\
	&   \push{\cX_2} \qw  & \ghost{T_2}& \push{\cY_2}  \qw & \qw  \\
} 
\quad = \quad 
\myQ{1}{0.7}{
	& \push{\cX_1}  \qw   & \gate{\tr}    \\
	& \Nmultigate{1}{s}  & \gate{\tr} &  \\
	& \Nghost{s} & \multigate{1}{T_2}  \\
	&   \push{\cX_2} \qw  & \ghost{T_2} & \push{\cY_2}  \qw & \qw \\
} 
\quad = \quad \myQ{1}{0.7}{
	& \push{\cX_1}  \qw   & \gate{\tr}    \\
	&   \push{\cX_2} \qw  & \gate{T'_2} & \push{\cY_2}  \qw & \qw \\
}$} \quad, 
\end{align}

and similarly with $\cY_2$ trashed in place of $\cY_1$. 

For some time, it was unknown whether the class of Bell-channels in $\QIT$ (with classical inputs and outputs) was restrained only by these two non-signalling conditions. The work of Refs. \cite{Cir80, PR94} settled this in the negative, and we now know that in fact the class of Bell-channels in $\QIT$ is highly complex (\cite{Brun14,Goh18}).

\end{Example} 

\begin{Example} (The PR Box.) \label{ex:PR}\\
	The concrete example given by S. Popescu and D. Rohrlich in Ref. \cite{PR94} of a non-signalling channel in $\CIT$, which cannot be realised as a Bell-channel in $\QIT$, was the classical channel $\scalemyQ{.8}{0.7}{0.5}{& \push{\{0,1\}} \qw& \multigate{1}{P} & \push{\{0,1\}} \qw & \qw \\ & \push{\{0,1\}} \qw& \ghost{P} & \push{\{0,1\}} \qw & \qw }$, determined by the probability distributions $(P^{x_1, x_2})_{x_1, x_2 \in \{0,1\}}$ given by 
	
	\begin{align}
		P^{x_1, x_2} (y_1, y_2) = \begin{cases}\frac{1}{2} &\text{for $y_1 \oplus y_2 = x_1 \cdot x_2$} \\ 
	0 &\text{for $y_1 \oplus y_2 \neq x_1 \cdot x_2$}  \end{cases},
	\end{align}
	
	with $\oplus$ denoting addition modulo $2$. It is easy to verify that each of the marginals equals the channel $\scalemyQ{.8}{0.7}{0.5}{ &\push{\{0,1\}} \qw & \gate{\tr} \\& \push{\{0,1\}} \qw & \gate{F} & \push{\{0,1\}} \qw& \qw}$, with $F$ the channel given by $F(k)= \frac{1}{2} \delta_0 + \frac{1}{2} \delta_1$ for $k=0,1$, so in particular $P$ is non-signalling. The reader is not expected to see why $P$ cannot be realised as a Bell-channel, but this is the result of Refs. \cite{Cir80,PR94} (it follows as $P$ violates Cirelson's bound mentioned in the general introduction). The channel $P$ has since become known in the folklore as the \emph{Popescu-Rohrlich Box}, or simply \emph{PR Box} (though it had in fact been identified earlier, e.g. by Cirelson himself (\cite{Cir93})). \end{Example}

\begin{Example} (Non-Signalling in $\Logic$.) \\
	Consider the thin theory of propositional logic, $\Logic$, from \cref{ex:Logic}. The channel $\scalemyQ{.7}{0.7}{0.5}{& \push{P_0} \qw & \multigate{1}{\phantom{X}} &  \push{P_1} \qw & \qw \\ & \push{Q_0} \qw & \ghost{\phantom{X}} &  \push{Q_1} \qw & \qw}$ asserts the relation $ P_0 \land Q_0 \succeq P_1 \land Q_1$ (i.e. "$P_0 \land Q_0$ implies $P_1 \land Q_1$"). To say that the channel is non-signalling from the $P_0$-port  to the $Q_1$-port would be to say that the marginal relation $\scalemyQ{.7}{0.7}{0.5}{& \push{P_0} \qw & \multigate{1}{\phantom{X}} &  \push{P_1} \qw & \qw \\ & \push{Q_0} \qw & \ghost{\phantom{X}} & \gate{\tr} }$ is of the form $\scalemyQ{.7}{0.7}{0.5}{& \push{P_0} \qw & \gate{\phantom{X}} &  \push{P_1} \qw & \qw \\ & \push{Q_0} \qw & \gate{\tr} &}$, or, what is actually equivalent, that "$P_0$ implies $P_1$". This of course does not generally follow from "$P_0 \land Q_0$ implies $P_1 \land Q_1$", so not all channels in the theory  $\Logic$ are non-signalling. 

\end{Example}

 Non-signalling as such shall not occupy us before \cref{chap:Causal}, but it is relevant in the  context of dilations because, due to our choice of defining them as two-sided, dilations and non-signalling are intimately related -- indeed, $T: \bbX \to \bbY$ is non-signalling from $\bbX_0$ to $\bbY_0$ if and only if it is the dilation of some channel $T': \bbX \setminus \bbX_0 \to \bbY_0$. This also means that any understanding we have of non-signalling channels can be transferred to an understanding about two-sided dilations. We end this section by discussing one such situation. \\ 
 
 It is obvious that a channel of the form

 	\begin{align} \label{eq:NS}
 \myQ{0.7}{0.5}{
 	& \push{\bbX \setminus \bbX_0}  \qw    &\multigate{1}{T_1}   & \qw &   \push{\bbY_0} \qw & \qw & \qw \\
 	&                   & \Nghost{T_1}  & \push{\bbH} \qw & \multigate{1}{T_2} & \\
 	& \qw &  \push{\bbX_0}  \qw & \qw &  \ghost{T_2}   &  \push{\bbY \setminus \bbY_0} \qw  & \qw  \\
 } 	\end{align}

is non-signalling from $\bbX_0$ to $\bbY_0$; trashing $\bbY \setminus \bbY_0$ results in trashing $\bbX_0$. More than 20 years ago, it was conjectured by DiVincenzo  (\cite{Beck01,Egg02}) that in $\QIT$ channels of the form \eqref{eq:NS} are in fact the only non-signalling channels. This conjecture was proven in Ref. \cite{Beck01} in a special case, and later in Ref. \cite{Egg02} for arbitrary channels. Phrased in terms of dilations, this means that in $\QIT$ every dilation of a channel  $\channel{\bbX}{T}{\bbY}$ must take the form 

	\begin{align}  \label{eq:dilNons}
\myQ{0.7}{0.5}{ & \push{\bbX}  \qw & \multigate{1}{L_0} & \qw & \push{\bbY} \qw & \qw &  \qw \\ & & \Nghost{L_0} &  \push{\bbE_0} \ww & \Nmultigate{1}{G}{\ww} & \push{\bbE} \ww & \ww\\
	& 	&  \push{\bbD} \ww &\ww & \Nghost{G}{\ww}} 
\end{align}

for some channels $L_0$ and $G$. Moreover, by trashing $\bbE$ we easily read off that $L_0$ is in this case a one-sided dilation of $T$. Conversely, the channel \eqref{eq:dilNons} defines a dilation of $T$ for any one-sided dilation $L_0$ and any channel $G$. \textbf{In other words, to determine the dilations of a channel in $\QIT$, we need only determine the one-sided dilations.} (In particular, this immediately hands us all dilations of states in $\QIT$, cf. \cref{ex:dilQIT}.) 

Let us name this phenomenon generally in DiVincenzo's honour:

\begin{Definition} (The DiVincenzo Property.) \label{def:DiVi}\\
	We say that a theory $\Theory$ \emph{has the DiVincenzo property} if every channel $\channel{\bbX}{T}{\bbY}$ which is non-signalling from $\bbX_0$ to $\bbY_0$ is of the form 
	
		\begin{align} 
	\myQ{0.7}{0.5}{
		& \push{\bbX \setminus \bbX_0}  \qw    &\multigate{1}{T_1}   & \qw &   \push{\bbY_0} \qw & \qw & \qw \\
		&                   & \Nghost{T_1}  & \push{\bbH} \qw & \multigate{1}{T_2} & \\
		& \qw &  \push{\bbX_0}  \qw & \qw &  \ghost{T_2}   &  \push{\bbY \setminus \bbY_0} \qw  & \qw  \\
	} 	\end{align}

	for some channels $T_1$ and $T_2$.
	
	\end{Definition}

\begin{Remark}
When checking for the DiVincenzo Property, it obviously suffices to consider the case where $\bbX$ and $\bbY$ are bipartite interfaces and $\bbX_0$ and $\bbY_0$ are simple (i.e. single-port) sub-interfaces. 

	\end{Remark}

\begin{Example} (Failure of the DiVincenzo Property.) \label{ex:DiViFail}\\
	Consider the thin theory $(\N, \cdot, 1, \geq)$ from \cref{ex:ThinNat}. The channel $\scalemyQ{.8}{0.7}{0.5}{& \push{7} \qw & \multigate{1}{\phantom{X}} & \push{4} \qw & \qw \\ & \push{2} \qw & \ghost{\phantom{X}} & \push{3} \qw & \qw}$ is non-signalling from the $2$-port to the $4$-port, since $7 \geq 4$. However, it cannot be written as $	\scalemyQ{.8}{0.7}{0.5}{
		& \push{7}  \qw    &\multigate{1}{\phantom{X}}   & \qw &   \push{4} \qw & \qw & \qw \\
		&                   & \Nghost{\phantom{X}}  & \push{z} \qw & \multigate{1}{\phantom{X}} & \\
		& \qw &  \push{2}  \qw & \qw &  \ghost{\phantom{X}}   &  \push{3} \qw  & \qw  \\
	} $ for any $z \in \N$, since $7 \geq 4 \cdot z$ forces $z=1$, which violates $2 \cdot z \geq 3$.
 
	\end{Example}

In the next section, we will see that the DiVincenzo property is very generic, and we will understand that it is a consequence of an information-theoretic principle governing dilations. As observed just above, this will greatly simplify the study of dilations. \\

We must begin, however, by giving shape to the intuition that some dilations can be \emph{constructed} from other dilations, a phenomenon which we also observed in connection with \cref{ex:dilSets}, \cref{ex:dilCIT}, and \cref{ex:dilQIT}.

\section{The Dilational Ordering}
\label{sec:DilOrd}

Dilations are not isolated islands -- some of them are connected by bridges. Ideally, these bridges should connect general two-sided dilations to general two-sided dilations, but it turns out that we will not be successful in attempting such a definition before introducing the theory of \emph{causal} channels in \cref{chap:Causal}. For now, we must restrict our attention to the case where at least the initial dilation is one-sided:

\begin{Definition} (The Dilational Ordering.) \label{def:DilOrd}\\
	Let $\channel{\bbX}{T}{\bbY}$ be a channel in $\Theory$, and denote by $\Dil{T}$ the class of all its dilations. The \emph{dilational ordering on $\Dil{T}$} is the relation $\der_T$ on $\Dil{T}$ given by 
	
		\begin{align} \label{eq:dilord}
	\begin{split}
	\scalemyQ{1}{0.7}{0.5}{& \push{\bbX} \qw &  \Nmultigate{1}{L}{\qw}   & \push{\bbY} \qw  & \qw  \\
		& & \Nghost{L}& \push{\bbE} \ww  & \ww
	} \der_T \scalemyQ{1}{0.7}{0.5}{& \push{\bbX} \qw &  \Nmultigate{1}{L'}{\qw}   & \push{\bbY} \qw  & \qw  \\
		&\push{\bbD'} \ww & \Nghost{L'}{\ww} & \push{\bbE'} \ww  & \ww
	} 
\; 	\Leftrightarrow \;  \exists G:  
	 \myQ{0.7}{0.5}{ & \push{\bbX} \qw &  \multigate{1}{L}  & \qw & \push{\bbY} \qw  & \qw  \\
		& &   \Nghost{L} & \push{\bbE} \ww & \Nmultigate{1}{G}{\ww} & \push{\bbE'} \ww & \ww & \\
	 &  \ww &  \push{\bbD'} \ww 	&  \ww & \Nghost{G}{\ww} & 
	} = \scalemyQ{1}{0.7}{0.5}{& \push{\bbX} \qw &  \Nmultigate{1}{L'}{\qw}   & \push{\bbY} \qw  & \qw  \\
		&\push{\bbD'} \ww & \Nghost{L'}{\ww} & \push{\bbE'} \ww  & \ww
	} \end{split}
	\end{align}
	
We say that \emph{$L'$ is derivable from $L$} if $ L \, \der_T L'$. 

\end{Definition}

\begin{Remark} (On the Nature of the Relation $\der_T$.) \\
	The domain of the relation $\der_T$ is somewhat awkward (since one dilation may be two-sided and the other not). If we restrict both dilations to be one-sided, however, $\der_T$ restricts to a reflexive and transitive relation, i.e. a pre-order. The reason that we do not confine ourselves to one-sided dilations in the definition is that we soon wish to give meaning to the idea that \myuline{every} dilation (even if two-sided), is derivable from a single fixed dilation.  
	\end{Remark}

\begin{Remark} (Notation -- On the Dependence of $\der_T$ on $T$.) \label{rem:DilDeponT}\\
We shall often drop the subscript from $\der_T$ and write simply $\der$. Note, however, that it \myuline{does} matter what the base interfaces $\bbX$ and $\bbY$ are; for example, dilations $L: \bbX \cup \bbD \to \bbY \cup \bbE$ of $T$ (with hidden interfaces $\bbD$ and $\bbE$) can also be seen as dilations of $\tr_\bbI$ (with hidden interfaces $\bbX \cup \bbD$ and $\bbY \cup \bbE$), but then of course the dilational ordering trivialises.
 
	\end{Remark}

\begin{Example} (Large Dilations.)\\
	\cref{ex:dilCIT} shows that for any state $\state{p}{X}$ in $\CIT$, the class of one-sided dilations of $\state{p}{X}$ has a $\der$-largest element, namely the dilation given by copying.  \cref{ex:dilQIT} demonstrates that this is also the case for states in $\QIT$, with a $\der$-largest element given by purification.
	\end{Example} 

\begin{Example} (DiVincenzo and Derivability.)\\
	In theories which have the DiVincenzo Property, any dilation $L$ of $T$ is derivable from a one-sided dilation $L_0$. In fact, this statement is equivalent to the DiVincenzo Property. 
	\end{Example}

\begin{Example} (Dilational Ordering in Cartesian Theories.)\\
	By \cref{ex:dilSets}, every dilation of a channel $\channel{\bbX}{T}{\bbY}$ is derivable from the dilation given by copying the input. 
	\end{Example}

\begin{Example} (Dilational Ordering in a Thin Theory.)\\
	Consider the channel $\channel{7}{\phantom{X}}{3}$ in the thin theory $(\N, \cdot, 1, \geq)$. Every one-sided dilation of this channel is derivable from the one-sided dilation $\oneext{7}{\phantom{X}}{3}{2}$. Some dilations of $\channel{7}{\phantom{X}}{3}$ , however, are not derivable from this dilation, for example the dilation $\twoext{7}{3}{\phantom{X}}{3}{7}$. 

	\end{Example}

\begin{Remark} (One-Sided Dilations as a Category.) \\
The class of one-sided dilations of $\channel{\bbX}{T}{\bbY}$ does not merely have the structure of a pre-order; in fact, the very definition of the dilational ordering reveals that the pre-order is the shadow of a \myuline{category} whose objects are the one-sided dilations of $T$ and whose morphisms from $\oneext{\bbX}{L}{\bbY}{\bbE}$ to $\oneext{\bbX}{L'}{\bbY}{\bbE'}$ are the channels $\scalemyQ{.8}{0.7}{0.5}{& \push{\bbE} \ww & \Ngate{G}{\ww} & \push{\bbE'} \ww & \ww}$ such that  $\scalemyQ{.8}{0.7}{0.5}{ & \push{\bbX} \qw &  \multigate{1}{L}  & \qw & \push{\bbY} \qw  & \qw  & \qw \\
	& &   \Nghost{L} & \push{\bbE} \ww & \Ngate{G}{\ww}& \push{\bbE'} \ww & \ww } = \scalemyQ{.8}{0.7}{0.5}{& \push{\bbX} \qw &  \Nmultigate{1}{L'}{\qw}   & \push{\bbY} \qw  & \qw  \\
	& & \Nghost{L'} & \push{\bbE'} \ww  & \ww
}$. We shall not have occasion to study this category in any detail. (Also, it seems to vanish in the elaboration of \cref{chap:Causal}.) \end{Remark}

It turns out that the dilational ordering is rather well-behaved in the theories of interest to us, confined by simple principles with a clear interpretation. Before we study those principles in \cref{subsec:Complete} and \cref{subsec:Localisable}, we will examine the possibility of a rather brutal collapse of the dilational ordering.

\subsection{Dilational Purity}
\label{subsec:Purity}

Any channel $\channel{\bbX}{T}{\bbY}$ has some dilations, namely the \emph{trivial dilations} of the form $\scalemyQ{.8}{0.7}{0.5}{& \push{\bbX} \qw &  \gate{T}   & \push{\bbY} \qw  & \qw  \\
& \push{\bbD }\ww & \Ngate{S}{\ww} & \push{\bbE} \ww & \ww	
}$. In the dilational ordering, the trivial dilations are precisely those that can be derived from the (very trivial) dilation  $\channel{\bbX}{T}{\bbY}$ itself. It may happen that there are no other dilations:

\begin{Definition} (Dilational Purity.)\label{def:DilPure} \\
A channel $\channel{\bbX}{T}{\bbY}$ is called \emph{dilationally pure} if every dilation of $T$ is of the  form $\scalemyQ{.8}{0.7}{0.5}{& \push{\bbX} \qw &  \gate{T}   & \push{\bbY} \qw  & \qw  \\
	& \push{\bbD }\ww & \Ngate{S}{\ww} & \push{\bbE} \ww & \ww	
}$ for some channel $S$. 
\end{Definition}

From an operational point of view, dilational purity of $T$ signifies that any side-computation in the environment of $T$ must be independent from it. We will see in the next section that all isometric channels in $\QIT$ are dilationally pure, and since identity channels are isometric this will immediately imply a rather strong version of the \emph{No Broadcasting Theorem} (\cite{Barn96}).

\begin{Example} (Dilational Purity in Thin Theories.) \label{ex:PureThin}\\
	In a thin theory, consider the identity channel $\channel{x}{\phantom{X}}{x}$ . Its dilations are $\twoext{x}{y}{\phantom{X}}{x}{z}$, with $x \star y \succeq x \star z$. If the monoid satisfies the cancellation law $x \star y \succeq x \star z \Leftarrow y \succeq z$, then these dilations factor as $\scalemyQ{.8}{0.7}{0.5}{& \push{y} \ww & \Ngate{\phantom{X}}{\ww}& \push{z} \ww & \ww \\ & \push{x} \qw &\gate{\phantom{X}} & \push{x} \qw & \qw}$, so the identity $\channel{x}{\phantom{X}}{x}$ is dilationally pure. 
	
\end{Example}

In $\CIT$ we have the following characterisation of the dilationally pure channels:

\begin{Prop} (Dilationally Pure Channels in $\CIT$.) \label{prop:PureinCIT}\\
	A channel in $\CIT$ is dilationally pure if and only if it is a probabilistically pure state.
\end{Prop}

\begin{proof} According to \cref{ex:dilCIT}, a probabilistically pure state has only trivial one-sided dilations.  The fact that a two-sided dilation must also be trivial can either be verified by a direct consideration or seen as a consequence of the DiVincenzo property for $\CIT$, which we shall prove in \cref{subsec:Localisable}. 
	
	Assume conversely that $\channel{\bbX}{T}{\bbY}$ is dilationally pure. Consider the copy channel $\scalemyQ{.8}{0.7}{0.5}{& & &  \qw  & \qw \\ & \push{\bbX} \qw & \ctrlo{-1} & \qw & \qw }$ from earlier.  Clearly, $\scalemyQ{.8}{0.7}{0.5}{& & &  \ww & \ww  \\ & \push{\bbX} \qw & \ctrlo{-1} & \gate{T} & \push{\bbY} \qw & \qw}$ is a dilation of $T$, so by dilational purity it must be of the form $\scalemyQ{.8}{0.7}{0.5}{& &  \Ngate{s} & \ww & \ww\\ & \push{\bbX} \qw & \gate{T} & \push{\bbY} \qw & \qw}$ for some state $s$. However, we now have

	\begin{align}
	\scalemyQ{1}{0.7}{0.5}{&  \push{\bbX} \qw & \gate{\id} & \push{\bbX} \ww & \ww & \ww}  = 	\scalemyQ{1}{0.7}{0.5}{& & &  \ww & \ww  \\ & \push{\bbX} \qw & \ctrlo{-1} & \gate{T} &  \gate{\tr}} 
	= 
	\scalemyQ{1}{0.7}{0.5}{& &   \Ngate{s} & \ww & \ww\\ & \push{\bbX} \qw & \gate{T} &  \gate{\tr}} = 	\scalemyQ{1}{0.7}{0.5}{&  \push{\bbX} \qw & \gate{\tr} & \Ngate{s} & \push{\bbX}\ww & \ww},
	\end{align}
	
	which is only possible if the system $X$ corresponding to $\bbX$ has size $\abs{X} =1$, so $T$ must be a state.\footnote{Ignoring that we strictly speaking defined states as having the particular domain $\triv$.} It is then an easy exercise to verify that a state  which is dilationally pure must be probabilistically pure; if it is not, then we obtain a non-trivial dilation by copying the output. \end{proof}

In the next section we shall be able to also give a complete characterisation of the dilationally pure channels in $\QIT$. For now, we content ourselves with a confinement:

\begin{Lem} (Necessary Condition for Dilational Purity in $\QIT$.) \label{lem:PureisIso}\\
	If a channel in $\QIT$ is dilationally pure, then it is isometric. 
\end{Lem}

\begin{proof}

It suffices to consider channels with simple input and output interfaces. Assume that $\channel{\cX}{\Lambda}{\cY}$ is a dilationally pure quantum channel, and let $(K_i)_{i \in I}$ be a Kraus representation of $\Lambda$, i.e. a family of linear operators $K_i : \cX \to \cY$ such that $\Lambda(A) = \sum_{i \in I} K_i A K^*_i$ for all $A \in \End{\cX}$. The channel $\oneext{\cX}{\Phi}{\cY}{\C^I}$ given by $\Phi(A) = \sum_{i \in I} K_i A K^*_i \otimes \ketbra{i}$ is then a dilation of $\channel{\cX}{\Lambda}{\cY}$ with hidden system $\C^I$, so by dilational purity we must have $\Phi(A) = \sum_{i \in I} K_i A K^*_i \otimes \varrho$ for some state $\varrho$ on $\C^I$. By equality of these two expressions we conclude that $K_{i_0} A K^*_{i_0} = \bra{i_0} \varrho \ket{i_0} \sum_{i \in I} K_i A K^*_i$ for any $i_0 \in I$, and since $1 = \tr(\varrho) = \sum_{i \in I} \bra{i} \varrho \ket{i}$ we can pick some $i_0 \in I$ with $\bra{i_0} \varrho \ket{i_0} > 0$, for which we consequently have

	\begin{align}
	\Lambda(A) = \sum_{i \in I} K_i A K^*_i = \frac{1}{\bra{i_0} \varrho \ket{i_0} } K_{i_0} A K^*_{i_0} 
	\end{align}
	
	for all $A \in \End{\cX}$. In other words, $\Lambda$ admits a Kraus representation using the \myuline{single} Kraus operator $K_{i_0} /\sqrt{\bra{i_0} \varrho \ket{i_0} }$. This operator must then be an isometry, so $\Lambda$ is isometric. 
\end{proof}

\begin{Remark} (Relation of Dilational Purity to other Purity Notions.) \label{rem:PurityNotions}\\
Whereas dilational purity is a well-defined concept in all theories that we consider, it does not generally make sense to speak of probabilistic purity as it requires a convex structure. \cref{prop:PureinCIT} shows that in theories where the concept does make sense, it may be distinct from dilational purity: Any deterministic function whose domain has at least two elements is probabilistically pure in $\CIT$, but not dilationally pure according to the proposition. (On the other hand, though we will not have occasion to be precise about this, it is easy to see that dilational purity implies probabilistic purity in theories where that concept does make sense, since to a non-trivial convex decomposition $\frac{1}{2} T_0 + \frac{1}{2} T_1$ we can associate the non-trivial dilation $\frac{1}{2} T_0  \otimes \delta_0+ \frac{1}{2} T_1 \otimes \delta_1$ which keeps as side-information a memory of which component was employed.)

The fact that these two purity notions do not coincide, also implies that dilational purity is distinct from the purity notion of Ref. \cite{Cunn17}, defined in terms of \emph{weak factorisation systems}, which reduces in $\CIT$ to probabilistic purity. 

If we restrict the condition in \cref{def:DilPure} to one-sided dilations then we obtain the purity notion proposed in Ref. \cite{Chir14pure}, but in general there is in fact a distinction between one- and two-sided purity.\footnote{For example, in the theory $\TREX$ consisting precisely of the surjective functions between finite sets (\cref{ex:TREX}), any one-sided dilation of a \myuline{bijection} is trivial, though a bijection generally has non-trivial two-sided dilations in $\TREX$.}
\end{Remark}

\section{Completeness and Localisability}
\label{sec:CompLoc}

In this section, the \emph{completeness} and \emph{localisability} principles are introduced, and we prove that they hold in many theories, in particular the two information theories $\CIT$ and $\QIT$. 

\subsection{Complete Dilations}
\label{subsec:Complete}

Dilational purity of a channel is an extreme condition, under which the dilational ordering implodes to a single level. In interesting theories, most channels will not be dilationally pure (cf. \cref{prop:PureinCIT} and \cref{lem:PureisIso}). They do, however, comply to another principle which is milder and still retains a remarkable simplicity:

\begin{Definition} (Complete Dilations. ) \\
Let $\channel{\bbX}{T}{\bbY}$ be a channel in $\Theory$, and let $\bfD \subseteq \Dil{T}$ be a class of dilations of $T$. We say that a (one-sided) dilation $K \in \bfD$ is \emph{complete for $\bfD$} if $K \der L$ for all $L \in \bfD$. A dilation $K \in \Dil{T}$ is called simply \emph{complete} if it is complete for $\Dil{T}$.	
\end{Definition}

\begin{Definition} (Complete Theories.) \label{def:Complete}\\
	A theory $\Theory$ is called \emph{complete} if every channel in $\Theory$ has a complete dilation. 
\end{Definition}

\begin{Example} (Completeness and Dilational Purity.) \label{ex:CompPure}\\
	If $\channel{\bbX}{T}{\bbY}$ is a dilationally pure channel in $\Theory$, then $\channel{\bbX}{T}{\bbY}$ is a complete dilation \myuline{of} \myuline{itself}; in fact, \emph{self-completeness} is equivalent to dilational purity.
\end{Example}

\begin{Example} (Trivial Completeness.) \label{ex:CompleteTrash}\\
	In any theory $\Theory$, the trash $\scalemyQ{.8}{0.7}{0.5}{& \push{\bbZ} \qw & \gate{\tr}}$ has a complete dilation, namely $\scalemyQ{.8}{0.7}{0.5}{& \push{\bbZ} \qw & \gate{\id}& \push{\bbZ} \ww & \ww}$.
\end{Example}

\begin{Example} (Completeness in Thin Theories.)\\
	The fact that a potential complete dilation is required to be one-sided means that every complete theory must have the DiVincenzo Property (\cref{def:DiVi}). Hence, it follows from \cref{ex:DiViFail} that the thin theory $(\N, \cdot, 1, \geq)$ is not complete. Some of its channels do have complete dilations, however. In fact, it is not difficult to see that the channel $\channel{x}{\phantom{X}}{y}$ has a complete dilation if and only if $x$ is divisible by $y$, in which case the dilation $\oneext{x}{\phantom{X}}{y}{x/y}$ is complete. 
\end{Example}

Though completeness might seem like a foolish mathematical fantasy rarely fulfilled in theories of real interest, the truth is quite the opposite. For example, we have already seen an argument to the effect that all cartesian theories are complete:

\begin{Thm} (Cartesian Theories are Complete.) \label{thm:CartisComp}\\
	Every cartesian theory $\Theory$ is complete. In fact, if $\channel{\bbX}{T}{\bbY}$ is a channel in $\Theory$ then the dilation given by copying the inputs, 
	
	\begin{align} \label{eq:CartComp}
\scalemyQ{1}{0.7}{0.5}{& &  & \ww & \push{\bbX} \ww & \ww \\ &  \\ & \push{\bbX} \qw & \ctrlo{-2} & \gate{T} & \push{\bbY} \qw & \qw}, 
\end{align} 
	
	is complete. 
\end{Thm}

\begin{proof}
	It suffices to show this in the case where the interfaces $\bbX$ and $\bbY$ are simple, but this is precisely what we did in \cref{ex:dilSets}.

	\end{proof}

	It is also the case that $\CIT$ and $\QIT$ are complete, but the most elegant way of reaching this conclusion is to start by proving completeness relative to \myuline{one-sided} dilations, and then lift those results in the next subsection.\\
	
	To ease language, let us say that a a dilation of $\channel{\bbX}{T}{\bbY}$ is \emph{one-sided-complete} if it is complete for one-sided dilations. Let us also say that a theory $\Theory$ is \emph{one-sided-complete} if every channel $\channel{\bbX}{T}{\bbY}$ in $\Theory$ has a one-sided-complete dilation.

\begin{Lem} (One-Sided-Completeness in $\CIT$.) \label{lem:CITOneSidedComp}\\
The theory $\CIT$ is one-sided-complete. Specifically, for a channel $\channel{\bbX}{T}{\bbY}$, the dilation given by copying both the inputs and outputs, 
	
		\begin{align}  \label{eq:CITcomp}
	\scalemyQ{1}{0.7}{0.5}{& &  & \ww & \ww & \ww & \ww \\ & & & & & \ww & \ww \\ & \push{\bbX} \qw & \ctrlo{-2} & \gate{T} & \ctrlo{-1} & \push{\bbY} \qw & \qw}, 
	\end{align} 
	
	is one-sided-complete.
	\end{Lem}

\begin{proof}
	In the case where $\bbX$ is trivial, this statement is exactly what was proved in \cref{ex:dilCIT}: The copy-dilation $\scalemyQ{.8}{0.7}{0.5}{&   & & \push{\bbY} \ww & \ww    \\ & \Ngate{p}  & \ctrlo{-1} & \push{\bbY} \qw & \qw  } $ is complete for one-sided dilations $\state{p}{\bbY}$,  by virtue of existence of conditional probabilities. A general classical channel $\channel{\bbX}{T}{\bbY}$ is simply a collection of states indexed by the input, so any one-sided dilation $\oneext{\bbX}{L}{\bbY}{\bbE}$ can be derived from the channel \eqref{eq:CITcomp}, since knowing a copy of the input $x$ essentially reduces the problem to the case of a single state (the details are left as an exercise).
	\end{proof}

\begin{Lem} (One-Sided-Completeness in $\QIT$.) \label{lem:QITOneSidedComp}\\
The theory $\QIT$ is one-sided-complete. Specifically, for a channel  $\channel{\bbX}{\Lambda}{\bbY}$,  any Stinespring dilation (i.e. any isometric dilation)
	
	\begin{align}
	\myQ{0.7}{0.5}{& & \Nmultigate{1}{\Sigma} & \push{\cE} \ww& \ww \\ & \push{\bbX} \qw & \ghost{\Sigma} & \push{\bbY} \qw & \qw}
	\end{align}
	
	of $\Lambda$ is  one-sided-complete.
\end{Lem}

\begin{proof}
	Again, we proved this in the case where $\Lambda$ is a state in \cref{ex:dilQIT}: Purifications are special cases of Stinespring dilations, and by the uniqueness of purifications it was demonstrated that any purification $\scalemyQ{.8}{0.7}{0.5}{& \Nmultigate{1}{\pi}  & \push{\cE} \ww & \ww    \\ & \Nghost{\pi}  & \push{\bbY} \qw & \qw }$ of a state $\state{\varrho}{\bbY}$ is a complete dilation. For general channels, Stinespring's Dilation Theorem (\cite{Stine55,NC02,Wat11}), which is covered in the preliminary section of the thesis, asserts that every quantum channel has an isometric one-sided dilation and that this dilation is unique up to a channel acting on the hidden interface. By an argument analogous to that of \cref{ex:dilQIT}, this implies that any one-sided dilation $\oneext{\bbX}{\Phi}{\bbY}{\bbE}$ of $\channel{\bbX}{\Lambda}{\bbY}$ can be derived from a Stinespring dilation.

\end{proof}

To conclude that the above dilations in $\CIT$ and $\QIT$ are not only one-sided-complete but in fact complete for all dilations, we will make use of the following observation:

\begin{Lem} (One-Sided-Completeness and DiVincenzo.) \label{lem:DiViComp}\\
	Let $\Theory$ be a theory. The following are equivalent: 
	
	\begin{enumerate}
		\item $\Theory$ is complete.
		\item $\Theory$ has the DiVincenzo property and $\Theory$ is one-sided-complete.
		\end{enumerate}
	
	In this case, any dilation of $\channel{\bbX}{T}{\bbY}$ which is complete for its one-sided dilations is in fact complete for all dilations. 
	\end{Lem}

\begin{proof}
	We have already observed that completeness implies the DiVincenzo property, so 2. clearly follows from 1. Conversely, assuming 2., if $\oneext{\bbX}{K}{\bbY}{\bbE}$ is complete for one-sided dilations of $\channel{\bbX}{T}{\bbY}$ then, by the DiVincenzo property, it is in fact complete for all dilations, so 1. follows, and so does the final statement in the lemma.
	
	\end{proof}

What remains in order to prove completeness is then only an argument to the effect that the DiVincenzo property holds in $\CIT$ and $\QIT$. This will be accomplished by relating the property to the principle of \emph{spatial localisability} which is introduced in the following subsection.

\subsection{Spatial and Temporal Localisability }
\label{subsec:Localisable}%

Completeness is a \emph{static} principle, in the sense that  it pertains to the dilational ordering for a fixed channel $T$. It is natural to ask how dilations behave under the \emph{dynamic} structure inherent in every theory: Parallel and serial composition. \\

Consider for example two channels $ \channel{\bbX_1}{T_1}{\bbY_1}$ and $ \channel{\bbX_2}{T_2}{\bbY_2}$. Clearly, for any one-sided dilations $\oneext{\bbX_1}{L_1}{\bbY_1}{\bbE_1}$ and  $\oneext{\bbX_2}{L_2}{\bbY_2}{\bbE_2}$ of $T_1$ and $T_2$ respectively, the channel $\scalemyQ{.8}{0.7}{0.5}{ & \push{\bbX_1}  \qw & \multigate{1}{L_1} & \push{\bbY_1} \qw & \qw \\ 
	&  & \Nghost{L_1} & \push{\bbE_1} \ww  & \ww   \\ 
	& & \Nmultigate{1}{L_2} & \push{\bbE_2} \ww & \ww \\ & \push{\bbX_2}  \qw & \ghost{L_2} & \push{\bbY_2} \qw  & \qw} $
is a dilation of their parallel composition $ \scalemyQ{.8}{0.7}{0.75}{ & \push{\bbX_1}  \qw & \gate{T_1} & \push{\bbY_1} \qw & \qw  \\ 
 & \push{\bbX_2}  \qw & \gate{T_2} & \push{\bbY_2} \qw  & \qw} $. In general, the parallel composition has other dilations than those of this form, indeed all the dilations \myuline{derivable} from dilations of this form. Such derivable dilations will be called \emph{spatially localisable} since, intuitively, the side-information corresponding to them can be `localised' as a combination of side-information from either $T_1$ or $T_2$:

\begin{Definition} (Spatial Localisability.) \label{def:SpatLoc}\\
Let  $ \channel{\bbX_1}{T_1}{\bbY_1}$ and $ \channel{\bbX_2}{T_2}{\bbY_2}$ be parallelly composable channels in $\Theory$. We say that a dilation of the parallel composition $ \scalemyQ{.8}{0.7}{0.75}{ & \push{\bbX_1}  \qw & \gate{T_1} & \push{\bbY_1} \qw & \qw  \\ 
	& \push{\bbX_2}  \qw & \gate{T_2} & \push{\bbY_2} \qw  & \qw} $ is \emph{spatially localisable w.r.t. $T_1$ and $T_2$} if it is of the form 

\begin{align} \label{eq:dilparallel}
\myQ{0.7}{0.5}{ & \push{\bbX_1}  \qw & \multigate{1}{L_1} & \push{\bbY_1} \qw & \qw \\ 
	&  & \Nghost{L_1} & \Nmultigate{2}{G}{\ww}   \\ 
	& & \push{\bbD} \ww& \Nghost{G}{\ww} & \push{\bbE} \ww &\ww  \\
	& & \Nmultigate{1}{L_2}  & \Nghost{G}{\ww} \\ & \push{\bbX_2}  \qw & \ghost{L_2} & \push{\bbY_2} \qw  & \qw} 
\end{align}

for some channel $G$ and some dilations $L_1$ of $T_1$ and $L_2$ of $T_2$.

	\end{Definition}

Obviously, there is a natural counterpart of localisability for serial compositions, which we name \emph{temporal localisability}:

\begin{Definition} (Temporal Localisability.)  \label{def:TempLoc}\\
	Let  $ \channel{\bbX}{T}{\bbY}$ and $ \channel{\bbY}{S}{\bbZ}$ be channels in $\Theory$. We say that a dilation of the serial composition $ \scalemyQ{.8}{0.7}{0.75}{ & \push{\bbX}  \qw & \gate{T} & \push{\bbY} \qw & \gate{S} & \push{\bbZ} \qw  & \qw} $ is \emph{temporally localisable w.r.t. $T$ and $S$} if it is of the form 
	
\begin{align}
\myQ{0.7}{0.5}{ & \push{\bbX}  \qw & \multigate{2}{L} &  \push{\bbY} \qw & \multigate{1}{M} &\push{\bbZ} \qw & \qw\\ 
	&  & \Nghost{L} & & \Nghost{M}& \Nmultigate{2}{G}{\ww}   \\
	&  & \Nghost{L}&\ww  & \ww & \Nghost{G}{\ww}& \push{\bbE}  \ww & \ww\\
	& & & & \push{\bbD} \ww& \Nghost{G}{\ww}  
}
\end{align}

for some channel $G$ and some dilations $L$ of $T$ and $M$ of $S$. 
	
\end{Definition}

By a somewhat awful abuse of language, we shall moreover use the following terminology:

\begin{Definition} (Localisability of a Theory.) \label{def:LocTheory}\\
	A theory $\Theory$ is called \emph{spatially localisable} if every one-sided dilation of a parallel composition is spatially localisable. 	A theory $\Theory$ is called \emph{temporally localisable} if every one-sided dilation of a serial composition is temporally localisable. We say  that $\Theory$ is \emph{localisable} if it is both spatially and temporally localisable.
	
		\end{Definition}

\begin{Remark}
	Observe that localisability is imposed on  \myuline{one-sided} dilations only. This provides a simpler interpretation of the requirement, and also makes the conditions easier to fulfil. (It will soon be clear, however, that the conditions automatically follow for two-sided dilations if they hold for the one-sided.) 

		\end{Remark}

It might seem a hopeless task to prove that a theory is spatially or temporally localisable. To our luck, however, we are mainly interested in theories which are one-sided-complete, and for those we have the following:

\begin{Lem} (Recharacterisation of Localisability.) \label{lem:RecharLoc}\\
	Suppose that $\Theory$ is one-sided-complete. Then, 
	
	\begin{itemize}
		\item $\Theory$ is spatially localisable if and only if for any channels $\channel{\bbX_1}{T_1}{\bbY_1}$ and $\channel{\bbX_2}{T_2}{\bbY_2}$, there exist one-sided-complete dilations $K_1$ of $T_1$ and $K_2$ of $T_2$, such that  $\scalemyQ{.8}{0.7}{0.5}{ & \push{\bbX_1}  \qw & \multigate{1}{K_1} & \push{\bbY_1} \qw & \qw \\
			&  & \Nghost{K_1} & \push{\bbE_1} \ww  & \ww   \\ 
			& & \Nmultigate{1}{K_2} & \push{\bbE_2} \ww & \ww \\ & \push{\bbX_2}  \qw & \ghost{K_2} & \push{\bbY_2} \qw  & \qw} $ is a one-sided-complete dilation of $ \scalemyQ{.8}{0.7}{0.75}{ & \push{\bbX_1}  \qw & \gate{T_1} & \push{\bbY_1} \qw & \qw  \\ 
			& \push{\bbX_2}  \qw & \gate{T_2} & \push{\bbY_2} \qw  & \qw} $.
		\item $\Theory$ is temporally localisable if and only if for any channels $\channel{\bbX}{T}{\bbY}$ and $\channel{\bbY}{S}{\bbZ}$, there exist one-sided-complete dilations $K$ of $T$ and $C$ of $S$, such that 	$	\scalemyQ{.8}{0.7}{0.5}{ & \push{\bbX}  \qw & \multigate{2}{K} &  \push{\bbY} \qw & \multigate{1}{C} &\push{\bbZ} \qw & \qw\\ 
			&  & \Nghost{K} & & \Nghost{C}& \ww   \\
			&  & \Nghost{K}&\ww  & \ww & \ww
		}$ is a one-sided-complete dilation of $ \scalemyQ{.8}{0.7}{0.75}{ & \push{\bbX}  \qw & \gate{T} & \push{\bbY} \qw & \gate{S} & \push{\bbZ} \qw  & \qw} $.
		\end{itemize}

	\end{Lem}

\begin{proof}
Obvious.
	\end{proof}

\begin{Remark} \label{rem:LocAnyComp}
	It is easy to see that if $\Theory$ is spatially [temporally] localisable, then  in fact the parallel [serial] composition of \myuline{any} complete dilations yields a complete dilation. 

	\end{Remark}

As if tailor-made, this lemma immediately implies that our pet theories are localisable: 

\begin{Thm} \label{thm:CartLoc}
	Every cartesian theory is localisable.
	\end{Thm}

\begin{proof}
	By  \cref{thm:CartisComp}, a cartesian theory is complete, in particular one-sided-complete, so we can use \cref{lem:RecharLoc}. However, it is obvious that the parallel compositions of the dilations \eqref{eq:CartComp}  obtained by copying the inputs from $T_1: \bbX_1 \to \bbY_1$ and $T_2: \bbX_2 \to \bbY_2$ yields the dilation obtained by copying the input from $T_1 \og T_2$, and it is also clear that the serial composition of the dilations obtained by copying from $T: \bbX \to \bbY$ and $S:\bbY \to \bbZ$ yields (a dilation from which we can derive) the dilation obtained by copying from $S \after T$.  The desired follows.
	\end{proof}

\begin{Thm} \label{thm:CITLoc} The theory $\CIT$ is localisable.

	\end{Thm}

\begin{proof}
	By  \cref{lem:CITOneSidedComp}, $\CIT$ is one-sided-complete, so again \cref{lem:RecharLoc} applies. The rest of the argument is exactly as in the previous proof, only now we have to copy the outputs as well as the inputs, cf. \cref{eq:CITcomp}. 
	\end{proof}

\begin{Thm} \label{thm:QITLoc} The theory $\QIT$ is localisable.

\end{Thm}

\begin{proof}
By  \cref{lem:QITOneSidedComp} $\QIT$ is one-sided complete, so we can use \cref{lem:RecharLoc} a third time.  The desired follows immediately, since the parallel and serial compositions of isometric channels are obviously isometric.\end{proof}

\vspace{.3cm}

The status of our dilational investigations is at this point as follows: Cartesian theories are complete and localisable; $\CIT$ and $\QIT$ are one-sided-complete and localisable; thin theories need not be complete. \\

What we shall prove now is that temporal localisability implies the existence of reversible dilations and that spatial localisability implies the DiVincenzo property. These two consequences will be used repeatedly and they also provide us with counterexamples to localisability. Moreover, as one-sided-completeness and completeness are by \cref{lem:DiViComp} equivalent under the DiVincenzo property, full completeness of $\CIT$ and $\QIT$ will follow.

\begin{Prop} (Temporal Localisability implies Reversible Dilations.) \label{prop:Reversibledilations} \\
If $\Theory$ is temporally localisable,  then every channel $\channel{\bbX}{T}{\bbY}$ has a one-sided dilation $\oneext{\bbX}{R}{\bbY}{\bbE}$ which is reversible.
\end{Prop}

\begin{proof}
	By \cref{lem:Trashes},

	\begin{align}
\myQ{0.7}{0.5}{& \push{\bbX} \qw & \gate{T} & \push{\bbY} \qw & \gate{\tr_\bbY}} = \myQ{0.7}{0.5}{& \push{\bbX} \qw  & \gate{\tr_\bbX}} .
\end{align}

Since $ \scalemyQ{.8}{0.7}{0.5}{& \push{\bbX} \qw  & \gate{\id_\bbX} & \push{\bbX} \ww & \ww}$ is a one-sided dilation of $\tr_\bbX$, we must by temporal localisability find one-sided dilations $R$ of $T$ and $M$ of $\tr_\bbX$ such that

\begin{align}
\myQ{0.7}{0.5}{ & \push{\bbX}  \qw & \multigate{2}{R} &  \push{\bbY} \qw & \multigate{1}{M} \\ 
	&  & \Nghost{R} & & \Nghost{M}& \Nmultigate{1}{G}{\ww}   \\
	&  & \Nghost{R}&\ww  & \push{\bbE}  \ww & \Nghost{G}{\ww}& \push{\bbX}  \ww & \ww  
} = \myQ{0.7}{0.5}{& \push{\bbX} \qw  & \gate{\id_\bbX} & \push{\bbX} \ww & \ww}
\end{align}

for some channel $G$. This identity implies in particular that $R$ is reversible. \end{proof}

\begin{Prop} (Spatial Localisability implies the DiVincenzo Property.) \label{prop:StructureNonSignalling} \\
If $\Theory$ is spatially localisable, then $\Theory$ has the DiVincenzo Property. That is, every channel $\channel{\bbX}{T}{\bbY}$ which is non-signalling from $\bbX_0$ to $\bbY_0$ is of the form 

\begin{align} \label{eq:SpatDiVi}
\myQ{0.7}{0.5}{
	& \push{\bbX \setminus \bbX_0}  \qw    &\multigate{1}{T_1}   & \qw &   \push{\bbY_0} \qw & \qw & \qw \\
	&                   & \Nghost{T_1}  & \push{\bbH} \qw & \multigate{1}{T_2} & \\
	& \qw &  \push{\bbX_0}  \qw & \qw &  \ghost{T_2}   &  \push{\bbY \setminus \bbY_0} \qw  & \qw  \\
} 	\end{align}

for some channels $T_1$ and $T_2$, or, equivalently, every dilation of a channel is derivable from a one-sided dilation. 
\end{Prop}

\begin{proof} Let $\channel{\bbX}{T}{\bbY}$ be non-signalling from $\bbX_0$ to $\bbY_0$. This means that there exists a channel $T'$ such that

		\begin{align}
	\myQ{0.7}{0.5}{& \push{\bbX \setminus \bbX_0} \qw &  \Nmultigate{1}{T}{\qw}   & \push{\bbY_0} \qw  & \qw  \\
		&\push{\bbX_0} \qw & \Nghost{T}{\qw} & \push{\bbY \setminus \bbY_0} \qw  & \Ngate{\tr}{\qw}
	}		
	\quad   = \quad 
	\myQ{0.7}{0.5}{& \push{\bbX \setminus \bbX_0} \qw &   \Ngate{T'}{\qw}    & \push{\bbY_0} \qw  & \qw  \\
		&\push{\bbX_0} \qw  & \Ngate{\tr}{\qw}
	}	\quad .
	\end{align}
	
	This, however, is to say that $	\scalemyQ{.8}{0.7}{0.5}{& \push{\bbX \setminus \bbX_0} \qw &  \Nmultigate{1}{T}{\qw}   & \push{\bbY_0} \qw  & \qw  \\
		&\push{\bbX_0} \qw & \Nghost{T}{\qw} & \push{\bbY \setminus \bbY_0} \ww  & \ww
	}	$ is a one-sided dilation of the parallel composition $	\scalemyQ{.8}{0.7}{0.5}{& \push{\bbX \setminus \bbX_0} \qw &   \Ngate{T'}{\qw}    & \push{\bbY_0} \qw  & \qw  \\
	&\push{\bbX_0} \qw  & \Ngate{\tr}{\qw}
}$. By spatial localisability, we thus find dilations $	\scalemyQ{.8}{0.7}{0.5}{& \push{\bbX \setminus \bbX_0} \qw &  \Nmultigate{1}{T_1}{\qw}   & \push{\bbY_0} \qw  & \qw  \\
& & \Nghost{T_1} & \push{\bbE_1} \ww  & \ww
}$ of $T'$ and $\scalemyQ{.8}{0.7}{0.5}{& \push{\bbX_0} \qw  & \gate{M} & \push{\bbE_2} \ww & \ww}$ of $\tr_{\bbX_0}$, such that 

\begin{align}
\scalemyQ{1}{0.7}{0.5}{& \push{\bbX \setminus \bbX_0} \qw &  \Nmultigate{1}{T}{\qw}   & \push{\bbY_0} \qw  & \qw  \\
	&\push{\bbX_0} \qw & \Nghost{T}{\qw} & \push{\bbY \setminus \bbY_0} \ww  & \ww
}	 = \scalemyQ{1}{0.7}{0.5}{& \push{\bbX \setminus \bbX_0} \qw &  \Nmultigate{1}{T_1}{\qw}   & \push{\bbY_0} \qw  & \qw  \\
& & \Nghost{T_1} & \push{\bbE_1} \ww  & \Nmultigate{1}{G}{\ww} 
 \\ & \push{\bbX_0} \qw & \gate{M} & \push{\bbE_2} \ww & \Nghost{G}{\ww} & \push{\bbY \setminus \bbY_0} \ww & \ww}
\end{align}

for some channel $G$. Merging $M$ and $G$ to form $T_2$ we obtain \cref{eq:SpatDiVi} (with $\bbH=\bbE_1$) as desired.

\end{proof}

As anticipated, we may now conclude full completeness of the information theories: 

\begin{Thm}(Completeness of $\CIT$.) \label{thm:CITComp} \\
	The theory $\CIT$ is complete. Specifically, for a channel $\channel{\bbX}{T}{\bbY}$, the dilation given by copying both the inputs and outputs, 
	
	\begin{align} 
	\scalemyQ{1}{0.7}{0.5}{& &  & \ww & \ww & \ww & \ww \\ & & & & & \ww & \ww \\ & \push{\bbX} \qw & \ctrlo{-2} & \gate{T} & \ctrlo{-1} & \push{\bbY} \qw & \qw}, 
	\end{align} 
	
	is complete.
	\end{Thm}

\begin{Thm} (Completeness of $\QIT$.) \label{thm:QITComp}\\
The theory $\QIT$ is complete. Specifically, for a channel  $\channel{\bbX}{\Lambda}{\bbY}$,  any Stinespring dilation (i.e. any isometric dilation)
	
	\begin{align}
	\myQ{0.7}{0.5}{& & \Nmultigate{1}{\Sigma} & \push{\cE} \ww& \ww \\ & \push{\bbX} \qw & \ghost{\Sigma} & \push{\bbY} \qw & \qw}
	\end{align}
	
 is complete.
\end{Thm}

\begin{proof}
	In both cases, the proof is by one-sided-completeness (\cref{lem:CITOneSidedComp},  \cref{lem:QITOneSidedComp}) and spatial localisability (\cref{thm:CITLoc}, \cref{thm:QITLoc}), using \cref{prop:StructureNonSignalling} in conjunction with \cref{lem:DiViComp}.  
\end{proof}

This also allows us to finally tie another loose end:
	
	\begin{Cor} (Dilationally Pure Channels in $\QIT$.)  \label{prop:PureEqualsIso}\\
A channel in $\QIT$ is dilationally pure if and only if it is isometric.  
		\end{Cor}
	
	\begin{proof}
By 	\cref{lem:PureisIso} any dilationally pure channel must be isometric, so it remains only to prove the converse. However, any isometric channel is by \cref{thm:QITComp} a complete dilation of itself and thus dilationally pure. 

	\end{proof}

Interestingly, this characterisation of the pure channels in $\QIT$ immediately implies a rather strong form of the No Broadcasting Theorem (\cite{Barn96}): Any dilation $\oneext{\cH}{L}{\cH}{\cE}$ of an identity $\channel{\cH}{\id}{\cH}$ is trivial, so in particular its $\cE$-marginal must trash the input $\cH$.\\

We shall return to such considerations in more detail in \cref{sec:Selfuniv}, but for now we end this section by recalling that we have ultimately proved completeness and localisability of $\CIT$, $\QIT$ and all cartesian theories. We have also seen that a thin theory need not be complete, and using \cref{prop:StructureNonSignalling} it need not by spatially localisable either, cf. \cref{ex:DiViFail} (it is also easy to find a direct example e.g. of a dilation of $\scalemyQ{.8}{0.7}{0.5}{& \push{3} \qw & \gate{\phantom{X}} & \push{2} \qw& \qw \\ & \push{3} \qw & \gate{\phantom{X}} & \push{2} \qw& \qw}$ which cannot be derived from a parallel composition of dilations). By \cref{prop:Reversibledilations} we can  similarly see that a thin theory need not be temporally localisable. The theory $\TREX$ also displays lack of localisability.

\section{Universal Dilations}
\label{sec:Universal}

Completeness and localisability give us leashes on the collections of dilations, but do not reveal what the pre-order $\der$ actually looks like among the one-sided dilations. This can be resolved if we slightly strengthen the notion of a complete dilation, introducing the idea of \emph{universal} dilations. \\

All of the complete theories we have seen so far turn out to admit universal dilations, but this has to be proved in each case, and is actually somewhat tricky for the theory $\QIT$. In fact, the result that $\QIT$ has universal dilations will constitute a generalisation of the injectivity of the Choi-Jamio\l{}kowski isomorphism (\cite{Jiang13,Pill67,Choi75,Jam72}) which apparently has not been observed before. 

Besides providing a firm grip on the dilational order, the existence of universal dilations implies a forceful \emph{contraction property} which will be instrumental in \cref{chap:Causal} (in the form of \cref{lem:UnivCont}).

\begin{Definition} (Universal Dilations.) \label{def:Univ} \\
	Let $\channel{\bbX}{T}{\bbY}$ be a channel in a theory $\Theory$. A one-sided dilation $\scalemyQ{.8}{0.7}{0.5}{& \push{\bbX}  \qw & \multigate{1}{U} & \push{\bbY} \qw & \qw \\ 
	& & \Nghost{U} & \push{\bbE_0} \ww  & \ww  }$ is called a \emph{universal dilation of $T$} if every dilation of $T$ is of the form 

	\begin{align}
\myQ{0.7}{0.5}{ & \push{\bbX}  \qw & \multigate{1}{U} & \qw & \push{\bbY} \qw & \qw &  \qw \\ & & \Nghost{U} &  \push{\bbE_0} \ww & \Nmultigate{1}{G}{\ww} & \push{\bbE} \ww & \ww\\
	& 	&  \push{\bbD} \ww &\ww & \Nghost{G}{\ww}} 
\end{align}

for a \myuline{unique} channel $G$. 	\end{Definition}

\begin{Definition}(Universal Theories.)\\
	A theory $\Theory$ is called \emph{universal} if every channel in $\Theory$ has a universal dilation.
	
	\end{Definition}

\begin{Example} (Trivial Universality.)\\
In \cref{ex:CompleteTrash}, we observed that $\scalemyQ{.8}{0.7}{0.5}{& \push{\bbZ} \qw & \gate{\id}& \push{\bbZ} \ww & \ww}$ is a complete dilation of the trash $\scalemyQ{.8}{0.7}{0.5}{& \push{\bbZ} \qw & \gate{\tr}}$ in any theory. In fact, the dilation is obviously universal. 
	\end{Example}

\begin{Example} (Universality and Dilational Purity.) \\
We observed in \cref{ex:CompPure} that a dilationally pure channel $\channel{\bbX}{T}{\bbY}$ is \emph{self-complete}, i.e. a complete dilation of itself. In fact, it is even \emph{self-\myuline{universal}}, since by the normality requirement (\cref{def:Normal}), the channel that derives a given trivial dilation is unique. 

	\end{Example}

We hasten to observe that cartesian theories and $\CIT$ are universal:

\begin{Thm} \label{thm:CartUniv} In a cartesian theory, every channel has a universal dilation. 
	
	\end{Thm}

\begin{proof}
	For any channel $\channel{\bbX}{T}{\bbY}$, the complete dilation $\scalemyQ{.8}{0.7}{0.5}{& &  & \ww & \push{\bbX} \ww & \ww  \\ & \push{\bbX} \qw & \ctrlo{-1} & \gate{T} & \push{\bbY} \qw & \qw}$ is in fact universal, since if  $\scalemyQ{.8}{0.7}{0.5}{& & & \ww & \Nmultigate{1}{G}{\ww}  & \ww \\& &  & \ww & \Nghost{G}{\ww} \\ & \push{\bbX} \qw & \ctrlo{-1} & \gate{T} & \push{\bbY} \qw & \qw} = \scalemyQ{.8}{0.7}{0.5}{& & & \ww& \Nmultigate{1}{G'}{\ww}  & \ww \\& &  & \ww & \Nghost{G'}{\ww} \\ & \push{\bbX} \qw & \ctrlo{-1} & \gate{T} & \push{\bbY} \qw & \qw} $ then it follows by trashing $\bbY$ that $\scalemyQ{.8}{0.7}{0.5}{ & \ww & \Nmultigate{1}{G}{\ww}  & \ww \\ & \ww & \Nghost{G}{\ww} \\ & \push{\bbX} \qw & \gate{\tr} } = \scalemyQ{.8}{0.7}{0.5}{ & \ww & \Nmultigate{1}{G'}{\ww}  & \ww \\ & \ww & \Nghost{G'}{\ww} \\ & \push{\bbX} \qw & \gate{\tr} } $ which by normality implies $G=G'$.

\end{proof}

\begin{Thm}\label{thm:CITUniv}  In $\CIT$, every channel has a universal dilation. 
	
\end{Thm}

\begin{proof}
Consider first a state $\state{p}{Y}$. The complete dilation $\scalemyQ{.8}{0.7}{0.5}{&   & & \push{Y} \ww & \ww    \\ & \Ngate{p}  & \ctrlo{-1} & \push{Y} \qw & \qw  }$  need not be complete, since, as discussed in \cref{ex:dilCIT}, the conditional distribution $g_x$ is only uniquely determined when $p(y)> 0$. However, this also means that to obtain a universal dilation, all we need to do is cut down the hidden system to the support of $p$, $\supp{p}$.  For a channel $\channel{X}{T}{Y}$ given by the distributions $(t_x)_{x \in X}$, we may similarly  turn the complete dilation $	\scalemyQ{.8}{0.7}{0.5}{& &  & \ww & \push{X} \ww & \ww & \ww \\ & & & & & \push{Y} \ww & \ww \\ & \push{X} \qw & \ctrlo{-2} & \gate{T} & \ctrlo{-1} & \push{Y} \qw & \qw}$ into a universal one by cutting down the hidden system $X \times Y$ to the subset $\{(x,y) \in X \times Y \mid y \in \supp{t_x}\}$. The details are left as exercise. 

\end{proof}

Before proving that also the theory $\QIT$ is universal, let us prove two general results about universal dilations and universal theories. \\

The first result classifies all universal and complete dilations in terms of a single universal dilation:

\begin{Prop} (One determines All.) \label{prop:OneDetAll}\\
Let $\scalemyQ{.8}{0.7}{0.5}{& \push{\bbX}  \qw & \multigate{1}{U} & \push{\bbY} \qw & \qw \\ 
	& & \Nghost{U} & \push{\bbE_0} \ww  & \ww  }$ be a universal dilation of $\channel{\bbX}{T}{\bbY}$. 
	
	\begin{enumerate}
		\item 
		For any isomorphism $\alpha$, $\scalemyQ{.8}{0.7}{0.5}{& \push{\bbX}  \qw & \multigate{1}{U} & \push{\bbY} \qw & \qw \\ 
			& & \Nghost{U} & \Ngate{\alpha}{\ww} & \ww  }$ is a universal dilation of $T$, and every universal dilation of $T$ is of this form for a unique isomorphism $\alpha$.
		\item For any reversible $R$, $\scalemyQ{.8}{0.7}{0.5}{& \push{\bbX}  \qw & \multigate{1}{U} & \push{\bbY} \qw & \qw \\ 
			& & \Nghost{U} & \Ngate{R}{\ww} & \ww  }$ is a complete dilation of $T$, and every complete dilation of $T$ is of this form for a unique reversible $R$. 
	\end{enumerate}
	
\end{Prop}

\begin{proof}
It is easy to see that the given channels	are universal, respectively complete, for any isomorphism $\alpha$, respectively reversible $R$. It thus suffices to prove the stated existence and uniqueness clauses. This is a standard `universal property in category'-argument. 

The key observation is that universal dilations $V$ have the property that if  $\scalemyQ{.8}{0.7}{0.5}{& \push{\bbX}  \qw & \multigate{1}{V} & \push{\bbY} \qw & \qw \\ 
	& & \Nghost{V} & \push{\bbE_0} \ww  & \ww  }= \scalemyQ{.8}{0.7}{0.5}{& \push{\bbX}  \qw & \multigate{1}{V} & \qw & \push{\bbY} \qw & \qw \\ 
	& & \Nghost{V} & \push{\bbE_0} \ww & \Ngate{H}{\ww} & \push{\bbE_0} \ww &\ww } $, then $H= \id_{\bbE_0}$; this is simply by the uniqueness clause in the definition of universality. We can use this observation as follows:

As for item 1., let $\scalemyQ{.8}{0.7}{0.5}{& \push{\bbX}  \qw & \multigate{1}{U'} & \push{\bbY} \qw & \qw \\ 
	& & \Nghost{U'} & \push{\bbE'_0} \ww  & \ww  }$ be any universal dilation of $T$. By universality of $U$, we must have 

\begin{align}
\scalemyQ{1}{0.7}{0.5}{& \push{\bbX}  \qw & \multigate{1}{U} & \push{\bbY} \qw & \qw \\ 
	& & \Nghost{U} & \Ngate{G}{\ww} & \ww  } = \scalemyQ{1}{0.7}{0.5}{& \push{\bbX}  \qw & \multigate{1}{U'} & \push{\bbY} \qw & \qw \\ 
	& & \Nghost{U'} & \ww  & \ww  }
\end{align}

for a unique channel $G$. Since however $U'$ is also universal, we similarly find $G'$ such that 

\begin{align}
\scalemyQ{1}{0.7}{0.5}{& \push{\bbX}  \qw & \multigate{1}{U'} & \push{\bbY} \qw & \qw \\ 
	& & \Nghost{U'} & \Ngate{G'}{\ww} & \ww  } = \scalemyQ{1}{0.7}{0.5}{& \push{\bbX}  \qw & \multigate{1}{U} & \push{\bbY} \qw & \qw \\ 
	& & \Nghost{U} &  \ww  & \ww  } \quad.
\end{align}

Together these two imply 

\begin{align}
\scalemyQ{.8}{0.7}{0.5}{& \push{\bbX}  \qw & \multigate{1}{U} & \qw & \push{\bbY} \qw & \qw \\ 
	& & \Nghost{U} & \Ngate{G}{\ww} & \Ngate{G'}{\ww} &\ww  } = \scalemyQ{.8}{0.7}{0.5}{& \push{\bbX}  \qw & \multigate{1}{U} & \push{\bbY} \qw & \qw \\ 
	& & \Nghost{U} & \ww  & \ww  } \quad \text{and} \quad \scalemyQ{.8}{0.7}{0.5}{& \push{\bbX}  \qw & \multigate{1}{U'} & \qw & \push{\bbY} \qw & \qw \\ 
	& & \Nghost{U'} & \Ngate{G'}{\ww} & \Ngate{G}{\ww} &\ww  } = \scalemyQ{.8}{0.7}{0.5}{& \push{\bbX}  \qw & \multigate{1}{U'} & \push{\bbY} \qw & \qw \\ 
	& & \Nghost{U'} & \ww  & \ww  } \quad,
\end{align}

from which we  by the initial observation conclude that $G' \after G = \id_{\bbE_0}$ and $G \after G' = \id_{\bbE'_0}$, i.e. the channel $G =: \alpha$ is an isomorphism with inverse $G'$. 

Item 2. is proved analogously.
	\end{proof}

The second result provides a recharacterisation of the dilational ordering in the presence of universal dilations. 

Let us denote by $\OneDil{T}$ the class of \myuline{one-sided} dilations of $T$ with \myuline{simple} hidden interface (any one-sided dilation is $\der$-equivalent to such a dilation, simply by merging systems in the environment). We have the following:  

\begin{Prop} (Structure of $(\OneDil{T}, \der)$.) \label{prop:Blackwell}\\
Suppose that $\channel{\bbX}{T}{\bbY}$ has a universal dilation for which the hidden system is $\cE_0$. Then, the pre-order $(\OneDil{T}, \der)$ is order-isomorphic to $(\up{Trans}_{\Theory}(\cE_0, -), \succeq)$, where  $\up{Trans}_{\Theory}(\cE_0, -)$ denotes the class of all transformations in $\Theory$ with domain $\cE_0$, and where $\succeq$ is the \emph{Blackwell order} given by

\begin{align}\channel{\cE_0}{G}{\cE}  \succeq  \channel{\cE_0}{G'}{\cE'}  \quad  \Leftrightarrow \quad \exists \channel{\cE}{M}{\cE'} :  \scalemyQ{.8}{0.7}{0.5}{& \push{\cE_0} \qw & \gate{G} & \push{\cE} \qw & \gate{M} & \push{\cE'} \qw & \qw }  = \channel{\cE_0}{G'}{\cE'} \quad.
\end{align} 
\end{Prop}

\begin{proof}
If $\scalemyQ{.8}{0.7}{0.5}{& \push{\bbX}  \qw & \multigate{1}{U} & \push{\bbY} \qw & \qw \\ 
	& & \Nghost{U} & \push{\cE_0} \ww  & \ww  }$ is a universal dilation of $T$ as asserted, then the map $\up{Trans}_{\Theory}(\cE_0, -) \to \OneDil{T}$ given by $G \mapsto \scalemyQ{1}{0.7}{0.5}{& \push{\bbX}  \qw & \multigate{1}{U} & \push{\bbY} \qw & \qw \\ 
	& & \Nghost{U} & \Ngate{G}{\ww} & \ww  } $ is an order-isomorphism. 
\end{proof}

\begin{Remark} (On Terminology.)\\
The name \emph{Blackwell order} was pointed out to me by Tobias Fritz; I have not subsequently been able to find a reference for this, but I have no reasons not to believe him. 

	\end{Remark}

The Blackwell order is non-trivial already in simple theories like $\FinSets^*$. The above result implies that intricacies of this order are mirrored in the dilational ordering.  For example, we can now prove that even for very simple channels in $\CIT$ the dilational ordering has infinitely many inequivalent levels: 

\begin{Example} (Infinite Descent in $(\OneDil{\id_{\{0,1\}}}, \der)$ in $\CIT$.)\\
	Consider the identity channel $\channel{\{0,1\}}{\id}{\{0,1\}}$ in $\CIT$. It has as universal dilation the copy-channel $\scalemyQ{.8}{0.7}{0.5}{& &  & \ww & \push{\{0,1\}} \ww & \ww  \\ & \push{\{0,1\}} \qw & \ctrlo{-1} & \gate{\id} & \push{\{0,1\}} \qw & \qw}$ whose hidden system is $\{0,1\}$. By \cref{prop:Blackwell}, the pre-order of its one-sided dilations is therefore isomorphic the the Blackwell order on $\up{Trans}_{\CIT}(\{0,1\}, -)$.

	Consider in particular the sequence of channels $(\channel{\{0,1\}}{G_n}{\{0,1\}})_{n \in \N_0}$ given by 
	
	\begin{align}
	 G_n(\delta_0)=\delta_0, \quad G_n(\delta_1)= \left(1-\frac{1}{2^n}\right) \delta_0 +  \frac{1}{2^n} \delta_1.
	 \end{align}
	 
	  Clearly, $G_0 = \id_{\{0,1\}}$ and $G_{n+1} = M \circ G_n$, with $M(\delta_0)= \delta_0$, $M(\delta_1)= \frac{1}{2} \delta_0 +\frac{1}{2} \delta_1 $. (Plainly speaking, $G_n$ represents $n$ iterations of a channel which always preserves $0$ but flips $1$ to $0$ with probability $1/2$.) As such, the sequence decreases according to the Blackwell order, 
	  
	  \begin{align}
	  G_0 \succeq G_1 \succeq G_2 \succeq \ldots.
	  \end{align}
	  
	   However, each inequality must be strict: If for some $n \in \N_0$ we had $G_{n+1} \succeq G_{n}$, i.e. if there were a channel $N: \{0,1\} \to \{0,1\}$ with $G_{n} = N \circ G_{n+1}$, then by evaluating this identity in $\delta_0$ and $\delta_1$ we would find
	  
	  \begin{align}
	  \delta_0 = N(\delta_0), \quad  \left(1-\frac{1}{2^n}\right)  \delta_0 + \frac{1}{2^n}\delta_1 = \left(1-\frac{1}{2^{n+1}}\right) N(\delta_0 )+  \frac{1}{2^{n+1}}N(\delta_1),
	  \end{align} 
	  
	  which leads to $N(\delta_1) = 2 \delta_1 - \delta_0$, contradicting that $N$ is a classical channel. 
	  
	  (Intuitively, it is clear that no channel $N$ could restore the damage inflicted by the erroneous channel $M$, and that is essentially what is asked for.)
\end{Example}

We end this section by demonstrating that $\QIT$ is universal: 

\begin{Thm} \label{thm:QITUniv} 
In $\QIT$, every channel has a universal dilation. 
	\end{Thm}

\begin{proof}
	By \cref{thm:QITComp}, every channel $\channel{\bbX}{\Lambda}{\bbY}$ in $\QIT$ has complete dilation given by a Stinespring dilation  $\oneext{\bbX}{\Sigma}{\bbY}{\cE}$. By \cref{lem:choigeneralised} below, any \myuline{minimal} Stinespring dilation moreover meets the uniqueness clause in the definition of universality and is thus a universal dilation. (A \emph{minimal} Stinespring dilation is one for which the system $\cE$ is as small as possible. For finite-dimensional spaces, this is simply a matter of choosing the dimension minimal, but in general it means that $\Sigma$ has full rank on $\cE$, in the sense that the smallest closed subspace which contains all local supports of the states $\{\Sigma(\varrho) \mid  \varrho \in \St{\cH_1}\}$, where $\cH_1$ is the total system corresponding to the interface $\bbX$, is all of $\cE$; compactly,\footnote{Here, the \emph{support} $\supp{P}$ of a positive linear operator $P$ means the range $\Im P$, or, equivalently, the orthogonal complement of $\ker{P}$.} $\displaystyle  \bigvee_{\varrho \in \St{\cH_1}} \supp{( \id_\cE \otimes \tr_{\cK_1})[\Sigma(\varrho)]}= \cE$.) \end{proof}

\begin{Lem} (Generalised Choi-Jamio\l{}kowski Isomorphism.) \label{lem:choigeneralised} \\
	Suppose that $
	\myQ{0.7}{0.7}{
		&                   & \Nmultigate{1}{\Sigma} & \push{\cE} \qw &\qw  \\
		&  \push{\cH_1}  \qw & \ghost{\Sigma} &  \push{\cK_1} \qw  & \qw  \\
	} 
	$ is an isometric channel in $\QIT$ (or $\QIT^\infty$) with full rank on system $\cE$, meaning that 
	
	\begin{align}
	\cE =  \bigvee_{\varrho \in \St{\cH_1}} \supp{( \id_\cE \otimes \tr_{\cK_1})[\Sigma(\varrho)]}.
	\end{align}
	
If $\scalemyQ{.8}{0.7}{0.5}{
		&           \push{\cH_2} \qw         & \multigate{1}{\Lambda} & \push{\cK_2} \qw & \qw  \\
		&  \push{\cE}  \qw & \ghost{\Lambda} &    \\
	} $ and $\scalemyQ{.8}{0.7}{0.5}{
		&           \push{\cH_2} \qw         & \multigate{1}{\Lambda'} & \push{\cK_2} \qw  & \qw \\
		&  \push{\cE}  \qw & \ghost{\Lambda'} &    \\
	} $ are quantum channels such that  
	
	\begin{align} \label{eq:choichannel} 	
	\myQ{0.7}{0.5}{
		& \push{\cH_2}  \qw    & \qw    & \qw & \multigate{1}{\Lambda}  &  \push{\cK_2} \qw & \qw \\
		&                   & \Nmultigate{1}{\Sigma} & \push{\cE} \qw& \ghost{\Lambda} & \\
		&  \push{\cH_1}  \qw & \ghost{\Sigma} & \qw & \qw &  \push{\cK_1} \qw  & \qw  \\
	} 
	\quad = 
	\quad 	
	\myQ{0.7}{0.5}{
		& \push{\cH_2}  \qw    & \qw    & \qw & \multigate{1}{\Lambda'}  &  \push{\cK_2} \qw & \qw \\
		&                   & \Nmultigate{1}{\Sigma} & \push{\cE} \qw& \ghost{\Lambda'} & \\
		&  \push{\cH_1}  \qw & \ghost{\Sigma} & \qw & \qw &  \push{\cK_1} \qw  & \qw  \\
	} 
	\quad,
	\end{align}
	
then it holds that $\Lambda= \Lambda'$.
	
\end{Lem}

\begin{Remark} (Relation to Ordinary Choi-Jamio\l{}kowski Isomorphism.)\\
	The injectivity of the ordinary Choi-Jamio\l{}kowski isomorphism (\cite{Jiang13,Pill67,Choi75,Jam72}) corresponds to the special case where $\cH_2 = \cH_1 = \C$, so that $\Sigma$ is simply a pure state on $\cE \otimes \cK_1$ with full rank on $\cE$ (usually taken to be maximally entangled).
\end{Remark}

\begin{Remark} (On Scope.) \\
Since the result holds as well in the case of infinite-dimensional Hilbert spaces, the theory $\QIT^\infty$ is universal by the same chain of arguments as used for $\QIT$.
	\end{Remark}

\begin{proof}
	I will give two different proofs of this statement. Note that, in any case, we may assume without loss of generality that $\cH_2=\C$ (so the $\cH_2$-wire can be omitted), since if the implication holds in that case then from \cref{eq:choichannel} we conclude that $
	\scalemyQ{.8}{0.7}{0.5}{
		&	\Ngate{\varrho}    &           \push{\cH_2} \qw         & \multigate{1}{\Lambda} & \push{\cK_2} \qw & \qw  \\
		& \qw 	&  \push{\cE}  \qw & \ghost{\Lambda} &    \\
	} =	\scalemyQ{.8}{0.7}{0.5}{
	&	\Ngate{\varrho}    &           \push{\cH_2} \qw         & \multigate{1}{\Lambda'} & \push{\cK_2} \qw & \qw  \\
	& \qw 	&  \push{\cE}  \qw & \ghost{\Lambda'} &    \\
}$ for all states $\varrho$ on $\cH_2$, which implies that $\Lambda = \Lambda'$ since quantum channels are determined by their action on product states. \textbf{Hence, we assume throughout that the domain of $\Lambda$ and $\Lambda'$ is the simple interface corresponding to system $\cE$.}\\
	
	The first proof looks simple, though it actually depends on some slightly tricky arguments detailed in the footnotes (one of which is invalid in the infinite-dimensional version of the statement). This proof was found in conversation with David P\'{e}rez-Garc\'{i}a when I visited Madrid, and it works essentially by reduction to the injectivity of the usual Choi-Jamio\l{}kowski isomorphism.
	
Let $\bistate{\phi}{\cH_1}{\cH_1}$ be a pure state with marginals of full rank. If we can argue that the pure state

\begin{align}
\myQ{0.7}{0.5}{& \Nmultigate{2}{\tilde{\phi}} & \push{\cE} \qw & \qw \\ 
& \Nghost{\tilde{\phi}} & \push{\cK_1} \qw & \qw \\
& \Nghost{\tilde{\phi}} & \push{\cH_1} \qw & \qw }
\quad := \quad 	\myQ{0.7}{0.5}{
	 & & & \Nmultigate{1}{\Sigma} & \push{\cE} \qw & \qw \\
	                  & \Nmultigate{1}{\phi} & \push{\cH_1} \qw &  \ghost{\Sigma} &\push{\cK_1} \qw & \qw  \\
   & \Nghost{\phi} & \qw &  \qw & \push{\cH_1} \qw  & \qw  \\
} \quad
\end{align}

has full rank on $\cE$, then the identity $\Lambda= \Lambda'$ follows from \cref{eq:choichannel} by pre-composing with $\phi$ and invoking injectivity of the usual Choi-Jamio\l{}kowski isomorphism. To show that $\tilde{\phi}$ has full rank on $\cE$ is to show that 

\begin{align}
\supp{( \id_\cE \otimes \tr_{\cK_1})[\Sigma(\tau)]} = \cE,
	\end{align}
	
	 where  $\tau := (\id_{\cH_1} \otimes \tr_{\cH_1})(\phi)$ is the marginal of $\phi$. To this end, let $\varrho$ be any state on $\cH_1$. Since $\tau$ has full rank on $\cH_1$ by assumption, there exists $p>0$ such that $p \, \varrho \leq \tau$.\footnote{This is by the spectral theorem, using that the eigenvalues of $\tau$ are lower bounded by a strictly positive constant. It generally fails in infinite-dimensional spaces; for example, if in $\ell^2(\N)$ the state $\tau$ is $\sum_{n=1}^\infty \frac{2}{3^n} \ketbra{e_n}$ and $\varrho$ is the pure state with vector representative $\sum_{n=1}^\infty \frac{1}{\sqrt{2^n}} \ket{e_n}$ (here, $(\ket{e_n})_{n \in \N}$ denotes an orthonormal basis), then the operator inequality $p \, \varrho \leq \tau$ implies $p \bra{e_m} \varrho \ket{e_m} \leq \bra{e_m} \tau \ket{e_m}$, i.e. $p/2^m \leq 2/3^m$ for all $m$, which forces $p \leq 0$.} This implies by positivity of the map $A \mapsto (\id_\cE \otimes \tr_{\cK_1})[\Sigma(A)]$ that 
	 
	 \begin{align}
	 p \, ( \id_\cE \otimes \tr_{\cK_1})[\Sigma(\varrho)] \leq ( \id_\cE \otimes \tr_{\cK_1})[\Sigma(\tau)],
	 \end{align}
	 
	  and this operator inequality in turn implies\footnote{To show this, we may equivalently show that the operator inequality $A \geq B \geq 0$ implies $\ker(A) \subseteq \ker(B)$. This can be proved in multiple ways, but the following elegant argument I owe to Lukas Schimmer: If $x \in \ker(A)$, then $0= \langle x, Ax \rangle \geq \langle x, Bx \rangle  \geq 0$, so $ \langle x, Bx \rangle  =0$. By the spectral theorem, $B= \sum_{j} \lambda_j P_{\lambda_j}$, where $\lambda_j$ is the $j$th eigenvalue of $B$ and $P_{\lambda_j}$ the orthogonal projection onto the corresponding eigenspace, and the identity $ \langle x, Bx \rangle  =0$ thus becomes $\sum_{j} \lambda_j \norm*{P_{\lambda_j} x}^2=0$. As the eigenvalues $\lambda_j$ are non-negative this implies that $P_{\lambda_j} x=0$ for any $j$ with $\lambda_j > 0$, but this means that $Bx = 0$, i.e. that $x \in \ker(B)$.} that 
	 
	 \begin{align}
 \supp{ p \, ( \id_\cE \otimes \tr_{\cK_1})[\Sigma(\varrho)} \subseteq  \supp{( \id_\cE \otimes \tr_{\cK_1})[\Sigma(\tau)]}.
	 \end{align}
	 
Since $p\neq 0$, the containment of supports holds also when $p$ is omitted from the left hand side, so as $\varrho$ was arbitrary we have

\begin{align}
\bigvee_{\varrho \in \St{\cH_1}} \supp{( \id_\cE \otimes \tr_{\cK_1})[\Sigma(\varrho)]} \subseteq \supp{( \id_\cE \otimes \tr_{\cK_1})[\Sigma(\tau)]},
\end{align}

and the full rank assumption on $\Sigma$ then yields the conclusion $\supp{( \id_\cE \otimes \tr_{\cK_1})[\Sigma(\tau)]} = \cE$, as desired. This concludes the first proof.\\

The second proof is the one I found originally. It proceeds in three steps to reduce to the special case where $\cK_2 = \cE$, $\Lambda = \id_\cE$ and $\Lambda'$ is a unitary conjugation $\cE$, and then handles this special case by combining a topological argument with a consideration about the cardinality of the spectrum of a unitary operator. \\

	\textbf{\myuline{Claim 1:}} \emph{We may assume that $\Lambda$ and $\Lambda'$ are both isometric.} Indeed, suppose that $\Lambda$ and $\Lambda'$ are general channels satisfying \eqref{eq:choichannel}. If $\hat{\Lambda}$ and $\hat{\Lambda}'$ are Stinespring dilations of $\Lambda$ and $\Lambda'$ respectively, then the channels

	\begin{align}	
	\myQ{0.7}{0.5}{
		& & & \Nmultigate{1}{\hat{\Lambda}} & \push{\cF} \ww & \ww \\
		&                   & \Nmultigate{1}{\Sigma} & \ghost{\hat{\Lambda}} &\push{\cK_2} \qw & \qw  \\
		&  \push{\cH_1}  \qw & \ghost{\Sigma} & \qw &  \push{\cK_1} \qw  & \qw  \\
	} 
	\quad \text{and} \quad 
	\myQ{0.7}{0.5}{
		& & & \Nmultigate{1}{\hat{\Lambda}'} & \push{\cF'} \ww & \ww \\
		&                   & \Nmultigate{1}{\Sigma} & \ghost{\hat{\Lambda}'} &\push{\cK_2} \qw & \qw  \\
		&  \push{\cH_1}  \qw & \ghost{\Sigma} & \qw &  \push{\cK_1} \qw  & \qw  \\
	} 
	\end{align}
	
	are Stinespring dilations of the same channel, so there exist isometric channels $\Gamma$ and $\Gamma'$ such that 
	
	\begin{align}	
	\myQ{0.7}{0.5}{
		& & & \Nmultigate{1}{\hat{\Lambda}} & \push{\cF} \ww & \Ngate{\Gamma}{\ww} & \push{\tilde{\cF}} \ww & \ww \\
		&                   & \Nmultigate{1}{\Sigma} & \ghost{\hat{\Lambda}} & \qw & \qw & \push{\cK_2} \qw & \qw \\
		&  \push{\cH_1}  \qw & \ghost{\Sigma} & \qw & \push{\cK_1} \qw  & \qw  & \qw &  \qw \\
	} 
	\quad = \quad 	
	\myQ{0.7}{0.5}{
		& & & \Nmultigate{1}{\hat{\Lambda}'} & \push{\cF'} \ww & \Ngate{\Gamma'}{\ww} & \push{\tilde{\cF}} \ww & \ww \\
		&                   & \Nmultigate{1}{\Sigma} & \ghost{\hat{\Lambda}'} & \qw & \qw & \push{\cK_2} \qw & \qw \\
		&  \push{\cH_1}  \qw & \ghost{\Sigma} & \qw & \push{\cK_1} \qw  & \qw  & \qw &  \qw \\
	} 
	\quad .
	\end{align}
	
If the implication holds for isometric channels, it follows that $\scalemyQ{.8}{0.7}{0.5}{
	& & & \Nmultigate{1}{\hat{\Lambda}} & \push{\cF} \ww & \Ngate{\Gamma}{\ww} & \push{\tilde{\cF}} \ww & \ww \\
	&                   & \push{\cE} \qw & \ghost{\hat{\Lambda}} & \qw & \qw & \push{\cK_2} \qw & \qw }=\scalemyQ{.8}{0.7}{0.5}{
	& & & \Nmultigate{1}{\hat{\Lambda}'} & \push{\cF'} \ww & \Ngate{\Gamma'}{\ww} & \push{\tilde{\cF}} \ww & \ww \\
	&                   & \push{\cE} \qw & \ghost{\hat{\Lambda}'} & \qw & \qw & \push{\cK_2} \qw & \qw }
$, and trashing $\tilde{\cF}$ we then conclude $\Lambda= \Lambda'$. \\

	\textbf{\myuline{Claim 2:}} \emph{We may assume that $\cK_2= \cE \otimes \cR$ for some system $\cR$, and that $\Lambda$ and $\Lambda'$ are both isometric, with $\Lambda$ of the form $\scalemyQ{0.7}{1}{0.7}{
		& \Ngate{\sigma} & \push{\cR} \qw & \qw \\
		& \qw & \push{\cE} \qw & \qw } $ for some pure state $\sigma$.}
	For this, first observe that we may by the previous claim assume both channels to be isometric. Clearly, we may also assume that $\cK_2 = \cE \otimes \cR$ for some system $\cR$, by possibly isometrically embedding the isometries into a larger space. Finally, any isometric conjugation from $\cE$ to $\cE \otimes \cR$ is of the form $\scalemyQ{.8}{0.7}{0.5}{
		& \Ngate{\sigma} & \push{\cR} \qw & \multigate{1}{\scrU} &\push{\cR} \qw & \qw \\
		& \qw & \push{\cE} \qw & \ghost{\scrU}  &\push{\cE} \qw & \qw } $ for some pure state $\sigma$ on $\cR$ and some unitary conjugation $\scrU$, so the desired follows by moving $\scrU$ to the other side of the identity, invoking injectivity in the special case, and moving $\scrU$ back. \\

	\textbf{\myuline{Claim 3:}} \emph{We may assume that $\cK_2= \cE$, that $\Lambda= \id_\cE$, and that $\Lambda'$ is a unitary conjugation on $\cE$.} %
		Indeed,  assume that the implication holds in this case. Let $\Lambda$ and $\Lambda'$ be of the form from the case in the previous claim -- i.e. $\cK_2= \cE \otimes \cR$ for some system $\cR$, $\Lambda$ isometric of the form $\scalemyQ{0.7}{1}{0.7}{
		& \Ngate{\sigma} & \push{\cR} \qw & \qw \\
		& \qw & \push{\cE} \qw & \qw }$ for some pure state $\sigma$, and $\Lambda' $ arbitrary isometric from $\cE$ to $\cE \otimes \cR$ -- and suppose that $\Lambda$ and $\Lambda'$ satisfy the identity
	
	\begin{align} \label{eq:assump}
	\myQ{0.7}{0.5}{
		& & & \Ngate{\sigma} &\push{\cR} \qw & \qw \\
		&                   & \Nmultigate{1}{\Sigma} & \qw &  \push{\cE} \qw & \qw  \\
		&  \push{\cH_1}  \qw & \ghost{\Sigma}  &  \push{\cK_1} \qw  & \qw  \\
	} 
	\quad = \quad 
	\myQ{0.7}{0.5}{
		& & & \Nmultigate{1}{\Lambda'}  &  \push{\cR} \qw & \qw \\
		&                   & \Nmultigate{1}{\Sigma}  & \ghost{\Lambda'}  & \push{\cE} \qw & \qw  \\
		&  \push{\cH_1}  \qw & \ghost{\Sigma}  &  \push{\cK_1} \qw  & \qw \\
	} 
	\quad ;
	\end{align}

	 we must show that $\Lambda = \Lambda'$. By tracing out system $\cK_1$ in \eqref{eq:assump}, we get

	\begin{align}
	\myQ{0.7}{0.5}{
		& & & \Ngate{\sigma} &\push{\cR} \qw & \qw \\
		&                   & \Nmultigate{1}{\Sigma} & \qw &  \push{\cE} \qw & \qw  \\
		&  \push{\cH_1}  \qw & \ghost{\Sigma}  &  \push{\cK_1} \qw  & \gate{\tr}  \\
	} 
	\quad = \quad 
	\myQ{0.7}{0.5}{
		& & & \Nmultigate{1}{\Lambda'}  &  \push{\cR} \qw & \qw \\
		&                   & \Nmultigate{1}{\Sigma}  & \ghost{\Lambda'}  & \push{\cE} \qw & \qw  \\
		&  \push{\cH_1}  \qw & \ghost{\Sigma}  &  \push{\cK_1} \qw  & \gate{\tr}  \\
	} 
	\quad ,
	\end{align}
	
	and by inserting various states $\varrho \in \St{\cH_1}$, this implies that 
	
	\begin{align} \label{eq:Wrho}
	\myQ{0.7}{0.5}{
		& & \Ngate{\sigma} &\push{\cR} \qw & \qw \\
		& \Ngate{\mu} & \qw &  \push{\cE} \qw & \qw  \\
	} 
	\quad = \quad 
	\myQ{0.7}{0.5}{
		& & &  \Nmultigate{1}{\Lambda'}  &  \push{\cR} \qw & \qw \\
		& \Ngate{\mu}  &\push{\cE} \qw &  \ghost{\Lambda'}  & \push{\cE} \qw & \qw  \\
	} 
	\end{align}
	
	for all states $\mu$ in the set $\scrS_0 := \{(\id_\cE \otimes \tr_{\cK_1})[\Sigma(\varrho)] \mid \varrho \in \St{\cH_1}\} \subseteq \St{\cE}$. The full rank assumption on $\Sigma$ does not not guarantee that $\scrS_0$ contains every state on $\cE$, but it does ensure that all of $\cE$ is covered by the supports, in the sense that 
	
\begin{align}	 \label{eq:mufullsupp}
\bigvee_{\mu \in \scrS_0} \up{supp}(\mu) = \cE .
\end{align}

Now, letting $\ket{\sigma} \in \cR$ be a vector representative of the pure state $\sigma$, and letting $W: \cE \to \cE \otimes \cR$ be an isometry representing $\Lambda'$ (i.e. $\Lambda'(A)= WAW^*$), \cref{eq:Wrho} implies that

	\begin{align} 
 \supp{\mu} \otimes \up{span}\{\ket{\sigma}\} = W(\text{supp}(\mu))
\end{align}

for all $\mu \in \scrS_0$. \cref{eq:mufullsupp} then yields 

	\begin{align} 
\cE  \otimes \up{span}\{\ket{\sigma}\} =		\bigvee_{\mu \in \scrS_0} W(\text{supp}(\mu)) = 	W \left( \bigvee_{\mu \in \scrS_0} \text{supp}(\mu)\right)= \up{Im}(W),
\end{align}

using linearity and continuity of $W$ for the middle equality. This identity, however, implies that $W=U \otimes\ket{\sigma}$ for some unitary operator $U: \cE \to \cE$, and thus $\Lambda' = \scrU \otimes \sigma$ for some unitary conjugation $\scrU$ on $\cE$. By the assumption that injectivity holds in the special unitary case, we must have $\scrU= \id_\cE$, and this implies that $\Lambda = \Lambda'$ as desired. \\

	\textbf{\myuline{Claim 4:}}  \emph{The implication of the lemma holds in the case described in the previous claim.} The assumption is that $\scrU$ is a unitary conjugation on $\cE$ such that
	
	\begin{align}	\label{eq:unitaryass}
	\myQ{0.7}{0.5}{
		&                   & \Nmultigate{1}{\Sigma} & \qw &  \push{\cE} \qw & \qw  \\
		&  \push{\cH_1}  \qw & \ghost{\Sigma}  &  \push{\cK_1} \qw  &\qw  & \qw \\
	} 
	\quad = \quad 	
	\myQ{0.7}{0.5}{
		&                   & \Nmultigate{1}{\Sigma}  & \push{\cE} \qw & \gate{\scrU}  & \push{\cE} \qw & \qw  \\
		&  \push{\cH_1}  \qw & \ghost{\Sigma}  & \qw &  \push{\cK_1} \qw  & \qw  & \qw \\
	} 
	\quad ,
	\end{align}
	
	and the objective is to show that $\scrU = \id_\cE$.

	Let $U: \cE \to \cE$ be a unitary operator which represents $\scrU$ (i.e. $\scrU(A) = UAU^*$). We must show that $U= \zeta_0 \, \bone_\cE$, where $\zeta_0 \in \C$ has modulus $1$ (here, $\bone_\cE$ denotes the identity operator on $\cE$). From \eqref{eq:unitaryass} it follows that
	
	\begin{align}	\label{eq:ChoiVary}
	\myQ{0.7}{0.5}{
		& \Nmultigate{1}{\Sigma(\psi)} & \qw &  \push{\cE} \qw & \qw  \\
		& \Nghost{\Sigma(\psi)}  &  \push{\cK_1} \qw  &\qw  & \qw \\
	} 
	\quad = \quad 
	\myQ{0.7}{0.5}{
		& \Nmultigate{1}{\Sigma(\psi)}  & \push{\cE} \qw & \gate{\scrU}  & \push{\cE} \qw & \qw  \\
		& \Nghost{\Sigma(\psi)}  & \qw &  \push{\cK_1} \qw  & \qw  & \qw  \\
	} 
	\quad 
	\end{align}
	
	for any pure state $\psi \in \St{\cH_1}$. This identity in turn lends itself to a use of the ordinary Choi-Jamio\l{}kowski isomorphism, or at least a variation of it: If $\ket{\phi} \in \cE \otimes \cK_1$ is a vector representative of the pure state $\phi := \Sigma(\psi)$, and if $\ket{\phi} = \sum_{j=1}^r \sqrt{p(j)} \ket{\phi^\cE(j)} \otimes \ket{\phi^{\cK_1}(j)}$ is a Schmidt decomposition with $p(j)>0$ for all $j=1, \ldots, r$, then \cref{eq:ChoiVary}, which asserts the equality of two pure state,  reads in terms of vector representatives 
	
	\begin{align}
	\sum_{j=1}^r \sqrt{p(j)} [U \ket{\phi^\cE(j)}] \otimes \ket{\phi^{\cK_1}(j)} = \zeta(\psi) \sum_{j=1}^r \sqrt{p(j)} \ket{\phi^\cE(j)} \otimes \ket{\phi^{\cK_1}(j)} 
	\end{align} 
	
for some (unique) phase $\zeta(\psi) \in \C$ with $\abs{\zeta(\psi)}=1$. As  $\left( \ket{\phi^{\cK_1}(j)} \right)_{j=1, \ldots, r}$ is an orthonormal system, this implies that $U \ket{\phi^\cE(j)} =  \zeta(\psi) \ket{\phi^\cE(j)}$ for all $j=1, \ldots, r$, or, equivalently, that 
	
	\begin{align} \label{Ueigen}
	U \ket{\phi} =  \zeta(\psi) \ket{\phi}
	\end{align}
	
	for all vectors $\ket{\phi}$ in the subspace 
	
	\begin{align} \label{eigenspace}
	\Span{\ket{\phi^\cE(j)} \mid j=1, \ldots, r} = \supp{(\id_\cE \otimes \tr_{\cK_1})(\phi)} = \supp{(\id_\cE \otimes \tr_{\cK_1})[\Sigma(\psi)]}.
	\end{align}
	
	Now, if $\phi= \Sigma(\psi)$ had full rank on $\cE$, i.e. if this span were all of $\cE$ (for some pure state $\psi$ on $\cH_1$), then we would be done, since \eqref{Ueigen} would then assert that $U$ acts as a multiple of the identity operator globally. In general, however, the full-rank assumption on $\Sigma$ merely implies that $\cE$ can be `patched up' from subspaces on which $U$ acts as (possibly different) multiples of the identity operator.
	We are saved by properties of a topological nature:

	Since $U$ acts as in \eqref{Ueigen} on the non-zero subspace \eqref{eigenspace}, the function $\zeta(\cdot)$ admits the explicit expression

	\begin{align} \label{eq:zeta}
	\zeta(\psi) = \tr \left( U (\id_\cE \otimes \tr_{\cK_1})[\Sigma(\psi)]\right)
	\end{align}  
	
	and for each $\psi$, $\zeta(\psi)$ is an eigenvalue of $U$ (whose eigenspace contains the subspace \eqref{eigenspace}). Now, the function $\zeta$ defined by \cref{eq:zeta}  is clearly continuous w.r.t. the topology induced by the trace-distance and the ordinary topology on $\C$. Also, the set of pure states on $\cH_1$ is path-connected with the topology induced by the trace-distance (meaning that for any pure states $\psi_0$ and $\psi_1$ on $\cH_1$, we can find a continuous map $\gamma :[0,1] \to \St{\cH_1}$ whose range contains only pure states, with $\gamma(0) = \psi_0$ and $\gamma(1) = \psi_1$). Therefore, the image of this set under $\zeta(\cdot)$ must be a non-empty path-connected subset of $\C$. However, every non-empty path-connected subset of $\C$ is either uncountable or consists of a single point, and since $\Im \zeta$ is a set of eigenvalues of $U$, it cannot be uncountable if $\cE$ is only finite-dimensional, or even separable. Consequently, $\Im \zeta$ contains only a single point, i.e. \myuline{$\zeta(\cdot)$ is constant}. With $\zeta_0$ denoting this constant value, we conclude that $U= \zeta_0 \bone_\cE$ globally, as desired. \end{proof}

\section{Purification}
\label{sec:Selfuniv}

The results of the previous sections imply that $\CIT$, $\QIT$ and all cartesian theories are localisable and universal. In particular, all of the dilational properties we have considered have been shared by the two information theories $\CIT$ and $\QIT$ alike. One dilational property, however, will set them apart, and that property is the topic of this section. \\

Picturesquely speaking, complete dilations are \emph{omniscient}: If $K$ is a complete dilation of $T$, then $K$ knows everything there is to know about $T$. Both $\CIT$ and $\QIT$ admit complete dilations. 

In $\QIT$, however, there exist complete dilations of $T$ which, in addition to knowing everything about $T$, also know everything there is to know \emph{about themselves}. Indeed, we have already seen that isometric channels in $\QIT$ are dilationally pure (\cref{prop:PureEqualsIso}), that is, \emph{self-complete}, and every quantum channel admits an isometric dilation, namely its Stinespring dilation. This phenomenon -- the fact that a quantum channel can be \emph{purified} -- abruptly terminates the process of forming dilations. \emph{Exhaustive knowledge is possible.}

In contrast, dilational purity -- or, self-completeness -- in $\CIT$ is a property reserved for the dull; any channel  which is not a pure state has non-trivial dilations (\cref{prop:PureinCIT}). As such, there is no ceiling to the formation of dilations: A complete dilation $K_0$ of a channel $T$ (given by copying all inputs and outputs) is not a complete dilation of itself; of course, $K_0$ too has a complete dilation, $K_1$, but $K_1$ is not a complete dilation of itself either -- and so it continues, ad infinitum, with complete dilations $K_2$ of $K_1$, $K_3$ of $K_2$, and so forth. \emph{There is always more to know}. \\

The goal of this section is to demonstrate that this distinction between $\CIT$ and $\QIT$ is not a randomly chosen one, but rather one that can be seen as responsible for \myuline{many} features which distinguish the two theories. As such, the significance of the isometric channels in the so-called `Church of the larger Hilbert space'\footnote{A phrase coined by John A. Smolin (\cite{Church}) about the possibility -- by virtue of Stinespring's dilation theorem -- of always  regarding a quantum channel as the marginal of an isometric channel into a larger space.} is not that they are reversible, or per se that they are isometric -- the significance is that \emph{in the church, (dilational) purity is obtained}. Jokingly, one might say that $\QIT$ is like Christianity whereas $\CIT$ is like Buddhism.   \\

`Purification' as an axiom has been considered in the literature before, with the profound conclusion that the property more or less characterises quantum theory uniquely (\cite{Chir10,Chir11}). However, there are differences in the statement of that principle, and the categorical framework used here is technically simpler and independent of probabilistic structure. Moreover, the main results of this chapter do not have counterparts in Refs. \cite{Chir10,Chir11}, and they are derived from fewer principles, none of which are not about dilations. (Further details are given in \cref{rem:PurePrinc}.)

\begin{Definition} (Purifiable Theories.) \label{def:SelfUniv}\\
	A theory $\Theory$ is called \emph{purifiable} if every channel in $\Theory$ has a  dilationally pure dilation.
	\end{Definition}

\begin{Example} (Purification in the Information Theories.)\\
	The theory $\QIT$ is purifiable, since Stinespring dilations are dilationally pure. The theory $\CIT$ is not purifiable, since the only channels which are dilationally pure are probabilistically pure states.
	\end{Example}

\begin{Example} (Purifiable Thin Theories.) \label{ex:ThinPurifiable}\\
	A thin theory described by the monoid $(M, \star, 1, \succeq)$ is purifiable if $\star$ satisfies the cancellation law $z \star u \succeq z \star v \Rightarrow u \succeq v$. Indeed, for any channel $\channel{x}{\phantom{X}}{y}$ the dilation $\twoext{x}{y}{\phantom{X}}{y}{x}$ is  dilationally pure by \cref{ex:PureThin} if $\star$ satisfies the cancellation law. Thus, for example, the thin theory $(\N, \cdot, 1, \geq)$ is purifiable. 
	\end{Example}

\begin{Remark} (Relation to the Purification Postulate of Refs. \cite{Chir10, Chir11}.) \label{rem:PurePrinc}\\
The main result of Ref. \cite{Chir11} is that, within a large ground class of theories (defined by a list of `standing assumptions'), five reasonable axioms determine a subclass of reasonable information theories, and an additional `Purification Postulate' (first introduced in Ref. \cite{Chir10}) uniquely identifies the theory $\QIT$. In our language, the Purification Postulate of Refs. \cite{Chir10,Chir11} is the requirement that every state has a dilationally pure one-sided dilation, and that any two such dilations with the same hidden interface are related by an isomorphism on the hidden interface (so-called `uniqueness' of purifications). This requirement is clearly from the same womb as that of \cref{def:SelfUniv}, but there are important differences. 

The Purification Postulate as formulated in Refs.  \cite{Chir10, Chir11} is about \emph{probabilistic} purity and as such does not a priori pertain to the general theories considered here; however, it can be reformulated equivalently in terms of dilational purity, so we may ignore that difference. 

More importantly, the Purification Postulate concerns only \myuline{states}, and though this seems to make it more general, the truth is actually the opposite:

Firstly, according to Thm. 15 of Ref. \cite{Chir10}, the Purification Postulate implies (within their large ground class of theories) that any channel $\channel{\bbX}{T}{\bbY}$ has a dilation of the form $\scalemyQ{.8}{0.7}{0.7}{&\push{\bbX} \qw & \multigate{1}{\alpha} & \push{\bbY} \qw & \qw \\ & \Ngate{\phi} & \ghost{\alpha} & \ww & \ww}$, where $\phi$ is a pure state and $\alpha$ an isomorphism. (This can be thought of as a generalisation of Stinespring's theorem, with $\alpha$ a unitary conjugation). By virtue of their `dynamically faithful states' (these are like the full rank states in the Choi-Jamio\l{}kowski isomorphism) one can easily show that the channel $\scalemyQ{.8}{0.7}{0.7}{&\push{\bbX} \qw & \multigate{1}{\alpha} & \push{\bbY} \qw & \qw \\ & \Ngate{\phi} & \ghost{\alpha} & \ww & \ww}$ is dilationally pure, by converting the argument to an argument about purity of states. Thus, within their ground class of theories, the Purification Postulate of Refs. \cite{Chir10, Chir11} is \textbf{a stronger requirement} than purifiability in the sense of \cref{def:SelfUniv}. (I do not know whether they are actually equivalent within this ground class; this depends on whether their uniqueness clause follows from purifiability in the sense of \cref{def:SelfUniv}.)

But secondly, not only is \cref{def:SelfUniv} less restrictive than the Purification Postulate, its effective scope is also larger: Indeed, the ground class of theories in Refs. \cite{Chir10,Chir11} assumes among other things that transformations are determined by their action on states (excluding thin theories and some cartesian theories), and that theories are non-deterministic (excluding cartesian theories), cf. Def. 2 in \cite{Chir10}. These standing assumptions are used repeatedly when deriving quantum-like consequences of the Purification Postulate, and this has the subtle side-effect that some theories not complying to the standing assumptions satisfy the Purification Postulate without being anything like quantum theory; for example, every cartesian theory, e.g. $\Sets^*$, satisfies the Purification Postulate, for the simple reason that its states are already pure. As such, purifiability in the sense of \cref{def:SelfUniv} seems more robust in that it does not accidentally include such theories.

As a final remark it should be noted that the uniqueness clause in the Purification Postulate can quite easily be used to show that the dilations $\scalemyQ{.8}{0.7}{0.7}{&\push{\bbX} \qw & \multigate{1}{\alpha} & \push{\bbY} \qw & \qw \\ & \Ngate{\phi} & \ghost{\alpha} & \ww & \ww}$ are \myuline{complete} (in the same way we demonstrated completeness of Stinespring dilations). Hence, the Purification Postulate ultimately implies both purifiability and completeness, whereas the framework of this chapter separates the two. 

\end{Remark}

In the remainder of the section, we state and prove  three general aspects of theories which are purifiable and comply to various subsets of our previous dilational principles. All of these aspects are traditionally considered `quantum', but the main point here is that they can be seen rather as consequences of a few simple and abstract principles about dilations. \\

The first result (\cref{prop:IsoSelfUniversal}) has to do with isomorphisms in a purifiable theory, and we prove in particular a rather simple recharacterisation of \myuline{universal} purifiable theories (\cref{prop:PureRechar}). %

Secondly, we prove a structure theorem for reversible channels in a universal, temporally localisable and purifiable theory (\cref{thm:StructureOfReversibles}). In the case of $\QIT$, it reproduces the not entirely trivial result that a quantum channel is reversible precisely if it is the tensoring with a (possibly mixed) ancillary state, followed by an isometric conjugation.

Finally, we discuss how purifiability entails (in conjunction with the other dilational principles) the notion of \emph{complementarity} between channels, generalising (by \cref{thm:Complementarity}) the concept of complementary quantum channels (\cite{Devetak05}). In particular, we recover an abstract version of the complementarity between reversible and `completely forgetful' channels (\cref{thm:InfoDist}). This complementarity will be generalised in \cref{chap:Metric} to an approximate setting.

	\subsection{Isomorphisms and Purifiability}
	\label{subsec:IsoPure}

We begin by proving that in a purifiable theory, isomorphisms not only have dilationally pure dilations, but must in fact already  themselves be dilationally pure: 

\begin{Prop} (Isomorphisms and Purifiability.) \label{prop:IsoSelfUniversal} \\
		If $\Theory$ is a purifiable theory, every isomorphism in $\Theory$ is dilationally pure.
\end{Prop}

\begin{Remark} (Complete Characterisation of Purifiable Thin Theories.)\\
 \cref{ex:ThinPurifiable} shows that for a thin theory $(M, \star, 1, \succeq)$, the cancellation law $x \star y \succeq x \star z \Rightarrow y \succeq z$ implies purifiability. \cref{prop:IsoSelfUniversal} shows the opposite implication, since to assert that the identity $x \succeq x$ is dilationally pure is precisely to assert that cancellation law.\end{Remark}

\begin{proof}

		Let us start by showing that any identity $\id_\cX$ is dilationally pure. By assumption, $\channel{\cX}{\id}{\cX}$ has a dilationally pure dilation, say $\twoext{\cX}{\bbD_0}{P}{\cX}{\bbE_0}$. Let $\twoext{\cX}{\bbD}{L}{\cX}{\bbE}$ be any dilation of $\channel{\cX}{\id}{\cX}$; we show that it is trivial.

 The channel

	\begin{align} \label{eq:PhiOmega}
\myQ{1}{0.7}{& \push{\bbD} \ww & \Nmultigate{1}{L}{\ww} &  \ww & \push{\bbE} \ww & \ww & \ww\\
	& \push{\cX}  \qw & \ghost{L} & \push{\cX} \qw &  \multigate{1}{P} & \push{\cX} \qw & \qw\\
	& & &\push{\bbD_0} \ww & \Nghost{P}{\ww} & \push{\bbE_0}\ww & \ww }
\end{align}
	
	is a dilation of $P$ with hidden interfaces $\bbD$ and $\bbE$ (trashing $\bbE$ yields $P \og \tr_\bbD$, since $L$ dilates $\id_\cX$). However, as $P$ is dilationally pure, this dilation must be trivial, i.e. of the form 
	
	\begin{align} \label{eq:sigmaOmega}
	\myQ{1}{0.7}{& \push{\bbD} \ww & \Ngate{M}{\ww} &  \push{\bbE} \ww & \ww\\
&	\push{\cX} \qw &  \multigate{1}{P} & \push{\cX} \qw & \qw\\
 &\push{\bbD_0} \ww & \Nghost{P}{\ww} & \push{\bbE_0}\ww & \ww }
	\end{align}
	
	for some channel $M$. Equating \eqref{eq:PhiOmega} with \eqref{eq:sigmaOmega} and trashing $\bbE_0$, we derive the identity
	 $\twoext{\cX}{\bbD}{L}{\cX}{\bbE}= 	\myQ{0.5}{0.5}{& \push{\bbD} \ww & \Ngate{M}{\ww} &  \push{\bbE} \ww & \ww\\
		& \push{\cX}  \qw  & \gate{\id_\cX} & \push{\cX} \qw & \qw}$, since $P$ dilates $\id_\cX$ and the theory is normal (\cref{def:Normal}). This demonstrates the desired.
	
	Now, if $\channel{\cX}{\alpha}{\cY}$ is an arbitrary isomorphism then 
	
	\begin{align} \label{eq:alphaalpha}
	\myQ{1}{0.7}{
		& \push{\cX}  \qw & \gate{\phantom{.}\alpha^{\phantom{!}}} & \push{\cY} \qw & \gate{\alpha^{-1}}   & \push{\cX} \qw & \qw} = \myQ{1}{0.7}{
		& \push{\cX}  \qw & \gate{\id_\cX} & \push{\cX} \qw & \qw } ,
	\end{align}
	
	so for any dilation $L$ of $\alpha$, dilational purity of $\id_\cX$ entails that
	
	\begin{align} 
	\myQ{1}{0.7}{
		& \push{\bbD} \ww & \Nmultigate{1}{L}{\ww} & \push{\bbE}\ww & \ww\\ 
		& \push{\cX}  \qw  & \ghost{\Phi} & \push{\cY} \qw & \gate{\alpha^{-1}} & \push{\cX} \qw & \qw } = 
	\myQ{1}{0.7}{& \push{\bbD} \ww& \Ngate{M}{\ww} &  \push{\bbE} \ww & \ww\\
		& \push{\cX}  \qw  & \gate{\id_\cX} & \push{\cX} \qw & \qw}
	\end{align}
	
	for some $M$. Composing with $\alpha$ on both sides gives $\twoext{\cX}{\bbD}{L}{\cY}{\bbE}= 	\myQ{0.5}{0.5}{&\push{\bbD} \ww & \Ngate{M}{\ww} &  \push{\bbE} \ww & \ww\\
		& \push{\cX}  \qw  & \gate{\alpha} & \push{\cY} \qw & \qw}$, and consequently $\alpha$ is dilationally pure.

\end{proof}

We immediately recover as a special case the \emph{No Broadcasting Theorem} (\cite{Barn96}):

\begin{Cor} (No Broadcasting.) \label{cor:NoBroadCast}\\
	If $\Theory$ is a purifiable theory, and if $\scalemyQ{.8}{0.7}{0.5}{& & \Nmultigate{1}{T} & \push{\cX} \qw & \qw \\ & \push{\cX} \qw & \ghost{T} & \push{\cX} \qw & \qw}$ is a channel for which one marginal is $\id_\cX$, then the other marginal must be of the form $\scalemyQ{.8}{0.7}{0.5}{& \push{\cX} \qw& \gate{\tr} & \Ngate{s} & \push{\cX} \qw & \qw}$ for some state $s$ on $\cX$.
	\end{Cor}

\begin{Remark}
In a purifiable thin theory $(M, \star, 1, \succeq)$ (i.e. one for which $\star$ satisfies the cancellation law), the `No Broadcasting' theorem simply says that if $x \succeq x \star x$, then $1  \succeq x$. 
\end{Remark}

We also recover the following non-trivial classification of isomorphisms in $\QIT$:

\begin{Cor} (Isomorphisms in $\QIT$.) \label{cor:IsosinQIT}\\
	A transformation in $\QIT$ is an isomorphism if and only if it is a unitary conjugation.
\end{Cor}

\begin{proof}
	It is clear that all unitary conjugations are isomorphisms. Conversely, since $\QIT$ is purifiable, any isomorphism in $\QIT$ is purifiable by \cref{prop:IsoSelfUniversal}, hence an isometric conjugation by \cref{prop:PureEqualsIso}. By a dimensional argument, an isometric conjugation has an inverse only if it is unitary. 
\end{proof}

Finally, we can augment the result to yield a surprising recharacterisation of purifiability for universal theories:

\begin{Prop} (Recharacterisation of Purifiable Theories.) \label{prop:PureRechar} \\
	A \myuline{universal} theory $\Theory$ is purifiable if and only if all identities $\id_\cX$ are dilationally pure. Moreover, in this case every universal dilation of any channel is dilationally pure. 
\end{Prop}

\begin{proof}
		The `only if'-direction follows from \cref{prop:IsoSelfUniversal}. As for the `if'-direction, assume that all identities in $\Theory$ are dilationally pure. Let $\channel{\bbX}{T}{\bbY}$ be a channel and let $\scalemyQ{.8}{0.7}{0.5}{& \push{\bbX}  \qw & \multigate{1}{U} & \push{\bbY} \qw & \qw \\ 
			& & \Nghost{U} & \push{\bbE_0} \ww  & \ww  }$ be a universal dilation of $T$.  We show that $U$ is dilationally pure. To this end, let $\scalemyQ{.8}{0.7}{0.5}{& \push{\bbX}  \qw & \multigate{2}{L} & \push{\bbY} \qw & \qw \\ 
			&& \Nghost{L} & \push{\bbE_0} \qw  & \qw   \\ 
		& \push{\bbD} \ww  &  \Nghost{L}{\ww}& \push{\bbE} \ww  & \ww  }$ be any dilation of $U$ . Since $L$ is a dilation \myuline{of $T$} (with hidden interfaces $\bbD$ and $\bbE_0 \cup \bbE$), we must have

	\begin{align} \label{eq:LUG}
	\scalemyQ{1}{0.7}{0.5}{& \push{\bbX}  \qw & \multigate{2}{L} & \push{\bbY} \qw & \qw \\ 
		&& \Nghost{L} & \push{\bbE_0} \qw  & \qw   \\ 
		& \push{\bbD} \ww  &  \Nghost{L}{\ww}& \push{\bbE} \ww  & \ww  } \quad = \quad 
\scalemyQ{1}{0.7}{0.5}{ & \push{\bbX}  \qw & \multigate{1}{U} & \qw & \push{\bbY} \qw & \qw &  \qw \\ & & \Nghost{U} &  \push{\bbE_0} \ww & \Nmultigate{1}{G}{\ww} & \push{\bbE_0} \qw & \qw\\
		& 	&  \push{\bbD} \ww &\ww & \Nghost{G}{\ww} &  \push{\bbE} \ww & \ww}
	\end{align}
	
	  for some channel $G$. As $L$ dilates $U$, however,
	  
	  	\begin{align}
	  \scalemyQ{1}{0.7}{0.5}{ & \push{\bbX}  \qw & \multigate{1}{U} & \qw & \push{\bbY} \qw & \qw &  \qw \\ & & \Nghost{U} &  \push{\bbE_0} \ww & \Nmultigate{1}{G}{\ww} & \push{\bbE_0} \qw & \qw\\
		& 	&  \push{\bbD} \ww &\ww & \Nghost{G}{\ww} &  \Ngate{\tr}{\ww} } 
	\quad = \quad  
	\scalemyQ{1}{0.7}{0.5}{& \push{\bbX}  \qw & \multigate{2}{L} & \push{\bbY} \qw & \qw \\ 
	  	&& \Nghost{L} & \push{\bbE_0} \qw  & \qw   \\ 
	  	& \push{\bbD} \ww  &  \Nghost{L}{\ww}& \Ngate{\tr}{\ww}    } 
  	\quad = \quad 
    \scalemyQ{1}{0.7}{0.5}{ & \push{\bbX}  \qw & \multigate{1}{U} & \qw & \push{\bbY} \qw & \qw &  \qw \\ & & \Nghost{U} &\push{\bbE_0} \ww  & \Ngate{\id_{\bbE_0}}{\ww} & \push{\bbE_0} \qw & \qw\\
  	& 	&  \push{\bbD} \ww &\ww & \Ngate{\tr}{\ww} }, 
	  \end{align}
	  
and comparing the first and last expression of this identity, the uniqueness clause in the definition of universality implies that $\scalemyQ{.8}{0.7}{0.5}{ &   \push{\bbE_0} \ww & \Nmultigate{1}{G}{\ww} & \push{\bbE_0} \qw & \qw\\
 	&  \push{\bbD} \ww  & \Nghost{G}{\ww} &   \Ngate{\tr}{\ww}} = \scalemyQ{.8}{0.7}{0.5}{&\push{\bbE_0} \ww  & \Ngate{\id_{\bbE_0}}{\ww} & \push{\bbE_0} \qw & \qw\\
 & 	&  \push{\bbD} \ww &\Ngate{\tr}{\ww} }$, i.e. $G$ is a dilation of $\id_{\bbE_0}$. But by assumption $\id_{\bbE_0}$ is dilationally pure, so must $G$ must be $\scalemyQ{.8}{0.7}{0.5}{&\push{\bbE_0} \ww  & \gate{\id_{\bbE_0}} & \push{\bbE_0} \qw & \qw\\
 &   \push{\bbD} \ww &\Ngate{M}{\ww} & \push{\bbE} \ww & \ww }$ for some $M$, and inserting this into \cref{eq:LUG} we see that $L$ is a trivial dilation of $U$. Hence, $U$ is dilationally pure as asserted.

\end{proof}

We will often use this result to ensure in a purifiable theory the existence of \myuline{one-sided} pure dilations (as every universal dilation is one-sided).

\subsection{Reversibles and Purifiability}
\label{subsec:RevPure}

By \cref{prop:IsoSelfUniversal}, every isomorphism in a purifiable theory is dilationally pure. The converse is not necessarily the case -- for example, a non-unitary isometric channel in $\QIT$ is not an isomorphism. It is, however, \myuline{reversible}, and as we shall now see, a few dilational axioms will guarantee in general that dilationally pure channels are reversible. In fact, we can use the dilationally pure channels in a purifiable theory to precisely understand the class of reversible channels:

\begin{Thm} (Structure of Reversible Channels.) \label{thm:StructureOfReversibles}\\
 	Let $\Theory$ be a universal, temporally localisable and purifiable theory. Then, a channel $\channel{\cX}{T}{\cY}$ in $\Theory$ is reversible if and only if it is of the form 
 	
 	\begin{align} \label{eq:RevChan}
 	\scalemyQ{1}{0.7}{0.7}{& \qw &\push{\cX} \qw & \multigate{1}{P} & \push{\cY} \qw & \qw \\ & \Ngate{r}  & \push{\cZ} \qw &  \ghost{P} }
 	\end{align}
 	
 	 for some state $r$ and some dilationally pure $P$.  
	\end{Thm}

\begin{proof}
Temporal localisability is only used to the effect that every channel has a reversible one-sided dilation (\cref{prop:Reversibledilations}).
	
If $P$ is dilationally pure, then any reversible dilation is trivial, so $P$ must already be reversible itself. Hence, it is clear that channels of the form \eqref{eq:RevChan} are reversible. The converse implication is a more intricate gymnastic exercise:

Assume that  $\channel{\cX}{T}{\cY}$ is reversible, and let $\channel{\cY}{T^-}{\cX}$ be a channel with 

\begin{align}\scalemyQ{1}{0.7}{0.5}{& \push{\cX} \qw & \gate{T} & \push{\cY} \qw & \gate{T^{-}} & \push{\cX} \qw & \qw} = \scalemyQ{1}{0.7}{0.5}{& \push{\cX} \qw & \gate{\id} & \push{\cX} \qw & \qw} \quad.
\end{align}

 Let $\oneext{\cX}{\breve{T}}{\cY}{\bbE}$ be a \myuline{universal} dilation of $T$; by \cref{prop:PureRechar} it is dilationally pure. Let moreover $\oneext{\cY}{R_{T^-}}{\cX}{\bbE^-}$ be a \myuline{reversible} dilation of $T^-$. The channel

 \begin{align} 
 \scalemyQ{1}{0.7}{0.5}{ & \push{\cX}  \qw & \multigate{2}{\breve{T}} &  \push{\cY} \qw & \multigate{1}{R_{T^-}} &\push{\cX} \qw & \qw\\ 
 	&  & \Nghost{L} & & \Nghost{R_{T^-}} & \push{\bbE^-}  \ww & \ww  \\
 	&  & \Nghost{\breve{T}}&\ww & \push{\bbE} \ww  & \ww}
 \end{align}
 
 is a dilation of $\channel{\cX}{\id}{\cX}$, and since $\Theory$ is purifiable, $\id_\cX$ is dilationally pure by \cref{prop:IsoSelfUniversal}, so 

\begin{align}
 \scalemyQ{1}{0.7}{0.5}{ & \push{\cX}  \qw & \multigate{2}{\breve{T}} &  \push{\cY} \qw & \multigate{1}{R_{T^-}} &\push{\cX} \qw & \qw\\ 
	&  & \Nghost{L} & & \Nghost{R_{T^-}} & \push{\bbE^-}  \ww & \ww  \\
	&  & \Nghost{\breve{T}}&\ww & \push{\bbE} \ww  & \ww} =  \scalemyQ{1}{0.7}{0.5}{& \push{\cX} \qw & \gate{\id} & \push{\cX} \qw & \qw \\ & & \Nmultigate{1}{s} & \push{\bbE^-} \ww & \ww \\ & & \Nghost{s} & \push{\bbE} \ww & \ww} 
\end{align}

for some state $s$. %
Now, pick a left-inverse $\tilde{P}$ of the reversible $R_{T^-}$, and observe that by composing both channels with $\tilde{P}$ we obtain the identity

\begin{align}
 \scalemyQ{.8}{0.7}{0.5}{ & \push{\cX}  \qw & \multigate{2}{\breve{T}} &  \push{\cY} \qw & \qw\\ 
	&  & \Nghost{L}  \\
	&  & \Nghost{\breve{T}}& \push{\bbE} \ww  & \ww} = \scalemyQ{1}{0.7}{0.5}{& \push{\cX} \qw  & \multigate{1}{\tilde{P}} & \push{\cY} \qw  & \qw\\  & \Nmultigate{1}{s} & \Nghost{\tilde{P}}{\ww}  \\  & \Nghost{s} & \push{\bbE} \ww & \ww} \quad.
\end{align}

If we knew that $\tilde{P}$ were dilationally pure, we could simply trash the system $\bbE$ and conclude the form \eqref{eq:RevChan}, but this conclusion is not within reach. The trick is to express $\scalemyQ{.8}{0.7}{0.5}{& \Nmultigate{1}{s}  & \ww  & \ww \\ & \Nghost{s}  & \push{\bbE} \ww & \ww }$ as $\scalemyQ{.8}{0.7}{0.5}{& \Nmultigate{1}{u}   & \Ngate{G}{\ww} & \push{\bbE^-} \ww & \ww    \\ & \Nghost{u}  & \push{\bbE} \ww & \ww  & \ww } $, where $u$ is a universal dilation of the marginal $\scalemyQ{.8}{0.7}{0.5}{& \Nmultigate{1}{s}  & \Ngate{\tr}{\ww}    \\ & \Nghost{s}  & \push{\bbE} \ww & \ww }$, thus obtaining

\begin{align} \label{eq:SuT}
\scalemyQ{1}{0.7}{0.5}{ & \push{\cX}  \qw & \multigate{2}{\breve{T}} &  \push{\cY} \qw & \qw\\ 
	&  & \Nghost{L}  \\
	&  & \Nghost{\breve{T}}& \push{\bbE} \ww  & \ww} = \scalemyQ{1}{0.7}{0.5}{& \push{\cX} \qw  & \multigate{1}{P} & \push{\cY} \qw  & \qw\\  & \Nmultigate{1}{u} & \Nghost{P}{\ww}  \\  & \Nghost{u} & \push{\bbE} \ww & \ww} \quad,
\end{align}

with $P$ the composition of $\tilde{P}$ with $G$. 
We can now argue that \myuline{$P$} is dilationally pure: Since $P$ has a complete (namely universal) one-sided dilation, it suffices to  argue that any \myuline{one-sided} dilation of $P$ is trivial. However, any one-sided dilation $ \scalemyQ{.8}{0.7}{0.5}{  & & \Nmultigate{2}{L} & \push{\bbG} \ww & \ww \\& \push{\cX} \qw  & \ghost{L} & \push{\cY} \qw  & \qw\\ & & \Nghost{L}{\ww} }$ of $P$ (with hidden interface $\bbG$) gives rise, by virtue of \cref{eq:SuT}, to a dilation of $\breve{T}$, and since the latter is dilationally pure, we must have    

\begin{align} 
 \scalemyQ{1}{0.7}{0.5}{  & & \Nmultigate{2}{L} & \push{\bbG} \ww & \ww \\& \push{\cX} \qw  & \ghost{L} & \push{\cY} \qw  & \qw\\  & \Nmultigate{1}{u} & \Nghost{L}{\ww}  \\  & \Nghost{u} & \push{\bbE} \ww & \ww} 
 \quad = \quad
 \scalemyQ{1}{0.7}{0.5}{& & \Ngate{t} & \push{\bbG} \ww & \ww \\& \push{\cX} \qw  & \multigate{1}{P} & \push{\cY} \qw  & \qw\\  & \Nmultigate{1}{u} & \Nghost{P}{\ww} \\  & \Nghost{u} & \push{\bbE} \ww & \ww} \quad
\end{align}

for some state $t$. By the universality property of $u$, however, this implies the identity $ \scalemyQ{.8}{0.7}{0.5}{  & & \Nmultigate{2}{L} & \push{\bbG} \ww & \ww \\& \push{\cX} \qw  & \ghost{L} & \push{\cY} \qw  & \qw\\ & & \Nghost{L}{\ww} } = \scalemyQ{.8}{0.7}{0.5}{  & & \Ngate{t} & \push{\bbG} \ww & \ww \\& \push{\cX} \qw  & \multigate{1}{P} & \push{\cY} \qw  & \qw\\ & & \Nghost{P}{\ww} } $, in other words, $P$ is dilationally pure, as desired. Trashing the system $\bbE$ in \cref{eq:SuT} finally yields the conclusion of the theorem.

	\end{proof}

The following follows immediately: 
\begin{Cor} (Reversibles in $\QIT$.) \label{cor:RevinQIT} \\A quantum channel $\channel{\cX}{\Lambda}{\cY}$ is reversible if and only if it is of the form $\scalemyQ{.8}{0.7}{0.7}{& \qw &\push{\cX} \qw & \multigate{1}{\Sigma} & \push{\cY} \qw & \qw \\ & \Ngate{\varrho}  & \push{\cZ} \qw &  \ghost{\Sigma} }$ for some state $\varrho$ and some isometric channel $\Sigma$. 
	\end{Cor}

\begin{proof}
Immediate from \cref{thm:StructureOfReversibles}, since by \cref{prop:IsoSelfUniversal} the dilationally pure channels in $\QIT$ are precisely the isometric channels. 
	\end{proof}

The proof of \cref{thm:StructureOfReversibles} required surprising assumptions additional to purifiability of the theory, namely the existence of reversible dilations and of universal dilations. The following is left unanswered:

\begin{OP}
Can the assumptions of \cref{thm:StructureOfReversibles} be weakened without compromising the conclusion?
	\end{OP}

\subsection{Complementarity}
\label{subsec:Complementarity}

If a channel knows everything about itself, does it necessarily know everything about any channel that it dilates? Moreover formally, if $\channel{\bbX}{T}{\bbY}$ is a channel and $\oneext{\bbX}{P}{\bbY}{\bbE}$ is dilationally pure, must then $P$ be a complete dilation of $T$?\\ 

This is not obvious. If true, though, it has a remarkable consequence for purifiable theories: In $\QIT$, Stinespring dilations creates a \emph{complementarity} between certain pairs of quantum channels (\cite{Devetak05}), and this is ultimately because it is a complete dilation of \myuline{any} channel that it dilates. Thus, an affirmative answer to the above question will foster a similar notion of complementarity in general. This will specialise in particular to a complementarity between reversible channels and \emph{completely forgetful} channels (that is, channels of the form $\scalemyQ{.8}{0.7}{0.5}{& \push{\bbX} \qw& \gate{\tr} & \Ngate{s} & \push{\bbY} \qw & \qw}$ for some state $s$) as known from the theory $\QIT$, and this immediately implies a rather long list of impossibility (`no go') theorems.\\

Let us start by answering the introductory question. It is a special case (corresponding to $K=L$) of the following more general question: If $L$ is a (one-sided) dilation of $T$ and $K$ is a complete dilation of $L$, is then $K$ a complete dilation of $T$?  \newpage

This is indeed true under suitable conditions:

\begin{Lem} (Completeness is Hereditary.) \label{lem:Hereditary} \\
Suppose that $\Theory$ is complete and temporally localisable. Let $\oneext{\bbX}{L}{\bbY}{\bbE}$ be a dilation of $\channel{\bbX}{T}{\bbY}$. If $\scalemyQ{.8}{0.7}{0.5}{& & \Nmultigate{2}{K} & \push{\bbG}\ww  & \ww \\ & & \Nghost{K} & \push{\bbE} \ww & \ww \\ & \push{\bbX} \qw& \ghost{K} & \push{\bbY} \qw & \qw}$ is a complete dilation of $L$, then $K$ is also a complete dilation of $T$. 

\end{Lem}

\begin{Remark}
The converse to this statement is true as well, but trivial: If $K$ is a complete dilation of $T$, then it is a complete dilation of $L$ simply for the reason that the dilations of $L$ form a sub-class of the dilations of $T$. What makes the above statement interesting is that, in general, $T$ has dilations which are not dilations of $L$.
	\end{Remark}

\begin{proof}
	By assumption, $\channel{\bbX}{T}{\bbY}$ has \myuline{some} complete dilation, say $\oneext{\bbX}{\hat{T}}{\bbY}{\hat{\bbE}}$, and it is enough to show that $K \der \hat{T}$. Since $\hat{T}$ is a complete dilation of $T$, there exists a channel $G$ such that 
	
	\begin{align}
	\myQ{0.7}{0.5}{& \push{\bbX}  \qw & \multigate{1}{\hat{T}} & \qw & \push{\bbY} \qw & \qw& \qw \\ 
		& & \Nghost{\hat{T}} & \push{\hat{\bbE}} \ww & \Ngate{G}{\ww} & \push{\bbE} \ww & \ww } =  	\myQ{0.7}{0.5}{& \push{\bbX}  \qw & \multigate{1}{L} & \push{\bbY} \qw & \qw \\ 
		& & \Nghost{L} &  \push{\bbE} \ww & \ww  } \quad .
	\end{align}

Now, by temporal localisability $G$ has a reversible dilation, say $R$, and since $	\scalemyQ{.8}{0.7}{0.5}{& \push{\bbX}  \qw & \multigate{1}{\hat{T}} & \qw & \push{\bbY} \qw & \qw& \qw \\ 
	& & \Nghost{\hat{T}} & \push{\hat{\bbE}} \ww & \Nmultigate{1}{R}{\ww} & \push{\bbE} \ww & \ww \\ & & & & \Nghost{R} & \push{\bbZ} \ww & \ww}$ is a dilation of $L$, we find a channel $F$ such that 

\begin{align}
	\myQ{0.7}{0.5}{& \push{\bbX}  \qw & \multigate{1}{\hat{T}} & \qw & \push{\bbY} \qw & \qw& \qw \\ 
	& & \Nghost{\hat{T}} & \push{\hat{\bbE}} \ww & \Nmultigate{1}{R}{\ww} & \push{\bbE} \ww & \ww \\ & & & & \Nghost{R} & \push{\bbZ} \ww & \ww} = \myQ{0.7}{0.5}{& \push{\bbX} \qw& \multigate{2}{K} & \push{\bbY} \qw & \qw \\ & & \Nghost{K} & \push{\bbE} \ww & \ww \\ & & \Nghost{K} & \push{\bbG}\ww  & \Ngate{F}{\ww} & \push{\bbZ} \ww & \ww} \quad,
\end{align}

by completeness of $K$. But now we can apply a left-inverse to $R$ on both sides, and the desired relation $K \der \hat{T}$ is evident. 
\end{proof}

\begin{Prop} (Pure Dilations are Complete.) \label{prop:PureareComp}\\
Suppose that $\Theory$ is complete and temporally localisable. If $\oneext{\bbX}{P}{\bbY}{\bbE}$ is a pure dilation of $\channel{\bbX}{T}{\bbY}$, then $P$ is a complete dilation of $T$. In particular, if two channels $\channel{\bbX}{T}{\bbY}$ and $\scalemyQ{.8}{0.7}{0.5}{& \push{\bbX} \qw & \gate{\tilde{T}}& \push{\bbE} \ww & \ww}$ have a common pure dilation, then this dilation is a complete dilation of them both. 
\end{Prop}

\begin{proof}
	Take $K=L=P$ in \cref{lem:Hereditary}.
\end{proof}

The point of \cref{prop:PureareComp} is that if we now define two channels $\channel{\bbX}{T}{\bbY}$ and $\scalemyQ{.8}{0.7}{0.5}{& \push{\bbX} \qw & \gate{\tilde{T}}& \push{\bbE} \ww & \ww}$ to be \emph{complementary} if they admit a common pure dilation $\oneext{\bbX}{P}{\bbY}{\bbE}$, then complementarity is `well-formed' in a sense which is not obvious from the definition itself:

\begin{Thm} (Complementarity.) \label{thm:Complementarity}\\
If $\scalemyQ{.8}{0.7}{0.5}{& \push{\bbX} \qw & \gate{\tilde{T}}& \push{\bbE} \ww & \ww}$ and $\scalemyQ{.8}{0.7}{0.5}{& \push{\bbX} \qw & \gate{\tilde{T}'}& \push{\bbE} \ww & \ww}$ are both complementary to $\channel{\bbX}{T}{\bbY}$, then they are equivalent in the Blackwell order, i.e.  there exist channels $G$ and $G'$ such that 

\begin{align}
\scalemyQ{1}{0.7}{0.5}{& \push{\bbX} \qw & \gate{\tilde{T}'}& \push{\bbE'} \ww & \ww} = \scalemyQ{1}{0.7}{0.5}{& \push{\bbX} \qw & \gate{\tilde{T}}& \push{\bbE} \ww & \Ngate{G}{\ww} & \push{\bbE'} \ww & \ww}
 \quad \text{and} \quad 
\scalemyQ{1}{0.7}{0.5}{& \push{\bbX} \qw & \gate{\tilde{T}}& \push{\bbE} \ww & \ww} = \scalemyQ{1}{0.7}{0.5}{& \push{\bbX} \qw & \gate{\tilde{T}'}& \push{\bbE'} \ww & \Ngate{G'}{\ww} & \push{\bbE} \ww & \ww}.
\end{align}

Moreover, complementarity is a \emph{duality}, in the sense that if a channel $T'$ is complementary to a channel complementary to $T$,  then $T'$  is equivalent to $T$ in the Blackwell order.
	\end{Thm}

\begin{proof}
	The first statement follows from \cref{prop:PureareComp}; the two pure dilations giving rise to $\tilde{T}$ and $\tilde{T}'$ are both complete, hence equivalent in the dilational order, which precisely implies equivalence of  $\tilde{T}$ and $\tilde{T}'$ in the Blackwell order. The second statement is a consequence of the first, since complementarity is clearly a symmetric relation. 
	\end{proof}

The point of \cref{thm:Complementarity} is that there is a strain, or balance, between channels $\channel{\bbX}{T}{\bbY}$ and their complementary channels $\scalemyQ{.8}{0.7}{0.5}{& \push{\bbX} \qw & \gate{\tilde{T}}& \push{\bbE} \ww & \ww}$.

In the case where the theory is furthermore universal, this strain implies a generalisation of the information-disturbance duality.

 Let us call a channel \emph{completely forgetful} if it is of the form  $\scalemyQ{.8}{0.7}{0.5}{& \push{\bbX} \qw& \gate{\tr} & \Ngate{s} & \push{\bbY} \qw & \qw}$ for some state $s$. It is clear that the property of being completely forgetful is invariant under equivalence in the Blackwell order, and so is the property of being reversible. We have the following:

\begin{Thm} (Duality between Reversible and Completely Forgetful Channels.) \label{thm:InfoDist}\\
Let $\Theory$ be a universal, localisable and purifiable theory. Then, a channel $\channel{\bbX}{T}{\bbY}$ is reversible if and only if the complementary channels $\scalemyQ{.8}{0.7}{0.5}{& \push{\bbX} \qw & \gate{\tilde{T}}& \push{\bbE} \ww & \ww}$ are completely forgetful, and vice versa.
\end{Thm}

\begin{proof}
One might prove this by restarting the argument from the proof of \cref{thm:StructureOfReversibles}, and we shall do so when proving its approximate generalisation (\cref{thm:AppInfoDist}) in \cref{chap:Metric}; however, for the sake of variation we instead use here an argument based on the statement of \cref{thm:StructureOfReversibles} itself. 

 If the channel $\channel{\bbX}{T}{\bbY}$ is reversible, then by \cref{thm:StructureOfReversibles} it is of the form $ 	\scalemyQ{.8}{0.7}{0.7}{& \qw &\push{\bbX} \qw & \multigate{1}{P} & \push{\bbY} \qw & \qw \\ & \Ngate{r}  & \push{\cZ} \qw &  \ghost{P} }$ for some pure channel $P$ and some state $r$. Choosing a pure dilation $v$ of $r$, the channel

\begin{align}
 \scalemyQ{1}{0.7}{0.5}{& \push{\bbX} \qw  & \multigate{1}{P} & \push{\bbY} \qw  & \qw\\  & \Nmultigate{1}{v} & \ghost{P}  \\  & \Nghost{v} & \push{\bbE} \ww & \ww}
\end{align} 

is then by localisability a pure dilation of $T$. Evidently, the corresponding complementary channel is then completely forgetful. 

Conversely, if $\channel{\bbX}{T}{\bbY}$ is completely forgetful, i.e. of the form $\scalemyQ{.8}{0.7}{0.5}{& \push{\bbX} \qw& \gate{\tr} & \Ngate{s} & \push{\bbY} \qw & \qw}$, then, letting $v$ be a pure dilation of $s$, $T$ has a pure dilation of the form 

\begin{align}
\scalemyQ{1}{0.7}{0.5}{& \push{\bbX} \qw  & \gate{\id} &  \push{\bbX} \ww  & \ww\\  & \Nmultigate{1}{v} & \push{\bbE} \ww & \ww  \\  & \Nghost{v} & \push{\bbY} \qw & \qw} \quad,
\end{align} 

for which the corresponding complementary channel is obviously reversible. 
	\end{proof}

Let me conclude this section by observing that the complementarity result framed by \cref{thm:InfoDist} implies not only the No Broadcasting Theorem (which we have already seen based on weaker assumptions), and hence the No Cloning Theorem (\cite{Woot82}) and the No Deletion Theorem (\cite{Pati00}), but also the more recent No Hiding (\cite{Braun07}) and No Masking (\cite{Modi18}) Theorems, which assert precisely the non-existence of pure channels both of whose marginals are completely forgetful.

\section{Summary and Outlook}
\label{sec:SummaryDilations}

In this chapter, we have seen a general theory of dilations. At the heart of this theory is the \emph{dilational ordering} (\cref{def:DilOrd}), which in particular facilitates the notions of \emph{complete} dilations (\cref{def:Complete}) and of \emph{spatial} and \emph{temporal localisability} (\cref{def:SpatLoc} and \cref{def:TempLoc}). 

These three \emph{dilational principles} have information-theoretic interpretations, and they formally have useful consequences (e.g. they imply the DiVincenzo Property and the existence of reversible dilations). They seem to hold in all theories that are remotely physical, e.g. in the information theories $\CIT$ and $\QIT$ (as well as $\QIT^\infty$) and in cartesian theories, but they are not derivable from the axioms which define a general theory, since they are violated for example in the thin theory $(\N, \cdot, 1, \geq)$. 

A strengthening of completeness in the guise of \emph{universality} (\cref{def:Univ}) holds in all of the above  complete theories, and universal dilations in principle allows us to fully characterise the dilational ordering among one-sided dilations (\cref{prop:Blackwell}). 

The theory $\QIT$ distinguishes itself from the theory $\CIT$ by being \emph{purifiable} (\cref{def:SelfUniv}), a notion which can be seen as generalising the `Purification Postulate' of Refs. \cite{Chir10,Chir11}. The purifiability principle implies, especially in conjunction with the other dilational principles, significant `quantum' features of a theory, most prominently the notion of complementarity and duality between reversible and completely forgetful channels (\cref{thm:InfoDist}).\\

One conclusion of this chapter rises above all, namely: \emph{It is possible to prove significant consequences from principles entirely about the structure of dilations.}

An obvious question for future work is whether even more properties can be  captured by principles about dilations. Another question is whether there are naturally occurring examples of `information theories' in which some of the dilational principles presented here fail. 

Thirdly, it is reasonable to investigate whether additional impossibility theorems known from quantum information theory can be derived using the principle of purifiability. In particular, some such impossibility results regard the non-existence of certain \emph{protocols}, for example the \emph{No Bit Commitment Theorem} (\cite{Mayers97, Lo97com}) and the \emph{insecurity of one-sided} (\cite{Lo97insec}) and \emph{two-sided} (\cite{Buhr12}) \emph{two-party computation}. Proof sketches lead me to believe that such results should be possible, but a proper coverage of impossibility theorems for protocols requires introducing an array of further concepts, and they must apparently be treated with the care of a soldier traversing a minefield (\cite{Bitcomreex}).

Finally, for the results of \cref{sec:Selfuniv} to be of interest beyond the virtue of simplicity, the following question is relevant: What are the theories besides $\QIT$ and $\QIT^\infty$ which are localisable, universal and purifiable?

\chapter{Metric Theories}
\label{chap:Metric}

{\centering
	\subsection*{§1. Introduction and Outline -- Comparison to Existing Literature.}}

The framework of theories as presented in \cref{chap:Theories} and \cref{chap:Dilations} is purely algebraic. All of its notions, primary and derived, are defined entirely in terms of the serial and parallel compositions in a theory $\Theory$. In this chapter, I present some initial thoughts on how to create a \emph{metric} version of this framework. That idea is not simply one of mathematical curiosity, but is physically natural too -- indeed, our two most important examples of theories, $\CIT$ and $\QIT$, are both endowed with natural metrics, namely the \emph{variational} (or \emph{statistical}) \emph{distance} (\cref{ex:d1CIT}) and the \emph{diamond-distance} (\cref{ex:dDQIT}), respectively.  

Whereas metric aspects of general theories have been considered before (e.g. in Ref. \cite{Chir10}), the defined metrics seem to be always grounded in the concept of \emph{optimal distinguishing probability}. As we have made no assumptions about probabilistic structure in the definition of a theory, an appropriate definition of \myuline{metric} theory also should not rely on such structure. Rather, we abstract from such `distinguishing  metrics' just two properties which can be phrased in terms of the serial and parallel composition in the theory. The main mantra of the ideas presented in this chapter, however, is the insistence that the metric structure comply to the architecture of \myuline{dilations}. This not only allows us to smoothly transfer results of \cref{chap:Dilations} to an approximate setting, but also contextualises sporadic definitions and observations in the literature, particularly of Refs. \cite{Toma10,Toma12} and Refs. \cite{KSW08math,KSW08phys}.  \\

\textbf{Topological Structure.} For the sake of completeness, we start in \cref{sec:Top}  by defining what one would call a \emph{topological theory}. It is more or less obvious how to do this: Simply posit that for any $\cX , \cY\in \Sys{\Theory}$, the set\footnote{We assume throughout this chapter that the category $\Theory$ is \emph{locally small}, i.e. that for any $\cX, \cY \in \Sys{\Theory}$ the class of transformations $\Trans{\Theory}{\cX}{\cY}$ is not a proper class, but merely a set. I know of no physically relevant example where this is not the case.} of transformations from $\cX$ to $\cY$, $\Trans{\Theory}{\cX}{\cY}$, is equipped with a topology $\scrT_{\cX, \cY}$, and that the various topologies cohere such that the serial and parallel compositions are continuous maps. For example, when $\Theory = \QIT$ there is a natural topology on the set of quantum channels $\Trans{\QIT}{\cH}{\cK}$, since it is a convex subset of a finite-dimensional vector space. In general, note that a topological structure on a theory entails in particular a topology on the set of states $\St{\cX} = \Trans{\Theory}{\triv}{\cX}$ for any system $\cX$. 

For reasons to come, it is actually more appropriate to define topologies and metrics on the sets of \myuline{channels}, $\Chan{\Theory}{\bbX}{\bbY}$, between interfaces $\bbX$ and $\bbY$, and this is what we shall do (\cref{def:TopTheory}); in principle, one could therefore have different topologies or metrics on $\Chan{\Theory}{\bbX}{\bbY}$  and $\Chan{\Theory}{\bbX'}{\bbY'}$, even if the total systems corresponding to $\bbX$ and $\bbX'$, respectively $\bbY$ and $\bbY'$, are the same,\footnote{And this might indeed be relevant if one were to generalise the theory to the causal setting of \cref{chap:Causal}.} but in the examples given in this chapter this will never be the case.\\

\textbf{Metric Structure and Compositional Compatibility.} In defining a \emph{metric theory}, we must clearly endow the sets of channels $\Chan{\Theory}{\bbX}{\bbY}$ with metrics, $d_{\bbX, \bbY}$, rather than topologies, $\scrT_{\bbX, \bbY}$. This is the topic of \cref{sec:Met}. However, whereas it is obvious in the case of topologies that the correct coherence condition across various channel-topologies is to require continuity of serial and parallel composition, the `correct' coherence conditions for metrics are not handed unambiguously to us -- for example, it is a priori unclear whether to impose on a metric $d$ in the case of serial composition that $d(S \after T, \tilde{S} \after \tilde{T}) \leq d(\tilde{S}, S)+ d(\tilde{T}, T)$ or that  $d(S \after T, \tilde{S} \after \tilde{T}) \leq \max\{d(\tilde{S}, S), d(\tilde{T}, T)\}$, or something else. 

The above notion of a topological theory can be seen as a special case of categorical \emph{enrichment} (\cite{MacLane}, \cite{Enriched}), where one basically replaces the sets of morphisms by objects from a category (e.g. topological spaces of morphisms). One might therefore suspect that the `mathematically correct' notion of a \myuline{metric} theory would similarly arise by enriching over a suitable category of metric spaces. F. W. Lawvere has argued that the most `natural' category of metric spaces is that of \emph{Lawvere metric spaces} and \emph{short maps} between them (\cite{Law73}); following this idea leads to the requirement  $d(S \after T, \tilde{S} \after \tilde{T}) \leq \max\{d(\tilde{S}, S), d(\tilde{T}, T)\}$. It turns out, however, that this cannot by `\myuline{physically} correct', since it implies for example that $d(T^n, \tilde{T}^n) \leq d(T, \tilde{T})$ for all $n \in \N$, for any channels  $T, \tilde{T}: \bbX \to \bbX$ (here, $T^n$ is the $n$-fold iterated serial composition), and it is easy to give examples in $\CIT$ with the total variation distance (or in $\QIT$ with the diamond-distance) for which $d(T, \tilde{T})<1$ whereas $d(T^n, \tilde{T}^n) \to 1$ for $n \to \infty$.\footnote{For instance, take $T: \{0,1\} \to \{0,1\}$ to be the identity and $\tilde{T}: \{0,1\} \to \{0,1\}$ given by $\tilde{T}(\delta_0) = \delta_0$, $\tilde{T}(\delta_1) = \frac{1}{2} \delta_0 + \frac{1}{2} \delta_1$.}

Rather, to state the correct conditions, we take inspiration from the metrics defined by optimal distinguishing probability, which suggest the requirement  $d(S \after T, \tilde{S} \after \tilde{T}) \leq d(\tilde{S}, S)+ d(\tilde{T}, T)$ (and a similar condition for parallel composition). Metrics on $\Theory$ which adhere to the two coherence conditions for serial and parallel composition will be called \emph{monotone} (\cref{def:MetTheory}). Both the variational metric $d_1$ on $\CIT$ and the diamond metric $d_\diamond$ on $\QIT$ are examples of monotone metrics. (On the other hand, the variational (trace) distance $d_1$ on $\QIT$ is not monotone, since, as is well-known, it fails to be invariant under parallel composition.) There are also more mathematical examples, however, which could not have been cast in terms of distinguishing probability, such as metrics deriving from operator norms in the cartesian theory $\Vect{\R}$ or $\Vect{\C}$ (\cref{ex:VectMet}).  \\

\textbf{Dilationality.} Monotonicity of a metric ensures compatibility with the serial and parallel composition in the theory, but it generally fails to guarantee an affirmative answer to the following problem:\footnote{The fact that we want to talk about dilations is why we want our metrics to be defined on sets of channels rather than sets of transformations.} Suppose that $T, \tilde{T}: \bbX \to \bbY$ are channels in a theory $\Theory$ and that $d$ is a monotone metric on $\Theory$; if $d(T,\tilde{T})$ is small, we tend to think of $T$ as being `close to' $\tilde{T}$. Does it follow that \myuline{dilations} of $T$ are close to dilations of $\tilde{T}$? More precisely, given a dilation $L$ of $T$, can we find a dilation $\tilde{L}$ of $\tilde{T}$ such that $d(L, \tilde{L}) \leq d(T, \tilde{T})$? (The inequality `$\geq$' will be automatic from monotonicity.)

In \cref{sec:Dilational}, we will name this property \emph{dilationality} (\cref{def:Dilationality}), and, importantly, it does constitute an additional requirement: Whereas monotone metrics in cartesian theories and in $\CIT$ are always dilational cf. \cref{ex:CITdilational}, ultimately due to the particular nature of complete dilations in these theories, it is known that e.g. the diamond-distance $d_\diamond$ on $\QIT$ is not dilational. Thus arises the question of how we might \myuline{construct} dilational, monotone metrics. 

The childishly naive guess of defining $\hat{d}(T, \tilde{T}) = \sup_{L} \inf_{\tilde{L}} d(L, \tilde{L})$ with $\tilde{L}$ ranging over dilations of $\tilde{T}$ and $L$ ranging over (compatible, i.e. with same hidden interfaces) dilations of $T$ does not necessarily produce a dilational metric $\hat{d}$. Indeed, $\hat{d}$ may not only fail to be symmetric (this issue could be ignored), but, more importantly, may even fail to be dilational; sure, it was defined so as to mimic dilationality of $d$, but the new standard that is must be held to -- namely dilationality \myuline{of $\hat{d}$} -- is distinct. To better come to terms with this issue, we derive a result (\cref{prop:GenUhl}) to the effect that dilationality of a (symmetric) metric $d$ can in a sufficiently nice theory be reformulated as the requirement that $d(T, \tilde{T}) =\inf_{K, \tilde{K}} d(K, \tilde{K})$, where $K, \tilde{K}$ range over \myuline{complete} dilations of $T, \tilde{T}$. This property will be termed the  \emph{Generalised Uhlmann Property}, as it imitates an infimum over Stinespring dilations, but rather using complete dilations which are more general. 

\cref{prop:GenUhl} could seem to suggest that a definition like $\hat{d}(T, \tilde{T}) = \inf_{K \tilde{K}} d(K,\tilde{K})$, which is now certainly symmetric, could yield a dilational metric $\hat{d}$, but, once again, the new standard that it must meet is rather $\hat{d}(T, \tilde{T}) = \inf_{K \tilde{K}} \hat{d}(K,\tilde{K})$; it appears as if the potential to always dilate further is dragged along wherever we go and hindering dilationality. What we need is to \emph{short circuit} this phenomenon, and we have already seen in \cref{chap:Dilations} what such a short circuit looks like:  \\

\textbf{Purified Distances.} Indeed, the trick is to use \myuline{pure} dilations in lifting a metric $d$ to a dilational metric $\breve{d}$, and this is possible in a \myuline{purifiable} theory, such as $\QIT$, as we will see in \cref{sec:PureUhlmann}. If we do this, then the self-completeness of pure dilations along with the hereditary nature of completeness  (as mixed in \cref{prop:PureareComp}) form the key to achieving dilationality. In particular, we obtain in the case of the diamond metric $d_\diamond$ on $\QIT$ a metric $P_\diamond := \breve{d}_\diamond$ which I will call the \emph{purified diamond-distance} and which generalises the purified distance for states as introduced in Refs. \cite{Toma10,Toma12}.  

In \cref{sec:PureUhlmann} we also prove (\cref{thm:AppInfoDist}) that a dilational metric in a purifiable theory implies a tight approximate converse to \cref{thm:InfoDist}, which can be seen as an abstract statement of the \emph{information-disturbance trade-off} of Ref. \cite{KSW08phys}. This conceptually has nothing to do with purified distances, but seems to fit in well at that point when we have established the existence of dilational metrics in purifiable theories. \\

\textbf{The Curious Case of $\QIT$.} In \cref{sec:BuresP}, we investigate the newly proposed diamond-distance $P_\diamond$ more closely. 

We start in \cref{subsec:PversD} by translating to $P_\diamond$ the continuity result from Refs. \cite{KSW08math,KSW08phys} which is there stated in terms of the so-called \emph{Bures distance} $\beta$ (as inspired by Ref. \cite{Bures69}), whose definition is based on a similar idea as $P_\diamond$. Specifically, we infer that the purified diamond-distance satisfies $d_\diamond \leq P_\diamond \leq \sqrt{2 \, d_\diamond}$, an inequality which was known in the case of states (\cite{Toma10,Toma12}) but which in the general case relies on the non-trivial result from Refs. \cite{KSW08math,KSW08phys} that $d_\diamond \leq \beta \leq \sqrt{2 \, d_\diamond}$.

So, what is the difference between $P_\diamond$ and $\beta$? The two quantities share their most important features (cf. \cref{rem:BPquali}), but the main argument in favour of $P_\diamond$ is that its definition is in terms of the diamond-distance (which is operational as optimal distinguishing probability) and thus quite theory-independent, whereas $\beta$ makes explicit and non-operational reference to the operator formalism of quantum information theory. In \cref{subsec:PversB} we tackle the question of determining a more quantitative relation between the two. We start by deriving two formulas which relate $P_\diamond$ and $\beta$ to two fidelity-like quantities (\cref{prop:fidelity}). Using a minimax-theorem, we can then link those two fidelity-like quantities (\cref{lem:Fakeit}), and from this ultimately derive a link between $P_\diamond$ and $\beta$ (\cref{thm:BP}). The quantitative difference between them is small for small values (in fact, $P_\diamond = \beta - \beta^3/8 + O(\beta^5)$ in the limit $\beta \to 0$), but for large values they can differ by a factor as large as $\sqrt{2}$.\\

{\centering
	\subsection*{§2. Contributions.}}

The original contributions of this chapter are the following:\\

\begin{itemize}
	\item Identifying a minimal framework for the discussion of topological (\cref{def:TopTheory}) and metric (\cref{def:MetTheory}) aspects of general theories in the sense of \cref{def:Theory}. 

	\item Articulating the property of \emph{dilationality} of a metric, and demonstrating its equivalence to a \emph{`Generalised Uhlmann property'} (\cref{prop:GenUhl}).
	
	\item Presenting a recipe for the construction in purifiable theories of dilational metrics called \emph{purified} (\cref{thm:Purified}), and observing their useful behaviour with regards to complementarity (\cref{prop:AppComp}). 
	
	\item Proving an approximate counterpart (\cref{thm:AppInfoDist}) to the duality between reversible and completely forgetful channels, which can be seen as an abstract statement of the \emph{information-disturbance trade-off} (\cite{KSW08phys}).

\item 	Defining in the specific theory $\QIT$ the \emph{purified diamond-distance} $P_\diamond$, which directly generalises purified distance between states of Refs. \cite{Toma10,Toma12}, and relates to the \emph{Bures distance} of Refs. \cite{KSW08math,KSW08phys,Bures69} but distinguishes itself by a more stable definition.  %
\end{itemize}

\newpage

\section{Topological Structure}
\label{sec:Top}

Let us begin by considering what would be a sensible notion of a \emph{topological theory}. Recall that a \myuline{topology} an a set $S$ is a system of subsets of $S$ which is closed under arbitrary unions and finite intersections, and which contains the sets $\emptyset$ and $S$. 

Let $\Int{\Theory}$ denote the class of interfaces in $\Theory$ and let $\Chan{\Theory}{\bbX}{\bbY}$ denote the set\footnote{Recall the assumption that $\Theory$ is locally small.} of channels in $\Theory$ from the interface $\bbX$ to the interface $\bbY$.

\begin{Definition} (Topological Theory.) \label{def:TopTheory}\\
	A \emph{topological structure} on a theory $\Theory$ is a collection $\scrT = (\scrT_{\bbX,\bbY})_{\bbX, \bbY \in \Int{\Theory}}$, where, for each pair of interfaces $\bbX, \bbY$ in $\Theory$, $\scrT_{\bbX,\bbY}$ is a topology on the set $\Chan{\Theory}{\bbX}{\bbY}$, such that the following compatibility conditions hold: 
	
		\begin{enumerate}
		\item For all interfaces $\bbX, \bbY, \bbZ$ in $\Theory$, the serial composition
		
		\begin{align}
		(S,T) \mapsto S \after T
		\end{align}
		
		 is continuous from $\Chan{\Theory}{\bbY}{\bbZ} \times \Chan{\Theory}{\bbX}{\bbY}$ to $\Chan{\Theory}{\bbX}{\bbZ}$ when the former is equipped with the product topology of $\scrT_{\bbY,\bbZ}$ and $\scrT_{\bbX,\bbY}$, and the latter with the topology $\scrT_{\bbX,\bbZ}$. 
		
		\item For all parallelly composable interfaces $\bbX_1, \bbX_2$ and $\bbY_1, \bbY_2$ in $\Theory$, the parallel composition
		
		\begin{align}
		(T_1,T_2) \mapsto T_1 \og T_2 
		\end{align}
		
		is continuous from $\Chan{\Theory}{\bbX_1}{\bbY_1} \times \Chan{\Theory}{\bbX_2}{\bbY_2}$ to $\Chan{\Theory}{\bbX_1 \cup \bbX_2}{\bbY_1 \cup \bbY_2}$ when the former is equipped with the product topology of $\scrT_{\bbX_1,\bbY_1}$ and $\scrT_{\bbX_2,\bbY_2}$, and the latter with the topology $\scrT_{\bbX_1 \cup \bbX_2, \bbY_1 \cup \bbY_2}$.

	\end{enumerate}

	A \emph{topological theory} $(\Theory, \scrT)$ is a theory $\Theory$ together with a topological structure $\scrT$.\footnote{Strictly speaking, if we work in a foundational framework where proper classes are predicates with existence only in the metalanguage, then there is formally no such thing as the `pair' $(\Theory, \scrT)$. We ignore this point.} 
\end{Definition}

Plainly speaking, what we have done is to replace in the definition of a theory the \myuline{sets} of channels with \myuline{topological spaces} of channels, and to require continuity of the two modes of composition in the theory. 

As mentioned in the introduction, the topologies $\scrT_{\bbX, \bbY}$ will in all our cases really be topologies $\scrT_{\cX, \cY}$ which depend only on the total systems $\cX$ and $\cY$ associated to the interfaces $\bbX$ and $\bbY$.  

\begin{Example} ($\QIT$ as a Topological Theory.)\\
	The theory $\QIT$ has an obvious topological structure: $\Chan{\QIT}{\bbX}{\bbY} \cong \Trans{\QIT}{\cX}{\cY}$ is a set of linear maps $\Lambda : \End{\cX} \to \End{\cY}$ between finite-dimensional vector spaces, and as such is naturally topologised. The continuity of the two modes of composition is clear. \end{Example}

\begin{Example} ($\CIT$ as a Topological Theory.)\\
	The theory $\CIT$ also has a natural topological structure; it can be most succinctly described as the subspace topology that arises from regarding $\CIT$ as a sub-theory of $\QIT$. This topology can also be given a more concrete description in terms of the natural topology on the sets of probability distributions.
		\end{Example} 

\newpage

\begin{Example} (Discrete Topological Theory.)\\
		Any theory $\Theory$ can be endowed with the \emph{discrete topological structure}, for which the topology $\scrT_{\bbX, \bbY}$ on $\Chan{\Theory}{\bbX}{\bbY}$ is simply the discrete topology, i.e. the topology consisting of \myuline{all} subsets of $\Chan{\Theory}{\bbX}{\bbY}$. In this case, the two continuity requirements are trivially satisfied. 
	
	\end{Example}

\begin{Example} (Thin Topological Theories.)\\
	Let $\Theory$ be a thin theory. For any interfaces $\bbX, \bbY$ in $\Theory$, the set $\Chan{\Theory}{\bbX}{\bbY}$ is either empty or contains a single element, in which case it can be topologised in only one way, namely with the discrete topology. Thus, topological structure is utterly uninteresting for thin theories. 
	\end{Example}

\begin{Example}  (Topologised Topology.)\\
	If we consider a sub-theory of the cartesian theory $\Top^*$ whose systems are sufficiently nice topological spaces (e.g. Hausdorff and locally compact), and if we equip the sets of continuous maps from $X$ to $Y$ with the  \emph{compact-open topology}, then both modes of composition are continuous and so we have a topological theory. (A sub-base for this topology is given by the collection of sets $\{f: X \to Y \mid f(K) \subseteq U\}$ with $K \subseteq X$ compact and $U \subseteq Y$ open.)
		\end{Example}

As a warm-up to metric theories, it is worth observing that, by properties of continuous maps, in a topological theory the map $T \mapsto S_0 \after T$ is continuous for any \myuline{fixed} channel $S_0$, as is the map $S \mapsto S \after T_0$ for any fixed $T_0$. It follows from this that also the swapping map $T_1 \og T_2 \mapsto T_2 \og T_1$ is continuous, since $T_2 \og T_1 =  \sigma_{\cY_1, \cY_2} \after (T_1 \og T_2) \after \sigma_{\cX_2, \cX_1}$ cf. \cref{def:SMC}. In short, all sensible operations involving the algebraic operations in $\Theory$ are continuous. \\

Let us to also note that, under condition 1., condition 2. in \cref{def:TopTheory} can be weakened to continuity, for all fixed $\bbZ_2$, of the map $T_1 \mapsto T_1 \og \id_{\bbZ_2}$. Indeed, from this follows continuity of $T_2 \mapsto \id_{\bbZ_1} \og T_2$ like above, and general parallel composition can then be realised by means of a serial composition, as $T_1 \og T_2 = (T_1 \og \id_{\bbY_2}) \after (\id_{\bbX_1} \og T_2)$.

\section{Metric Structure}

\label{sec:Met}

Though it is possible and potentially beneficial to work with very general kinds of metrics (e.g. those in the sense of Lawvere, which need neither be symmetric nor non-degenerate), we shall content ourselves with the usual definition of a metric:

\begin{Definition} (Metric.) \\
Let $S$ be a set. A map $d: S \times S \to [0,\infty]$ is called a \emph{metric on $S$} if it satisfies the following four requirements: 

\begin{itemize}
		\item (Nullity.) For all $x \in S$, $d(x,x)=0$
		\item (Non-Degeneracy.) For all $x,y \in S$, $d(x,y)=d(y,x)=0$ implies $x=y$
		\item (Symmetry.)  For all $x,y \in S$, $d(x,y)=d(y,x)$.
		\item (Triangle Inequality.) For all $x,y,z \in S$, $d(x,y) \leq d(x,z)+d(z,y)$. 
	\end{itemize}

(By symmetry, non-degeneracy can be weakened to $d(x,y)=0 \Rightarrow x=y$.)
\end{Definition}

Most of our metrics will moreover be \emph{normalised}, meaning that $\sup_{x, y \in S} d(x,y) = 1$ (for example, this is the case for the metrics derived from optimal distinguishing probability), but we need not impose this.

\begin{Definition} (Monotone Metrics on a Theory.) \label{def:MetTheory}\\
	A \emph{metric} on a theory $\Theory$ is a collection $d = (d_{\bbX,\bbY})_{\bbX, \bbY \in \Int{\Theory}}$, where, for each pair of interfaces $\bbX, \bbY$ in $\Theory$, $d_{\bbX,\bbY}$ is a metric on the set $\Chan{\Theory}{\bbX}{\bbY}$. A metric is called

	\begin{enumerate}
		\item \emph{serially monotone}, if 

		\begin{align} \label{eq:sumser}
		d_{\bbX, \bbZ}(S \after T, \tilde{S} \after \tilde{T}) \leq d_{\bbX, \bbY}(T, \tilde{T})+ d_{\bbY, \bbZ}(S, \tilde{S})
		\end{align}

	for all channels $T, \tilde{T}: \bbX \to \bbY$ and $S,\tilde{S}: \bbY \to \bbZ$;

		\item  \emph{parallelly invariant}, if 
		
		\begin{align} \label{eq:parinv}
				d_{\bbX \cup \bbZ, \bbY \cup \bbW}(T \og T_0, \tilde{T} \og T_0) = d_{\bbX, \bbY}(T,\tilde{T})  %
			\end{align}

			for all channels $T, \tilde{T}: \bbX \to \bbY$ and $T_0: \bbZ\to \bbW$.

	\end{enumerate}

A metric is called \emph{monotone} if it is both serially monotone and parallelly invariant.\footnote{The word `monotone' seems to not do justice to the requirement of parallel \myuline{invariance}, but that requirement is in fact equivalent (in the presence of serial monotonicity) to the `monotonicity' requirement that results from replacing in \cref{eq:parinv} the equality with `$\leq$', provided the theory has states on every system. We shall use the term generally even so.} 

A \emph{metric theory} $(\Theory, d)$ is a theory $\Theory$ together with a monotone metric $d$.\end{Definition}

\begin{Remark} (Notational Convention I.)\\
	From now on, we always omit the subscripts on $d$ indicating the interfaces, and write e.g. $d(T, \tilde{T})$ rather than $d_{\bbX, \bbY}(T, \tilde{T})$ for channels $T, \tilde{T}: \bbX \to \bbY$. 
\end{Remark}

\begin{Remark} (Notational Convention II.) \\
We will in this chapter move quite liberally between the algebraic notation in terms of operations `$\after$' and `$\og$' and the pictorial notation, depending on the situation at hand. For reference, note that the conditions of serial monotonicity and parallel invariance are pictorially stated as 
  	\begin{align} 
  	\begin{split}
d(\scalemyQ{.8}{0.7}{0.5}{ & \push{\bbX} \qw &  \gate{T} & \push{\bbY} \qw & \gate{S} & \push{\bbZ} \qw & \qw}, \scalemyQ{.8}{0.7}{0.5}{ & \push{\bbX} \qw &  \gate{\tilde{T}} & \push{\bbY} \qw & \gate{\tilde{S}}  & \push{\bbZ} \qw & \qw}) & \leq d(\channel{\bbX}{T}{\bbY}, \channel{\bbX}{\tilde{T}}{\bbY}) \\
 & + d(\channel{\bbY}{S}{\bbZ}, \channel{\bbY}{\tilde{S}}{\bbZ}),
 \end{split}
\end{align}

respectively

\begin{align}
	d( \scalemyQ{.8}{0.7}{0.5}{ & \push{\bbX} \qw & \gate{T} & \push{\bbY} \qw & \qw \\ & \push{\bbZ} \qw & \gate{T_0} & \push{\bbW} \qw & \qw },   \scalemyQ{.8}{0.7}{0.5}{ & \push{\bbX} \qw & \gate{\tilde{T}} & \push{\bbY} \qw & \qw \\ & \push{\bbZ} \qw & \gate{T_0} & \push{\bbW} \qw & \qw }) =d(\channel{\bbX}{T}{\bbY}, \channel{\bbX}{\tilde{T}}{\bbY}) . 
	\end{align}
\end{Remark}

Like for topological theories, it will in our examples always be the case that the metric is really defined on the set of \myuline{transformations}, rather than channels. 

As such, we will see shortly that the \emph{variational distance} and the \emph{diamond-distance} are monotone metrics on the theories $\CIT$ and $\QIT$, respectively; this can be realised already now by the acquainted reader, since both of these metrics are defined using optimal distinguishing probability. However, the conditions of serial monotonicity and parallel invariance abstract away just the most essential features of such metrics, and they consequently accommodate examples which have nothing to do with distinguishing probability: 

\begin{Example} (Operator Norms.) \label{ex:VectMet}\\
	Recall from \cref{ex:VectCart} the cartesian theory whose systems are vector spaces over $k$ and whose transformations are $k$-linear operators, composing serially by functional composition and parallelly by the direct sum, $\oplus$. Suppose that $k=\R$ or $k=\C$, and consider instead the cartesian theory $\Vect{k}^{\norm{\cdot}}$ whose systems are \myuline{normed} vector spaces and whose transformations $A: (V, \norm{\cdot}_V) \to (W, \norm{\cdot}_W)$ are linear operators with operator norm $\norm{A}_\infty \leq 1$. The serial and parallel compositions are as in $\Vect{k}$, with the additional detail that the norm on $V_1 \oplus V_2$ is given by $\max\{\norm{\cdot}_{V_1}, \norm{\cdot}_{V_2} \}$. 
	
	Now, consider the metric $d$ given on the set of transformations from $(V, \norm{\cdot}_V)$ to $(W, \norm{\cdot}_W)$ by  $d(A, \tilde{A}) = \norm{A-\tilde{A}}_\infty$. It is easy to verify that $d$ is serially monotone (using the fact that all allowed transformations have operator norm at most $1$) and that $d$ is parallelly invariant (since $A \oplus A_0 - \tilde{A} \oplus A_0 = (A-\tilde{A}) \oplus 0$), so $d$ defines a monotone metric on the theory $\Vect{k}^{\norm{\cdot}}$.
	
	\end{Example}

The following observation is often helpful in demonstrating that a given metric is serially monotone and parallelly invariant: 

\begin{Lem} (Recharacterisation of Monotone Metrics.) \label{lem:MetSimpl}\\
	A metric $d$ on $\Theory$ is monotone (i.e. serially monotone and parallelly invariant) if and only if the following three conditions hold:

\begin{itemize}
	\item $d$ is \textbf{monotone under serial pre-composition}, meaning that 
	
	\begin{align}
	d(S_0 \after T, S_0 \after \tilde{T}) \leq d(T,\tilde{T})
	\end{align}
	
	for all channels $T, \tilde{T}: \bbX \to \bbY$ and $S_0 : \bbY \to \bbZ$;
	
	\item $d$ is \textbf{monotone under serial post-composition}, meaning that 
	
	\begin{align}
	d(S \after T_0, \tilde{S} \after T_0) \leq d(S,\tilde{S})
	\end{align}
	
	for all channels $S, \tilde{S}: \bbY \to \bbZ$ and $T_0: \bbX\to \bbY$;
	
	\item $d$ is \textbf{parallelly invariant under identities}, meaning that 
	
	\begin{align} \label{eq:parmon}
d(T \og \id_{\bbZ}, \tilde{T} \og \id_{\bbZ}) =d(T,\tilde{T})
\end{align}

for all interfaces $\bbZ$ and all channels $T, \tilde{T}: \bbX \to \bbY$.

	\end{itemize}

\end{Lem}

\begin{proof}
	To see that all three conditions must hold if $d$ is a monotone metric, simply observe that they are special instances of the general conditions (e.g. letting $S=\tilde{S}=S_0$ and using nullity of the metric). 
	
	Conversely, if $d$ is monotone under serial pre- and post-composition then 
	
		\begin{align}
		\begin{split}
	d(S \after T, \tilde{S} \after \tilde{T})& \leq 	d(S \after T, \tilde{S} \after T) + 	d(\tilde{S} \after T, \tilde{S} \after \tilde{T}) \\
	&\leq d(S,\tilde{S}) + d(T, \tilde{T})
	\end{split}
	\end{align}
	
	by the triangle inequality, so $d$ is monotone under serial composition. If in addition $d$ is invariant under parallel identities, then 
	
		\begin{align}
		\begin{split}
	d(T_1 \og T_2, \tilde{T}_1 \og \tilde{T}_2) &= 	d([T_1 \og \id_{\bbY_2}] \after[\id_{\bbX_1}\og T_2] , [\tilde{T}_1 \og \id_{\bbY_2}] \after[\id_{\bbX_1}\og \tilde{T}_2]) \\
	 & \leq 	d(T_1 \og \id_{\bbY_2},  \tilde{T}_1 \og \id_{\bbY_2} )+ d(\id_{\bbX_1}\og T_2 ,  \id_{\bbX_1}\og \tilde{T}_2) \\
	&= d(T_1,  \tilde{T}_1 )+ d( T_2, \tilde{T}_2) .
	\end{split}
	\end{align}

	\end{proof}

Whereas the conditions of \cref{lem:MetSimpl} are often easier to verify than those of \cref{def:MetTheory} (the reader may revisit \cref{ex:VectMet} in this light), a further simplification in the form of \cref{lem:MetSimpl2} more clearly yields the argument that the usual metrics in $\CIT$ and $\QIT$ are monotone. Before observing this result, however, let us recall the definition of these metrics (for more details, the reader may consult Refs. \cite{NC02,Wat11}):

\begin{Example} (The Variational Distance $d_1$ on $\CIT$.) \label{ex:d1CIT}\\
Consider the theory $\CIT$. On $\St{Y}$, the set of probability distributions on $Y$, let $d_1$ be the metric given by 

\begin{align}
d_1(p, q) = \frac{1}{2} \sum_{y \in Y} \abs{p(y)-q(y)}.
\end{align}

The metric $d_1$ is called the \emph{total variation distance}, or simply the \emph{variational distance}. It can be shown that $d_1(p,q)= \max_{A \subseteq Y} \abs{p(A)-q(A)}$, where we use the notation $r(A)$ as shorthand for $\sum_{y \in A} r(y)$, and as such $d_1(p,q)$ quantifies the largest possible difference in probability that $p$ and $q$ assign to any subset of $Y$. It can also be shown that $\frac{1+d_1(p,q)}{2}$ is the optimal probability of distinguishing between the probability distributions $p$ and $q$ based on an outcome drawn according to either $p$ or $q$, with equal probability. 

We can easily extend $d_1$ to a metric on general  classical channels, by defining, for $T=(t_x)_{x \in X} : X \to Y$ and $\tilde{T}=(\tilde{t}_x)_{x \in X} : X \to Y$, 

\begin{align}
d_1(T, \tilde{T}) = \max_{x \in X} d_1(t_x, \tilde{t}_x).
\end{align}

As such, $d_1$ provides a metric on the theory $\CIT$. It is not eye-catching that $d_1$ is serially monotone and parallelly invariant (or, equivalently, satisfies the conditions of \cref{lem:MetSimpl}). A convexity argument shows however that $d_1(T, \tilde{T}) = \sup_{p \in \St{X}} d_1(T \circ p, \tilde{T} \circ p)$, and more generally that $d_1(T, \tilde{T}) =  \sup_{p \in \St{X \times R}} d_1([T \otimes \id_R] \circ p, [\tilde{T} \otimes \id_R] \circ p)$ for any system $R$, which implies one of the conditions stated in our next result, \cref{lem:MetSimpl2}; as the other condition can also quite easily be demonstrated, that result will imply the desired. Thus, $(\CIT, d_1)$ is a metric theory.
	\end{Example}

\begin{Example} (The Diamond-Distance $d_\diamond$ on $\QIT$.) \label{ex:dDQIT}\\
Consider the theory $\QIT$. On $\St{\cK}= \scrD(\cK)$, the set of density matrices on $\cK$, let $d_1$ be the metric given by 

\begin{align}
d_1(\varrho, \sigma) = \frac{1}{2} \norm{\varrho-\sigma}_1,
\end{align}

where $\norm{\cdot}_1$ denotes the trace-norm on $\End{\cK}$. This metric, called \emph{trace distance}, is denoted by the same symbol as the variational distance in \cref{ex:d1CIT}, since under the natural embedding of $\CIT$ in $\QIT$ it is easily seen to reproduce $d_1$ as defined in $\CIT$. It can be shown that also in $\QIT$ the quantity $\frac{1+d_1(\varrho, \sigma)}{2}$ is the optimal probability of distinguishing between the states $\varrho$ and $\sigma$ if one of these is given at random with equal probability. 

If we were to define the trace-distance between general quantum channels $\Lambda, \tilde{\Lambda}: \cH \to \cK$ as $d_1(\Lambda, \tilde{\Lambda}) = \sup_{\varrho \in \St{\cH}} d_1(\Lambda_1(\varrho), \tilde{\Lambda}(\varrho))$, then $d_1$ would be a metric on $\QIT$ which is serially monotone. As is well-known, however, this metric is not parallelly invariant.\footnote{See e.g. Ref. \cite{Wat11}, where the diamond norm is referred to as the \emph{completely bounded} norm.} The fix is to define for channels instead the \emph{diamond-distance} $d_\diamond$ given by

\begin{align}
d_\diamond(\Lambda, \tilde{\Lambda}) = \sup_{\cR \in \Sys{\QIT}} \sup_{\varrho \in \St{\cH \otimes \cR}} d_1([\Lambda \otimes \id_\cR] (\varrho) ,  [\tilde{\Lambda} \otimes \id_\cR] (\varrho))
\end{align}  

for $\Lambda, \tilde{\Lambda}: \cH \to \cK$. Again, by \cref{lem:MetSimpl2}, $d_\diamond$ will a be serially monotone and parallelly invariant, so we have a metric theory $(\QIT, d_\diamond)$.

\end{Example}

The above two examples illustrate that naturally occurring metrics in a theory are often constructed on the basis of metrics on the sets of states. The following result will be applicable in such circumstances:

\begin{Lem} (Sufficient Conditions for Monotonicity.) \label{lem:MetSimpl2}\\
	Suppose that $d$ is	a metric on $\Theory$ which is \textbf{state-determined}, meaning that for all channels $T, \tilde{T}: \bbX \to \bbY$,

		\begin{align}
	d(T, \tilde{T}) = \sup_{\cR, s  }  d\left(\scalemyQ{.8}{0.7}{0.5}{& \Nmultigate{1}{s} &\push{\bbX} \qw &  \gate{T} & \push{\bbY}\qw & \qw \\ & \Nghost{s} & \qw & \push{\cR} \qw & \qw }, \scalemyQ{.8}{0.7}{0.5}{& \Nmultigate{1}{s} &\push{\bbX} \qw &  \gate{\tilde{T}} & \push{\bbY}\qw & \qw \\ & \Nghost{s} & \qw & \push{\cR} \qw & \qw } \right).
	\end{align}
	
	Then, $d$ is monotone (i.e. serially monotone and parallelly invariant) if and only if $d$ is \textbf{state-monotone}, meaning that
				
				\begin{align}
				d(\scalemyQ{.8}{0.7}{0.5}{& \Ngate{r} & \push{\bbX} \qw & \gate{T} & \push{\bbY} \qw & \qw}, \scalemyQ{.8}{0.7}{0.5}{& \Ngate{\tilde{r}} & \push{\bbX} \qw & \gate{T} & \push{\bbY} \qw & \qw}) \leq d(\state{r}{\bbX}, \state{\tilde{r}}{\bbX})
				\end{align}
				
				for all channels $T: \bbX \to \bbY$ and all states $r, \tilde{r}$ on $\bbX$.

\end{Lem}

\begin{Remark} (Metrics form Distinguishing Probability.)  \label{rem:OptDist}\\
	Any metric which like $d_1$ and $d_\diamond$ quantifies optimal distinguishing probability will be (quite evidently) state-determined and state-monotone. Therefore, any such metric is a monotone metric in the sense of \cref{def:MetTheory}.
	\end{Remark}

\begin{proof}
	
	It is clear that state-monotonicity follows from general serial monotonicity. Assume conversely that $d$ is state-determined and state-monotone. We need to verify the conditions in \cref{lem:MetSimpl}, but it suffices to prove parallel \myuline{monotonicity} rather than parallel invariance, since there are states on every system in $\Theory$. 
	
	The monotonicity under parallel composition with identities is built into the condition of being state-determined. So is monotonicity under serial post-composition, since the supremum is after composition with $T_0$ confined to a smaller set. Finally, monotonicity under serial pre-composition follows from $d$ being both state-determined and state-monotone. 

\end{proof}

The fact that $d_1$ on $\CIT$ and $d_\diamond$ on $\QIT$ are state-monotone are proved in most introductory texts, see e.g. Ref. \cite{NC02}. (Alternatively, the reader might find the demonstration to be a nice exercise.) Thus, we confirm that indeed $d_1$ is a monotone metric on $\CIT$, and $d_\diamond$ a monotone metric on $\QIT$.

\section{Dilationality and the Generalised Uhlmann Property}
\label{sec:Dilational}

Thus far, we have stated a general definition of metric theories and provided two important examples, $(\CIT,d_1)$ and $(\QIT, d_\diamond)$. The usefulness of the monotonicity properties is likely well-known to anyone who has experience with these two examples, but one additional property of metrics is desirable, and that property has do with dilations:\\

Suppose that $\channel{\bbX}{T}{\bbY}$ and $\channel{\bbX}{\tilde{T}}{\bbY}$ are channels in $\Theory$, and that $d$ is a monotone metric on $\Theory$. If $\twoext{\bbX}{\bbD}{L}{\bbY}{\bbE}$ and $\twoext{\bbX}{\bbD}{\tilde{L}}{\bbY}{\bbE}$ are \emph{compatible} (meaning with the same hidden interfaces) dilations of $T$ and $\tilde{T}$ respectively, then

\begin{align}
\begin{split}
d(\channel{\bbX}{T}{\bbY},\channel{\bbX}{\tilde{T}}{\bbY})  &= d\left( \scalemyQ{.8}{0.7}{0.5}{ & \push{\bbD} \ww & \Ngate{\tr}{\ww} \\ & \push{\bbX} \qw & \gate{T} & \push{\bbY} \qw & \qw }, \scalemyQ{.8}{0.7}{0.5}{ & \push{\bbD} \ww & \Ngate{\tr}{\ww} \\ & \push{\bbX} \qw & \gate{\tilde{T}} & \push{\bbY} \qw & \qw }\right) \\
& = d\left( \scalemyQ{.8}{0.7}{0.5}{ & \push{\bbD} \ww & \Nmultigate{1}{L}{\ww}  & \Ngate{\tr}{\ww}\\ & \push{\bbX} \qw & \ghost{L} & \push{\bbY} \qw & \qw }, \scalemyQ{.8}{0.7}{0.5}{ & \push{\bbD} \ww & \Nmultigate{1}{\tilde{L}}{\ww}  & \Ngate{\tr}{\ww}\\ & \push{\bbX} \qw & \ghost{\tilde{L}} & \push{\bbY} \qw & \qw }\right)  \\
& \leq d\left(\twoext{\bbX}{\bbD}{L}{\bbY}{\bbE}, \twoext{\bbX}{\bbD}{\tilde{L}}{\bbY}{\bbE}\right)
\end{split}
\end{align}

by parallel invariance and serial monotonicity of $d$. In other words, \emph{distance never decreases under dilations}. Of course, even when $T$ and $\tilde{T}$ are identical, the dilations $L$ and $\tilde{L}$ might be far from each other; what is desirable, however, is that for any dilation $L$ of $T$ we can \myuline{find} some compatible dilation $\tilde{L}$ of $\tilde{T}$ such that $d(T, \tilde{T}) = d(L, \tilde{L})$. 

Somewhat more forgivingly, we may ask that merely $d(T, \tilde{T}) = \inf_{\tilde{L} } d(L, \tilde{L})$, where the infimum is over dilations of $\tilde{T} $ compatible with $L$. The universal quantification over $L$ can be then replaced be a supremum for a closed statement:

\begin{Definition} (Dilationality.) \label{def:Dilationality}\\
A monotone metric $d$  on $\Theory$ is called \emph{dilational} if for all channels $T, \tilde{T}: \bbX \to \bbY$, it holds that
	
	\begin{align} \label{eq:dila}
	d(T, \tilde{T}) = \sup_{L \in \Dil{T}} \inf_{\tilde{L} \in \Dil{\tilde{T}}} d(L, \tilde{L}),
	\end{align}
	
	where the inner infimum is over dilations $\tilde{L}$ \emph{compatible with $L$}, i.e. with the same hidden interfaces as $L$. 
	\end{Definition}

\begin{Example} (Dilationality of the Discrete Metric.) \\
	Consider on any theory $\Theory$ the discrete metric $d_0$, for which $d_0(T,\tilde{T})=1$ if $T \neq \tilde{T}$. The metric $d_0$ is trivially dilational.
\end{Example}

Before presenting more interesting examples of dilational metrics, we demonstrate a number of properties surrounding dilationality. (Though I encourage the reader already at this point to ponder why a monotone metric in a cartesian theory, or in $\CIT$, is necessarily dilational.)\newpage

First, let us observe the following useful fact (which has nothing to do with dilationality):

\begin{Lem} (Monotonicity of Derivability.) \label{lem:FundDil}\\
Let $d$ be a monotone metric on $\Theory$. Suppose that $T, \tilde{T}: \bbX \to \bbY$ are channels, and let $L \in \Dil{T}$ and $\tilde{L} \in \Dil{\tilde{T}}$ be compatible one-sided dilations. For any dilation $L' \red L$ of $T$, there exists a compatible dilation $\tilde{L}' \red \tilde{L}$ such that $d(L', \tilde{L}') \leq d(L, \tilde{L})$. 
	\end{Lem}

\begin{proof}
	Since $L' \red L$, we find a channel $G$ such that  $ \scalemyQ{.8}{0.7}{0.5}{& \push{\bbX} \qw &  \Nmultigate{1}{L'}{\qw}   & \push{\bbY} \qw  & \qw  \\
		&\push{\bbD'} \ww & \Nghost{L'}{\ww} & \push{\bbE'} \ww  & \ww
	} = \scalemyQ{.8}{0.7}{0.5}{ & \push{\bbX} \qw &  \multigate{1}{L}  & \qw & \push{\bbY} \qw  & \qw  \\
		& &   \Nghost{L} & \Nmultigate{1}{G}{\ww} & \push{\bbE'} \ww & \ww & \\
		&   \push{\bbD'} \ww 	&  \ww & \Nghost{G}{\ww} & 
	} $. But then defining $\tilde{L}'$ to be $\scalemyQ{.8}{0.7}{0.5}{ & \push{\bbX} \qw &  \multigate{1}{\tilde{L}}  & \qw & \push{\bbY} \qw  & \qw  \\
	& &   \Nghost{\tilde{L}} & \Nmultigate{1}{G}{\ww} & \push{\bbE'} \ww & \ww & \\
	&   \push{\bbD'} \ww 	&  \ww & \Nghost{G}{\ww} & 
} $, we have $d(L', \tilde{L}') \leq  d(L, \tilde{L})$ by serial and parallel monotonicity of $d$. 	\end{proof}

Among other things, \cref{lem:FundDil} has as consequence that the supremum in the condition \eqref{eq:dila} can be restricted to one-sided dilations if every dilation is derivable from a one-sided dilation:

\begin{Prop}  (Restricting to One-Sided Dilations.) \label{rem:MetDilaOne}\\
Suppose that $\Theory$ has the DiVincenzo Property (e.g. by being spatially localisable). Then, a monotone metric $d$ on $\Theory$ is dilational precisely if for all channels $T, \tilde{T}: \bbX \to \bbY$, and for all \myuline{one-sided} dilations $L$ of $T$, it holds that 

	\begin{align} 
d(T, \tilde{T}) = \inf_{\tilde{L} } d(L, \tilde{L}),
\end{align}

where the infimum is over all one-sided dilations $\tilde{L}$ of $\tilde{T}$ which are compatible with $L$. 
	\end{Prop}

\begin{proof} Obvious from \cref{lem:FundDil}.\end{proof}

We will often use this result more or less implicitly. \\

\begin{Remark} (On the Importance of Dilationality for the Establishment of Dilations.) \\
One of the useful consequences of dilationality has to the with the very establishment of the concept of a dilation. 

In the operational narrative, we have pretended that by interaction with a pair of open interfaces $\bbX$ and $\bbY$ we can certify exactly the occurrence of a given channel $\channel{\bbX}{T_0}{\bbY}$. In reality, we can of course, even under an i.i.d. assumption, only verify that we interact with a channel $T$ \myuline{close} to $T_0$. Now, the agent (or, `Nature') who implements $T$ is really employing some channel $\scalemyQ{.8}{0.7}{0.5}{& \push{\bbX} \qw &  \Nmultigate{1}{M}{\qw}   & \push{\bbY} \qw  & \qw  \\
	&\push{\bbD} \ww & \Nghost{M}{\ww} & \push{\bbE} \ww  & \ww}$ between the interfaces accessible to us and additional interfaces in the environment which are inaccessible to us. Not only have we pretended that $T=T_0$, we have also pretended that $M$ must in fact be a dilation of $T_0$. In reality, however, the channel we `see' when $M$ is employed is the channel $\scalemyQ{.8}{0.7}{0.5}{& \push{\bbX} \qw &  \Nmultigate{1}{M}{\qw}   & \push{\bbY} \qw  & \qw  \\
	&\push{\bbD} \ww & \Nghost{M}{\ww} & \push{\bbE} \ww  & \Ngate{\tr}{\ww}}$, and we control only the input to the interface $\bbX$. By doing this, we may (under an i.d.d. assumption) certify that

\begin{align} \label{eq:dilclose}
\scalemyQ{1}{0.7}{0.5}{& \push{\bbX} \qw &  \Nmultigate{1}{M}{\qw}   & \push{\bbY} \qw  & \qw  \\
	&\push{\bbD} \ww & \Nghost{M}{\ww} & \push{\bbE} \ww  & \Ngate{\tr}{\ww}} \approx^d_\varepsilon 	\myQ{0.7}{0.5}{& \push{\bbX} \qw &   \Ngate{T_0}{\qw}    & \push{\bbY} \qw  & \qw  \\
	&\push{\bbD} \ww  & \Ngate{\tr}{\ww}
},
\end{align}

where `$\approx^d_\varepsilon$' signifies that the two channels are at most $\varepsilon$ apart w.r.t. $d$. Hence, what we \myuline{actually} need for all of our mathematical pretendings to be operationally justifiable is that $T_0$ have a two-sided dilation which is at most $\varepsilon$ apart from $M$ w.r.t. $d$. 

Dilationality hands us precisely this feature: If $d$ is dilational, then according to \cref{eq:dilclose}, some one-sided dilation of $\scalemyQ{.8}{0.7}{0.5}{& \push{\bbX} \qw &   \Ngate{T_0}{\qw}    & \push{\bbY} \qw  & \qw  \\
	&\push{\bbD} \ww  & \Ngate{\tr}{\ww}
}$ will be $\varepsilon$-close to $M$, and this dilation will be a dilation of $T_0$ when including the input interface $\bbD$.

(It is worth observing that another way of phrasing this discussion is by saying that dilationality guarantees that a channel which is almost non-signalling is close to a channel which is perfectly non-signalling.)

\end{Remark}

Now, it is known that the dilational property fails for the diamond-distance $d_\diamond$ in $\QIT$ already in the case of distances between states. This is one of the reasons for the introduction of the `purified distance' (\cite{Toma10,Toma12}), and will ultimately also be the reason for us to introduce a generalisation of this distance to arbitrary channels in $\QIT$, the \emph{purified diamond-distance} (\cref{def:PurifDiam}). \\

In contrast, the dilational property is satisfied for the variational $d_1$ on $\CIT$, and will in fact be satisfied for \myuline{any} monotone metric on $\CIT$. To better understand the mechanism behind this, it is helpful to observe the following recharacterisation of dilationality, which is also of independent interest: 

\begin{Prop} (Dilationality means the Generalised Uhlmann Property.) \label{prop:GenUhl} \\
Suppose that $\Theory$ is complete and localisable, and suppose that $d$ is a monotone metric on $\Theory$. Then, the following are equivalent: 
	
	\begin{enumerate}
		\item The metric $d$ is dilational, that is, for any channels $T, \tilde{T}: \bbX \to \bbY$, and any one-sided dilation $L$ of $T$, it holds that
		
			\begin{align} \label{eq:dilational}
		d\left( \channel{\bbX}{T}{\bbY}, \channel{\bbX}{\tilde{T}}{\bbY}\right) = \inf_{ \tilde{L}} d\left(\oneext{\bbX}{L}{\bbY}{\bbE}, \oneext{\bbX}{\tilde{L}}{\bbY}{\bbE} \right),
		\end{align}
		
		where the infimum is over all dilations $\tilde{L}$ of $\tilde{T}$, compatible with $L$.

		\item The metric $d$ has the \emph{Generalised Uhlmann Property}, that is, for any channels $T, \tilde{T}: \bbX \to \bbY$, it holds that

			\begin{align} \label{eq:uhlmann}
		d\left( \channel{\bbX}{T}{\bbY}, \channel{\bbX}{\tilde{T}}{\bbY}\right) = \inf_{K, \tilde{K}} d\left(\oneext{\bbX}{K}{\bbY}{\bbE_0}, \oneext{\bbX}{\tilde{K}}{\bbY}{\bbE_0} \right),
		\end{align} 
		
		where the infimum is over all compatible pairs of complete dilations $K$ of $T$ and $\tilde{K}$ of $\tilde{T}$.

	\end{enumerate} 
\end{Prop}

\begin{proof}
By the monotonicity properties, the inequalities `$\leq$' are clear in both \eqref{eq:dilational} and \eqref{eq:uhlmann}. Hence, for each implication, it suffices to show the inequality `$\geq$'. 
	
First assume that $d$ has the Generalised Uhlmann Property. Let $T, \tilde{T}: \bbX \to \bbY$ be channels and let $L$ be a dilation of $T$. Given $\varepsilon > 0$, pick, according to the Generalised Uhlmann Property, compatible complete dilations $K, \tilde{K}$  such that
	
	\begin{align}
	d(K, \tilde{K}) < d(T, \tilde{T}) + \varepsilon.
	\end{align}
	
Now, since by completeness $L \red K$, it follows from \cref{lem:FundDil}	that

	\begin{align}
\inf_{\tilde{L}} d(L, \tilde{L}) \leq d(K, \tilde{K}) < d(T, \tilde{T}) + \varepsilon,
\end{align}

where the infimum is over dilations $\tilde{L}$ of $\tilde{T}$, compatible with $L$. Since $\varepsilon > 0$ was arbitrary the desired inequality follows. 

The converse implication is more interesting. Suppose that $d$ is dilational, and let $T, \tilde{T} : \bbX\to \bbY$ be channels. Now, choose some complete dilation $K_0$ of $T$. Given $\varepsilon > 0$, pick, according to dilationality, a dilation $\tilde{L}$ of $\tilde{T}$ such that

	\begin{align} \label{eq:tri1}
	d(K_0, \tilde{L}) < d(T, \tilde{T}) + \varepsilon.
	\end{align}
	
Now, by symmetry of $d$ we may switch the order of arguments in the condition of dilationality. Therefore, if we pick a complete dilation $\tilde{K}$ of $\tilde{L}$, we find by dilationality some dilation $K$ of $K_0$, compatible with $\tilde{K}$, such that 
	
	\begin{align} \label{eq:tri2}
	d(K,  \tilde{K}) < d(K_0, \tilde{L}) + \varepsilon.
	\end{align}
	
	Combining Eqs. \eqref{eq:tri1} and \eqref{eq:tri2} yields
	
	\begin{align}
	d(K, \tilde{K}) < d(T,  \tilde{T}) + 2 \varepsilon,
	\end{align}
	
	and so we are done if we can argue that $K$ and $\tilde{K}$ are complete dilations of $T$ and $\tilde{T}$, respectively. 	
	
	However,  $K$ dilates $K_0$ which was chosen complete for $T$, and therefore $K$ is trivially a complete dilation of $T$. More interestingly, $\tilde{K}$ is a complete dilation of $\tilde{L}$, which dilates $\tilde{T}$, and is therefore by the fact that completeness is \myuline{hereditary} (\cref{lem:Hereditary}) a complete dilation of $\tilde{T}$. The desired follows again since $\varepsilon> 0$ was arbitrary.

\end{proof}

\begin{Example} (Dilationality in $\CIT$.) \label{ex:CITdilational}\\
In $\CIT$, a complete dilation $K_T$ of $T$ can be canonically obtained by copying inputs and outputs (\cref{thm:CITComp}). Since $K_T$ is given from $T$ by tensoring with identities and serially composing with surrounding channels, we must have $d(T, \tilde{T}) = d(K_T, K_{\tilde{T}})$ for any monotone metric $d$. Consequently, $d$ trivially has the Generalised Uhlmann Property, and $d$ is therefore necessarily dilational. In particular, the variational distance $d_1$ on $\CIT$ is dilational.  
	\end{Example}

By a completely similar argument, any monotone metric on a cartesian theory is dilational. 

In general, however, complete dilations are not computable simply by pre- and post-composing cleverly with channels -- for example, this is not the case in $\QIT$ where completeness goes through Stinespring dilations. A natural question is thus the following: Given a monotone metric $d$ on a theory $\Theory$,  can we construct from $d$ a monotone metric $\hat{d}$ which is also \myuline{dilational}? 

\cref{prop:GenUhl} suggests that the quantity

\begin{align} \label{eq:Uhlatt}
\hat{d}(T, \tilde{T}) = \inf_{K, \tilde{K}} d(T, \tilde{T})
\end{align} 

could be a sensible candidate (where the infimum ranges over compatible pairs of complete dilations). Though $\hat{d}$ can be proven to be a metric under circumstances which ensure the triangle inequality, and proven to be also monotone, it seems to me that (as mentioned in the introduction) dilationality could well fail for $\hat{d}$, although I know of no specific counterexample.

\begin{OP} (Construction of Dilational Metrics.) \label{op:Dhat}\\
	Suppose that $\Theory$ is complete and localisable and that $d$ is a monotone metric on $\Theory$. Is the quantity $\hat{d}$ given by \cref{eq:Uhlatt} a monotone dilational metric on $\Theory$? If not, are there tractable additional conditions which ensure that it is?  
	\end{OP}

\section{Purified Dilationality -- the Uhlmann Property}
\label{sec:PureUhlmann}

In this section, we shall see that \myuline{purifiability} allows us to squeeze out a dilational, monotone metric starting from one which is just monotone. This is expressed by \cref{thm:Purified}. We will also prove an approximate version of the duality of \cref{thm:InfoDist} which applies to any dilational monotone metric in a purifiable theory.\\

More precisely, let us assume from now on that the theory $\Theory$ is localisable, universal and purifiable. In particular, every channel $T$  in $\Theory$ has a \myuline{one-sided} pure dilation $\Sigma$ by \cref{prop:PureRechar}. (We will in the remainder of the chapter use the notation `$\Sigma$' rather than `$P$' for pure dilations, since $P$ will be soon used for something else; in $\QIT$, the pure dilations $\Sigma$ are Stinespring dilations.)

 Let us also make the technical assumption that if $T, \tilde{T} : \bbX \to \bbY$ are channels, then there exist pure one-sided dilations $\Sigma, \tilde{\Sigma}: \bbX \to \bbY \cup \bbE$ which are \myuline{compatible}, i.e. have the same hidden interface. This is true, for example, in $\QIT$, and more generally whenever the theory $\Theory$ has a pure state on every system.\footnote{If $\Sigma_0: \bbX \to \bbY \cup \bbE_0$ is a pure dilation of $T$ and $\tilde{\Sigma}_0 : \bbX \to \bbY \cup \tilde{\bbE}_0$ a pure dilation of $\tilde{T}$, then we obtain compatible pure dilations $\Sigma, \tilde{\Sigma}: \bbX\to \bbY \cup \bbE_0 \cup \tilde{\bbE}_0$ by parallelly composing with any pure states on the appropriate systems.}  \\

 We make the following definition:

\begin{Definition} (Purified Distance.) \\
	Let $d$ be a monotone metric on $\Theory$.	We define the \emph{purified $d$-distance} between channels $\channel{\bbX}{T}{\bbY}$ and $\channel{\bbX}{\tilde{T}}{\bbY}$ by
	
	\begin{align} \label{eq:Dpurified}
	\breve{d}(\channel{\bbX}{T}{\bbY},\channel{\bbX}{\tilde{T}}{\bbY}) = \inf_{\Sigma, \tilde{\Sigma}}  d \left(\oneext{\bbX}{\Sigma}{\bbY}{\bbE} , \oneext{\bbX}{\tilde{\Sigma}}{\bbY}{\bbE}\right),
	\end{align}	
	
	where the infimum is over all compatible pure one-sided dilations $\Sigma$ of $T$ and $\tilde{\Sigma}$ of $\tilde{T}$.
	
\end{Definition}

Observe that in the special case where $\Theory = \QIT$, the infimum is over compatible \myuline{Stinespring} dilations of the channels. Taking in this case $d= d_\diamond$ (the diamond-distance) is our most important example, and we will give it a special name:

\begin{Definition} (Purified Diamond-Distance.) \label{def:PurifDiam}\\
	The \emph{purified diamond-distance in $\QIT$} is the quantity $P_\diamond := \breve{d}_\diamond$. 
	\end{Definition}

In \cref{sec:BuresP} we shall discuss how to calculate $P_\diamond(T, \tilde{T})$ and relate it to the so-called \emph{Bures distance} of Refs. \cite{KSW08math, Bures69}. For the moment, let us simply observe that in the case where the channels are states, the quantity $P_\diamond(\varrho, \tilde{\varrho})$ is already known in the literature as the \emph{purified distance} (\cite{Toma10, Toma12}).\\

Right now, we are confronted with the much more pressing problem of demonstrating that in general the purified distance $\breve{d}$ defines a monotone metric, so that indeed its name is sensible. Moreover -- and this was the reason for its introduction --  it is \myuline{dilational}:

\begin{Thm} (Properties of the Purified Distance.)\\ \label{thm:Purified}
Assume that $\Theory$ is localisable, universal and purifiable.\footnote{And satisfies the technical assumption about existence of compatible pure dilations.} For any monotone metric $d$ on $\Theory$, the purified distance $\breve{d}$ defines a monotone, dilational metric on $\Theory$. We moreover have the inequality $d(T, \tilde{T}) \leq \breve{d}(T, \tilde{T})$ with equality when $T$ and $\tilde{T}$ are dilationally pure.  

\end{Thm} 

\begin{proof}

The inequality $d(T, \tilde{T}) \leq \breve{d}(T, \tilde{T})$ is clear from the definition of $\breve{d}$ by monotonicity of $d$. It is moreover obvious that we have the reverse inequality (and hence equality) when the channels $T$ and $\tilde{T}$ are dilationally pure. 

	As for the claim that $\breve{d}$ is a metric, symmetry follows from symmetry of $d$, and non-degeneracy from non-degeneracy of $d$ along with the inequality $d \leq \breve{d}$. It thus only remains to show the triangle inequality, which is also the hardest part. To this end, let $T_1, T_2, T_3 : \bbX \to \bbY$ be channels, and assume to arrive at a contradiction that $ \breve{d}(T_1, T_2) + \breve{d}(T_2, T_3)< \breve{d}(T_1, T_3) $. Then, we can pick by definition of $\breve{d}$ pure dilations $\Sigma_1$ and $\Sigma_2$ of $T_1$ and $T_2$, respectively, and pure dilations $\Sigma'_2$ and $\Sigma_3$ of $T_2$ and $T_3$, respectively, such that 
	
	\begin{align} \label{eq:contra}
	d(\Sigma_1, \Sigma_2) + d(\Sigma'_2, \Sigma_3) < \breve{d}(T_1, T_3) .
	\end{align} 
	
	Now, if it were the case that $\Sigma_2 = \Sigma'_2$ then the triangle inequality for $d$ would imply that $d(\Sigma_1, \Sigma_3)$ is a lower bound to the left hand side, in conflict with the definition of $\breve{d}(T_1, T_3)$. The challenge is, however, that this need not be the case. In $\QIT$, we can apply isometries to `align' $\Sigma_2$ and $\Sigma'_2$, and the general remedy is indeed a generalisation of this trick.
	
	First, pick a universal dilation $U_2$ of $T_2$. Then there exist channels $V$ and $V'$ such that $\Sigma_2$ equals $\scalemyQ{.8}{0.7}{0.5}{& \push{\bbX}  \qw & \multigate{1}{U_2} & \push{\bbY} \qw & \qw \\ 
		& & \Nghost{U_2} & \Ngate{V}{\ww} & \ww  }$  and $\Sigma'_2$ equals $\scalemyQ{.8}{0.7}{0.5}{& \push{\bbX}  \qw & \multigate{1}{U_2} & \push{\bbY} \qw & \qw \\ 
		& & \Nghost{U_2} & \Ngate{V'}{\ww} & \ww  }$. By universality, $V$ and $V'$ must moreover be dilationally pure since $\Sigma_2$ and $\Sigma'_2$ are, so by \cref{lem:PureAlign} below we find pure channels $\tilde{V}$ and $\tilde{V}'$ such that $ \scalemyQ{.8}{0.7}{0.5}{& \gate{V} & \gate{\tilde{V}}  & \qw} = \scalemyQ{.8}{0.7}{0.5}{& \gate{V'} & \gate{\tilde{V}'}  & \qw} $, whence we actually have $\scalemyQ{.8}{0.7}{0.5}{& \push{\bbX}  \qw & \multigate{1}{\Sigma_2} & \push{\bbY} \qw & \qw \\ 
		& & \Nghost{\Sigma_2} & \Ngate{\tilde{V}}{\ww} & \ww  } =\scalemyQ{.8}{0.7}{0.5}{& \push{\bbX}  \qw & \multigate{1}{\Sigma'_2} & \push{\bbY} \qw & \qw \\ 
		& & \Nghost{V'} & \Ngate{\tilde{V}'}{\ww} & \ww  }$. In other words, though $\Sigma_2$ and $\Sigma'_2$ are not necessarily identical, we can apply pure channels in the environment and thereby align them. This now implies by the triangle inequality for $d$ that 
	
	\begin{align} \begin{split}
	d\left( \scalemyQ{.8}{0.7}{0.5}{& \push{\bbX}  \qw & \multigate{1}{\Sigma_1} & \push{\bbY} \qw & \qw \\ 
		& & \Nghost{\Sigma_1} & \Ngate{\tilde{V}}{\ww} & \ww  },  \scalemyQ{.8}{0.7}{0.5}{& \push{\bbX}  \qw & \multigate{1}{\Sigma_3} & \push{\bbY} \qw & \qw \\ 
		& & \Nghost{\Sigma_3} & \Ngate{\tilde{V}'}{\ww} & \ww  }\right) \leq \;   
	&d\left( \scalemyQ{.8}{0.7}{0.5}{& \push{\bbX}  \qw & \multigate{1}{\Sigma_1} & \push{\bbY} \qw & \qw \\ 
		& & \Nghost{\Sigma_1} & \Ngate{\tilde{V}}{\ww} & \ww  },  \scalemyQ{.8}{0.7}{0.5}{& \push{\bbX}  \qw & \multigate{1}{\Sigma_2} & \push{\bbY} \qw & \qw \\ 
		& & \Nghost{\Sigma_2} & \Ngate{\tilde{V}}{\ww} & \ww  }\right) \\
	&+ d\left( \scalemyQ{.8}{0.7}{0.5}{& \push{\bbX}  \qw & \multigate{1}{\Sigma'_2} & \push{\bbY} \qw & \qw \\ 
		& & \Nghost{\Sigma'_2} & \Ngate{\tilde{V}'}{\ww} & \ww  },  \scalemyQ{.8}{0.7}{0.5}{& \push{\bbX}  \qw & \multigate{1}{\Sigma_3} & \push{\bbY} \qw & \qw \\ 
		& & \Nghost{\Sigma_3} & \Ngate{\tilde{V}'}{\ww} & \ww  }\right),
	\end{split}
	\end{align}
	
	and because the channels $\tilde{V}$ and $\tilde{V}'$ are reversible, monotonicity of $d$ implies that the two distances on the right hand side coincide with $d(\Sigma_1, \Sigma_3)$ and $d(\Sigma'_2, \Sigma_3)$, respectively, so by \cref{eq:contra} we actually have
	
	\begin{align} 
	d\left( \scalemyQ{.8}{0.7}{0.5}{& \push{\bbX}  \qw & \multigate{1}{\Sigma_1} & \push{\bbY} \qw & \qw \\ 
		& & \Nghost{\Sigma_1} & \Ngate{\tilde{V}}{\ww} & \ww  },  \scalemyQ{.8}{0.7}{0.5}{& \push{\bbX}  \qw & \multigate{1}{\Sigma_3} & \push{\bbY} \qw & \qw \\ 
		& & \Nghost{\Sigma_3} & \Ngate{\tilde{V}'}{\ww} & \ww  }\right) < \breve{d}(T_1, T_3). 
	\end{align}
	
	Finally, however, as $\tilde{V}$ and $\tilde{V}'$ are pure, the two dilations $\scalemyQ{.8}{0.7}{0.5}{& \push{\bbX}  \qw & \multigate{1}{\Sigma_1} & \push{\bbY} \qw & \qw \\ 
		& & \Nghost{\Sigma_1} & \Ngate{\tilde{V}}{\ww} & \ww  }$ and $\scalemyQ{.8}{0.7}{0.5}{& \push{\bbX}  \qw & \multigate{1}{\Sigma_3} & \push{\bbY} \qw & \qw \\ 
		& & \Nghost{\Sigma_3} & \Ngate{\tilde{V}'}{\ww} & \ww  }$ are (by localisability) pure dilations of $T_1$ and $T_3$, respectively, so \myuline{this} contradicts the definition of $\breve{d}(T_1, T_3)$. This altogether demonstrates the triangle inequality. (Incidentally, the specialisation of this proof to the case of states yields a different proof for the triangle inequality of the purified distance than the one given in Ref. \cite{Toma12}.)
	
	To show that $\breve{d}$ is serially monotone, it suffices to observe that by localisability the serial composition of pure dilations is a pure dilation and then use serial monotonicity of $d$. Parallel invariance of $\breve{d}$ is proved by using parallel invariance of $d$ and the fact that identities are dilationally pure (\cref{prop:IsoSelfUniversal}).
	
	Lastly, to see the $\breve{d}$ is dilational, we may by \cref{prop:GenUhl} equivalently demonstrate that $\breve{d}$ has the Generalised Uhlmann Property. The inequality $\breve{d}(T, \tilde{T}) \leq \inf_{K, \tilde{K}} \breve{d}(K, \tilde{K})$ holds simply because $d$ is monotone; the non-trivial inequality is the converse, $\breve{d}(T, \tilde{T}) \geq \inf_{K, \tilde{K}} \breve{d}(K, \tilde{K})$, and it owes its validity to the hereditary property of completeness in the guise of \cref{prop:PureareComp}. Indeed, we have 
	
	\begin{align}
	\breve{d}(T, \tilde{T}) = \inf_{\Sigma, \tilde{\Sigma}} d(\Sigma, \tilde{\Sigma}) =  \inf_{\Sigma, \tilde{\Sigma}} \breve{d}(\Sigma, \tilde{\Sigma}) 
	\end{align} 
	
	by definition of $\breve{d}$ and the introductory observation, and because every pure dilation of a channel is \myuline{complete} (\cref{prop:PureareComp}), the above quantity is trivially an upper bound to $\inf_{K, \tilde{K}} \breve{d}(K, \tilde{K})$.

\end{proof}

\begin{Lem} (Pure Channels can be Purely Aligned.) \label{lem:PureAlign}\\
	Suppose that $\Theory$ is localisable, universal and purifiable. Then for any  pure channels  $\channel{\cX}{V}{\cY}$ and $\channel{\cX}{V'}{\cY'}$, there exist pure channels $\channel{\cY}{\tilde{V}}{\cZ}$ and $\channel{\cY'}{\tilde{V}'}{\cZ}$ such that 
	\begin{align}
	\myQ{0.7}{0.5}{& \push{\cX} \qw & \gate{V} & \push{\cY} \qw & \gate{\tilde{V}} & \push{\cZ} \qw & \qw} = 	\myQ{0.7}{0.5}{& \push{\cX} \qw & \gate{V'} & \push{\cY'} \qw & \gate{\tilde{V}'} & \push{\cZ} \qw & \qw} \quad.
	\end{align}
\end{Lem}

\begin{proof}
	
By \cref{thm:StructureOfReversibles}, $V$ and $V'$ are reversible, hence we can find $S$ and $S'$ with
	
	\begin{align}
	\myQ{0.7}{0.5}{& \push{\cX} \qw & \gate{V} & \push{\cY} \qw & \gate{S} & \push{\cX} \qw & \qw} = \myQ{0.7}{0.5}{ & \push{\cX} \qw & \gate{\id} & \push{\cX} \qw & \qw } = 	\myQ{0.7}{0.5}{& \push{\cX} \qw & \gate{V'} & \push{\cY'} \qw & \gate{S'} & \push{\cX} \qw & \qw} \quad.
	\end{align} 
	
Now, $S$ and $S'$ have universal dilations, say $U$ and $U'$, which by \cref{prop:PureRechar} are dilationally pure, so the channels 

	\begin{align}
\myQ{0.7}{0.5}{& \push{\cX} \qw & \gate{V}  & \multigate{1}{U} & \push{\cX} \qw & \qw \\ & & &  \Nghost{U} & \ww & \ww}  , \quad \myQ{0.7}{0.5}{& \push{\cX} \qw & \gate{V'} & \multigate{1}{U'} & \push{\cX} \qw & \qw \\ & & &  \Nghost{U'} & \ww & \ww} 
\end{align} 

are pure dilations of $\channel{\cX}{\id}{\cX}$. Since identities are dilationally pure by \cref{prop:IsoSelfUniversal}, these dilations must consequently take the form

\begin{align}
	\myQ{0.7}{0.5}{& \push{\cX} \qw &  \gate{\id} & \push{\cX} \qw & \qw \\ & &  \Ngate{t} & \ww& \ww }, \quad 
	\myQ{0.7}{0.5}{& \push{\cX} \qw &  \gate{\id} & \push{\cX} \qw & \qw \\ & &  \Ngate{t'} & \ww& \ww },
\end{align}

respectively, for pure states $t$ and $t'$. Now, \myuline{those} two channels can clearly be purely aligned, simply by tensoring with $t'$ and $t$, respectively. Pre-composing with $U$ and $U'$ we altogether obtain pure channels $\tilde{V}$ and $\tilde{V}'$ which align $V$ and $V'$, as desired.

\end{proof}

\begin{Cor} The purified diamond-distance $P_\diamond$ is a monotone, dilational metric on the theory $\QIT$.
	
	\end{Cor}

It is quite easy to identify \myuline{the} characteristic property of purified distances:\footnote{We will not need this result, I simply produce it for the reader to gain further intuition.}

\begin{Prop} (Purified means Uhlmann.)\\
	Suppose that $\Theory$ is a localisable, universal and purifiable theory, and suppose that $D$ is a monotone metric on $\Theory$. Then the following are equivalent:
	
	\begin{enumerate}
		\item $D = \breve{d}$ for some monotone metric $d$.
	\item $D$ has the \emph{Uhlmann Property}, that is, for any channels $T, \tilde{T}: \bbX \to \bbY$, it holds that

		\begin{align}
		D\left( \channel{\bbX}{T}{\bbY}, \channel{\bbX}{\tilde{T}}{\bbY}\right) = \inf_{\Sigma, \tilde{\Sigma}} D\left(\oneext{\bbX}{\Sigma}{\bbY}{\bbE}, \oneext{\bbX}{\tilde{\Sigma}}{\bbY}{\bbE} \right),
		\end{align} 
		
		where the infimum is over all compatible pairs of pure one-sided dilations $\Sigma$ of $T$ and $\tilde{\Sigma}$ of $\tilde{T}$.

		\end{enumerate}
	
\end{Prop}

\begin{proof}
	If $D = \breve{d}$, then $D(T, \tilde{T}) = \inf_{\Sigma, \tilde{\Sigma}} d(\Sigma, \tilde{\Sigma}) = \inf_{\Sigma, \tilde{\Sigma}} \breve{d}(\Sigma, \tilde{\Sigma}) =  \inf_{\Sigma, \tilde{\Sigma}} D(\Sigma, \tilde{\Sigma}) $ since $\breve{d}(\Sigma, \tilde{\Sigma}) = d(\Sigma, \tilde{\Sigma})$ by the statement in \cref{thm:Purified}. If conversely $D$ has the Uhlmann property, then obviously $D = \breve{D}$.
	\end{proof}

Now that we have established the purified distance $\breve{d}$ as a monotone, dilational metric, and understood its characteristic trait, let us observe what could be an important quality of it, namely the fact that it tightly captures the approximate theory of complementary channels. 

Recall from \cref{sec:Selfuniv} that the channels $\channel{\bbX}{T}{\bbY}$ and $\scalemyQ{.8}{0.7}{0.5}{& \push{\bbX} \qw & \gate{T^c}& \push{\bbE} \ww & \ww}$ are said to be \emph{complementary} if they admit a common pure dilation $\oneext{\bbX}{\Sigma}{\bbY}{\bbE}$. We have the following:

\begin{Prop} (Approximate Complementarity.) \label{prop:AppComp}\\
Suppose that $D=\breve{d}$ is a purified distance, or, equivalently, has the Uhlmann property. For any channels $\channel{\bbX}{T}{\bbY}$ and $\channel{\bbX}{\tilde{T}}{\bbY}$, it holds that

\begin{align}
 \inf_{T^c, \tilde{T}^c}\breve{d}(\scalemyQ{.8}{0.7}{0.5}{& \push{\bbX} \qw & \gate{T^c}& \push{\bbE} \ww & \ww}, \scalemyQ{.8}{0.7}{0.5}{& \push{\bbX} \qw & \gate{\tilde{T}^c}& \push{\bbE} \ww & \ww})  \leq  \breve{d}(\channel{\bbX}{T}{\bbY}, \channel{\bbX}{\tilde{T}}{\bbY}) ,
\end{align}

where the infimum is over all pairs of channels $T^c$ complementary to $T$ and $\tilde{T}^c$ complementary to $\tilde{T}$, with the same codomain. 

\end{Prop}

\begin{proof}
Simply observe that by monotonicity of $\breve{d}$ under marginalisation we have $\breve{d}(T^c, \tilde{T}^c) \leq \breve{d}(\Sigma, \tilde{\Sigma})$ for pure dilations $\Sigma$ and $\tilde{\Sigma}$ which witness the complementarity, so $\inf_{T^c, \tilde{T}^c} \breve{d}(T^c, \tilde{T}^c) \leq \inf_{\Sigma, \tilde{\Sigma}} \breve{d}(\Sigma, \tilde{\Sigma}) = \breve{d}(T, \tilde{T})$ as desired.	\end{proof}

This result immediately implies an approximate version of \cref{thm:InfoDist}, in the sense that a channel $T$ is close to a reversible channel if and only if its complementary channel $T^c$ is close to being completely forgetful. \\

We can, however, improve on this observation by discarding \cref{prop:AppComp} and using instead mere dilationality, and with this I will conclude the section.  The improvement will reside in the fact that there is a natural notion of a channel $T$ being `approximately reversible', and this notion is weaker than $T$ being approximately equal to a channel which is genuinely reversible.\footnote{To be fair, it constitutes an improvement only in one direction of the bi-implication.} \\

The improved result can be seen as an abstract statement of the `information-disturbance trade off' of Ref. \cite{KSW08phys}:

\begin{Thm} \scalebox{0.95}{(Approximate Duality between Reversible and Completely Forgetful Channels.)}\label{thm:AppInfoDist}\\
	Let $\Theory$ be a universal, localisable and purifiable theory, and let $D$ be a monotone and dilational metric on $\Theory$. Then, for any channel $\channel{\cX}{T}{\cY}$ and any complementary channel $\scalemyQ{.8}{0.7}{0.5}{& \push{\cX} \qw & \gate{T^c}& \push{\cE} \ww & \ww}$, we have 
		
		\begin{align} \label{eq:EqInfoDist}
		\inf_{T^-}  D(\scalemyQ{.8}{0.7}{0.5}{& \push{\cX} \qw & \gate{T} & \push{\cY} \qw & \gate{T^{-}} & \push{\cX} \qw & \qw}, \scalemyQ{.8}{0.7}{0.5}{& \push{\cX} \qw & \gate{\id} & \push{\cX} \qw & \qw} ) \, = \, \inf_{ s} D(\scalemyQ{.8}{0.7}{0.5}{& \push{\cX} \qw & \gate{T^c}& \push{\cE} \ww & \ww}, \scalemyQ{.8}{0.7}{0.5}{& \push{\cX} \qw & \gate{\tr} & \Ngate{s} & \push{\cE} \ww & \ww }),
		\end{align}
		
		where the first infimum is over all channels $\channel{\cY}{T^-}{\cX}$ and the second infimum over all states $\scalemyQ{.8}{0.7}{0.5}{& \Ngate{s} & \push{\cE} \ww & \ww}$. 

	\end{Thm}

\begin{proof}

Fix a channel $\channel{\cY}{T^-}{\cX}$ and define  

\begin{align}
\varepsilon := D(\scalemyQ{.8}{0.7}{0.5}{& \push{\cX} \qw & \gate{T} & \push{\cY} \qw & \gate{T^{-}} & \push{\cX} \qw & \qw}, \scalemyQ{.8}{0.7}{0.5}{& \push{\cX} \qw & \gate{\id} & \push{\cX} \qw & \qw} ).
 \end{align}

As in the proof of \cref{thm:StructureOfReversibles}, let $\oneext{\cX}{\breve{T}}{\cY}{\breve{\cE}}$ be a universal  (and hence by \cref{prop:PureRechar} pure) dilation of $T$. By dilationality of $D$, we find for each $n \in \N$ a state $s'_n$ on $\breve{\cE}$ such that

\begin{align}
D\left( \scalemyQ{1}{0.7}{0.5}{ & \push{\cX}  \qw & \multigate{1}{\breve{T}} &  \push{\cY} \qw & \gate{T^-} &\push{\cX} \qw & \qw\\ 
	&  & \Nghost{\breve{T}}&\ww & \push{\breve{\cE}} \ww  & \ww} ,  \scalemyQ{1}{0.7}{0.5}{& \push{\cX} \qw & \gate{\id} & \push{\cX} \qw & \qw \\ & & \Ngate{s'_n} & \push{\breve{\cE}} \ww & \ww} \right) \leq  \varepsilon + \frac{1}{n} =: \varepsilon_n
\end{align}

(since $\id_\cX$ is dilationally pure). Now, trashing $\cX$ yields on the left hand side \myuline{some} channel $\scalemyQ{.8}{0.7}{0.5}{& \push{\cX} \qw & \gate{T'}& \push{\breve{\cE}} \ww & \ww}$ complementary to $T$, and on the right hand side the completely forgetful channel $\scalemyQ{.8}{0.7}{0.5}{& \push{\cX} \qw & \gate{\tr} & \Ngate{s'_n} & \push{\breve{\cE}} \ww & \ww }$ which by monotonicity is $\varepsilon_n$-close to $\scalemyQ{.8}{0.7}{0.5}{& \push{\cX} \qw & \gate{T'}& \push{\breve{\cE}} \ww & \ww}$ w.r.t. $D$. By \cref{thm:Complementarity}, the \myuline{specific} complementary channel $\scalemyQ{.8}{0.7}{0.5}{& \push{\cX} \qw & \gate{T^c} & \push{\cE} \ww & \ww}$ must equal $\scalemyQ{.8}{0.7}{0.5}{& \push{\cX} \qw & \gate{T'} & \push{\breve{\cE}} \ww & \Ngate{G}{\ww} & \push{\cE} \ww & \ww}$ for some channel $G$, and by monotonicity it is then $\varepsilon_n$-close w.r.t. $D$ to the completely forgetful channel $\scalemyQ{.8}{0.7}{0.5}{& \push{\cX} \qw & \gate{\tr} & \Ngate{s_n} & \push{\cE} \ww & \ww }$, where $\state{s_n}{\cE} = \scalemyQ{.8}{0.7}{0.5}{ & \Ngate{s'_n} & \push{\breve{\cE}} \ww & \Ngate{G}{\ww} & \push{\cE} \ww & \ww }$. In other words, 

\begin{align}
 D(\scalemyQ{.8}{0.7}{0.5}{& \push{\cX} \qw & \gate{T^c}& \push{\cE} \ww & \ww}, \scalemyQ{.8}{0.7}{0.5}{& \push{\cX} \qw & \gate{\tr} & \Ngate{s_n} & \push{\cE} \ww & \ww }) \leq \varepsilon_n
\end{align}

for each $n \in \N$. Since $\varepsilon_n \to \varepsilon=D(\scalemyQ{.8}{0.7}{0.5}{& \push{\cX} \qw & \gate{T} & \push{\cY} \qw & \gate{T^{-}} & \push{\cX} \qw & \qw}, \scalemyQ{.8}{0.7}{0.5}{& \push{\cX} \qw & \gate{\id} & \push{\cX} \qw & \qw} )$ for $n \to \infty$ and since $T^-$ was arbitrary, the inequality `$\geq$' holds in \cref{eq:EqInfoDist}. 

For the converse inequality, fix a state $\scalemyQ{.8}{0.7}{0.5}{& \Ngate{s} & \push{\cE} \ww & \ww}$ and define 

\begin{align}
\varepsilon := D(\scalemyQ{.8}{0.7}{0.5}{& \push{\cX} \qw & \gate{T^c}& \push{\cE} \ww & \ww}, \scalemyQ{.8}{0.7}{0.5}{& \push{\cX} \qw & \gate{\tr} & \Ngate{s} & \push{\cE} \ww & \ww }) .
\end{align}

Letting $\scalemyQ{.8}{0.7}{0.5}{& \Nmultigate{1}{\breve{s}} & \push{\cE} \ww & \ww \\ & \Nghost{\breve{s}} & \push{\cZ} \qw & \qw }$ be a pure dilation of $s$ (with hidden system $\cZ$), we find by dilationality of $D$ for each $n \in \N$ that 

\begin{align}
D\left(\scalemyQ{1}{0.7}{0.5}{ 
	&  & \Nmultigate{2}{L^c_n}& \push{\cE} \ww  & \ww \\ & \push{\cX}  \qw & \ghost{L^c_n} &  \push{\cZ} \qw & \qw \\  & & \Nghost{L^c_n} &  \push{\cX} \qw & \qw} , \scalemyQ{1}{0.7}{0.5}{& & \Nmultigate{1}{\breve{s}} & \push{\cE} \ww & \ww \\ & & \Nghost{\breve{s}} & \push{\cZ} \qw & \qw \\ & \push{\cX} \qw & \gate{\id} & \push{\cX} \qw & \qw  }\right) \leq  \varepsilon + \frac{1}{2n}
\end{align}

for some dilation $L^c_n$ of $T^c$ (with hidden system $\cZ \og \cX$). The channel on the right is dilationally pure, so picking pure dilations $P_n$ of $L^c_n$ (with hidden systems $\cZ_n$) there exist states $t_n$ on $\cZ_n$ such that 

\begin{align}
D\left(\scalemyQ{1}{0.7}{0.5}{ 
	&  & \Nmultigate{3}{P_n}& \push{\cE} \ww  & \ww \\ & \push{\cX}  \qw & \ghost{P_n} &  \push{\cZ} \qw & \qw \\  & & \Nghost{P_n} &  \push{\cX} \qw & \qw \\& & \Nghost{P_n} & \push{\cZ_n} \qw & \qw} , \scalemyQ{1}{0.7}{0.5}{& & \Nmultigate{1}{\breve{s}} & \push{\cE} \ww & \ww \\ & & \Nghost{\breve{s}} & \push{\cZ} \qw & \qw \\ & \push{\cX} \qw & \gate{\id} & \push{\cX} \qw & \qw   \\ & & \Ngate{t_n} & \push{\cZ_n}\qw & \qw}\right) \leq  \varepsilon + \frac{1}{n} = \varepsilon_n,
\end{align}

by dilationality of $D$. Trashing $\cE$ now yields on the left hand side a channel $\channel{\cX}{T^{cc}_n}{\cY_n}$ (with $\cY_n = \cZ \og \cX \og \cZ_n$) complementary to $T^c$ and on the right hand side a channel $\channel{\cX}{R_n}{\cY_n}$ which is evidently reversible, and such that $D(T^{cc}_n, R_n) \leq \varepsilon_n$. Choosing $R^{-}_n$ with $\scalemyQ{.8}{0.7}{0.5}{& \push{\cX} \qw & \gate{R_n} & \gate{R^{-}_n} & \push{\cX} \qw & \qw}= \channel{\cX}{\id}{\cX}$, monotonicity of $D$ yields

\begin{align}
D\left(\scalemyQ{.8}{0.7}{0.5}{& \push{\cX} \qw & \gate{T^{cc}_n } & \gate{R^{-}_n} & \push{\cX} \qw & \qw},  \channel{\cX}{\id}{\cX} \right) \leq \varepsilon_n, 
\end{align}

and since by \cref{thm:Complementarity} $\channel{\cX}{T^{cc}_n}{\cY_n}$ is of the form $\scalemyQ{.8}{0.7}{0.5}{& \push{\cX} \qw & \gate{T} & \push{\cY} \qw & \gate{G_n} & \push{\cY_n} \qw & \qw}$ for some channel $G_n$, the above implies that

\begin{align}
D(\scalemyQ{.8}{0.7}{0.5}{& \push{\cX} \qw & \gate{T} & \push{\cY} \qw & \gate{T^{-}_n} & \push{\cX} \qw & \qw}, \scalemyQ{.8}{0.7}{0.5}{& \push{\cX} \qw & \gate{\id} & \push{\cX} \qw & \qw} ) \leq \varepsilon_n
\end{align}

with $\channel{\cY}{T^-_n}{\cX} = \scalemyQ{.8}{0.7}{0.5}{& \push{\cY} \qw & \gate{G_n} & \gate{R^-_n} & \push{\cX} \qw & \qw}$. As $\varepsilon_n \to \varepsilon=   D(\scalemyQ{.8}{0.7}{0.5}{& \push{\cX} \qw & \gate{T^c}& \push{\cE} \ww & \ww}, \scalemyQ{.8}{0.7}{0.5}{& \push{\cX} \qw & \gate{\tr} & \Ngate{s} & \push{\cE} \ww & \ww })$
for $n \to \infty$ and as $s$ was arbitrary, the inequality `$\leq$' in \cref{eq:EqInfoDist} follows.

This altogether finishes the proof.

\end{proof}

The significance of \cref{thm:AppInfoDist} is that, when measured using an appropriate notion of distance -- namely a monotone, dilational metric -- the duality of \cref{thm:InfoDist} tightens under approximations. By an interpretation similar to the one usually employed in the case of $\QIT$, we might say that the degree to which information is preserved  by a channel $T$ (i.e. the degree to which it can be reversed) equals the degree to which it leaks no information to the environment.

\section{The Purified Diamond-Distance $P_\diamond$}
\label{sec:BuresP}

In this section, we cast our attention on the \emph{purified diamond-distance} which we defined above as the metric $P_\diamond := \breve{d}_\diamond$ in $\QIT$ (\cref{def:PurifDiam}). 

First, we will observe (owing fully to the result of Refs. \cite{KSW08math,KSW08phys}) that $P_\diamond$ can be bounded non-trivially in terms of $d_\diamond$ (\cref{thm:KSW}). Then, we will compare more systematically the metric $P_\diamond$ to the \emph{Bures distance} $\beta$ used in Refs. \cite{KSW08math,KSW08phys}, in particular establishing a quantitative relationship between the two (\cref{thm:BP}).\\

Recall that the purified diamond-distance between quantum channels $\Lambda_1, \Lambda_2$ was defined as

	\begin{align}  \label{eq:purediamond}
	P_\diamond(\Lambda_1, \Lambda_2) = \inf_{\Sigma_1, \Sigma_2} d_\diamond (\Sigma_1, \Sigma_2),
	\end{align}	
	
where the infimum is over all compatible Stinespring dilations $\Sigma_1$ of $\Lambda_1$ and $\Sigma_2$ of $\Lambda_2$.\footnote{I will use a notation with number indices ($\Sigma_1, \Sigma_2$) rather than with tildes ($\Sigma, \tilde{\Sigma}$) in this section for better legibility.} \\

On the other hand, the so-called \emph{Bures-distance} (\cite{KSW08math, KSW08phys}, generalising \cite{Bures69}) between $\Lambda_1$ and $\Lambda_2$, is given by $\beta(\Lambda_1, \Lambda_2) = \inf_{S_1, S_2} \norm{S_1-S_2}_\infty$, where the infimum is again over compatible Stinespring dilations of $\Lambda_1, \Lambda_2$, but this time in terms of isometries $S_j$ that represent the isometric dilations $\Sigma_j$ (i.e. $\Sigma_j(A)=S_j A S^*_j$). The isometry representing a given isometric channel is unique up to a phase, and therefore we can equivalently define $\beta$ as 

\begin{align} \label{eq:bures}
\beta(\Lambda_1, \Lambda_2) = \inf_{\Sigma_1, \Sigma_2} d_\infty(\Sigma_1, \Sigma_2), 
\end{align}

with $d_\infty$ given by

\begin{align}
d_\infty(\Sigma_1, \Sigma_2) := \inf_{\lambda \in \bbT} \norm{S_1-\lambda S_2}_\infty
\end{align}

 for isometric channels $\Sigma_j(\cdot)=S_j (\cdot) S^*_j$, with $\bbT$ denoting the unit circle in $\C$. The quantity $d_\infty(\Sigma_1, \Sigma_2)$ is independent of the choice of representatives $S_1$ and $S_2$, and the definition \eqref{eq:bures} is more easily compared to \eqref{eq:purediamond}.\\

 \begin{Remark} (Qualitative Comparison of $P_\diamond$ and $\beta$.) \label{rem:BPquali}\\
 Already before we quantitatively compare $P_\diamond$ and $\beta$, a more qualitative comparison is possible and appropriate: 
 
 It can be checked (by more or less the same arguments as used in the proof of \cref{thm:Purified}, but specialised to isometric channels in $\QIT$), that $\beta$ defines a monotone and dilational metric on $\QIT$, even one which has the Uhlmann Property. This set of properties (which $P_\diamond$ also has) therefore does not give a useful evaluation of the two against each other. For this reason, it also probably does not matter much in $\QIT$ which metric is used. However, the purified diamond-distance $P_\diamond$ is defined in terms of dilations and the operational metric $\delta_\diamond$, whereas the Bures distance $\beta$ lacks a similar operational definition. Indeed, its definition is derived from operator algebra, and therefore contingent on a particular formalism of quantum information theory. It should also be noted that $P_\diamond$ restricts in the case of states to the well-known purified distance (\cite{Toma10,Toma12}), whereas the Bures distance does not. It is finally worth remarking that, somewhat curiously, while $P_\diamond$ always agrees with $d_\diamond$ on isometric channels, $\beta$ and $d_\infty$ need not agree on isometric channels (cf. \cref{rem:Bdinf}).\end{Remark}

 \subsection{$P_\diamond$ versus $d_\diamond$}
 \label{subsec:PversD}

The weightiest result about the purified distance $P_\diamond$ is a rip-off from Refs. \cite{KSW08math,KSW08phys} (whose formulation is in terms of the Bures distance $\beta$). Whereas dilationality can act as a supply of `magic' in manipulations (as exemplified by \cref{thm:AppInfoDist}), this result will allow us to connect the magic of $P_\diamond$ to the more mundane world of $d_\diamond$:

 \begin{Thm} (Equivalence of $d_\diamond$ and $P_\diamond$.) \label{thm:KSW}\\
 	For any quantum channels $\Lambda_1, \Lambda_2$, we have the inequalities 
 	
 	\begin{align} \label{purified}
 	d_\diamond(\Lambda_1, \Lambda_2) \leq P_\diamond(\Lambda_1, \Lambda_2) \leq 	\sqrt{2 \, d_\diamond(\Lambda_1, \Lambda_2) }.
 	\end{align}
 	
 \end{Thm}
 
 \begin{Remark} 
 	When $\Lambda_1$ and $\Lambda_2$ are states, so that $P_\diamond$ is the purified distances for states, this result specialises to Lem. 6 in Ref. \cite{Toma10} (or Prop. 3.3 in Ref. \cite{Toma12}).
 	
 \end{Remark}
 
 \begin{proof}
 	The inequality $d_\diamond \leq P_\diamond$ follows from \cref{thm:Purified} (and its proof was easy). The hard part is the inequality $P_\diamond \leq \sqrt{2 \, d_\diamond}$. 
 	
 	The main result (Thm. 1) of Ref. \cite{KSW08math} states in our language that 

 	\begin{align}
 	\frac{\norm{\Lambda_1-\Lambda_2}_{\diamond}}{\sqrt{\norm{\Lambda_1}_{\diamond}}+ \sqrt{\norm{\Lambda_2}_{\diamond}}} \leq \beta(\Lambda_1, \Lambda_2) \leq \sqrt{\norm{\Lambda_1-\Lambda_2}_{\diamond}},
 	\end{align}
 	
 where $\norm{\cdot}_\diamond$ is the diamond norm and $\beta$ the Bures distance.	Now, any quantum channel has diamond-norm $1$, and the diamond-norm is related to our diamond-\myuline{distance} by a factor of $2$, so the above is the statement that 
 	
 	\begin{align} \label{bures}
 	d_\diamond(\Lambda_1, \Lambda_2) \leq \beta(\Lambda_1, \Lambda_2) \leq \sqrt{2 \, d_\diamond(\Lambda_1,\Lambda_2)}.
 	\end{align}

 To finish, it therefore suffices to prove $P_\diamond(\Lambda_1, \Lambda_2) \leq \beta(\Lambda_1, \Lambda_2)$, and by Eqs. \eqref{eq:purediamond} and \eqref{eq:bures} this will follow if $d_\diamond(\Sigma_1, \Sigma_2) \leq d_\infty(\Sigma_1, \Sigma_2)$ for any isometric channels $\Sigma_1$ and $\Sigma_2$. In particular, it suffices to show that for any isometries $S_1, S_2: \cH \to \cK$, we have $d_\diamond(\Sigma_1, \Sigma_2) \leq \norm{S_1-S_2}_\infty$, where $\Sigma_j $ denotes the channel $A \mapsto S_j A S^*_j$. 
 
 But this follows from elementary calculations: For any state $\varrho \in \scrD(\cH \otimes \cH)$, and any linear operators $A, B : \cH \otimes \cH \to \cL$, we have 
 	
 	\begin{align}
 	\begin{split}
 	\norm{A \varrho A^* - B \varrho B^*}_1 & \leq \norm{(A -B)\varrho A^*}_1 + \norm{ B \varrho (A^*-B^*)}_1  \\ & \leq  \norm{A -B}_\infty \norm{\varrho}_1 \norm{A^*}_\infty + \norm{ B}_\infty \norm{ \varrho}_1 \norm{ A^*-B^*}_\infty \\
 	&=  \norm{A -B}_\infty ( \norm{A}_\infty + \norm{ B}_\infty  ),
 	\end{split}
 	\end{align}
 	
 	so in particular, for isometries $S_1, S_2$,
 	
 	\begin{align}
 	\begin{split}
 	2 \, d_1((\Sigma_1\otimes \id_\cH)( \varrho),(\Sigma_2\otimes \id_\cH)( \varrho) ) 
 	&= \norm{(S_1\otimes \bone_\cH)\varrho(S^*_1\otimes \bone_\cH)-(S_2\otimes \bone_\cH)\varrho(S^*_2\otimes \bone_\cH)}_1  \\ 
 	& \leq  \norm{S_1 \otimes \bone_\cH -S_2 \otimes \bone_\cH}_\infty ( \norm{S_1 \otimes \bone_\cH}_\infty + \norm{ S_2\otimes \bone_\cH}_\infty ) \\
 	&= \norm{S_1  -S_2}_\infty ( 1+ 1),
 	\end{split}
 	\end{align}
 	
 	from which it follows that 
 	
 	\begin{align}
 	d_\diamond(\Sigma_1, \Sigma_2) = \sup_{\varrho \in \scrS(\cH \otimes \cH)} d_1((\Sigma_1\otimes \id_\cH)( \varrho),(\Sigma_2\otimes \id_\cH)( \varrho) ) \leq \norm{S_1  -S_2}_\infty,
 	\end{align}
 	
 	as desired. 
 \end{proof}

\cref{thm:KSW} implies that we may (if we so desire) reformulate the result of \cref{thm:AppInfoDist}, which applies to the dilational metric $D=P_\diamond$, as a result about \myuline{diamond}-distances, though in doing so we of course lose a square root (the resulting bound is one of the main results of Ref. \cite{KSW08phys}). 

\cref{thm:KSW} also implies more abstract features, such as the non-trivial fact that the topologies induced by $d_\diamond$ and $P_\diamond$ coincide. (In particular, since sets of channels are compact w.r.t. this topology, the infimum in the dilationality condition for $P_\diamond$ is always attained.)\\

In translating the proof of Refs. \cite{KSW08math,KSW08phys} from $\beta$ to $P_\diamond$, we demonstrated the inequality $P_\diamond \leq \beta$. In the next and final subsection, we will examine the relationship between these two metrics more carefully.

\subsection{$P_\diamond$ versus $\beta$}
\label{subsec:PversB}

The complications of calculating $P_\diamond$ and $\beta$ is two-fold; both of determining an infimum over Stinespring dilations, and in turn of determining the $d_\diamond$-distance, respectively $d_\infty$-distance, itself between Stinespring dilations. For the latter distances, we have the following formulas:

\begin{Prop} (Calculating $d_\diamond$- and $d_\infty$-distances between Isometric Channels.) \label{prop:fidelity}\\
		Let $\Sigma_1, \Sigma_2: \cH \to \cK$ be isometric quantum channels. Define the \emph{fidelity}, respectively \emph{fake fidelity}, between $\Sigma_1$ and $\Sigma_2$ as
	
	\begin{align} \label{eq:Fid}
	F(\Sigma_1, \Sigma_2) &= \inf_{\varrho \in \scrD(\cH)} \abs{\tr(S^*_1  S_2 \varrho)}, \; \text{respectively}\\ \label{eq:FakeFid}
	\textit{FF}(\Sigma_1, \Sigma_2) &= \sup_{\lambda \in \bbT} \, \inf_{\varrho \in \scrD(\cH)} \Re [\lambda  \tr(S^*_1  S_2 \varrho)],
	\end{align}
	
	where $S_1, S_2$ represent $\Sigma_1, \Sigma_2$ in the sense that $\Sigma_j = S_j(\cdot) S^*_j$. It then holds that 
	
	\begin{align} \label{eq:FormFid}
	d_\diamond(\Sigma_1, \Sigma_2) &= \sqrt{1-F(\Sigma_1, \Sigma_2)^2}, \; \text{and} \\ \label{eq:FormFakeFid}
d_\infty(\Sigma_1, \Sigma_2) &= \sqrt{2- 2 \, \textit{FF}(\Sigma_1, \Sigma_2)}. 
	\end{align}

\end{Prop} 

\begin{Remark} (Relation to Fidelity between States.)\\
When the domain $\cH$ is trivial, $\Sigma_1$ and $\Sigma_2$ are pure states on $\cK$, say $\psi_1$ and $\psi_2$, so the infima in Eqs. \eqref{eq:Fid} and \eqref{eq:FakeFid} are over a one-element set, and both fidelities reduce to the ordinary fidelity for states, $\abs{\braket{\psi_1}{\psi_2}}$. This means in particular that we have an equational relationship for states $\varrho_1, \varrho_2$ between $\beta$ and $P_\diamond$, given by 
	
	\begin{align} \label{statecorrespondence}
	\sqrt{1-P_\diamond(\varrho_1, \varrho_2)^2} = 1 - \frac{\beta(\varrho_1, \varrho_2)^2}{2},  \quad \text{or} \quad  P_\diamond(\varrho_1, \varrho_2) = \beta(\varrho_1, \varrho_2) \sqrt{1- \frac{\beta(\varrho_1, \varrho_2)^2}{4}}.
	\end{align}
	
	It is not clear that this relationship should hold for general channels, since the two fidelities might differ, but we shall  demonstrate that this is in fact the case (\cref{thm:BP}).
\end{Remark}

\begin{proof}
	
	As for $d_\diamond$, we have
	
	\begin{align}
	d_\diamond(\Sigma_1, \Sigma_2)  = \sup_{\text{$\psi \in \scrD(\cH \otimes \cH)$ pure}} d_1((\Sigma_1 \otimes \id_\cH)(\psi), (\Sigma_2 \otimes \id_\cH)(\psi)),
	\end{align}

	since by convexity the supremum over all states is attained already on the set of pure states. Using the well-known formula (see e.g. p. 415 in \cite{NC02}) $d_1(\phi_1, \phi_2) = \sqrt{1-\abs{\braket{\phi_1}{\phi_2}}^2}$ for pure states $\phi_1$ and $\phi_2$, we find
	
	\begin{align}
	\begin{split}
	d_1((\Sigma_1 \otimes \id_\cH)(\psi), (\Sigma_2 \otimes \id_\cH)(\psi)) &= \sqrt{1-\abs{\bra{\psi}(S^*_1 S_2 \otimes \bone_\cH )\ket{\psi}}^2} \\ 
	&= \sqrt{1-\abs{ \tr[(S^*_1 S_2 \otimes \bone_\cH )\psi]}^2} \\
	& =\sqrt{1-\abs{ \tr[S^*_1 S_2  \pi_1 (\psi)]}^2},
	\end{split}
	\end{align}
	
	where $\pi_1(\psi) := [\id_\cH \otimes \tr](\psi)$ signifies the first marginal of $\psi$. As $\psi$ ranges over all pure states on $\cH \otimes \cH$, the quantity $\pi_1(\psi)$ ranges precisely over all states $\varrho$ on $\cH$. Therefore, 
	
	\begin{align}
	d_\diamond(\Sigma_1, \Sigma_2) = \sup_{\varrho \in \scrD(\cH)} \sqrt{1- \abs{\tr(S^*_1 S_2 \varrho)}^2},
	\end{align}
	
	from which the formula \eqref{eq:FormFid} follows. 
	
	As for $d_\infty$, we find (using the formula $\norm{\ket{\phi_1}-\ket{\phi_2}}^2= \norm{\ket{\phi_1}}^2+\norm{\ket{\phi_2}}^2 - 2 \Re (\braket{\phi_1}{\phi_2})$), that

	\begin{align}
	\begin{split}
	\norm{S_1- \lambda S_2}^2_\infty &= \sup_{\ket{\psi} \in \cH, \norm{\ket{\psi}}=1} \norm{S_1 \ket{\psi}- \lambda S_2 \ket{\psi}}^2 = \sup_{\ket{\psi} \in \cH, \norm{\ket{\psi}}=1} \big( 1+1 - 2 \Re(\lambda \bra{\psi} S^*_1 S_2 \ket{\psi}) \big) \\
	&=\sup_{\text{$\psi \in \scrD(\cH)$ pure}} \big( 2 - 2 \Re(\lambda \tr( S^*_1 S_2 \psi) \big)  = \sup_{\varrho \in \scrD(\cH)} \big( 2 - 2 \Re[\lambda \tr( S^*_1 S_2 \varrho) ] \big) ,
	\end{split}
	\end{align}
	
	where the last equality is due to convex-linearity of the map $\varrho \mapsto \Re[\lambda \tr( S^*_1 S_2 \varrho)]$. Consequently,  
	
	\begin{align}
	d_\infty(\Sigma_1, \Sigma_2) = \inf_{\lambda \in \bbT} \norm{S_1- \lambda S_2}_\infty = \inf_{\lambda \in \bbT} \sup_{\varrho \in \scrD(\cH)} \sqrt{ 2 - 2 \Re[\lambda \tr(  S^*_1 S_2 \varrho)]},
	\end{align}
	
	from which the formula \eqref{eq:FormFakeFid} follows. This finishes the proof.
	
\end{proof}

From \cref{prop:fidelity} follow the formulas 

\begin{align}
	P_\diamond(\Lambda_1, \Lambda_2) &= \sqrt{1- \sup_{\Sigma_1, \Sigma_2}F(\Sigma_1, \Sigma_2)^2}, 
	\end{align}
	\begin{align}
\beta(\Lambda_1, \Lambda_2) &= \sqrt{2- 2\, \sup_{\Sigma_1, \Sigma_2} \textit{FF}(\Sigma_1, \Sigma_2)},
\end{align}

where the suprema range over pairs of compatible Stinespring dilations $\Sigma_1$ of ${\Lambda_1}$ and $\Sigma_2$ of $\Lambda_2$. It is not evident from those formulas how $P_\diamond$ and $\beta$ might be related, since we lack a relation between the fidelity and the fake fidelity, except for the obvious inequality $\textit{FF}(\Sigma_1, \Sigma_2) \leq F(\Sigma_1, \Sigma_2)$ (which, incidentally, yields a different argument for the inequality $d_\diamond(\Sigma_1, \Sigma_2) \leq d_\infty(\Sigma_1, \Sigma_2)$ used in the proof of \cref{thm:KSW}).

We will establish in a moment that $\sup_{\Sigma_1, \Sigma_2}F(\Sigma_1, \Sigma_2)= \sup_{\Sigma_1, \Sigma_2}FF(\Sigma_1, \Sigma_2)$, but it is worth observing that the identity $F(\Sigma_1, \Sigma_2)= FF(\Sigma_1, \Sigma_2)$ does \myuline{not} necessarily hold; in fact, the fake fidelity $FF(\Sigma_1, \Sigma_2)$ can even be negative (hence its name). To gain a feeling for this, and for the two fidelities in general, we will consider below an example in which $\Sigma_1$ and $\Sigma_2$ are unitary conjugations. \\

A key observation in analysing the fidelities is that for any linear operator $A: \cH \to \cH$, the set 

\begin{align} \label{eq:CA}
C(A) := \{\tr(A\varrho ) \mid \varrho \in \scrD(\cH)\},
\end{align}

known as the \emph{numerical range of $A$}, is a convex compact subset of $\C$ (contained in the disc of radius $\norm{A}_\infty$), since it is a continuous linear image of the convex compact set $\scrD(\cH)$. In particular, this holds for the linear operator $A= S^*_1S_2$. \\

The following (rather long) example exhibits a calculation of the fidelity and fake fidelity in the case where $\Sigma_1$ and $\Sigma_2$ are unitary. It can be skipped without losing coherence. \\

\begin{Example} (Fidelity and Fake Fidelity between Unitary Conjugations.) \label{ex:UnitaryDistance}\\
	Suppose that $\Sigma_1$ and $\Sigma_2$ are \myuline{unitary} conjugations. Let $S_1$ and $S_2$ be some choice of unitaries representing $\Sigma_1$ and $\Sigma_2$, respectively. Then, the operator $S^*_1 S_2$ is unitary, and thus admits a basis of eigenvectors. This means that for any unit vector $\ket{\psi} \in \cH$, the quantity $\tr(S^*_1 S_2 \ketbra{\psi})= \bra{\psi} S^*_1 S_2 \ket{\psi}$ is a convex combination of eigenvalues of $S^*_1 S_2$, and any particular eigenvalue is obtained by a suitable choice of $\ket{\psi}$. It follows that $C(S^*_1 S_2)$ as given by \cref{eq:CA} coincides with $\up{Conv}(\textup{spec}(S^*_1 S_2))$, the convex hull of the spectrum of $S^*_1 S_2$.  
	
\textbf{Calculating the Fidelity.} Now, the fidelity $F(\Sigma_1, \Sigma_2)$ is geometrically the distance from the origin $0 \in \C$ to the set $C(S^*_1 S_2)$. We can determine this distance by simple considerations. All of the eigenvalues of $S^*_1S_2$ will lie on the unit circle in $\C$; let $\gamma(\Sigma_1, \Sigma_2) \in [0, 2 \pi)$ denote the length of the smallest closed arc containing all of the eigenvalues. (Equivalently, $\gamma(\Sigma_1, \Sigma_2) = 2 \pi - \bar{\gamma}(\Sigma_1, \Sigma_2)$, where $\bar{\gamma}(\Sigma_1, \Sigma_2)$ is the length of the largest of the finitely many open arcs into which the unit circle is divided by the eigenvalues.) Observe that this quantity indeed depends only on the channels $\Sigma_1$ and $\Sigma_2$, and not on the chosen representatives $S_1$ and $S_2$, since another choice will merely have the effect of rotating the set of eigenvalues around the origin. It turns out that $F(\Sigma_1, \Sigma_2)$ is a function of $\gamma(\Sigma_1, \Sigma_2)$ alone:

	If $\gamma(\Sigma_1, \Sigma_2) \geq \pi$, no straight line can strictly separate $\up{spec}(S^*_1 S_2)$ from the point $0$, and by a standard result in convex analysis, we must therefore have $0 \in C(S^*_1 S_2) $; consequently, $F(\Sigma_1, \Sigma_2)=0$. 
	
	If $\gamma(\Sigma_1, \Sigma_2) < \pi$, all of the eigenvalues lie on one side of some straight line through $0$. Now, if $C(S^*_1 S_2)$ consists of a single point (i.e. if $\gamma(\Sigma_1, \Sigma_2)=0$), we clearly have $F(\Sigma_1, \Sigma_2)=1$. If $C(S^*_1 S_2)$ does not consist of sa single point, the infimal distance from $0$ to a point in $C(S^*_1 S_2) $ must be attained on one of those faces of the polytope $C(S^*_1 S_2)$ which is a straight line segment between two distinct eigenvalues. In fact, by simple trigonometry, it must be attained precisely in the midpoint of such a line segment. It is intuitively obvious that the two eigenvalues, for which this midpoint is closest to $0$, are the ones which have the entire arc of length $\gamma(\Sigma_1, \Sigma_2)$ between them, say $\lambda_a$ and $\lambda_b$. We can also see this formally: By definition of $\gamma(\Sigma_1, \Sigma_2)$, any two distinct eigenvalues are separated by an arc of length $\alpha \leq \gamma(\Sigma_1, \Sigma_2) (< \pi)$. For a pair of points on the circle separated by an arc of length $\alpha \in [0, \pi]$, the distance from $0$ to the midpoint of the line segment between them is given by $\cos(\alpha/2)$. Since this expression decreases as $\alpha$ increases, it follows that indeed the desired infimal distance is $\cos(\gamma(\Sigma_1, \Sigma_2)/2)$.
	
	Summarising these findings, the fidelity is given by 
	
	\begin{align} \label{unitaryfidelity}
	F(\Sigma_1, \Sigma_2) = \begin{cases}  
	\cos\left( \frac{\gamma(\Sigma_1, \Sigma_2)}{2}\right)  & \text{if $\gamma(\Sigma_1, \Sigma_2) \in [0, \pi)$} \\  0  & \text{if $\gamma(\Sigma_1, \Sigma_2) \in [\pi,  2\pi)$} \end{cases} = \cos( \gamma(\Sigma_1, \Sigma_2)/2)^+,
	\end{align}
	
	where, as usual, $u^+ := \max\{0,u\}$ for $u \in \R$.

\textbf{Calculating the Fake Fidelity.}	As for the fake fidelity, the analysis is also rather easily carried out using elementary arguments. $\textit{FF}(\Sigma_1, \Sigma_2)$ is by definition the minimal real part of a point in the rotated set $\lambda C(S^*_1 S_2)$, for a choice of $\lambda$ which gives the largest possible such minimal real part. Again, this quantity turns out to depend only on $\gamma(\Sigma_1, \Sigma_2)$:
	
	If $\gamma(\Sigma_1, \Sigma_2) < \pi$, then, as observed before, all of $\up{spec}(S^*_1 S_2)$ will lie on one side of some straight line through $0$. It is intuitively clear, that the largest possible smallest real part of a rotation of $C(S^*_1 S_2)$ is in this case coincident with the distance from $C(S^*_1 S_2)$ to $0$, i.e. equals $F(\Sigma_1, \Sigma_2) (=\cos( \gamma(\Sigma_1, \Sigma_2)/2))$, but a formal argument is also rather simple: $\textit{FF}(\Sigma_1, \Sigma_2) \leq F(\Sigma_1, \Sigma_2)$ always, and $\textit{FF}(\Sigma_1, \Sigma_2) \geq F(\Sigma_1, \Sigma_2)$ follows by choosing $\lambda \in \bbT$ such that the straight line through $0$ and some $0$-nearest point $z_0 \in C(S^*_1 S_2)$ becomes horizontal; $z_0$ will then also be a point with minimal real part, since if $z' \in C(S^*_1 S_2)$ had strictly smaller real part, then some point on the line segment between $z_0$ and $z'$  (which by convexity belongs to $C(S^*_1 S_2)$) would be strictly closer to $0$ than $z_0$ is, contradicting its choice. 
	
	If, on the other hand, $\gamma(\Sigma_1, \Sigma_2)\geq \pi$, a new situation emerges. In this case, the rotation with the largest possible minimal real part is the one which positions the arc of length $\gamma(\Sigma_1, \Sigma_2)$ such that its mid-point is the point $1 \in \C$, or, equivalently, positions the eigenvalue-free arc of length $\bar{\gamma}(\Sigma_1, \Sigma_2)$ such that its mid-point is $-1 \in \C$ (thus with its endpoints having identical real parts). To see this, simply note that if some choice of $\lambda$ yielded a strictly larger minimal real part, say $r \in (-1,0]$, then the rotated set of eigenvalues, $\lambda \,  \up{spec}(S^*_1 S_2)$, would be contained in the half-plane to the right of the straight line $\Re(z)=r$, and this would imply that the eigenvalue-free arc to the left of this line had length strictly greater than $\bar{\gamma}(\Sigma_1, \Sigma_2)$, contradicting its definition. Having established this, we see by easy trigonometric considerations that the relevant minimal real part is $\cos\left(\frac{\gamma(\Sigma_1, \Sigma_2)}{2}\right)$, so that altogether we conclude that, globally for $\gamma(\Sigma_1, \Sigma_2) \in [0, 2 \pi)$, we have 
	
	\begin{align} \label{unitaryfakefidelity}
	\textit{FF}(\Sigma_1, \Sigma_2)  = \cos( \gamma(\Sigma_1, \Sigma_2)/2).
	\end{align}

\textbf{Comparison.} We observe the relationship $F(\Sigma_1, \Sigma_2) = \max\{0, FF(\Sigma_1, \Sigma_2)\}$, which implies that for $\gamma(\Sigma_1, \Sigma_2) \in [0, \pi)$ the two fidelities coincide. However, by choosing $\Sigma_1$ and $\Sigma_2$ appropriately, we can design the value of $\gamma(\Sigma_1, \Sigma_2)$ at will -- e.g. letting $S_1 = \bone_\cH$ and letting $S_2$ have eigenvalues which are spread out suitably along the unit circle. (Note, though, that since $S^*_1  S_2$ has at most $\dim \cH$ distinct eigenvalues, the arc length $\gamma(\Sigma_1, \Sigma_2)$ is at most $2 \pi - \frac{2\pi}{\dim \cH}$). In particular, by choosing $\gamma(\Sigma_1, \Sigma_2) > \pi$ the fidelity $F(\Sigma_1, \Sigma_2)$ will be constantly $0$, whereas the fake fidelity $FF(\Sigma_1, \Sigma_2)$ approaches $-1$ as $\gamma(\Sigma_1, \Sigma_2)$ approaches $2 \pi$.\end{Example}

The relationship $F(\Sigma_1, \Sigma_2) = \max\{0, FF(\Sigma_1, \Sigma_2)\}$ observed in \cref{ex:UnitaryDistance} for unitary channels turns out to be no coincidence. To see this, we use a minimax-theorem:

\begin{Lem} (Fake it Till You Make it. \label{lem:Fakeit})\\
For any isometric quantum channels $\Sigma_1, \Sigma_2: \cH \to \cK$, it holds that 

\begin{align} \label{eq:fakeit}
F(\Sigma_1, \Sigma_2) = \max \{0, FF(\Sigma_1, \Sigma_2)\}. 
\end{align}

In particular, $FF(\Sigma_1, \Sigma_2) = F(\Sigma_1, \Sigma_2)$ as soon as $FF(\Sigma_1, \Sigma_2)  \geq 0$.

\end{Lem}

\begin{proof}
Let $S_1$ and $S_2$ be isometries representing $\Sigma_1$ and $\Sigma_2$, respectively. As already observed, the set 

\begin{align}
C(S^*_1S_2) := \{\tr(S^*_1S_2\varrho) \mid \varrho \in \scrD(\cH)\}
\end{align}

is a convex compact subset of $\C$. Now, we can write the two fidelities as 

\begin{align} \label{eq:FidRe}
F(\Sigma_1, \Sigma_2) &=  \inf_{z \in C(S^*_1S_2)} \abs{z}, \\ \label{eq:FakeFidRe}
FF(\Sigma_1, \Sigma_2) &=   \sup_{\lambda \in \bbT} \inf_{z \in C(S^*_1S_2)} \Re(\lambda z).
\end{align}

Clearly, $\abs{z} = \sup_{\alpha \in \bbD} \Re(\alpha z)$ for any $z \in \C$, where $\bbD := \{\alpha \in \C \mid \abs{\alpha} \leq 1\}$ denotes the closed unit disc in $\C$, so we really have 

\begin{align}
F(\Sigma_1, \Sigma_2) =  \inf_{z \in C(S^*_1S_2)} \sup_{\alpha\in \bbD} \Re( \alpha z). 
\end{align}

Now, the map $(z, \alpha) \mapsto \Re(\alpha z)$ is convex-linear in both of its arguments, and both of the sets $C(S^*_1 S_2)$ and $\bbD$ are convex and compact. Therefore, the order of optimisation can be interchanged by (e.g.) von Neumann's minimax-theorem, so in fact

\begin{align}
F(\Sigma_1, \Sigma_2) = \sup_{\alpha \in \bbD} \inf_{z \in C(S^*_1S_2)}  \Re(\alpha z) = \sup_{r \in [0,1]} \sup_{\lambda \in \bbT} \inf_{z \in C(S^*_1S_2)}  \Re(r \lambda z) = \sup_{r \in [0,1]} r \cdot FF(\Sigma_1, \Sigma_2).
\end{align}

The formula \eqref{eq:fakeit} now follows, since if $FF(\Sigma_1, \Sigma_2) \leq  0$ the last supremum above equals $0$, and if $FF(\Sigma_1, \Sigma_2) >0$ it equals $FF(\Sigma_1, \Sigma_2)$.

	\end{proof}

As demonstrated by \cref{ex:UnitaryDistance}, the condition $FF(\Sigma_1, \Sigma_2) \geq 0$ is not a void one. It turns out, however, that we nevertheless obtain a functional relationship between $P_\diamond$ and $\beta$, and with this result we conclude the section:

\begin{Thm}($P_\diamond$ versus $\beta$.) \label{thm:BP}\\
	For any quantum channels $\Lambda_1, \Lambda_2$, it holds that

\begin{align} \label{eq:supagree}
	\sup_{\Sigma_1, \Sigma_2}F(\Sigma_1, \Sigma_2)= \sup_{\Sigma_1, \Sigma_2}FF(\Sigma_1, \Sigma_2),
	\end{align}
	
	where the suprema range over pairs of compatible Stinespring dilations $\Sigma_1$ of ${\Lambda_1}$ and $\Sigma_2$ of $\Lambda_2$. In particular, we have the relationship 
	
	\begin{align} \label{eq:pb1}
	\sqrt{1-P_\diamond(\Lambda_1, \Lambda_2)^2} = 1 - \frac{\beta(\Lambda_1, \Lambda_2)^2}{2},
	\end{align}
	
	or 
	
		\begin{align} \label{eq:pb2}
P_\diamond(\Lambda_1, \Lambda_2) =\beta(\Lambda_1, \Lambda_2) \sqrt{1 - \frac{\beta(\Lambda_1, \Lambda_2)^2}{4}}.
	\end{align}

	\end{Thm}

\begin{Remark} (Ranges and Asymptotics.) \label{rem:RangeAsymp} \\
\cref{thm:BP} shows in particular that as $P_\diamond$ goes from $0$ to $1$, $\beta$ goes from $0$ to $\sqrt{2}$, and vice versa. Moreover, it shows that the ratio $\beta/ P_\diamond$ approaches $1$ when they are near $0$ (i.e. $\beta$ and $P_\diamond$ are asymptotically equal near $0$), whereas it approaches $\sqrt{2}$ when they are near their respective maximal values. The approximation $P_\diamond \approx \beta$ near $0$ is even correct to quadratic order ($P_\diamond = \beta - \beta^3/8 + O(\beta^5)$ in the limit $\beta \to 0$). 
\end{Remark}

\begin{Remark} ($\beta$ versus $d_\infty$.) \label{rem:Bdinf} \\
	We have previously observed that $P_\diamond$ and $d_\diamond$ agree on isometric channels. Note that this is \myuline{not} the case for $\beta$ and $d_\infty$: By \cref{ex:UnitaryDistance}, $FF(\Sigma_1, \Sigma_2)$ can range from $-1$ to $1$, so (by \cref{prop:fidelity}) $d_\infty(\Sigma_1, \Sigma_2)$ can range from $0$ to $2$. On the other hand, by \cref{thm:BP} $\beta(\Sigma_1, \Sigma_2)$ only ranges from $0$ to $\sqrt{2}$ as $P_\diamond(\Sigma_1, \Sigma_2)$ ranges from $0$ to $1$, ultimately due to the collapse of the \myuline{infimum} over fake fidelities to the infimum over (non-fake) fidelities.
	
	\end{Remark}

\begin{proof}
	
	The relations \eqref{eq:pb1} and \eqref{eq:pb2} follow from \cref{eq:supagree} by \cref{prop:fidelity}, so we need only show \cref{eq:supagree}. Since $F(\Sigma_1, \Sigma_2) \geq FF(\Sigma_1, \Sigma_2)$ always, the inequality `$\geq$' is clear so it suffices to argue that 	$\sup_{\Sigma_1, \Sigma_2}F(\Sigma_1, \Sigma_2) \leq  \sup_{\Sigma_1, \Sigma_2}FF(\Sigma_1, \Sigma_2)$.
	
	Suppose first that 	$\sup_{\Sigma_1, \Sigma_2}F(\Sigma_1, \Sigma_2)= 0$. It is sufficient to find $\Sigma'_1$, $\Sigma'_2$ such that $FF(\Sigma'_1, \Sigma'_2) = 0$. However, starting from \myuline{any} (compatible) Stinespring dilations $\Sigma_1, \Sigma_2$, the channels $\Sigma'_1 = \Sigma_1 \otimes \psi_1$ and $\Sigma'_2 = \Sigma_2 \otimes \psi_2$ are (compatible) Stinespring dilations for any choice of pure states $\psi_1$ and $\psi_2$ on the same system. In particular, if $\psi_1$ and $\psi_2$ are chosen perfectly distinguishable ($\braket{\psi_1}{\psi_2} = 0$), then 
	
	\begin{align}
	FF(\Sigma'_1, \Sigma'_2) = FF(\Sigma_1 \otimes \psi_1, \Sigma_2 \otimes \psi_2)  = \sup_{\lambda \in \bbT} \, \inf_{\varrho \in \scrD(\cH)} \Re [\lambda  \tr( \braket{\psi_1}{\psi_2} S^*_1  S_2 \varrho)] = 0,
	\end{align}  

as desired.\footnote{In this argument lies incidentally the reason why it is possible to have $\beta(\Sigma_1, \Sigma_2)< d_\infty(\Sigma_1, \Sigma_2)$ for isometric channels $\Sigma_1$ and $\Sigma_2$.}

Next, suppose that $\sup_{\Sigma_1, \Sigma_2}F(\Sigma_1, \Sigma_2)> 0$. We show that for any $\Sigma_1, \Sigma_2$ with $F(\Sigma_1, \Sigma_2) > 0$, we must have $FF(\Sigma_1, \Sigma_2) \geq 0$ and thus by \cref{lem:Fakeit} $FF(\Sigma_1, \Sigma_2) = F(\Sigma_1, \Sigma_2)$; this will imply that  $ \sup_{\Sigma_1, \Sigma_2}FF(\Sigma_1, \Sigma_2) \geq \sup_{\Sigma_1, \Sigma_2}F(\Sigma_1, \Sigma_2)$. To this end, observe that the assumption $F(\Sigma_1, \Sigma_2)> 0$ implies by \cref{eq:FidRe} that $0 \notin C(S^*_1 S_2)$. As the set $C(S^*_1 S_2) \subseteq \C$ is convex, it must therefore be entirely contained in the half-space defined by some straight line in $\C$ through $0$. This means, however, that for some $\lambda \in \bbT$, the rotated set $\lambda C(S^*_1 S_2)$ is contained in the specific half-space $\{z \in \C \mid \Re(z) \geq 0\}$, and consequently we must by \cref{eq:FakeFidRe} have $FF(\Sigma_1, \Sigma_2) \geq 0$, as claimed. \end{proof}

\section{Summary and Outlook}
\label{sec:SummaryMetric}

In this chapter, we have seen a general definition of topological and metric theories. We have discussed the properties of \emph{monotonicity} (\cref{def:MetTheory}) and \emph{dilationality} (\cref{def:Dilationality}) of a metric, and reformulated the latter in terms of a `Generalised Uhlmann Property' (\cref{prop:GenUhl}). This reformulation reveals in particular why monotone metrics in $\CIT$ (and cartesian theories) are bound to be automatically dilational, a phenomenon which might explain why the property of dilationality was only properly installed in metrics on $\QIT$ fairly recently (\cite{Toma10}). 

In \cref{sec:PureUhlmann}, we saw how the concept of purifiability can be used to overcome the challenges that a priori surround the construction of a (monotone) dilational metric from a monotone one, and this led to the idea of \emph{purified distances}. In particular, the \emph{purified diamond-distance}, $P_\diamond$, was introduced, a metric which has to the best of my knowledge not been considered before, but whose generalisation from the purified distance for states (\cite{Toma10,Toma12}) is fairly straightforward. In \cref{sec:BuresP}, we compared the metric $P_\diamond$ to the Bures distance $\beta$ of Refs. \cite{KSW08math,KSW08phys}, with the main conclusion that the two metrics share their most characteristic properties but differ in their relation to the formalism of quantum information theory. We also found an explicit quantitative relation between $P_\diamond$ and $\beta$ (\cref{thm:BP}), which entails that they are asymptotically equal when they are small, but can deviate for larger values.\\

 \cref{chap:Metric} leaves open the following problems:

\begin{enumerate}
	\item Can dilational metrics be constructed by a general scheme from monotone metrics $d$ on theories which are not necessarily purifiable? For example, does \cref{op:Dhat} have an affirmative answer?
	
	\item Does \cref{thm:KSW} generalise to give a non-trivial bound on $\breve{d}$ in terms of $d$ for general monotone metrics $d$ on purifiable theories? Can for example arguments of Refs. \cite{KSW08math,KSW08phys} (which are based on a minimax-theorem) be somehow abstracted to the general setting?
	\end{enumerate}

However, the most important sense in which \cref{chap:Metric} is open-ended is probably by virtue of circumstances which we shall only come to see in the last two chapters of the thesis: Namely that it is unclear how to appropriately adapt the metric theory to the setting of \emph{causal channels} and \emph{causal dilations} of \cref{chap:Causal}, and by extension that of quantum self-testing in \cref{chap:Selftesting} (these issues will be commented briefly in due time).

\chapter{Contractible Theories and Causal Dilations}
\label{chap:Causal}

{\centering
	\subsection*{§1. Introduction and Outline.}}

So far, we have considered channels $T: \bbX \to \bbY$ in a theory as more or less equivalent to their underlying transformations $T: \cX \to \cY$, and as such as processes which produce on a given input from the system $\cX$ an output in the system $\cY$. This perception, however, is too crude to adequately model a real physical device with input interface $\bbX$ and output interface $\bbY$. Indeed, in such a device it may be the case that some ports in $\bbY$ deliver an output already when a strict subset of the ports in $\bbX$ have been fed with an input. We shall encode this idea in a \emph{causal specification}, a map $\scrC$ which associates to each output port $\sfy \in \ports{\bbY}$ a set $\scrC(\sfy) \subseteq \ports{\bbX}$, thought of as the \emph{causes of $\sfy$}, namely the precise set of ports which require input before the output at $\sfy$ is available. A pair $(T, \scrC)$ consisting of a channel $T$ and a causal specification $\scrC$ will be called a \emph{causal channel}.

Causal channels force us to re-examine our understanding of dilations. If the ports in the accessible interfaces take part in an intricate causal relationship, then the ports in the \myuline{hidden} interfaces are involved in that relationship too. This leads to the notion of a \emph{causal dilation} of $(T, \scrC): \bbX \to \bbY$, which is simply a causal channel $(L, \scrE): \bbX \cup \bbD \to \bbY \cup \bbE$ whose $\bbY$-marginal factors  as $(T, \scrC) \og \tr_\bbD$ (in a sense which suitably accounts for causality). A causal dilation is meant to capture a `causally structured side-computation' in the presence of $(T, \scrC)$; if we understand the causal dilations of $(T, \scrC)$, we understand every way in which it may be immersed in its environment. \\

In a sense this chapter improves the causality-free theory of dilations introduced in \cref{chap:Dilations}. This is not to say that the theory of that chapter is rendered obsolete, merely that it is realised as the special case of a more general and stable framework. As often, greater generality does not necessarily entail greater versions of the same results, but rather entails \myuline{different} types of results, due to a shift in focus. In particular, having introduced the causal version of the dilational ordering (\cref{def:CausDilOrd}) it will quickly become clear that its higher complexity makes it much harder to confine by dilational axioms. For example, the existence of \emph{complete} (causal) dilations will cease to be a property of the theory in question, and instead becomes a property of the (causal) \myuline{channel} in question. In fact, the existence of a complete dilation will be coined as \emph{rigidity} of the channel (\cref{def:Rigidity}), and as we shall see in \cref{chap:Selftesting} this property is more or less the hallmark of quantum self-testing. \\

\textbf{Causal Channels and Constructibility.} We start in \cref{sec:CausChan} by defining the concept of a \emph{causal channel}, consisting of a channel $T$ together with a \emph{causal specification} $\scrC$. The most obvious way in which a causal specification emerges, is when the channel is represented by means of diagram like the following (each wire representing a simple interface):

\begin{align}   \label{eq:causex}
\scalemyQ{1}{0.7}{0.5}{
	&	&  &  & \Nmultigate{2}{ T_1} & \qw & \qw & \qw & \sfy_1\\
	& \sfx_1\quad 	&  \qw  & \qw & \ghost{ T_1} &  \multigate{2}{T_2} &  \qw & \qw & \sfy_2 \\
	&	\sfx_2 \quad& \qw  &  \multigate{1}{T_3} & \ghost{T_1} &  \Nghost{T_2} & \qw & \qw & \sfy_3\\
	&	& \Nmultigate{2}{T_4} &  \ghost{T_3} & \qw & \ghost{T_2} \\
	&	& \Nghost{T_4} & \qw & \qw & \qw & \qw & \qw & \sfy_4\\
	&	& \Nghost{T_2} & \qw & \multigate{1}{T_5}  \\
	& \sfx_3\quad	&   \qw  & \qw &\ghost{T_5} & \qw &   \qw & \qw & \sfy_5  \\
} \quad .
\end{align} 

The channel $T$ is the total channel resulting from the appropriate composition of $T_1, \ldots, T_5$, and the causal specification $\scrC$ is the one that associates to $\sfy_j$ the set of those $\sfx_i$ from which a directed path from left to right leads to $\sfy_j$. For example, $\scrC(\sfy_2)= \scrC(\sfy_3)=\{\sfx_1, \sfx_2\}$ and $\scrC(\sfy_4) = \emptyset$. By properties of the trashes, $\scrC$ encodes a collection of \emph{non-signalling conditions} to which $T$ complies; for any subset $\sfJ \subseteq \ports{\bbY}$, $T$ is necessarily non-signalling from the interface $\bbX \lvert_{\ports{\bbX} \setminus \scrC(\sfJ)}$ (the `non-causes' of $\sfJ$) to the interface $\bbY \lvert_\sfJ$. This observation tempts us to define, abstractly, a \emph{causal channel} as any pair $(T, \scrC): \bbX \to \bbY$ with $T: \bbX \to \bbY$ a channel and $\scrC$ an additive function from subsets of $\ports{\bbY}$ to subsets of $\ports{\bbX}$, such that $T$ adheres to the non-signalling conditions suggested by $\scrC$.\footnote{Importantly, $T$ might enjoy further non-signalling conditions than those implied by $\scrC$, and as such the specification $\scrC$ cannot be derived from knowledge of $T$ but rather formalises how causality is \myuline{perceived} (cf. \cref{ex:onetimepad}).} If a causal channel can be realised using a diagram as above, we will call it \emph{constructible}, and though the constructible causal channels will be the most important, not all imaginable causal channels are constructible (e.g. the Popescu-Rohrlich box from \cref{ex:PR} is inconstructible in $\QIT$ when equipped with its natural causal specification). Causal channels can be composed serially and parallelly, with causal specifications composing in the obvious way, and this leads to the \emph{category of interfaces and causal channels} (\cref{def:ICC}), which will constitute our final model for physical devices. \\

\textbf{Notions of Contraction.} Whereas there is nothing inconsistent about reducing the information of a network of channels like \eqref{eq:causex} to the information of the causal channel $(T, \scrC)$, it is utterly unobvious that this reduction is reasonable. Specifically, it is unclear that we do not thereby lose the possibility of forming \emph{contractions} of the network, that is, feeding an output to one of the inputs as long as this does not create any cycles in the network. For example, it would seem as if we should be able to feed the output at $\sfy_4$ to any of the input ports $\sfx_1, \sfx_2$ or $\sfx_3$, or that of $\sfy_5$ to $\sfx_2$ or $\sfx_3$ (provided that the corresponding systems match), but it is not trivial that the resulting causal channel can be obtained from knowledge of $(T, \scrC)$ alone. Indeed, not only does the skeleton of the network (which we will call a \emph{stencil}) not determine the channels $T_1, T_2, \ldots, T_5$ uniquely, it might also be that several different stencils can be used to represent $(T, \scrC)$. In \cref{sec:Contractions} we will see that one \myuline{can} define an unambiguous notion of contraction, provided that the underlying theory $\Theory$ has universal dilations (\cref{thm:StdCont}). This is one of the main results of this chapter, and I consider it quite remarkable. In the case of $\QIT$, the statement follows already by the work of Ref. \cite{Chir09combs}, but the proof given here is conceptually simpler and evidently has much larger scope, as it encompasses all universal theories (in particular $\CIT$ and all cartesian theories).  

The operation we obtain from the concrete contraction of wires will be called the  \emph{standard notion of contraction}. It turns out that, similarly to the way in which causal channels can be abstracted beyond constructible ones, the standard notion of contraction can be extended to abstract \emph{notions of contraction}, operations defined axiomatically on causal channels which swallow (`contract')  pairs of input and output ports and yield causal channels between the remaining interfaces. This is beneficial because, as detailed in the beginning of \cref{subsec:ContGeneral}, our construction of the standard notion of contraction will be sensitive to somewhat inessential particularities. The abstract notion is more stable, and also more widely applicable, e.g. in a thin theory a contraction is essentially a \myuline{cancellation} in the corresponding monoid. Finally, abstract notions of contraction will be realised to be a generalisation of \emph{traces} in symmetric monoidal categories as introduced by Ref. \cite{JSV96}.\footnote{Admittedly, I only became aware of this work a short time ago.} 

A notion of contraction, abstract or standard, is not merely additional to the operations of serial and parallel composition in a theory; it actually obviates the serial composition: If $\channel{\bbX}{(T, \scrC)}{\bbY}$ and $\channel{\bbY}{(S, \scrD)}{\bbZ}$ are causal channels, then their serial composition can be constructed from their parallel composition $\scalemyQ{.8}{0.7}{0.5}{& \push{\bbX} \qw  & \gate{(T,\scrC)} & \push{\bbY} \qw & \qw \\ & \push{\bbY} \qw & \gate{(S,\scrD)} & \push{\bbZ} \qw & \qw }$ simply by contracting the interface $\bbY$. This not only conceptually simplifies the complexity of operations in a theory, but is also technically simplifying in proofs. \\

\textbf{Climax: A Hierarchy of Causal Dilations.} With causal channels and contractions in place, we can in \cref{sec:CausDilations} elevate the theory of dilations to its full potential. \emph{Causal dilations} of $\channel{\bbX}{(T, \scrC)}{\bbY}$ are simply causal channels $ \scalemyQ{.8}{0.7}{0.5}{& \push{\bbX} \qw &  \Nmultigate{1}{(L, \scrE)}{\qw}   & \push{\bbY} \qw  & \qw  \\
	&\push{\bbD} \ww & \Nghost{(L, \scrE)}{\ww} & \push{\bbE} \ww  & \ww
}		$ such that $\scalemyQ{.8}{0.7}{0.5}{& \push{\bbX} \qw &  \Nmultigate{1}{(L, \scrE)}{\qw}   & \push{\bbY} \qw  & \qw  \\
	&\push{\bbD} \ww & \Nghost{(L, \scrE)}{\ww} & \push{\bbE} \ww  & \Ngate{\tr}{\ww}
}		
= 
\scalemyQ{.8}{0.7}{0.5}{& \push{\bbX} \qw &   \Ngate{(T, \scrC)}{\qw}    & \push{\bbY} \qw  & \qw  \\
	&\push{\bbD} \ww  & \Ngate{\tr}{\ww}
}$, a condition which is visually similar to that of dilations in \cref{chap:Dilations}, but is now a condition among \myuline{causal} channels (\cref{def:CausDil}). Again, the interpretation is that the interfaces $\bbX$ and $\bbY$ are \emph{open} or \emph{accessible} for us to probe, establishing that we interact with $(T, \scrC)$, and the dilation 
$(L, \scrE)$ signifies a way in which the functionality described by $(T, \scrC)$ is actually causally immersed in a larger environment, whose interfaces $\bbD$ and $\bbE$ are inaccessible to us. The \emph{causal-dilational order} which we will synonymously refer to as \emph{derivability in the environment} (\cref{def:CausDilOrd}) rectifies the dilational ordering of \cref{chap:Dilations}. It formalises the intuition that some dilations can be derived from others by operations in the environment; this time, however, we keep track of the causal order and we have at our disposal not only serial and parallel composition, but also contractions. This finally makes it possible to treat also two-sided dilations in a satisfactory manner. Precisely, we will render a dilation $ \scalemyQ{.8}{0.7}{0.5}{& \push{\bbX} \qw &  \Nmultigate{1}{(L, \scrE)}{\qw}   & \push{\bbY} \qw  & \qw  \\
	&\push{\bbD} \ww & \Nghost{(L, \scrE)}{\ww} & \push{\bbE} \ww  & \ww
}		$ \emph{equivalent} to the dilations $\scalemyQ{.8}{0.7}{0.5}{& \push{\bbX} \qw &  \Nmultigate{1}{(L, \scrE)}{\qw}   & \push{\bbY} \qw  & \qw  \\
&\push{\bbD} \ww & \Nghost{(L, \scrE)}{\ww} & \push{\bbE} \ww  &  \ww \\
& \push{\bbA} \ww & \Ngate{(G, \scrB)}{\ww} & \push{\bbB} \ww& \ww
} $, with $(G, \scrB)$ an arbitrary causal channel; a dilation $(L', \scrE')$ is then \emph{derivable} from $(L, \scrE)$, if a dilation equivalent to $(L, \scrE)$ is can be contracted along some ports to yield a dilation equivalent to $(L', \scrE')$. The intuition is that, in the environment, it is possible to construct a network of channels which when coupled suitably to the dilation $(L, \scrE)$ yields the dilation $(L', \scrE')$. 

The definition of this relation raises an interesting question: Is it necessarily the case that the contraction of ports in a causal dilation of $(T, \scrC)$ yields another causal dilation of $(T, \scrC)$? Though this is not formally implied by the definition, a result of this kind is desirable as it consolidates the concept of causal dilations. It is also operationally relevant, since, if false, the agents controlling the environment are not free to perform operations on the hidden interfaces without the risk that these become detectable at the open interfaces. That it should actually be the case is, however, far from obvious (and as demonstrated by \cref{ex:NearDil} it is almost false), and the most interesting result of  \cref{sec:CausDilations} is therefore the proof that causal dilations \myuline{are} actually stable under contractions in the environment (\cref{thm:Contractionsofdilations}).

 \cref{sec:CausDilations} also contains results to the effect that causal dilations are stable under parallel compositions and contraction of the \myuline{open} interfaces (in combination, these two results subsume serial compositions of open interfaces); those results (\cref{thm:ParStab} and \cref{thm:ContStab}) are of course also necessary for the concept of causal dilations to be well-behaved, but they are expectable and less subtle. \\

\textbf{Completeness, Rigidity and Density Theorems.} In stark contrast to the dilational ordering that arose in the causality-free setting of \cref{chap:Dilations}, axiomatic regularity of the causal-dilational ordering seems to escape into thin air. In \cref{sec:Density} we will scratch the surface, by mostly focusing on special cases. The goal is once again to understand the large dilations w.r.t. the causal-dilational ordering, and we do this by means of \emph{density theorems}. A class $\bfD$ of causal dilations of $(T, \scrC)$ is said to be \emph{dense} if every dilation of $(T, \scrC)$ is derivable from some dilation in $\bfD$. A density theorem asserts the density of a particular class of dilations. Obviously, the smaller the class, the better the theorem. 

The ultimate density theorem asserts the density of a class containing just a single dilation $(K, \scrF)$, that is, it asserts that the dilation $(K, \scrF)$ is a \emph{complete} causal dilation. In general, however, causal channels will not have complete dilations, and the various dilations in the dense class will reflect genuinely different ways of implementing $(T, \scrC)$ across the hidden interfaces. In fact, if $(T, \scrC)$ has a complete dilation we will call it \emph{rigid} (\cref{def:Rigidity}), and in \cref{chap:Selftesting} we will see the relation of this notion to quantum self-testing. 

In a cartesian theory, every causal channel turns out to be rigid (\cref{thm:CartRigid}), but already in the theory $\CIT$ rigidity fails in the simplest cases -- for example, we will finally see why the `bit refreshment' channel from the general introduction is not rigid. Indeed, density theorems for $\CIT$ (\cref{thm:UniBell} and \cref{thm:BiBell}) suggest that to study the causal-dilational ordering in $\CIT$ is essentially to study the various convex decompositions of channels, and in particular rigidity occurs if certain decompositions are unique (\cref{cor:UniRigidCIT} and \cref{cor:BiRigidCIT}). 

{\centering
	\subsection*{§2. Comparison to Existing Literature.}}

\textbf{On Causality.} The modelling of causality constitutes a vivid field of study in quantum foundations. Whereas much modern interest is directed towards finding a framework of `indefinite causal structure' in which causality  is malleable and even quantumly super-posable (\cite{Hardy07, Oresh12, Chir13, Bruk14}), we consider exclusively the setting of fixed causal structures, e.g. as represented by the network \eqref{eq:causex}. The main technical obstacle in treating such networks is that it is cumbersome -- boarding unfeasible -- to keep track of the individual channels of the network and their intricate connections. This problem was solved in the theory $\QIT$ by the framework of \emph{quantum combs} (\cite{Chir09combs}), which demonstrated that a network of quantum channels can be summarised by a linear operator bearing witness only to the open ports that enter and exit the network. This linear operator is obtained from the network by means of the Choi-Jamio\l{}kowski isomorphism, and connecting two networks can be achieved by a binary operation among these linear operators. 

The framework proposed here -- namely the approach of summarising the network by a pair $(T, \scrC)$ --  was developed also with the motivation of reducing complexity. In comparison to Ref. \cite{Chir09combs}, it distinguishes itself by the following traits: \\

\begin{enumerate}
	\item A network such as that in \cref{eq:causex} is summarised not by means of a linear operator which happens to exist by virtue of the Choi-Jamio\l{}kowski representation, but rather by its actual input-output behaviour as a channel, $T$, along with a specification, $\scrC$, of which inputs cause which outputs. This is conceptually and mathematically less awkward, and it broadens the scope to arbitrary theories.
	
	\item The connection of two or more networks is defined not by a binary operation, but rather is derived from parallel composition along with an operation on a single network, namely \emph{contraction}. The existences of these two operations are equivalent (since a contraction of a network can be viewed as coupling it to one consisting of identity channels), but for our purposes the approach in terms of contractions supports a cleaner presentation and method of proof.

	\item The well-definedness of the contraction operation  (or, equivalently, the composition of networks) is proved not using features of the Choi-Jamio\l{}kowski representation, but rather employing universal dilations (\cref{def:Univ}). As such, the approach immediately lends itself to all universal theories.  
	\end{enumerate}

There is also a difference in spirit between the approach here and that of Ref. \cite{Chir09combs}. Indeed, the main objective of the latter work is that of determining what is the class of transformations that a given network can undergo by combining it with other networks. This starts with the observation that quantum channels are the admissible transformation of \myuline{states}, and that networks with one `open slot' (a so-called \emph{$1$-comb}) are the admissible transformations of \myuline{channels}; the authors then proceed to define an entire hierarchy of \emph{$(n+1)$-combs} which act on \emph{$n$-combs}. In contrast, the approach taken here is rooted in the viewpoint that all these transformations can be realised in a flat, non-hierarchical structure, in which \emph{contraction} is a simple operation performed in a single network.

Ref. \cite{Kiss17} has also introduced a framework for arguing about causality in general categorical terms; it too is based on the hierarchical viewpoint of Ref. \cite{Chir09combs} (as distilled in Ref. \cite{Peri17}). Putting that difference aside, it compares to the framework introduced in this chapter by being more general in that it is able to treat indefinite causal structures as mentioned above, but less general in that  it requires (among other things) transformations in the category to be determined by their action on states, and the category itself to be compact closed. In fact, this assumption is used to establish essentially an equivalent of the Choi-Jamio\l{}kowski representation which then facilitates a treatment analogous to that of Refs. \cite{Chir09combs,Peri17}.  

Finally, all of the above comparisons pertain to the \emph{constructible causal channels} and the \emph{standard notion of contraction} only. As already mentioned, the framework introduced here accommodates causal channels of a more abstract character, and notions of contraction which are not necessarily physically realisable but more mathematical. It is my hope that this yields a more stable theory which might find applications elsewhere. \\

\textbf{On the Term `Causal Channel'.} Ref. \cite{Beck01} and off-spring work (e.g. Ref.\cite{Egg02}) define the term `causal channel' in a setting which can be seen as an extremely special case of ours: Namely, a bipartite-bipartite quantum channel	$\myQ{0.7}{0.5}{& \push{\cX_1} \qw & \multigate{1}{\Lambda} & \push{\cY_1} \qw & \qw \\ & \push{\cX_2} \qw & \ghost{\Lambda} & \push{\cY_2} \qw & \qw }$ is called \emph{causal} if it is non-signalling from $\sfx_1$ to $\sfy_2$ and from $\sfx_2$ to $\sfy_1$. In our language, this is simply to say that $\Lambda$ is compatible with the specification $\scrC$ given by $\scrC(\sfy_1)=\{\sfx_1\}$ and $\scrC(\sfy_2)=\{\sfx_2\}$.\\

 \vspace{.2cm}
 
 \textbf{Contractions and Traced Monoidal Categories.} Ref. \cite{JSV96} introduced the notion of a \emph{traced (symmetric) monoidal category}, namely a (symmetric) monoidal category equipped with an additional operation on channels generalising the trace in the symmetric monoidal category $(\Vect{k}, \otimes, k)$. More precisely, a trace in the sense of Ref. \cite{JSV96} maps channels $\scalemyQ{.8}{0.7}{0.7}{& \push{X} \qw & \multigate{1}{T}  & \push{Y} \qw & \qw \\ & \push{Z} \qw & \ghost{T} & \push{Z} \qw & \qw}$ to channels $\channel{X}{T'}{Y}$ by means of an operation which coheres to certain natural conditions. It turns out what is defined in this chapter as \emph{notions of contraction} (\cref{def:AbsCont}) can be seen a generalisation of such traces, as detailed in \cref{rem:Traces}. The generalisation consists in not requiring total domain for this operation, reflecting the situation that not necessarily any pair of wires is contractible just because they correspond to the same system. (For example, the output at $Z$ might require the input at $Z$ to be given first.) This generalisation may seem like an obvious one, but subtlety resides in the fact that those coherence conditions must now also specify the relationship between domains. (In a similar fashion, the generalisation from bounded linear operators to unbounded operators cannot be done mindlessly, but must consider what should be e.g. the domain of the operator $A+B$ given the domains of $A$ and $B$.) \\

{\centering
	\subsection*{§3. Contributions.}}

The original contributions of this chapter are the following:\\

\begin{enumerate}
	
	\item Defining the notion of a \emph{causal channel} (\cref{def:CausChan}), in particular providing a general and conceptually simple way of thinking about complicated networks of channels in a theory as 
	so-called \emph{constructible causal channels} (\cref{def:Constructible}).
	\item Establishing the operation of \emph{contractions} of causal channels (\cref{thm:StdCont}) which offer a technically simple alternative to other frameworks (\cite{Chir09combs, Kiss17}), and abstracting it to \emph{notions of contraction}, which generalise traces in a symmetric monoidal category as introduced in Ref. \cite{JSV96} (cf. \cref{rem:Traces}).
	\item Introducing the concept of \emph{causal dilations} (\cref{def:CausDil}) of a causal channel, which model the various possible environments in which the channel may be causally immersed, and defining a version of the dilational ordering for causal channel, \emph{derivability (in the environment)}, which models the relative strength of such dilations (\cref{def:CausDilOrd}). A number of stability results about causal dilations and derivability ordering is proved, most surprisingly the fact that the property of being a dilation is preserved under derivability (\cref{thm:Contractionsofdilations}).

\item Giving a general definition of \emph{rigidity} of a causal channel in terms of complete causal dilations (\cref{def:Rigidity}), and establishing \emph{density theorems} in some example theories (\cref{sec:Density}), in particular clarifying the meaning of rigidity in cartesian theories and in the classical information theory $\CIT$.

\end{enumerate}

{\centering
	\subsection*{§4. The Metric Aspect.}}

 In \cref{chap:Metric} was rolled out a general theory of metric structure in theories. As with dilations, it is only reasonable to revise this theory in light of the shift from channels to causal channels. I shall not take on such a revision here, but rather leave it for future work. It seems that there there are basically two challenges to overcome.  %

First of all, whereas it is sensible to leave the property of \emph{parallel invariance} of metrics unchanged in the causal setting, the \emph{serial monotonicity} should be upgraded to monotonicity under \myuline{contractions}; if we take seriously the idea that a contraction is a valid operation that can be performed, and if a metric is to abstractly quantify distinguishability between two causal channels, this is only logical. It is probably easy to define such a contractually monotone metric by forming a supremum over various  contractions, but it would seem difficult to calculate.\footnote{Note that e.g. the variational distance $d_1$ on $\CIT$ is not contractually monotone already.} A related idea has also been considered in Ref. \cite{Christ09}, and in Ref. \cite{Chir09combs} in the concept of \emph{comb distance} (herein, a formula is derived). 

Secondly, since the concept of dilations has changed, so should the concept of \myuline{dilationality} for a metric on causal channels. We do not have an equivalent of \cref{prop:GenUhl} to guide us, since we generally lack complete causal dilations, and it is not obvious how (or even in what sense) dilationality can be achieved.

\newpage

\section{Causal Channels and Constructibility}
\label{sec:CausChan}
In this section, we define \emph{causal channels} and consider some basic examples. An especially tangible and important class of causal channels is that of \emph{constructible causal channels}, which is meant to formalise the class of causal channels which can actually be `built' in the real world. 

\subsection{Causal Specifications and Causal Channels}

A causal channel $(T, \scrC): \bbX \to \bbY$ consists of two parts, a channel $T: \bbX \to \bbY$ and a causal specification $\scrC$. The channel represents the input-output behaviour without any regards to causality, and the specification specifies the causal relationships between the ports of $\bbX$ and $\bbY$.\\

\begin{Definition} (Causal Specifications.)\\
	Let $\bbX$ and $\bbY$ be interfaces in $\Theory$. A \emph{causal specification from $\bbX$ to $\bbY$} is a map $\scrC: \pow{\ports{\bbY}} \to \pow{\ports{\bbX}}$ such that
		
		\begin{itemize}
			\item  $\scrC(\emptyset) = \emptyset$ and 
			\item $\scrC(\sfJ_1 \cup \sfJ_2) = \scrC(\sfJ_1) \cup \scrC(\sfJ_2)$ for any $\sfJ_1, \sfJ_2 \subseteq \ports{\bbY}$.
			\end{itemize}
			 We write $\scrC: \bbX \to \bbY$ to denote that $\scrC$ is a causal specification from $\bbX$ to $\bbY$.
	\end{Definition}

\begin{Remark} (Specifying a Specification.) \\
	The \emph{additivity requirement} $\scrC(\sfJ_1 \cup \sfJ_2) = \scrC(\sfJ_1) \cup \scrC(\sfJ_2)$ implies that $\scrC$ is determined by its value on single ports, $\scrC(\{\sfy\})$. We tend to write $\scrC(\sfy)$ rather than $\scrC(\{\sfy\})$, and by additivity we thus have $\scrC(\sfJ) = \bigcup_{\sfy \in \sfJ} \scrC(\sfy)$ for any $\sfJ \subseteq \ports{\bbY}$. 

	\end{Remark}

Given an interface $\bbZ$, the subsets of $\ports{\bbZ}$ are in natural correspondence with the sub-interfaces of $\bbZ$. To ease notation, we might therefore sometimes think of a causal specification as being defined on sub-interfaces rather than subsets of port names, as in the following definition:

\begin{Definition} (Causal Channels.) \label{def:CausChan}\\
	Let $T: \bbX\to \bbY$ be a channel in $\Theory$, and $\scrC: \bbX \to \bbY$ a causal specification. $T$ and $\scrC$ are said to be \emph{compatible} if for any sub-interface $\bbY_0 \subseteq \bbY$, $T$ is non-signalling from $\bbX \setminus \scrC(\bbY_0)$ to $\bbY_0$, i.e. there exists a channel $T' : \scrC(\bbY_0) \to \bbY_0$ such that 
	
		\begin{align}
	\myQ{1}{0.7}{& \push{ \scrC(\bbY_0)} \qw &  \Nmultigate{1}{T}{\qw}   & \push{\bbY_0} \qw  & \qw  \\
		&\push{\bbX \setminus \scrC(\bbY_0)} \qw & \Nghost{T}{\qw} & \push{\bbY \setminus \bbY_0} \qw  & \Ngate{\tr}{\qw}
	}		
	\quad   = \quad 
	\myQ{1}{0.7}{& \push{\scrC(\bbY_0)} \qw &   \Ngate{T'}{\qw}    & \push{\bbY_0} \qw  & \qw  \\
		&\push{\bbX \setminus \scrC(\bbY_0)} \qw  & \Ngate{\tr}{\qw}
	}	\quad .
	\end{align}
	
	A \emph{causal channel from $\bbX$ to $\bbY$} is a pair $(T, \scrC)$ consisting of a channel $T: \bbX \to \bbY$ and a compatible causal specification $\scrC: \bbX\to \bbY$. 
\end{Definition}

To verify that a channel $T$ is compatible with a causal specification $\scrC$, one must in principle check $2^\abs{\bbY}$ non-signalling conditions, one for each sub-interface of $\bbY$ (though they are always trivial for the sub-interfaces $\bbI$ and $\bbY$). If the theory $\Theory$ has states on every system, it actually suffices to check the condition for simple sub-interfaces, reducing the number to $\abs{\bbY}$, but this can still be cumbersome. An important class of examples of causal channels is therefore those which are visually presented in such a way that the non-signalling conditions are obvious:

\begin{Example} (A Generic Causal Channel.) \label{ex:Generic} \\
		In any theory $\Theory$, the channel $T$ given by 
	
\begin{align}  
\scalemyQ{1}{0.7}{0.5}{
		&  &  & & &  \Nmultigate{2}{ T_1} & \qw & \qw & \push{\cY_1} \qw & \qw & \\
	 	& \push{\cX_1} \qw  & \qw & \qw& \qw &  \ghost{ T_1} & \push{\cZ} \qw &  \multigate{2}{T_2} &   \push{\cY_2} \qw & \qw  \\
	& \push{\cX_2}\qw & \qw  &  \multigate{1}{T_3} &  \push{\cW} \qw & \ghost{T_1} & &  \Nghost{T_2} & \push{\cY_3}\qw & \qw \\
		& \Nmultigate{2}{T_4} &  \push{\cU} \qw & \ghost{T_3} & \qw & \push{\cV} \qw & \qw & \ghost{T_2} \\
		& \Nghost{T_4} & \qw & \qw & \qw & \qw& \qw  & \qw &\push{\cY_4} \qw & \qw \\
		& \Nghost{T_2} & \push{\cS} \qw & \multigate{1}{T_5}  \\
	 	&    \push{\cX_3} \qw  & \qw &\ghost{T_5} & \qw & \qw & \qw& \qw &   \push{\cY_5}  \qw &  \qw   \\
} 
\end{align}  

for suitably composable channels $T_1, \ldots, T_5$ is compatible with the causal specification $\scrC$ that arises from ancestry in the network, when thinking of it as a directed (from left to right) graph. Precisely, $\scrC$ is given by $\scrC(\sfy_j)=\{\sfx_1, \sfx_2\}$ for $j=1,2,3$, $\scrC(\sfy_4)=\emptyset$ and $\scrC(\sfy_5) = \{\sfx_3\}$. The non-signalling conditions follow from properties of the trashes (\cref{lem:Trashes}), which in each case eliminate those input ports which are not ancestral to the non-trashed output ports. 
\end{Example}

 \begin{Example} (Bell-Channels.) \\
Recall from \cref{ex:Bellch} that a \emph{(bipartite) Bell-channel} is a channel $T$ of the form  

	\begin{align}  	
\myQ{0.7}{0.5}{
		& \push{\cX_\sfA}  \qw  & \multigate{1}{ T_\sfA} & \push{\cY_\sfA}  \qw & \qw  \\
		& \Nmultigate{1}{s}  & \ghost{T_\sfA} &  \\
		& \Nghost{s} & \multigate{1}{T_\sfB}  \\
&   \push{\cX_\sfB} \qw  & \ghost{T_\sfB} & \push{\cY_\sfB}  \qw & \qw  \\
} 
\quad.
\end{align}
	
We may augment it to a causal channel by defining the specification $\scrC$ again by ancestry in the network: $\scrC(\emptyset)=\emptyset$,  $\scrC(\sfy_\sfA)= \{\sfx_\sfA\}$, $\scrC(\sfy_\sfB)= \{\sfx_\sfB\}$ and $\scrC(\{\sfy_\sfA, \sfy_\sfB\}\}) = \{\sfx_\sfA, \sfx_\sfB\}$. 
\end{Example}

It is important to realise that a channel $T$ may enjoy non-signalling conditions which are \myuline{additional} to those implied by a given compatible specification $\scrC$. As such, the causal specification is an integral part of a causal channel $(T, \scrC)$ and cannot be derived from $T$ itself: 

\begin{Example} (Primitive Causal Channels.) \label{ex:PrimChan}\\
\myuline{Any} channel $T: \bbX \to \bbY$ is compatible with the \emph{primitive specification} $\scrC_0 : \bbX \to \bbY$, according to which $\scrC_0(\sfJ) =\ports{\bbX}$ for all non-empty subsets $\sfJ \subseteq \ports{\bbY}$. A causal channel $(T, \scrC_0)$ is called \emph{primitive} if it is equipped with the primitive specification $\scrC_0$. Intuitively, the primitive specification expresses that the all ports of $\bbX$ must be fed with an input before any outputs appear at $\bbY$. This causal specification reflects how we implicitly thought of channels in all preceding chapters. 
\end{Example}

 \begin{Example} (One-Time Pad.) \label{ex:onetimepad}\\
	Let $P$ be the channel in $\mathbf{CIT}$ given by 
	
	\begin{align}   	
	\myQ{0.7}{0.5}{
		& \sfm \hspace{1cm} & \qw & \push{\{0,1\}}  \qw  & \multigate{1}{ \up{XOR}} & \push{\{0,1\}}  \qw & \qw  & \sfc\\
		&	& \Nmultigate{1}{\kappa}  &  \push{\{0,1\}}  \qw   & \ghost{\up{XOR}} &  \\
		&	& \Nghost{\kappa} &  \push{\{0,1\}}  \qw  & \qw & \qw  & \qw & \sfk\\
	} 
	\quad,
	\end{align}
	
	where $\kappa$ is the state on $\{0,1\} \times \{0,1\}$ given by $\kappa= \frac{1}{2} \delta_0 \otimes \delta_0 + \frac{1}{2} \delta_1 \otimes \delta_1$ (i.e. $\kappa$ is two copies of a uniformly random bit), and where $\up{XOR}$ is the `exclusively OR', i.e. the deterministic function which adds the inputs modulo $2$. The channel $P$ represents the so-called \emph{one-time pad}, a cryptographic device used for encryption (see any introductory book on cryptography, e.g. \cite{Crypt}). In the one-time pad, a key bit is chosen at random and copied; one copy of the key bit is kept in memory (the $\sfk$-port) for later decryption, and the other copy is used to decide on whether or not to apply a bit flip on the message bit coming in at the $\sfm$-port, thus obtaining an encrypted cipher bit at the $\sfc$-port. Clearly, we want to equip $P$ with the causal specification that arises from ancestry: $\sfm$ is a cause of $\sfc$, and $\sfk$ has no causes. And indeed, this specification is guaranteed to be compatible with $P$: Trashing the $\sfc$-port results in a trash on the $\sfm$-port.
	
	\emph{However}, the entire point of the one-time pad is that trashing the \myuline{$\sfk$-port} also results in a trash of the $\sfm$-port, that is, to an individual who does not know the key bit, the cipher bit will look completely random, independently of the message bit that was to be encrypted. Thus, the channel $P$ is non-signalling from the input port to either of its output ports. In fact, the channel $P$ is even symmetric under permutation of the two output ports: One cannot tell which bit is the key bit, and which is the cipher bit. Asymmetry between the ports only arises in the presence of a chosen causal specification, and that specification is not derivable from the input-output behaviour of the channel $P$ itself. 

\end{Example}

What the above examples demonstrate is that causal specifications are not merely about non-signalling conditions, but rather about how causality is \emph{perceived}. 

Before proceeding, let us consider three more examples of causal channels.

 \begin{Example} (`Linked' Parallel Composition.) \label{ex:linkedproduct} \\
Let $\channel{\cX_1}{T_1}{\cY_1}$ and $\channel{\cX_2}{T_2}{\cY_2}$ be channels in $\Theory$, and consider their parallel composition 

\begin{align} \label{eq:unlinked}
\myQ{0.7}{0.5}{& \push{\cX_1} \qw & \multigate{1}{T} & \push{\cY_1} \qw & \qw \\ & \push{\cX_2} \qw & \ghost{T} & \push{\cY_2} \qw & \qw } \quad := \myQ{0.7}{0.5}{& \push{\cX_1} \qw & \gate{T_1} & \push{\cY_1} \qw & \qw \\ & \push{\cX_2} \qw & \gate{T_2} & \push{\cY_2} \qw & \qw } \quad.
\end{align}

Of course, we can equip $T$ with the causal $\scrC$ specification that arises from the ancestry on the right hand side ($\scrC(\sfy_1)= \{\sfx_1\}$ and $\scrC(\sfy_2)= \{\sfx_2\}$), but $T$ is also compatible with another causal specification, namely $\scrC'$ given by $\scrC'(\sfy_1)= \{\sfx_1\}$ and $\scrC'(\sfy_2)=\{\sfx_1, \sfx_2\}$. (In general, enlarging the cause sets of a compatible specification yields another compatible specification.) According to the specification $\scrC'$, the output at $\sfy_1$ occurs as soon as $\sfx_1$ has been fed with an input, but the output at $\sfy_2$ requires both $\sfx_1$ and $\sfx_2$ to be given an input before it shows any output. 

Though $(T, \scrC)$ is not the causal channel whose causality derives from ancestry in the network from \eqref{eq:unlinked}, it does arrive from the ancestry in a different network, namely

\begin{align} \label{eq:linked}
\myQ{0.7}{0.5}{
	& \push{\cX_1}  \qw    &\multigate{1}{T_1}   & \qw &   \push{\cY_1} \qw & \qw & \qw \\
	&                   & \Nghost{T_1}  & \push{\triv} \qw & \multigate{1}{T_2} & \\
	& \qw &  \push{\cX_2}  \qw & \qw &  \ghost{T_2}   &  \push{\cY_2} \qw  & \qw  \\
} 	\quad,
\end{align}

where $\triv$ is the trivial system. In effect, the trivial system acts to \emph{stall} the execution of $T_2$ until $T_1$ has been applied. These two alternative depictions would not have been clear without the pedantic discussion about interfaces and channels versus systems and transformations in \cref{subsec:Inter}. \end{Example}

 \begin{Example} (States and Trashes as Causal Channels.) \label{ex:statetrashcaus}\\
 	If a causal $(T, \scrC): \bbX \to \bbY$ has trivial output interface, $\bbY= \bbI$, the specification can only be the one given by $\scrC(\emptyset)=\emptyset$, and its channel $T$ must be $\tr_\bbX$. Similarly, when its input interface is trivial, $\bbX = \bbI$, its specification must be given by $\scrC(\sfJ)= \emptyset$ for all $\sfJ \subseteq \ports{\bbY}$ and its channel $T$ must be a state. In short, trashes and states have unique causal specifications when thought of as channels to, respectively from, the trivial interface. (Incidentally, these unique specifications are in both cases primitive.)
 	
 	However, we need not think of a trash as mapping to the trivial interface, nor of a state as mapping from it. For instance, we can write a trash $\tr_\cZ$ as 
 	
 	\begin{align}
 	\myQ{0.7}{0.5}{& \push{\cZ} \qw & \gate{\tr} & \push{\triv} \qw& \qw}
 	\end{align}
 	
 	and give it the (primitive) causal specification $\scrT$ according to which $\scrT(\sfone) = \sfz$. As such, the output port becomes an \emph{indicator} for whether or not something was trashed. Similarly, a state $s$ on $\cY$ can be represented by the channel 
 	
 		\begin{align}
 	\myQ{0.7}{0.5}{& \push{\triv} \qw & \gate{s} & \push{\cY} \qw& \qw},
 	\end{align}
 	
 	and when endowed with the (primitive) specification $\scrS$ given by $\scrS(\sfy)= \sfone$, the input port becomes an \emph{activator} for the state, releasing it at the output port. 
 	
 	Of course, indicators and activators can act in intricate ways, indicating or activating only a subset of ports. Also, they can be mounted to any sorts of channels, not just trashes and states (indeed the stalling phenomenon from \cref{ex:linkedproduct} can be seen as an example of this).

\end{Example}

\begin{Example} (The Popescu-Rohrlich Box as a Causal Channel.) \label{ex:PRCaus}\\
Recall the PR Box from \cref{ex:PR}, the classical channel $\scalemyQ{.8}{0.7}{0.5}{& \sfx_\sfA \quad & \push{\{0,1\}} \qw& \multigate{1}{P} & \push{\{0,1\}} \qw & \qw & \sfy_\sfA\\ & \sfx_\sfB \quad & \push{\{0,1\}} \qw& \ghost{P} & \push{\{0,1\}} \qw & \qw & \sfy_\sfB}$ determined by the probability distributions $(P^{x_\sfA, x_\sfB})_{x_\sfA, x_\sfB \in \{0,1\}}$ given by 
	
	\begin{align}
	P^{x_\sfA, x_\sfB} (y_\sfA, y_\sfB) = \begin{cases}\frac{1}{2} &\text{for $y_\sfA \oplus y_\sfB = x_\sfA \cdot x_\sfB$} \\ 
	0 &\text{for $y_\sfA \oplus y_\sfB \neq x_\sfA \cdot x_\sfB$}  \end{cases},
	\end{align}
	
	with $\oplus$ denoting addition modulo $2$. We saw there that this channel was non-signalling from $\sfx_\sfA$ to $\sfy_\sfB$ and from $\sfx_\sfB$ to $\sfy_\sfB$, and as such it is compatible with a \emph{local} causal specification $\scrC$, namely the one given by $\scrC(\sfy_i)=\{\sfx_i\}$ for $i= \sfA, \sfB$. Thus, we have a causal channel $(P, \scrC)$.
	
	\end{Example}

\subsection{Constructible Causal Channels}
\label{subsec:Constructible}

All of the above examples of causal channels, with the exception of the PR box in \cref{ex:PRCaus}, were presented by means of a network of channels which defined simultaneously the channel and its causal specification. We now discuss how to formalise this idea in the concept of a \emph{constructible} causal channel. \\ %

To exemplify, consider the depiction

\begin{align}  
\scalemyQ{1}{0.7}{0.5}{
	&  &  & & &  \Nmultigate{2}{ T_1} & \qw & \qw & \push{\cY_1} \qw & \qw & \\
	& \push{\cX_1} \qw  & \qw & \qw& \qw &  \ghost{ T_1} & \push{\cZ} \qw &  \multigate{2}{T_2} &   \push{\cY_2} \qw & \qw  \\
	& \push{\cX_2}\qw & \qw  &  \multigate{1}{T_3} &  \push{\cW} \qw & \ghost{T_1} & &  \Nghost{T_2} & \push{\cY_3}\qw & \qw \\
	& \Nmultigate{2}{T_4} &  \push{\cU} \qw & \ghost{T_3} & \qw & \push{\cV} \qw & \qw & \ghost{T_2} \\
	& \Nghost{T_4} & \qw & \qw & \qw & \qw& \qw  & \qw &\push{\cY_4} \qw & \qw \\
	& \Nghost{T_2} & \push{\cS} \qw & \multigate{1}{T_5}  \\
	&    \push{\cX_3} \qw  & \qw &\ghost{T_5} & \qw & \qw & \qw& \qw &   \push{\cY_5}  \qw &  \qw   \\
} 
\end{align}

from \cref{ex:Generic}. It is comprised of two pieces of information. One of them is purely graph-theoretic, namely the \emph{stencil} 

\begin{align} \label{eq:stencil}
\scalemyQ{1}{0.7}{0.5}{
	&	&  &  & \Nmultigate{2}{ \phantom{X}} & \qw & \qw & \qw & \bullet \\
	& \bullet  \quad 	&  \qw  & \qw & \ghost{ \phantom{X}} &  \multigate{2}{\phantom{X}} &  \qw & \qw & \bullet  \\
	&	  \bullet   \quad& \qw  &  \multigate{1}{\phantom{X}} & \ghost{\phantom{X}} &  \Nghost{\phantom{X}} & \qw & \qw &  \bullet\\
	&	& \Nmultigate{2}{\phantom{X}} &  \ghost{\phantom{X}} & \qw & \ghost{\phantom{X}} \\
	&	& \Nghost{\phantom{X}} & \qw & \qw & \qw & \qw & \qw &  \bullet \\
	&	& \Nghost{\phantom{X}} & \qw & \multigate{1}{\phantom{X}}  \\
	& 	  \bullet   \quad	&   \qw  & \qw &\ghost{\phantom{X}} & \qw &   \qw & \qw &  \bullet
}  \quad
\end{align}
 (directed from left to right). The other piece of information rests upon the theory $\Theory$, namely the \emph{filling}, which assigns to each wire a simple interface in $\Theory$ and to each box a suitably compatible channel between the adjacent interfaces.\\
 
 A general stencil can be described by a \emph{directed acyclic graph} (for short, \emph{DAG}), in which we distinguish two kinds of vertices; \emph{boxes}, which are vertices to be filled with a channel, and \emph{ports} (represented in \eqref{eq:stencil} by bullets):

 \begin{Definition} (Stencils.) \label{def:Stencil} \\
 	A \emph{stencil} is a triple $(G, W_\up{in}, W_\up{out})$, where\footnote{Recall that a \emph{source} in a directed graph is a vertex with no incoming edges, and that a \emph{sink} is a vertex with no outgoing edges. An \emph{isolated vertex} is a vertex which is both a source and a sink.}

 	\begin{itemize}
 		\item[-] $G$ is a finite, non-empty DAG with no isolated vertices;
 		\item[-] $W_\up{in}$ is a set of edges in $G$, each of which comes from a source in $G$ that has no other outgoing edges;  %
 		\item[-] $W_\up{out}$ is a set of edges in $G$, each of which go to a sink in $G$ which has no other incoming edges.
 	\end{itemize}
 	
 	An edge in $G$ is called a \emph{wire}. The edges in $W_\up{in}$ are called \emph{input wires} and the edges in $W_\up{out}$ are called \emph{output wires}. A vertex which is the source of a wire in $W_\up{in}$, or the sink of a wire in $W_\up{out}$, is called a \emph{port}, and every other vertex in $G$ is called a \emph{box}.
 	
 \end{Definition}

The reader is encouraged to consider how these  notions pan out in the example illustrated by \cref{eq:stencil}. \\

By abuse of notation, we use the symbol $G$ to represent the whole stencil $(G, W_\up{in}, W_\up{out})$, and we write $\cW(G)$, $\cB(G)$, $\cW_\up{in}(G)$ and $\cW_\up{out}(G)$ and  to denote respectively the sets of wires, boxes, input wires and output wires in the stencil.

Let us also denote, for a box $b \in \cB(G)$, by $\up{In}(b)$ and $\up{Out}(b)$ the set of wires which go to, respectively from, $b$. Fillings of a stencil are now straightforward to define:

\begin{Definition} (Stencil Fillings.)\\
Let $\Theory$ be a theory. A \emph{$\Theory$-filling of the stencil $G$} is a pair $\fF= ((\bbZ_w)_{w \in \cW(G)}, (T_b)_{b \in \cB(G)} )$, where
	
	\begin{enumerate}
		\item $(\bbZ_w)_{w \in \cW(G)}$ is a collection of simple interfaces in $\Theory$, indexed by the wires $w \in \cW(G)$; we require that $\bbZ_w$ and $\bbZ_{w'}$ can only be identical if a path in $G$ leads from $w$ to $w'$, or from $w'$ to $w$. 
		\item $(T_b)_{b \in \cB(G)}$ is a collection of channels in $\Theory$, indexed by the boxes $b \in \cB(G)$, such that $T_b$ is a channel from the interface $\bigcup_{w \in \up{In}(b)} \bbZ_w$ to the interface $\bigcup_{w \in \up{Out}(b)} \bbZ_w$. 
	\end{enumerate}

	\end{Definition}

\begin{Remark}
	The requirement about distinctness of the simple interfaces $\bbZ_w$ is so as to ensure that on the one hand identical interfaces are never composed in parallel (that was explicitly forbidden), while on the other hand an input wire and output wire may correspond to the same interface. 
	\end{Remark}

Now, it is intuitively clear that a stencil $G$ together with a filling $\fF$ defines in a natural way a total channel

\begin{align} 
\fF[G] : \bigcup_{w \in \cW_\up{in}(G)} \bbZ_w \to \bigcup_{w \in \cW_\up{out}(G)} \bbZ_w
\end{align}

 from the interface of input wires to the interface of output wires. We will call $\fF[G]$ the \emph{value of $\fF$ on $G$}. The formal groundwork needed for this construction, however, is tedious, and the reader is referred to the original work of Ref. \cite{JS91} for the details.\\  %

We can now give a more precise definition of what is meant by a constructible causal channel. For vertices $v'$ and $v$ in a directed graph $G$, it is customary to write $v' \to v$ if there is a path in $G$ from $v'$ to $v$. We shall write also $w' \to w$ for \myuline{wires} (edges) $w'$ and $w$, if there is a path in $G$ which includes the wire $w'$ before the wire $w$ (or if $w'=w$). As such, for any output wire $w \in \cW_\up{out}(G)$, the set

\begin{align} \label{eq:specDAG}
\scrA_G(w) := \{w' \in \cW_\up{in}(G) \mid w' \to w\}
\end{align}

is the set of input wires ancestral to $w$. 

\begin{Definition} (Stencil-Representability.) \\
	Let $G$ be a stencil. We say that a causal channel $(T, \scrC): \bbX \to \bbY$ in $\Theory$ is \emph{representable on $G$} if there exists a  $\Theory$-filling $\fF$ of $G$ such that $(T, \scrC) = (\fF[G], \scrC_G)$, where $\fF[G]$ is the value of $\fF$ on $G$, and where $\scrC_G$ is the causal specification given by ancestry in $G$, i.e. $\scrC_G(\bbZ_w) = \bigcup_{w' \in \scrA_G(w)} \bbZ_{w'}$ for $w \in \cW_\up{out}(G)$, with $\scrA_G$ as in \cref{eq:specDAG}. 
	\end{Definition}

\begin{Definition} (Constructible Causal Channels.) \label{def:Constructible}\\
A causal channel is called \emph{constructible} if it is representable on some stencil. 
\end{Definition}

It is not a priori clear that there even exist examples of causal channels which are \myuline{not} constructible. Observe however the following: If a channel $\scalemyQ{.8}{0.7}{0.5}{& \push{\cX_\sfA} \qw & \multigate{1}{T} & \push{\cY_\sfA} \qw & \qw \\ & \push{\cX_\sfB} \qw & \ghost{T} & \push{\cY_\sfB} \qw & \qw } $ is compatible with the \emph{local} causal specification $\scrC$ given by $\scrC(\sfy_i)= \{\sfx_i\}$, and if $(T, \scrC)$ is constructible, then $T$ must be a Bell-channel, i.e. of the form $\scalemyQ{.8}{0.7}{0.5}{
		& \push{\cX_\sfA}  \qw  & \multigate{1}{T_\sfA} & \push{\cY_\sfA}  \qw & \qw   \\
		& \Nmultigate{1}{s}  & \ghost{T_\sfA} &  \\
		& \Nghost{s} & \multigate{1}{T_\sfB}  \\
	 &   \push{\cX_\sfB} \qw  & \ghost{T_\sfB} & \push{\cY_\sfB}  \qw & \qw \\
} $. For now, I leave a formal graph-theoretic argument to the reader (since we will see in the next subsection an alternative proof technique by an \emph{induction principle}, cf. \cref{ex:IncRevis}), but the idea is basically that if $(T, \scrC)$ is representable on $G$ and if $G$ is chosen to have a minimal number of boxes among all such stencils, then one can argue by contradiction that $G$ must be either

\begin{align}
	\myQ{0.7}{0.5}{
	&  \qw  &  \qw  & \qw  \\
	& & & \\
	&   \qw  &  \qw & \qw \\
} , \quad 	
\myQ{0.7}{0.5}{
	&  \qw  &   \qw  & \qw \\
		& & & \\
	&   \qw  & \gate{\phantom{X}} & \qw \\
} , \quad
\myQ{0.7}{0.5}{
	&   \qw  & \gate{\phantom{X}} & \qw \\
		& & & \\
		&  \qw  &   \qw  & \qw \\
} , \quad \text{or} \quad
\myQ{0.7}{0.5}{
	 	&  \qw  & \multigate{1}{ \phantom{X}} &  \qw  & \qw \\
		& \Nmultigate{1}{\phantom{X}}  & \ghost{\phantom{X}} &  \\
		& \Nghost{\phantom{X}} & \multigate{1}{\phantom{X}}  \\
	 &   \qw  & \ghost{\phantom{X}} & \qw & \qw \\
} \quad,
\end{align}

and in either case $T$ is a Bell-channel. \\

This observation immediately shows that the classic counterexamples to being a Bell-channel are in fact counterexamples to constructibility:

\begin{Example} (Inconstructibility of the Popescu-Rohrlich Box in $\QIT$.) \\
The PR box is inconstructible in $\QIT$ when given the local specification from \cref{ex:PRCaus}. This is because, as mentioned in \cref{ex:PR}, it follows from the work of Cirelson (\cite{Cir80}) that the PR box does not have the form of a Bell-channel in $\QIT$.	\end{Example}

\begin{Example} (Inconstructibility of the CHSH-Behaviour in $\CIT$.) \label{ex:CHSH}\\
	As mentioned in \cref{ex:Bellch}, the work of Ref. \cite{Bell64}, as simplified in Ref. \cite{CHSH69}, gives an example of a Bell-channel in $\QIT$ which has classical inputs and outputs but is not a Bell-channel, i.e. is not constructible as a causal channel, in $\CIT$.

 Specifically, it is the causal channel given by 	
 
 \begin{align} \label{eq:CHSH}
 \myQ{0.7}{0.5}{
 		& \push{\C^{X_\sfA}}  \qw  & \qw & \multigate{1}{ \Lambda_\sfA} & \push{\C^{Y_\sfA}}  \qw & \qw   \\
 		& \Nmultigate{1}{\psi}  & \push{\C^2} \qw & \ghost{\Lambda_\sfA} &  \\
 		& \Nghost{\psi} &  \push{\C^2} \qw & \multigate{1}{\Lambda_\sfB}  \\
  &  \push{\C^{X_\sfB}}   \qw  & \qw & \ghost{\Lambda_\sfB} & \push{\C^{Y_\sfB}}  \qw & \qw  \\
 } 
 \quad,
 \end{align}
 
with $X_\sfA=X_\sfB=\{0,1\}$ and $Y_\sfA=Y_\sfB=\{+1,-1\}$, where $\psi$ is the maximally entangled state on $\C^2\otimes \C^2$  represented by the vector $\ket{\psi} = \frac{\ket{0} \otimes \ket{0}+\ket{1} \otimes \ket{1}}{\sqrt{2}}$, and where $\Lambda_i$ is the ensemble of projective measurements $\Pi^{x_i}_i$ on $\C^2$ (indexed by $x_i \in X_i$) for which $\Pi^{x_\sfA}_\sfA(\pm 1)$ are the two projections corresponding to the orthonormal basis $(\ket{0}, \ket{1})$ or $\left(\frac{\ket{0}+\ket{1}}{\sqrt{2}}, \frac{\ket{0}-\ket{1}}{\sqrt{2}} \right)$ depending on whether $x_\sfA=0$ or $x_\sfA=1$, respectively, and for which $\Pi^{x_\sfB}_\sfB(\pm 1)$ are the projections corresponding to the orthonormal basis $\left(\frac{ \cos(\pi/8)\ket{0}+\sin(\pi/8)\ket{1}}{\sqrt{2}}, \frac{\sin(\pi/8)\ket{0}-\cos(\pi/8)\ket{1}}{\sqrt{2}} \right)$ or $\left(\frac{ \cos(\pi/8)\ket{0}-\sin(\pi/8)\ket{1}}{\sqrt{2}}, \frac{\sin(\pi/8)\ket{0}+\cos(\pi/8)\ket{1}}{\sqrt{2}} \right)$ depending on whether $x_\sfB=0$ or $x_\sfB=1$, respectively. One can check that the resulting channel \eqref{eq:CHSH} is the classical channel $P=(P^{x_\sfA, x_\sfB})_{x_\sfA, x_\sfB \in \{0,1\}}$ given by

\begin{align}
P^{x_\sfA, x_\sfB}(y_\sfA, y_\sfB) =\frac{1}{4} + \frac{y_\sfA y_\sfB}{4}  \frac{(-1)^{x_\sfA \cdot x_\sfB}}{  \sqrt{2}},
\end{align}

and it can be verified by evaluation against a convex-linear functional (known as the \emph{CHSH game}) that this channel is not the convex combination of products of deterministic functions, hence not a Bell-channel in $\CIT$.
	\end{Example}

More recent work (\cite{Cola18,Cola20}) implies in a similar way that some constructible causal channels in $\QIT^\infty$ are inconstructible in $\QIT$.\\

The interested reader may prove as exercise that the phenomenon of inconstructibility is not bound to occur in every theory:\footnote{It is easiest to start with a small example, e.g. by proving that a bipartite channel with local causal specification is a Bell-channel (which in the case of cartesian theories means that it factors).} 

\begin{Thm} \label{thm:CartAllCons}
If $\Theory$ is a cartesian theory, every causal channel in $\Theory$ is constructible.  
	\end{Thm}

\subsection{The Category of Causal Channels -- Relative Constructibility}

Recall that a \emph{primitive} causal channel $(T, \scrC_0): \bbX \to \bbY$ is one whose specification renders every port in $\bbX$ a cause of any port in $\bbY$:  $\scrC_0(\sfJ) = \ports{\bbX}$ for all non-empty $\sfJ  \subseteq \ports{\bbY}$.  This notion of causality reflects how we (implicitly) thought of channels in the preceding chapters. If we think about it, what we did just above when defining constructible causal channels was \myuline{again} to think of each channel filling a box in the stencil as a primitive causal channel. Through an intuitive notion of how those causal channels should compose, a total causal channel thus emerged from the stencil and its filling.

It is natural and beneficial to explicitly articulate the notions of parallel and serial composition of causal channels which gives substance to this intuition. Effectively, we thereby achieve a \emph{new} (partial) symmetric monoidal category:

 \begin{Definition} ($\ICC{\Theory}$.) \label{def:ICC}\\
 	The \emph{category of interfaces and causal channels in $\Theory$}, denoted $\ICC{\Theory}$, is the partial\footnote{In the same sense as discussed towards the end of \cref{subsec:Inter}.} symmetric monoidal category which has as its objects the interfaces in $\Theory$, and as morphisms from $\bbX$ to $\bbY$ the causal channels from $\bbX$ to $\bbY$. Its serial and parallel composition is given as follows:
 		\begin{itemize} 		
 		\item The serial composition of $(T, \scrC): \bbX \to \bbY$ with $(S, \scrD): \bbY \to \bbZ$ is given by $(S \after T, \scrD \scrC)$, where $\scrD \scrC: \bbX \to \bbZ$ is the causal specification given by the functional composition $\scrC \circ \scrD$, i.e. 
 		
 		\begin{align}
 		(\scrD \scrC)(\sfJ) = \scrC(\scrD(\sfJ))
 		\end{align}
 		
 		 for $\sfJ\subseteq \ports{\bbZ}$.\footnote{This may look slightly confusing. The issue is ultimately that causal specifications have a `contravariant' nature reversing the order in their definition, e.g. a causal specification from $\bbX$ to $\bbY$ is a map from $\pow{\ports{\bbY}}$ to $\pow{\ports{\bbX}}$.} The identity on the interface $\bbX$ is given by $(\id_\bbX, \scrI_\bbX)$, where $\scrI_\bbX(\sfJ)=\sfJ$ for all $\sfJ \subseteq \ports{\bbX}$.
 		
 		\item If the underlying channels are pairwise parallelly composable, then $(T_1, \scrC_1): \bbX_1 \to \bbY_1$ and $(T_2, \scrC_2): \bbX_2 \to \bbY_2$ are parallelly composable, and the parallel composition is $(T_1 \og T_2, \scrC_1 \og \scrC_2)$, where  $\scrC _1 \og \scrC_2: \bbX_1 \og \bbX_2 \to \bbY_1 \og \bbY_2$ is the causal specification given by
 		
 		 \begin{align}
 		 (\scrC _1 \og\scrC_2)(\sfJ_1 \cup \sfJ_2) = \scrC_1(\sfJ_1) \cup \scrC_2(\sfJ_2)
 		 \end{align} 
 		 
 		 for $\sfJ_1 \subseteq \ports{\bbX_1}$ and $\sfJ_2 \subseteq \ports{\bbX_2}$. 
 	\end{itemize} \end{Definition}

\begin{Remark}
	One must of course check the appropriate compatibility requirements (e.g. if $T$ is compatible with $\scrC$ and $S$ with $\scrD$, then $S \after T$ is compatible with $\scrD\scrC$), but they are both obvious. 
	\end{Remark}

From now on, we will work in the category $\ICC{\Theory}$. In fact, the remainder of the thesis is to a great extent about revisiting and modifying the theory of dilations to the category of interfaces and causal channels, $\ICC{\Theory}$,  in place of the category of interfaces and channels, $\IC{\Theory}$. We will use the same sort of pictorial syntax, writing e.g. $\channel{\bbX}{(T, \scrC)}{\bbY}$ to denote the causal channel $(T, \scrC): \bbX \to \bbY$. \\

Let us re-examine in this fresh light the concept of constructible causal channels. With the general composition of causal channels at our disposal, we can of course build networks where the constituents are not channels, but \myuline{causal} channels: From now on, we will thus assume a filling $\fF$ of a stencil $G$ to consist of simple interfaces $(\bbZ_w)_{w \in \cW(G)}$ and \myuline{causal} channels $((T_b, \scrC_b))_{b \in \cB(G)}$, and we will denote by $\fF[G]$ the total \myuline{causal} channel that arises from filling the stencil $G$ according to $\fF$.\footnote{The theorem of Ref. \cite{JS91} that yielded a well-defined value $\fF[G]$ still applies, since what we have done is effectively to substitute the category for a different category.}  

\begin{Definition} (Relative Constructibility.) \label{def:RelConst}\\
	Let $\bfP$ be a class of causal channels in $\Theory$. We say that a causal channel $(T, \scrC)$ is \emph{constructible from $\bfP$} if it is of the form $\fF[G]$ for some stencil $G$ and some filling $\fF$, all of whose causal channels belong to $\bfP$.
	\end{Definition}

The notion of (absolute) constructibility is obviously recovered as follows:

\begin{Prop} (Absolute Constructibility as Special Relative Constructibility.)\\
	A causal channel is constructible in the sense of \cref{def:Constructible} precisely if it is constructible from primitive causal channels in the sense of \cref{def:RelConst}.
	\end{Prop}

The notion of relative constructibility can be recast as a minimality notion. Let us denote by $\up{Const}(\bfP)$ the class of causal channels constructible from a given class $\bfP$. Let us say that a class of causal channels $\bfC$ is \emph{constructibly closed} if $\up{Const}(\bfC) \subseteq \bfC$. 

One can prove (by induction on the complexity of DAGs) that a class is constructibly closed if and only if it contains identities between simple interfaces and is closed under serial and parallel composition. With this equivalence in mind, it is a relatively simple exercise to show the following: 

\begin{Thm} (Principle of Induction.)\label{thm:Induc}\\
	The class $\up{Const}(\bfP)$ is the smallest class of causal channels which contains $\bfP$ and is constructibly closed. More precisely,\footnote{The precision here is quite subtle: A standard foundation for mathematics will \myuline{not} abstractly prove the existence of a `smallest class' of channels subject to some requirements, since it is generally too large to be a set. As such, \cref{thm:Induc} is really a so-called \emph{theorem \myuline{schema}}, one theorem for each possible class $\bfC$.} $\up{Const}(\bfP)$ contains $\bfP$ and is constructibly closed, and if $\bfC$ is any class of causal channels which contains $\bfP$ and identities between simple interfaces, and which is closed under serial and parallel composition, then $\up{Const}(\bfP) \subseteq \bfC$. 
\end{Thm}

To showcase the induction technique, we prove the following result which is often useful:

\begin{Lem} (Factorisation of Constructible Channels.) \label{lem:CnstrFact}\\
	If $\channel{\bbX}{(T, \scrC)}{\bbY}$ is constructible, then for any sub-interface $\bbY_0 \subseteq \bbY$, there exist constructible causal channels $(T_1, \scrC_1)$ and $(T_2, \scrC_2)$ such that 
	
	\begin{align} \label{eq:ConsFac}
\myQ{0.7}{0.5}{& \push{\bbX} \qw & \gate{(T, \scrC)} & \push{\bbY} \qw & \qw} 
\quad = \quad 	
\myQ{0.7}{0.5}{
		& \push{\scrC(\bbY_0)}  \qw    &\multigate{1}{(T_1, \scrC_1)}   & \qw &   \push{\bbY_0} \qw & \qw & \qw \\
		&                   & \Nghost{(T_1, \scrC_1)}  & \push{\bbH} \qw & \multigate{1}{(T_2, \scrC_2)} & \\
		& \qw &  \push{\bbX \setminus \scrC(\bbY_0)}  \qw & \qw &  \ghost{(T_2, \scrC_2)}   &  \push{\bbY \setminus \bbY_0} \qw  & \qw  \\
	} \quad.	\end{align}

\end{Lem}

\begin{Remark} Though visually similar, this statement has nothing to do with the DiVincenzo Property (\cref{def:DiVi}), as clarified by the proof. It holds regardless of whether the theory $\Theory$ has the property or not, for it is ultimately a graph-theoretic statement. Rather, the DiVincenzo Property would be the statement that if  $\scalemyQ{.8}{0.7}{0.5}{& \push{\cX_1} \qw & \multigate{1}{(T, \scrC)} & \push{\cY_1} \qw & \qw \\ & \push{\cX_2} \qw & \ghost{(T, \scrC)} & \push{\cY_2} \qw & \qw } $ is a causal channel for which $\scrC(\{\sfy_1\})=\{\sfx_1\}$, $\scrC(\{\sfy_2\})=\{\sfx_1, \sfx_2\}$, then $(T, \scrC)$ is constructible. In other words, it would say that constructibility follows from a property of the causal specification alone. (\cref{lem:CnstrFact} would then provide a factorisation.)	\end{Remark}

\begin{proof}
	We prove the result by induction on constructible channels.

	Let $\pi((T, \scrC))$ be the predicate, ranging over all causal channels $(T, \scrC): \bbX \to \bbY$ in $\Theory$, which asserts that for all sub-interfaces $\bbY_0 \subseteq \bbY$ there exist constructible channels $(T_1, \scrC_1)$ and $(T_2, \scrC_2)$ satisfying \eqref{eq:ConsFac}. Given a causal channel $(T, \scrC)$, the statement $\pi((T, \scrC))$ is either true or false. What we wish to prove is that it is true for every constructible channel $(T, \scrC)$. Consider the class of causal channels

	\begin{align}
	\bfC := \{(T, \scrC) \mid \pi((T, \scrC))\}
	\end{align}
	
	for which $\pi((T, \scrC))$ is true. By virtue of \cref{thm:Induc}, we can show that $\bfC$ contains all constructible channels by showing that $\bfC$ contains every primitive causal channel and that $\bfC$ is closed under serial and parallel composition (whenever these are defined). 
	
	It is easy to see that $\bfC$ contains all primitive channels: If $(T, \scrC)$ is primitive and $\bbY_0 \subseteq \bbY$, then either $\bbY_0= \bbI$ or $\scrC(\bbY_0) = \bbX$; in either case, there is really nothing to show. More precisely, in the case $\bbY_0= \bbI$ we may take $\bbH= \bbI$, $(T_1, \scrC_1) = \id_\bbI$ and $(T_2, \scrC_2) = (T, \scrC)$, and in the case $\scrC(\bbY_0) = \bbX$ we may take $\bbH = \bbY \setminus \bbY_0$, $(T_1, \scrC_1) = (T, \scrC)$ and $(T_2, \scrC_2) = \id_{\bbY \setminus \bbY_0}$ (with the identity specification $\scrI_{\bbY \setminus \bbY_0}$).
	
	It is also easy to see that $\bfC$ is closed under parallel composition. Let $(T, \scrC): \bbX \to \bbY$ and $(S, \scrD): \bbZ \to \bbW$ belong to $\bfC$ and be parallelly composable. Consider the parallel composition $(T, \scrC) \og (S, \scrD): \bbX  \cup \bbZ \to \bbY \cup \bbW$. Any sub-interface $\bbA$ of the output interface $\bbY \cup\bbW$ is of the form $\bbY_0 \cup \bbW_0$ for sub-interfaces $\bbY_0 \subseteq \bbY$ and $\bbW_0 \subseteq \bbW$, and thus
	
	\begin{align}
	(\scrC \og \scrD)(\bbA)= (\scrC \og \scrD)(\bbY_0 \cup \bbW_0) = \scrC(\bbY_0) \cup \scrD(\bbW_0).
	\end{align}  
	
	Now, because $(T, \scrC)$ and $(S, \scrD)$ belong to $\bfC$ we may write 
	
		\begin{align} 
	\myQ{0.7}{0.5}{& \push{\bbX} \qw & \gate{(T, \scrC)} & \push{\bbY} \qw & \qw} 
	\quad = \quad 	
	\myQ{0.7}{0.5}{
		& \push{\scrC(\bbY_0)}  \qw    &\multigate{1}{(T_1, \scrC_1)}   & \qw &   \push{\bbY_0} \qw & \qw & \qw \\
		&                   & \Nghost{(T_1, \scrC_1)}  & \push{\bbH} \qw & \multigate{1}{(T_2, \scrC_2)} & \\
		& \qw &  \push{\bbX \setminus \scrC(\bbY_0)}  \qw & \qw &  \ghost{(T_2, \scrC_2)}   &  \push{\bbY \setminus \bbY_0} \qw  & \qw  \\
	} \quad	\end{align}

for constructible channels $(T_1, \scrC_1)$ and $(T_2, \scrC_2)$, and 

	\begin{align} 
\myQ{0.7}{0.5}{& \push{\bbZ} \qw & \gate{(S, \scrD)} & \push{\bbW} \qw & \qw} 
\quad = \quad 	
\myQ{0.7}{0.5}{
	& \push{\scrD(\bbW_0)}  \qw    &\multigate{1}{(S_1, \scrD_1)}   & \qw &   \push{\bbW_0} \qw & \qw & \qw \\
	&                   & \Nghost{(S_1, \scrD_1)}  & \push{\bbK} \qw & \multigate{1}{(S_2, \scrD_2)} & \\
	& \qw &  \push{\bbZ \setminus \scrD(\bbW_0)}  \qw & \qw &  \ghost{(S_2, \scrD_2)}   &  \push{\bbW \setminus \bbW_0} \qw  & \qw  \\
} \quad	\end{align}
	
for constructible channels $(S_1, \scrD_1)$ and $(S_2, \scrD_2)$. By parallelly composing and merging the causal channels pairwise, we obtain the desired form of $(T, \scrC) \og (S, \scrD)$.

	The fact that $\bfC$ is closed under serial composition is proved similarly, using the induction hypothesis for each constituent. The details are left as exercise. 
\end{proof}

We end this section by observing another factorisation result (valid for arbitrary causal channels) which is often useful, and by employing \cref{lem:CnstrFact} to give an alternative analysis of constructible channels with local specifications.

\begin{Lem} (Extracting Trashes from a Causal Channel.) \label{lem:ExtractTriv}\\
Every causal channel $\channel{\bbX}{(T, \scrC)}{\bbY}$ can be written in the form 

	\begin{align}
\myQ{0.7}{0.5}{&\push{\bbX' } \qw &  \gate{(T', \scrC')}& \qw &  \push{\bbY} \qw & \qw \\ & \push{\bbX \setminus \bbX'} \qw & \gate{\tr}  },
\end{align}

with $\bbX' \subseteq \bbX$ and $\scrC'(\ports{\bbY})= \ports{\bbX'}$.
\end{Lem}

\begin{proof}
Let $\bbX'$ be the sub-interface of $\bbX$ defined by $\ports{\bbX'}= \scrC(\ports{\bbY})$. The desired factorisation follows from the non-signalling conditions implied by $\scrC$ (and $\scrC'$ is given by $\scrC'(\sfJ)= \scrC(\sfJ)$ for $\sfJ \subseteq \ports{\bbY}$).	

\end{proof}

\begin{Example} (Inconstructibility Revisited.) \label{ex:IncRevis}\\
	Suppose that $\scalemyQ{.8}{0.7}{0.5}{& \push{\cX_1} \qw & \multigate{1}{(T, \scrC)} & \push{\cY_1} \qw & \qw \\ & \push{\cX_2} \qw & \ghost{(T, \scrC)} & \push{\cY_2} \qw & \qw } $ is constructible, with $\scrC$ the local specification given by  $\scrC(\sfy_1)= \{\sfx_1\}$ and $\scrC(\sfy_2)= \{\sfx_2\}$. Then, by \cref{lem:CnstrFact}, we can write e.g.

\begin{align} \label{eq:NSfact}
	\myQ{0.7}{0.5}{& \push{\cX_1} \qw & \multigate{1}{(T, \scrC)} & \push{\cY_1} \qw & \qw \\ & \push{\cX_2} \qw & \ghost{(T, \scrC)} & \push{\cY_2} \qw & \qw } 
	\quad = \quad 	
	\myQ{0.7}{0.5}{
		& \push{\cX_1}  \qw    &\multigate{1}{(T_1, \scrC_1)}   & \qw &   \push{\cY_1} \qw & \qw & \qw \\
		&                   & \Nghost{(T_1, \scrC_1)}  & \push{\bbH} \qw & \multigate{1}{(T_2, \scrC_2)} & \\
		& \qw &  \push{\cX_2}  \qw & \qw &  \ghost{(T_2, \scrC_2)}   &  \push{\cY_2} \qw  & \qw  \\
	} 
\end{align}

for constructible channels $(T_1, \scrC_1)$ and $(T_2, \scrC_2)$. Now, however, observe that there can be no port $\sfh \in \ports{\bbH}$ for which $\sfh \in \scrC_2(\sfy_2)$ and $\sfx_1 \in \scrC_1(\sfh)$, since such a port would provide a causal link effectuating (by the definition of composition) that $\sfx_1 \in \scrC(\sfy_2)$. Consequently, every $\sfh \in \ports{\bbH}$ either satisfies $\sfh \notin \scrC_2(\sfy_2)$ or $\scrC_1(\sfh)=\emptyset$. By \cref{lem:ExtractTriv}, we may assume, after possibly extracting a trash from $(T_2, \scrC_2)$, that there are only ports of the latter kind.
Applying \cref{lem:CnstrFact} to $(T_1, \scrC_1)$, we can thus write 

\begin{align}
	\myQ{0.7}{0.5}{
	& \push{\cX_1}  \qw    &\multigate{1}{(T_1, \scrC_1)}   &   \push{\cY_1} \qw & \qw  \\
	&                   & \Nghost{(T_1, \scrC_1)}  & \push{\bbH} \qw & \qw
} \quad = \quad =  
	\myQ{0.7}{0.5}{
	& \push{\cX_1}  \qw    &\qw & \multigate{1}{(T'_1, \scrC'_1)}   &   \push{\cY_1} \qw & \qw  \\
	&         \Nmultigate{1}{s}          & \push{\bbG} \qw & \ghost{(T_1, \scrC_1)}  \\
	& \Nghost{s} & \push{\bbH} \qw & \qw& \qw
}, 
\end{align}

which ultimately implies that $(T, \scrC)$ is necessarily a Bell-channel. Hence, we confirm that for example the PR box considered as a causal channel (\cref{ex:PRCaus}) is inconstructible in the theory $\QIT$. 	\end{Example}

Note, importantly, that \emph{by the DiVincenzo Property} of $\QIT$, the \myuline{channel} $\scalemyQ{.8}{0.7}{0.5}{& \push{\cX_1} \qw & \multigate{1}{T} & \push{\cY_1} \qw & \qw \\ & \push{\cX_2} \qw & \ghost{T} & \push{\cY_2} \qw & \qw } $ which underlies the PR box is of the form $\scalemyQ{.8}{0.7}{0.5}{
	& \push{\cX_1}  \qw    &\multigate{1}{T_1}   & \qw &   \push{\cY_1} \qw & \qw  \\
	&                   & \Nghost{T_1}  & \multigate{1}{T_2} & \\
	&  \push{\cX_2}  \qw & \qw &  \ghost{T_2}   &  \push{\cY_2} \qw  & \qw  \\
} $ for \myuline{channels} $T_1$ and $T_2$; as a \myuline{causal} channel, however, it is not of the form \eqref{eq:NSfact} for any \myuline{causal} channels $(T_1, \scrC_1)$ and $(T_2, \scrC_2)$.

\section{Notions of Contraction}
\label{sec:Contractions}

The class of causal channels admits an operation additional to those of serial and parallel composition. That this \myuline{should} be the case is imminent when we look at stencil-representations of causal channels, but that it \myuline{is} the case is by no means obvious. \\

Consider our generic stencil, filled with causal channels:

\begin{align} \label{eq:GenCaus}
\myQ{1}{0.5}{
	&	&  &  & \Nmultigate{2}{ (T_1, \scrC_1)} & \qw & \qw & \qw & \sfy_1\\
	& \sfx_1\quad 	&  \qw  & \qw & \ghost{ (T_1, \scrC_1)} &  \multigate{2}{(T_2, \scrC_2)} &  \qw & \qw & \sfy_2 \\
	&	\sfx_2 \quad& \qw  &  \multigate{1}{(T_3, \scrC_3)} & \ghost{(T_1, \scrC_1)} &  \Nghost{(T_2, \scrC_2)} & \qw & \qw & \sfy_3\\
	&	& \Nmultigate{2}{(T_4, \scrC_4)} &  \ghost{(T_3, \scrC_3)} & \qw & \ghost{(T_2, \scrC_2)} \\
	&	& \Nghost{(T_4, \scrC_4)} & \qw & \qw & \qw & \qw & \qw & \sfy_4\\
	&	& \Nghost{(T_2, \scrC_2)} & \qw & \multigate{1}{(T_5, \scrC_5)}  \\
	& \sfx_3\quad	&   \qw  & \qw &\ghost{(T_5, \scrC_5)} & \qw &   \qw & \qw & \sfy_5  \\
}  \quad. 
\end{align}

As outlined in the previous section, the stencil $G$ with its filling $\fF$ defines a total causal channel, $(T, \scrC) := \fF[G]$. Now, if the system at $\sfy_4$ matches the system at $\sfx_3$, it should intuitively be possible to \emph{contract} those two wires, effectively feeding the output at $\sfy_4$ to the input port $\sfx_3$. Similarly for $\sfy_3$ into $\sfx_3$, or $\sfy_5$ into $\sfx_2$. (On the other hand, an insertion of $\sfy_2$ into $\sfx_2$ is not a priori sensible, as the circuit would thereby acquire a cycle.) 

Clearly, any of the total causal channels which would result from such a contracted diagram can be determined from $G$ and $\fF$. But might it be determinable from $(T, \scrC)$ alone?  \\

If the answer to this question were `no', we would in a sense have done a lousy job in stating the very definition of a causal channel; for the concept to be of use, it must reflect all operational aspects of what it aims to model, and certainly contraction is such an aspect. However, it is not at all clear that the violent reduction of $(\fF, G)$ to $\fF[G]$ should spare the life of the possibility to contract. In fact, there are not only different fillings on the same stencil which will reproduce a given causal channel $(T, \scrC)$, there might also be different \myuline{stencils} on which $(T, \scrC)$ is representable. What we ask is ultimately that the contraction be independent of any details of the representation whatsoever.  \\

In \cref{subsec:ContStandard}, we will see that if the underlying theory $\Theory$ has universal dilations (cf. \cref{def:Univ}), we \myuline{can} introduce unambiguous contractions, definable from $(T, \scrC)$ alone. This result is one of the technical highlights of the chapter, and one that consolidates and justifies the definition of a causal channel. %

In \cref{subsec:ContGeneral}, we will then discuss why and how one might wish to define abstract notions of contraction, also in theories where the operational narrative of connecting wires does not really make sense. (For example, \emph{cancellation} in a monoid, i.e. the act of forming the relation $x \succeq y$ from the relation $x \star z \succeq y \star z$ can be seen as the result of contraction.) Such abstract notions of contraction might be seen as generalisation of the \emph{traces} in symmetric monoidal categories as introduced in Ref. \cite{JSV96}.

\subsection{The Standard Notion of Contraction}
\label{subsec:ContStandard}
Let $\fF$ be a filling (with causal channels) of a stencil $G$. Let $(T, \scrC): \bbX \to \bbY$ be the resulting causal channel, i.e. the value $\fF[G]$. The above discussion alluded to the contraction of a single pair of ports, but we might as well contract several pairs of ports at once. It is notationally convenient that the ports in each pair we wish to contract is given the same name (their systems must match anyway) -- for instance, in \eqref{eq:GenCaus} we would require $\sfy_4=\sfx_3$ for the contraction of $\sfy_4$ with $\sfx_3$. Hence, what we will the define is the notion of \emph{contracting $(T, \scrC)$ along $\bbP$}, where $\bbP$ is a an \myuline{interface} which is a sub-interface of both $\bbX$ and $\bbY$. 

First, we will define what it means for a common sub-interface of $\bbX$ and $\bbY$ to be \emph{seemingly contractible}. This will be simply a matter of whether there is a stencil-representation of $(T, \scrC)$ which renders $\bbP$ contractible based on ancestry in the stencil, that is, which does not acquire cycles when a contraction of the wires is forced. Having done this, we will then show that in universal theories the seemingly contractible interfaces \myuline{can} actually be contracted in a well-defined manner.\\ %

Given a stencil $G$ and a bijection $h: P_\up{out} \to P_\up{in}$, where $P_\up{in} \subseteq \cW_\up{in}(G)$ and $P_\up{out} \subseteq \cW_\up{out}(G)$, let us say that \emph{$G$ is contractible w.r.t. $h$} if the directed graph that results from adjoining each wire $w \in P_\up{out}$ with the wire $h(w) \in P_\up{in}$ remains acyclic. If $\fF$ is a filling of $G$ and $\bbP$ a common sub-interface of the input and output interfaces $\bbX$ and $\bbY$, there is a natural bijection $h^\fF_\bbP$ from the output wires $w$ corresponding to $\sfp \in \ports{\bbP}$ (on the output side) to the input wires $w'=h^\fF_\bbP(w)$ corresponding to $\sfp \in \ports{\bbP}$ (on the input side).

\begin{Definition} (Seemingly Contractible Interfaces.) \label{def:seemCont}\\
	Let $\channel{\bbX}{(T, \scrC)}{\bbY}$ be a causal channel, and let $\bbP$ be a common sub-interface of $\bbX$ and $\bbY$. If $(\fF, G)$ is a stencil-representation of $(T, \scrC)$, we say that \emph{$\bbP$ is contractible in $(\fF,G)$} if $G$ is contractible w.r.t. $h^\fF_\bbP$. We say that $\bbP$ is \emph{seemingly contractible in $(T, \scrC)$}, or that $(T, \scrC)$ is \emph{seemingly contractible along $\bbP$}, if there exists a stencil-representation $(\fF, G)$ of $(T, \scrC)$ in which $\bbP$ is contractible.\end{Definition}

\begin{Example} (Seeming Contractibility in a Generic Causal Channel.) \label{ex:ContGeneric}\\
	Suppose that $(T, \scrC)$ admits the stencil-representation 

\begin{align}
\myQ{1}{0.5}{
	&	&  &  & \Nmultigate{2}{ (T_1, \scrC_1)} & \qw & \qw & \qw & \sfy_1\\
	& \sfp_0\quad 	&  \qw  & \qw & \ghost{ (T_1, \scrC_1)} &  \multigate{2}{(T_2, \scrC_2)} &  \qw & \qw & \sfy_2 \\
	&	\sfp_1 \quad& \qw  &  \multigate{1}{(T_3, \scrC_3)} & \ghost{(T_1, \scrC_1)} &  \Nghost{(T_2, \scrC_2)} & \qw & \qw & \sfp_2\\
	&	& \Nmultigate{2}{(T_4, \scrC_4)} &  \ghost{(T_3, \scrC_3)} & \qw & \ghost{(T_2, \scrC_2)} \\
	&	& \Nghost{(T_4, \scrC_4)} & \qw & \qw & \qw & \qw & \qw & \sfp_0\\
	&	& \Nghost{(T_2, \scrC_2)} & \qw & \multigate{1}{(T_5, \scrC_5)}  \\
	& \sfp_2 \quad	&   \qw  & \qw &\ghost{(T_5, \scrC_5)} & \qw &   \qw & \qw & \sfp_1  \\
}  \quad.
\end{align}

In this representation, the sub-interfaces with port sets $\{\sfp_0\}$, $\{\sfp_1\}$, $\{\sfp_2\}$, $\{\sfp_0, \sfp_1\}$ and $\{\sfp_0, \sfp_2\}$ are all contractible. Hence, those sub-interfaces are seemingly contractible in $(T, \scrC)$. 

It might be that the sub-interface $\{\sfp_1, \sfp_2\}$ is also seemingly contractible in $(T, \scrC)$, but this is not witnessed by the above representation and would require the existence of \myuline{another} stencil-representation of $(T, \scrC)$, in which both $\sfp_1$ and $\sfp_2$ can be contracted without creating cycles. In such a representation, some of the first-mentioned sub-interfaces (e.g. $\{\sfp_0, \sfp_1\}$ or $\{\sfp_0, \sfp_2\}$) might cease to be contractible -- in fact, it is unclear whether there exists a single representation of $(T, \scrC)$ in which every seemingly contractible interface is contractible (and I do not know the answer to this question).\end{Example}

\begin{Example} (Obvious Seeming Contractibility.) \\
The trivial interface $\bbI$ is seemingly contractible in every causal channel $(T, \scrC)$. It is similarly clear that if $\bbP$ is seemingly contractible, then so is any sub-interface $\bbP_0 \subseteq \bbP$.\end{Example}

\begin{Example} (Seeming Contractibility versus the Causal Specification.) \label{ex:ContversusCaus}\\
Seeming contractibility can sometimes be excluded on the basis of the causal specification: If $\bbP$ is seemingly contractible in $(T, \scrC)$, then necessarily  $\sfp \notin \scrC(\sfp)$ for all $\sfp \in \ports{\bbP}$. (Indeed, if $\sfp \in \scrC(\sfp)$ and $(T, \scrC)=\fF[G]$, then there must be a path in $G$ leading from the input port $\sfp$ to the output port $\sfp$; thus a cycle would form if those two ports were contracted.) More generally, define on $\ports{\bbP}$ the transitive relation $<^{\bbP}_\scrC$ by letting $\sfp <^{\bbP}_\scrC \sfq$ if and only if there is a sequence $\sfp_0, \sfp_1, \ldots,\sfp_n \in \ports{\bbP}$, $n \geq 1$, such that  $\sfp_0 = \sfp$, $\sfp_n= \sfq$ and $\sfp_{k-1}\in \scrC(\sfp_{k})$ for all $k=1, \ldots, n$; if $\bbP$ is seemingly contractible in $(T, \scrC)$, then $<^{\bbP}_\scrC$ must be irreflexive on $\ports{\bbP}$, i.e. there can be no $\sfp \in \ports{\bbP}$ with $\sfp <^{\bbP}_\scrC \sfp$.

Since intuitively the relationship $\sfp \nless^{\bbP}_\scrC \sfp$ signifies that the input $\sfp$ is not in the causal past of the output $\sfp$, one might naively think that an interface $\bbP$ is seemingly contractible in $(T, \scrC)$ \myuline{if} and only if $<^{\bbP}_\scrC$ is irreflexive on $\ports{\bbP}$. This fails, however, even for the simple interfaces:  If a simple interface with port $\sfp$ is seemingly contractible, then, by merging components in a stencil-representation that witnesses this, we see that $(T, \scrC)$ admits a stencil-representation of the specific form

\begin{align} \label{eq:contcan}
\scalemyQ{.8}{0.7}{0.5}{
	&  \qw    &\multigate{4}{(T_1, \scrC_1)}   & \qw &    \qw & \push{\sfp} \\
		&   &\Nghost{(T_1, \scrC_1)}   \\
	&        \vdots          & \Nghost{(T_1, \scrC_1)}  &  \qw & \multigate{4}{(T_2, \scrC_2)} &  \qw & \qw \\
		&              & \Nghost{(T_1, \scrC_1)}  & \vdots & \Nghost{(T_2, \scrC_2)}  \\
			&       \qw       & \ghost{(T_1, \scrC_1)}  & \qw & \ghost{(T_2, \scrC_2)}  & \vdots\\
		&                   &  &  & \Nghost{(T_2, \scrC_2)} &  \\
&	& \sfp \hspace{1cm}  & \qw &  \ghost{(T_2, \scrC_2)}   &  \qw  & \qw  \\
} 	\quad . 
\end{align}

Thus, for example, if we consider any of our inconstructible channels from \cref{subsec:Constructible}, e.g. the PR box $\scalemyQ{.8}{0.7}{0.5}{& \sfx_\sfA \quad & \push{\{0,1\}} \qw & \multigate{1}{(T, \scrC)} &\push{\{0,1\} }\qw & \qw& \sfy_\sfA \\ & \sfx_\sfB \quad& \push{\{0,1\} }\qw & \ghost{(T, \scrC)} & \push{\{0,1\}} \qw & \qw & \sfy_\sfB} $ and if we pair the ports `oppositely' (letting $\sfp := \sfx_\sfA = \sfy_\sfB$ and $\sfq := \sfx_\sfB = \sfy_\sfA$),  then we have e.g. $\sfp\notin \scrC(\sfp)$, but the interface with port $\sfp$ is \myuline{not} seemingly contractible, since, as we saw in \cref{ex:PRCaus}, the PR box admits no representation of the form \eqref{eq:contcan}.

Note, however, that if $(T, \scrC)$ is \myuline{constructible}, then $\scrC$ alone determines seeming contractibility of $\bbP$. Indeed, if $<^{\bbP}_\scrC$ is irreflexive on $\ports{\bbP}$, then $\bbP$ must be contractible in any stencil-representation of $(T, \scrC)$ whose filling consists of primitive causal channels, since there is in such a representation a path from $\sfx$ to $\sfy$ if and only if $\sfx \in \scrC(\sfy)$. 

\end{Example}

Now, let us consider the problem of actually forming contractions of seemingly contractible interfaces, independently of the representations. A simple instance of this problem is the following: Suppose that 

\begin{align}  \label{eq:simpContAss}
\scalemyQ{.8}{0.7}{0.5}{
	&  \qw    &\multigate{4}{(T_1, \scrC_1)}   & \qw &    \qw &\sfp \\
	&   &\Nghost{(T_1, \scrC_1)}   \\
	&        \vdots          & \Nghost{(T_1, \scrC_1)}  &  \qw & \multigate{4}{(T_2, \scrC_2)} &  \qw & \qw \\
	&              & \Nghost{(T_1, \scrC_1)}  & \vdots & \Nghost{(T_2, \scrC_2)}  \\
	&       \qw       & \ghost{(T_1, \scrC_1)}  & \qw & \ghost{(T_2, \scrC_2)}  & \vdots\\
	&                   &  &  & \Nghost{(T_2, \scrC_2)} &  \\
	&	& \sfp \hspace{1cm}  & \qw &  \ghost{(T_2, \scrC_2)}   &  \qw  & \qw  \\
} 	\quad  = \quad 
\scalemyQ{.8}{0.7}{0.5}{
	&  \qw    &\multigate{4}{(T'_1, \scrC'_1)}   & \qw &    \qw & \sfp \\
	&   &\Nghost{(T'_1, \scrC'_1)}   \\
	&        \vdots          & \Nghost{(T'_1, \scrC'_1)}  &  \qw & \multigate{4}{(T'_2, \scrC'_2)} &  \qw & \qw \\
	&              & \Nghost{(T'_1, \scrC'_1)}  & \vdots & \Nghost{(T'_2, \scrC'_2)}  \\
	&       \qw       & \ghost{(T'_1, \scrC'_1)}  & \qw & \ghost{(T'_2, \scrC'_2)}  & \vdots\\
	&                   &  &  & \Nghost{(T'_2, \scrC'_2)} &  \\
	&	& \sfp \hspace{1cm}  & \qw &  \ghost{(T'_2, \scrC'_2)}   &  \qw  & \qw  \\
}  \quad  .
\end{align}

Is it necessarily the case that 

\begin{align}  \label{eq:simpContCon}
\scalemyQ{.8}{0.7}{0.5}{
	&  \qw    &\multigate{4}{(T_1, \scrC_1)}  &  \push{\sfp} \qw &  \multigate{4}{(T_2, \scrC_2)} &  \qw & \qw  \\
	&   &\Nghost{(T_1, \scrC_1)}  & & \Nghost{(T_2, \scrC_2)}  \\
	&        \vdots          & \Nghost{(T_1, \scrC_1)}  &  \qw & \ghost{(T_2, \scrC_2)} &  \vdots \\
	&              & \Nghost{(T_1, \scrC_1)}  & \vdots & \Nghost{(T_2, \scrC_2)}  \\
	&       \qw       & \ghost{(T_1, \scrC_1)}  & \qw & \ghost{(T_2, \scrC_2)}  &\qw & \qw\\
} 	\quad = \quad 
\scalemyQ{.8}{0.7}{0.5}{
	&  \qw    &\multigate{4}{(T'_1, \scrC'_1)}  &  \push{\sfp} \qw &  \multigate{4}{(T'_2, \scrC'_2)} &  \qw & \qw  \\
	&   &\Nghost{(T'_1, \scrC'_1)}  & & \Nghost{(T'_2, \scrC'_2)}  \\
	&        \vdots          & \Nghost{(T'_1, \scrC'_1)}  &  \qw & \ghost{(T'_2, \scrC'_2)} &  \vdots \\
	&              & \Nghost{(T'_1, \scrC'_1)}  & \vdots & \Nghost{(T'_2, \scrC'_2)}  \\
	&       \qw       & \ghost{(T'_1, \scrC'_1)}  & \qw & \ghost{(T'_2, \scrC'_2)}  &\qw& \qw \\
}  \quad ?
\end{align}

(Here, the  wire-pairs corresponding to $\sfp$ have been contracted while all others remain as before.) In fact, we can ask an even simpler question: Disregarding the causal specifications altogether, is it even the case that the identity

\begin{align}
	\myQ{0.7}{0.5}{
	& \push{\cX}  \qw    &\multigate{1}{T_1}   & \qw &   \push{\cW} \qw & \qw & \qw \\
	&                   & \Nghost{T_1}  & \push{\cZ} \qw & \multigate{1}{T_2} & \\
	& \qw &  \push{\cW}  \qw & \qw &  \ghost{T_2}   &  \push{\cY} \qw  & \qw  \\
} 	
=
\myQ{0.7}{0.5}{
	& \push{\cX}  \qw    &\multigate{1}{T'_1}   & \qw &   \push{\cW} \qw & \qw & \qw \\
	&                   & \Nghost{T'_1}  & \push{\cZ'} \qw & \multigate{1}{T'_2} & \\
	& \qw &  \push{\cW}  \qw & \qw &  \ghost{T'_2}   &  \push{\cY} \qw  & \qw  \\
} 	
\end{align}

between \myuline{channels} implies the identity 

\begin{align}
\myQ{0.7}{0.5}{
& \push{\cX}  \qw    &\multigate{1}{T_1}   & \push{\cW} \qw &   \multigate{1}{T_2} & \push{\cY} \qw & \qw \\
&                 & \Nghost{T_1}  & \push{\cZ} \qw &  \ghost{T_2}
} 	
=
\myQ{0.7}{0.5}{
& \push{\cX}  \qw    &\multigate{1}{T'_1}   & \push{\cW} \qw &   \multigate{1}{T'_2} & \push{\cY} \qw & \qw \\
&                 & \Nghost{T'_1}  & \push{\cZ'} \qw &  \ghost{T'_2}
} 	\quad ?
\end{align}

The latter identity asserts the equality of the serial compositions $T_2 \after T_1$ and $T'_2 \after T'_1$, whereas the former asserts the equality of a `partial' serial composition, as if interrupted in the midst of composing. What we are asking is thus \emph{whether it is possible to finish a half-hearted serial composition}. \\

It turns out that it is enough to solve that problem: If we can answer the latter question affirmatively, then we obtain independence of the stencil-representations all the way up to general contractions,  essentially because any contraction can be decomposed into a sequence of contractions of this kind. The most interesting part is thus the result that the interrupted serial compositions can be finished without knowing the individual components. 

To show this, an old acquaintance comes to rescue:

\begin{Lem} (Interrupted Serial Compositions can be Completed in Universal Theories.) \label{lem:UnivCont}\\
	Suppose that $\Theory$ is a \myuline{universal} theory. If $\onetwochannel{\cX}{T_1}{\cW}{\cZ}$, $\twoonechannel{\cZ}{\cW}{T_2}{\cY}$ and $\onetwochannel{\cX}{T'_1}{\cW}{\cZ'}$, $\twoonechannel{\cZ'}{\cW}{T'_2}{\cY}$ are pairs of channels such that
	
	\begin{align} \label{eq:contAss}
	\myQ{0.7}{0.5}{
		& \push{\cX}  \qw    &\multigate{1}{T_1}   & \qw &   \push{\cW} \qw & \qw & \qw \\
		&                   & \Nghost{T_1}  & \push{\cZ} \qw & \multigate{1}{T_2} & \\
		& \qw &  \push{\cW}  \qw & \qw &  \ghost{T_2}   &  \push{\cY} \qw  & \qw  \\
	} 	
	=
	\myQ{0.7}{0.5}{
		& \push{\cX}  \qw    &\multigate{1}{T'_1}   & \qw &   \push{\cW} \qw & \qw & \qw \\
		&                   & \Nghost{T'_1}  & \push{\cZ'} \qw & \multigate{1}{T'_2} & \\
		& \qw &  \push{\cW}  \qw & \qw &  \ghost{T'_2}   &  \push{\cY} \qw  & \qw  \\
	} 	,
	\end{align}
	
	then 
	
	\begin{align}
	\myQ{0.7}{0.5}{
		& \push{\cX}  \qw    &\multigate{1}{T_1}   & \push{\cW} \qw &   \multigate{1}{T_2} & \push{\cY} \qw & \qw \\
		&                 & \Nghost{T_1}  & \push{\cZ} \qw &  \ghost{T_2}
	} 	
	=
	\myQ{0.7}{0.5}{
		& \push{\cX}  \qw    &\multigate{1}{T'_1}   & \push{\cW} \qw &   \multigate{1}{T'_2} & \push{\cY} \qw & \qw \\
		&                 & \Nghost{T'_1}  & \push{\cZ'} \qw &  \ghost{T'_2}
	} 	.
	\end{align}
\end{Lem}

\begin{proof}
	By trashing $\cY$ in \cref{eq:contAss}, we see that 
	
	\begin{align} 
	\myQ{0.7}{0.5}{
		& \push{\cX}  \qw    &\multigate{1}{T_1}   & \qw &   \push{\cW} \qw & \qw & \qw \\
		&                   & \Nghost{T_1}  & \push{\cZ} \qw& \gate{\tr}
	} =  	\myQ{0.7}{0.5}{
		& \push{\cX}  \qw    &\multigate{1}{T'_1}   & \qw &   \push{\cW} \qw & \qw & \qw \\
		&                   & \Nghost{T'_1}  & \push{\cZ'} \qw & \gate{\tr}
	} 	.
	\end{align}
	
	Call this channel $\channel{\cX}{T}{\cW}$, and let  $\scalemyQ{.8}{0.7}{0.5}{& \push{\cX}  \qw & \multigate{1}{U} & \push{\cW} \qw & \qw \\ 
		& & \Nghost{U} & \push{\cE_0} \ww  & \ww  }$ be a universal dilation of $T$. Evidently, both $T_1$ and $T'_1$ are one-sided dilations of $T$, so, by completeness of $U$, we find $G$ and $G'$ with 
	
	\begin{align} \label{eq:T1T2}
	\scalemyQ{.8}{0.7}{0.5}{& \push{\cX}  \qw & \multigate{1}{U} & \push{\cW} \qw & \qw \\ 
		& & \Nghost{U} & \Ngate{G}{\ww} & \qw  } = \scalemyQ{.8}{0.7}{0.5}{& \push{\cX}  \qw & \multigate{1}{T_1} & \push{\cY} \qw & \qw \\ 
		& & \Nghost{T_1} & \push{\cZ} \qw  & \qw  } \quad \text{and} \quad \scalemyQ{.8}{0.7}{0.5}{& \push{\cX}  \qw & \multigate{1}{U} & \push{\cW} \qw & \qw \\ 
		& & \Nghost{U} & \Ngate{G'}{\ww} & \qw  } = \scalemyQ{.8}{0.7}{0.5}{& \push{\cX}  \qw & \multigate{1}{T'_1} & \push{\cY} \qw & \qw \\ 
		& & \Nghost{T'_1} & \push{\cZ'} \qw  & \qw  } \quad. 
	\end{align}
	
	Plugging this into \cref{eq:contAss} yields
	
	\begin{align}
	\myQ{0.7}{0.5}{
		& \push{\cX}  \qw    &\multigate{1}{U}   & \qw &   \push{\cW} \qw & \qw & \qw \\
		&                   & \Nghost{U}  & \Ngate{G}{\ww} & \multigate{1}{T_2} & \\
		& \qw &  \push{\cW}  \qw & \qw &  \ghost{T_2}   &  \push{\cY} \qw  & \qw  \\
	} 	
	=
	\myQ{0.7}{0.5}{
		& \push{\cX}  \qw    &\multigate{1}{U}   & \qw &   \push{\cW} \qw & \qw & \qw \\
		&                   & \Nghost{U}  & \Ngate{G'}{\ww} & \multigate{1}{T'_2} & \\
		& \qw &  \push{\cW}  \qw & \qw &  \ghost{T'_2}   &  \push{\cY} \qw  & \qw  \\
	} \quad, 
	\end{align}
	
so universality of $U$ implies that  $	\scalemyQ{.8}{0.7}{0.5}{& \Ngate{G}{\ww} & \multigate{1}{T_2} & \\
 &  \push{\cW}  \qw  &  \ghost{T_2}   &  \push{\cY} \qw  & \qw  
} = \scalemyQ{.8}{0.7}{0.5}{& \Ngate{G'}{\ww} & \multigate{1}{T'_2} & \\
&  \push{\cW}  \qw  &  \ghost{T'_2}   &  \push{\cY} \qw  & \qw  
} $.  By \cref{eq:T1T2}, we must then have

		\begin{align}
	\begin{split}
	\myQ{0.7}{0.5}{
		& \push{\cX}  \qw    &\multigate{1}{T_1}   & \push{\cW} \qw &   \multigate{1}{T_2} & \push{\cY} \qw & \qw \\
		&                 & \Nghost{T_1}  & \push{\cZ} \qw &  \ghost{T_2}
	} 	
	=& \myQ{0.7}{0.5}{
		& \push{\cX}  \qw    &\multigate{1}{U}   & \push{\cW} \qw &   \multigate{1}{T_2} & \push{\cY} \qw & \qw \\
		&                 & \Nghost{U}  & \gate{G} &  \ghost{T_2}
	} 	\\ &= 
	\myQ{0.7}{0.5}{
		& \push{\cX}  \qw    &\multigate{1}{U}   & \push{\cW} \qw &   \multigate{1}{T'_2} & \push{\cY} \qw & \qw \\
		&                 & \Nghost{U}  & \gate{G'} &  \ghost{T'_2}
	} 	= 
	\myQ{0.7}{0.5}{
		& \push{\cX}  \qw    &\multigate{1}{T'_1}   & \push{\cW} \qw &   \multigate{1}{T'_2} & \push{\cY} \qw & \qw \\
		&                 & \Nghost{T'_1}  & \push{\cZ'} \qw &  \ghost{T'_2}
	} 	,
	\end{split}
	\end{align}

as desired. \end{proof}

Let us now argue  that \cref{lem:UnivCont} implies that contractions are generally independent of stencil-representations. \\

The first thing to realise is that \cref{lem:UnivCont} entails the implication of \cref{eq:simpContCon} by  \cref{eq:simpContAss}. It follows clearly from the lemma that the \myuline{channels} on each side will be the same, so only the causal specifications need to be accounted for. However, we can think of causal specifications as \emph{transformations} in a theory $\Theory'$ in which the \emph{systems} are interfaces, with serial and parallel composition given as in \cref{def:ICC}. This theory can be regarded a sub-theory of $\Sets^*$ since causal specifications are ultimately functions between sets,\footnote{Note that though $\ports{\bbZ}$ may be empty, the set $\pow{\ports{\bbZ}}$ is always non-empty, and $\pow{\ports{\bbZ_1 \cup \bbZ_2}} \cong \pow{\ports{\bbZ_1}} \times \pow{\ports{\bbZ_2}}$ when $\bbZ_1$ and $\bbZ_2$ are parallelly composable (i.e. disjoint).} and as the theory $\Sets^*$ is cartesian \cref{lem:UnivCont} itself applies to show that an interrupted serial composition \myuline{of causal specifications} can be completed, which exactly yields the desired. 

Next, we must realise that the implication of  \cref{eq:simpContCon} by  \cref{eq:simpContAss} is enough to show general independence of the stencil-representation. More precisely, given a stencil-representation $(\fF, G)$ of $(T, \scrC): \bbX \to \bbY$ in which $\bbP \subseteq \bbX \cap \bbY$ is contractible, let us denote by $(G)_\bbP$ the stencil that arises when contracting $G$ along the wires corresponding $\bbP$; the boxes in $(G)_{\bbP}$ are the same as those in $G$ so we can consider $\fF$ as a filling of $(G)_{\bbP}$, and the value $\fF[(G)_{\bbP}] $ is precisely the causal channel  from  $\bbX \setminus \bbP$ to $\bbY \setminus \bbP$ that corresponds to contracting $\bbP$ in the representation $(\fF, G)$. What we must show is that $\fF[(G)_{\bbP}] = \fF'[(G')_{\bbP}]$ whenever $(\fF, G)$ and $(\fF', G')$ are stencil-representations of $(T, \scrC)$ in which $\bbP$ is contractible. This can be done by an induction argument.  Let $P(n)$, with $n \in \N_0$, be the statement that for every causal channel $(T, \scrC)$ and every seemingly contractible interface interface $\bbP$ of size $\abs{\bbP}=n$, we have $\fF[(G)_{\bbP}] = \fF'[(G')_{\bbP}]$ whenever $(\fF, G)$ and $(\fF', G')$ are stencil-representations of $(T, \scrC)$ in which $\bbP$ is contractible. Then, $P(0)$ is true by definition of what a stencil-representation is. And if $P(n)$ is true, then $P(n+1)$ follows. Indeed, in any given stencil-representation the total contraction can be executed by first contracting a single port, say $\sfp \in \ports{\bbP}$, and then the remaining $n$ ports. But by merging channels, the contraction of a single pair of ports always takes the form of going from \eqref{eq:simpContAss} to \eqref{eq:simpContCon}, so for any two stencil-representations $(\fF, G)$ and $(\fF', G')$, we have $\fF[(G)_{\sfp}]= \fF'[(G')_\sfp] =: (S, \scrD)$ and the induction hypothesis then applies to the two stencil-representations $(\fF, (G)_\sfp)$ and $(\fF', (G')_\sfp)$ of $(S, \scrD)$ and yields the desired.

All in all, we have proved the following:

\begin{Thm} (The Standard Notion of Contraction.) \label{thm:StdCont}\\
Let $\Theory$ be a universal theory. If $(T, \scrC): \bbX \to \bbY$ is a causal channel in $\Theory$ which is seemingly contractible along $\bbP \subseteq \bbX \cap \bbY$, then $\fF[(G)_{\bbP}] = \fF'[(G')_{\bbP}]$ for all stencil-representations $(\fF, G)$ and $(\fF', G')$ of $(T, \scrC)$ in which $\bbP$ is contractible. 
\end{Thm}

\subsection{General Notions of Contraction}
\label{subsec:ContGeneral}

In order to use contractions effectively, it is desirable to have a more abstract reformulation of the concept. \\

Ideally, we would like to posit the existence of \emph{contraction maps} $\fC^{\bbX \to \bbY}_\bbP$, which render certain causal channels $(T, \scrC): \bbX \to \bbY$ \emph{contractible along $\bbP$}, and map them to causal channels $\fC^{\bbX \to \bbY}_\bbP((T, \scrC)): \bbX \setminus \bbP \to \bbY \setminus \bbP$, their \emph{contractions along $\bbP$}. In the case of the standard contraction, `contractibility along $\bbP$' would mean simply seeming contractibility along $\bbP$, and the `contraction' would be simply the contraction in any stencil-representation. Abstractly, this interpretation must be enforced by subjecting the maps $\fC^{\bbX \to \bbY}_\bbP$ to certain \emph{coherence conditions}, regulating how contractibility of given causal channels affects that of others, and how their contractions relate to each other.

Such a reformulation is not only desirable on the account of swifter applicability, but also because the standard notion of contraction is somewhat fickle. Consider for example the bipartite causal channels $\scalemyQ{.8}{0.7}{0.5}{& \push{\cX_\sfA} \qw & \multigate{1}{(T, \scrC)} & \push{\cY_\sfA} \qw & \qw \\ & \push{\cX_\sfB} \qw & \ghost{(T, \scrC)} & \push{\cY_\sfB} \qw & \qw }$ in $\QIT$, with the local causal specification given by $\scrC(\sfy_i)=\{\sfx_i\}$. Suppose moreover that the output interface labelled by $\sfA$ matches the input interface labelled by $\sfB$, i.e. that $\sfy_\sfA = \sfx_\sfB$ and $\cY_\sfA = \cX_\sfB$. As we saw in \cref{ex:ContversusCaus}, this common sub-interface is seemingly contractible if and only if $(T, \scrC)$ factors as $\scalemyQ{.8}{0.7}{0.5}{
	& \push{\cX_\sfA}  \qw    &\multigate{1}{(T_1, \scrC_1)}   & \qw &\push{\cY_\sfA}  \qw & \qw\\
	&                   & \Nghost{(T_1, \scrC_1)}  & \multigate{1}{(T_2, \scrC_2)} & \\
	&\push{\cX_\sfB}  \qw &    \qw &  \ghost{(T_2, \scrC_2)}   &  \push{\cY_\sfB} \qw  & \qw  \\
}$. However, as discussed earlier, some causal channels in $\QIT$ admit such a factorisation with $(T_1, \scrC_1)$ and $(T_2, \scrC_2)$ \myuline{in $\QIT^\infty$}, even though they have no such factorisation in $\QIT$. In other words, by slightly enlarging the theory, some channels become constructible, and (in the present case) indeed \myuline{contractible},\footnote{Note that the contraction will belong to $\QIT$ even if $T_1$ and $T_2$ do not, since the systems $\cX_i$ and $\cY_i$ are finite-dimensional.} even though they were not covered by the standard notion of contraction in $\QIT$. 

Finally, abstraction is of general interest as it often allows us to see more clearly what are the important features of a given concept. As such, a rather mathematical example of notions of contraction will be \emph{cancellations} in a thin theory whose corresponding monoid satisfies a cancellation law.\\

It turns out that a reformulation of the precise content of \cref{thm:StdCont} in terms of contraction maps might not be within scope. The reason is that one of the properties that one would like an abstract notion of contraction to have (and which seems necessary to recover the physical contraction of wires, cf. the proof of \cref{prop:AbsStd} below) is `transitivity' of contractibility, namely: If $\bbP_1$ is contractible in $(T, \scrC): \bbX \to \bbY$ and if $\bbP_2$ is contractible in the resulting contraction, $\fC^{\bbX \to \bbY}_{\bbP_1}((T, \scrC))$, then the total interface $\bbP_1 \cup \bbP_2$ is contractible in $(T, \scrC)$. That this property should hold for \myuline{seeming} contractibility is however not obvious, and in fact I do not know whether it is true or false;\footnote{It seems related to the problem raised in \cref{ex:ContGeneric} about whether there always exists a stencil-representation $(\fF,G)$ of $(T, \scrC)$ such that if $\bbP$ is seemingly contractible in $(T, \scrC)$ then $\bbP$ is contractible in the specific representation $(\fF,G)$.} it is stated as an open problem below. 

Instead, what we shall do in providing an abstract reformulation of \cref{thm:StdCont} is to restrict its content to \myuline{constructible} causal channels, since the transitivity property can be proven for those. More general notions of contraction might work for larger collections of causal channels:

\begin{Definition} (Schemes of Causal Channels.) \label{def:Scheme} \\
	Let $\Theory$ be a theory. A \emph{scheme of causal channels in $\Theory$} is a class $\bfE$ of causal channels in $\Theory$ which is closed under serial and parallel compositions, and which contains every constructible causal channel in $\Theory$. 
\end{Definition} 

Physically, a scheme $\bfE$ of causal channels reflects a vision of which causal channels in $\Theory$ we imagine occurring in the world. Mathematically, a scheme is just an aggregate for defining notions of contraction. The largest scheme is $\bfE = \ICC{\Theory}$, the class of all causal channels, and the smallest scheme is $\bfE = \Cns{\Theory}$, the class of constructible causal channels; the latter is the most important example. 

We will often call a causal channel belonging to $\bfE$ simply an \emph{$\bfE$-channel}. 
For interfaces $\bbX$, $\bbY$ in $\Theory$, let us denote by $\bfE(\bbX, \bbY)$ the class of $\bfE$-channels from $\bbX$ to $\bbY$.

\begin{Definition} (Notions of Contraction.) \label{def:AbsCont}\\
Let $\bfE$ be a scheme of causal channels in $\Theory$.  A \emph{notion of contraction in $\bfE$} is a collection $\fC$ of partially defined \emph{contraction maps}, indexed by interfaces $\bbX, \bbY$ and common sub-interfaces $\bbP \subseteq \bbX \cap \bbY$,

	\begin{align}
	\fC^{\bbX\to \bbY}_\bbP: \up{Dom}(\fC^{\bbX\to \bbY}_\bbP) \subseteq \bfE(\bbX, \bbY) \to \bfE(\bbX \setminus \bbP, \bbY \setminus \bbP),
	\end{align}
	
	for which the causal channels in the domain $\up{Dom}(	\fC^{\bbX\to \bbY}_\bbP)$ are called \emph{contractible along $\bbP$}, and which are subject to the following five conditions (abbreviating $	\fC^{\bbX\to \bbY}_\bbP$ by $\fC_\bbP$ when $\bbX$ and $\bbY$ are implicit):
	
	\begin{enumerate}
		
		\item \textbf{Soundness.} If  $\channel{\bbX}{(T, \scrC)}{\bbY}$ and $\channel{\bbY}{(S, \scrD)}{\bbZ}$ are parallelly composable $\bfE$-channels, then their parallel composition is contractible along $\bbY$, and the contraction equals the serial composition, 

		\begin{align}
		\fC_\bbY
		\left(\scalemyQ{.8}{0.7}{0.5}{& \push{\bbX} \qw & \gate{(T, \scrC)} & \push{\bbY} \qw & \qw \\ & \push{\bbY} \qw & \gate{(S, \scrD)} & \push{\bbZ} \qw & \qw }\right) = \scalemyQ{.8}{0.7}{0.5}{& \push{\bbX} \qw & \gate{(T, \scrC)} & \push{\bbY} \qw & \gate{(S, \scrD)} & \push{\bbZ} \qw & \qw}.
		\end{align}

		\item  \textbf{Coherence of Nested Contraction.} An $\bfE$-channel  $\scalemyQ{.8}{0.7}{0.5}{& \push{\bbX} \qw & \multigate{2}{(T, \scrC)} & \push{\bbY} \qw & \qw \\ & \push{\bbP_1} \qw & \ghost{(T, \scrC)} & \push{\bbP_1} \qw & \qw \\ & \push{\bbP_2} \qw & \ghost{(T, \scrC)} & \push{\bbP_2} \qw & \qw}$ is contractible along $\bbP_1 \cup \bbP_2$ if and only if it is contractible along $\bbP_1$ and the resulting contraction is contractible along $\bbP_2$. Moreover, in this case the successive contraction coincides with the total contraction,

		\begin{align}
		\fC_{\bbP_2} \left(\scalemyQ{.8}{0.7}{0.5}{& \push{\bbX} \qw & \multigate{1}{\fC_{\bbP_1}((T, \scrC))} & \push{\bbY} \qw & \qw \\ & \push{\bbP_2} \qw & \ghost{\fC_{\bbP_1}((T, \scrC))} & \push{\bbP_2} \qw & \qw } \right) 
		= 
		\fC_{\bbP_1 \cup \bbP_2} \left( \scalemyQ{.8}{0.7}{0.5}{& \push{\bbX} \qw & \multigate{2}{(T, \scrC)} & \push{\bbY} \qw & \qw \\ & \push{\bbP_1} \qw & \ghost{(T, \scrC)} & \push{\bbP_1} \qw & \qw \\ & \push{\bbP_2} \qw & \ghost{(T, \scrC)} & \push{\bbP_2} \qw & \qw}\right).
		\end{align}
		
		\item \textbf{Freeness of Non-Contracted Interfaces.} If $\scalemyQ{.8}{0.7}{0.5}{& \push{\bbX} \qw & \multigate{1}{(T, \scrC)} & \push{\bbY} \qw & \qw \\ & \push{\bbP} \qw & \ghost{(T, \scrC)} & \push{\bbP} \qw & \qw }$  is contractible along $\bbP$, then for any $\bfE$-channels $\channel{\bbX'}{(S_1, \scrD_1)}{\bbX}$ and $\channel{\bbY}{(S_2, \scrD_2)}{\bbY'}$ the causal channel $\scalemyQ{.8}{0.7}{0.5}{& \push{\bbX'} \qw & \gate{(S_1, \scrD_1)} &  \multigate{1}{(T, \scrC)} & \gate{(S_2, \scrD_2)} & \push{\bbY'} \qw & \qw \\  & & \push{\bbP} \qw & \ghost{(T, \scrC)} &   \push{\bbP} \qw & \qw }$ is contractible along $\bbP$, and contraction commutes with the processing of the non-contracted interfaces, 

		\begin{align}
		\fC_\bbP \left(  \scalemyQ{.8}{0.7}{0.5}{& \push{\bbX'} \qw & \gate{(S_1, \scrD_1)} &  \multigate{1}{(T, \scrC)} & \gate{(S_2, \scrD_2)} & \push{\bbY'} \qw & \qw \\ & & \push{\bbP} \qw & \ghost{(T, \scrC)} &   \push{\bbP} \qw & \qw } \right) 
		=  \scalemyQ{.8}{0.7}{0.5}{& \push{\bbX'} \qw & \gate{(S_1, \scrD_1)} &  \gate{	\fC_\bbP((T, \scrC))} & \gate{(S_2, \scrD_2)} & \push{\bbY'} \qw & \qw }.
		\end{align}

		\item \textbf{Freeness of Parallel Composition.} If $\scalemyQ{.8}{0.7}{0.5}{& \push{\bbX} \qw & \multigate{1}{(T, \scrC)} & \push{\bbY} \qw & \qw \\ & \push{\bbP} \qw & \ghost{(T, \scrC)} & \push{\bbP} \qw & \qw }$ and $\channel{\bbZ}{(S, \scrD)}{\bbW}$ are parallelly composable $\bfE$-channels, then $\scalemyQ{.8}{0.7}{0.5}{& \push{\bbX} \qw & \multigate{1}{(T, \scrC)} & \push{\bbY} \qw & \qw \\ & \push{\bbP} \qw & \ghost{(T, \scrC)} & \push{\bbP} \qw & \qw }$ is contractible along $\bbP$ if and only if $\scalemyQ{.8}{0.7}{0.5}{& \push{\bbX} \qw & \multigate{1}{(T, \scrC)} & \push{\bbY} \qw & \qw \\ & \push{\bbP} \qw & \ghost{(T, \scrC)} & \push{\bbP} \qw & \qw  \\ & \push{\bbZ} \qw & \gate{(S, \scrD)} & \push{\bbW} \qw & \qw}$ is contractible along $\bbP$, and in that case contraction commutes with the parallel composition, 

		\begin{align}
		\fC_\bbP \left( \scalemyQ{.8}{0.7}{0.5}{& \push{\bbX} \qw & \multigate{1}{(T, \scrC)} & \push{\bbY} \qw & \qw \\ & \push{\bbP} \qw & \ghost{(T, \scrC)} & \push{\bbP} \qw & \qw  \\ & \push{\bbZ} \qw & \gate{(S, \scrD)} & \push{\bbW} \qw & \qw} \right)  =\scalemyQ{.8}{0.7}{0.5}{& \push{\bbX} \qw & \gate{	\fC_\bbP((T, \scrC))} & \push{\bbY} \qw & \qw \\ & \push{\bbZ} \qw & \gate{(S, \scrD)} & \push{\bbW} \qw & \qw }.
		\end{align}

		\item \textbf{Well-Foundedness.} If $\channel{\bbX}{(T, \scrC)}{\bbY}$ is contractible along $\bbP \subseteq \bbX \cap \bbY$, then $\sfp \notin \scrC(\sfp)$ for all $\sfp \in \ports{\bbP}$.
		
	\end{enumerate}
	
\end{Definition}

To digest this definition, several observations are in order.

\begin{Remark} (On the Significance of Well-Foundedness.) \\
	In comparison to the remaining conditions, Well-Foundedness might seem a strange condition to highlight. However, we will use it at a crucial point in the proof of \cref{thm:Contractionsofdilations}.
\end{Remark}

\begin{Remark} (On the Statement and Significance of Soundness.) \label{rem:SoundSigni} \\
	The conditions have not been stated in their weakest form under their mutual presence. For example, we could have restricted the statement of the Soundness condition to the case where both $(T, \scrC)$ and $(S, \scrD)$ are identity channels between simple interfaces. Since a general identity channel $(\id_\bbY, \scrI_\bbY)$ can be realised as a parallel composition of identity channels between simple interfaces, Freeness of Parallel Composition along with Coherence of Nested Contraction would then extend the scope to this case, and by Freeness of Non-Contracted Interfaces the case of arbitrary $(T, \scrC)$ and $(S, \scrD)$ would follow. 
	
		In fact, using a suitable combination of conditions it follows even more generally that a channel of the form $\scalemyQ{.8}{0.5}{0.5}{
		& \push{\bbX}  \qw    &\multigate{1}{(T_1, \scrC_1)}   & \qw &   \push{\bbP} \qw & \qw & \qw \\
		&                   & \Nghost{(T_1, \scrC_1)}  & \push{\bbH} \qw & \multigate{1}{(T_2, \scrC_2)} & \\
		& \qw &  \push{\bbP}  \qw & \qw &  \ghost{(T_2, \scrC_2)}   &  \push{\bbY} \qw  & \qw  \\
	}$ is contractible along $\bbP$ with contraction given by $\scalemyQ{.8}{0.5}{0.5}{
		& \push{\bbX}  \qw    &\multigate{1}{(T_1, \scrC_1)}   & \push{\bbP} \qw &   \multigate{1}{(T_2, \scrC_2)} & \push{\bbY} \qw & \qw \\
		&                 & \Nghost{(T_1, \scrC_1)}  & \push{\bbH} \qw &  \ghost{(T_2, \scrC_2)}
	} 	$, and this requirement would thus have been yet another equivalent way of stating the Soundness condition. \end{Remark}

	Whatever its guise, the point of the Soundness condition is to anchor any notion of contraction in terms of actually contracting wires in stencil-representations, as expressed by the following: 
	
	\begin{Prop} (Abstract Notions of Contractions Generalise the Standard Notion.) \label{prop:AbsStd}\\
		Let $\fC$ be a notion of contraction in $\bfE$. Suppose that $(T, \scrC): \bbX \to \bbY$ is a causal channel which has a stencil-representation $(\fF, G)$ using a filling with $\bfE$-channels, and that $\bbP \subseteq \bbX \cap \bbY$ is contractible in the representation $(\fF, G)$  in the sense of \cref{def:seemCont}. Then, $(T, \scrC)$ is contractible along $\bbP$ according to $\fC$ (i.e. $(T, \scrC) \in \up{Dom}(\fC^{\bbX \to \bbY}_\bbP)$) and the contraction $\fC_\bbP((T, \scrC))$ is the causal channel that arises from contracting $\bbP$ in the stencil-representation $(\fF, G)$.
		
		\end{Prop}
	
	\begin{proof}
The statement is proved by induction on $n = \abs{\ports{\bbP}}$. 

For $n=0$, i.e. $\bbP=\bbI$, what we must show is simply that $\bbI$ is contractible in $(T, \scrC)$ according to $\fC$, with $\fC_\bbI((T, \scrC))= (T, \scrC)$. But by Soundness, the causal channel $\id_\bbI$ is contractible along $\bbI$ with contraction $\id_\bbI$, and by Freeness of Parallel Composition the desired then follows.

As for the induction step, suppose the statement is always true when $\abs{\ports{\bbP}}=n$. If $\abs{\ports{\bbP'}} = n+1$, then by the induction hypothesis and the `if'-direction in Coherence of Nested Contraction it suffices to show that a single port $\sfp \in \ports{\bbP'}$ is contractible according to $\fC$, and that the contraction is given by contracting $\sfp$ in any stencil-representation. In other words, the induction step is really the case $n=1$. But in the case $n=1$, any contraction in a stencil-representation takes the form of going from  $\scalemyQ{.8}{0.7}{0.5}{
	& \push{\bbX}  \qw    &\multigate{1}{(T_1, \scrC_1)}   & \qw &   \push{\sfp} \qw & \qw & \qw \\
	&                   & \Nghost{(T_1, \scrC_1)}  & \push{\bbH} \qw & \multigate{1}{(T_2, \scrC_2)} & \\
	& \qw &  \push{\sfp}  \qw & \qw &  \ghost{(T_2, \scrC_2)}   &  \push{\bbY} \qw  & \qw  \\
}$ to $\scalemyQ{.8}{0.7}{0.5}{
	& \push{\bbX}  \qw    &\multigate{1}{(T_1, \scrC_1)}   & \push{\sfp} \qw &   \multigate{1}{(T_2, \scrC_2)} & \push{\bbY} \qw & \qw \\
	&                 & \Nghost{(T_1, \scrC_1)}  & \push{\bbH} \qw &  \ghost{(T_2, \scrC_2)}
} 	$, and by \cref{rem:SoundSigni} such a contraction is indeed rendered possible by $\fC$, and yields the correct contraction. 

		\end{proof}
	
	An immediate corollary of \cref{prop:AbsStd} is that contractions of seemingly contractible interfaces are independent of the chosen representation. However,  as mentioned above, it is not clear that this statement can replace \cref{thm:StdCont}. The problem is essentially that `the standard notion of contraction' might not literally be a notion of contraction in the sense of \cref{def:AbsCont}. More precisely, if in a universal theory $\Theory$ we take $\up{Dom}(\fC^{\bbX \to \bbY}_\bbP)$ to be the class of all causal channels $(T, \scrC): \bbX \to \bbY$ in which $\bbP$ is seemingly contractible, and if we define for that class the contraction $\fC_\bbP((T, \scrC)): \bbX \setminus \bbP \to \bbY\setminus \bbP$ as the contraction in any valid representation, then, whereas the conditions 1., 3., 4., 5. and the `only if'-direction in 2.  can be rather easily proved, I do not know a proof in condition 2. of the `if'-direction:
	
	\begin{OP} (Existence of Notions of Contraction.) \label{op:Contractions}\\
		Suppose that $\Theory$ is universal. Does there exist a notion of contraction in $\bfE = \ICC{\Theory}$, the scheme of all causal channels?
		\end{OP}

What we \myuline{can} obtain is a notion of contraction for \myuline{constructible} channels, and with this we will content ourselves: 

\begin{Thm} (The Standard Notion of Contraction in $\Cns{\Theory}$.) \label{thm:StContisAbsCont}\\
	Suppose that $\Theory$ is a universal theory. Then there exists a notion of contraction in the scheme $\up{Cons}(\Theory)$ of constructible causal channels in $\Theory$.
	\end{Thm}

\begin{proof} %

	We simply define $(T, \scrC): \bbX\to \bbY$ to be an element of the domain $\up{Dom}(\fC^{\bbX \to \bbY}_\bbP)$ (i.e. `contractible along $\bbP$') precisely if $(T, \scrC)$ is seemingly contractible along $\bbP$ in the sense of \cref{def:seemCont}. For $(T, \scrC) \in \up{Dom}(\fC^{\bbX \to \bbY}_\bbP)$, we define the image $\fC^{\bbX \to \bbY}_\bbP((T, \scrC))$ as the contraction in any stencil-representation; by \cref{thm:StdCont}, this is independent of the chosen representation. We have already seen in \cref{ex:ContversusCaus} that 5. (Well-Foundedness) holds, and 1. (Soundness) and 3. (Freeness of Non-Contracted Interfaces) are elementary to verify. The `only if'-directions in 2. (Coherence of Nested Contraction) and 4. (Freeness of Parallel Composition) are clear, and it is elementary that in this case the relevant contractions coincide. The `if'-direction in 4. follows by observing that a stencil-representation of the parallel composition $(T, \scrC) \og (S, \scrD)$ which witnesses seeming contractibility of $\bbP$ factors to witness seeming contractibility of $(T, \scrC)$.
	
	Thus, only the `if'-direction in 2. remains. So far, we have not used a single time that the considered channels are constructible, but we need that now. 
	
	Suppose that $\bbP_1$ is seemingly contractible in $(T, \scrC)$, and that $\bbP_2$ is seemingly contractible in $\fC_{\bbP_1}((T, \scrC))$. We must show that $\bbP_1 \cup \bbP_2$ is seemingly contractible in $(T, \scrC)$. To this end, observe that by constructibility we can pick a stencil-representation $(\fF, G)$ of $(T, \scrC)$ whose filling has only \myuline{primitive} causal channels. By the observation in \cref{ex:ContversusCaus}, such a representation correctly decides contractibility of sub-interfaces, so $\bbP_1$ is contractible in $(\fF, G)$ and contracting $\bbP_1$ yields the representation $(\fF, (G)_{\bbP_1})$ of $\fC_{\bbP_1}((T, \scrC))$. This representation also has only primitive causal channels; thus, again, it correctly decides contractibility of sub-interfaces. In particular, since $\bbP_2$ is seemingly contractible in $\fC_{\bbP_1}((T, \scrC))$, it must be contractible in the representation $(\fF, (G)_{\bbP_1})$; but by definition of what this means -- namely, that $(G)_{\bbP_1}$ does not acquire cycles when contracting $\bbP_2$ -- we see that all of $\bbP_1 \cup \bbP_2$ can be contracted in $(\fF, G)$ without creating cycles. Consequently, $\bbP_1 \cup \bbP_2$ is seemingly contractible in $(T, \scrC)$, as desired.
\end{proof}

In general, it is natural to sum up a notion of contraction along with its scheme and underlying theory:

\begin{Definition} (Contractible Theories.) \\
	A \emph{contractible theory} is a triple $(\Theory, \bfE, \fC)$, where $\Theory$ is a theory, $\bfE$ is a scheme of causal channels in $\Theory$, and $\fC$ is a notion of contraction in $\bfE$. 
\end{Definition}

A point which is both conceptually and technically simplifying is the following:

\begin{Remark} (Serial Composition as a Derived Notion.) \label{rem:SerDerived}\\
If $(\Theory, \bfE, \fC)$ is a contractible theory, then the parallel composition  in $\Theory$ together with $\fC$ actually \emph{defines} the serial composition in $\Theory$, by the Soundness condition for $\fC$. As such, we can think of serial composition as a derived notion, and we shall often do so in the following. 
\end{Remark}

It is important to stress that even though schemes were introduced above mainly to define notions of contraction, they are relevant even if the open problem \ref{op:Contractions} has an affirmative answer. Indeed, if $\fC$ is a notion of contraction in all of $\ICC{\Theory}$, it makes sense to talk about schemes $\bfE$ which are closed under the contraction $\fC$ as models of \emph{those causal channels which are physically implementable}. This will be significant  in particular when we introduce the causal-dilational ordering in \cref{sec:CausDilations}.  \\

Let me conclude the subsection by mentioning a relation between abstract notions of contraction and so-called \emph{traces} in symmetric monoidal categories:

\begin{Remark} (Relation to Traces in Symmetric Monoidal Categories.) \label{rem:Traces}\\
Generalising the ordinary concept of trace in the symmetric monoidal category $(\Vect{k}, \otimes, k)$, Ref. \cite{JSV96} introduced the concept of \emph{(abstract) traces} in general (symmetric)	monoidal categories, operations which map transformations $T: \cX \og \cZ \to \cY \og \cZ$ in the category to transformations $T' : \cX \to \cY$ (thus `swallowing' the object $\cZ$), subjected to certain coherence conditions. Though I was not aware of this work when first developing the theory presented here, there are extremely close connections between abstract traces and abstract notions of contraction. The connection is probably most easily seen by means of the simplification presented in Ref. \cite{Hase09} of the original conditions for traces. 

Specifically, the \emph{Yanking} condition of Ref. \cite{Hase09} corresponds to the Soundness condition in \cref{def:AbsCont}, whereas  the \emph{Tightening} and \emph{Superposing} conditions correspond to the Freeness of Non-Contracted Interfaces and Freeness of Parallel Composition, respectively. The \emph{Vanishing} condition corresponds to Coherence under Nested Contraction. The Well-Foundedness condition has no equivalent, since causality is absent in the framework of traced categories. Likewise, \myuline{any} system is considered contractible, so the statements in \cref{def:AbsCont} about contractibility have no equivalents in that framework either (as we will see in \cref{sec:CausDilations}, those conditions have been carefully chosen).  In fact, a trace in a symmetric monoidal category can be seen as a notion of contraction for which causality (and the condition of Well-Foundedness) is completely disregarded and for which every channel $T: \bbX \to \bbY$ is contractible along any common sub-interface $\bbP \subseteq \bbX\cap \bbY$.  \end{Remark}

\section{Causal Dilations}
\label{sec:CausDilations}

Now that we have introduced the concept of causal channels and notions of contraction, we are ready to introduce a causal version of the theory of \cref{chap:Dilations}. Throughout, $(\Theory, \bfE, \fC)$ will denote a fixed contractible theory. The narrative is that $\bfE$ defines the collection of causal channels which we imagine to be physically possible (the reader is free to think of the example $\bfE = \Cns{\Theory}$, the constructible channels, since this will anyway be the only example we ultimately care about). Whenever we speak of `contractibility' we will mean contractibility according to $\fC$, and implicitly assume that the involved channels are $\bfE$-channels.\footnote{As such, it would often be really  nice if we could take $\bfE= \ICC{\Theory}$, i.e. if there were an affirmative answer to the open problem \ref{op:Contractions}.}  \\

In \cref{subsec:CausDilBasic}, we define the concept of \emph{causal dilations}, consider a lot of examples, and prove three general stability results about causal dilations, most significantly \cref{thm:Contractionsofdilations} which asserts that causal dilations are stable under contractions in the environment. 

Then, in \cref{subsec:CausalDilOrd}, we introduce the causal version of the dilational ordering; this \emph{causal-dilational ordering} will depend on the scheme $\bfE$, since it reflects that one dilation can be derived from another using $\bfE$-channels in the environment. We then show some basic \emph{composability} results, regarding how the causal-dilational ordering interplays with various compositions.

\subsection{Causal Dilations -- Definition and Basic Properties}
\label{subsec:CausDilBasic}

The concept of a causal dilation can in principle be very compactly defined: It is simply the notion of dilation obtained by replacing channels by causal channels. 

More explicitly, recall that the trash channel $\tr_\bbZ$ can be equipped with a unique causal specification, and that when we write $\id_\bbZ$ as a causal channel, we mean really $(\id_\bbZ, \scrI_\bbZ)$, where $\scrI_\bbZ$ is the specification given by $\scrI_\bbZ(\sfJ)=\sfJ$.

\begin{Definition} (Causal Dilations.) \label{def:CausDil}\\
	Let $(T, \scrC): \bbX \to \bbY$ be a causal channel in $\Theory$. A \emph{(causal) dilation of $(T, \scrC)$} is a causal channel $(L, \scrE): \bbX \og \bbD  \to \bbY \og \bbE$ such that $(\id_\bbY \og \tr_\bbE) \after (L, \scrE) = (T, \scrC) \og \tr_\bbD$, with composition in the sense of \cref{def:ICC}. In pictures,

		\begin{align} \label{eq:causdil}
	\myQ{0.7}{0.5}{& \push{\bbX} \qw &  \Nmultigate{1}{(L, \scrE)}{\qw}   & \push{\bbY} \qw  & \qw  \\
		&\push{\bbD} \ww & \Nghost{(L, \scrE)}{\ww} & \push{\bbE} \ww  & \Ngate{\tr}{\ww}
	}		
	\quad   = \quad 
	\myQ{0.7}{0.5}{& \push{\bbX} \qw &   \Ngate{(T, \scrC)}{\qw}    & \push{\bbY} \qw  & \qw  \\
		&\push{\bbD} \ww  & \Ngate{\tr}{\ww}
	}	\quad .
	\end{align}

\end{Definition}

\begin{Remark} (Terminology.) \\
	We will often simply say that $(L, \scrE)$ is a \emph{dilation} of $(T, \scrC)$ (rather than a \emph{\myuline{causal} dilation}), since by including the causal specifications there is no risk of confusion. 
	\end{Remark}

It is instructive to write out explicitly what the condition \eqref{eq:causdil} says about the channels and specifications individually. It is simply the following:  

	\begin{align} 
\myQ{0.7}{0.5}{& \push{\bbX} \qw &  \Nmultigate{1}{L}{\qw}   & \push{\bbY} \qw  & \qw  \\
	&\push{\bbD} \ww & \Nghost{L}{\ww} & \push{\bbE} \ww  & \Ngate{\tr}{\ww}
}		
\quad   = \quad 
\myQ{0.7}{0.5}{& \push{\bbX} \qw &   \Ngate{T }{\qw}    & \push{\bbY} \qw  & \qw  \\
	&\push{\bbD} \ww  & \Ngate{\tr}{\ww}
}	\quad  \text{and}  \quad \scrE \lvert_{\ports{\bbY}} = \scrC \quad;
\end{align}

in other words, $L$ is a dilation of $T$ and $\scrE(\sfJ) = \scrC(\sfJ)$ for all $\sfJ\subseteq \ports{\bbY}$. The requirement on the specifications expresses that in the causal channel $(L, \scrE)$, every set of ports in the accessible output interface $\bbY$ has the precise same causes as they do in the causal channel $(T, \scrC)$. As such, the specification $\scrE$ is completely determined by its restriction to $\ports{\bbE}$, i.e. by the cause sets $\scrE(\sfJ) \subseteq \ports{\bbX \cup \bbD}$ for $\sfJ \subseteq \ports{\bbE}$. In particular, though there may exist intricate causal relationships between the outputs in $\bbE$ and the inputs in $\bbX \cup \bbD$, no port in $\bbY$ can have causes in $\bbD$.\footnote{One could certainly speculate whether this restriction is fair, but it turns out that the dilation concept which results from its absence is not only much more complicated, but also ill-behaved, cf. \cref{ex:NearDil}. The requirement might be physically motivated on the same ground that we require non-signalling from $\bbD$ to $\bbY$ in the concept of a dilation: An implementer of a channel who wants to give the impression that we are interacting with $(T, \scrC)$ through the accessible interface could not exploit functionality which would ruin this impression.} \\

\begin{Example} (Primitive Dilations.) \label{ex:primdil}\\
	If $\channel{\bbX}{(T, \scrC)}{\bbY}$ is any causal channel and $L$ is a dilation of $T$ as a channel, then the causal channel $\twoext{\bbX}{\bbD}{(L, \scrE)}{\bbY}{\bbE}$ whose specification $\scrE$ is determined by $\scrE(\bbE_0) = \bbX \cup \bbD$ for any non-trivial sub-interface $\bbE_0 \subseteq \bbE$, is a causal dilation of $(T, \scrC)$. Intuitively, it corresponds to thinking of the hidden outputs as requiring all inputs in $\bbX \cup \bbD$ to be fed. We will call such a dilation a \emph{primitive causal dilation}, as it is the natural extension of primitive specifications in the sense of \cref{ex:primdil} to the realm of dilations. (Note, however, that $(L, \scrE)$ itself need not be a primitive causal channel; it is primitive if and only if $(T, \scrC)$ is.) 	\end{Example}

\begin{Example} (All Causal Dilations of a Trash.) \label{ex:causexttrash}\\
Consider the channel $\scalemyQ{.8}{0.7}{0.5}{& \push{\bbX} \qw & \gate{\tr}}$, equipped with its unique causal specification. We can characterise all its causal dilations. Evidently, they are simply the causal channels $\scalemyQ{.8}{0.7}{0.5}{& \push{\bbX} \qw & \multigate{1}{(L, \scrE)} \\ & \push{\bbD} \ww & \Nghost{(L, \scrE)}{\ww} & \push{\bbE} \ww & \ww}$, with $\bbD$ and $\bbE$ arbitrary. These can be written in the form

\begin{align}
\myQ{0.7}{0.5}{ & \push{\bbX} \qw & \gate{(\id_\bbX, \scrI_\bbX)} &  \push{\bbX} \ww & \Nmultigate{1}{(L, \scrE)}{\ww} \\ & & & \push{\bbD} \ww & \Nghost{(L, \scrE)}{\ww} & \push{\bbE} \ww & \ww} \quad, 
	\end{align}

and though this observation is trivial, it suggests that the particular causal dilation $\scalemyQ{.8}{0.7}{0.5}{ & \push{\bbX} \qw & \gate{(\id_\bbX, \scrI_\bbX)} &  \push{\bbX} \ww & \ww}$ should be rendered a \emph{complete causal dilation}, in an extension of the completeness notion to causal dilations. 

The reader is encouraged to consider similarly what are all the causal dilations of $\scalemyQ{.8}{0.7}{0.5}{& \push{\bbX} \qw & \gate{\tr}}$ if we replace the trivial output interface $\bbI$ by a non-trivial interface in which some of the ports act as `indicators' of its use, as in \cref{ex:statetrashcaus}. \end{Example}%

\begin{Example} (All Causal Dilations of a State.) \label{ex:causdilstate}\\ 
	Consider a state $\state{s}{\bbY}$, equipped with its unique causal specification. If $\state{s}{\bbY}$ as a channel in $\Theory$ has a complete one-sided dilation $\scalemyQ{.8}{0.7}{0.5}{& \Nmultigate{1}{c} & \push{\bbY} \qw & \qw  \\ & \Nghost{c} & \push{\bbE_0} \ww & \ww}$, then every dilation (disregarding causality) is of the form $\scalemyQ{.8}{0.7}{0.5}{& \Nmultigate{1}{c} & \push{\bbY} \qw & \qw  \\ & \Nghost{c} & \Nmultigate{1}{G}{\ww}  \\ & \push{\bbD} \ww & \Nghost{G}{\ww} & \push{\bbE} \ww}$ for some channel $G$. Now, if $(L, \scrE)$ is a \myuline{causal} dilation of $s$ then necessarily $\scrE(\bbY) = \bbI$ (i.e. no port in $\bbY$ has any causes whatsoever), so it is easy to see that $(L, \scrE)$ must in fact be of the form $\scalemyQ{.8}{0.7}{0.5}{& \Nmultigate{1}{c} & \push{\bbY} \qw & \qw  \\ & \Nghost{c} & \Nmultigate{1}{(G, \scrB)}{\ww}  \\ & \push{\bbD} \ww & \Nghost{(G, \scrB)}{\ww} & \push{\bbE} \ww}$ for some \myuline{causal} channel $(G, \scrB)$. As such, we have a `complete' causal dilation, like in the previous example.\end{Example}

\begin{Example} (Causal-Dilational Purity?)\\
		Recall that a channel is called dilationally pure if every dilation is obtained by parallel composition (\cref{def:DilPure}).  One might be tempted to similarly call dilationally pure a \myuline{causal} channel $\channel{\bbX}{(T, \scrC)}{\bbY}$ with the property that every causal dilation is of the form $\scalemyQ{.8}{0.7}{0.5}{& \push{\bbX} \qw & \gate{(T, \scrC)} & \push{\bbY} \qw & \qw \\ & \push{\bbD} \ww & \Ngate{(S, \scrD)}{\ww} & \push{\bbE} \ww & \ww }$. This, however, is quickly realised to be a rather dull notion: As soon as $\bbX \neq \bbI$, there exists a dilation which extracts as side-information \emph{the fact that some port in $\bbX$ was fed with an input}, i.e. we have the dilation $\scalemyQ{.8}{0.7}{0.5}{& \push{\bbX} \qw & \multigate{1}{(T, \scrC')} & \push{\bbY} \qw & \qw\\ & & \Nghost{(T, \scrC')} & \push{\triv} \ww & \ww}$ with $\scrC'(\sfone) = \ports{\bbX}$, and this dilation does not factor by parallel composition. (If $\bbX$ has several ports, there are other non-trivial dilations as well, since we may have indicators for the different subsets of ports in $\bbX$.) As such, the `dilationally pure' causal  channels would be exactly those causal channels with $\bbX = \bbI$ and for which the underlying channel is a dilationally pure state.

	\end{Example}

\begin{Example} (Causal Dilations of the Bit Refreshment.) \label{ex:bitrefresh}\\
Consider the `bit refreshment' from the general introduction to the thesis, i.e. the causal channel $\channel{\{0,1\}}{(T, \scrC_0)}{\{0,1\}}$ in $\CIT$ with primitive specification $\scrC_0$ and with $T(b) = r$ for any input $b$, where $r \in \St{\{0,1\}}$ is the uniformly random bit. Pictorially, we can represent $(T, \scrC_0)$ as 

\begin{align}
\myQ{0.7}{0.5}{& \push{2} \qw & \gate{\tr} &\push{\triv} \qw & \gate{r} & \push{2} \qw & \qw} \quad,
\end{align}

abbreviating the system $\{0,1\}$ as `$2$', and with the understanding that each component is given its primitive specification. One causal dilation of this channel is 

\begin{align} \label{eq:dil1}
\myQ{0.7}{0.5}{& \qw& \push{2} \qw & \multigate{1}{\id} & \ww &  \push{2} \ww & \ww \\ && & \Nghost{\id} & \push{\triv} \qw & \multigate{1}{\id} &\push{2} \qw  & \qw \\&
	& & & \push{2} \qw & \ghost{\id} \\ & \Ngate{r}   & \push{2} \ww &  \Nctrlo{-1}{\ww}  & \push{2} \ww & \ww} \quad,
\end{align}

each component with its primitive specification, one identity thus stalling the other. It corresponds to a scenario in which the agents controlling the environment draw a random bit, and upon our input to the open interface gives us a copy of that bit, while storing our input in their memory. Precisely, the specification $\scrE$ of this dilation is such that the two upper output ports have the input as a cause, whereas the lower output port has no causes. This reflects the fact that the random bit can be drawn in advance of seeing our input. 

Another causal dilation is given by

\begin{align} \label{eq:dil2}
\myQ{0.7}{0.5}{& &  &  & \ww &  \push{2} \ww & \ww \\ & \qw& \push{2} \qw & \ctrlo{-1} & \push{2} \qw & \multigate{1}{\up{XOR}} &\push{2} \qw  & \qw \\&
	& & & \push{2} \qw & \ghost{\up{XOR}} \\ & \Ngate{r}   & \push{2} \ww &  \Nctrlo{-1}{\ww}  & \push{2} \ww & \ww} \quad
\end{align}

(again equipping each component with its primitive specification), corresponding to a scenario in which the agents draw a random bit and use it to decide whether or not to give us back as output our original input, or to flip it. 

Now, disregarding causality altogether, both of the channels \eqref{eq:dil1} and \eqref{eq:dil2} are easily seen to be complete dilations of $T$, and as such they are equivalent in the dilational ordering of \cref{sec:DilOrd}. \emph{However}, what we shall do in a moment is to introduce a causal version of this dilational ordering, and it will \myuline{not} be possible to go from one \myuline{causal} dilation to the other in a way that respects the causality: As causal dilations, \eqref{eq:dil1} and \eqref{eq:dil2} are simply \myuline{different}, formalising the intuition that the side-information in one dilation (pre-existing knowledge of which bit will be given as output) is information about something entirely different than the side-information  in the other (pre-existing knowledge of whether or not the input will be flipped). 	\end{Example} 

\begin{Example} (Acausal Side-Information.) \label{ex:Acausal}\\
Let $\scalemyQ{.8}{0.7}{0.5}{& \push{\bbX} \qw &  \Nmultigate{1}{(L, \scrE)}{\qw}   & \push{\bbY} \qw  & \qw  \\
		&\push{\bbD} \ww & \Nghost{(L, \scrE)}{\ww} & \push{\bbE} \ww  & \ww}
	$ be a causal dilation of $\channel{\bbX}{(T, \scrC)}{\bbY}$. A sub-interface $\bbE_0 \subseteq \bbE$ is said to be \emph{acausal} if $\scrE(\bbE_0) \subseteq \bbD$, or, equivalently, if $\scrE(\bbE_0) \cap \bbX = \bbI$. The output at an acausal interface $\bbE_0$ represents the information that can be available at the hidden interface \myuline{before} the open interface $\bbX$ of the channel has been used. For instance, in \cref{ex:bitrefresh} the copies in the environment of the random bits were acausal. 

We will say a dilation $(L, \scrE)$ \emph{has no acausal side-information} if $\bbI$ is the only acausal sub-interface of $\bbE$, i.e. if $\scrE(\sfe) \cap \ports{\bbX} \neq \emptyset$ for all $\sfe \in \ports{\bbE}$. In such a dilation, no hidden outputs are available before the accessible interface has been used. 

On the other hand, a dilation $(L, \scrE)$ for which \myuline{all} of $\bbE$ is acausal is called simply an \emph{acausal dilation}. Every trivial dilation $\scalemyQ{.8}{0.7}{0.5}{& \push{\bbX} \qw & \gate{(T, \scrC)} & \push{\bbY} \qw & \qw \\ & \push{\bbD} \ww & \Ngate{(S, \scrD)}{\ww} & \push{\bbE} \ww & \ww }$ is acausal, but there may be other acausal dilations. For example, if $\channel{\bbX}{T}{\bbY}$ can be represented as $\scalemyQ{.8}{0.7}{0.5}{& \push{\bbX} \qw & \multigate{1}{\tilde{T}} & \push{\bbY}\qw & \qw \\ & \Ngate{p} & \ghost{\tilde{T}}}$ for some channel $\tilde{T}$ and some state $p$ (e.g. in $\CIT$ or $\QIT$, if $T$ admits a convex decomposition with weights described by $p$), then for any dilation $\scalemyQ{.8}{0.7}{0.5}{& \Nmultigate{1}{t} & \push{\bbH} \qw& \qw \\ & \Nghost{t} & \push{\bbE}\ww & \ww}$ of $\state{p}{\bbH}$ the causal channel $\scalemyQ{.8}{0.7}{0.5}{& \push{\bbX} \qw & \qw &  \multigate{1}{(\tilde{T}, \tilde{\scrC})} & \push{\bbY}\qw & \qw \\ & \Nmultigate{1}{t} & \push{\bbH} \qw &\ghost{(\tilde{T}, \tilde{\scrC})} \\ & \Nghost{t} & \ww&  \push{\bbE} \ww & \ww}$, with $\tilde{\scrC}$ given by $\tilde{\scrC}(\bbY_0) = \scrC(\bbY_0) \cup \bbH$ for $\bbY_0 \subseteq \bbY$ non-trivial, is an acausal dilation of $(T, \scrC)$, assuming that $\tilde{T}$ is compatible with $\tilde{\scrC}$. In general (for example, if $\tilde{T}$ and $p$ constitute a non-trivial convex decomposition of $T$) this dilation does not factor as a trivial dilation.

As we will see in \cref{chap:Selftesting}, quantum self-testing ensures that all acausal dilations of certain channels are trivial, and this has the interpretation that any randomness produced by those channels is \emph{fresh}, independent of pre-existing randomness (\cref{prop:SecSelfTest}). By the preceding considerations, it also implies that the channel at the open interface has only trivial convex decompositions, i.e. is extremal (\cref{prop:SecExt}).

 \end{Example}

\begin{Example} (Causal Dilations of a Quantum Measurement?) \label{ex:MeasDil}\\
	Consider in $\QIT$ the decoherence channel $\channel{\C^2}{\Delta}{\C^2}$ given by $\Delta(A) = \ketbra{0}A\ketbra{0}+ \ketbra{1}A\ketbra{1}$ for $A \in \End{\C^2}$. The channel $\Delta$ models a measurement in the computational basis (\cite{NC02}). Let us equip $\Delta$ with the primitive causal specification $\scrC_0$ (this is the only compatible specification anyway). As in \cref{ex:primdil}, we obtain of course for any dilation $\oneext{\C^2}{\Phi}{\C^2}{\bbE}$ of $\Delta$ as a channel, a causal dilation by simply equipping $\Phi$ with its primitive specification. 
	
	There is, however, another dilation, namely one which is acausal: Based on the convex decomposition $\Delta = \frac{1}{2} \id_{\C^2} + \frac{1}{2} Z$, where $Z$ is the channel given by conjugation by the Pauli $z$-unitary $\sigma_z := \left(\begin{array}{cc} 1 & 0 \\ 0 & -1 \end{array} \right)$, we can write $\channel{\C^2}{\Delta}{\C^2}$ as $\scalemyQ{.8}{0.7}{0.5}{& \qw & \push{\C^2} \qw & \multigate{1}{\Gamma} & \push{\C^2} \qw & \qw \\ & \Ngate{\tau}  & \push{\C^2} \qw & \ghost{\Gamma}}$, where $\tau := \frac{1}{2} \ketbra{0}+\frac{1}{2} \ketbra{1} $ is the fully mixed state on $\C^2$, and where $\Gamma$ measures the lower system to determine whether to apply $\id_{\C^2}$ or $Z$ to the upper system. But then, letting $K: \C^2 \to \C^2 \otimes \C^2$ denote the embedding in $\QIT$ of the classical copy-channel,  i.e. the quantum channel given by $K(A) = \bra{0}A \ket{0} \ketbra{0} \otimes \ketbra{0}+ \bra{1}A \ket{1} \ketbra{1} \otimes \ketbra{1}$, the state $\scalemyQ{.8}{0.7}{0.5}{& \Ngate{\tau}  & \multigate{1}{K}  & \qw\\ & & \Nghost{K} &  \ww }$ is a dilation of $\tau$ (corresponding to a copy), and following \cref{ex:Acausal} we thus obtain the causal dilation  
	
	\begin{align} \label{eq:measdil}
	\myQ{0.7}{0.5}{& \qw & \push{\C^2} \qw & \multigate{1}{\Gamma} & \push{\C^2} \qw & \qw \\ & \Ngate{\tau}  & \multigate{1}{K}  & \ghost{\Gamma} \\&  & \Nghost{K} & \push{\C^2} \ww & \ww}
	\end{align}
	
of $\channel{\C^2}{\Delta}{\C^2}$, where each component is equipped with its primitive specification. In equations, the channel \eqref{eq:measdil} is given by $A \mapsto \frac{1}{2}  A \otimes \ketbra{0} + \frac{1}{2} \sigma_z A \sigma^*_z \otimes \ketbra{1}$. 

This dilation is strange in the context of measurements since it is acausal, and thus represents side-information available before the execution of the measurement; the dilation formalises the curious circumstance that a quantum measurement can -- mathematically -- be implemented by tossing a fair coin and using the outcome to decide whether to apply a Pauli $z$-conjugation or do nothing to the system at hand. There are other such strange causal dilations, indeed one for each convex decomposition of $\Delta$ (for example, we also have $\Delta = \frac{1}{2} X + \frac{1}{2}Y$, where $X$ is $Y$ are conjugations by the Pauli $x$- and $y$-unitaries, respectively.)

 It would seem that there is no sensible way in which these strange dilations really reflect the nature of quantum measurements, and I do not know how to interpret them other than by refuting them as valid dilations when thinking of $\Delta$ as a measurement. As we will see in \cref{chap:Selftesting}, the strangeness is not purely philosophical, but constitutes a mathematical nuisance when establishing the precise connection between quantum self-testing and the theory of causal dilations.
	
	\end{Example}

We now show three stability results about causal dilations, two of which are simple to prove and the third of which is highly subtle, representing perhaps the most surprising result about causal dilations at all. \\

The first result is immediate from the definition, like it was in the causality-free setting: 

\begin{Prop} (Stability of Causal Dilations under Parallel Composition.) \label{prop:ParaStab}\\
Suppose that$\twoext{\bbX_1}{\bbD_1}{(L_1, \scrE_1)}{\bbY_1}{\bbE_1}$ is a causal dilation of $\channel{\bbX_1}{(T_1, \scrC_1)}{\bbY_1}$ and $\twoext{\bbX_2}{\bbD_2}{(L_2, \scrE_2)}{\bbY_2}{\bbE_2}$ is a causal dilation of $\channel{\bbX_2}{(T_2, \scrC_2)}{\bbY_2}$. Then,

	\begin{align}
	\myQ{0.7}{0.5}{ & \push{\bbX_1}  \qw & \multigate{1}{(L_1, \scrE_1)} & \push{\bbY_1} \qw & \qw \\ 
		& \push{\bbD_1} \ww & \Nghost{(L_1, \scrE_1)}{\ww} & \push{\bbE_1} \ww  & \ww   \\ 
		& \push{\bbD_2} \ww& \Nmultigate{1}{(L_2, \scrE_2)}{\ww} & \push{\bbE_2} \ww & \ww \\ & \push{\bbX_2}  \qw & \ghost{(L_2, \scrE_2)} & \push{\bbY_2} \qw  & \qw} 
	\end{align} 
	
	is a causal dilation of $\scalemyQ{.8}{0.7}{0.5}{& \push{\bbX_1} \qw & \gate{(T_1, \scrC_1)} & \push{\bbY_1} \qw & \qw \\ & \push{\bbX_2} \qw & \gate{(T_2, \scrC_2)} & \push{\bbY_2} \qw & \qw }$.
	\end{Prop}

The next obvious result to prove would be that also the serial composition of causal dilations is a causal dilation, and this is immediate from the definition too. However, as discussed earlier, serial composition can be derived from parallel composition and contraction, so given \cref{prop:ParaStab} it is stronger to state a stability result about \myuline{contractions} in the accessible interface:

\begin{Prop} (Stability of Causal Dilations under Accessible Contractions.)  \label{prop:ContStab}\\
Suppose that $\twoext{\bbX}{\bbD}{(L, \scrE)}{\bbY}{\bbE}$ is a causal dilation of $\channel{\bbX}{(T, \scrC)}{\bbY}$, and that $\bbP$ is a common sub-interface of $\bbX$ and $\bbY$. Then, $\bbP$ is contractible in $(T, \scrC)$ if and only if it is contractible in $(L, \scrE)$, and in that case the contraction

\begin{align}
\myQ{0.7}{0.5}{	& \push{\bbD} \ww& \Nmultigate{1}{\contr{(L, \scrE)}{\bbP}}{\ww} & \push{\bbE} \ww & \ww \\ & \push{\bbX \setminus \bbP}  \qw & \ghost{\contr{(L, \scrE)}{\bbP}} & \push{\bbY \setminus \bbP} \qw  & \qw} 
\end{align} 

is a causal dilation of the contraction $\channel{\bbX \setminus \bbP}{\contr{(T, \scrC)}{\bbP}}{\bbY \setminus \bbP}$.
	\end{Prop}

\begin{proof}
Consider the causal channel

\begin{align} \label{eq:diltrash}
\myQ{0.7}{0.5}{& \push{\bbE} \ww&  \Ngate{\tr}{\ww} \\	& \push{\bbD} \ww& \Nmultigate{1}{(L, \scrE)}{\ww} & \push{\bbE} \ww & \ww \\ & \push{\bbX}  \qw & \ghost{(L, \scrE)} & \push{\bbY} \qw  & \qw} \quad.
\end{align} 

By Soundness it is contractible along $\bbE$, and its contraction along $\bbE$ is given by

\begin{align} \label{eq:dilcont}
\myQ{0.7}{0.5}{& \push{\bbD} \ww& \Ngate{\tr}{\ww} \\	& \push{\bbX}  \qw & \gate{(T, \scrC)} & \push{\bbY} \qw  & \qw} \quad,
\end{align} 

since $(L, \scrE)$ dilates $(T, \scrC)$. Now, by Freeness of Parallel Composition, $\bbP$ is contractible in $(T, \scrC)$ if and only if it is contractible in the channel \eqref{eq:dilcont}. But by Nested Contraction, $\bbP$ is contractible in \eqref{eq:dilcont} if and only if it is contractible in \eqref{eq:diltrash}. And this again is the case if and only if $\bbP$ is contractible in $(L, \scrE)$. Moreover, in the affirmative case, Nested Contraction implies that the order of contraction of along $\bbP$ and $\bbE$ can be interchanged, so 

\begin{align} 
\myQ{0.7}{0.5}{& \push{\bbE} \ww&  \Ngate{\tr}{\ww} \\	& \push{\bbD} \ww& \Nmultigate{1}{\contr{(L, \scrE)}{\bbP}}{\ww} & \push{\bbE} \ww & \ww \\ & \push{\bbX \setminus \bbP}  \qw & \ghost{\contr{(L, \scrE)}{\bbP}} & \push{\bbY \setminus \bbP} \qw  & \qw} \quad
\end{align} 

contracts along $\bbE$ to yield

\begin{align} 
\myQ{0.7}{0.5}{& \push{\bbD} \ww& \Ngate{\tr}{\ww} \\	& \push{\bbX \setminus \bbP}  \qw & \gate{\contr{(T, \scrC)}{\bbP}} & \push{\bbY \setminus \bbP} \qw  & \qw} \quad,
\end{align} 

which by Soundness is precisely to say that $\contr{(L, \scrE)}{\bbP}$ is a dilation of $\contr{(T, \scrC)}{\bbP}$.
	\end{proof}

The third and final result is the most surprising. It also concerns contractions, but in the \myuline{hidden} rather than the accessible interface, and I should like to stress two points which serve to illustrate that it is non-trivial.\\

First of all, the result implies a surprising modularity property: If the causal channel $(T_1, \scrC_1)$ is implemented by means of the dilation $(L_1, \scrE_1)$, and, meanwhile, $(T_2, \scrC_2)$ is implemented by means of the dilation $(L_2, \scrE_2)$, and maybe $(T_3, \scrC_3)$, $(T_4, \scrC_4)$, and so forth, are also implemented, some in parallel with each other, and others in series (or by contraction), then \cref{prop:ParaStab} and \cref{prop:ContStab} imply that the total network comprised by all of the dilations, no matter how complicated it might be, is one big causal dilation of the channel connecting the accessible interfaces. The result we are now about to prove then shows that this remains true even if the implementer should choose to connect in all sorts of obscure ways the hidden wires of this dilation; in other words, the implementer is free to do anything in the environment and we will not be able to detect this at the accessible interfaces. 

Secondly, as the following example demonstrates, the result is dangerously close to being false, bearing witness to the fact that the definition of causal dilations was delicately chosen:

\begin{Example} (Hidden Contractions of Near-Dilations may Cease to be Near-Dilations.)\label{ex:NearDil}\\
Suppose we had defined the concept of a causal dilation slightly more liberally: Let us say that $(L, \scrE)$ is a \emph{near-dilation} of $(T, \scrC)$, if $L$ is a dilation of $T$ and $\scrE(\sfJ) \cap \ports{\bbX} = \scrC(\sfJ)$ for all $\sfJ\subseteq \ports{\bbY}$ (rather than $\scrE(\sfJ) = \scrC(\sfJ)$ for all $\sfJ\subseteq \ports{\bbY}$). The requirement is loosened so that while the causes of $ \sfy \in \ports{\bbY}$ \myuline{within $\bbX$} remain as prescribed by $\scrC$, $\sfy$ may have causes in $\bbD$. (Still, however, $L$ will be non-signalling from $\bbD$ to $\bbY$ as a channel, since $L$ is a dilation of $T$). The concept of near-dilations is unstable under hidden contractions:

Consider the bit refreshment causal channel $\myQ{0.7}{0.5}{& \push{2} \qw & \gate{\tr} &\push{\triv} \qw & \gate{r} & \push{2} \qw & \qw}$ from \cref{ex:bitrefresh}, and consider the causal channel $(L, \scrE)$ given by

	\begin{align} \label{eq:neardil}
	\myQ{0.7}{0.5}{
		& \push{2}  \ww   & \Nmultigate{1}{\up{XOR}}{\ww} & \push{2}  \qw & \qw \\
		& \Nmultigate{1}{\kappa}   & \ghost{\up{XOR}} &  \\
		& \Nghost{\kappa}   & \multigate{1}{\up{XOR}} \\
		& \push{2}  \qw  & \ghost{\up{XOR}} & \push{2} \ww & \ww \\
	} \quad, 
\end{align}

where $\kappa$ denotes two copies of the uniformly random bit, i.e. the state $\scalemyQ{.8}{0.7}{0.5}{& & &  \qw \\& \Ngate{r} & \ctrlo{-1} & \qw }$, and where each component is equipped with its primitive specification. By the encrypting property of the one-time pad (\cref{ex:onetimepad}), the channel $L$ dilates $T$. Moreover, the specification $\scrE$ is easily seen to have the property required for $(L, \scrE)$ to be a near-dilation of $(T, \scrC)$. If we give the input and output ports in the hidden interface of $(L, \scrE)$ the same name, then the pair is contractible, and the resulting contraction is 

	\begin{align}
\myQ{0.7}{0.5}{
	& \Nmultigate{1}{\kappa}   & \qw & \push{2} \qw&  \multigate{2}{\up{XOR}} &  \\
	& \Nghost{\kappa}   & \multigate{1}{\up{XOR}} & & \Nghost{\up{XOR}} \\
	& \push{2}  \qw  & \ghost{\up{XOR}} & \push{2} \ww & \Nghost{\up{XOR}}{\ww} & \push{2} \qw & \qw \\
} = \myQ{0.7}{0.5}{
& \push{2} \qw & \gate{\id} & \push{2} \qw & \qw
} \quad,  
\end{align}

which suddenly is no longer a near-dilation of $(T, \scrC)$, but rather of the identity $\id_{\{0,1\}}$. Plainly speaking, though the implementer of $(T, \scrC)$ achieves the correct input-output behaviour on the accessible interface with the causal channel \eqref{eq:neardil}, the implementer is not free to connect at will the various wires of the hidden interfaces without destroying the correctness of this behaviour. \end{Example}

\begin{Thm} (Hidden Contractions of Causal Dilations are Causal Dilations.) \label{thm:Contractionsofdilations}\\
Let $\channel{\bbX}{(T, \scrC)}{\bbY}$ be a causal channel. Suppose that $\twoext{\bbX}{\bbD}{(L, \scrE)}{\bbY}{\bbE}$ is a causal dilation of $\channel{\bbX}{(T, \scrC)}{\bbY}$, and that $\bbQ \subseteq \bbD \cap \bbE $ is contractible. Then the contraction

\begin{align}
\myQ{0.7}{0.5}{	& \push{\bbD \setminus \bbQ} \ww& \Nmultigate{1}{\contr{(L, \scrE)}{\bbQ}}{\ww} & \push{\bbE \setminus \bbQ} \ww & \ww \\ & \push{\bbX }  \qw & \ghost{\contr{(L, \scrE)}{\bbQ}} & \push{\bbY } \qw  & \qw} 
\end{align} 

is a causal dilation of $\channel{\bbX}{(T, \scrC)}{\bbY}$.
\end{Thm}

\begin{proof}
	By induction and Coherence of Nested Contraction, it suffices to prove the statement in the case where $\bbQ$ is a simple interface, i.e. has only a single port $\sfq$. 
	
	Under this assumption, consider the causal channel 
	
	\begin{align} \label{eq:parttrash}
	\myQ{0.7}{0.5}{	& \push{\bbQ} \ww& \Nmultigate{2}{(L, \scrE)}{\ww} & \push{\bbQ} \ww & \ww \\& \push{\bbD \setminus \bbQ} \ww& \Nghost{(L, \scrE)}{\ww} & \push{\bbE \setminus \bbQ} \ww & \Ngate{\tr}{\ww}\\ & \push{\bbX }  \qw & \ghost{(L, \scrE)} & \push{\bbY } \qw  & \qw} \quad.
	\end{align} 
	
	By Freeness of Non-Contracted Interfaces, this causal channel is contractible along $\bbQ$, and its contraction is  
	
	\begin{align} \label{eq:partcont}
\fC_{\bbQ}\left(	\myQ{0.7}{0.5}{	& \push{\bbQ} \ww& \Nmultigate{2}{(L, \scrE)}{\ww} & \push{\bbQ} \ww & \ww \\& \push{\bbD \setminus \bbQ} \ww& \Nghost{(L, \scrE)}{\ww} & \push{\bbE \setminus \bbQ} \ww & \Ngate{\tr}{\ww}\\ & \push{\bbX }  \qw & \ghost{(L, \scrE)} & \push{\bbY } \qw  & \qw}\right) \quad = \quad 	\myQ{0.7}{0.5}{	& \push{\bbD \setminus \bbQ} \ww& \Nmultigate{1}{\contr{(L, \scrE)}{\bbQ}}{\ww} & \push{\bbE \setminus \bbQ} \ww & \Ngate{\tr}{\ww} \\ & \push{\bbX }  \qw & \ghost{\contr{(L, \scrE)}{\bbQ}} & \push{\bbY } \qw  & \qw} 
\quad	.
	\end{align}
	
	Notice that the right hand side is precisely the thing to consider if we wish to show that $\contr{(L, \scrE)}{\bbQ}$ is a dilation of $(T, \scrC)$. Hence, let us compute this contraction in a different way. 
 
 Observe the following about the causal specification of the channel \eqref{eq:parttrash}: The \myuline{input} port $\sfq \in \ports{\bbQ}$ is not a cause of $\bbY$, since $\sfq \notin \scrE(\bbY)$ by virtue of $(L, \scrE)$ being a causal dilation of $(T, \scrC)$; on the other hand, $\sfq$ is not a cause of the output port $\sfq$ either, since this would contradict Well-Foundedness by virtue of $\bbQ$ being contractible. By additivity of causal specifications, we thus conclude that the input interface $\bbQ$ is not a cause of $\bbY \cup \bbQ$, i.e. not a cause of any outputs whatsoever. But this implies (by \cref{lem:ExtractTriv}) that \eqref{eq:parttrash} must factor as
 
 	\begin{align}  \label{eq:insight}
 \myQ{0.7}{0.5}{	& \push{\bbQ} \ww& \Nmultigate{2}{(L, \scrE)}{\ww} & \push{\bbQ} \ww & \ww \\& \push{\bbD \setminus \bbQ} \ww& \Nghost{(L, \scrE)}{\ww} & \push{\bbE \setminus \bbQ} \ww & \Ngate{\tr}{\ww}\\ & \push{\bbX }  \qw & \ghost{(L, \scrE)} & \push{\bbY } \qw  & \qw} \quad = \quad  \myQ{0.7}{0.5}{& \push{\bbQ} \ww& \Ngate{\tr}{\ww}\\	& \push{\bbD \setminus \bbQ} \ww& \Nmultigate{1}{(S, \scrD)}{\ww} & \push{\bbQ} \ww & \ww \\ & \push{\bbX }  \qw & \ghost{(S, \scrD)} & \push{\bbY } \qw  & \qw}
 \end{align} 
 
 for some causal channel $(S, \scrD)$. This identity allows us to determine the contraction \eqref{eq:partcont} differently.
 
By Soundness we see from \eqref{eq:insight} that
 
 	\begin{align} \label{eq:altsound}
 \fC_{\bbQ}\left(	\myQ{0.7}{0.5}{	& \push{\bbQ} \ww& \Nmultigate{2}{(L, \scrE)}{\ww} & \push{\bbQ} \ww & \ww \\& \push{\bbD \setminus \bbQ} \ww& \Nghost{(L, \scrE)}{\ww} & \push{\bbE \setminus \bbQ} \ww & \Ngate{\tr}{\ww}\\ & \push{\bbX }  \qw & \ghost{(L, \scrE)} & \push{\bbY } \qw  & \qw}\right) \quad = \quad  \myQ{0.7}{0.5}{& \push{\bbD \setminus \bbQ} \ww& \Nmultigate{1}{(S, \scrD)}{\ww} & \push{\bbQ} \ww & \Ngate{\tr}{\ww} \\ & \push{\bbX }  \qw & \ghost{(S, \scrD)} & \push{\bbY } \qw  & \qw} \quad.
 \end{align}
 
We also see from \eqref{eq:insight} that

	\begin{align} 
\scalemyQ{.9}{0.7}{0.5}{& \push{\bbQ} \ww& \Ngate{\tr}{\ww}\\	& \push{\bbD \setminus \bbQ} \ww& \Nmultigate{1}{(S, \scrD)}{\ww} & \push{\bbQ} \ww & \Ngate{\tr}{\ww} \\ & \push{\bbX }  \qw & \ghost{(S, \scrD)} & \push{\bbY } \qw  & \qw} 
\quad = \quad  
\scalemyQ{.9}{0.7}{0.5}{	& \push{\bbQ} \ww& \Nmultigate{2}{(L, \scrE)}{\ww} & \push{\bbQ} \ww & \Ngate{\tr}{\ww} \\& \push{\bbD \setminus \bbQ} \ww& \Nghost{(L, \scrE)}{\ww} & \push{\bbE \setminus \bbQ} \ww & \Ngate{\tr}{\ww}\\ & \push{\bbX }  \qw & \ghost{(L, \scrE)} & \push{\bbY } \qw  & \qw} 
 \quad = \quad 
 \scalemyQ{.9}{0.7}{0.5}{& \push{\bbQ} \ww& \Ngate{\tr}{\ww}\\	& \push{\bbD \setminus \bbQ} \ww& \Ngate{\tr}{\ww}\\& \push{\bbX }  \qw & \gate{(T, \scrC)} & \push{\bbY } \qw  & \qw} \quad,
\end{align} 

as $(L, \scrE)$ is a dilation of $(T, \scrC)$. Hence, $\scalemyQ{.8}{0.5}{0.5}{& \push{\bbD \setminus \bbQ} \ww& \Nmultigate{1}{(S, \scrD)}{\ww} & \push{\bbQ} \ww & \Ngate{\tr}{\ww} \\ & \push{\bbX }  \qw & \ghost{(S, \scrD)} & \push{\bbY } \qw  & \qw} = \scalemyQ{.8}{0.5}{0.5}{	& \push{\bbD \setminus \bbQ} \ww& \Ngate{\tr}{\ww}\\& \push{\bbX }  \qw & \gate{(T, \scrC)} & \push{\bbY } \qw  & \qw}$ by normality of the theory. Combining this with \cref{eq:altsound} and \cref{eq:partcont}, we finally conclude that 

	\begin{align} 
 	\myQ{0.7}{0.5}{	& \push{\bbD \setminus \bbQ} \ww& \Nmultigate{1}{\contr{(L, \scrE)}{\bbQ}}{\ww} & \push{\bbE \setminus \bbQ} \ww & \Ngate{\tr}{\ww} \\ & \push{\bbX }  \qw & \ghost{\contr{(L, \scrE)}{\bbQ}} & \push{\bbY } \qw  & \qw} 
\quad	= \quad \myQ{0.7}{0.5}{	& \push{\bbD \setminus \bbQ} \ww& \Ngate{\tr}{\ww}\\& \push{\bbX }  \qw & \gate{(T, \scrC)} & \push{\bbY } \qw  & \qw} \quad,
\end{align}

which shows that $\contr{(L, \scrE)}{\bbQ}$ is a dilation of $(T, \scrC)$, as desired. \end{proof}

\subsection{The Causal-Dilational Ordering}
\label{subsec:CausalDilOrd}
Now that we have seen a number of examples of causal dilations, we are ready to introduce the causal version of \emph{derivability}, namely the dilational ordering of \cref{chap:Dilations}. As outlined in the introduction to the chapter, this order is meant to formalise the idea that some causal dilations can be `constructed in the environment' from others. Since we now have contractions in the bag of possible `constructions' which can take place in the environment, we can treat two-sided dilations properly, but must also slightly rethink what a sensible notion of derivability should be. \\

Given a dilation $\scalemyQ{.8}{0.7}{0.5}{& \push{\bbX} \qw &  \Nmultigate{1}{(L, \scrE)}{\qw}   & \push{\bbY} \qw  & \qw  \\
	&\push{\bbD} \ww & \Nghost{(L, \scrE)}{\ww} & \push{\bbE} \ww  & \ww
}$ of $\channel{\bbX}{(T, \scrC)}{\bbY}$, we would like to render it equivalent to any dilation of the form $\scalemyQ{.8}{0.7}{0.5}{& \push{\bbX} \qw &  \Nmultigate{1}{(L, \scrE)}{\qw}   & \push{\bbY} \qw  & \qw  \\
&\push{\bbD} \ww & \Nghost{(L, \scrE)}{\ww} & \push{\bbE} \ww  &  \ww \\
& \push{\bbA} \ww & \Ngate{(G, \scrB)}{\ww} & \push{\bbB} \ww& \ww
} $.\footnote{That the latter should be deemed \myuline{derivable} from the former is clear, since the latter dilation represents the independent execution of a causal channel $(G, \scrB)$ in the environment. That the former should be derivable from the latter, however, is ultimately a choice which reflects a convention. For example, it is not obvious that $\scalemyQ{.8}{0.7}{0.5}{& \push{\bbX} \qw &  \Nmultigate{1}{(L, \scrE)}{\qw}   & \push{\bbY} \qw  & \qw  \\
	&\push{\bbD} \ww & \Nghost{(L, \scrE)}{\ww} & \push{\bbE} \ww  & \ww
}$ can be `constructed' from $\scalemyQ{.8}{0.7}{0.5}{& \push{\bbX} \qw &  \Nmultigate{1}{(L, \scrE)}{\qw}   & \push{\bbY} \qw  & \qw  \\
&\push{\bbD} \ww & \Nghost{(L, \scrE)}{\ww} & \push{\bbE} \ww  &  \ww \\
& \push{\bbA} \ww & \Ngate{\tr}{\ww} } $; if the theory $\Theory$ has states, we can construct $(L, \scrE)$ by inserting a state into $\bbA$, but if there are no states on $\bbA$ there is no obvious `construction' in the environment which yields $(L, \scrE)$. } We would also like to render a dilation $(L', \scrE')$ derivable from a dilation $(L, \scrE)$, if $(L', \scrE')$ results from the contraction of a sub-interface of the hidden interface in $(L, \scrE)$.  These two ideas are combined in \cref{def:CausDilOrd} below.\\

As we want to speak of contraction, we need to restrict attention to $\bfE$-channels from now on. (As mentioned earlier, the reader is free to consider for concreteness the case where $\bfE$ is the class of constructible causal channels in $\Theory$.) Given an $\bfE$-channel $(T, \scrC)$, let us denote by $\CDil{(T, \scrC)}{\bfE}$ the class of causal dilations of $(T, \scrC)$ which are $\bfE$-channels.

\begin{Definition} (The Causal-Dilational Ordering.) \label{def:CausDilOrd} \\
Let $\channel{\bbX}{(T, \scrC)}{\bbY}$ be an $\bfE$-channel. We denote by $\cder^\bfE_{(T, \scrC)}$ the relation on $\CDil{(T, \scrC)}{\bfE}$ given by 

\begin{align}
\scalemyQ{1}{0.7}{0.5}{& \push{\bbX} \qw &  \Nmultigate{1}{(L, \scrE)}{\qw}   & \push{\bbY} \qw  & \qw  \\
	&\push{\bbD} \ww & \Nghost{(L, \scrE)}{\ww} & \push{\bbE} \ww  & \ww
} \cder^\bfE_{(T, \scrC)}\scalemyQ{1}{0.7}{0.5}{& \push{\bbX} \qw &  \Nmultigate{1}{(L', \scrE')}{\qw}   & \push{\bbY} \qw  & \qw  \\
	&\push{\bbD'} \ww & \Nghost{(L', \scrE')}{\ww} & \push{\bbE'} \ww  & \ww
}
\end{align}

 if and only if there exists, possibly after renaming the ports in $\bbD$ and $\bbE$, $\bfE$-channels $\scalemyQ{.8}{0.7}{0.5}{& \push{\bbA} \ww & \Ngate{(G, \scrB)}{\ww} & \push{\bbB} \ww& \ww}$ and $\scalemyQ{.8}{0.7}{0.5}{& \push{\bbA'} \ww & \Ngate{(G', \scrB')}{\ww} & \push{\bbB'} \ww& \ww}$ such that 

\begin{align} \label{eq:defsim}
\fC_{\bbQ} \left( \myQ{0.7}{0.5}{& \push{\bbX} \qw &  \Nmultigate{1}{(L, \scrE)}{\qw}   & \push{\bbY} \qw  & \qw  \\
	&\push{\bbD} \ww & \Nghost{(L, \scrE)}{\ww} & \push{\bbE} \ww  &  \ww \\
	& \push{\bbA} \ww & \Ngate{(G, \scrB)}{\ww} & \push{\bbB} \ww& \ww
} \right) = \myQ{0.7}{0.5}{& \push{\bbX} \qw &  \Nmultigate{1}{(L', \scrE')}{\qw}   & \push{\bbY} \qw  & \qw  \\
&\push{\bbD'} \ww & \Nghost{(L', \scrE')}{\ww} & \push{\bbE'} \ww  &  \ww \\
& \push{\bbA'} \ww & \Ngate{(G', \scrB')}{\ww} & \push{\bbB'} \ww& \ww
} 
\end{align}

for some contractible common sub-interface $\bbQ$ of $\bbD \cup \bbA$ and $\bbE \cup \bbB$. We say in this case that \emph{$(L', \scrE')$ is derivable from $(L, \scrE)$}.

	\end{Definition}

Before exemplifying derivability, some observations are in order: 

\begin{Prop} (Derivability is a Pre-Order.) \label{prop:CDerPreorder}\\
	The derivability relation $\cder^\bfE_{(T, \scrC)}$ is a pre-order (i.e. is reflexive and transitive) on the class $\CDil{(T, \scrC)}{\bfE}$.
\end{Prop}

\begin{proof}
	The relation is reflexive by taking $(G, \scrB) = (G', \scrB')$ and $\bbQ = \bbI$ in the condition \eqref{eq:defsim}.  As for transitivity, suppose that $(L, \scrE) \cder (L', \scrE')$ and $(L', \scrE') \cder (L'', \scrE'')$. If $(G_1, \scrB_1)$, works on  the left-hand side of \eqref{eq:defsim} to realise the relation $(L, \scrE) \cder (L', \scrE')$,  and if $(G_2, \scrB_2)$ works on the left-hand side of \eqref{eq:defsim} to realise the relation $(L', \scrE') \cder (L'', \scrE'')$,  then the parallel composition $(G, \scrB) := (G_1, \scrB_1) \og (G_2, \scrB_2)$ works to realise the relation $(L, \scrE) \cder (L'', \scrE'')$, using Freeness of Parallel Composition and Coherence of Nested Contraction. (The intuition is obvious: If $(L', \scrE')$ can be constructed from $(L, \scrE)$ and $(L'', \scrE'')$ from $(L', \scrE')$, then by successive construction we can reach $(L'', \scrE'')$ from $(L, \scrE)$.) The details are left as exercise.
\end{proof}

\begin{Remark} (On the `Renaming of Ports' in \cref{def:CausDilOrd}.)\\
The reader may wonder why it the renaming of ports cannot simply be absorbed into the action of $(G, \scrB)$; this is ultimately due to our choice of formally defining contractibility for \myuline{common} sub-interfaces (rather than, say, by means of merely a bijective correspondence between sub-interfaces), and it is illustrated in \cref{ex:primsim}. 
	\end{Remark}

\begin{Remark} (On the Significance of $(G', \scrB')$ in Condition \eqref{eq:defsim}.) \label{rem:GprimeTriv}\\
We will shortly give a simplifying reformulation of the condition of derivability (\cref{lem:DerSimpl}). Let us already now observe, however, that we may without loss of generality assume that $\bbB'= \bbI$ and that $(G', \scrB') =\tr_{\bbA'} $; indeed, $\scalemyQ{.8}{0.7}{0.5}{& \push{\bbX} \qw &  \Nmultigate{1}{(L', \scrE')}{\qw}   & \push{\bbY} \qw  & \qw  \\
	&\push{\bbD'} \ww & \Nghost{(L', \scrE')}{\ww} & \push{\bbE'} \ww  &  \ww \\
	& \push{\bbA'} \ww & \Ngate{\tr}{\ww} 
} $ is derivable from $\scalemyQ{.8}{0.7}{0.5}{& \push{\bbX} \qw &  \Nmultigate{1}{(L', \scrE')}{\qw}   & \push{\bbY} \qw  & \qw  \\
	&\push{\bbD'} \ww & \Nghost{(L', \scrE')}{\ww} & \push{\bbE'} \ww  &  \ww \\
	& \push{\bbA'} \ww & \Ngate{(G', \scrB')}{\ww} & \push{\bbB'} \ww& \ww
} $, by parallelly composing the latter with $\scalemyQ{.8}{0.7}{0.5}{& \push{\bbB'} \ww & \Ngate{\tr}{\ww}}$ and contracting along $\bbB'$, using Soundness. The formulation \eqref{eq:defsim} was merely  chosen for the sake of symmetric appearance. 
	\end{Remark}

\begin{Remark} (Notation -- On the Dependence of $\cder^\bfE_{(T, \scrC)}$ on $(T, \scrC)$ and $\bfE$.) \\
	Like the dilational ordering $\der_T$ in the causality-free case, the relation $\cder^\bfE_{(T, \scrC)}$ depends only on $(T, \scrC)$ through the dependence on the open interfaces $\bbX$ and $\bbY$ (cf. \cref{rem:DilDeponT}). On the other hand, there is a strong dependence of $\cder^\bfE_{(T, \scrC)}$ on $\bfE$, since the channel $(G, \scrB)$ used to construct $(L', \scrE')$ in the environment must be an $\bfE$-channel. For example, in $\QIT$, if $\bfE$ is the constructible class, any Bell-channel can be used for $(G, \scrB)$, whereas the inconstructible PR box cannot. 
	
	In practice, we will abbreviate $\cder^\bfE_{(T, \scrC)}$ by $\cder^\bfE$, or even by $\cder$ if $\bfE$ is clear from the context.
	\end{Remark}

The canonical examples of derivability are given by serial composition in the environment, and by contraction in the environment:

 \begin{Example} (Derivation by Serial Composition.) \label{ex:primsim} \label{ex:DerBySer}\\
	Let $\scalemyQ{.8}{0.7}{0.5}{& \push{\bbX} \qw &  \Nmultigate{1}{(L, \scrE)}{\qw}   & \push{\bbY} \qw  & \qw  \\
		&\push{\bbD} \ww & \Nghost{(L, \scrE)}{\ww} & \push{\bbE} \ww  & \ww
	}$ be a dilation of $\channel{\bbX}{(T, \scrC)}{\bbY}$. For any causal channel  $\scalemyQ{.8}{0.7}{0.5}{& \push{\bbE} \ww & \Ngate{(G,\scrB)}{\ww} & \push{\bbE'} \ww & \ww}$, the dilation $\scalemyQ{.8}{0.7}{0.5}{& \push{\bbX} \qw &  \Nmultigate{1}{(L, \scrE)}{\qw}   & \push{\bbY} \qw  & \qw  \\
	&\push{\bbD} \ww & \Nghost{(L, \scrE)}{\ww} & \push{\bbE} \ww  & \Ngate{(G, \scrB)}{\ww} & \push{\bbE'} \ww & \ww
}$ can be derived from $(L, \scrE)$, as witnessed by the parallel composition 

\begin{align}
\myQ{0.7}{0.5}{& \push{\bbX} \qw &  \Nmultigate{1}{(L, \scrE)}{\qw}   & \push{\bbY} \qw  & \qw  \\
	&\push{\bbD} \ww & \Nghost{(L, \scrE)}{\ww} & \push{\bbE} \ww  &  \ww \\
	& \push{\bbE} \ww & \Ngate{(G, \scrB)}{\ww} & \push{\bbE'} \ww& \ww
} 
\end{align}

which by Soundness is contractible along $\bbE$ with the desired contraction. Note, however, that we might need to intermittently rename the ports in $\bbE$ so as to make them differ from those of $\bbE'$, $\bbD$ and $\bbX$, in order for the channels to be parallelly composable in the first place (according to our requirement of distinct port names in a parallel composition). 

This example serves to illustrate how the causal-dilational ordering generalises the one of \cref{chap:Dilations} (note that $(G, \scrB)$ could easily have had an additional input interface $\bbD'$).

\end{Example}

\begin{Example} (Derivation by Contraction.) \label{ex:DerByCont}\\
	Let $\scalemyQ{.8}{0.7}{0.5}{& \push{\bbX} \qw &  \Nmultigate{1}{(L, \scrE)}{\qw}   & \push{\bbY} \qw  & \qw  \\
	&\push{\bbD} \ww & \Nghost{(L, \scrE)}{\ww} & \push{\bbE} \ww  & \ww
}$ be a dilation of $\channel{\bbX}{(T, \scrC)}{\bbY}$, and let $\scalemyQ{.8}{0.7}{0.5}{& \push{\bbY} \qw &  \Nmultigate{1}{(M, \scrF)}{\qw}   & \push{\bbZ} \qw  & \qw  \\
&\push{\bbE} \ww & \Nghost{(M, \scrF)}{\ww} & \push{\bbK} \ww  & \ww
}$ be a dilation of $\channel{\bbY}{(S, \scrD)}{\bbZ}$ whose hidden input interface matches the hidden output interface of $(L, \scrE)$. Of course, the causal channel

\begin{align} \label{eq:dilstart}
\myQ{0.7}{0.5}{& \push{\bbX} \qw &  \Nmultigate{1}{(L, \scrE)}{\qw}   &  \qw & \push{\bbY} \qw  & \qw &  \qw  & \multigate{1}{(M, \scrF)} & \push{\bbZ} \qw & \qw  \\
	&\push{\bbD} \ww & \Nghost{(L, \scrE)}{\ww} & \push{\bbE} \ww  & \ww & & \push{\bbE} \ww& \Nghost{(M, \scrF)}{\ww} & \push{\bbK} \ww & \ww
}
\end{align}

is a causal dilation of $\scalemyQ{.8}{0.7}{0.5}{& \push{\bbX} \qw & \gate{(T, \scrC)} & \gate{(S, \scrD)} & \push{\bbZ} \qw & \qw}$. (A better two-dimensional drawing of the channel \eqref{eq:dilstart} would have the $\bbE$-wires prolonged, crossing each other, so as to display all the input interfaces to the far left, and all the output interfaces to the far right.) The interface $\bbE$ is contractible, and the contraction is given by 

\begin{align} \label{eq:dilfromcont}
\myQ{0.7}{0.5}{& \push{\bbX} \qw &  \Nmultigate{1}{(L, \scrE)}{\qw}   &   \push{\bbY} \qw   & \multigate{1}{(M, \scrF)} & \push{\bbZ} \qw & \qw  \\
	&\push{\bbD} \ww & \Nghost{(L, \scrE)}{\ww}  & \push{\bbE} \ww& \Nghost{(M, \scrF)}{\ww} & \push{\bbK} \ww & \ww
} \quad;
\end{align}

consequently, \eqref{eq:dilfromcont} is dilation derivable from the dilation \eqref{eq:dilstart}. 

This derivation could \myuline{not} have been realised by `building clever circuits around' the original dilation; at some point a contraction is needed, and ordinary serial and parallel composition will not accomplish it. A similar example can be based on the parallel composition of two dilations, rather than a serial. These two examples illustrate why we could not properly have treated a dilational ordering among two-sided dilations in \cref{chap:Dilations} before introducing the formalism of causal channels and contractions.

	\end{Example}

We shall now restate the definition of derivability in simpler terms, effectively demonstrating that \cref{ex:DerBySer} and \cref{ex:DerByCont} are indicative of what general derivability looks like, namely, a combination of serial composition and contraction of the inaccessible interface. We have the following:

\begin{Lem} (Simplified Statement of Derivability.)\label{lem:DerSimpl}
\\
Let $\scalemyQ{.8}{0.7}{0.5}{& \push{\bbX} \qw &  \Nmultigate{1}{(L, \scrE)}{\qw}   & \push{\bbY} \qw  & \qw  \\
	&\push{\bbD} \ww & \Nghost{(L, \scrE)}{\ww} & \push{\bbE} \ww  & \ww
}$ and $\scalemyQ{.8}{0.7}{0.5}{& \push{\bbX} \qw &  \Nmultigate{1}{(L', \scrE')}{\qw}   & \push{\bbY} \qw  & \qw  \\
		&\push{\bbD'} \ww & \Nghost{(L', \scrE')}{\ww} & \push{\bbE'} \ww  & \ww
	}$ be causal dilations of $(T, \scrC)$, and assume that the naming of ports is such that the indicated interfaces are sufficiently disjoint (for example, all pairwise disjoint). Then, $(L, \scrE) \cder^\bfE (L', \scrE')$ if and only if there exists an $\bfE$-channel $(G, \scrB)$ such that

	\begin{align}  \label{eq:CD}
\fC_{\bbD} \left( \myQ{0.7}{0.5}{& \push{\bbX} \qw &  \Nmultigate{1}{(L, \scrE)}{\qw}   & \push{\bbY} \qw  & \qw  \\
	&\push{\bbD} \ww & \Nghost{(L, \scrE)}{\ww} & \push{\bbE} \ww  &   \Nmultigate{2}{(G, \scrB)}{\ww} & \push{\bbD} \ww& \ww  \\
&&	& \push{\bbD'} \ww & \Nghost{(G, \scrB)}{\ww} & \push{\bbE'} \ww& \ww \\	&& &\push{\bbA'} \ww & \Nghost{(G, \scrB)}{\ww} 
} \right) = \myQ{0.7}{0.5}{& \push{\bbX} \qw &  \Nmultigate{1}{(L', \scrE')}{\qw}   & \push{\bbY} \qw  & \qw  \\
	&\push{\bbD'} \ww & \Nghost{(L', \scrE')}{\ww} & \push{\bbE'} \ww  &  \ww \\
	& \push{\bbA'} \ww & \Ngate{\tr}{\ww} 
} \quad,
\end{align}

with $\bbD$ contractible. (If $\Theory$ has a state on every system, we may additionally take $\bbA' = \bbI$.)
	\end{Lem}

\begin{proof}
	The `if'-direction is clear. For the  `only if'-direction, assume $(G, \scrB)$ and $(G', \scrB')$ are given such that \eqref{eq:defsim} holds. Recall from \cref{rem:GprimeTriv} that we may assume without loss of generality that $\bbB'= \bbI$ and $(G', \scrB') = \tr_{\bbA'}$, since we may always (possibly after renaming inaccessible ports) parallelly compose with $\tr_{\bbB'}$ and contract along $\bbB'$ using Soundness.  (If the theory $\Theory$ has states on every system, we may here similarly assume that $\bbA'=\bbI$, so that altogether the channel $(G', \scrB')$ does not appear at all on the right hand side of  \cref{eq:defsim}.) 
	
We thus have a causal channel $(G, \scrB)$ and a contractible sub-interface $\bbQ$, such that 
	
	\begin{align} 
	\fC_{\bbQ} \left( \myQ{0.7}{0.5}{& \push{\bbX} \qw &  \Nmultigate{1}{(L, \scrE)}{\qw}   & \push{\bbY} \qw  & \qw  \\
		&\push{\bbD} \ww & \Nghost{(L, \scrE)}{\ww} & \push{\bbE} \ww  &  \ww \\
		& \push{\bbA} \ww & \Ngate{(G, \scrB)}{\ww} & \push{\bbB} \ww& \ww
	} \right) = \myQ{0.7}{0.5}{& \push{\bbX} \qw &  \Nmultigate{1}{(L', \scrE')}{\qw}   & \push{\bbY} \qw  & \qw  \\
		&\push{\bbD'} \ww & \Nghost{(L', \scrE')}{\ww} & \push{\bbE'} \ww  &  \ww \\
		& \push{\bbA'} \ww & \Ngate{\tr_{\bbA'}}{\ww} 
	} 
	\end{align}

	(the naming assumption means that a potential renaming of hidden ports can be absorbed into $(G, \scrB)$). Now, observe that we can always enlarge the contracted interface $\bbQ$ so as to ensure that $\bbQ \supseteq \bbD \cup \bbE$, by replacing $(G, \scrB)$ with $(G, \scrB) \og \id_{(\bbD \cup \bbE) \setminus \bbQ}$ and redefining $\bbQ$ as $\bbQ \cup [(\bbD \cup \bbE ) \setminus \bbQ]$; indeed, contracting the ports in $(\bbD \cup \bbE) \setminus \bbQ$ with identities makes no difference, by Soundness. Under this modification, we actually have

		\begin{align}  \label{eq:Qsplit}
	\fC_{\bbQ} \left( \myQ{0.7}{0.5}{& \push{\bbX} \qw &  \Nmultigate{1}{(L, \scrE)}{\qw}   & \push{\bbY} \qw  & \qw  \\
		&\push{\bbD} \ww & \Nghost{(L, \scrE)}{\ww} & \push{\bbE} \ww  &  \ww \\ & \push{\bbE} \ww & \Nmultigate{2}{(G, \scrB)}{\ww} & \push{\bbD} \ww & \ww \\
		& \push{\bbQ_\bbA} \ww & \Nghost{(G, \scrB)}{\ww} & \push{\bbQ_\bbB} \ww& \ww \\
		& \push{\bbA \setminus \bbQ_\bbA} \ww & \Nghost{(G, \scrB)}{\ww} & \push{\bbB \setminus \bbQ_\bbB} \ww& \ww 
	} \right) = \myQ{0.7}{0.5}{& \push{\bbX} \qw &  \Nmultigate{1}{(L', \scrE')}{\qw}   & \push{\bbY} \qw  & \qw  \\
		&\push{\bbD'} \ww & \Nghost{(L', \scrE')}{\ww} & \push{\bbE'} \ww  &  \ww \\
		& \push{\bbA'} \ww & \Ngate{\tr_{\bbA'}}{\ww} 
	} ,
	\end{align}
	
where $\bbQ_\bbA = \bbQ \cap \bbA$ and $\bbQ_\bbB = \bbQ \cap \bbB$, and where $\bbQ = \bbD \cup \bbE \cup \bbQ_\bbA = \bbD \cup \bbE \cup \bbQ_\bbB $. This last identity implies that $\bbQ_\bbA = \bbQ_\bbB =: \bbQ_0$, and by Nested Contraction we can start by contracting the interface $\bbQ_0$ and may thus assume without loss of generality that actually $\bbQ_\bbA = \bbQ_\bbB= \bbI$, and thus $\bbQ = \bbD \cup \bbE$. But then, by comparing the remaining interfaces on each side of \eqref{eq:Qsplit}, we must have $\bbA = \bbA \setminus \bbQ_\bbA = \bbD' \cup \bbA'$ and $\bbB =  \bbB \setminus \bbQ_\bbB = \bbE'$, so that \cref{eq:Qsplit} reads

	\begin{align} \label{eq:QDE}
\fC_{\bbD \cup \bbE} \left( \myQ{0.7}{0.5}{& \push{\bbX} \qw &  \Nmultigate{1}{(L, \scrE)}{\qw}   & \push{\bbY} \qw  & \qw  \\
	&\push{\bbD} \ww & \Nghost{(L, \scrE)}{\ww} & \push{\bbE} \ww  &  \ww \\ & \push{\bbE} \ww & \Nmultigate{2}{(G, \scrB)}{\ww} & \push{\bbD} \ww & \ww \\
	& \push{\bbD'} \ww & \Nghost{(G, \scrB)}{\ww} & \push{\bbE'} \ww& \ww \\
	& \push{\bbA'} \ww & \Nghost{(G, \scrB)}{\ww} 
} \right) = \myQ{0.7}{0.5}{& \push{\bbX} \qw &  \Nmultigate{1}{(L', \scrE')}{\qw}   & \push{\bbY} \qw  & \qw  \\
	&\push{\bbD'} \ww & \Nghost{(L', \scrE')}{\ww} & \push{\bbE'} \ww  &  \ww \\
	& \push{\bbA'} \ww & \Ngate{\tr_{\bbA'}}{\ww} 
} \quad.
\end{align}

It now follows by Nested Contraction that we can start in \cref{eq:QDE} by contracting $\bbE$, and by Soundness that this contraction is given by serial composition, so we conclude \cref{eq:CD} as desired.

	\end{proof}

\begin{Remark} (On Interpretation.)\\
Intuitively, \cref{lem:DerSimpl} is not surprising. It merely says that any construction in the environment can be realised by adjoining all hidden outputs and inputs of $(L, \scrE)$ to a network in the environment (one of them by a serial composition, the other necessarily by a contraction). The statement could equally well have been formulated using a serial composition on the $\bbD$-interface followed by a contraction of the $\bbE$-interface.	\end{Remark}

\begin{Remark} (Derivations from One-Sided Dilations.)\\
	\cref{lem:DerSimpl} implies that any dilation derivable from a \myuline{one-sided} dilation can be obtained by a serial composition with a causal channel $(G, \scrB)$. As such, when restricting to one-sided dilations  the causal-dilational ordering looks similar to the dilational ordering of \cref{chap:Dilations}.
	
	\end{Remark}

We end this subsection by observing two results to the effect that derivability interacts sensibly with parallel composition and contraction in the accessible interface (thus subsuming serial composition). Both results are easy to prove, boiling down to axiomatic gymnastics of contractions, similar to that of earlier proofs. They can be seen as companions to \cref{prop:ParaStab} and \cref{prop:ContStab}.

\begin{Thm} (Parallel Composability of $\cder^\bfE$.) \label{thm:ParStab} \\
Let $\twoext{\bbX_j}{\bbD_j}{(L_j, \scrE_j)}{\bbY_j}{\bbE_j}$ and $\twoext{\bbX_j}{\bbD'_j}{(L'_j, \scrE'_j)}{\bbY_j}{\bbE'_j}$ be causal dilations of $\channel{\bbX_j}{(T_j, \scrC_j)}{\bbY_j}$ for $j=1,2$, such that

\begin{align}
\scalemyQ{.8}{0.7}{0.5}{& \push{\bbX_1} \qw &  \Nmultigate{1}{(L_1, \scrE_1)}{\qw}   & \push{\bbY_1} \qw  & \qw  \\
	&\push{\bbD_1} \ww & \Nghost{(L_1, \scrE_1)}{\ww} & \push{\bbE_1} \ww  & \ww
} \cder^\bfE \scalemyQ{.8}{0.7}{0.5}{& \push{\bbX_1} \qw &  \Nmultigate{1}{(L'_1, \scrE'_1)}{\qw}   & \push{\bbY_1} \qw  & \qw  \\
&\push{\bbD'_1} \ww & \Nghost{(L'_1, \scrE'_1)}{\ww} & \push{\bbE'_1} \ww  & \ww
} 
\; \text{and} \;
\scalemyQ{.8}{0.7}{0.5}{& \push{\bbX_2} \qw &  \Nmultigate{1}{(L_2, \scrE_2)}{\qw}   & \push{\bbY_2} \qw  & \qw  \\
	&\push{\bbD_2} \ww & \Nghost{(L_2, \scrE_2)}{\ww} & \push{\bbE_2} \ww  & \ww
} \cder^\bfE \scalemyQ{.8}{0.7}{0.5}{& \push{\bbX_2} \qw &  \Nmultigate{1}{(L'_2, \scrE'_2)}{\qw}   & \push{\bbY_2} \qw  & \qw  \\
	&\push{\bbD'_2} \ww & \Nghost{(L'_2, \scrE'_2)}{\ww} & \push{\bbE'_2} \ww  & \ww
} .
\end{align}

Then, it holds that

	\begin{align}
	\scalemyQ{1}{0.7}{0.5}{ & \push{\bbX_1}  \qw & \multigate{1}{(L_1, \scrE_1)} & \push{\bbY_1} \qw & \qw \\ 
		& \push{\bbD_1} \ww & \Nghost{(L_1, \scrE_1)}{\ww} & \push{\bbE_1} \ww  & \ww   \\ 
		& \push{\bbD_2} \ww& \Nmultigate{1}{(L_2, \scrE_2)}{\ww} & \push{\bbE_2} \ww & \ww \\ & \push{\bbX_2}  \qw & \ghost{(L_2, \scrE_2)} & \push{\bbY_2} \qw  & \qw}  \cder^\bfE
		\scalemyQ{1}{0.7}{0.5}{ & \push{\bbX'_1}  \qw & \multigate{1}{(L'_1, \scrE'_1)} & \push{\bbY'_1} \qw & \qw \\ 
		& \push{\bbD'_1} \ww & \Nghost{(L'_1, \scrE'_1)}{\ww} & \push{\bbE'_1} \ww  & \ww   \\ 
		& \push{\bbD'_2} \ww& \Nmultigate{1}{(L'_2, \scrE'_2)}{\ww} & \push{\bbE'_2} \ww & \ww \\ & \push{\bbX'_2}  \qw & \ghost{(L'_2, \scrE'_2)} & \push{\bbY'_2} \qw  & \qw}   \quad. 
	\end{align}
	 
	\end{Thm}

\begin{proof}
The proof uses Coherence of Nested Contraction and Freeness of Parallel Composition and is left as a straightforward exercise. 

	\end{proof}

\begin{Thm} (Contractual Composability of $\cder^\bfE$.) \label{thm:ContStab}\\
Let $\twoext{\bbX}{\bbD}{(L, \scrE)}{\bbY}{\bbE}$ and  $\twoext{\bbX}{\bbD'}{(L', \scrE')}{\bbY}{\bbE'}$ be causal dilations of $\channel{\bbX}{(T, \scrC)}{\bbY}$, such that 

\begin{align}
\scalemyQ{1}{0.7}{0.5}{& \push{\bbX} \qw &  \Nmultigate{1}{(L, \scrE)}{\qw}   & \push{\bbY} \qw  & \qw  \\
	&\push{\bbD} \ww & \Nghost{(L, \scrE)}{\ww} & \push{\bbE} \ww  & \ww
} \cder^\bfE\scalemyQ{1}{0.7}{0.5}{& \push{\bbX} \qw &  \Nmultigate{1}{(L', \scrE')}{\qw}   & \push{\bbY} \qw  & \qw  \\
	&\push{\bbD'} \ww & \Nghost{(L', \scrE')}{\ww} & \push{\bbE'} \ww  & \ww
} \quad . 
\end{align}

Then, a sub-interface $\bbP \subseteq \bbX \cap \bbY$ is contractible in $(L, \scrE)$ if and only if it is contractible in $(L', \scrE')$, and in that case

\begin{align}
\scalemyQ{1}{0.7}{0.5}{& \push{\bbX \setminus \bbP} \qw &  \Nmultigate{1}{\fC_\bbP((L, \scrE))}{\qw}   & \push{\bbY \setminus \bbP} \qw  & \qw  \\
	&\push{\bbD} \ww & \Nghost{\fC_\bbP((L, \scrE))}{\ww} & \push{\bbE} \ww  & \ww
} \cder^\bfE\scalemyQ{1}{0.7}{0.5}{& \push{\bbX \setminus \bbP} \qw &  \Nmultigate{1}{\fC_\bbP((L', \scrE'))}{\qw}   & \push{\bbY \setminus \bbP} \qw  & \qw  \\
	&\push{\bbD'} \ww & \Nghost{\fC_\bbP((L', \scrE'))}{\ww} & \push{\bbE'} \ww  & \ww
} \quad . 
\end{align}
	 
\end{Thm}

\begin{Remark}
	Observe that, as usual, \cref{thm:ContStab} about contractions subsumes serial composition, given \cref{thm:ParStab} about parallel composition.
	\end{Remark}

\begin{proof}
	Contractions now occur on two distinct levels; on the explicitly stated level of contraction in the accessible interfaces $\bbX$ and $\bbY$, and on the level of contractions in the environment implicit within the definition of the pre-order $\cder^\bfE$.

	If $(G, \scrB)$ realises the relation $(L, \scrE) \cder^\bfE (L', \scrE')$, then the desired follows by Freeness under Parallel Composition and Coherence of Nested Contraction. Again, the details are left as exercise. (The statement about equivalence of contractibility follows since contractibility of $\bbP$ in either dilation is equivalent to contractibility of $\bbP$ in $(T, \scrC)$, by \cref{prop:ContStab}.)

\end{proof}

\section{Density Theorems and Rigidity}
\label{sec:Density}

Now that we have understood the basic properties of causal dilations and the pre-order of derivability, it is natural to start executing the same program as for dilations in \cref{chap:Dilations}, namely, that of uncovering the structure of causal dilations of a given causal channel $(T, \scrC)$. 

This time, however, the problem at hand is much more complex, and we will not be as successful in doing this. In this section, I will only make some initial observations about the structure of causal dilations, and almost exclusively in specific examples. Whether a more systematic theory is possible is unclear, and it is probably the most important problem left open from the work of this thesis. \\

Again, we work in a fixed contractible theory $(\Theory, \fC, \bfE)$. First, let us transfer to the setting of causal dilations the idea of \emph{completeness}, which was one of the most powerful concepts of \cref{chap:Dilations}:

\begin{Definition} (Causal Completeness.) \\
	Let $\channel{\bbX}{(T, \scrC)}{\bbY}$ be an $\bfE$-channel, and let $\bfD \subseteq \CDil{(T, \scrC)}{\bfE}$ be a class of dilations of $(T, \scrC)$. A dilation $(K, \scrF) \in \bfD$ is called \emph{(causally) complete for $\bfD$}, if $(K, \scrF)\cder^\bfE (L, \scrE)$ for all  $(L, \scrE) \in \bfD$. A dilation  $(K, \scrF) \in \CDil{(T, \scrC)}{\bfE}$ is called simply \emph{(causally) complete} if it is complete for $ \CDil{(T, \scrC)}{\bfE}$.	
	\end{Definition} 

\begin{Example}
	By \cref{ex:causexttrash}, both trashes and states have complete causal dilations. 
	\end{Example}

In \cref{chap:Dilations}, completeness was ubiquitous and its absence confined to obscure examples. \emph{\myuline{Causal}} completeness, on the other hand, turns out to be a scarce resource, at least in the information theories $\CIT$ and $\QIT$. The reality seems to be that causal completeness should no longer be regarded as a property of a (contractible) theory, cf. the idea of `complete theories' in \cref{chap:Dilations}, but rather as a property of certain causal \myuline{channels} in the theory:

\begin{Definition} (Rigidity.)  \label{def:Rigidity}\\
An $\bfE$-channel $\channel{\bbX}{(T, \scrC)}{\bbY}$ is said to be \emph{rigid} if it has a complete causal dilation. More generally, we call $(T, \scrC)$ \emph{rigid relative to $\bfD \subseteq \CDil{(T, \scrC)}{\bfE}$} if $(T, \scrC)$ has a causal dilation which is complete for $\bfD$.
\end{Definition}

Rigidity signifies that by locking the input-output behaviour $(T, \scrC)$ between the accessible interfaces, there is a single `worst case' causal dilation, formalising the strongest possible causal side-computations the environment can perform during our interaction with the causal channel $(T, \scrC)$. At this point, it is not clear that this property has anything to do with rigidity as it is understood in quantum self-testing (except, perhaps, for similarity in spirit), but the relationship will be uncovered in \cref{chap:Selftesting}. Essentially, `rigidity' as used in the field of quantum self-testing will correspond to rigidity relative to a certain class of dilations, but not relative to all; that some dilations have to be disregarded slightly disturbs the simplicity, but it is intimately related to the fact that measurement channels have strange dilations (\cref{ex:MeasDil}).\\

How should we tackle the problem of clarifying the structure of dilations if we cannot generally rely on completeness?

 As a substitute for complete (i.e. $\cder^\bfE$-largest) dilations of $(T, \scrC)$, one might look for $\cder^\bfE$-\myuline{maximal}\footnote{See the Preliminary section of the thesis for basic terminology of pre-orders.} dilations, and try to prove results to the effect that any dilation is derivable from some maximal dilation. Whereas this could indeed be successful, pre-orders in general need not exhibit such universal boundedness by maximal elements.\footnote{It seems to me hard to find principles which in this context facilitate an application of Zorn's Lemma.} 
 
 As such, the best we can do is to identify a \emph{dense} class of dilations, that is, a class\footnote{This class has conceptually nothing to do with the class in the definition of relative completeness and rigidity, though we denoted it too by the symbol `$\bfD$'. Observe, however, that if $\bfD$ is dense in $\CDil{(T, \scrC)}{\bfE}$ and $(K, \scrF)$ is complete for $\bfD$, then $(K, \scrF)$ is in fact complete for all of $\CDil{(T, \scrC)}{\bfE}$.} $\bfD$ of dilations such that for any dilation $(L', \scrE')$, there exists $(L, \scrE) \in \bfD$ with $(L, \scrE) \cder^\bfE (L', \scrE')$. The smaller we can make the dense class, the better we have understood the large elements in the causal-dilational ordering, that is, the better we have understood the abilities of a potentially adversarial environment of the channel $(T, \scrC)$.  \\

We will use the term \emph{density theorem} about a result which identifies a certain class of dilations as dense. A \emph{completeness theorem} would be an extreme density theorem, asserting that a class $\{(K, \scrF)\}$  consisting of a single dilation is dense.

To make any sort of progress, we will from now on restrict attention to constructible channels, i.e. we will fix $\bfE = \Cns{\Theory}$ and take $\fC$ to be the standard notion of contraction. This is not only a mathematical convenience, but is also, at least in the case of the information theories $\CIT$ and $\QIT$, physically reasonable.\footnote{Otherwise, even a causal channel as simple as the bit refreshment channel has as among its dilation the PR box, whose physical existence is questionable (\cite{vanD13}).} We will write simply $\cder$ in place of $\cder^{\Cns{\Theory}}$.\\

First, we show a general density theorem which holds by virtue of the constructibility restriction. The most important case of this result is that of Bell-channels. Then, we observe that  causal channels in cartesian theories are always rigid. Finally, we examine density and rigidity for Bell-channels in $\CIT$.

\subsection{... in General}
\label{subsec:DenseGeneral}

In approaching a general density theorem, the first observation to make is that, by \cref{lem:CnstrFact}, every constructible dilation of a causal channel $\channel{\bbX}{(T, \scrC)}{\bbY}$ is of the form 

\begin{align}
\scalemyQ{1}{0.7}{0.5}{& \push{\bbX} \qw &  \multigate{1}{(L, \scrE)}   & \qw  & \push{\bbY} \qw  & \qw  \\
	& & \Nghost{(L, \scrE)} & \push{\bbE} \ww  &  \Nmultigate{1}{(G, \scrB)}{\ww}  & \\ & & \push{\bbD'} \ww  &\ww &  \Nghost{(G, \scrB)}{\ww} & \push{\bbE'} \ww & \ww
}
\end{align}

 with $(L, \scrE)$ and $(G, \scrB)$ constructible. This implies that the class of \myuline{one-sided} constructible dilations of $(T, \scrC)$ is dense. As such, the density question is reduced to finding a class which is dense \emph{within} the class of one-sided dilations. 
 
 A first step in this direction is to clarify the complexity of the various possible hidden output interfaces $\bbE$. The ports $\sfe \in \ports{\bbE}$ distinguish themselves by their sets of causes $\scrE(\sfe) \subseteq \ports{\bbX}$. These cause-sets encode which ports in the open interface $\bbX$ have to be fed with an input before side-information becomes available to the environment. 

\begin{Example} (One-sided Dilations of a General Causal Channel.) \label{ex:expports}\\
		Let  $\oneext{\bbX}{(L, \scrE)}{\bbY}{\bbE}$ be a one-sided dilation of $\channel{\bbX}{(T, \scrC)}{\bbY}$. If two ports $\sfe, \sfe' \in \ports{\bbE}$ have the same causes, $\scrE(\sfe) = \scrE(\sfe')$, then they can by merged in the environment to a single port, and this merging can be reversed also by a causal channel in the environment. In other words, $(L, \scrE)$ is $\cder$-equivalent to a one-sided dilation for which all hidden ports have distinct sets of causes. Consequently, a dense class of dilations is given by those one-sided dilations with (at most) $2^\abs{\bbX}$ hidden ports, one for each possible set of causes in the open interface $\bbX$.

	\end{Example}

A priori, the size of $\bbE$ might not be further reducible than suggested by the exponential bound $2^\abs{\bbX}$. But sometimes we can reduce it by virtue of (constructibility and) details of the causal specification $\scrC$. The following example not only serves to illustrate this point, but will also be extremely important for establishing the relation to quantum self-testing in \cref{chap:Selftesting}:

\begin{Example} (Constructible Dilations of a Bipartite Bell-Channel.) \label{ex:BellDil}\\
	Consider a bipartite Bell-channel, i.e. a constructible causal channel $\scalemyQ{.8}{0.7}{0.5}{& \push{\cX_\sfA} \qw & \multigate{1}{(T, \scrC)} & \push{\cY_\sfA} \qw & \qw \\ & \push{\cX_\sfB} \qw & \ghost{(T, \scrC)} & \push{\cY_\sfB} \qw & \qw }$ with $\scrC(\sfy_\sfA)=\{\sfx_\sfA\}$ and $\scrC(\sfy_\sfB)= \{\sfx_\sfB\}$. By \cref{ex:expports}, any constructible dilation of $(T, \scrC)$ is derivable from a one-sided constructible causal dilation with four hidden ports, corresponding to the cause-sets  $\emptyset, \{\sfx_\sfA\}, \{\sfx_\sfB\}$ and $\{\sfx_\sfA, \sfx_\sfB\}$. It turns out that already the dilations with \myuline{three} types of hidden output ports -- corresponding to cause-sets $\emptyset$, $\{\sfx_\sfA\}$ and $\{{\sfx_\sfB}\}$ -- suffice to form a dense class. (As such, side-information with cause-set $\{\sfx_\sfA, \sfx_\sfB\}$ can always be derived from side-information with cause-sets $\{\sfx_\sfA\}$ and $\{\sfx_\sfB\}$.) The argument is essentially graph-theoretic:
	
Define the \emph{stencil-complexity} of a constructible channel to be the minimal number of boxes required in a stencil-representation whose filling uses only primitive causal channels. Given any constructible dilation $(L', \scrE')$ of $(T, \scrC)$, let $(L, \scrE) \cder (L', \scrE')$ be a one-sided constructible dilation with smallest possible stencil-complexity. Let $(\fF,G)$ be a stencil-representation of $(L, \scrE)$ witnessing this stencil-complexity. We aim to show that $(L, \scrE)$ has no hidden port with cause set $\{\sfx_\sfA, \sfx_\sfB\}$. This will prove the desired, since the remaining ports can then be merged according to causes. 

To arrive at a contradiction, assume that for some $\sfe \in \bbE$ we have $\scrE(\sfe)=\{\sfx_\sfA, \sfx_\sfB\}$. Then in the stencil $G$, the port $\sfe$ must be the child\footnote{Recall that a vertex $v$ in a directed graph is a \emph{child} of the vertex $v'$ if there exists an edge from $v'$ to $v$.} of some box, say $b$. Moreover, both $\sfx_\sfA$ and $\sfx_\sfB$ must be ancestral to $b$, and therefore $\sfx_\sfA$ and $\sfx_\sfB$ are ancestral to any \myuline{descendant} of $b$. In particular, any output port that descends from $b$ has cause-set $\{\sfx_\sfA, \sfx_\sfB\}$. But by virtue of the specification $\scrC$, this implies that $b$ cannot have $\sfy_\sfA$ or $\sfy_\sfB$ as descendants, so the ports which descend from $b$ must correspond to a sub-interface $\bbE_0$ of $\bbE$. Pictorially, the situation now looks as follows:

\begin{align}
\myQ{0.7}{0.5}{& \push{\bbX} \qw &  \multigate{1}{(L, \scrE)}   & \push{\bbY} \qw  & \qw  \\
	& & \Nghost{(L, \scrE)} & \push{\bbE} \ww & \ww
} = \myQ{0.7}{0.5}{& \push{\bbX} \qw &  \multigate{2}{(L_0, \scrE_0)}   & \push{\bbY} \qw  & \qw  \\
& & \Nghost{(L_0, \scrE_0)} &\push{\bbE \setminus \bbE_0} \ww  & \ww\\ &  & \Nghost{(L_0, \scrE_0)} & \push{\bbH} \ww & \Ngate{T_b}{\ww} & \push{\bbE_0} \ww & \ww
} ,
\end{align}

where $\scalemyQ{.8}{0.7}{0.5}{& \Ngate{T_b} }$ denotes the box $b$ filled with its primitive channel $T_b$, and where $(L_0, \scrE_0)$ abbreviates the remaining part of the representation $(\fF, G)$, that results from disregarding the box $b$. Now, however, if we remove the box $b$ we are left with a one-sided constructible dilation $(L_0, \scrE_0) \cder (L, \scrE)$ of strictly lower stencil-complexity than $(L, \scrE)$, contradicting the choice of $(L, \scrE)$. Consequently, there cannot be any ports $\sfe \in \ports{\bbE}$ with $\scrE(\sfe)= \{\sfx_\sfA, \sfx_\sfB\}$, and the desired follows: Any constructible dilation of $(T, \scrC)$ derives from a constructible dilation $\scalemyQ{.8}{0.7}{0.5}{& \push{\cX_\sfA} \qw & \multigate{4}{(L, \scrE)} & \push{\cY_\sfA} \qw & \qw \\& & \Nghost{(L, \scrE)} & \push{\cE_\sfA} \ww & \ww \\& & \Nghost{(L, \scrE)} & \push{\cE_0} \ww & \ww \\& & \Nghost{(L, \scrE)} & \push{\cE_\sfB} \ww & \ww
	\\& \push{\cX_\sfB} \qw & \ghost{(L, \scrE)} & \push{\cY_\sfB} \qw & \qw}$, where $\scrE(\sfe_0)= \emptyset$, $\scrE(\sfe_\sfA)=\{\sfx_\sfA\}$ and $\scrE(\sfe_\sfB)=\{\sfx_\sfB\}$.

\end{Example}

It is clear that \cref{ex:BellDil} can be easily generalised to the case of a multipartite Bell-channel, i.e. a constructible causal channel $\channel{\bbX}{(T, \scrC)}{\bbY}$ with $\ports{\bbX} = \{\sfx_1, \ldots, \sfx_n\}$ and $\ports{\bbY}= \{\sfy_1, \ldots, \sfy_n\}$, and for which the specification $\scrC$ is given by $\scrC(\sfy_j)= \{\sfx_j\}$ for $j=1, \ldots, n$. As dense class of dilations we thus obtain those one-sided constructible dilations with $n+1 = \abs{\bbX}+1$ hidden ports, one port for each $j=1, \ldots, n$ with cause-set $\{\sfx_j\}$, and one port which is acausal i.e. has cause-set $\emptyset$. This is a tremendous improvement over the $2^\abs{\bbX}$ ports that would a priori be expected.\\

In fact, the graph-theoretic argument we gave can even be generalised beyond Bell-channels without much difficulty. 

Let $\channel{\bbX}{(T, \scrC)}{\bbY}$ be an arbitrary  constructible causal channel, but let us suppose for simplicity that $\bbY \neq \bbI$ and $\scrC(\bbY)= \bbX$. (Minding \cref{lem:ExtractTriv}, this is to say that no trash can be factored out of $(T, \scrC)$.) Let us call a set of ports $\sfP \subseteq \ports{\bbX}$ a \emph{minorant of $\scrC$ given $\sfy$} if $\sfP\subseteq \scrC(\sfy)$. Let

 \begin{align}
 \up{Min}_\scrC := \{\sfP \subseteq \ports{\bbX} \mid \exists \sfy \in \ports{\bbY}: \sfP \subseteq \scrC(\sfy)\}
 \end{align}
 
 be the collection of minorants of $\scrC$ (given some $\sfy$). If $\scrC$ is a primitive causal specification ($\scrC(\sfy) = \ports{\bbX}$ for all $\sfy \in \ports{\bbY}$), then $ \up{Min}_\scrC $ is the collection of all subsets of $\ports{\bbX}$, so $\abs{ \up{Min}_\scrC }= 2^\abs{\bbX}$. At the other extreme, if $\scrC(\sfy)$ contains just a single port $\sfx_\sfy \in \ports{\bbX}$ for every $\sfy \in \ports{\bbY}$ (like for the Bell-channels), then $ \up{Min}_\scrC $ contains precisely the singletons $\{\sfx_\sfy\}$ along with the empty set $\emptyset$, so $\abs{ \up{Min}_\scrC } =  \abs{\bbX}+1$ (since $\sfy \mapsto \sfx_\sfy$ is surjective by the requirement $\scrC(\bbY)= \bbX$).

We have the following:

\begin{Thm} (Density of Minorantal One-Sided Dilations.) \label{thm:Minorantal}\\
Let $\channel{\bbX}{(T, \scrC)}{\bbY}$ be a constructible causal channel with $\bbY \neq \bbI$ and $\scrC(\bbY)= \bbX$. Then the class of constructible one-sided dilations $\oneext{\bbX}{(L, \scrE)}{\bbY}{\bbE}$ for which $\bbE$ contains (at most) $\abs{\up{Min}_\scrC}$ ports, one port $\sfe$ with $\scrE(\sfe) = \sfP$ for each minorant $\sfP\in \up{Min}_\scrC$, is a dense class of dilations.

\end{Thm}

\begin{Remark} For the case $\bbY= \bbI$ we already have a density theorem in the form of \cref{ex:causexttrash}, indeed, the trashes have complete causal dilations. For $\bbY \neq \bbI$ but $\scrC(\bbY) \subsetneq \bbX$, one can factor out a maximal trash from $(T, \scrC)$ and prove that the dense class of \cref{thm:Minorantal} can be combined in parallel with those complete dilations and yield a dense class. 
	\end{Remark}

\begin{proof}

The proof closely follows that given in \cref{ex:BellDil}. %

For any constructible dilation $(L', \scrE')$ of $(T, \scrC)$, let $(L, \scrE) \cder (L', \scrE')$ be a one-sided constructible dilation with minimal stencil-complexity, and let $(\fF, G)$ be a stencil-representation witnessing this. We prove that $\scrE(\sfe) \in \up{Min}_\scrC$ for every $\sfe \in \ports{\bbE}$, and this is enough. 

Any port $\sfe \in \ports{\bbE}$ is the child of some box $b$ in the stencil $G$ (if it were the child of an input port $\sfx$, then the wire $\scalemyQ{.8}{0.7}{0.5}{& \sfx \quad & \qw & \sfe}$ would be disconnected from the rest of $G$, so $(T, \scrC)$, which appears by trashing all the ports in $\bbE$, would have $\tr_\sfx$ as a factor, contradicting $\scrC(\bbX)= \bbY$). By the exact same argument as given in \cref{ex:BellDil}, $b$ cannot only have descendants in $\ports{\bbE}$, since this would allow us to remove the box $b$ and obtain a constructible dilation $(L_0, \scrE_0) \cder (L, \scrE)$ of strictly lower stencil-complexity. Hence, $b$ must have some $\sfy \in \ports{\bbY}$ as descendant. But then every ancestor of $b$ is also an ancestor of $\sfy$, so every ancestor \myuline{of $\sfe$} is an ancestor of $\sfy$. In particular, the cause-set $\scrE(\sfe)$ is contained in the cause-set $\scrE(\sfy) = \scrC(\sfy)$, which is precisely to say that $\scrE(\sfe) \in \up{Min}_\scrC$. 	\end{proof}

\cref{thm:Minorantal} is neat, in particular as it applies to Bell-channels in the form of \cref{ex:BellDil}, but it does not really bring us much closer to what causal dilations look like. The point is, however, that in certain cases we can now use the knowledge of the cause-sets $\scrE(\sfe)$ to dessicate the dilations $(L, \scrE)$ even further, by giving a concrete stencil-representation:

\begin{Thm} (General Density Theorem for Bell-Channels, Part I.) \label{thm:BellDense} \\
Let $\scalemyQ{.8}{0.7}{0.5}{& \push{\cX_\sfA} \qw & \multigate{1}{(T, \scrC)} & \push{\cY_\sfA} \qw & \qw \\ & \push{\cX_\sfB} \qw & \ghost{(T, \scrC)} & \push{\cY_\sfB} \qw & \qw }$ be a bipartite Bell-channel. Then, the causal dilations of the form 

\begin{align} \label{eq:belldil}
\myQ{0.7}{0.5}{
	&  \push{\cX_\sfA}  \qw& \multigate{1}{ L_\sfA} &  \push{\cY_\sfA} \qw & \qw  \\
	& \Nmultigate{2}{t}  & \ghost{L_\sfA} &  \push{\cE_\sfA} \ww & \ww\\
	& \Nghost{t} & \ww & \push{\cE_0} \ww & \ww\\
	& \Nghost{t} & \multigate{1}{L_\sfB} &  \push{\cE_\sfB}\ww & \ww  \\
	&   \push{\cX_\sfB} \qw & \ghost{L_\sfB} & \push{\cY_\sfB} \qw & \qw  \\
} \quad ,
\end{align}

where each component is given its primitive specification, constitute a dense class for constructible dilations of $(T, \scrC)$. 

\end{Thm}

\begin{Remark} (General Bell-Channels.)\\
	It should be clear how to generalise this statement to multipartite Bell-channels. In the case of a unipartite Bell-channel (that is, in the case of a primitive causal channel $\channel{\cX}{(T, \scrC)}{\cY}$), the dense class is comprised by causal dilations of the form $\scalemyQ{.8}{0.7}{0.5}{&
		  \push{\cX}  \qw & \multigate{1}{ L} &  \push{\cY} \qw & \qw  \\
		& \Nmultigate{1}{t}  &  \ghost{L} &  \push{\cE} \ww & \ww\\
		& \Nghost{t} & \ww & \push{\cE_0} \ww & \ww	}$.
	\end{Remark}

\begin{proof}
	We know from \cref{ex:BellDil} that the constructible dilations $\scalemyQ{.8}{0.7}{0.5}{& \push{\cX_\sfA} \qw & \multigate{4}{(L, \scrE)} & \push{\cY_\sfA} \qw & \qw \\& & \Nghost{(L, \scrE)} & \push{\cE_\sfA} \ww & \ww \\& & \Nghost{(L, \scrE)} & \push{\cE_0} \ww & \ww \\& & \Nghost{(L, \scrE)} & \push{\cE_\sfB} \ww & \ww
		\\& \push{\cX_\sfB} \qw & \ghost{(L, \scrE)} & \push{\cY_\sfB} \qw & \qw}$, where $\scrE(\sfe_0)= \emptyset$, $\scrE(\sfe_\sfA)=\{\sfx_\sfA\}$ and $\scrE(\sfe_\sfB)=\{\sfx_\sfB\}$, form a dense class. However, using \cref{lem:CnstrFact} such a causal channel must be of the form $\scalemyQ{.8}{0.7}{0.5}{& \push{\cX_\sfA} \qw & \multigate{3}{(S_\sfA, \scrD_\sfA)} & \qw & \push{\cY_\sfA} \qw & \qw \\& & \Nghost{(S_\sfA, \scrD_\sfA)} & \ww &  \push{\cE_\sfA} \ww & \ww \\& & \Nghost{(S_\sfA, \scrD_\sfA)} &\ww & \push{\cE_0} \ww & \ww \\& & \Nghost{(S_\sfA, \scrD_\sfA)} & \push{\bbH} \qw  & \multigate{1}{(L_\sfB, \scrF_\sfB)} & \push{\cE_\sfB} \ww & \ww
		\\& \push{\cX_\sfB} \qw & \qw & \qw & \ghost{(L_\sfB, \scrF_\sfB)} & \push{\cY_\sfB} \qw & \qw}$ for constructible $(S_\sfA, \scrD_\sfA)$ and $(L_\sfB, \scrF_\sfB)$. Since $\sfx_\sfA \notin \scrE(\{\sfy_\sfB, \sfe_\sfB\})$, we may assume as in \cref{ex:IncRevis} without loss of generality that $\scrD_\sfA(\ports{\bbH}) = \emptyset$ (if $\ports{\bbH}$ had $\scrD_\sfA$-causes, these would propagate to $\{\sfy_\sfB, \sfe_\sfB\}$, unless trashed in $(L_\sfB, \scrF_\sfB)$).  Using \cref{lem:CnstrFact} again, $(S_\sfA, \scrD_\sfA)$ must therefore be of the form$\scalemyQ{.8}{0.7}{0.5}{
		&  \push{\cX_\sfA}  \qw&  \multigate{1}{ (L_\sfA, \scrF_\sfA)} &  \push{\cY_\sfA} \qw & \qw  \\
		& \Nmultigate{2}{t}  &  \ghost{(L_\sfA, \scrF_\sfA)} &  \push{\cE_\sfA} \ww & \ww\\
		& \Nghost{t} &  \push{\cE_0} \ww & \ww\\
		& \Nghost{t}   &  \push{\bbH}\qw &\qw }$ for some state $t$ and some constructible $(L_\sfA, \scrF_\sfA)$. Altogether, $(L, \scrE)$ is now of the form $\scalemyQ{.8}{0.7}{0.5}{
		&  \push{\cX_\sfA}  \qw& \multigate{1}{ (L_\sfA, \scrF_\sfA)} &  \push{\cY_\sfA} \qw & \qw  \\
		& \Nmultigate{2}{t}   & \ghost{(L_\sfA, \scrF_\sfA)} &  \push{\cE_\sfA} \ww & \ww\\
		& \Nghost{t} & \push{\cE_0} \ww & \ww\\
		& \Nghost{t}  & \multigate{1}{(L_\sfB, \scrF_\sfB)} &  \push{\cE_\sfB}\ww & \ww  \\
		&   \push{\cX_\sfB} \qw & \ghost{(L_\sfB, \scrF_\sfB)} & \push{\cY_\sfB} \qw & \qw  \\
	} $,  with $\sfx_i \in \scrF_i(\sfy_i)$ and $\sfx_i \in \scrF_i(\sfe_i)$. Now, however, the total causal specification that results from this is precisely the same as when $\scrF_\sfA$ and $\scrF_\sfB$ are substituted by primitive specifications. Hence, we can make this substitution and achieve the desired.  
	\end{proof}

The significance of \cref{thm:BellDense} is the following:  A channel of the form \eqref{eq:belldil} is a dilation of $(T, \scrC)$ if and only if $t$ is a dilation a state $s$ and the channels $L_i$ are dilations of channels $T_i$, such that 

\begin{align}
\myQ{0.7}{0.5}{
	&  \push{\cX_\sfA}  \qw& \multigate{1}{T_\sfA} &  \push{\cY_\sfA} \qw & \qw  \\
	& \Nmultigate{1}{s}  & \ghost{T_\sfA} \\
	& \Nghost{s} & \multigate{1}{T_\sfB}   \\
	&   \push{\cX_\sfB} \qw & \ghost{T_\sfB} & \push{\cY_\sfB} \qw & \qw  \\
} =  \myQ{0.7}{0.5}{& \push{\cX_\sfA} \qw & \multigate{1}{(T, \scrC)} & \push{\cY_\sfA} \qw & \qw \\ & \push{\cX_\sfB} \qw & \ghost{(T, \scrC)} & \push{\cY_\sfB} \qw & \qw } \quad.
\end{align}

Though it was of course already clear that $(T, \scrC)$ must have this form (since it is a Bell-channel), it was \myuline{not} clear that any causal dilation of $(T, \scrC)$ must derive from a dilation obtained by simply dilating every component in a triple of possible components. That result also brings us for the first time promisingly close to the concepts of quantum self-testing: The various triples $(T_\sfA, T_\sfB,s)$ will correspond essentially to \emph{quantum strategies} which produce a given input-output behaviour. \\

Now, in a general theory it seems that we cannot achieve a better density theorem (in the case of Bell-channels) than \cref{thm:BellDense}. In specific theories, however, certain features often make an improvement possible. As such, it is essential to have a criterion which expresses when two dilations of the form \eqref{eq:belldil} are related in the pre-order $\cder$. Indeed, if one is derivable from another, then the weaker one can be removed from the class without losing density. We have the following:

\begin{Thm} (General Density Theorem for Bell-Channels, Part II.) \label{thm:BellDenseII} \\
Let $\scalemyQ{.8}{0.7}{0.5}{& \push{\cX_\sfA} \qw & \multigate{1}{(T, \scrC)} & \push{\cY_\sfA} \qw & \qw \\ & \push{\cX_\sfB} \qw & \ghost{(T, \scrC)} & \push{\cY_\sfB} \qw & \qw }$ be a bipartite Bell-channel. Then, two causal dilations of the form \eqref{eq:belldil} satisfy $\scalemyQ{.8}{0.7}{0.5}{
	&  \push{\cX_\sfA}  \qw& \multigate{1}{ L_\sfA} &  \push{\cY_\sfA} \qw & \qw  \\
	& \Nmultigate{2}{t}  & \ghost{L_\sfA} &  \push{\cE_\sfA} \ww & \ww\\
	& \Nghost{t} & \ww & \push{\cE_0} \ww & \ww\\
	& \Nghost{t} & \multigate{1}{L_\sfB} &  \push{\cE_\sfB}\ww & \ww  \\
	&   \push{\cX_\sfB} \qw & \ghost{L_\sfB} & \push{\cY_\sfB} \qw & \qw  \\
} \cder \scalemyQ{.8}{0.7}{0.5}{
&  \push{\cX_\sfA}  \qw& \multigate{1}{ L'_\sfA} &  \push{\cY_\sfA} \qw & \qw  \\
& \Nmultigate{2}{t'}  & \ghost{L'_\sfA} &  \push{\cE'_\sfA} \ww & \ww\\
& \Nghost{t'} & \ww & \push{\cE'_0} \ww & \ww\\
& \Nghost{t'} & \multigate{1}{L'_\sfB} &  \push{\cE'_\sfB}\ww & \ww  \\
&   \push{\cX_\sfB} \qw & \ghost{L'_\sfB} & \push{\cY_\sfB} \qw & \qw  \\
}$ if and only if there exist channels $F$, $G_\sfA$ and $G_\sfB$ such that 	

\begin{align}  \label{eq:ThinDense}
\scalemyQ{.8}{0.7}{0.5}{
	&   \push{\cX_\sfA} \qw  & \multigate{1}{L_\sfA}  &   \qw &  \push{\cY_\sfA}  \qw &  \qw  & \qw \\
	& \Nmultigate{4}{t}  & \ghost{L_\sfA} &  \push{\cE_\sfA} \ww &  \Nmultigate{1}{G_\sfA}{\ww} & \push{\cE'_\sfA} \ww & \ww \\
	& \Nghost{t}   &  & \Nmultigate{2}{F} & \Nghost{G_\sfA}{\ww} &  \\  
	& \Nghost{t}  & \push{\cE_0} \ww &   \Nghost{F}{\ww}  & \ww & \push{\cE'_0} \ww & \ww \\
	& \Nghost{t}  &  & \Nghost{ F} &   \Nmultigate{1}{G_\sfB}{\ww}   \\
	& \Nghost{t}   & \multigate{1}{L_\sfB} &  \push{\cE_\sfB} \ww & \Nghost{G_\sfB}{\ww} & \push{\cE'_\sfB}\ww & \ww \\ 
	&   \push{\cX_\sfB} \qw  & \ghost{L_\sfB}   & \qw &  \push{\cY_\sfB} \qw &   \qw & \qw} =\scalemyQ{.8}{0.7}{0.5}{
	& \push{\cX_\sfA}\qw  & \multigate{1}{ L'_\sfA} &  \push{\cY_\sfB}  \qw & \qw  \\
	& \Nmultigate{2}{t'}  & \ghost{L'_\sfA} &  \push{\cE'_\sfA} \ww & \ww\\
	& \Nghost{t'} & \ww & \push{\cE'_0} \ww & \ww\\
	& \Nghost{t'} & \multigate{1}{L'_\sfB} & \push{\cE'_\sfB} \ww & \ww  \\
	&  \push{\cX_\sfA} \qw  & \ghost{L'_\sfB} &  \push{\cY_\sfB} \qw & \qw  \\
}  \quad. 
\end{align}

	\end{Thm}

\begin{Remark} (General Bell-channels.)\\
	Again, it should be clear how to generalise the statement to multipartite Bell-channels. In the case of a unipartite Bell-channel (i.e. of a primitive causal channel $\channel{\cX}{(T, \scrC)}{\cY}$), we have $\scalemyQ{.8}{0.7}{0.5}{
		&  \push{\cX_\sfA}  \qw & \multigate{1}{ L} &  \push{\cY_\sfA} \qw & \qw  \\
		& \Nmultigate{1}{t}   & \ghost{L} &  \push{\cE} \ww & \ww\\
		& \Nghost{t} & \ww & \push{\cE_0} \ww & \ww\\
	} \cder \scalemyQ{.8}{0.7}{0.5}{
		&  \push{\cX_\sfA}  \qw & \multigate{1}{ L'} &  \push{\cY_\sfA} \qw & \qw  \\
		& \Nmultigate{1}{t'}   & \ghost{L'} &  \push{\cE'} \ww & \ww\\
		& \Nghost{t'} & \ww & \push{\cE'_0} \ww & \ww\\
	}$ if and only if there exist channels $G$ and $F$ such that

\begin{align}
\scalemyQ{.8}{0.7}{0.5}{
	&  \push{\cX_\sfA}  \qw & \multigate{1}{ L} &  \qw & \push{\cY_\sfA} \qw & \qw  \\
	& \Nmultigate{2}{t}   & \ghost{L} &  \push{\cE} \ww & \Nmultigate{1}{G}{\ww} & \push{\cE'} \ww & \ww \\
	& \Nghost{t} &   & \Nmultigate{1}{F} &   \Nghost{G}{\ww} \\
	& \Nghost{t} & \push{\cE_0} \ww & \Nghost{F}{\ww} & \ww &  \push{\cE'_0} \ww & \ww 
}  \quad = \quad 
 \scalemyQ{.8}{0.7}{0.5}{
	&  \push{\cX_\sfA}  \qw & \multigate{1}{ L'} &  \push{\cY_\sfA} \qw & \qw  \\
	& \Nmultigate{1}{t'}   & \ghost{L'} &  \push{\cE'} \ww & \ww\\
	& \Nghost{t'} & \ww & \push{\cE'_0} \ww & \ww\\
} \quad.
\end{align}	\end{Remark}

\begin{proof}
By \cref{lem:DerSimpl}, the derivability relation holds if and only if there exists a constructible causal channel $(G, \scrB)$ such that

\begin{align} 
\scalemyQ{1}{0.7}{0.5}{
	& \push{\cX_\sfA}\qw  & \multigate{1}{ L_\sfA} &  \push{\cY_\sfB}  \qw & \qw  \\
	& \Nmultigate{2}{t}  & \ghost{L_\sfA} &  \push{\cE_\sfA} \ww & \Nmultigate{2}{(G, \scrB)}{\ww} & \push{\cE'_\sfA} \ww & \ww \\
	& \Nghost{t} & \ww & \push{\cE_0} \ww & \Nghost{(G, \scrB)}{\ww} & \push{\cE'_0} \ww & \ww \\
	& \Nghost{t} & \multigate{1}{L_\sfB} & \push{\cE_\sfB} \ww & \Nghost{(G, \scrB)}{\ww} & \push{\cE'_\sfB} \ww & \ww   \\
	&  \push{\cX_\sfA} \qw  & \ghost{L_\sfB} &  \push{\cY_\sfB} \qw & \qw  \\
}  =\scalemyQ{1}{0.7}{0.5}{
	& \push{\cX_\sfA}\qw  & \multigate{1}{ L'_\sfA} &  \push{\cY_\sfB}  \qw & \qw  \\
	& \Nmultigate{2}{t'}  & \ghost{L'_\sfA} &  \push{\cE'_\sfA} \ww & \ww\\
	& \Nghost{t'} & \ww & \push{\cE'_0} \ww & \ww\\
	& \Nghost{t'} & \multigate{1}{L'_\sfB} & \push{\cE'_\sfB} \ww & \ww  \\
	&  \push{\cX_\sfA} \qw  & \ghost{L'_\sfB} &  \push{\cY_\sfB} \qw & \qw  \\
}  \quad. 
\end{align}

Since the causal specifications on each side must match, we conclude that $\scrB(\sfe'_0) \subseteq \{\sfe_0\}$, $\scrB(\sfe'_\sfA) \subseteq \{\sfe_0, \sfe_\sfA\}$ and $\scrB(\sfe'_\sfB) \subseteq \{\sfe_0, \sfe_\sfB\}$. Now, however, it can be demonstrated by an argument slightly more general than in the proof of \cref{thm:BellDense} (when arguing about the structure of $(L, \scrE)$) that if a constructible $(G, \scrB)$ has a specification with those properties, then it is necessarily of the form represented in \cref{eq:ThinDense}. (Importantly, we rely in this proof on the fact that we fixed the scheme $\bfE$ of allowed causal channels to be the constructible causal channels.)
	\end{proof}

\subsection{... in Cartesian Theories}

In cartesian theories, the above observations are not particularly useful, except perhaps in the case of Bell-channels. However, it is quite easy to see that, regardless of whether the channel $(T, \scrC)$ is a Bell-channel or not, it has a causally complete dilation: 

\begin{Thm} (Completeness and Rigidity in Cartesian Theories.) \label{thm:CartRigid}\\
	Assume that $\Theory$ is a cartesian theory, and let $\channel{\bbX}{(T, \scrC)}{\bbY}$ be any (constructible) causal channel in $\Theory$. Then $(T, \scrC)$ is rigid, and a complete causal dilation is given by 
	
	\begin{align} \label{eq:dilcopycaus}
	\myQ{0.7}{0.5}{&& &  \ww  & \ww & \ww  \\
		& && &  \vdots \\ &&& \\ &  & & & \ww & \ww \\& \push{\cX_1} \qw  & \ctrlo{-4} & \qw &  \multigate{2}{(T, \scrC)} \\ & \vdots & & &  \Nghost{(T, \scrC)} & \push{\bbY} \qw& \qw \\ & \push{\cX_n} \qw & \qw & \ctrlo{-3} & \ghost{(T, \scrC)}} \quad,
	\end{align}

where the individual copy channels are given their primitive causal specifications.  (Observe in this regard that the copy channels are formally also boxes in the stencil.)

\end{Thm}

\begin{Remark}
	Note that a restriction to constructible channels in $\Theory$ is void, since by \cref{thm:CartAllCons} all causal channels in a cartesian theory are constructible. 
	\end{Remark}

\begin{proof}
It suffices to show that any one-sided causal dilation of $(T, \scrC)$ can be derived from the dilation \eqref{eq:dilcopycaus}. If however $(L, \scrE)$ is a one-sided dilation, then the underlying channel $L$ can be derived by completeness of copying for \myuline{channels} (\cref{thm:CartisComp}), so only the causal specification $\scrE$ remains to be accounted for. But we can argue abstractly as earlier, when we discussed the standard notion of contraction, by noting that the collection of causal specifications themselves form a cartesian theory themselves, that $\scrE$ is as such a dilation of $\scrC$, and that the causal specification of the channel \eqref{eq:dilcopycaus} is a complete dilation of the specification $\scrC$. More down-to-earth arguments are also possible, and left as exercise. 
	\end{proof}

\subsection{... in $\CIT$}
\label{subsec:DenseCIT}

In the theory $\CIT$ things are much more complicated than in cartesian theories, and here the general discussion of \cref{subsec:DenseGeneral} can be put to use. We will restrict attention entirely to Bell-channels.\\ %

Let us start by considering the case of a unipartite Bell-channel. Put differently, this is really to study the causal dilations of a channel between simple interfaces, $\channel{X}{(T, \scrC)}{Y}$, where $\scrC$ is the primitive specification given by $\scrC(\sfy)=\{\sfx\}$. (Note that this scenario includes the `bit refreshment' channel of \cref{ex:bitrefresh}.) The analysis of this simple case is already rather involved, culminating with \cref{thm:UniBell}.

By \cref{thm:BellDense}, any constructible dilation of $(T, \scrC)$ is derivable from a dilation of the form 

\begin{align} \label{eq:densestart}
\myQ{0.7}{0.5}{
	& \qw &  \push{X}  \qw & \multigate{1}{ L} &  \push{Y} \qw & \qw  \\
	& \Nmultigate{1}{t}  & \push{R} \qw & \ghost{L} &  \push{E} \ww & \ww\\
	& \Nghost{t} & \ww & \ww & \push{E_0} \ww & \ww\\
} \quad ,
\end{align}

where the state $t$ and the channel $L$ are given their primitive specifications. Due to special features of $\CIT$, this density statement can be improved; we shall now argue that effectively any randomness involved in $L$ can be pushed to $E_0$ as acausal side-information, thus moving up higher in the causal-dilational ordering and thereby narrowing the dense class:

Every classical channel is a convex combination of deterministic channels (functions). In particular, this holds for the channel $\scalemyQ{.8}{0.7}{0.5}{	&  \push{X}  \qw& \multigate{1}{ T} &  \push{Y} \qw & \qw  \\
	& \push{R} \qw & \ghost{T}} := \scalemyQ{.8}{0.7}{0.5}{	&  \push{X}  \qw& \multigate{1}{ L} &  \push{Y} \qw & \qw  \\
	& \push{R} \qw & \ghost{L} &  \push{E} \ww & \Ngate{\tr}{\ww}}$; but this is to say that 

\begin{align} \label{eq:convdec}
\myQ{0.7}{0.5}{
		&  \push{X}  \qw& \multigate{1}{ T} &  \push{Y} \qw & \qw  \\
	& \push{R} \qw & \ghost{T}
} \quad = \quad 
\myQ{0.7}{0.5}{
	&\qw 	&  \push{X}  \qw&  \multigate{2}{f} &  \push{Y} \qw & \qw  \\
	& \qw 	& \push{R} \qw & \ghost{f} & \\
	& \Ngate{q} & \push{H}\qw & \ghost{f}
} \quad 
\end{align}

for some deterministic channel (function) $f$ and some state $q$ on some system $H$. Forgetting for a moment about causality, $L$ is evidently a dilation of the channel \eqref{eq:convdec}, so by completeness and localisability of $\CIT$ (\cref{thm:CITComp} and \cref{thm:CITLoc}) the channel $L$ can be derived (in the dilational ordering of \cref{chap:Dilations}) from the dilation given by copying $q$ and the inputs to $f$, that is, there exists a channel $G$ such that 

\begin{align}
\myQ{0.7}{0.5}{
	&  \push{X}  \qw& \multigate{1}{L} &  \push{Y} \qw & \qw  \\
	& \push{R} \qw & \ghost{L} & \push{E} \ww & \ww
} \quad = \quad 
\myQ{0.7}{0.5}{
	& \qw & \push{X} \qw &\qw 	& \ctrlo{3} &  \multigate{2}{f} &  \push{Y} \qw & \qw  \\
	& \qw & \push{R} \qw 	& \ctrlo{3} & \qw & \ghost{f} & \\
	& \Ngate{q} & \ctrlo{3} & \qw & \qw  & \ghost{f} \\
	& 	& & &&  \Nmultigate{2}{G}{\ww} \\
	& 	& &&  \ww & \Nghost{G}{\ww} & \push{E} \ww & \ww\\
	& 	& & \ww & \ww & \Nghost{G}{\ww}
} \quad .
\end{align}

(Clearly, we need not include two copies of $q$; one suffices.) Turning causality back on, we see that by giving each component on the right hand side its primitive specification, $L$ indeed gains the primitive specification, so in summary we can rewrite \eqref{eq:densestart} as 

\begin{align}
\myQ{0.7}{0.5}{
&	& \qw &  \push{X}\qw &\qw 	& \ctrlo{3} &  \multigate{2}{f} &  \push{Y} \qw & \qw  \\
&  \Nmultigate{5}{t}	& \qw & \push{R} \qw 	& \ctrlo{3} & \qw & \ghost{f} & \\
&\Nghost{t}	& \Ngate{q} & \ctrlo{3} & \qw & \qw  & \ghost{f} \\
&\Nghost{t}		& 	& & &&  \Nmultigate{2}{G}{\ww} \\
&\Nghost{t}		& 	& &&  \ww & \Nghost{G}{\ww} & \push{E} \ww & \ww\\
&\Nghost{t}		& 	& & \ww & \ww & \Nghost{G}{\ww}\\
& \Nghost{t}	& \ww & \ww & \ww & \push{E_0} \ww & \ww & \ww
} \quad ,
\end{align}

with each component given its primitive specification. Now, by simply removing $G$ we obviously rise in the causal-dilational ordering -- in fact, we generically rise \myuline{strictly}, since, importantly, the copy of $q$ transits from contributing to the side-information available after the input to $\sfx$ (as stalled by $G$) to \myuline{acausal} side-information (cf. \cref{ex:Acausal}). A similar thing can be said about the copy of the $R$-system, and by merging the three acausal hidden outputs in the environment, we conclude in summary that the causal dilation \cref{eq:densestart} is derivable from a causal dilation of the form 

\begin{align}
\myQ{0.7}{0.5}{
	&	& \qw &  \push{X}\qw &\qw 	& \ctrlo{2} &  \multigate{1}{f} &  \push{Y} \qw & \qw  \\
& 	&  \Nmultigate{2}{\tilde{t}}	&  \push{R} \qw 	& \qw &  \qw & \ghost{f} & \\
& 	&\Nghost{\tilde{t}}	& & & & \push{X} \ww& \ww&  \\
&	& \Nghost{\tilde{t}}	& \ww & \ww & \ww & \push{\tilde{E}_0} \ww & \ww 
} \quad ,
\end{align}

so such dilations form a dense class. \\

One further improvement is possible by noting that since $\tilde{t}$ is a dilation of its $R$-marginal, it can be derived from a copy of that marginal by completeness of copying (\cref{thm:CITComp}). Hence, we have altogether proved the following density theorem for constructible dilations:

\begin{Thm} (Preliminary Density Theorem for Unipartite Bell-Channels in $\CIT$.)\label{thm:UniDensePrel}\\
Let $\channel{X}{(T, \scrC)}{Y}$ be a primitive causal channel in $\CIT$ between simple interfaces. Then the class of causal dilations of the form 

\begin{align} \label{eq:unidense}
\myQ{0.7}{0.5}{
	&	& \qw &  \push{X}\qw &\qw 	& \ctrlo{2} &  \multigate{1}{f} &  \push{Y} \qw & \qw  \\
	& 	&  \Ngate{p}	&  \push{R} \qw 	& \ctrlo{2} &  \qw & \ghost{f} & \\
	& 	&	& & & & \push{X} \ww& \ww&  \\
		& 	&	& & & \ww & \push{R} \ww& \ww&  \\
} \quad ,
\end{align}

where $p$ is a state on some system $R$ and $f$ a deterministic channel (function), and where each component is given its primitive causal specification, is a dense class of dilations.
\end{Thm}

\cref{thm:BellDenseII} tells us when two dilations of the form \eqref{eq:unidense} are related in the causal-dilational ordering, and will allow us to investigate whether a further thinning of this dense class is possible. But another question merits attention first: How can we interpret the dilations \eqref{eq:unidense}? \\

This question is actually easy to answer. The dilations \eqref{eq:unidense} are indexed by pairs $(f,p)$, and to understand which pairs give rise to a dilation we simply need to observe that the channel \eqref{eq:unidense} constitutes a dilation of $(T, \scrC)$ precisely when $\scalemyQ{.8}{0.7}{0.5}{& \push{X} \qw & \multigate{1}{f} & \push{Y} \qw & \qw \\ & \Ngate{p} & \ghost{f} } = \channel{X}{T}{Y}$. If we consider the function $f: X \times R \to Y$ equivalently as a family of functions $f_r := f(\cdot, r) :X \to Y $ indexed by $r \in R$, then this requirement is the condition 

\begin{align}
\sum_{r \in R} p(r) f_r = T,
\end{align}

in other words, \emph{the dilations \eqref{eq:unidense} correspond to convex decompositions of $T$ into deterministic functions}. 

In this light, \cref{thm:UniDensePrel} is perhaps not so surprising -- what it says, basically, is that any causal dilation of $T$ can be thought of as deriving from one in which the agent in the environment draws a random $r \in R$ according to the distribution $p$, keeps a copy of $r$ in memory while using another copy to choose a function $f_r$ to apply to our input, copies our input $x \in X$ to memory and gives us the value $f_r(x)$ as output. If we think about it, this exactly fits the scheme of the two causal dilations of the bit refreshment channel from \cref{ex:bitrefresh}; the two dilations correspond to the convex decompositions $\frac{1}{2} f_0 + \frac{1}{2} f_1$, where $f_k: \{0,1\} \to \{0,1\}$ is the function that is constantly $k$, respectively $\frac{1}{2} \id_{\{0,1\}}  + \frac{1}{2} \up{NOT}$, where $\up{NOT}: \{0,1\} \to \{0,1\}$ is the bit flip. \\

This understanding of the nature of the dilations also gives a hint on how to thin the dense class further. First of all, we may clearly assume that $p$ has full rank on $R$, i.e. $p(r)> 0$ for all $r \in R$. Secondly, any convex decomposition $(f,p)$ of $T$ can intuitively be reduced to one for which $f_r \neq f_{r'}$ whenever $r \neq r'$; this comes about by merging any instances of $r,r'$ which violate this. Formally, we replace $R$ by the set of equivalence classes $R / \sim$ under the equivalence relation $r \sim r' \Leftrightarrow f_r = f_{r'}$, we replace $(f_r)_{r \in R}$ by $(\tilde{f}_{[r]})_{[r] \in R / \sim}$, well-defined by $\tilde{f}_{[r]}=f_r$, and we replace $p$ by $\tilde{p}: R / \sim \, \to [0,1]$ given by $\tilde{p}([r]) = \sum_{r' \in [r]} p(r)$; the original dilation \eqref{eq:unidense} can be derived from the new dilation given by $(\tilde{f}, \tilde{p})$, since we can draw a value of $r$ in the environment conditional on its equivalence class. 
In summary, the dense class of \cref{thm:UniDensePrel} can be thinned to those dilations \eqref{eq:unidense} for which the pair $(f,p)$ correspond to a \myuline{proper} convex decomposition of $T$, that is, one for which $p(r)> 0$ for all $r$ and $f_r \neq f_{r'}$ for $r \neq r'$. \myuline{That} dense class essentially cannot be thinned any further, however, for what we have done is to identify a class of inequivalent maximal causal dilations:

\begin{Thm} (Ultimate Density Theorem for Unipartite Bell-Channels in $\CIT$.) \label{thm:UniBell}	\\
	Let $\channel{X}{(T, \scrC)}{Y}$ be a primitive causal channel in $\CIT$ between simple interfaces. Then the class of causal dilations of the form 
	
	\begin{align} \label{eq:ultunidense}
	\myQ{0.7}{0.5}{
		&	& \qw &  \push{X}\qw &\qw 	& \ctrlo{2} &  \multigate{1}{f} &  \push{Y} \qw & \qw  \\
		& 	&  \Ngate{p}	&  \push{R} \qw 	& \ctrlo{2} &  \qw & \ghost{f} & \\
		& 	&	& & & & \push{X} \ww& \ww&  \\
		& 	&	& & & \ww & \push{R} \ww& \ww&  \\
	} \quad ,
	\end{align}
	
where $p$ is a full-rank state on some system $R$, and where $f:X \times R \to Y$  is a function with $f_r \neq f_{r'}$ for $r \neq r'$, is a dense class of dilations. Moreover, each such dilation is $\cder$-maximal, and the dilations given by $(f,p)$ and $(f', p')$ are $\cder$-equivalent if and only they are related by a relabelling of the elements in $R$, i.e. there exists a bijection $h: R \to R'$ such that $p(r) = p'(h(r))$ and $f_r=f'_{h(r)}$ for all $r \in R$. 
\end{Thm}

\begin{Remark} (The Dilational Ordering versus the Causal-Dilational Ordering.) \label{rem:CompleteWithout}\\
	It is worth observing that if we forget about causality, then the dilations \eqref{eq:ultunidense} are all complete dilations of the underlying channel $T$. In particular, all these dilations are \myuline{equivalent} under the dilational ordering of \cref{chap:Dilations}, that is, there always exist channels relating them in the environment. The significance of the \myuline{causal-}dilational ordering, which renders some of them inequivalent, is that those channels in the environment are required to take a specific form so as to preserve causality.
\end{Remark}

\begin{proof}
	We have already argued that the class is dense. It remains to show maximality of each dilation, and the criterion for equivalence. 
	
Assume we are given two comparable dilations within the class, say $	\scalemyQ{.8}{0.7}{0.5}{
	&	& \qw &  \push{X}\qw &\qw 	& \ctrlo{2} &  \multigate{1}{f} &  \push{Y} \qw & \qw  \\
	& 	&  \Ngate{p}	&  \push{R} \qw 	& \ctrlo{2} &  \qw & \ghost{f} & \\
	& 	&	& & & & \push{X} \ww& \ww&  \\
	& 	&	& & & \ww & \push{R} \ww& \ww&  \\
} \cder \scalemyQ{.8}{0.7}{0.5}{
&	& \qw &  \push{X}\qw &\qw 	& \ctrlo{2} &  \multigate{1}{f'} &  \push{Y} \qw & \qw  \\
& 	&  \Ngate{p'}	&  \push{R'} \qw 	& \ctrlo{2} &  \qw & \ghost{f'} & \\
& 	&	& & & & \push{X} \ww& \ww&  \\
& 	&	& & & \ww & \push{R'} \ww& \ww&  \\
}$. We show that the relabelling criterion is satisfied. This will prove all the desired statements (since the relabelling condition is clearly sufficient for equivalence). 

First, note that \cref{thm:BellDenseII} implies the existence of channels $F$ and $G$ such that 

\begin{align} \label{eq:maxiproof}
	\scalemyQ{1}{0.7}{0.5}{
	&	& \qw &  \push{X}\qw &\qw 	& \ctrlo{2} &  \multigate{1}{f} &  \push{Y} \qw & \qw  \\
	& 	&  \Ngate{p}	&  \push{R} \qw 	& \ctrlo{3} &  \qw & \ghost{f} & \\
	& 	&	& & & & \push{X} \ww& \ww & \Nmultigate{1}{G}{\ww}&  \push{X} \ww & \ww \\
	&&& & & & \Nmultigate{1}{F} & \push{Z} \ww & \Nghost{G}{\ww}  \\
	& 	&	& & & \push{R} \ww& \Nghost{F}{\ww}&  \ww &  \push{R'} \ww & \ww\\
} =  \scalemyQ{1}{0.7}{0.5}{
	&	& \qw &  \push{X}\qw &\qw 	& \ctrlo{2} &  \multigate{1}{f'} &  \push{Y} \qw & \qw  \\
	& 	&  \Ngate{p'}	&  \push{R'} \qw 	& \ctrlo{2} &  \qw & \ghost{f'} & \\
	& 	&	& & & & \push{X} \ww& \ww&  \\
	& 	&	& & & \ww & \push{R'} \ww& \ww&  \\
} \quad.
\end{align}

By trashing the hidden copy of $X$, we see that 

\begin{align} \label{eq:Xtrashed}
\scalemyQ{1}{0.7}{0.5}{
	& \qw &  \push{X}\qw &\qw  &  \multigate{1}{f} &  \push{Y} \qw & \qw  \\
		&  \Ngate{p}	&  \push{R} \qw 	& \ctrlo{1}  & \ghost{f}  \\
		& & && \push{R} \ww& \Ngate{H}{\ww}&  \push{R'} \ww & \ww}=\scalemyQ{1}{0.7}{0.5}{
		& \qw &  \push{X}\qw &\qw  &  \multigate{1}{f'} &  \push{Y} \qw & \qw  \\
		&  \Ngate{p'}	&  \push{R'} \qw 	& \ctrlo{1}  & \ghost{f'}  \\
		& & &&  \ww&  \push{R'} \ww & \ww} \quad,
\end{align}

where $H$ is the marginal of $F$. Let us prove that this identity can only hold if $H =:h$ is in fact a deterministic bijection from $R$ to $R'$; this implies the desired. 

Given $r \in R$ and $r' \in R'$, let us write $\up{P}(H(r)=r')$ for the `probability that $H$ maps $r$ to $r'$', that is, for the quantities defined by $H(\delta_r)= \sum_{r' \in R'} \up{P}(H(r)=r') \delta_{r'}$. Algebraically, \cref{eq:Xtrashed} is then the identity $ \sum_{r' \in R'}  \sum_{r \in R} p(r) \up{P}(H(r)=r')\,  f_r \otimes \delta_{r'} = \sum_{r' \in R'} p'(r') \, f'_{r'} \otimes \delta_{r'}$.

 Comparing terms, this identity is equivalent the identities

\begin{align} \label{eq:condmixt}
f'_{r'} = \sum_{r \in R} \frac{p(r)}{p'(r')} \up{P}(H(r)=r') \, f_r  \quad \text{for all $r' \in R'$}.
\end{align}

(These have the following interpretation: $f'_{r'}$ is the probabilistic mixture of the $f_r$'s, with weights determined by the conditional distribution of $r$ given $r'=H(r)$.) But now the desired follows: For any fixed $r' \in R'$, extremality of the channel $f'_{r'}$ and the fact that $p(r)> 0$ for all $r \in R$ implies by \cref{eq:condmixt} that  $\up{P}(H(r)=r')  \in \{0,1\}$ for all $r \in R$, and that there must be a unique (since the $f_r$'s are all distinct) $r$ with $\up{P}(H(r)=r')  =1$. This is precisely to say that $H =: h$ is deterministic, injective and (since $r'$ was arbitrary) surjective. 

	\end{proof}

\cref{thm:UniBell} ends our analysis of the unipartite case, and the conclusion is clear: Every causal dilation of $(T, \scrC)$ is derivable from a maximal causal dilation, and the maximal causal dilations correspond precisely to the distinct convex decompositions of $T$ into deterministic functions. As such, \emph{the study of causal dilations of $(T, \scrC)$ is the study of convex decompositions of $T$.}

The following consequence is immediate:

\begin{Cor} (Rigidity of Unipartite Bell-Channels in $\CIT$.) \label{cor:UniRigidCIT}\\
A primitive causal channel between simple interfaces $\channel{X}{(T, \scrC)}{Y}$ in $\CIT$ is rigid if and only if $T$ admits a unique convex decomposition into functions from $X$ to $Y$. 	
	\end{Cor}

\vspace{.1cm}

\begin{Example} (Extremal implies Rigid.)\\
If $\channel{X}{(T, \scrC)}{Y}$ is deterministic, it is rigid.	\end{Example}

\begin{Example} (Rigid does not imply Extremal.)\\
	If $\abs{X}=1$, then $\channel{X}{(T, \scrC)}{Y}$ corresponds to a state on $Y$, which is released upon giving an input (`pushing a button'). Any state in $\CIT$ has a unique convex decomposition into deterministic functions (pure states), hence every such causal channel is rigid. In particular, extremality of $T$ is not necessary for rigidity.  \end{Example}

\begin{Example} (Non-Rigidity of Bit Refreshment.) \label{ex:NonrigidBitRefresh} \\
The bit refreshment dilations from \cref{ex:bitrefresh} are maximal and inequivalent, as they correspond to the distinct convex decompositions $\frac{1}{2} f_0 + \frac{1}{2} f_1$ and $\frac{1}{2} \id_{\{0,1\}} + \frac{1}{2} \up{NOT}$, as described earlier. Hence, the bit refreshment channel is \myuline{not} rigid. 	\end{Example}

\vspace{.2cm}
The non-rigidity of the bit refreshment channel has another peculiar consequence: 

\begin{Remark} (Rigidity is Non-Composable -- Temporal Localisability fails for Causal Dilations.)\\
Consider the two causal channels $\channel{\{0,1\}}{\tr}{\triv}$ and $\channel{\triv}{r}{\{0,1\}}$ in $\CIT$, both equipped with their primitive specifications. By \cref{ex:causexttrash} and \cref{ex:causdilstate}, both of them have complete causal dilations (so they are rigid). However, their serial composition is the bit refreshment channel, which is not rigid; in particular, the composition of the complete causal dilations for $\channel{\{0,1\}}{\tr}{\triv}$ and $\channel{\triv}{r}{\{0,1\}}$ is not a complete causal dilation of the bit refreshment, that is, the principle of temporal localisability fails for causal channels in $\CIT$. 
	\end{Remark}

Now, let us move on to the multipartite case. The analysis that led us in the unipartite case to the preliminary density theorem \cref{thm:UniDensePrel} generalises without difficulty (for simplicity, it is stated merely for the bipartite case):

\begin{Thm} (Preliminary Density Theorem for Bipartite Bell-Channels in $\CIT$.) \label{thm:BiBell}	\\
	Let $\scalemyQ{.8}{0.7}{0.5}{& \push{X_\sfA} \qw & \multigate{1}{(T, \scrC)} & \push{Y_\sfA} \qw & \qw \\ & \push{X_\sfB} \qw & \ghost{(T, \scrC)} & \push{Y_\sfB} \qw & \qw }$ be a bipartite Bell-channel in $\CIT$. Then the class of causal dilations of the form 
	
	\begin{align}  \label{eq:bidense}
	\myQ{0.7}{0.5}{& &  & \ww & \push{X_\sfA} \ww & \ww\\
		&  \push{X_\sfA}  \qw& \ctrlo{-1} & \multigate{1}{ f_\sfA} &  \push{Y_\sfA} \qw & \qw  \\
		& \Nmultigate{3}{p}  & \ctrlo{1} & \ghost{f_\sfA} \\
		& \Nghost{p} &   & \ww & \push{R_\sfA} \ww & \ww\\
			& \Nghost{p} & & \ww & \push{R_\sfB} \ww & \ww\\
		& \Nghost{p} & \ctrlo{-1} & \multigate{1}{f_\sfB}   \\
		&   \push{X_\sfB} \qw & \ctrlo{1} & \ghost{f_\sfB} & \push{Y_\sfB} \qw & \qw  \\
		& & & \ww &  \push{X_\sfB} \ww & \ww 
	} \quad ,
	\end{align}
	
	where $p$ is a state on some system $R_\sfA \times R_\sfB$, where $f_\sfA$ and $f_\sfB$ are deterministic channels (functions), and where each component is given its primitive causal specification, is a dense class of dilations.

\end{Thm}

Clearly, we may in \cref{thm:BiBell} moreover restrict to triples $(f_\sfA, f_\sfB, p)$ for which the state $p$ has \myuline{locally} full rank, meaning that the $\sfA$- and $\sfB$-marginals of $p$ both have full rank. However, the rest of the analysis that led us in the unipartite case to the improved density theorem \cref{thm:UniBell} is not immediately transferable. In particular, that $p$ has locally full rank does not imply that $p$ itself has full rank, indeed there can be strong correlations between the $\sfA$- and $\sfB$-parts of the state.

In fact, I will leave open the problem of finding an equivalent to \cref{thm:UniBell}, characterising the $\cder$-maximal causal dilations and $\cder$-equivalence among them. The following improvement of \cref{thm:BellDenseII} (which appeared in disguise in the proof of \cref{thm:UniBell} in the unipartite case) might be helpful in this regard:

\begin{Thm} (Relations within the Dense Class for Bipartite Bell-Channels.)\\
		Let $\scalemyQ{.8}{0.7}{0.5}{& \push{X_\sfA} \qw & \multigate{1}{(T, \scrC)} & \push{Y_\sfA} \qw & \qw \\ & \push{X_\sfB} \qw & \ghost{(T, \scrC)} & \push{Y_\sfB} \qw & \qw }$ be a bipartite Bell-channel in $\CIT$. Then, two causal dilations of the form \eqref{eq:bidense} satisfy $\scalemyQ{.8}{0.7}{0.5}{& &  & \ww & \push{X_\sfA} \ww & \ww\\
		&  \push{X_\sfA}  \qw& \ctrlo{-1} & \multigate{1}{ f_\sfA} &  \push{Y_\sfA} \qw & \qw  \\
		& \Nmultigate{3}{p}  & \ctrlo{1} & \ghost{f_\sfA} \\
		& \Nghost{p} &   & \ww & \push{R_\sfA} \ww & \ww\\
		& \Nghost{p} & & \ww & \push{R_\sfB} \ww & \ww\\
		& \Nghost{p} & \ctrlo{-1} & \multigate{1}{f_\sfB}   \\
		&   \push{X_\sfB} \qw & \ctrlo{1} & \ghost{f_\sfB} & \push{Y_\sfB} \qw & \qw  \\
		& & & \ww &  \push{X_\sfB} \ww & \ww 
	} \quad  \cder  \quad 	\scalemyQ{.8}{0.7}{0.5}{& &  & \ww & \push{X_\sfA} \ww & \ww\\
	&  \push{X_\sfA}  \qw& \ctrlo{-1} & \multigate{1}{ f'_\sfA} &  \push{Y_\sfA} \qw & \qw  \\
	& \Nmultigate{3}{p'}  & \ctrlo{1} & \ghost{f'_\sfA} \\
	& \Nghost{p'} &   & \ww & \push{R'_\sfA} \ww & \ww\\
	& \Nghost{p'} & & \ww & \push{R'_\sfB} \ww & \ww\\
	& \Nghost{p'} & \ctrlo{-1} & \multigate{1}{f'_\sfB}   \\
	&   \push{X_\sfB} \qw & \ctrlo{1} & \ghost{f'_\sfB} & \push{Y_\sfB} \qw & \qw  \\
	& & & \ww &  \push{X_\sfB} \ww & \ww 
} $ if and only if there exists a channel $F$ such that 

\begin{align} \label{eq:BiRec}
\myQ{0.7}{0.5}{
	&  \push{X_\sfA}  \qw& \qw & \multigate{1}{ f_\sfA} &  \push{Y_\sfA} \qw & \qw  \\
	& \Nmultigate{3}{p}  & \ctrlo{1} & \ghost{f_\sfA} \\
	& \Nghost{p} &    & \push{R_\sfA} \ww & \Nmultigate{1}{F}{\ww} & \ww \\
	& \Nghost{p} &  & \push{R_\sfB} \ww & \Nghost{F}{\ww} & \ww\\
	& \Nghost{p} & \ctrlo{-1} & \multigate{1}{f_\sfB}   \\
	&   \push{X_\sfB} \qw & \qw  & \ghost{f_\sfB} & \push{Y_\sfB} \qw & \qw } \quad = \quad 	\myQ{0.7}{0.5}{
	&  \push{X_\sfA}  \qw& \qw & \multigate{1}{ f'_\sfA} &  \push{Y_\sfA} \qw & \qw  \\
	& \Nmultigate{3}{p'}  & \ctrlo{1} & \ghost{f'_\sfA} \\
	& \Nghost{p'} &   & \ww & \push{R'_\sfA} \ww & \ww\\
	& \Nghost{p'} & & \ww & \push{R'_\sfB} \ww & \ww\\
	& \Nghost{p'} & \ctrlo{-1} & \multigate{1}{f'_\sfB}   \\
	&   \push{X_\sfB} \qw & \qw & \ghost{f'_\sfB} & \push{Y_\sfB} \qw & \qw 
} \quad .
\end{align}
	
	\end{Thm}

\begin{proof}
	The `only if'-direction is clear from \cref{thm:BellDenseII}, by trashing the hidden ports corresponding to $X_\sfA$ and $X_\sfB$ (this is also what we did in the proof of \cref{thm:UniBell}). The `if'-direction follows simply by noting that the identity $\scalemyQ{.8}{0.7}{0.5}{& &  & \ww & \push{X_\sfA} \ww & \ww\\
		&  \push{X_\sfA}  \qw& \ctrlo{-1} & \multigate{1}{ f_\sfA} &  \push{Y_\sfA} \qw & \qw  \\
		& \Nmultigate{3}{p}  & \ctrlo{1} & \ghost{f_\sfA} \\
		& \Nghost{p} &    & \push{R_\sfA} \ww & \Nmultigate{1}{F}{\ww} & \ww \\
	& \Nghost{p} &  & \push{R_\sfB} \ww & \Nghost{F}{\ww} & \ww\\
		& \Nghost{p} & \ctrlo{-1} & \multigate{1}{f_\sfB}   \\
		&   \push{X_\sfB} \qw & \ctrlo{1} & \ghost{f_\sfB} & \push{Y_\sfB} \qw & \qw  \\
		& & & \ww &  \push{X_\sfB} \ww & \ww 
	} =	\scalemyQ{.8}{0.7}{0.5}{& &  & \ww & \push{X_\sfA} \ww & \ww\\
		&  \push{X_\sfA}  \qw& \ctrlo{-1} & \multigate{1}{ f'_\sfA} &  \push{Y_\sfA} \qw & \qw  \\
		& \Nmultigate{3}{p'}  & \ctrlo{1} & \ghost{f'_\sfA} \\
		& \Nghost{p'} &   & \ww & \push{R'_\sfA} \ww & \ww\\
		& \Nghost{p'} & & \ww & \push{R'_\sfB} \ww & \ww\\
		& \Nghost{p'} & \ctrlo{-1} & \multigate{1}{f'_\sfB}   \\
		&   \push{X_\sfB} \qw & \ctrlo{1} & \ghost{f'_\sfB} & \push{Y_\sfB} \qw & \qw  \\
		& & & \ww &  \push{X_\sfB} \ww & \ww 
	} $ follows from \eqref{eq:BiRec}. 

	\end{proof}

The fact that we have not obtained a multipartite counterpart to \cref{thm:UniBell} in principle also leaves open the problem of finding a characterisation of \myuline{rigidity} in the multipartite case (i.e. an equivalent to \cref{cor:UniRigidCIT}). We do, however, have the following, which for simplicity is again only stated in the bipartite case:

\begin{Cor} (Sufficient Condition for Rigidity of Bipartite Bell-Channels in $\CIT$.) \label{cor:BiRigidCIT}\\
	Let $\scalemyQ{.8}{0.7}{0.5}{& \push{X_\sfA} \qw & \multigate{1}{(T, \scrC)} & \push{Y_\sfA} \qw & \qw \\ & \push{X_\sfB} \qw & \ghost{(T, \scrC)} & \push{Y_\sfB} \qw & \qw }$ be a bipartite Bell-channel in $\CIT$. If $T$ admits a unique convex decomposition into deterministic product channels, then $(T, \scrC)$ is rigid. 
	
	Explicitly, $(T, \scrC)$ is rigid if there exists a unique probability density $p: Y^{X_\sfA}_\sfA \times Y^{X_\sfB}_\sfB \to [0,1]$ (here, $Y^{X_i}_i$ denotes the set of functions from $X_i$ to $Y_i$) such that 
	
	\begin{align} \label{eq:UniqueDec}
T = 	\sum_{ g_\sfA, g_\sfB} p(g_\sfA, g_\sfB) \; g_\sfA \times g_\sfB .
	\end{align}
	
	\end{Cor}

\begin{proof}
By an argument similar to that used in the unipartite case, we may restrict the dense class of dilations \eqref{eq:bidense} to ensure that $f_\sfA(\cdot, r_\sfA) \times f_\sfB(\cdot, r_\sfB) \neq f_\sfA(\cdot, r'_\sfA) \times f_\sfB(\cdot, r'_\sfB) $ whenever the pairs $(r_\sfA, r_\sfB)$ and $(r'_\sfA, r'_\sfB)$ are distinct. However, the uniqueness of $p$ in the representation \eqref{eq:UniqueDec} means that there is, up to relabelling, only one dilation of this kind, so it must be a complete dilation. 	\end{proof}

\begin{Example} (Bipartite Rigidity without Marginal Rigidity.) \label{ex:BipartRigidCIT}\\
	The bipartite Bell-channel $\scalemyQ{.8}{0.7}{0.5}{& \push{\{0,1\}} \qw & \multigate{1}{(T, \scrC)} & \push{\{0,1\}} \qw & \qw \\ & \push{\{0,1\}} \qw & \ghost{(T, \scrC)} & \push{\{0,1\}} \qw & \qw }$ with $T= \frac{1}{2} \id_{\{0,1\}} \otimes \up{NOT} + \frac{1}{2} \up{NOT} \otimes \id_{\{0,1\}} $ is rigid. Both of its marginals are $\frac{1}{2} \id_{\{0,1\}} + \frac{1}{2} \up{NOT}  = \frac{1}{2} f_0 + \frac{1}{2} f_1$, i.e. they are the bit refreshment channel, which is not rigid. Hence, the rigidity of $(T, \scrC)$ resides in a sense in the correlation between the local behaviours. 
	\end{Example}

\vspace{.2cm}

\begin{OP} (Rigidity of Multipartite Bell-Channels.) \label{op:BiRigidCITop}\\
 \cref{cor:BiRigidCIT} says that the uniqueness of convex decomposition into deterministic product channels implies rigidity. Is the converse true? 
	\end{OP}  

\begin{OP} (An Ultimate Density Theorem for Multipartite Bell-Channels.) \label{op:Ultimatedensity}\\
For a bipartite (or generally multipartite) Bell-channel $(T, \scrC)$, is there an equivalent to \cref{thm:UniBell}, i.e. a dense class of $\cder$-maximal causal dilations, and what is a criterion for $\cder$-equivalence among such dilations? 	
	\end{OP}

\begin{OP} (Analysis of Other Causal Channels.) \label{op:CITotherchannels}\\
	Is it possible to give an analysis of the causal dilations of a constructible causal channel in $\CIT$ which is not a Bell-channel? 
	\end{OP}

\section{Summary and Outlook}
\label{sec:SummaryCausal}

In this chapter, we have defined \emph{causal channels} (\cref{def:CausChan}) and \emph{causal dilations} (\cref{def:CausDil}), both of which have been heavily exemplified. The result that ultimately justifies our definition of a causal channel is the fact that \emph{contractions} can be well-defined from knowledge of a channel and its causal specification alone (\cref{thm:StdCont}). An especially important class of causal channels are the \emph{constructible} channels (\cref{def:Constructible}), which model physically realisable causal channels. 

We have also seen how one may define \emph{notions of contraction} in general (\cref{def:AbsCont}), though the simplicity of this idea is slightly contaminated by the fact that we had to introduce \emph{schemes} of causal channels (\cref{def:Scheme}), which physically represent `all sensible causal channels' (e.g. the constructible ones), but mathematically are primarily needed so as to facilitate the \emph{Coherence of Nested Contraction} (as for the constructible scheme in \cref{thm:StContisAbsCont}). 

Given (a scheme and) a notion of contraction, it is possible to define the \emph{causal-dilational ordering} (\cref{def:CausDilOrd}), which is our final model for derivability among various dilations, hinted at already in the general introduction to the thesis. We have proved a handful of stability results about the causal-dilational ordering (\cref{thm:Contractionsofdilations}, \cref{thm:ParStab}, \cref{thm:ContStab}) which all serve to consolidate the concepts of causal dilations and derivability. 

Finally, we introduced the idea of \emph{density theorems} and \emph{rigidity} (\cref{sec:Density}), and have seen this exemplified most interestingly in the theory $\CIT$, where in particular we saw that the bit refreshment channel is an example of non-rigidity since it allows two distinct convex decompositions (\cref{ex:NonrigidBitRefresh}). \\

Several problems remain open for future investigation:

\begin{enumerate}
	
\item \cref{op:Contractions}: Given a universal theory, does there exist a notion of contraction in the scheme of all causal channels?

\item Can we find substitutes for the universal dilations of \cref{chap:Dilations} so as to nail down more precisely what the causal-dilational ordering looks like, similar to what we did in \cref{prop:Blackwell}?
	
	\item In particular, how can we further the understanding of density theorems and rigidity in $\CIT$, cf. \cref{op:BiRigidCITop}, \cref{op:Ultimatedensity} and \cref{op:CITotherchannels}?  
	
	\item Can we develop a metric version of the theory of causal channels and causal dilations, cf. the discussion in §4 of the prelude to this chapter? 
\end{enumerate}

\chapter {Rigidity and Quantum Self-Testing}
\label{chap:Selftesting}

{\centering
	\subsection*{§1. Introduction and Outline.}}

We concluded the preceding chapter by examining the causal-dilational ordering in $\CIT$ in a few special cases. The main lesson was that the dilational structure of \cref{chap:Dilations} is fundamentally altered when causality is taken into account. In particular, the completeness principle no longer reigns globally, but rather becomes a volatile feature which depends on specifics of the causal channel in question. This phenomenon persists in the theory $\QIT$, where it connects to the well-established field of quantum self-testing. That connection is the topic of this final chapter of the thesis. 

(The reader is recommended at this point to review the material in the preliminary section of the thesis if it is unknown.)  \\

Let us recall from the general introduction that \emph{quantum self-testing} (\cite{MY98,MY04,MYS12}) pertains to the following scenario:\footnote{We restrict for simplicity to the bipartite case.} Imagine that we interact with two separated computing devices, labelled by $\sfA$ and $\sfB$. Device $i \in \{\sfA, \sfB\}$ accepts inputs from a finite set $X_i$ and gives outputs from a finite set $Y_i$. By probing the devices many times, we may (under an i.i.d. assumption) establish the statistical input-output behaviour of the devices. These statistics are summarised by a collection of probability distributions $(P^x)_{x \in X}$ on the set of output pairs, $Y:= Y_\sfA \times Y_\sfB$, indexed by the set of input pairs, $X:= X_\sfA \times X_\sfB$. In general, the sets $X_\sfA, X_\sfB, Y_\sfA$ and $Y_\sfB$ are said to define a \emph{(bipartite) Bell-scenario}, and a collection of probability distributions $(P^x)_{x \in X}$ is called a \emph{behaviour}\footnote{Some authors use the term \emph{correlation}. The term `behaviour' is due to Cirelson (\cite{Cir93}).} for that Bell-scenario. Now assume that the devices work by sharing a bipartite quantum state $\varrho \in \St{\cH_\sfA \otimes \cH_\sfB}$, and device $i$ performing, on input $x_i \in X_i$, a projective quantum measurement to produce an outcome $y_i \in Y_i$, the measurement being described by the PVM $(\Pi^{x_i}_i(y_i))_{y_i \in Y_i}$. By the formalism of quantum theory, the input-output behaviour $P$ is given by the Born rule, $P^{x_\sfA, x_\sfB}(y_\sfA, y_\sfB) = \tr([\Pi^{x_\sfA}_\sfA(y_\sfA) \otimes \Pi^{x_\sfB}_\sfB(y_\sfB) ]\varrho)$. We then ask the following: \\

\emph{Can the state and measurements be deduced from the input-output behaviour $P$?}\\

A configuration of state and measurements, $(\varrho, \Pi_\sfA, \Pi_\sfB)$, is called a \emph{(tensor-product\footnote{There is in the infinite-dimensional case a more general notion of \emph{quantum commuting strategies} (\cite{Dyke16}). These are usually not considered in quantum self-testing, and neither shall we.}) quantum strategy}. Clearly, there is always more than one quantum strategy which produces a given input-output behaviour, since e.g. a local unitary rotation of state and measurements leaves the behaviour invariant. As such, taking the above question literally its answer is always in the negative. For some behaviours, however, it turns out that the strategy is `essentially unique', in the sense that every strategy $(\varrho, \Pi_\sfA, \Pi_\sfB)$ with behaviour $P$ is `reducible' to a fixed canonical strategy $(\tilde{\varrho}, \tilde{\Pi}_\sfA, \tilde{\Pi}_\sfB)$, where the state $\tilde{\varrho}=: \tilde{\psi}$ is pure.\footnote{In fact, many treatments also take the state $\varrho$ to be pure as well, as we did for simplicity in the general introduction. This, however, is unjustified as one should quantify universally over general strategies.}  \myuline{This} is the phenomenon of quantum self-testing. 

To be precise, we say that \emph{$(\varrho, \Pi_\sfA, \Pi_\sfB)$ is reducible to $(\tilde{\psi},\tilde{\Pi}_\sfA, \tilde{\Pi}_\sfB)$},  if there exist Hilbert spaces $\cH^\res_\sfA, \cH^\res_\sfB$ (called \emph{residual spaces}), isometries $W_i : \cH_i \to \tilde{\cH}_i \otimes \cH^\res_i$, and a pure state $\psi^\res$ on  $\cH^\res_\sfA \otimes \cH^\res_\sfB \otimes \cP$ such that 

\begin{align} \label{eq:attl}
(W \Pi^x(y) \otimes \bone_{\cP}) \ket{\psi} = \tilde{\Pi}^x(y) \tilde{\ket{\psi}} \otimes \ket{\psi^\res} \quad \text{for all $x \in X$, $y \in Y$},
\end{align}

where $W= W_\sfA \otimes W_\sfB$, $\Pi^x(y)= \Pi^{x_\sfA}_\sfA(y_\sfA) \otimes \Pi^{x_\sfB}_\sfB(y_\sfB)$, $\tilde{\Pi}^x(y)= \tilde{\Pi}^{x_\sfA}_\sfA(y_\sfA) \otimes \tilde{\Pi}^{x_\sfB}_\sfB(y_\sfB)$ and where $\psi$ is a purification of $\varrho$ with purifying system $\cP$.\footnote{It is easy to see that the choice of purification does not matter, since we may change the state $\psi^\res$ accordingly. The choices of vector representatives do not matter either, since we may absorb a potential phase into one of the isometries $W_i$.} (This definition, which includes a purification of the state $\varrho$, is modelled on that of Refs. \cite{SB19,RUV13}.)

Intuitively, reducibility expresses that the strategy $(\varrho, \Pi_\sfA, \Pi_\sfB)$ can be locally embedded into larger spaces and thereby realised as the strategy $(\tilde{\psi}, \tilde{\Pi}_\sfA, \tilde{\Pi}_\sfB)$, augmented by a \emph{residual state} $\psi^\res$ which is left unmeasured. It is obvious that such a relation between strategies implies that their behaviours are the same -- it is the converse which is interesting. One says that \emph{$P$ self-tests the strategy $( \tilde{\psi},\tilde{\Pi}_\sfA, \tilde{\Pi}_\sfB)$}  if every strategy $(\varrho,\Pi_\sfA, \Pi_\sfB)$ with behaviour $P$ is reducible to $(\tilde{\psi}, \tilde{\Pi}_\sfA, \tilde{\Pi}_\sfB)$.  Occasionally, one calls a behaviour $P$ \emph{rigid} if it self-tests \myuline{some} strategy, but we will refer to this as \emph{rigidity a.t.t.l.} (\emph{according to the literature}), so as to avoid confusion with rigidity in the sense of \cref{def:Rigidity} with which we ultimately want to link it.    \\

Whereas the standard definition of self-testing reviewed above is mathematically unambiguous, its operational significance is unclear. The best way to expose this is by observing that is not at all obvious how one would formulate self-testing in a theory distinct from quantum theory -- indeed, the reducibility condition is cast in a formalism specific to quantum theory, in terms of operators on Hilbert spaces. In particular, the projections $\Pi^{x_i}_i(y_i)$ are merely mathematical representations of certain measurement channels, guaranteed to exist by abstract theorems (as summarised in the preliminary section of the thesis); in fact, a general measurement channel is equivalent to a POVM, and whereas the reduction to PVMs is often justified in the literature on self-testing by reference to Naimark's theorem, the meaning of this reduction remains somewhat obscure.\footnote{The reduction is often said to be `without loss of generality' -- whereas this is of course perfectly legitimate in cases where the objective is to prove a particular theorem using the self-testing phenomenon (as in e.g. Ref. \cite{RUV13}), a statement of this kind is mathematically nonsensical in the absence of such a target theorem.}  Moreover, these unclarities are amplified when considering the fact that there could be ways for the two computing devices to locally establish an output on a given input which are more intricate than locally measuring a state;\footnote{For example, locally throwing a die and using the outcome to establish which of several shared states to use in a long sequence of various operations which ultimately make up an output, etc.}  though it seems intuitively clear that we can standardise the form of more general `strategies' to the form above, the meaning of the components deteriorates in that process. \\

In this chapter, we will see a fundamentally different way of looking at quantum self-testing. The idea is to consider the behaviour $P=(P^x)_{x \in X}$ not as a collection of probability distributions, but as a \myuline{causal} \myuline{channel} $\scalemyQ{.8}{0.7}{0.5}{& \push{\cX_\sfA} \qw & \multigate{1}{(P, \scrC)} & \push{\cY_\sfA} \qw & \qw \\ & \push{\cX_\sfB} \qw & \ghost{(T, \scrC)} & \push{\cY_\sfB} \qw & \qw }$ in $\QIT$, with classical inputs on $\cX_i := \C^{X_i}$ and classical outcomes on $\cY_i := \C^{Y_i}$, and where $\scrC$ is the local causal specification that stems from the devices being separated. It is in fact easy to see that that the behaviours that can be realised by quantum strategies are precisely the Bell-channels in $\QIT$.

The key is then to relate the causal dilations of this channel to the (tensor-product) quantum strategies as ordinarily perceived, and to relate the reducibility criterion \eqref{eq:attl} to the causal-dilational ordering. As such, self-testing will be recast operationally in terms of causally structured side-computations which the two computing devices may perform during our interaction with them. \\

\emph{As in \cref{sec:Density}, we will assume throughout that the scheme $\bfE$  is that of constructible causal channels in $\QIT$. In particular, all causal dilations will be constructible and derivability ($\cder$) will always refer to derivability using constructible channels.} \\

\textbf{Density in Purifiable Theories.} We start in \cref{sec:DensePure} by making a few observations about the causal-dilational ordering in general purifiable theories (of which $\QIT$ is an example), in particular establishing a density theorem (\cref{thm:BellDenseSelf}) which slightly improves on \cref{thm:BellDense} and \cref{thm:BellDenseII}. The density theorem says in the special case of $\QIT$ that \emph{causal Stinespring dilations} form a dense class of dilations of a given Bell-channel. More precisely, in the case of a bipartite Bell-channel $\scalemyQ{.8}{0.7}{0.5}{& \push{\cX_\sfA} \qw & \multigate{1}{(P, \scrC)} & \push{\cY_\sfA} \qw & \qw \\ & \push{\cX_\sfB} \qw & \ghost{(T, \scrC)} & \push{\cY_\sfB} \qw & \qw }$, these are the causal dilations of the form 

	\begin{align} \label{eq:CausStine}
\myQ{0.7}{0.5}{
	&  \push{\cX_\sfA}  \qw& \multigate{1}{ \Sigma_\sfA} &  \push{\cY_\sfA} \qw & \qw  \\
	& \Nmultigate{2}{\pi}  & \ghost{\Sigma_\sfA} &  \push{\cE_\sfA} \ww & \ww\\
	& \Nghost{\pi} & \ww & \push{\cE_0} \ww & \ww\\
	& \Nghost{\pi} & \multigate{1}{\Sigma_\sfB} &  \push{\cE_\sfB}\ww & \ww  \\
	&   \push{\cX_\sfB} \qw & \ghost{\Sigma_\sfB} & \push{\cY_\sfB} \qw & \qw  \\
},
\end{align}

with $\pi$ a pure state and with $\Sigma_\sfA, \Sigma_\sfB$ isometric quantum channels. It is (some of) those dilations which we will ultimately identify as representing the traditional quantum strategies.

 In \cref{subsec:NonsigIso} I will give an abstract recharacterisation of the channels \eqref{eq:CausStine}, to the effect that an arbitrary isometric channel is of the form \eqref{eq:CausStine} precisely if it satisfies the correct non-signalling conditions (\cref{thm:StructfromNS}). This implies in particular that a multipartite channel is the behaviour of a (tensor-product) quantum strategy if and only if it admits a multipartite Stinespring dilation which is non-signalling among the different parties (\cref{cor:QBRechar}). This result is quite surprising, since there are certainly channels which \myuline{themselves} are non-signalling without arising from quantum strategies, e.g. the PR box. As such, though the results presented in \cref{subsec:NonsigIso} will not be put to use in the remainder of the chapter (in fact \cref{subsec:NonsigIso} can be skipped without diminishing coherence in reading), they have been included because of their independent interest and connections to the work of Refs. \cite{PR94,Beck01,Egg02}, which first contemplated the relationship between non-signalling and structural representations. 
 
In \cref{subsec:Collapse} we improve in purifiable theories on \cref{thm:BellDenseII}, giving a criterion (\cref{lem:CderRechar}) for derivability among causal dilations in the dense class of dilations \eqref{eq:CausStine}; this will bring us closer to the language of quantum self-testing, in particular by emergence of the residual state $\psi^\res$. The result also implies a partial `collapse' of the causal-dilational ordering (\cref{thm:Collapse}), which in particular entails that the existence of a complete dilations with no acausal side-information is equivalent to all of the dilations \eqref{eq:CausStine} being $\cder$-equivalent. We will ultimately prove that this circumstance occurs for quantum self-testing, and this can be seen as justifying the intuition of some authors in the field who use the term `equivalence' about the reducibility criterion \eqref{eq:attl}, even though that criterion itself does not define an equivalence relation. (It also implies, however, that the condition \eqref{eq:attl} must be something slightly different from our notion of $\cder$-derivability, as detailed below.) \\

\textbf{The Bridge to Quantum Self-Testing.} In \cref{sec:SelfTest}, we relate the framework proposed in this thesis to the standard framework of quantum self-testing. Specifically, the goal is to  link rigidity and complete dilations of the causal behaviour channel $(P, \scrC)$ to the standard notion of self-testing in terms of $P$. Basically, we face two challenges:

First, we must identify within the framework of causal dilations precisely what are the entities that correspond to quantum strategies for $P$ in the usual sense. Then, we must rephrase in our language what the reducibility criterion \eqref{eq:attl} means in terms of these entities. 

\emph{ \textbf{Strategies and Classically Bound Dilations.}} First, every ordinary quantum strategy $(\varrho, \Pi_\sfA, \Pi_\sfB)$ will give rise to a causal dilation of the form \eqref{eq:CausStine}; indeed, letting $\scalemyQ{.8}{0.7}{0.5}{& \push{\C^{X_i}} \qw  & \multigate{1}{\Lambda_i} & \push{\C^{Y_i}} \qw & \qw \\ & \push{\cH_i} \qw & \ghost{\Lambda_i}}$ denote the ensemble of projective measurements corresponding to $\Pi_i$ (cf. the definition in the preliminary section of the thesis), we obtain a dilation \eqref{eq:CausStine} by taking $\pi = \psi$ a purification of $\varrho$ and $\Sigma_i = \hat{\Lambda}_i$ a Stinespring dilation of $\Lambda_i$. As it turns out, however, some dilations of the form \eqref{eq:CausStine} will \myuline{not} arise in this way from quantum strategies, and this is ultimately due to the fact that measurements can be dilated in peculiar ways, cf. \cref{ex:MeasDil}. In the present context, those strange dilations reflect the fact (see \cref{ex:StrangeRel}) that it is mathematically possible to choose a state $\varrho$ and channels $\Lambda_\sfA$ and $\Lambda_\sfB$ such that even though the channel

\begin{align}
\myQ{0.7}{0.5}{
	& \qw &  \push{\cX_\sfA}  \qw& \multigate{1}{ \Lambda_\sfA} &  \push{\cY_\sfA} \qw & \qw  \\
	& \Nmultigate{1}{\varrho}   & \push{\cH_\sfA} \qw & \ghost{\Lambda_\sfA} \\
	& \Nghost{\varrho} & \push{\cH_\sfB} \qw & \multigate{1}{\Lambda_\sfB}  \\
	&  \qw &  \push{\cX_\sfB} \qw & \ghost{\Lambda_\sfB} & \push{\cY_\sfB} \qw & \qw  \\
}  
\end{align}

behaves as $\scalemyQ{.8}{0.7}{0.5}{& \push{\cX_\sfA} \qw & \multigate{1}{(P, \scrC)} & \push{\cY_\sfA} \qw & \qw \\ & \push{\cX_\sfB} \qw & \ghost{(T, \scrC)} & \push{\cY_\sfB} \qw & \qw }$, with classical inputs and outputs, the channels $\Lambda_\sfA$ and $\Lambda_\sfB$ are not themselves measurement channels, and thus the `strategy' defined by the triple $(\varrho, \Lambda_\sfA, \Lambda_\sfB)$ has no counterpart in the ordinary framework of self-testing. We will resolve this basically by explicitly eliminating such `bad' dilations, but providing the interpretation (\cref{prop:RecharCB}) that it corresponds to restricting the class of imaginable dilations in a sensible way, namely to dilations which will be called \emph{classically bound} (\cref{def:ClassBound}).

\emph{\textbf{Reducibility and Local Derivability.}} Secondly, as for the reducibility relation \eqref{eq:attl}, there seems to be something fundamentally off with the \myuline{direction} of the relation, when compared to the philosophy that underlies the framework of causal dilations. Indeed, rigidity in terms of causal dilations means the existence of a $\cder$-\myuline{largest} dilation, whereas quantum self-testing would seem to assert the existence of a \myuline{smallest} (w.r.t. reducibility) strategy. The resolution to this paradox is offered by the collapse of the causal-dilational ordering as expressed by \cref{cor:AcausalColl} mentioned above.  More precisely, what we will show (\cref{thm:Bridge}) is that if $(\psi, \hat{\Lambda}_\sfA, \hat{\Lambda}_\sfB)$ defines a causal Stinespring dilation that corresponds to the strategy $(\varrho, \Pi_\sfA, \Pi_\sfB)$ and if $(\tilde{\psi}, \hat{\tilde{\Lambda}}_\sfA, \hat{\tilde{\Lambda}}_\sfB)$ defines one that corresponds to $(\tilde{\psi}, \tilde{\Pi}_\sfA, \tilde{\Pi}_\sfA)$, then the reducibility condition \eqref{eq:attl} holds if and only if there exist channels $\Gamma_\sfA$ and $\Gamma_\sfB$ such that

	\begin{align}
\myQ{0.7}{0.5}{
	& \push{\cX_\sfA}  \qw  & \multigate{1}{ \hat{\Lambda}_\sfA} & \push{\cY_\sfA}  \qw & \qw  \\
	& \Nmultigate{2}{\psi}  & \ghost{\hat{\Lambda}_\sfA} &  \Ngate{\Gamma_\sfA}{\ww} & \ww\\
	& \Nghost{\psi}  & \Ngate{\tr}{\ww} \\
	& \Nghost{\psi} & \multigate{1}{\hat{\Lambda}_\sfB} & \Ngate{\Gamma_\sfB}{\ww} & \ww \\
	&   \push{\cX_\sfB} \qw  & \ghost{\hat{\Lambda}_\sfB} & \push{\cY_\sfB}  \qw & \qw  
} 
\quad =  \quad 
\myQ{0.7}{0.5}{
	& \push{\cX_\sfA}  \qw  & \multigate{1}{ {\hat{\tilde{\Lambda}}}_\sfA} & \push{\cY_\sfA}  \qw & \qw  \\
	& \Nmultigate{2}{\tilde{\psi}}  & \ghost{{\hat{\tilde{\Lambda}}}_\sfA} &  \ww\\
	& \Nghost{\tilde{\psi}}  \\
	& \Nghost{\tilde{\psi}} & \multigate{1}{{\hat{\tilde{\Lambda}}}_\sfB} & \ww  \\
	&   \push{\cX_\sfB} \qw  & \ghost{{\hat{\tilde{\Lambda}}}_\sfB} & \push{\cY_\sfB}  \qw & \qw  
}   \quad.
\end{align}

(The key to proving this is a technical result, \cref{lem:technical}.) We may call the above relation \emph{local derivability}, since it asserts the $\cder$-derivability of one dilation by another using a product of three channels in the environment. (Though this condition is somewhat theory-independent, it is not purely operational, since causal Stinespring dilations do not seem to admit an operational definition.) Using the above-mentioned collapse-corollary, however, we will show (\cref{cor:RigidSelftest}) that this condition can be rephrased as rigidity of $(P, \scrC)$ relative to classically bound dilations witnessed by a complete dilation with no acausal  side-information, \myuline{along} with the existence of a simple representative in the $\cder$-equivalence class of causal Stinespring dilations, namely the one corresponding to the canonical strategy $(\tilde{\psi},\tilde{\Pi}_\sfA, \tilde{\Pi}_\sfA)$. Conjecturing that such a simple representative always exists (\cref{conj:Selftesting}) suggests a fully operational view on quantum self-testing.  \\

\textbf{Some Perks of a Reformulation.} The recasting of quantum self-testing in the language of causal dilations not only points towards generalisations to other theories, but also loosens the original theory from linear operators and thus facilitates a new way of understanding some well-known consequences of quantum self-testing. 

For example, it is easy to prove that self-testing implies the production of genuine randomness in a Bell-experiment (\cref{prop:SecSelfTest}) and that extremality of a behaviour is a necessary condition for quantum self-testing (\cref{prop:SecExt}). The former fact connects to the work of Ref. \cite{FFW11}, and the latter (though seemingly common knowledge for many years) was first proven formally in Ref. \cite{Goh18}.   

It will also be clear from the reformulation why the canonical state and the canonical measurements can always be locally extracted from the state and measurements of arbitrary strategies. Extractibility of the state can be proven in a few lines using the original formulation, and is also easily proved in the reformulation using local derivability, though by a completely different argument (\cref{prop:StateExtract}). Extractibility of the measurements does not seem to have been proved before in the formulation presented here (\cref{prop:MeasExtract}), but relates to some authors' alternative phrasing of self-testing (see e.g. Ref. \cite{Kan17} and the discussion in Ref. \cite{SB19}). %

{\centering
	\subsection*{§2. Further Notes on the Existing Literature.}}

\textbf{Quantum Self-Testing as Ordinarily Perceived.} The concept of quantum self-testing is widely recognised  as being introduced in Refs. \cite{MY98} and \cite{MY04} by D. Mayers and A. Yao, with Ref. \cite{MY04} establishing much of the current terminology. However, Ref. \cite{Ek91} by Artur Ekert contains already on an informal level some core ideas, in particular pointing towards the field of \emph{device-independent cryptography}, and the mathematical results that some behaviours can be achieved by essentially unique quantum strategies were independently reported already in Refs. \cite{SW87} and \cite{PR92}, though with a foundational rather than cryptographic perspective.

The general mathematical definition of self-testing took its modern standard form in Ref. \cite{MYS12}, and it is the one employed in the vast majority of contemporary expositions on the subject. There are authors who, when proving rigidity results, seek alternative formulations, but to the best of my knowledge all such are either designed for specialised situations (e.g. \cite{RUV13}, Def. 5.5), or, though generally applicable, capture only certain aspects of the commonly used definition (e.g. \cite{Kan17}). As such, a general operational definition of quantum self-testing has not been attempted before. \\

{\centering
	\subsection*{§3. Contributions.}}

The original contributions of this chapter are the following:\\

\begin{enumerate}

	\item Proving a density theorem for Bell-channels in purifiable theories (\cref{thm:BellDenseSelf}) and providing and abstract characterisation of elements in the dense class, implying in particular a surprising new characterisation of tensor-product quantum behaviours in terms of non-signalling properties of their Stinespring dilations (\cref{cor:QBRechar}).
	
	\item Establishing a formal connection between the conventional definition of quantum self-testing (\cite{MY98,MY04,MYS12, SB19}) and rigidity in the sense of \cref{chap:Causal} (\cref{thm:Bridge}, \cref{cor:RigidSelftest} and \cref{conj:Selftesting}). 
	
	\item Giving simple proofs for known implications of quantum self-testing, including the necessity of extremality of the behaviour (\cref{prop:SecSelfTest} and \cref{prop:SecExt}), and local extractibility of the canonical state (\cref{prop:StateExtract}) and the canonical measurements (\cref{prop:MeasExtract}).

	\end{enumerate}

The contribution mentioned in 2. is in a sense the most important, as it was the original motivation for the thesis project itself. It points towards a purely operational interpretation of quantum self-testing, entailing in particular that

\begin{itemize}
	\item the origin of \emph{quantum strategies} (\cref{def:StratAttl}) is explained, including the validity of restricting to projective measurements;
	\item the nature of the reducibility relation (\cref{def:Redattl}) is clarified, by exhibiting it as a strong version of derivability in the environment; 
	\item the notions of reducibility and self-testing are put into a context of general theories;
\item quantum self-testing is seen to imply rigidity in the sense of causal dilations, a fully operational notion, and a converse to this implication is conjectured. 
	\end{itemize}

{\centering
	\subsection*{§4. The Metric Aspect.}}

We will not touch at the concept of \emph{robust} self-testing, an approximate version of the self-testing story in which the exact equality of behaviours and exact fulfilment of the reducibility condition are relaxed (\cite{MYS12}). It seems quite obvious that an understanding of robustness in the framework of causal dilations hinges upon a metric version of \cref{chap:Causal}, whose challenges have already been discussed. 

\newpage

\section{Density Considerations in Purifiable Theories}
\label{sec:DensePure}

Suppose that $\scalemyQ{.8}{0.7}{0.5}{& \push{\cX_\sfA} \qw & \multigate{1}{(T, \scrC)} & \push{\cY_\sfA} \qw & \qw \\ & \push{\cX_\sfB} \qw & \ghost{(T, \scrC)} & \push{\cY_\sfB} \qw & \qw }$ is a bipartite Bell-channel in a purifiable theory $\Theory$ where every channel has a pure \myuline{one-sided} dilation (e.g. by virtue of $\Theory$ being universal, cf. \cref{prop:PureRechar}). We have the following improvement of the general density theorem (\cref{thm:BellDense}), which as usual generalises without effort to the multipartite (or unipartite) case: 

\begin{Thm} (Density Theorem for Bell-Channels in Purifiable Theories.) \label{thm:BellDenseSelf}\\
The causal dilations of the form 
	
	\begin{align} \label{eq:StineImp}
	\myQ{0.7}{0.5}{
		&  \push{\cX_\sfA}  \qw& \multigate{1}{ \Sigma_\sfA} &  \push{\cY_\sfA} \qw & \qw  \\
		& \Nmultigate{2}{\pi}  & \ghost{\Sigma_\sfA} &  \push{\cE_\sfA} \ww & \ww\\
		& \Nghost{\pi} & \ww & \push{\cE_0} \ww & \ww\\
		& \Nghost{\pi} & \multigate{1}{\Sigma_\sfB} &  \push{\cE_\sfB}\ww & \ww  \\
		&   \push{\cX_\sfB} \qw & \ghost{\Sigma_\sfB} & \push{\cY_\sfB} \qw & \qw  \\
	} \quad, 
	\end{align}
	
where $\Sigma_\sfA, \Sigma_\sfB$ and $\pi$ are dilationally pure,	constitute a dense class for the constructible dilations of $(T, \scrC)$. Moreover, $\scalemyQ{.8}{0.7}{0.5}{
		&  \push{\cX_\sfA}  \qw& \multigate{1}{ \Sigma_\sfA} &  \push{\cY_\sfA} \qw & \qw  \\
		& \Nmultigate{2}{\pi}  & \ghost{\Sigma_\sfA} &  \push{\cE_\sfA} \ww & \ww\\
		& \Nghost{\pi} & \ww & \push{\cE_0} \ww & \ww\\
		& \Nghost{\pi} & \multigate{1}{\Sigma_\sfB} &  \push{\cE_\sfB}\ww & \ww  \\
		&   \push{\cX_\sfB} \qw & \ghost{\Sigma_\sfB} & \push{\cY_\sfB} \qw & \qw  \\
	} \cder \scalemyQ{.8}{0.7}{0.5}{
		&  \push{\cX_\sfA}  \qw& \multigate{1}{ \Sigma'_\sfA} &  \push{\cY_\sfA} \qw & \qw  \\
		& \Nmultigate{2}{\pi'}  & \ghost{\Sigma'_\sfA} &  \push{\cE'_\sfA} \ww & \ww\\
		& \Nghost{\pi'} & \ww & \push{\cE'_0} \ww & \ww\\
		& \Nghost{\pi'} & \multigate{1}{\Sigma'_\sfB} &  \push{\cE'_\sfB}\ww & \ww  \\
		&   \push{\cX_\sfB} \qw & \ghost{\Sigma'_\sfB} & \push{\cY_\sfB} \qw & \qw  \\
	}$ if and only if there exist channels  $\Gamma_\sfA$, $\Gamma_\sfB$ and a dilationally pure channel $\Phi$ such that 	
	
	\begin{align} \label{eq:primerel}
	\scalemyQ{1}{0.7}{0.5}{
		&   \push{\cX_\sfA} \qw  & \multigate{1}{\Sigma_\sfA}  &   \qw &  \push{\cY_\sfA}  \qw &  \qw  & \qw \\
		& \Nmultigate{4}{\pi}  & \ghost{\Sigma_\sfA} &  \push{\cE_\sfA} \ww &  \Nmultigate{1}{\Gamma_\sfA}{\ww} & \push{\cE'_\sfA} \ww & \ww \\
		& \Nghost{\pi}   &  & \Nmultigate{2}{\Phi} & \Nghost{\Gamma_\sfA}{\ww} &  \\  
		& \Nghost{\pi}  & \push{\cE_0} \ww &   \Nghost{\Phi}{\ww}  & \ww & \push{\cE'_0} \ww & \ww \\
		& \Nghost{\pi}  &  & \Nghost{ \Phi} &   \Nmultigate{1}{\Gamma_\sfB}{\ww}   \\
		& \Nghost{\pi}   & \multigate{1}{\Sigma_\sfB} &  \push{\cE_\sfB} \ww & \Nghost{\Gamma_\sfB}{\ww} & \push{\cE'_\sfB}\ww & \ww \\ 
		&   \push{\cX_\sfB} \qw  & \ghost{\Sigma_\sfB}   & \qw &  \push{\cY_\sfB} \qw &   \qw & \qw} =\scalemyQ{1}{0.7}{0.5}{
		& \push{\cX_\sfA}\qw  & \multigate{1}{ \Sigma'_\sfA} &  \push{\cY_\sfB}  \qw & \qw  \\
		& \Nmultigate{2}{\pi'}  & \ghost{\Sigma'_\sfA} &  \push{\cE'_\sfA} \ww & \ww\\
		& \Nghost{\pi'} & \ww & \push{\cE'_0} \ww & \ww\\
		& \Nghost{\pi'} & \multigate{1}{\Sigma'_\sfB} & \push{\cE'_\sfB} \ww & \ww  \\
		&  \push{\cX_\sfA} \qw  & \ghost{\Sigma'_\sfB} &  \push{\cY_\sfB} \qw & \qw  \\
	} 
	\end{align}

\end{Thm}

\begin{proof}
	The density statements follow from \cref{thm:BellDense}, simply by additionally forming for each component a pure dilation, thus moving further up in the $\cder$-ordering. As for the characterisation of the derivability relation $\cder$, that follows directly from the statements in \cref{thm:BellDenseII}, observing that $\Phi$ can be taken dilationally pure by choosing a pure dilation and including a trash in one of the channels $\Gamma_i$ if necessary. 
	\end{proof}

As mentioned in the introduction, a \emph{(tensor-product) quantum strategy} will be encoded in the isometric channel \eqref{eq:StineImp} in the theory $\QIT$. This isometric channel will reveal all relevant information about the strategy with regards to self-testing. We will call it a \emph{causal Stinespring dilation} of $(T, \scrC)$. In a general purifiable theory, the channels \eqref{eq:StineImp} are pure if the theory is localisable and we may thus more generally call them \emph{pure causal dilations}.

 As was the case in $\CIT$ (cf. \cref{rem:CompleteWithout}), the channels of the dense class are complete dilations of $T$ if we disregard causality.\footnote{This is no coincidence; in a complete and localisable theory, \cref{thm:BellDense} can always be such improved, by picking complete dilations of each component.} Hence, the dilations \eqref{eq:StineImp} are all equivalent by means of a channel acting on the inaccessible interface; the intricacy arises because for two channels in the dense class to be related in the \myuline{causal}-dilational order, they must be related by a channel with a particular structure as prescribed by \cref{eq:primerel}.

\subsection{An Abstract Characterisation of Pure Causal Dilations}
\label{subsec:NonsigIso}

In \cref{chap:Dilations}, we saw that many theories have the DiVincenzo Property, which allows one to conclude a that a channel $\scalemyQ{.8}{0.7}{0.5}{& \push{\cX_\sfA} \qw & \multigate{1}{T} & \push{\cY_\sfA} \qw & \qw \\ & \push{\cX_\sfB} \qw & \ghost{T} & \push{\cY_\sfB} \qw & \qw }$ has the structure $	\scalemyQ{.8}{0.7}{0.5}{
	& \push{\cX_\sfA}  \qw    &\multigate{1}{T_1}   & \qw &   \push{\cY_\sfA} \qw & \qw  \\
	&                   & \Nghost{T_1}   & \multigate{1}{T_2} & \\
&  \push{\cX_\sfB}  \qw & \qw &  \ghost{T_2}   &  \push{\cY_\sfB} \qw  & \qw  
} $ merely provided that it satisfies the necessary non-signalling condition -- in the case of $\QIT$, this statement was raised and illuminated in Ref. \cite{Beck01}, and then fully proved in Ref. \cite{Egg02}. 

On the other hand, it has been known since the work of Ref. \cite{PR94} that more intricate structures generally can \myuline{not} be concluded based on fulfilment of non-signalling conditions; for example, not every channel $\scalemyQ{.8}{0.7}{0.5}{& \push{\cX_\sfA} \qw & \multigate{1}{T} & \push{\cY_\sfA} \qw & \qw \\ & \push{\cX_\sfB} \qw & \ghost{T} & \push{\cY_\sfB} \qw & \qw }$ satisfying the same non-signalling conditions as the channel

\begin{align} \label{eq:Localis}
 	\myQ{0.7}{0.5}{
 &  \push{\cX_\sfA}  \qw& \multigate{1}{ T_\sfA} &  \push{\cY_\sfA} \qw & \qw  \\
	& \Nmultigate{1}{s}   & \ghost{T_\sfA} \\
	& \Nghost{s} & \multigate{1}{T_\sfB}  \\
&  \push{\cX_\sfB} \qw & \ghost{T_\sfB} & \push{\cY_\sfB} \qw & \qw  \\
}  \quad
\end{align}

(i.e. compatible with the local causal specification $\scrC$ given by $\scrC(\sfy_i)=\{\sfx_i\}$) is of that form. In the language of \cref{chap:Causal}, not every causal channel with local specification is constructible. 

What we will prove in the present subsection is that, under mild dilational assumptions, every \myuline{pure} channel which satisfies the necessary non-signalling conditions is of the form \eqref{eq:Localis}. In fact, we have the following:

\begin{Thm} (Structure from Non-Signalling.) \label{thm:StructfromNS}\\
	Suppose that $\Theory$ is localisable, and that there is a state on every system in $\Theory$. Then, a dilationally pure channel 
	
	\begin{align}
	\myQ{0.7}{0.5}{& \push{\cX_\sfA} \qw & \multigate{2}{T} & \push{\bbY_\sfA} \qw & \qw \\ & & \Nghost{T} & \push{\bbZ} \qw & \qw \\ & \push{\cX_\sfB} \qw & \ghost{T} & \push{\bbY_\sfB} \qw & \qw } 
\end{align}

is of the form 

\begin{align} \label{eq:LocForm}
\myQ{0.7}{0.5}{
	&  \push{\cX_\sfA}  \qw& \multigate{1}{ T_\sfA} &  \push{\bbY_\sfA} \qw & \qw  \\
	& \Nmultigate{2}{s}  & \ghost{T_\sfA} \\
	& \Nghost{s} & \qw & \push{\bbZ} \qw & \qw\\
	& \Nghost{s} & \multigate{1}{T_\sfB}   \\
	&   \push{\cX_\sfB} \qw & \ghost{T_\sfB} & \push{\bbY_\sfB} \qw & \qw  \\
}
\end{align}

if and only if it is compatible with the causal specification $\scrC$ given by $\scrC(\ports{\bbZ})= \emptyset$, $\scrC(\ports{\bbY_\sfA})= \{\sfx_\sfA\}$ and $\scrC(\ports{\bbY_\sfA})= \{\sfx_\sfB\}$. If $\Theory$ is moreover universal and purifiable, then the components $T_\sfA$, $T_\sfB$ and $s$ may be taken pure in that case. 
	\end{Thm}

\begin{Remark} (More Parties.) \\
	The statement and proof generalise to the arbitrary multipartite setting by induction. In fact, the proof can be seen as the induction step executed in going from the case of unipartite to bipartite Bell-scenarios. 
\end{Remark} 

\begin{Remark} (Approximate Generalisation.) \\
	If $D$ is a dilational metric on $\Theory$ (cf. \cref{def:Dilationality}), it is easy to see from the proof that the first statement in \cref{thm:StructfromNS} admits an approximate generalisation, such that if $T$ is $\varepsilon$-close to being suitably non-signalling, then $T$ is $3 \varepsilon$-close to a channel of the form \eqref{eq:LocForm}.
	
	\end{Remark}

\begin{proof}
	The `only if'-direction is clear. As for the `if'-direction, first observe that, since $\Theory$ is spatially localisable, $\Theory$ has the DiVincenzo Property (\cref{prop:StructureNonSignalling}). Since $T$ is compatible with $\scrC$, the non-signalling property implied by the condition $\scrC(\ports{\bbZ} \cup \{\sfy_\sfB\})=\{\sfx_\sfB\}$ therefore yields that $T$ is of the form  
	
	\begin{align} \label{eq:form1} 	
	\myQ{0.7}{0.5}{
		& \push{\cX_\sfA}  \qw    & \qw    & \multigate{1}{T_\sfA}  &  \push{\bbY_\sfA} \qw & \qw \\
		&                   & \Nmultigate{2}{S_\sfB}  &  \ghost{T_\sfA} & \\
		&                   & \Nghost{S_\sfB} & \qw &  \push{\bbZ} \qw& \qw\\
		&  \push{\cX_\sfB}  \qw & \ghost{\Xi_\sfB}  & \qw &  \push{\bbY_\sfB} \qw  & \qw  \\
	} 	
	\end{align}
	
	for some channels $S_\sfB$ and $T_\sfA$. If we can show that $S_\sfB$ is of the form $\scalemyQ{.8}{0.7}{0.5}{
		   & \Nmultigate{2}{s}  &  \qw  \\
		                & \Nghost{s}  &\qw & \push{\bbZ}  \qw&  \qw\\
		               & \Nghost{s}  & \multigate{1}{T_\sfB} &  & \\
		&  \push{\cX_\sfB}  \qw & \ghost{T_\sfB} &  \push{\bbY_\sfB} \qw  & \qw  \\
	}$, then we are done proving the first statement in the theorem. To this end, however, observe that because of the non-signalling property implied by the condition $\scrC(\ports{\bbZ} \cup \{\sfy_\sfA\})=\{\sfx_\sfA\}$, the channel $T$ is also of the form

	\begin{align} \label{eq:form2} 	
	\myQ{0.7}{0.5}{
		& \push{\cX_\sfA}  \qw     & \multigate{2}{S'_\sfA}  & \qw &   \push{\bbY_\sfA} \qw & \qw \\
		&                 & \Nghost{S'_\sfA}  & \qw&\push{\bbZ}  \qw&  \qw\\
		&                 & \Nghost{S'_\sfA}  & \multigate{1}{T'_\sfB} &  & \\
		&  \push{\cX_\sfB}  \qw &\qw &  \ghost{T'_\sfB} &  \push{\bbY_\sfB} \qw  & \qw  \\
	} 
	\end{align}
	
	for some channels $S'_\sfA$ and $T'_\sfB$. In general, the equality of the two forms \eqref{eq:form1} and \eqref{eq:form2} of $T$ is all that can be concluded on the basis of compatibility with the specification $\scrC$. If we could somehow reverse $T_\sfA$ and move it to the other side of this equality, effectively isolating $S_\sfB$, we would have the desired form of $S_\sfB$, but this is not possible in general. \emph{However}, since $T$ is dilationally pure, we can squeeze out this opportunity: If we pick (by \cref{prop:Reversibledilations}) a reversible \myuline{dilation} $R_\sfA$ of $T_\sfA$, then dilational purity of $T$ implies that  
	
	\begin{align}
		\myQ{0.7}{0.5}{
			& & & \Nmultigate{2}{R_\sfA} & \ww & \ww\\
		& \push{\cX_\sfA}  \qw    & \qw    & \ghost{R_\sfA}  &  \push{\bbY_\sfA} \qw & \qw \\
		&                   & \Nmultigate{2}{S_\sfB}  &  \ghost{R_\sfA} & \\
		&                   & \Nghost{S_\sfB} & \qw &  \push{\bbZ} \qw& \qw\\
		&  \push{\cX_\sfB}  \qw & \ghost{\Xi_\sfB}  & \qw &  \push{\bbY_\sfB} \qw  & \qw  \\
	} 	=  	\myQ{0.7}{0.5}{& & \Ngate{t} & \ww & \ww \\ & \push{\cX_\sfA} \qw & \multigate{2}{T} & \push{\bbY_\sfA} \qw & \qw \\ & & \Nghost{T} & \push{\bbZ} \qw & \qw \\ & \push{\cX_\sfB} \qw & \ghost{T} & \push{\bbY_\sfB} \qw & \qw }  =  \myQ{0.7}{0.5}{
& & \Ngate{t} & \ww & \ww \\ 	& \push{\cX_\sfA}  \qw     & \multigate{2}{S'_\sfA}  & \qw &   \push{\bbY_\sfA} \qw & \qw \\
	&                 & \Nghost{S'_\sfA}  & \qw&\push{\bbZ}  \qw&  \qw\\
	&                 & \Nghost{S'_\sfA}  & \multigate{1}{T'_\sfB} &  & \\
	&  \push{\cX_\sfB}  \qw &\qw &  \ghost{T'_\sfB} &  \push{\bbY_\sfB} \qw  & \qw  \\
} 
	\end{align}
	
	for some state $t$, and applying an inverse\footnote{We might pick $R_\sfA$ to be dilationally pure, and for this reason reversible; in $\QIT$, this corresponds to picking $R_\sfA$ isometric, i.e. of `going to the church of the larger Hilbert space'. As such, the following joke may help to remember the idea behind the proof of \cref{thm:StructfromNS}: \emph{Why is the church of the larger Hilbert space better than the church of Catholicism? Because, while in the Catholic church your sins can be forgiven, in the church of the larger Hilbert space your sins can be \myuline{undone}}. } $R^{-}_\sfA$ to $R_\sfA$ on both sides yields the identity
	
		\begin{align}
	\myQ{0.7}{0.5}{
		& \push{\cX_\sfA}  \qw    & \qw    &   \qw & \qw \\
		&                   & \Nmultigate{2}{S_\sfB}  & \qw & \qw \\
		&                   & \Nghost{S_\sfB} & \qw &  \push{\bbZ} \qw& \qw\\
		&  \push{\cX_\sfB}  \qw & \ghost{\Xi_\sfB}  & \qw &  \push{\bbY_\sfB} \qw  & \qw  \\
	} 	=  	 \myQ{0.7}{0.5}{
		& & \Ngate{t} & \Nmultigate{1}{R^{-}_\sfA}{\ww} & \push{\cX_\sfA} \qw & \qw  \\ 	& \push{\cX_\sfA}  \qw     & \multigate{2}{S'_\sfA}  & \ghost{R^{-}_\sfA} &   \qw & \qw \\
		&                 & \Nghost{S'_\sfA}  & \qw&\push{\bbZ}  \qw&  \qw\\
		&                 & \Nghost{S'_\sfA}  & \multigate{1}{T'_\sfB} &  & \\
		&  \push{\cX_\sfB}  \qw &\qw &  \ghost{T'_\sfB} &  \push{\bbY_\sfB} \qw  & \qw  \\
	} .
	\end{align}
	
	Now, by finally inserting an arbitrary state as input to $\cX_\sfA$ and  trashing it, we obtain 
	
		\begin{align} \label{eq:SBform}
	\myQ{0.7}{0.5}{
		&                   & \Nmultigate{2}{S_\sfB}  & \qw & \qw \\
		&                   & \Nghost{S_\sfB} & \qw &  \push{\bbZ} \qw& \qw\\
		&  \push{\cX_\sfB}  \qw & \ghost{\Xi_\sfB}  & \qw &  \push{\bbY_\sfB} \qw  & \qw  \\
	} 	=  	\myQ{0.7}{0.5}{
	& \Nmultigate{2}{s}  &  \qw   \\
	& \Nghost{s}  &\qw & \push{\bbZ}  \qw&  \qw\\
	& \Nghost{s}  & \multigate{1}{T'_\sfB} &  & \\
	&  \push{\cX_\sfB}  \qw & \ghost{T'_\sfB} &  \push{\bbY_\sfB} \qw  & \qw  \\
}
	\end{align}
	
for some state $s$. Letting $T_\sfB = T'_\sfB$, we conclude that $T$ is of the form \cref{eq:LocForm}, as desired.

To prove the second statement in the theorem, first observe that we may without loss of generality assume that $S_\sfB$ in \cref{eq:form1} is dilationally pure. Indeed, by redefining $T_\sfA$ to include a trash, we may replace $S_\sfB$ with a universal dilation of $S_\sfB$, and by \cref{prop:PureRechar} this dilation is dilationally pure. By the same argument, the state $s$ in \eqref{eq:SBform} may be taken to be pure and a universal dilation of $\scalemyQ{.8}{0.7}{0.5}{	& \Nmultigate{2}{s}  &  \qw  & \qw \\
& \Nghost{s}  & \push{\bbZ}  \qw&  \qw\\
& \Nghost{s}  & \gate{\tr}  }$ without loss of generality. Then, however,  \cref{eq:SBform} implies by purity of $S_\sfB$ and universality of the dilation $s$ that $T_\sfB (=T'_\sfB)$ is dilationally pure. Inserting \eqref{eq:SBform} into \cref{eq:form1}, and once again replacing $s$ by a universal dilation $\scalemyQ{.8}{0.7}{0.5}{	& \Nmultigate{2}{s'}  &  \ww  & \ww \\
& \Nghost{s'}  & \push{\bbZ}  \qw&  \qw\\
& \Nghost{s'}  & \qw & \qw  }$  of $\scalemyQ{.8}{0.7}{0.5}{	& \Nmultigate{2}{s}  & \gate{\tr}   \\
& \Nghost{s}  & \push{\bbZ}  \qw&  \qw\\
& \Nghost{s}  & \qw & \qw }$ then similarly implies (since $\scalemyQ{.8}{0.7}{0.5}{
& \Nmultigate{2}{s'}  &  \ww  & \ww \\
& \Nghost{s'}  &\qw & \push{\bbZ}  \qw&  \qw\\
& \Nghost{s'}  & \multigate{1}{T_\sfB} &  & \\
&  \push{\cX_\sfB}  \qw & \ghost{T_\sfB} &  \push{\bbY_\sfB} \qw  & \qw  \\
}$ is a universal dilation, $T_\sfB$ being pure and reversible) that the modified $T_\sfA$ is dilationally pure.

	\end{proof}

\cref{thm:StructfromNS} immediately implies a recharacterisation  in terms of non-signalling of the channels in the dense class of \cref{thm:BellDenseSelf}. In a purifiable theory such as $\QIT$, however, it also directly yields a surprising recharacterisation of the entire class of Bell-channels in terms of non-signalling in their pure dilations:\\

Given an arbitrary bipartite channel $\scalemyQ{.8}{0.7}{0.5}{& \push{\cX_\sfA} \qw & \multigate{1}{T} & \push{\cY_\sfA} \qw & \qw \\ & \push{\cX_\sfB} \qw & \ghost{T} & \push{\cY_\sfB} \qw & \qw }$, let us say that $T$ is \emph{purely non-signalling} if $T$ has a dilationally pure bipartite dilation  $\scalemyQ{.8}{0.7}{0.5}{& \push{\cX_\sfA} \qw & \multigate{3}{\Sigma} & \push{\cY_\sfA} \qw & \qw \\  &  & \Nghost{\Sigma} & \push{\bbE_\sfA} \ww & \ww\\  &  & \Nghost{\Sigma} & \push{\bbE_\sfB} \ww & \ww\\ & \push{\cX_\sfB} \qw & \ghost{P} & \push{\cY_\sfB} \qw & \qw }$ which is compatible with the specification $\scrC$ given by $\scrC(\ports{\bbE_i} \cup \{\sfy_i\})=\{\sfx_i\}$ for $i = \sfA, \sfB$. This requirement does not imply that \myuline{every} pure bipartite dilation will be non-signalling, but merely says there exist ones which are.\footnote{In general, there will be ones which are not non-signalling, since given any one that is non-signalling we can  swap $\bbE_\sfA$ with $\bbE_\sfB$ and generically obtain one that is not.} In the specific theory $\QIT$, the condition is a matter of whether there exists a bipartite \myuline{Stinespring} dilation which is non-signalling, i.e. whether $T$ can be implemented in a non-signalling way `in the Church of the larger Hilbert space'.
	
	We have the following:
	
	\begin{Cor} (Purely Non-Signalling means Bell.) \label{cor:QBRechar}\\
		Let $\Theory$ be a universal, localisable and purifiable theory. Then, an arbitrary bipartite channel  $\scalemyQ{.8}{0.7}{0.5}{& \push{\cX_\sfA} \qw & \multigate{1}{T} & \push{\cY_\sfA} \qw & \qw \\ & \push{\cX_\sfB} \qw & \ghost{T} & \push{\cY_\sfB} \qw & \qw }$ is purely non-signalling if and only if it is a Bell-channel, i.e. of the form 	
		
		\begin{align} \label{eq:BellC}
		\myQ{0.7}{0.5}{
			&  \push{\cX_\sfA}  \qw& \multigate{1}{ T_\sfA} &  \push{\cY_\sfA} \qw & \qw  \\
			& \Nmultigate{1}{s}   & \ghost{T_\sfA} \\
			& \Nghost{s} & \multigate{1}{T_\sfB}  \\
			&  \push{\cX_\sfB} \qw & \ghost{T_\sfB} & \push{\cY_\sfB} \qw & \qw  \\
		}.
	\end{align}
		
		\end{Cor}
	
	\begin{Remark} Again, the generalisation to the multipartite case is straightforward. \end{Remark}
	
	\begin{proof}
	First recall that every channel in $\Theory$ has a one-sided pure dilation, by the last statement in \cref{prop:PureRechar}. 
		
		The `if'-direction is not difficult (the reader is encouraged to think about the special case of $\QIT$, where pure one-sided dilations mean Stinespring dilations): If $T$ is a Bell-channel, we may take it to be of the form \eqref{eq:BellC} with $s$ pure. By dilating $T_\sfA$ and $T_\sfB$ to pure channels, we thus obtain a pure (by localisability) bipartite dilation of $T$ which evidently has the required non-signalling properties, so $T$ is purely non-signalling. 
		
		The `only if'-direction is the surprise: Given a purely non-signalling channel $T$, any pure bipartite dilation which witnesses this must by \cref{thm:StructfromNS} (with $\bbZ = \bbI$) factor as a Bell-channel; but then, trashing the hidden interfaces gives the Bell-structure \eqref{eq:BellC} for $T$ itself.	\end{proof}

\begin{Cor}
In $\QIT$, there exist non-signalling channels $\scalemyQ{.8}{0.7}{0.5}{& \push{\cX_\sfA} \qw & \multigate{1}{T} & \push{\cY_\sfA} \qw & \qw \\ & \push{\cX_\sfB} \qw & \ghost{T} & \push{\cY_\sfB} \qw & \qw }$ which are not purely non-signalling. For example, the PR box is non-signalling but not purely non-signalling. 
	\end{Cor}

\cref{cor:QBRechar} gives a novel characterisation of the \myuline{tensor-product} quantum behaviours in $\QIT$, i.e. those that arise as Bell-channels. It is unexpected because, a priori, there is no reason to think that `non-signalling in the Church' should favour tensor-product behaviours over more general behaviours. By previous observations, \cref{cor:QBRechar} applies not only to the theory $\QIT$ but also the theory $\QIT^\infty$. In that theory, a slightly more general model for the implementable behaviours is that arising from so-called \emph{commuting-operator strategies} (\cite{Dyke16}); however, \cref{cor:QBRechar} means that the behaviours from this class which are not already tensor-product behaviours do not admit non-signalling Stinespring dilations. (The approximate generalisation of \cref{cor:QBRechar} even implies, in light of the recently established fact that there exist commuting-operator behaviours which are not even close to being tensor-product behaviours (\cite{MIP20}), that some commuting-operator behaviours are not even close to having non-signalling Stinespring dilations.) \\ 

It is also worth observing (still in $\QIT^\infty$) that if the systems $\cX_i$ and $\cY_i$ are finite-dimensional, then a slightly more detailed book-keeping in the proof of \cref{cor:QBRechar} will reveal that the bipartite system $\cH_\sfA \otimes \cH_\sfB$ corresponding to the state $s$ can be taken finite-dimensional if and only if the hidden systems in the non-signalling Stinespring dilation can be taken finite-dimensional.  Since there are Bell-channels in $\QIT^\infty$ which cannot be realised using a state on a finite-dimensional system, this means that some bipartite channels $\scalemyQ{.8}{0.7}{0.5}{& \push{\cX_\sfA} \qw & \multigate{1}{T} & \push{\cY_\sfA} \qw & \qw \\ & \push{\cX_\sfB} \qw & \ghost{T} & \push{\cY_\sfB} \qw & \qw }$ in $\QIT^\infty$ have a non-signalling bipartite Stinespring dilation with infinite-dimensional hidden systems, but no one with finite-dimensional hidden systems.

\subsection{A Partial Collapse of the Causal-Dilational Ordering}
\label{subsec:Collapse}

 It is of interest to better understand the relation \eqref{eq:primerel} of derivability between the dilations of \cref{thm:BellDenseSelf}. As it turns out, purifiability of a theory entails a recharacterisation of this relation:

\begin{Lem} (Recharacterisation of $\cder$ among Pure Causal Dilations.) \label{lem:CderRechar} \\
Suppose that $\Theory$ is purifiable and localisable, and let $\scalemyQ{.8}{0.7}{0.5}{& \push{\cX_\sfA} \qw & \multigate{1}{(T, \scrC)} & \push{\cY_\sfA} \qw & \qw \\ & \push{\cX_\sfB} \qw & \ghost{(T, \scrC)} & \push{\cY_\sfB} \qw & \qw }$ be a bipartite Bell-channel in $\Theory$. Then, two pure causal dilations satisfy 
	$\scalemyQ{.8}{0.7}{0.5}{
		&  \push{\cX_\sfA}  \qw& \multigate{1}{ \Sigma_\sfA} &  \push{\cY_\sfA} \qw & \qw  \\
		& \Nmultigate{2}{\pi}  & \ghost{\Sigma_\sfA} &  \push{\cE_\sfA} \ww & \ww\\
		& \Nghost{\pi} & \ww & \push{\cE_0} \ww & \ww\\
		& \Nghost{\pi} & \multigate{1}{\Sigma_\sfB} &  \push{\cE_\sfB}\ww & \ww  \\
		&   \push{\cX_\sfB} \qw & \ghost{\Sigma_\sfB} & \push{\cY_\sfB} \qw & \qw  \\
	} \cder \scalemyQ{.8}{0.7}{0.5}{
		&  \push{\cX_\sfA}  \qw& \multigate{1}{ \Sigma'_\sfA} &  \push{\cY_\sfA} \qw & \qw  \\
		& \Nmultigate{2}{\pi'}  & \ghost{\Sigma'_\sfA} &  \push{\cE'_\sfA} \ww & \ww\\
		& \Nghost{\pi'} & \ww & \push{\cE'_0} \ww & \ww\\
		& \Nghost{\pi'} & \multigate{1}{\Sigma'_\sfB} &  \push{\cE'_\sfB}\ww & \ww  \\
		&   \push{\cX_\sfB} \qw & \ghost{\Sigma'_\sfB} & \push{\cY_\sfB} \qw & \qw  \\
	}$ if and only if there exist dilationally pure channels  $\hat{\Gamma}_\sfA$, $\hat{\Gamma}_\sfB$, $\Phi$ and a dilationally pure state $\pi^\res$, such that
	\begin{align} \label{eq:cderpure}
\scalemyQ{.8}{0.7}{0.5}{
	&   \push{\cX_\sfA} \qw  & \multigate{1}{\Sigma_\sfA}  &   \qw &  \push{\cY_\sfA}  \qw &  \qw  & \qw \\
	& \Nmultigate{4}{\pi}  & \ghost{\Sigma_\sfA} &  \push{\cE_\sfA} \ww &  \Nmultigate{1}{\hat{\Gamma}_\sfA}{\ww} & \push{\cE'_\sfA} \ww & \ww \\
	& \Nghost{\pi}   &  & \Nmultigate{2}{\Phi} & \Nghost{\hat{\Gamma}_\sfA}{\ww} & \dw  & \dw \\  
	& \Nghost{\pi}  & \push{\cE_0} \ww &   \Nghost{\Phi}{\ww}  & \ww & \push{\cE'_0} \ww & \ww \\
	& \Nghost{\pi}  &  & \Nghost{ \Phi} &   \Nmultigate{1}{\hat{\Gamma}_\sfB}{\ww} & \dw & \dw  \\
	& \Nghost{\pi}   & \multigate{1}{\Sigma_\sfB} &  \push{\cE_\sfB} \ww & \Nghost{\hat{\Gamma}_\sfB}{\ww} & \push{\cE'_\sfB}\ww & \ww \\ 
	&   \push{\cX_\sfB}  \qw  & \ghost{\Sigma_\sfB}   & \qw &   \push{\cY_\sfB}\qw &   \qw & \qw} =  \scalemyQ{.8}{0.7}{0.5}{
	&   \push{\cX_\sfA} \qw  & \multigate{1}{\Sigma'_\sfA}  &   \qw &  \push{\cY_\sfA}  \qw &  \qw  & \qw \\
	& \Nmultigate{4}{\pi'}  & \ghost{\Sigma'_\sfA} & \ww &   \push{\cE'_\sfA} \ww & \ww &  \Nmultigate{1}{\id}{\ww} & \push{\cE'_\sfA} \ww & \ww \\
	& \Nghost{\pi'}   & &  & & \Nmultigate{2}{\pi^\res}  & \Nghost{\id}{\dw} & \dw & \dw \\  
	& \Nghost{\pi'}  & \push{\cE'_0} \ww &  \ww & &  \Nghost{\pi^\res}   \\
	& \Nghost{\pi'}  & &  & & \Nghost{ \pi^\res} &  \Nmultigate{1}{\id}{\dw} & \dw & \dw  \\
	& \Nghost{\pi'}   & \multigate{1}{\Sigma'_\sfB} &  \ww & \push{\cE'_\sfB} \ww & \ww & \Nghost{\id}{\ww} & \push{\cE'_\sfB}\ww & \ww \\ 
	&   \push{\cX_\sfB}  \qw  & \ghost{\Sigma'_\sfB}   & \qw &  \push{\cY_\sfB} \qw &   \qw & \qw}  \quad,
\end{align}
	
 where each channel is given its primitive specification (the identities thus stalling the outputs), and where some of the wires of the hidden interfaces are drawn dotted and unlabelled for the sake of clearer perception. 
 
The channel $\Phi$ may be taken identical to that of \cref{eq:primerel}. %
	
	\end{Lem}

\begin{Remark} (Interpretation.)\\
	The main difference between the above condition and the original condition \eqref{eq:primerel} is that all involved channels above are dilationally pure. In $\QIT$, one would say that \eqref{eq:cderpure} is the `purified' version of \eqref{eq:primerel}, or that it corresponds to viewing the condition \eqref{eq:primerel} `in the Church of the larger Hilbert space'. The surprise of \cref{lem:CderRechar} is that even in this purified view the condition is rather simple, with no shenanigans playing out on the right hand side -- only a harmless \emph{residual} state $\pi^\res$ is adjoined. 
	\end{Remark}

\begin{proof}
	It is clear that \eqref{eq:cderpure} implies \eqref{eq:primerel}, simply by trashing all systems corresponding to dotted wires. Conversely, if \eqref{eq:primerel} holds as an equation between channels, then by forming dilationally pure dilations $\hat{\Gamma}_\sfA$ of $\Gamma_\sfA$ and $\hat{\Gamma}_\sfB$ of $\Gamma_\sfB$, dilational purity of the right hand side of \eqref{eq:primerel} implies the identity  \eqref{eq:cderpure} for \myuline{some} state $\pi^\res$. That $\pi^\res$ must in fact be dilationally pure follows from dilational purity of the left hand side of \eqref{eq:cderpure}. 
	
	(We use in this proof that compositions of pure channels are pure; this hinges on localisability.)
	\end{proof}

\begin{Thm} (Partial Collapse.) \label{thm:Collapse}\\
A pure causal dilation of $(T, \scrC)$ is derivable from a given pure causal dilation $\scalemyQ{.8}{0.7}{0.5}{
	&  \push{\cX_\sfA}  \qw& \multigate{1}{ \Sigma_\sfA} &  \push{\cY_\sfA} \qw & \qw  \\
	& \Nmultigate{2}{\pi}  & \ghost{\Sigma_\sfA} &  \push{\cE_\sfA} \ww & \ww\\
	& \Nghost{\pi} & \ww & \push{\cE_0} \ww & \ww\\
	& \Nghost{\pi} & \multigate{1}{\Sigma_\sfB} &  \push{\cE_\sfB}\ww & \ww  \\
	&   \push{\cX_\sfB} \qw & \ghost{\Sigma_\sfB} & \push{\cY_\sfB} \qw & \qw  \\
}$ if and only if it is $\cder$-\myuline{equivalent} to a dilation of the form

  \begin{align} \label{eq:collapse}
  \scalemyQ{1}{0.7}{0.5}{
  	&   \push{\cX_\sfA} \qw  & \multigate{1}{\Sigma_\sfA}  &   \qw &  \push{\cY_\sfA}  \qw &  \qw  & \qw \\
  	& \Nmultigate{4}{\pi}  & \ghost{\Sigma_\sfA} &  \push{\cE_\sfA} \ww & \ww &  \Nmultigate{1}{\id}{\ww} & \push{\cE_\sfA \og \cG_\sfA} \ww & \ww \\
  	& \Nghost{\pi}   &  & \Nmultigate{2}{\Phi} & \push{\cG_\sfA} \ww & \Nghost{\id}{\ww} &  \\  
  	& \Nghost{\pi}  & \push{\cE_0} \ww &   \Nghost{\Phi}{\ww}  & \ww & \push{\cE'_0} \ww & \ww \\
  	& \Nghost{\pi}  &  & \Nghost{ \Phi} &  \push{\cG_\sfB} \ww &  \Nmultigate{1}{\id}{\ww}   \\
  	& \Nghost{\pi}   & \multigate{1}{\Sigma_\sfB} &  \push{\cE_\sfB} \ww & \ww & \Nghost{\id}{\ww} & \push{\cE_\sfB \og \cG_\sfB}\ww & \ww \\ 
  	&   \push{\cX_\sfB}  \qw  & \ghost{\Sigma_\sfB}   & \qw &  \push{\cY_\sfB} \qw &   \qw & \qw} 
\end{align}

for some dilationally pure channel $\Phi$, where each channel is given its primitive specification (the identities thus stalling the $\cG_i$-outputs). In fact, the channel $\Phi$ can be taken to be the one that derives it according to condition \eqref{eq:primerel}.
	\end{Thm}

\begin{Remark} The significance of this theorem is that within the class of pure causal dilations, the pre-order $\cder$ implodes (up to equivalence) to simple modifications which redistribute the acausal side-information of $\cE_0$ to acausal side-information in $\cE'_0$ along with side-information in $\cG_\sfA$ and $\cG_\sfB$, by means of the pure channel $\Phi$.	\end{Remark}

\begin{proof}
	It is clear that the dilation \eqref{eq:collapse} is derivable from the dilation $\scalemyQ{.8}{0.7}{0.5}{
		&  \push{\cX_\sfA}  \qw& \multigate{1}{ \Sigma_\sfA} &  \push{\cY_\sfA} \qw & \qw  \\
		& \Nmultigate{2}{\pi}  & \ghost{\Sigma_\sfA} &  \push{\cE_\sfA} \ww & \ww\\
		& \Nghost{\pi} & \ww & \push{\cE_0} \ww & \ww\\
		& \Nghost{\pi} & \multigate{1}{\Sigma_\sfB} &  \push{\cE_\sfB}\ww & \ww  \\
		&   \push{\cX_\sfB} \qw & \ghost{\Sigma_\sfB} & \push{\cY_\sfB} \qw & \qw  \\
	}$, so any $\cder$-equivalent dilation must be as well. Conversely, we find for any derivable dilation $\scalemyQ{.8}{0.7}{0.5}{
	&  \push{\cX_\sfA}  \qw& \multigate{1}{ \Sigma'_\sfA} &  \push{\cY_\sfA} \qw & \qw  \\
	& \Nmultigate{2}{\pi'}  & \ghost{\Sigma'_\sfA} &  \push{\cE'_\sfA} \ww & \ww\\
	& \Nghost{\pi'} & \ww & \push{\cE'_0} \ww & \ww\\
	& \Nghost{\pi'} & \multigate{1}{\Sigma'_\sfB} &  \push{\cE'_\sfB}\ww & \ww  \\
	&   \push{\cX_\sfB} \qw & \ghost{\Sigma'_\sfB} & \push{\cY_\sfB} \qw & \qw  \\
}$ by \cref{lem:CderRechar} a pure state $\pi^\res$ and pure channels $\hat{\Gamma}_\sfA, \hat{\Gamma}_\sfB$ and $\Phi$ such that \eqref{eq:cderpure} holds. As such, it is also derivable from the dilation \eqref{eq:collapse}. But it is in fact $\cder$-equivalent to it: By \cref{thm:StructureOfReversibles}, the pure channels $\hat{\Gamma}_\sfA$ and $\hat{\Gamma}_\sfB$ are \myuline{reversible}, so by applying left-inverses $\tilde{\Gamma}_\sfA$ and $\tilde{\Gamma}_\sfB$, with their primitive specifications, we obtain the identity 

	\begin{align} 
  \scalemyQ{.8}{0.7}{0.5}{
	&   \push{\cX_\sfA} \qw  & \multigate{1}{\Sigma_\sfA}  &   \qw &  \push{\cY_\sfA}  \qw &  \qw  & \qw \\
	& \Nmultigate{4}{\pi}  & \ghost{\Sigma_\sfA} &  \push{\cE_\sfA} \ww & \ww &  \Nmultigate{1}{\id}{\ww} & \push{\cE_\sfA \og \cG_\sfA} \ww & \ww \\
	& \Nghost{\pi}   &  & \Nmultigate{2}{\Phi} & \push{\cG_\sfA} \ww & \Nghost{\id}{\ww} &  \\  
	& \Nghost{\pi}  & \push{\cE_0} \ww &   \Nghost{\Phi}{\ww}  & \ww & \push{\cE'_0} \ww & \ww \\
	& \Nghost{\pi}  &  & \Nghost{ \Phi} &  \push{\cG_\sfB} \ww &  \Nmultigate{1}{\id}{\ww}   \\
	& \Nghost{\pi}   & \multigate{1}{\Sigma_\sfB} &  \push{\cE_\sfB} \ww & \ww & \Nghost{\id}{\ww} & \push{\cE_\sfB \og \cG_\sfB}\ww & \ww \\ 
&	 \push{\cX_\sfB}    \qw  & \ghost{\Sigma_\sfB}   & \qw &  \push{\cY_\sfB} \qw &   \qw & \qw} 
 =
 \scalemyQ{.8}{0.7}{0.5}{
 	&   \push{\cX_\sfA} \qw  & \multigate{1}{\Sigma'_\sfA}  &   \qw &  \push{\cY_\sfA}  \qw &  \qw  & \qw \\
 	& \Nmultigate{4}{\pi'}  & \ghost{\Sigma'_\sfA} & \ww &   \push{\cE'_\sfA} \ww & \ww &  \Nmultigate{1}{\tilde{\Gamma}_\sfA}{\ww} & \push{\cE_\sfA \og \cG_\sfA} \ww & \ww \\
 	& \Nghost{\pi'}   & &  & & \Nmultigate{2}{\pi^\res}  & \Nghost{\tilde{\Gamma}_\sfA}{\dw}  \\  
 	& \Nghost{\pi'}  & \push{\cE'_0} \ww &  \ww & &  \Nghost{\pi^\res}   \\
 	& \Nghost{\pi'}  & &  & & \Nghost{ \pi^\res} &  \Nmultigate{1}{\tilde{\Gamma}_\sfB}{\dw}   \\
 	& \Nghost{\pi'}   & \multigate{1}{\Sigma'_\sfB} &  \ww & \push{\cE'_\sfB} \ww & \ww & \Nghost{\tilde{\Gamma}_\sfB}{\ww} & \push{\cE_\sfB \og \cE_\sfB}\ww & \ww \\ 
 	&   \push{\cX_\sfB}  \qw  & \ghost{\Sigma'_\sfB}   & \qw &  \push{\cY_\sfB} \qw &   \qw & \qw} 
\quad ,
\end{align}

showing that the converse derivability holds as well, as desired. (Plainly speaking, we have managed to transfer to the other side of the equality symbol the circuitry in the environment that derives one dilation from the other.) 
	\end{proof}

Recall from \cref{ex:Acausal} the notion of a dilation having \emph{no acausal side-information}. If a pure causal dilation $\scalemyQ{.8}{0.7}{0.5}{
	&  \push{\cX_\sfA}  \qw& \multigate{1}{ \Sigma_\sfA} &  \push{\cY_\sfA} \qw & \qw  \\
	& \Nmultigate{2}{\pi}  & \ghost{\Sigma_\sfA} &  \push{\cE_\sfA} \ww & \ww\\
	& \Nghost{\pi} & \ww & \push{\cE_0} \ww & \ww\\
	& \Nghost{\pi} & \multigate{1}{\Sigma_\sfB} &  \push{\cE_\sfB}\ww & \ww  \\
	&   \push{\cX_\sfB} \qw & \ghost{\Sigma_\sfB} & \push{\cY_\sfB} \qw & \qw  \\
}$ has $\cE_0=\triv$, then it is clearly $\cder$-equivalent to a dilation with no acausal side-information, so we might as well use the term about that dilation itself. In this case (i.e. when $\cE_0 = \triv$), the dilation \eqref{eq:collapse} is $\cder$-equivalent to it for any $\Phi$, and this gives us the following two corollaries to \cref{thm:Collapse}, which will be important in the context of quantum self-testing:

\begin{Cor} (Derivability Collapses to Equivalence.)\label{cor:AcausalColl} \\
If $\scalemyQ{.8}{0.7}{0.5}{
	&  \push{\cX_\sfA}  \qw& \multigate{1}{ \Sigma_\sfA} &  \push{\cY_\sfA} \qw & \qw  \\
	& \Nmultigate{2}{\pi}  & \ghost{\Sigma_\sfA} &  \push{\cE_\sfA} \ww & \ww\\
	& \Nghost{\pi} & \ww & \push{\triv} \ww & \ww\\
	& \Nghost{\pi} & \multigate{1}{\Sigma_\sfB} &  \push{\cE_\sfB}\ww & \ww  \\
	&   \push{\cX_\sfB} \qw & \ghost{\Sigma_\sfB} & \push{\cY_\sfB} \qw & \qw  \\
}$ is a pure causal dilation of $(T, \scrC)$ with no acausal side-information, then the pure causal dilations which are derivable from it are precisely those which are $\cder$-equivalent to it. %
	\end{Cor}

\begin{proof}Obvious from \cref{thm:Collapse}. \end{proof}

\begin{Cor} \label{cor:AcRigid}
The following are equivalent:

\begin{enumerate}
	\item $(T, \scrC)$ has a complete dilation with no acausal side-information.
	\item All pure causal dilations of $(T, \scrC)$ are $\cder$-equivalent. 
\end{enumerate}

Moreover, in this case any pure causal dilation is complete.

	\end{Cor}

\begin{proof}
	If 2. holds, then the dense class collapses to a single level; hence, any pure causal dilation is a complete dilation of $(T, \scrC)$. Clearly, we find among the pure causal dilations one with $\cE_0 = \triv$, simply by picking anyone and applying some $\Phi$ which merges $\cE_0$ with $\cE_\sfA$. %
	
	If 1. holds, then any given complete dilation with no acausal side-information can be derived from a pure complete dilation, by density of pure dilations. We may moreover assume without loss of generality that this pure dilation has $\cE_0= \triv$. In other words, $(T, \scrC)$ has a \myuline{pure} causal dilation which is complete and has no acausal side-information. In particular, it must be possible to derive any pure causal dilation from it, so \cref{cor:AcausalColl} implies that 2. holds. 
	\end{proof}

What we now aim to get at, is that when $(T, \scrC)= (P, \scrC)$ encodes the behaviour in a Bell-scenario, quantum self-testing is essentially equivalent to the two conditions of \cref{cor:AcRigid}. There are, however, two subtleties related to this statement.\\
 
 First, it turns out that we have to consider rigidity relative to a \myuline{sub-class} of possible dilations, because some dilations of $(P, \scrC)$ will not be reflected in the space of ordinary quantum strategies; this is related to the peculiarities of dilating measurements, cf. \cref{ex:MeasDil}. 
 
 Secondly, in the usual formulation of self-testing a specific strategy $\tilde{S}$ is singled out as canonical, and the relations that other strategies $S$ bear to it is not a symmetric one, and thus cannot possibly be $\cder$-\myuline{equivalence}; what is actually going on is that $\tilde{S}$ is smallest in the $\cder$-equivalence class, w.r.t. a pre-order (namely, reducibility) which refines the pre-order $\cder$.

\section{Rigidity in $\QIT$ --  Quantum Self-Testing}
\label{sec:SelfTest}

Consider a bipartite Bell-channel $\scalemyQ{.8}{0.7}{0.5}{& \push{\cX_\sfA} \qw & \multigate{1}{(P, \scrC)} & \push{\cY_\sfA} \qw & \qw \\ & \push{\cX_\sfB} \qw & \ghost{(P, \scrC)} & \push{\cY_\sfB} \qw & \qw }$ in $\QIT$.
 Suppose that the systems $\cX_i$ and $\cY_i$ are embeddings of classical systems, $\cX_i = \C^{X_i}$ and $\cY_i = \C^{Y_i}$, for finite sets $X_i$ and $Y_i$. Suppose moreover that the channel $P$ is classical on its interfaces, i.e. that

\begin{align}
	\myQ{0.7}{0.5}{& \gate{\Delta_{X_\sfA}} & \multigate{1}{(P, \scrC)} & \gate{\Delta_{Y_\sfA}}& \qw \\ & \gate{\Delta_{X_\sfB}} & \ghost{(P, \scrC)} & \gate{\Delta_{Y_\sfB}} & \qw }  \quad = \quad 
	\myQ{0.7}{0.5}{& \qw & \multigate{1}{(P, \scrC)}  & \qw & \qw \\ &  \qw & \ghost{(P, \scrC)} & \qw & \qw} \quad,  
	\end{align}
	
	where $\Delta_Z$ is the decoherence channel on $\C^Z$ given by $\Delta_Z(A) = \sum_{z \in Z} \ketbra{z}A \ketbra{z}$. As discussed earlier, $P$ can then be thought of as a channel in $\CIT$, determined by the states 
	
	\begin{align}
	P^{(x_\sfA, x_\sfB)}  := P(\ketbra{x_\sfA} \otimes \ketbra{x_\sfB} ),  \quad x_\sfA \in X_\sfA, x_\sfB \in X_\sfB, 
	\end{align}
	
	which are classical states on $\C^{Y_\sfA} \otimes \C^{Y_\sfB} \cong \C^{Y_\sfA \times Y_\sfB}$, that is, probability distributions on $Y_\sfA \times Y_\sfB $. In the traditional language of self-testing, we are given a \emph{behaviour} or \emph{correlation} $(P^{(x_\sfA, x_\sfB)} )_{x_\sfA \in X_\sfA, x_\sfB \in X_\sfB}$ for the bipartite Bell-scenario with input sets $X_\sfA, X_\sfB$ and output sets $Y_\sfA, Y_\sfB$. Let us denote by $X:= X_\sfA \times X_\sfB$ and $Y:= Y_\sfA \times Y_\sfB$ the total input and output sets, and let us generically write $x$ and $y$ for the tuples $(x_\sfA, x_\sfB)$ and $(y_\sfA, y_\sfB)$, respectively. 

By \cref{thm:BellDenseSelf}, the isometric channels

\begin{align} \label{eq:BellPure}
	\myQ{0.7}{0.5}{
	& \qw &  \push{\cX_\sfA}  \qw& \multigate{1}{ \Sigma_\sfA} &  \push{\cY_\sfA} \qw & \qw  \\
	& \Nmultigate{2}{\pi}   & \push{\cH_\sfA} \qw & \ghost{\Sigma_\sfA} &  \push{\cE_\sfA} \ww & \ww\\
	& \Nghost{\pi} & \ww & \push{\cE_0} \ww & \ww\\
	& \Nghost{\pi} & \push{\cH_\sfB} \qw & \multigate{1}{\Sigma_\sfB} &  \push{\cE_\sfB}\ww & \ww  \\
	&  \qw &  \push{\cX_\sfB} \qw & \ghost{\Sigma_\sfB} & \push{\cY_\sfB} \qw & \qw  \\
}
\end{align}

form a dense class for constructible causal dilations of $(P, \scrC)$. The goal is now to build our way from the channels \eqref{eq:BellPure} to the traditional (tensor-product) quantum strategies which realise the input-output behaviour $(P^x)_{x \in X}$, and to connect rigidity of the causal channel $(P, \scrC)$ to the traditional definition of quantum self-testing. 

\subsection{The Standard Definition of Quantum Self-Testing}

Let us begin by recalling the traditional notions from the literature (recall in particular that measurements are in this traditional conception taken to be projective):

\begin{Definition} (Quantum Strategies.) \label{def:StratAttl}\\
	A \emph{(finite-dimensional tensor-product) quantum strategy} is a triple $(\varrho, \Pi_\sfA, \Pi_\sfB)$, where $\varrho$ is a state on some bipartite finite-dimensional system $\cH_\sfA \otimes \cH_\sfB$, and where, for $i=\sfA, \sfB$,  $\Pi_i = (\Pi^{x_i}_i)_{x_i \in X_i}$ is a collection of PVMs on $\cH_i$, that is, families $\Pi^{x_i}_i = (\Pi^{x_i}_i(y_i))_{y_i \in Y_i}$ of orthogonal projections on $\cH_i$ with $\sum_{y_i \in Y_i} \Pi^{x_i}_i(y_i) = \bone_{\cH_i}$.\end{Definition}

Given a quantum strategy $(\varrho, \Pi_\sfA, \Pi_\sfB)$, we will generically denote the system $\cH_\sfA \otimes \cH_\sfB$ by $\cH$, and for $x=(x_\sfA, x_\sfB)$ and $y=(y_\sfA, y_\sfB)$ the tensor-product projection $\Pi^{x_\sfA}_\sfA(y_\sfA) \otimes \Pi^{x_\sfB}_\sfB(y_\sfB)$ by $\Pi^x(y)$.

\begin{Definition} (Behaviour of a Quantum Strategy.) \\
The \emph{behaviour of the quantum strategy $(\varrho, \Pi_\sfA, \Pi_\sfB)$} is the behaviour $P=(P^x)_{x \in X}$ given by 

\begin{align}
 P^x(y) = \tr(\Pi^x(y) \varrho) = \tr([\Pi^{x_\sfA}_\sfA(y_\sfA) \otimes \Pi^{x_\sfB}_\sfB(y_\sfB)] \varrho ) 
\end{align}

for $x=(x_\sfA, x_\sfB) \in X$ and $y=(y_\sfA, y_\sfB) \in Y$.

\end{Definition}

Observe that the collection $\Pi_i = (\Pi^{x_i}_i)_{x_i \in X_i}$ of projective measurements can be packed (cf. the observations in the preliminary section of the thesis) in the \emph{ensemble} $\scalemyQ{.8}{0.7}{0.5}{& \push{\C^{X_i}} \qw  & \multigate{1}{\Lambda_i} & \push{\C^{Y_i}} \qw & \qw \\ & \push{\cH_i} \qw & \ghost{\Lambda_i}}$, and as such the behaviour $P$ is nothing but the channel 

\begin{align}
\myQ{0.7}{0.5}{
	& \qw &  \push{\C^{X_\sfA}}  \qw& \multigate{1}{ \Lambda_\sfA} &  \push{\C^{Y_\sfA}} \qw & \qw  \\
	& \Nmultigate{1}{\varrho}   & \push{\cH_\sfA} \qw & \ghost{\Lambda_\sfA} \\
	& \Nghost{\varrho} & \push{\cH_\sfB} \qw & \multigate{1}{\Lambda_\sfB}  \\
	&  \qw &  \push{\C^{X_\sfB}} \qw & \ghost{\Lambda_\sfB} & \push{\C^{Y_\sfB}} \qw & \qw  \\
}  \quad. 
\end{align}

A behaviour $P$ is often called a \emph{quantum behaviour} (or, \emph{quantum correlation}) if it is the behaviour of some quantum strategy, and in our terminology the quantum behaviours are thus simply the Bell-channels in $\QIT$. 

\begin{Definition} (Reducibility of Strategies.) \label{def:Redattl}\\
	Let $(\varrho, \Pi_\sfA, \Pi_\sfB)$ and $(\tilde{\psi}, \tilde{\Pi}_\sfA, \tilde{\Pi}_\sfB)$ be two quantum strategies, with $\tilde{\psi}$ a pure state. We say that \emph{$(\varrho, \Pi_\sfA, \Pi_\sfB)$ is reducible to $(\tilde{\psi}, \tilde{\Pi}_\sfA, \tilde{\Pi}_\sfB)$} if there exists a purification $\psi \in \St{\cH_\sfA \otimes \cH_\sfB \otimes \cP}$ of $\varrho$, systems $\cH^\res_\sfA$ and $\cH^\res_\sfB$, a pure state $\psi^\res$ on $\cH^\res_\sfA \otimes \cH^\res_\sfB \otimes \cP$ and isometries $W_i: \cH_i  \to \tilde{\cH}_i \otimes \tilde{\cH}^\res_i$ such that

	\begin{align} \label{eq:RedAttl}
	[	W _\sfA \Pi^{x_\sfA}_\sfA(y_\sfA) \otimes W_\sfB \Pi^{x_\sfB}_\sfB(y_\sfB)   \otimes \bone_\cP ] \ket{\psi} = [ \tilde{\Pi}^{x_\sfA}_\sfA(y_\sfA) \otimes  \tilde{\Pi}^{x_\sfB}_\sfB(y_\sfB)] \tilde{\ket{\psi}} \otimes \ket{\psi^\res} \; \text{for all $x \in X, y \in Y$},
	\end{align}
	
	or, with $W= W_\sfA \otimes W_\sfB$, more compactly
	
	\begin{align} \label{eq:RedAttlComp}
	[	W \Pi^{x}(y)   \otimes \bone_\cP ] \ket{\psi} =  \tilde{\Pi}^{x}(y) \tilde{\ket{\psi}} \otimes \ket{\psi^\res} \quad \text{for all $x \in X, y \in Y$},
	\end{align}

 where, by usual  abuse of notation, we write $\ket{\cdot}$ to denote vector representatives of pure states.\footnote{The condition is not dependent on the choice of vector representatives, since a phase may be absorbed into one of the isometries $W_i$.}\end{Definition}

\begin{Remark} (Terminology and Scope.) \label{rem:Reducibility}\\
	The term \emph{reducible} is not standard in the literature, indeed the condition is rarely separated from the self-testing definition and explicitly named. The domain of the reducibility relation is slightly odd, since $\varrho$ can be arbitrary but $\tilde{\psi}$ is assumed pure, but the relation restricts to a pre-order on the class of pure-state strategies. Alternatively, it could be made a pre-order on the class of all strategies by rephrasing it to quantify also over purifications of potentially mixed state $\tilde{\varrho}$ of the strategy $(\tilde{\varrho}, \tilde{\Pi}_\sfA, \tilde{\Pi}_\sfB)$. This notion is apparently never considered in the literature on self-testing. 
	\end{Remark}

\begin{Definition} (Quantum Self-Testing and Rigidity According to the Literature.) \label{def:SelftestAttl}\\
	Suppose that $(\tilde{\psi}, \tilde{\Pi}_\sfA, \tilde{\Pi}_\sfB)$ is a quantum strategy with behaviour $P$, whose state $\tilde{\psi}$ is pure. We say that \emph{$P$ self-tests the strategy  $(\tilde{\psi}, \tilde{\Pi}_\sfA, \tilde{\Pi}_\sfB)$} if every strategy $(\varrho, \Pi_\sfA, \Pi_\sfB)$ with behaviour $P$ is reducible to $(\tilde{\psi}, \tilde{\Pi}_\sfA, \tilde{\Pi}_\sfB)$. 
	
	A behaviour $P$ is called \emph{rigid according to the literature} (for short, \emph{rigid a.t.t.l.}) if $P$ self-tests some quantum strategy. 	\end{Definition}

\begin{Remark} (To be Pure or Not to be Pure.) \label{rem:PureNotPure}\\
In \cref{def:SelftestAttl} we quantify over strategies whose state $\varrho$ may be non-pure. In much work, this quantification is (seemingly unconsciously) restricted to strategies for which the state $\varrho$ is pure, thus obtaining an a priori weaker condition. Later results will imply (cf. \cref{rem:PurevsMixed}) that if the isometries $W_i$ in condition \eqref{eq:RedAttl} can be chosen independently of $\varrho$ (only depending on the PVMs $\Pi^{x_i}_i$), and if the behaviour $P$ is extremal in the convex set of quantum behaviours, then these two definitions are actually equivalent. However, I do not know whether this is generally the case, and the question apparently has not been considered in the literature. The strong version of quantifying over all strategies agrees with the definition that appears in the recent review article \cite{SB19}, and is also used in Ref. \cite{RUV13}.
	\end{Remark}

\begin{Example} (Rigidity a.t.t.l. of the CHSH-Behaviour.)\\
Recall the CHSH-behaviour from \cref{ex:CHSH}. This behaviour self-tests the strategy which we explicitly used to define it in \cref{ex:CHSH}. Historically, the CHSH-behaviour was the first behaviour to be proven rigid a.t.t.l. (\cite{SW87, PR92, Cir93}), though the terminology of self-testing was only coined later (\cite{MY98, MY04}), in the context of a slightly different behaviour. The CHSH-behaviour remains the quintessential example of quantum self-testing, and there are by now several different proofs for its rigidity (\cite{MYS12, Miller13, SB19}).
	\end{Example}

\subsection{Classically Bound Dilations}
\label{subsec:ClasBound}

Consider again the channel \eqref{eq:BellPure} from the dense class of causal Stinespring dilations of the Bell-channel $(P, \scrC)$, i.e. the channel 

\begin{align} \label{eq:BellPure2}
\myQ{0.7}{0.5}{
	& \qw &  \push{\cX_\sfA}  \qw& \multigate{1}{ \Sigma_\sfA} &  \push{\cY_\sfA} \qw & \qw  \\
	& \Nmultigate{2}{\pi}   & \push{\cH_\sfA} \qw & \ghost{\Sigma_\sfA} &  \push{\cE_\sfA} \ww & \ww\\
	& \Nghost{\pi} & \ww & \push{\cE_0} \ww & \ww\\
	& \Nghost{\pi} & \push{\cH_\sfB} \qw & \multigate{1}{\Sigma_\sfB} &  \push{\cE_\sfB}\ww & \ww  \\
	&  \qw &  \push{\cX_\sfB} \qw & \ghost{\Sigma_\sfB} & \push{\cY_\sfB} \qw & \qw  \\
}. 
\end{align}
 When analysing density in $\CIT$, we were successful by introducing the \myuline{marginals} of the individual components,

\begin{align}
\myQ{0.7}{0.5}{& \push{\cX_i} \qw & \multigate{1}{\Lambda_i} & \push{\cY_i} \qw & \qw \\ 
	& \push{\cH_i} \qw & \ghost{\Lambda_i} } := 	\myQ{0.7}{0.5}{
	&  \push{\cX_i}  \qw & \multigate{1}{\Sigma_i} &  \push{\cY_i} \qw & \qw  \\
	& \push{\cH_i} \qw & \ghost{\Sigma_i} &  \push{\cE_i} \ww & \Ngate{\tr}{\ww}
} \quad \text{and} \quad 
\myQ{0.7}{0.5}{& \Nmultigate{1}{\varrho} & \push{\cH_\sfA} \qw & \qw \\ & \Nghost{\varrho} & \push{\cH_\sfB} \qw & \qw } := 	\myQ{0.7}{0.5}{& \Nmultigate{2}{\pi}  & \push{\cH_\sfA} \qw & \qw \\
	& \Nghost{\pi} & \push{\cE_0} \ww & \Ngate{\tr}{\ww}\\
	& \Nghost{\pi}  & \push{\cH_\sfB} \qw & \qw 
} \quad,
\end{align}

so that 

\begin{align} \label{eq:BellBhv}
\myQ{0.7}{0.5}{& \push{\cX_\sfA} \qw & \multigate{1}{P} & \push{\cY_\sfA} \qw & \qw \\ & \push{\cX_\sfB} \qw & \ghost{P} & \push{\cY_\sfB} \qw & \qw } =  	\myQ{0.7}{0.5}{
	& \qw &  \push{\cX_\sfA}  \qw& \multigate{1}{ \Lambda_\sfA} &  \push{\cY_\sfA} \qw & \qw  \\
	& \Nmultigate{1}{\varrho}   & \push{\cH_\sfA} \qw & \ghost{\Lambda_\sfA} \\
	& \Nghost{\varrho} & \push{\cH_\sfB} \qw & \multigate{1}{\Lambda_\sfB}  \\
	&  \qw &  \push{\cX_\sfB} \qw & \ghost{\Lambda_\sfB} & \push{\cY_\sfB} \qw & \qw  \\
}  \quad.
\end{align}

If we can understand the triples $(\varrho, \Lambda_\sfA, \Lambda_\sfB)$ which satisfy \cref{eq:BellBhv}, then we understand the possible triples $(\pi,\Sigma_\sfA, \Sigma_\sfB)$ defining the dilations \eqref{eq:BellPure2}, for these arise by Stinespring dilating the components of the former triples. And now, for the first time in our story, quantum self-testing finally meets causal dilations: Indeed, as we observed above, any triple $(\varrho, \Lambda_\sfA, \Lambda_\sfB)$ that corresponds to an ordinary quantum strategy (i.e. for which $\Lambda_\sfA$ and $\Lambda_\sfB$ are ensembles of projective measurements) is a triple that satisfies \cref{eq:BellBhv}; hence, any ordinary quantum strategy defines a causal Stinespring dilation \eqref{eq:BellPure2}. The challenge is now that there might be \myuline{other} triples $(\varrho, \Lambda_\sfA, \Lambda_\sfB)$ satisfying \cref{eq:BellBhv} than those arising from such strategies.\\

First of all, there might of course be triples for which $\Lambda_\sfA$ and $\Lambda_\sfB$ correspond to ensembles of measurements which are not projective. This is not problematic, since, as we will show in due time, any causal Stinespring dilation corresponding to such a POVM-strategy is \myuline{derivable} from one corresponding to a PVM-strategy, i.e. the dense class can be further thinned -- for this, we use Naimark's theorem.\footnote{\myuline{This} is the correct way of stating that the measurements in a `quantum strategy' can without loss of generality be assumed to be projective.}

A much more peculiar problem, however, is that even though the channel $P$ has classical inputs and outputs, the channels $\Lambda_\sfA$ and $\Lambda_\sfB$ actually need not be measurement ensembles: 

\begin{Example} (Strange Realisations of the CHSH-Behaviour.) \label{ex:StrangeRel}\\
Consider the strategy for the CHSH-behaviour which we defined in \cref{ex:CHSH}. Or, more generally, consider for the Bell-scenario with $X_\sfA = X_\sfB = \{0,1\}$ and $Y_\sfA = Y_\sfB=\{+1,-1\}$ \myuline{any}  strategy $(\tilde{\psi}, \tilde{\Pi}_\sfA, \tilde{\Pi}_\sfB)$ for which $\tilde{\psi}$ is a pure state on $\C^2 \otimes \C^2$, and all projections $\tilde{\Pi}^{x_i}_i(\pm 1)$ have rank one (i.e. project onto $1$-dimensional subspaces). The channel

\begin{align}  \label{eq:canonical}
  	\myQ{0.7}{0.5}{
	& \qw &  \push{\C^{X_\sfA}}  \qw& \multigate{1}{ \tilde{\Lambda}_\sfA} &  \push{\C^{Y_\sfA}} \qw & \qw  \\
	& \Nmultigate{1}{\tilde{\psi}}   & \push{\C^2} \qw & \ghost{\tilde{\Lambda}_\sfA} \\
	& \Nghost{\tilde{\psi}} & \push{\C^2} \qw & \multigate{1}{\tilde{\Lambda}_\sfB}  \\
	&  \qw &  \push{\C^{X_\sfB}} \qw & \ghost{\tilde{\Lambda}_\sfB} & \push{\C^{Y_\sfB}} \qw & \qw  \\
}  
\end{align}

is the behaviour of this strategy, and it has classical inputs and outputs.
Now, however, observe the following: Every projective rank-one measurement on $\C^2$ can be realised as a unitary conjugation followed by a measurement in the computational basis $(\ket{0}, \ket{1})$; in other words, there exists an ensemble of unitary conjugations $(\tilde{\scrU}^{x_i}_i)_{x_i \in X_i }$ on $\C^2$, such that, with $[\tilde{\scrU}]_i$ denoting the channel that encodes this ensemble, we have $\scalemyQ{.8}{0.7}{0.5}{& \push{\C^{X_i}} \qw & \multigate{1}{\tilde{\Lambda}_i} & \push{\C^{Y_i}} \qw & \qw \\ &\push{\C^2} \qw & \ghost{\tilde{\Lambda}_i}} = \scalemyQ{.8}{0.7}{0.5}{& \push{\C^{X_i}} \qw & \multigate{1}{[\tilde{\scrU}]_i}& \push{\C^2} \qw & \gate{\Delta} & \push{\C^{Y_i}} \qw & \qw \\ &\push{\C^2} \qw & \ghost{[\tilde{\scrU}]_i }}$, where $\Delta$ is the measurement in the computational basis of $\C^2$ (under some identification of $\C^{Y_i}$ with $\C^2$). As we have seen in \cref{ex:MeasDil}, the channel $\Delta$ can be written as a convex combination of unitary conjugations, that is, $\scalemyQ{.8}{0.7}{0.5}{& \push{\C^2} \qw & \gate{\Delta} & \push{\C^{Y_i}} \qw & \qw} = \scalemyQ{.8}{0.7}{0.5}{& \push{\C^2} \qw & \multigate{1}{[\mathscr{V}]_i} & \push{\C^{Y_i}} \qw & \qw \\ &\Ngate{\tau_i} & \ghost{[\mathscr{V}]_i}}$ for some classical state $\tau_i$ on $\C^{m_i}$ (e.g. with $m_i=2$), and some ensemble of unitaries $[\mathscr{V}]_i$. In total, we can therefore rewrite the channel \eqref{eq:canonical} as

\begin{align} \label{eq:strange}
	\myQ{0.7}{0.5}{
	& \qw &  \push{\C^{X_\sfA}}  \qw& \multigate{1}{ [\tilde{\scrU}]_\sfA} &  \push{\C^2} \qw & \multigate{2}{[\mathscr{V}]_\sfA} & \push{\C^{Y_\sfA}} \qw & \qw  \\
	& \Nmultigate{3}{\tilde{\psi}}   &  \push{\C^2} \qw & \ghost{[\tilde{\scrU}]_\sfA} & & \Nghost{[\mathscr{V}]_\sfA}  \\
		& \Nghost{\tilde{\psi}} & \Ngate{\tau_\sfA}& \push{\C^{m_\sfA}}\qw& \qw& \ghost{[\mathscr{V}]_\sfA}\\
			& \Nghost{\tilde{\psi}} & \Ngate{\tau_\sfB}& \push{\C^{m_\sfB}}\qw& \qw& \multigate{2}{[\mathscr{V}]_\sfB} \\
	& \Nghost{\tilde{\psi}} & \push{\C^2} \qw & \multigate{1}{[\tilde{\scrU}]_\sfB} & & \Nghost{[\mathscr{V}]_\sfB} \\
	&  \qw &  \push{\C^{X_\sfB}} \qw & \ghost{[\tilde{\scrU}]_\sfB} &  \push{\C^2} \qw & \ghost{[\mathscr{V}]_\sfB} & \push{\C^{Y_\sfA}} \qw & \qw  \\
}  \quad . \end{align}

Now, if we regroup components so as to make a new triple $(\varrho, \Lambda_\sfA, \Lambda_\sfB)$ for which $\varrho= \tilde{\psi} \otimes \tau_\sfA \otimes \tau_\sfB$ and $\Lambda_i$ is the channel given by the composition of $[\tilde{\scrU}]_i$ and $[\mathscr{V}]_i$, then the channel $\Lambda_i$ will generally not have classical outputs. In fact, on input $\ketbra{x_i}$ from $\C^{X_i}$ and $\ketbra{k}$ from $\C^{m_i}$, the channel acts as a unitary conjugation from $\C^2$ to $\C^{Y_i} \cong \C^2$. The resulting triple $(\varrho, \Lambda_\sfA, \Lambda_\sfB)$ has no chance of being interpreted as a strategy in the traditional sense -- it corresponds to a `strategy' in which the outcomes are not produced by measurements, but by unitarily conjugating a state $\varrho$ whose conjugates \myuline{happen} to be classical states on $\C^{Y_\sfA} \otimes \C^{Y_\sfB}$.   \end{Example} 

\vspace{.1cm}

The issue illustrated by \cref{ex:StrangeRel} is not merely a formal nuisance, but has operational ramifications as well. Indeed, the representation \eqref{eq:strange} gives a convex decomposition of e.g. the CHSH-behaviour (interpreting the classical distribution $\tau_\sfA \otimes \tau_\sfB$ as the weights), and it is a convex decomposition into channels which individually map every classical input $\ketbra{x_\sfA} \otimes \ketbra{x_\sfB}$ to a \myuline{pure} state; since the CHSH-behaviour itself yields a non-pure state on any classical input, these pure states cannot all be identical, so if we dilate $\tau_\sfA \otimes \tau_\sfB$ by forming a copy to obtain an acausal dilation of the CHSH-behaviour (as in \cref{ex:Acausal}), then this acausal  dilation \emph{does not factor}. In other words, it is possible to have side-information correlated with the outputs, before the inputs have been given, a circumstance which drastically contradicts the usual conception of self-testing (\cite{SB19}). The twist is, of course, that the outputs provided by the strange strategy of \cref{ex:StrangeRel} are not classical, and thus cannot be interpreted in the usual framework of self-testing. The unwelcome acausal dilation arising from this is in turn related to the strange dilations of measurements which we saw in \cref{ex:MeasDil}. \\

In this subsection, we will eliminate this problem, more or less by forcing it away.  (We must also enforce classicality of the inputs, but though we ultimately pick the same solution, this problem is of a different character -- indeed, it is the honest users of the open interfaces, and not the potentially malicious agents implementing the channel, who provide the inputs.)

The solution we choose, though not overly elegant, is essentially to dismiss the dilations that arise from the strange strategies of \cref{ex:StrangeRel}. That this should be sensible relies on the fact that it is the honest users of the channel who control the open interfaces; as such they may well choose to only provide classical inputs, and to \myuline{measure} the outputs received from the channels. Under this course of action, the class of dilations which we deemed reasonable would shrink:

\begin{Definition} (Classical Dilations.)\\
	Let $\scalemyQ{.8}{0.7}{0.5}{& \push{\cX_\sfA} \qw & \multigate{1}{(P, \scrC)} & \push{\cY_\sfA} \qw & \qw \\ & \push{\cX_\sfB} \qw & \ghost{(P, \scrC)} & \push{\cY_\sfB} \qw & \qw }$ be a Bell-channel with classical inputs and outputs. A causal dilation $\scalemyQ{.8}{0.7}{0.5}{& \push{\cX_\sfA} \qw & \multigate{2}{(L, \scrE)} & \push{\cY_\sfA} \qw & \qw \\ & \push{\cX_\sfB} \qw & \ghost{(L, \scrE)} & \push{\cY_\sfB} \qw & \qw \\
		& \push{\bbD} \ww & \Nghost{(L, \scrE)}{\ww} & \push{\bbE} \ww & \ww}$ of $(P, \scrC)$ is called \emph{classical} if

\begin{align} \label{eq:ClasDil}
\myQ{0.7}{0.5}{& \gate{\Delta_{X_\sfA}} & \multigate{2}{(L, \scrE)} &\gate{\Delta_{Y_\sfA}} & \qw \\ &\gate{\Delta_{X_\sfB}}  & \ghost{(L, \scrE)} & \gate{\Delta_{Y_\sfB}}  & \qw \\
	& \push{\bbD} \ww & \Nghost{(L, \scrE)}{\ww} & \push{\bbE} \ww & \ww} = \myQ{0.7}{0.5}{&  \qw & \multigate{2}{(L, \scrE)} &  \qw & \qw \\ & \qw & \ghost{(L, \scrE)} & \qw & \qw \\
	& \push{\bbD} \ww & \Nghost{(L, \scrE)}{\ww} & \push{\bbE} \ww & \ww} \quad.
\end{align}

	\end{Definition}

Recall from \cref{def:Rigidity} the concept of \emph{rigidity relative to $\bfD$}, namely the notion that a class of causal dilations $\bfD$ has a $\cder$-largest element.

Now, it \myuline{is} possible to prove that rigidity a.t.t.l. of $P$ (\cref{def:SelftestAttl}) implies rigidity of $(P, \scrC)$ relative to the class of classical dilations in our sense. However, we shall eventually prove a different result, and there are two reasons for that.\\

First of all, the above implication is not optimal: Self-testing in the sense of \cref{def:SelftestAttl} implies rigidity relative to a slightly larger class of dilations than the classical ones, and up to some small caveats the converse is true as well. On the other hand, I do not know that rigidity relative to classical dilations should conversely imply self-testing in the usual sense.

Secondly, though the problem exposed by \cref{ex:StrangeRel} is ultimately a child of the strange dilations of measurements (\cref{ex:MeasDil}), there seems to be something odd about the assumption that we can ignore \myuline{any} dilations of the measurements that are performed by the honest users (cf. the condition \eqref{eq:ClasDil}). After all, this apparently goes against the very idea of asking about the possible dilations that the accessible channel could have come from. More concretely: The devices which allegedly execute our measurements could \myuline{themselves} be manufactured by malicious agents. \\

Fortunately, these two issues are solved by the same move, namely, by not assuming that absolutely no information leaks to the environment from those measurements, but only restraining the allowed dilations of the measurements so as to exclude the strange acausal dilations. What we must require is that only \myuline{primitive} dilations\footnote{That is, dilations with no acausal side-information.} of a measurement count as reasonable (cf. \cref{ex:primdil}): 

\begin{Definition} (Classically Bound Dilations.) \label{def:ClassBound}\\
		Let $\scalemyQ{.8}{0.7}{0.5}{& \push{\cX_\sfA} \qw & \multigate{1}{(P, \scrC)} & \push{\cY_\sfA} \qw & \qw \\ & \push{\cX_\sfB} \qw & \ghost{(P, \scrC)} & \push{\cY_\sfB} \qw & \qw }$ be a Bell-channel with classical inputs and outputs. A causal dilation $\scalemyQ{.8}{0.7}{0.5}{& \push{\cX_\sfA} \qw & \multigate{2}{(L, \scrE)} & \push{\cY_\sfA} \qw & \qw \\ & \push{\cX_\sfB} \qw & \ghost{(L, \scrE)} & \push{\cY_\sfB} \qw & \qw \\
		& \push{\bbD} \ww & \Nghost{(L, \scrE)}{\ww} & \push{\bbE} \ww & \ww}$ of $(P, \scrC)$ is called \emph{classically bound} if there exists a causal dilation $(\check{L}, \check{\scrE})$ of $(P, \scrC)$, such that 

	\begin{align} \label{eq:CB}
	\myQ{0.7}{0.5}{& \Nmultigate{1}{\overline{\Delta}_{X_\sfA}} & \ww &  &\Nmultigate{1}{\overline{\Delta}_{Y_\sfA}} & \ww \\ 
		& \ghost{\Delta_{X_\sfA}} & \qw & \multigate{3}{(\check{L}, \check{\scrE})} &\ghost{\Delta_{Y_\sfA}} & \qw \\
			& \Nmultigate{1}{\overline{\Delta}_{X_\sfB}} & \ww &  \Nghost{(\check{L}, \check{\scrE})} &\Nmultigate{1}{\overline{\Delta}_{Y_\sfB}} & \ww \\ &\ghost{\overline{\Delta}_{X_\sfB}}  & \qw & \ghost{(\check{L}, \check{\scrE})} & \ghost{\overline{\Delta}_{Y_\sfB}}  & \qw \\
		& \push{\check{\bbD}} \ww &\ww &  \Nghost{(\check{L}, \check{\scrE})}{\ww} & \push{\check{\bbE}} \ww & \ww} \cder \myQ{0.7}{0.5}{& \push{\cX_\sfA} \qw & \multigate{2}{(L, \scrE)} & \push{\cY_\sfA} \qw & \qw \\ & \push{\cX_\sfB} \qw & \ghost{(L, \scrE)} & \push{\cY_\sfB} \qw & \qw \\
		& \push{\bbD} \ww & \Nghost{(L, \scrE)}{\ww} & \push{\bbE} \ww & \ww} \quad
	\end{align}
	
	for some primitive dilations $\overline{\Delta}_{X_i}$ of $\Delta_{X_i}$ and $\overline{\Delta}_{Y_i}$ of $\Delta_{Y_i}$.

	\end{Definition}

\begin{Remark} 
	The reason for allowing the dilation $(\check{L}, \check{\scrE})$ to be distinct from $(L, \scrE)$ is that if we force  $(\check{L}, \check{\scrE})= (L, \scrE)$, it does not seem obvious that any dilation \myuline{derivable} from a classically bound dilation is itself classically bound, and this property is desirable for the interpretation of classically bound dilations as `the allowed ones'.
	\end{Remark}

Every classical dilation is classically bound, as witnessed by choosing $(\check{L}, \check{\scrE})= (L, \scrE)$, $\overline{\Delta}_{X_i} = \Delta_{X_i}$ and $\overline{\Delta}_{Y_i} = \Delta_{Y_i}$. Classical boundedness is however slightly less restrictive than classicality. In fact we have the following result, which expresses that we have achieved exactly the desired:

\begin{Prop} (Concrete Recharacterisation of Classical Boundedness.) \label{prop:RecharCB}\\
The classically bound dilations of $(P, \scrC)$ are precisely those which are derivable from a dilation of the form 

	\begin{align} \label{eq:MeasStine}
\myQ{0.7}{0.5}{
	& \qw &  \push{\cX_\sfA}  \qw& \multigate{1}{ \hat{\Lambda}_\sfA} &  \push{\cY_\sfA} \qw & \qw  \\
	& \Nmultigate{2}{\psi }  & \push{\cH_\sfA} \qw & \ghost{\hat{\Lambda}_\sfA} &  \push{\hat{\cE}_\sfA} \ww & \ww\\
	& \Nghost{\psi} & \ww & \push{\cE_0} \ww & \ww\\
	& \Nghost{\psi} & \push{\cH_\sfB} \qw & \multigate{1}{\hat{\Lambda}_\sfB} &  \push{\hat{\cE}_\sfB}\ww & \ww  \\
	&  \qw &  \push{\cX_\sfB} \qw & \ghost{\hat{\Lambda}_\sfB} &\push{\cY_\sfB} \qw& \qw   \\
} \quad,
\end{align}

where $\psi$ is a purification of a state $\varrho$, and where $\hat{\Lambda}_\sfA$ and $\hat{\Lambda}_\sfB$ are Stinespring dilations of measurement ensembles $\Lambda_\sfA$ and $\Lambda_\sfB$ such that $\scalemyQ{.8}{0.7}{0.5}{& \push{\cX_\sfA} \qw & \multigate{1}{P} & \push{\cY_\sfA} \qw & \qw \\ & \push{\cX_\sfB} \qw & \ghost{P} & \push{\cY_\sfB} \qw & \qw } =  	\scalemyQ{.8}{0.7}{0.5}{
	& \qw &  \push{\cX_\sfA}  \qw& \multigate{1}{ \Lambda_\sfA} &  \push{\cY_\sfA} \qw & \qw  \\
	& \Nmultigate{1}{\varrho}   & \push{\cH_\sfA} \qw & \ghost{\Lambda_\sfA} \\
	& \Nghost{\varrho} & \push{\cH_\sfB} \qw & \multigate{1}{\Lambda_\sfB}  \\
	&  \qw &  \push{\cX_\sfB} \qw & \ghost{\Lambda_\sfB} & \push{\cY_\sfB} \qw & \qw  \\
}$. 
\end{Prop}

\begin{proof}
Let us start by observing that the dilations of the form

	\begin{align}  \label{eq:CBdense}
	\myQ{0.7}{0.5}{
		& \Nmultigate{1}{\hat{\Delta}_{X_\sfA}}& \ww & & & \Nmultigate{1}{\hat{\Delta}_{Y_\sfA}} & \ww \\
		& \ghost{\hat{\Delta}_{X_\sfA}} &  \push{\cX_\sfA}  \qw& \multigate{1}{ \Sigma_\sfA} &  \push{\cY_\sfA} \qw & \ghost{\hat{\Delta}_{Y_\sfA}} & \qw  \\
		& \Nmultigate{2}{\pi}   & \push{\cH_\sfA} \qw & \ghost{\Sigma_\sfA} &  \push{\cE_\sfA} \ww & \ww\\
		& \Nghost{\pi} & \ww & \push{\cE_0} \ww & \ww\\
		& \Nghost{\pi} & \push{\cH_\sfB} \qw & \multigate{1}{\Sigma_\sfB} &  \push{\cE_\sfB}\ww & \ww  \\
		&  \multigate{1}{\hat{\Delta}_{X_\sfB}} &  \push{\cX_\sfB} \qw & \ghost{\Sigma_\sfB} &\push{\cY_\sfB} \qw& \multigate{1}{\hat{\Delta}_{Y_\sfB}} & \qw \\
		&\Nghost{\hat{\Delta}_{X_\sfB}} &\ww & & & \Nghost{\hat{\Delta}_{Y_\sfB}} & \ww
	} \quad,
	\end{align}
	
	where $\pi$, $\Sigma_\sfA$ and $\Sigma_\sfB$ are isometric, and where $\hat{\Delta}_{X_i}$ are Stinespring dilations of $\Delta_{X_i}$ and $\hat{\Delta}_{Y_i}$ of $\Delta_{Y_i}$, constitute a dense class within the class of classically bound dilations. First, any classically bound dilation $(L, \scrE)$ can be derived from such a dilation: Pick a dilation $(\check{L}, \check{\scrE})$ such that 
	
	\begin{align}
		\myQ{0.7}{0.5}{& \Nmultigate{1}{\hat{\Delta}_{X_\sfA}} & \ww &  &\Nmultigate{1}{\hat{\Delta}_{Y_\sfA}} & \ww \\ 
		& \ghost{\Delta_{X_\sfA}} & \qw & \multigate{3}{(\check{L}, \check{\scrE})} &\ghost{\Delta_{Y_\sfA}} & \qw \\
		& \Nmultigate{1}{\hat{\Delta}_{X_\sfB}} & \ww &  \Nghost{(\check{L}, \check{\scrE})} &\Nmultigate{1}{\hat{\Delta}_{Y_\sfB}} & \ww \\ &\ghost{\hat{\Delta}_{X_\sfB}}  & \qw & \ghost{(\check{L}, \check{\scrE})} & \ghost{\hat{\Delta}_{Y_\sfB}}  & \qw \\
		& \push{\check{\bbD}} \ww &\ww &  \Nghost{(\check{L}, \check{\scrE})}{\ww} & \push{\check{\bbE}} \ww & \ww} 
	\cder 
	\myQ{0.7}{0.5}{& \push{\cX_\sfA} \qw & \multigate{2}{(L, \scrE)} & \push{\cY_\sfA} \qw & \qw \\ & \push{\cX_\sfB} \qw & \ghost{(L, \scrE)} & \push{\cY_\sfB} \qw & \qw \\
		& \push{\bbD} \ww & \Nghost{(L, \scrE)}{\ww} & \push{\bbE} \ww & \ww}
	\end{align}
	
	holds (observe that in \cref{eq:CB} we may always without loss of generality take the dilations of the measurements to be Stinespring dilations, by completeness). Then, by the density theorem \cref{thm:BellDenseSelf}, we find a pure causal dilation such that 
	
	\begin{align}
		\myQ{0.7}{0.5}{
		&  \push{\cX_\sfA}  \qw& \multigate{1}{ \Sigma_\sfA} &  \push{\cY_\sfA} \qw & \qw  \\
		& \Nmultigate{2}{\pi}  & \ghost{\Sigma_\sfA} &  \push{\cE_\sfA} \ww & \ww\\
		& \Nghost{\pi} & \ww & \push{\cE_0} \ww & \ww\\
		& \Nghost{\pi} & \multigate{1}{\Sigma_\sfB} &  \push{\cE_\sfB}\ww & \ww  \\
		&   \push{\cX_\sfB} \qw & \ghost{\Sigma_\sfB} & \push{\cY_\sfB} \qw & \qw  \\
	} \cder 
\myQ{0.7}{0.5}{& \push{\cX_\sfA} \qw & \multigate{2}{(\check{L}, \check{\scrE})} & \push{\cY_\sfA} \qw & \qw \\ & \push{\cX_\sfB} \qw & \ghost{(L, \scrE)} & \push{\cY_\sfB} \qw & \qw \\
	& \push{\check{\bbD}} \ww & \Nghost{(\check{L}, \check{\scrE})}{\ww} & \push{\check{\bbE}} \ww & \ww} \quad,
\end{align}
	
and	by our coherence theorems from \cref{chap:Causal} it follows that
	
		\begin{align}
		\myQ{0.7}{0.5}{
		& \Nmultigate{1}{\hat{\Delta}_{X_\sfA}}& \ww & & & \Nmultigate{1}{\hat{\Delta}_{Y_\sfA}} & \ww \\
		& \ghost{\hat{\Delta}_{X_\sfA}} &  \push{\cX_\sfA}  \qw& \multigate{1}{ \Sigma_\sfA} &  \push{\cY_\sfA} \qw & \ghost{\hat{\Delta}_{Y_\sfA}} & \qw  \\
		& \Nmultigate{2}{\pi}   & \push{\cH_\sfA} \qw & \ghost{\Sigma_\sfA} &  \push{\cE_\sfA} \ww & \ww\\
		& \Nghost{\pi} & \ww & \push{\cE_0} \ww & \ww\\
		& \Nghost{\pi} & \push{\cH_\sfB} \qw & \multigate{1}{\Sigma_\sfB} &  \push{\cE_\sfB}\ww & \ww  \\
		&  \multigate{1}{\hat{\Delta}_{X_\sfB}} &  \push{\cX_\sfB} \qw & \ghost{\Sigma_\sfB} &\push{\cY_\sfB} \qw& \multigate{1}{\hat{\Delta}_{Y_\sfB}} & \qw \\
		&\Nghost{\hat{\Delta}_{X_\sfB}} &\ww & & & \Nghost{\hat{\Delta}_{Y_\sfB}} & \ww
	}  \cder 
	\myQ{0.7}{0.5}{& \Nmultigate{1}{\hat{\Delta}_{X_\sfA}} & \ww &  &\Nmultigate{1}{\hat{\Delta}_{Y_\sfA}} & \ww \\ 
	& \ghost{\Delta_{X_\sfA}} & \qw & \multigate{3}{(\check{L}, \check{\scrE})} &\ghost{\Delta_{Y_\sfA}} & \qw \\
	& \Nmultigate{1}{\hat{\Delta}_{X_\sfB}} & \ww &  \Nghost{(\check{L}, \check{\scrE})} &\Nmultigate{1}{\hat{\Delta}_{Y_\sfB}} & \ww \\ &\ghost{\hat{\Delta}_{X_\sfB}}  & \qw & \ghost{(\check{L}, \check{\scrE})} & \ghost{\hat{\Delta}_{Y_\sfB}}  & \qw \\
	& \push{\check{\bbD}} \ww &\ww &  \Nghost{(\check{L}, \check{\scrE})}{\ww} & \push{\check{\bbE}} \ww & \ww}   \quad;
	\end{align} 
	
derivability of $(L, \scrE)$ from \eqref{eq:CBdense} is then implied by  transitivity of $\cder$. Secondly, the dilations \eqref{eq:CBdense} are themselves classically bound: This is simply by virtue of the fact that, for any set $Z$,  we have $\Delta_{Z} \circ \Delta_{Z} = \Delta_{Z}$ so there there exists a channel $\Gamma_Z$ such that  

\begin{align}
\myQ{0.7}{0.5}{ &  \qw & \multigate{2}{\hat{\Delta}_{Z}}  & \multigate{1}{\hat{\Delta}_{Z}} & \qw & \qw\\ 
	&  & \Nghost{\hat{\Delta}_{Z}}  & \Nghost{\hat{\Delta}_{Z}}& \Nmultigate{1}{\Gamma_Z}{\ww}  &  \ww & \ww \\
	&  & \Nghost{\hat{\Delta}_{Z}}& \ww  &  \Nghost{\Gamma_Z}{\ww}
} = \myQ{0.7}{0.5}{  & \multigate{1}{\hat{\Delta}_{Z}} & \qw & \qw \\ 
&  \Nghost{\hat{\Delta}_{Z}} & \ww & \ww
},
	\end{align}
	
as the left-most composition is a Stinespring (hence complete) dilation of $\Delta_{Z}$. Altogether, the dilations \eqref{eq:CBdense} therefore constitute a dense class within classically bound dilations, as asserted.

Next, note that if a dilation is derivable from a classically bound one, then it must itself be classically bound (this is obvious from the definition). Hence, the classically bound dilations are \myuline{exactly} those that can be derived from a dilation of the form \eqref{eq:CBdense}. 

With this in place it is however easy to reach the desired conclusion, by observing that the class of channels $\scalemyQ{.8}{0.7}{0.5}{
&	 \Nmultigate{1}{\hat{\Delta}_{X_i}}  &\ww  & & &  \Nmultigate{1}{\hat{\Delta}_{Y_i}} & \ww \\
 & \ghost{\hat{\Delta}_{X_i}}& \push{\cX_i} \qw & \multigate{1}{ \Sigma_i} &  \push{\cY_i} \qw & \ghost{\hat{\Delta}_{Y_i}} & \qw  \\ & 
	   & \push{\cH_i} \qw & \ghost{\Sigma_i} &  \push{\cE_i} \ww & \ww}$, with $\Sigma_i$ an arbitrary isometric channel, is equivalent (by means of channels acting in the environment) to the class of channels $\scalemyQ{.8}{0.7}{0.5}{
	   &  \push{\cX_i} \qw & \multigate{1}{\hat{\Lambda}_i} &  \push{\cY_i} \qw & \qw \\
	   & \push{\cH_i} \qw & \ghost{\hat{\Lambda}_i} &  \push{\hat{\cE}_i} \ww & \ww}$, with $\hat{\Lambda}_i$ the Stinespring dilation of a measurement ensemble $\Lambda_i$.
	\end{proof}

 \cref{prop:RecharCB} implies in particular that the class of dilations \eqref{eq:MeasStine} is \myuline{dense} in the class of classically bound dilations, so rigidity relative to classically bound dilations can be decided from the former class. One last stroke is needed to go from general measurement ensembles to \myuline{projective} measurement ensembles, which then link directly to the usual definition of quantum strategies. This reduction amounts to the following density theorem:  %

\begin{Thm} (Density of PVM-Dilations.) \label{thm:PVMdense}\\ 
	The dilations of $(P, \scrC)$ of the form

		\begin{align}  \label{eq:PVMdense}
	\myQ{0.7}{0.5}{
		& \qw &  \push{\cX_\sfA}  \qw& \multigate{1}{ \hat{\Lambda}_\sfA} &  \push{\cY_\sfA} \qw & \qw  \\
		& \Nmultigate{2}{\psi }  & \push{\cH_\sfA} \qw & \ghost{\hat{\Lambda}_\sfA} &  \push{\hat{\cE}_\sfA} \ww & \ww\\
		& \Nghost{\psi} & \ww & \push{\cE_0} \ww & \ww\\
		& \Nghost{\psi} & \push{\cH_\sfB} \qw & \multigate{1}{\hat{\Lambda}_\sfB} &  \push{\hat{\cE}_\sfB}\ww & \ww  \\
		&  \qw &  \push{\cX_\sfB} \qw & \ghost{\hat{\Lambda}_\sfB} &\push{\cY_\sfB} \qw& \qw   \\
	} \quad,
	\end{align}

	where $\psi$ is a purification of a state $\varrho$, and where $\hat{\Lambda}_\sfA$ and $\hat{\Lambda}_\sfB$ are Stinespring dilations of \myuline{projective} measurement ensembles $\Lambda_\sfA$ and $\Lambda_\sfB$ such that $\scalemyQ{.8}{0.7}{0.5}{& \push{\cX_\sfA} \qw & \multigate{1}{P} & \push{\cY_\sfA} \qw & \qw \\ & \push{\cX_\sfB} \qw & \ghost{P} & \push{\cY_\sfB} \qw & \qw } =  	\scalemyQ{.8}{0.7}{0.5}{
		& \qw &  \push{\cX_\sfA}  \qw& \multigate{1}{ \Lambda_\sfA} &  \push{\cY_\sfA} \qw & \qw  \\
		& \Nmultigate{1}{\varrho}   & \push{\cH_\sfA} \qw & \ghost{\Lambda_\sfA} \\
		& \Nghost{\varrho} & \push{\cH_\sfB} \qw & \multigate{1}{\Lambda_\sfB}  \\
		&  \qw &  \push{\cX_\sfB} \qw & \ghost{\Lambda_\sfB} & \push{\cY_\sfB} \qw & \qw  \\
	}$, is dense in the class of constructible classically bound dilations of $(P, \scrC)$.

		\end{Thm}
	
	\begin{Remark} (Re-Recharacterisation of Classically Bound Dilations.)\\
		Obviously, a slightly stronger statement is also true, namely that a dilation is classically bound \myuline{if} and only if it can be derived from a dilation of the form \eqref{eq:PVMdense}.
		\end{Remark}
	
	\begin{Remark} (Relation to the Usual Reduction.) \\
		\cref{thm:PVMdense} is a precise mathematical statement. In contrast, the standard comment in the literature in motivating the definition of quantum strategies (\cref{def:StratAttl}) -- namely, that measurements can `without loss of generality' be assumed to be projective -- is in most contexts where it is used not a precise mathematical statement. Indeed, a `without loss of generality'-claim is formally the claim that the statement $\forall s \in S : Q(s)$ follows from the statement $\forall s \in S_0 : Q(x)$, where $S_0 \subseteq S$, and in the absence of such a predicate $Q$ it is meaningless. (To some authors, the predicate $Q(s)$ seems to be implicitly `the strategy $s$ is reducible to the canonical strategy', but this of course is meaningless too, since the relation of being reducible to the canonical strategy has not even been \emph{defined} for strategies which do not have projective measurements.)
		\end{Remark}
	
		\begin{proof}
		We have already seen in \cref{prop:RecharCB} that the class is dense when $\Lambda_\sfA, \Lambda_\sfB$ range over measurement ensembles, so we need only prove that any such dilation can be derived from one in which the measurements are projective. 
		
		By the well-known theorem of Naimark (see the preliminary section), any measurement on $\cH$ with outcomes in $Y$, $\channel{\cH}{M}{\C^Y}$, can be written as $\scalemyQ{.8}{0.7}{0.5}{& \qw &  \push{\cH} \qw & \multigate{1}{M^{\up{Nai}}} & \push{\C^Y} \qw & \qw \\ & \Ngate{\phi^{\up{Nai}}} & \push{\cK^{\up{Nai}}} \qw & \ghost{M^{\up{Nai}}}}$, where  $\phi^{\up{Nai}}$ is a pure state on some system $\cK^{\up{Nai}}$, and where $M^{\up{Nai}}$ is a projective measurement on $\cH \otimes \cK^{\up{Nai}}$. By an inductive argument we can easily extend this to ensembles: If 	$
		\scalemyQ{.8}{0.7}{1}{
			& \push{\C^{X_i}}  \qw  & \multigate{1}{\Lambda_i}   \\
			& \push{\cH_i} \qw & \ghost{\Lambda_i} & \qw & \push{\C^{Y_i}}  \qw & \qw  \\
		}$ is an ensemble of measurements, it can be written as	$	\scalemyQ{.8}{0.7}{0.5}{
			& \qw      & \push{\C^{X_i}}  \qw  & \multigate{2}{\Lambda^\up{Nai}_i}    \\
			& \qw & \push{\cH_i} \qw & \ghost{\Lambda^\up{Nai}_i} & \qw & \push{\C^{Y_i}}  \qw & \qw  \\
			& \Ngate{\phi^\up{Nai}_i} & \push{\cK^\up{Nai}_i} \qw & \ghost{\Lambda^\up{Nai}_i} &
		}$, with $\phi^\up{Nai}_i$ a pure state and $\Lambda^\up{Nai}_i$ an ensemble of projective measurements. Then, however, if $\hat{\Lambda}^\up{Nai}_i$ is a Stinespring dilation of $\Lambda^\up{Nai}_i$, the channel $\scalemyQ{.8}{0.7}{0.5}{
		& \qw      & \push{\C^{X_i}}  \qw  & \multigate{2}{\hat{\Lambda}^\up{Nai}_i}    \\
		& \qw & \push{\cH_i} \qw & \ghost{\hat{\Lambda}^\up{Nai}_i} & \push{\C^{Y_i}}  \qw & \qw  \\
		& \Ngate{\phi^\up{Nai}_i} & \push{\cK^\up{Nai}_i} \qw & \ghost{\hat{\Lambda}^\up{Nai}_i} &
	\push{\cE_i} \ww & \ww}$ is a Stinespring dilation of $\Lambda_i$, so the desired follows by redefining the state $\varrho$ as $\varrho \otimes \phi^\up{Nai}_\sfA \otimes \phi^\up{Nai}_\sfB$.
	\end{proof}

\cref{thm:PVMdense} concludes our strive for locating quantum strategies within the framework of causal dilations: They correspond to the channels \eqref{eq:PVMdense}, forming a dense class of classically bound dilations. Whereas it would evidently have been cleaner if that class were dense in the class of \myuline{all} (constructible) dilations, the reader should think of this result as explaining the origin of the traditional quantum strategies. This understanding will be solidified and improved when in the next subsection we see how the relation between strategies used in the conventional definition of self-testing can be re-expressed in terms of operations on the inaccessible interface of \eqref{eq:PVMdense}.

\subsection{The Bridge to Quantum Self-Testing}
\label{subsec:Bridge}

Suppose that $(\varrho, \Pi_\sfA, \Pi_\sfB)$ is a quantum strategy, and that $\Lambda_\sfA$ and $\Lambda_\sfB$ are the associated measurement ensembles. What does the channel \eqref{eq:PVMdense}, i.e. the channel	

\begin{align}  \label{eq:SHat}
	\myQ{0.7}{0.5}{
	& \qw &  \push{\cX_\sfA}  \qw& \multigate{1}{ \hat{\Lambda}_\sfA} &  \push{\cY_\sfA} \qw & \qw  \\
	& \Nmultigate{2}{\psi }  & \push{\cH_\sfA} \qw & \ghost{\hat{\Lambda}_\sfA} &  \push{\hat{\cE}_\sfA} \ww & \ww\\
	& \Nghost{\psi} & \ww & \push{\cE_0} \ww & \ww\\
	& \Nghost{\psi} & \push{\cH_\sfB} \qw & \multigate{1}{\hat{\Lambda}_\sfB} &  \push{\hat{\cE}_\sfB}\ww & \ww  \\
	&  \qw &  \push{\cX_\sfB} \qw & \ghost{\hat{\Lambda}_\sfB} &\push{\cY_\sfB} \qw& \qw   \\
}
\end{align}

look like? That of course depends on the concrete choice of purification and Stinespring dilations, but by the completeness properties of \cref{chap:Dilations}, all choices are equivalent by means of channels acting locally on the inaccessible systems. Therefore,  we may fix them at our convenience without compromising density of the dilation class. In equations, the channel $\scalemyQ{.8}{0.7}{1}{
	& \push{\cX_i}  \qw  & \multigate{1}{\Lambda_i}  & \push{\cY_i}  \qw & \qw \\
	& \push{\cH_i} \qw & \ghost{\Lambda_i}    
}$ which encodes the measurement ensemble $\Pi_i$ is given by 

\begin{align}
\Lambda_i(A \otimes B) = \sum_{y_i \in Y_i , x_i \in X_i } \tr[\Pi^{x_i}(y_i) A] \ketbra{y_i} \bra{x_i} B \ket{x_i} \quad \text{for $A \in \End{\cH_i}$, $B \in \End{\cX_i}$}.
\end{align}

A particularly nice choice of Stinespring dilation of this channel is the isometric channel  $\scalemyQ{.8}{0.7}{1}{
	& \push{\cX_i}  \qw  & \multigate{1}{\hat{\Lambda}_i}  & \push{\cY_i}  \qw & \qw \\
	& \push{\cH_i} \qw & \ghost{\hat{\Lambda}_i} & \push{\cH_i \otimes \cX_i} \ww & \ww   
}$ corresponding to the isometry $S_i : \cH_i \otimes \cX_i \to \cH_i \otimes \cX_i \otimes \cY_i$ given by\footnote{Observe that if the measurements $\Pi^{x_i}_i$ had not been projective but rather general POVMs, $ E^{x_i}_i = (E^{x_i}_i(y_i))_{y_i \in Y_i}$, then we would have had to include additionally a copy of $\cY_i$ in the inaccessible system, and the Stinespring isometry $S_i$ would instead have been $\sum_{x_i \in X_i , y_i \in Y_i } \sqrt{E^{x_i}_i(y_i)}  \otimes \ket{x_i} \otimes \ket{y_i} \otimes  \ket{y_i} \otimes \bra{x_i}$.}

\begin{align}
	S_i = \sum_{x_i \in X_i , y_i \in Y_i } \Pi^{x_i}_i(y_i)  \otimes \ket{x_i} \otimes \ket{y_i} \otimes \bra{x_i}.
	\end{align}

With this choice, the isometric channel \eqref{eq:SHat} corresponds to the isometry $S: \cX \to \cH \otimes \cX \otimes \cY$ given by  

\begin{align}
 S= \sum_{x \in X, y \in Y}[ \Pi^{x_\sfA}_\sfA(y_\sfA) \otimes \Pi^{x_\sfB}_\sfB(y_\sfB)  \otimes \bone_{\cE_0}] \ket{\psi} \otimes \ket{x} \otimes \ket{y} \otimes \bra{x},
\end{align}

where $\cX = \cX_\sfA \otimes \cX_\sfB$, $\cY= \cY_\sfA \otimes \cY_\sfB$ (and as usual $\cH = \cH_\sfA \otimes \cH_\sfB$), and where $\ket{x} = \ket{(x_\sfA, x_\sfB)}= \ket{x_\sfA} \otimes \ket{x_\sfB}$ and $\ket{y} = \ket{(y_\sfA, y_\sfB)}= \ket{y_\sfA} \otimes \ket{y_\sfB}$. (Observe that $\hat{\cE}_i = \cH_i \otimes \cX_i$.)\\

Qualitatively, this is the pivotal moment in our storyline: We now see the quantities from the reducibility condition \eqref{eq:RedAttl} emerge in the guise of a causal Stinespring dilation of the behaviour channel. The fundamental link between the reducibility criterion and these casual dilations is contained in the following result:

\begin{Thm} (Reducibility among Quantum Strategies in Terms of PVM-Dilations.)\label{thm:Bridge} 
	Let $(\tilde{\psi}, \tilde{\Pi}_\sfA, \tilde{\Pi}_\sfB)$ be a quantum strategy for which the state $\tilde{\psi}$ is pure and has locally full rank. Then, a quantum strategy $(\varrho, \Pi_\sfA, \Pi_\sfB)$ is reducible to $(\tilde{\psi}, \tilde{\Pi}_\sfA, \tilde{\Pi}_\sfB)$ in the sense of \cref{def:Redattl} if and only if there exist channels $\Gamma_A, \Gamma_B$ such that 
	
	\begin{align} \label{eq:locder}
\myQ{0.7}{0.5}{
	& \push{\cX_\sfA}  \qw  & \multigate{1}{ \hat{\Lambda}_\sfA} & \push{\cY_\sfA}  \qw & \qw  \\
	& \Nmultigate{2}{\psi}  & \ghost{\hat{\Lambda}_\sfA} &  \Ngate{\Gamma_\sfA}{\ww} & \ww\\
		& \Nghost{\psi}  & \Ngate{\tr}{\ww} \\
	& \Nghost{\psi} & \multigate{1}{\hat{\Lambda}_\sfB} & \Ngate{\Gamma_\sfB}{\ww} & \ww \\
	&   \push{\cX_\sfB} \qw  & \ghost{\hat{\Lambda}_\sfB} & \push{\cY_\sfB}  \qw & \qw  
} 
	\quad =  \quad 
\myQ{0.7}{0.5}{
	& \push{\cX_\sfA}  \qw  & \multigate{1}{ {\hat{\tilde{\Lambda}}}_\sfA} & \push{\cY_\sfA}  \qw & \qw  \\
	& \Nmultigate{2}{\tilde{\psi}}  & \ghost{{\hat{\tilde{\Lambda}}}_\sfA} &  \ww\\
		& \Nghost{\tilde{\psi}}  \\
	& \Nghost{\tilde{\psi}} & \multigate{1}{{\hat{\tilde{\Lambda}}}_\sfB} & \ww  \\
	&   \push{\cX_\sfB} \qw  & \ghost{{\hat{\tilde{\Lambda}}}_\sfB} & \push{\cY_\sfB}  \qw & \qw  
}   \quad,
	\end{align}
	
	where $\hat{\Lambda}_i$ and $\hat{\tilde{\Lambda}}_i$ are Stinespring dilations of the corresponding measurement ensembles $\Lambda_i$ and $\tilde{\Lambda_i}$, respectively, and where $\psi$ is a purification of $\varrho$. 
	
\end{Thm}

\begin{Remark} (On the Full-Rank Assumption.)\\
	The technical full-rank assumption is, to the best of my knowledge, satisfied in all known examples of quantum self-testing. Also, it is only needed for the `if'-direction of the statement. 
	\end{Remark}

\begin{Remark} (Reducibility of Pure-State Strategies versus General Strategies.) \label{rem:PurevsMixed} \\
	From \cref{thm:Bridge}, we may argue that in order to establish self-testing it is often enough to quantify over pure-state strategies (cf. \cref{rem:PureNotPure}):
	
Suppose that every pure-state strategy $(\psi, \Pi_\sfA, \Pi_\sfB)$ with behaviour $P$ is reducible to $(\tilde{\psi}, \tilde{\Pi}_\sfA, \tilde{\Pi}_\sfB)$, and that this is witnessed by isometries $W_i$ which do not depend on $\psi$. (This is often the case, cf. the so-called \emph{swap-method} (\cite{SB19}), which produces isometries $W_i$ depending only on $\Pi_i$.) By the proof of \cref{thm:Bridge}, the channels $\Gamma_i$ are simply marginals of the isometric conjugations by $W_i$, so $\Gamma_i$ can then be chosen independently of $\psi$. Suppose moreover that the behaviour $P$ is extremal (this is true, for example, if the reducibility of pure-state strategies can be established merely based on the value of the behaviour in a so-called \emph{non-local game} (\cite{SB19})). Now, if $(\varrho, \Pi_\sfA, \Pi_\sfB)$ is a \myuline{general} strategy with behaviour $P$, then, writing $\varrho = \sum_{k=1}^n p_k \psi_k$ by the spectral theorem, with $\psi_1, \ldots, \psi_n$ pure and $p_1, \ldots, p_n > 0$, we conclude  by extremality that all of the pure-state strategies $(\psi_k, \Pi_\sfA, \Pi_\sfB)$ have behaviour $P$. We thus find (by the state-independence assumption) channels $\Gamma_\sfA, \Gamma_\sfB$ such that $\scalemyQ{.8}{0.7}{0.5}{
	& \push{\cX_\sfA}  \qw  & \multigate{1}{ \hat{\Lambda}_\sfA} & \push{\cY_\sfA}  \qw & \qw  \\
	& \Nmultigate{1}{\psi_k}  & \ghost{\hat{\Lambda}_\sfA} &  \Ngate{\Gamma_\sfA}{\ww} & \ww\\
	& \Nghost{\psi_k} & \multigate{1}{\hat{\Lambda}_\sfB} & \Ngate{\Gamma_\sfB}{\ww} & \ww \\
	&   \push{\cX_\sfB} \qw  & \ghost{\hat{\Lambda}_\sfB} & \push{\cY_\sfB}  \qw & \qw  
} 
=
\scalemyQ{.8}{0.7}{0.5}{
	& \push{\cX_\sfA}  \qw  & \multigate{1}{ {\hat{\tilde{\Lambda}}}_\sfA} & \push{\cY_\sfA}  \qw & \qw  \\
	& \Nmultigate{1}{\tilde{\psi}}  & \ghost{{\hat{\tilde{\Lambda}}}_\sfA} &  \ww\\
	& \Nghost{\tilde{\psi}} & \multigate{1}{{\hat{\tilde{\Lambda}}}_\sfB} & \ww  \\
	&   \push{\cX_\sfB} \qw  & \ghost{{\hat{\tilde{\Lambda}}}_\sfB} & \push{\cY_\sfB}  \qw & \qw  
}   $ for all $k=1, \ldots, n$, and hence by forming the convex combination on each side we have $\scalemyQ{.8}{0.7}{0.5}{
& \push{\cX_\sfA}  \qw  & \multigate{1}{ \hat{\Lambda}_\sfA} & \push{\cY_\sfA}  \qw & \qw  \\
& \Nmultigate{1}{\varrho}  & \ghost{\hat{\Lambda}_\sfA} &  \Ngate{\Gamma_\sfA}{\ww} & \ww\\
& \Nghost{\varrho} & \multigate{1}{\hat{\Lambda}_\sfB} & \Ngate{\Gamma_\sfB}{\ww} & \ww \\
&   \push{\cX_\sfB} \qw  & \ghost{\hat{\Lambda}_\sfB} & \push{\cY_\sfB}  \qw & \qw  
} 
=
\scalemyQ{.8}{0.7}{0.5}{
& \push{\cX_\sfA}  \qw  & \multigate{1}{ {\hat{\tilde{\Lambda}}}_\sfA} & \push{\cY_\sfA}  \qw & \qw  \\
& \Nmultigate{1}{\tilde{\psi}}  & \ghost{{\hat{\tilde{\Lambda}}}_\sfA} &  \ww\\
& \Nghost{\tilde{\psi}} & \multigate{1}{{\hat{\tilde{\Lambda}}}_\sfB} & \ww  \\
&   \push{\cX_\sfB} \qw  & \ghost{{\hat{\tilde{\Lambda}}}_\sfB} & \push{\cY_\sfB}  \qw & \qw  
}   $, which is exactly the condition \eqref{eq:locder}, so $(\varrho, \Pi_\sfA, \Pi_\sfB)$ is reducible to $(\tilde{\psi}, \tilde{\Pi}_\sfA, \tilde{\Pi}_\sfB)$. 

In summary, self-testing relative to pure-state strategies along with extremality of the behaviour implies, under the state-independence assumption, the full self-testing condition. (The converse holds as well, since the full self-testing condition always implies extremality of the behaviour, as detailed in \cref{subsec:SecExt}.) 
	\end{Remark}

\vspace{.2cm}

\begin{proof}[Proof of \cref{thm:Bridge}] 
	By \cref{lem:CderRechar}, the condition \eqref{eq:locder} is equivalent to the existence of isometric channels $\hat{\Gamma}_\sfA, \hat{\Gamma}_\sfB$ and a pure state $\psi^\res$ on $\cH^\res_\sfA \otimes \cH^\res_\sfB \otimes {\cE_0}$, such that 
	
	\begin{align} \label{eq:LocPure}
	\myQ{0.7}{0.5}{
		& \push{\cX_\sfA}  \qw  & \multigate{1}{ \hat{\Lambda}_\sfA} & \qw &  \push{\cY_\sfA}  \qw & \qw  \\
		& \Nmultigate{4}{\psi}  & \ghost{\hat{\Lambda}_\sfA}  & \push{\cH_\sfA \otimes \cX_\sfA} \ww &  \Nmultigate{1}{\hat{\Gamma}_\sfA}{\ww} & \push{\tilde{\cH}_\sfA \otimes \cX_\sfA}  \ww & \ww \\
		& \Nghost{\psi}& & & \Nghost{\hat{\Gamma}_\sfA} & \push{\cH^\res_\sfA} \dw & \dw\\
		& \Nghost{\psi}& \ww & \ww & \ww & \push{{\cE_0}} \ww & \ww\\
		&\Nghost{\psi} & & & \Nmultigate{1}{\hat{\Gamma}_\sfB} &   \push{\cH^\res_\sfB} \dw & \dw\\
		& \Nghost{\psi} & \multigate{1}{\hat{\Lambda}_\sfB} & \push{\cH_\sfB \otimes \cX_\sfB} \ww &  \Nghost{\hat{\Gamma}_\sfB}{\ww} & \push{\tilde{\cH}_\sfB \otimes \cX_\sfB}  \ww & \ww  \\
		&   \push{\cX_\sfB} \qw  & \ghost{\hat{\Lambda}_\sfB} & \qw & \push{\cY_\sfB}  \qw & \qw  \\
	} 
	 =  
	\myQ{0.7}{0.5}{
		& \push{\cX_\sfA}  \qw  & \multigate{1}{ \hat{\tilde{\Lambda}}_\sfA} & \push{\cY_\sfA}  \qw & \qw  \\
		& \Nmultigate{4}{\tilde{\psi}}  & \ghost{\hat{\tilde{\Lambda}}_\sfA} & \push{\tilde{\cH}_\sfA \otimes \cX_\sfA}  \ww & \ww \\
		& \Nghost{\tilde{\psi}} &  \Nmultigate{2}{\psi^\textup{res}} &   \push{\cH^\res_\sfA} \dw & \dw \\
		&	\Nghost{\psi'} & \Nghost{\psi^\textup{res}} & \push{{\cE_0}}\ww & \ww  \\ 
		& \Nghost{\psi'} &  \Nghost{\psi^\textup{res}} &   \push{\cH^\res_\sfB} \dw & \dw \\
		& \Nghost{\tilde{\psi}} & \multigate{1}{\hat{\tilde{\Lambda}}_\sfB}  & \push{\tilde{\cH}_\sfB \otimes \cX_\sfB}  \ww & \ww \\
		&   \push{\cX_\sfB} \qw  & \ghost{\hat{\tilde{\Lambda}}_\sfB} & \push{\cY_\sfB}  \qw & \qw  \\
	} 
	\quad ,
	\end{align}

with our convenient choice of Stinespring dilations $\hat{\Lambda}_i$ and $\hat{\tilde{\Lambda}}_i$ from above. \cref{eq:LocPure} is an identity between two isometric channels, and can thus be rephrased as an identity between the two isometries which represent them (possibly absorbing a phase). Let $V_i :   \cH_i  \otimes \cX_i \to  \tilde{\cH}_i \otimes \cH^\res_i \otimes \cX_i$ be an isometry representing the channel $\hat{\Gamma}_i$, and let $\ket{\psi}, \tilde{\ket{\psi}}$ and $\ket{\psi^\res}$ represent the states. Then,  \cref{eq:LocPure} is equivalent to the equation

\begin{align}
\begin{split}
&\sum_{x \in X, y \in Y} [V^{x_\sfA}_\sfA \Pi^{x_\sfA}_\sfA(y_\sfA) \otimes V^{x_\sfB}_\sfB \Pi^{x_\sfB}_\sfB(y_\sfB)  \otimes \bone_{\cE_0} ] \ket{\psi}  \otimes \ket{y} \otimes \bra{x} \\
& = \sum_{x \in X, y \in Y} [\tilde{\Pi}^{x_\sfA}_\sfA(y_\sfA) \otimes  \tilde{\Pi}^{x_\sfB}_\sfB(y_\sfB)  ] \tilde{\ket{\psi}} \otimes \ket{\psi^\res} \otimes \ket{x}  \otimes \ket{y} \otimes \bra{x},
\end{split}
\end{align}

where $V^{x_i}_i : \cH_i \to  \tilde{\cH}_i \otimes \cH^\res_i \otimes \cX_i$ denotes the isometry $V_i (\bone_{\cH_i} \otimes \ket{x_i})$. Now, since the system $(\ket{y} \otimes \bra{x})_{x \in X, y \in Y}$ is orthonormal, this identity can be rephrased component-wise as

\begin{align} \label{eq:Vq}
\forall x \in X, y \in Y : \quad  [V^x \Pi^x(y)  \otimes \bone_{\cE_0} ] \ket{\psi} = \tilde{\Pi}^{x}(y)  \tilde{\ket{\psi}}\otimes \ket{\psi^\res} \otimes \ket{x} ,
\end{align}

with the abbreviations $V^x = V^{x_\sfA}_\sfA \otimes V^{x_\sfB}_\sfB$, and, as usual, $\Pi^x(y) = \Pi^{x_\sfA}_\sfA(y_\sfA) \otimes \Pi^{x_\sfB}_\sfB(y_\sfB)$, $\tilde{\Pi}^x(y) = \tilde{\Pi}^{x_\sfA}_\sfA(y_\sfA) \otimes \tilde{\Pi}^{x_\sfB}_\sfB(y_\sfB)$. In comparison, the reducibility condition (\cref{eq:RedAttlComp}) reads

\begin{align} \label{eq:Redu}
\forall x \in X, y \in Y : \quad  [W \Pi^x(y)  \otimes \bone_{\cE_0} ] \ket{\psi} = \tilde{\Pi}^{x}(y)  \tilde{\ket{\psi}} \otimes \ket{\psi^\res},
\end{align}

with $W= W_\sfA \otimes W_\sfB$, for some isometries $W_i : \cH_i \to \tilde{\cH}_i \otimes \cH^\res_i$ and some pure state $\psi^\res$. If \cref{eq:Redu} holds, then \cref{eq:Vq} follows by taking $V_i = W_i \otimes \bone_{\cX_i}$, so that $V^{x_i}_i = W_i \otimes \ket{x_i}$. Hence, it is clear that \eqref{eq:locder} follows from the reducibility condition. That conversely \cref{eq:Redu} follows from \cref{eq:Vq} is not obvious, and it is the content of \cref{lem:technical} below (using the assumption that $\tilde{\psi}$ has locally full rank). Hence, the reducibility condition follows from \eqref{eq:locder}, finishing the proof. \end{proof}

The main technicalities of the above proof reside in \cref{lem:technical} below. We will postpone this lemma to the end of the subsection, however, in order to first discuss the consequences of \cref{thm:Bridge} and phrase it more transparently in our language. \\

When $P$ self-tests $(\tilde{\psi}, \tilde{\Pi}_\sfA, \tilde{\Pi}_\sfB)$, \cref{thm:Bridge} apparently says that the causal dilation corresponding to $(\tilde{\psi}, \tilde{\Pi}_\sfA, \tilde{\Pi}_\sfB)$ can be derived from the causal dilation corresponding to any strategy $(\varrho, \Pi_\sfA, \Pi_\sfB)$. This, of course, is the \emph{wrong way around} compared to our conception of rigidity. The point is, however, that as in the proof of \cref{thm:Collapse}, the purified relation \eqref{eq:LocPure} allows us to \emph{move the channels $\hat{\Gamma}_i$ to the other side}, thus realising also that the dilation corresponding to $(\varrho, \Pi_\sfA, \Pi_\sfB)$ is derivable from the dilation corresponding to $(\tilde{\psi}, \tilde{\Pi}_\sfA, \tilde{\Pi}_\sfB)$. As such, self-testing really implies that all dilations corresponding to strategies are $\cder$-\myuline{equivalent}  (as in \cref{cor:AcausalColl}).\\

Consequently, rigidity a.t.t.l. implies in our language rigidity relative to classically bound dilations, but in the rather strong sense that all pure causal dilations are $\cder$-equivalent. In this light, there is nothing special about the `canonical' strategy $(\tilde{\psi}, \tilde{\Pi}_\sfA, \tilde{\Pi}_\sfB)$; any representative of the equivalence class will be a complete dilation of $(P, \scrC)$. However, \cref{eq:locder} expresses that the dilation corresponding to the strategy $(\tilde{\psi}, \tilde{\Pi}_\sfA, \tilde{\Pi}_\sfB)$ is particularly simple, because it can be derived from the others using \emph{local} operations in the environment. \\

More precisely, suppose we define the following pre-order on the class of pure causal dilations:

\begin{Definition} (Local Derivability.)\\
	Let $(P, \scrC)$ be a Bell-channel. For two pure causal dilations we write 
	
	\begin{align}
	\scalemyQ{1}{0.7}{0.5}{
	&  \push{\cX_\sfA}  \qw& \multigate{1}{ \Sigma_\sfA} &  \push{\cY_\sfA} \qw & \qw  \\
	& \Nmultigate{2}{\pi}  & \ghost{\Sigma_\sfA} &  \push{\cE_\sfA} \ww & \ww\\
	& \Nghost{\pi} & \ww & \push{\cE_0} \ww & \ww\\
	& \Nghost{\pi} & \multigate{1}{\Sigma_\sfB} &  \push{\cE_\sfB}\ww & \ww  \\
	&   \push{\cX_\sfB} \qw & \ghost{\Sigma_\sfB} & \push{\cY_\sfB} \qw & \qw  \\
} \cder_{loc.}\scalemyQ{1}{0.7}{0.5}{
	&  \push{\cX_\sfA}  \qw& \multigate{1}{ \Sigma'_\sfA} &  \push{\cY_\sfA} \qw & \qw  \\
	& \Nmultigate{2}{\pi'}  & \ghost{\Sigma'_\sfA} &  \push{\cE'_\sfA} \ww & \ww\\
	& \Nghost{\pi'} & \ww & \push{\cE'_0} \ww & \ww\\
	& \Nghost{\pi'} & \multigate{1}{\Sigma'_\sfB} &  \push{\cE'_\sfB}\ww & \ww  \\
	&   \push{\cX_\sfB} \qw & \ghost{\Sigma'_\sfB} & \push{\cY_\sfB} \qw & \qw  \\
}
\end{align}

 and say that the latter is \emph{locally derivable} from the former if there exist channels  $\Gamma_\sfA$, $\Gamma_\sfB$ and $\Gamma_0$ such that

	\begin{align}
\scalemyQ{1}{0.7}{0.5}{
	& \push{\cX_\sfA}\qw  & \multigate{1}{ \Sigma_\sfA} &  \push{\cY_\sfB}  \qw & \qw  \\
	& \Nmultigate{2}{\pi}  & \ghost{\Sigma_\sfA} &  \push{\cE_\sfA} \ww & \Ngate{\Gamma_\sfA}{\ww} & \push{\cE'_\sfA} \ww & \ww \\
	& \Nghost{\pi} & \ww & \push{\cE_0} \ww & \Ngate{\Gamma_0}{\ww} & \push{\cE'_0} \ww & \ww\\
	& \Nghost{\pi} & \multigate{1}{\Sigma_\sfB} & \push{\cE_\sfB} \ww & \Ngate{\Gamma_\sfB}{\ww} & \push{\cE'_\sfB} \ww & \ww \\
	&  \push{\cX_\sfA} \qw  & \ghost{\Sigma_\sfB} &  \push{\cY_\sfB} \qw & \qw  \\
} =\scalemyQ{1}{0.7}{0.5}{
	& \push{\cX_\sfA}\qw  & \multigate{1}{ \Sigma'_\sfA} &  \push{\cY_\sfB}  \qw & \qw  \\
	& \Nmultigate{2}{\pi'}  & \ghost{\Sigma'_\sfA} &  \push{\cE'_\sfA} \ww & \ww\\
	& \Nghost{\pi'} & \ww & \push{\cE'_0} \ww & \ww\\
	& \Nghost{\pi'} & \multigate{1}{\Sigma'_\sfB} & \push{\cE'_\sfB} \ww & \ww  \\
	&  \push{\cX_\sfA} \qw  & \ghost{\Sigma'_\sfB} &  \push{\cY_\sfB} \qw & \qw  \\
} .
\end{align}

\end{Definition}

Then, the pre-order $\cder_{loc.}$ refines the pre-order $\cder$, and we have the following which will constitute our best relation between self-testing (that is, rigidity a.t.t.l.) and rigidity in our sense:

\begin{Cor} (Rigidity versus Rigidity a.t.t.l.) \label{cor:RigidSelftest}\\
	Let $\scalemyQ{.8}{0.7}{0.5}{& \push{\cX_\sfA} \qw & \multigate{1}{(P, \scrC)} & \push{\cY_\sfA} \qw & \qw \\ & \push{\cX_\sfB} \qw & \ghost{(P, \scrC)} & \push{\cY_\sfB} \qw & \qw }$ be bipartite Bell-channel, and let $(\tilde{\psi}, \tilde{\Pi}_\sfA, \tilde{\Pi}_\sfB)$ be a quantum strategy with behaviour $P$ for which the state $\tilde{\psi}$ is pure and has locally full rank. Then, the following are equivalent:
	
	\begin{enumerate} 
		\item $P$ self-tests $(\tilde{\psi}, \tilde{\Pi}_\sfA, \tilde{\Pi}_\sfB)$.
		\item All the causal Stinespring dilations of $(P, \scrC)$ corresponding to quantum strategies are $\cder$-equivalent, and $(\tilde{\psi}, \tilde{\Pi}_\sfA, \tilde{\Pi}_\sfB)$ is a $\cder_{loc.}$-smallest element of this equivalence class.
		\item $(P, \scrC)$ is rigid relative to classically bound dilations and has a complete dilation with no acausal side-information.  Moreover, the $\cder$-equivalence class of causal Stinespring dilations corresponding to quantum strategies has a $\cder_{loc.}$-smallest element, namely the dilation corresponding to $(\tilde{\psi}, \tilde{\Pi}_\sfA, \tilde{\Pi}_\sfB)$.
	\end{enumerate}
\end{Cor}

\begin{proof}
	The equivalence of 1. and 2. follows from \cref{thm:Bridge} and the considerations that follow. The equivalence of 2. and 3. follows by \cref{cor:AcausalColl} (or rather, an adaption of this corollary to the class of classically bound dilations).
	\end{proof}

Now, of the statements appearing in \cref{cor:RigidSelftest} only the statement \emph{`$(P, \scrC)$ is rigid relative to classically bound dilations and has a complete dilation with no acausal side-information.'} is of a purely operational nature. Indeed, as far as I can see, the causal Stinespring dilations themselves a priori have no operational interpretation. 

Consider therefore the following question:

\begin{OP} (Existence of Simple Representatives.) \label{op:SimpleRep}\\
	Does every $\cder$-equivalence class of [classically bound] causal Stinespring dilations contain a  $\cder_{loc.}$-smallest element? 
\end{OP}

If it has an affirmative answer, then, by \cref{cor:RigidSelftest}, quantum self-testing can be given a purely operational formulation. Indeed, the following will be true (assuming that also the full-rank condition on the state is always satisfied): 

\begin{Conj} (Equivalence of Quantum Self-Testing with Acausal Rigidity.) \label{conj:Selftesting}\\
Given a bipartite Bell-channel $\scalemyQ{.8}{0.7}{0.5}{& \push{\cX_\sfA} \qw & \multigate{1}{(P, \scrC)} & \push{\cY_\sfA} \qw & \qw \\ & \push{\cX_\sfB} \qw & \ghost{(P, \scrC)} & \push{\cY_\sfB} \qw & \qw }$, the following are equivalent: 
	
	\begin{enumerate}
		\item $P$ is rigid a.t.t.l., i.e. self-tests (in the sense of \cref{def:SelftestAttl}) some strategy  $(\tilde{\psi}, \tilde{\Pi}_\sfA, \tilde{\Pi}_\sfB)$.
		\item $(P, \scrC)$ is rigid relative to classically bound dilations, and has a complete dilation with no acausal side-information. 
		\end{enumerate}
	
	\end{Conj}

On the other hand, there seems to me to be no good reasons why it should not be possible to have rigidity by means of a complete dilation with non-trivial acausal side-information:

\begin{OP} (Rigidity Beyond the Usual Assumptions of Quantum Self-Testing.)\\ \label{op:CausalRigidity}
Does there exist a Bell-channel $(P, \scrC)$ which is rigid relative to classically bound dilations, but for which no complete dilation has no acausal side-information? 
\end{OP}

We end the subsection by establishing the missing technical ingredient for the proof of \cref{thm:Bridge}:

\begin{Lem} (Technical Result for \cref{thm:Bridge}). \label{lem:technical} \\
	Let $(\varrho, \Pi_\sfA, \Pi_\sfB)$ and $(\tilde{\psi}, \tilde{\Pi}_\sfA, \tilde{\Pi}_\sfB)$ be quantum strategies, for which $\tilde{\psi}$ is a pure state with locally full rank. Suppose that there exists a purification $\psi$ of $\varrho$ with purifying space $\cP$, a pure state $\psi^\res$ on $\cH^\res_\sfA \otimes \cH^\res_\sfB \otimes \cP$, and isometries $V^{x_i}_i: \cH_i \into \tilde{\cH}_i \otimes \cH^\res_i \otimes \cX_i$, such that, with $V^{x} = V^{x_\sfA}_\sfA \otimes V^{x_\sfB}_\sfB$,
	
	\begin{align} \label{eq:qdependent}
[	V^x \Pi^x(y) \otimes \bone_\cP] \ket{\psi} = \tilde{\Pi}^x(y) \tilde{\ket{\psi}} \otimes \ket{\psi^\res} \otimes \ket{x} \; \text{	for all $x \in X, y \in Y$. }
	\end{align}

Then, there exist $x_i$-independent isometries $W_i :  \cH_i \into \tilde{\cH}_i \otimes \cH^\res_i$ such that, with $W= W_\sfA \otimes W_\sfB$,

	\begin{align} \label{eq:qindependent}
	[W \Pi^x(y) \otimes \bone_\cP] \ket{\psi} = \tilde{\Pi}^x(y) \tilde{\ket{\psi}} \otimes \ket{\psi^\res} \; \text{	for all $x \in X, y \in Y$. }
	\end{align} 
	
In other words, $(\varrho, \Pi_\sfA, \Pi_\sfB)$ is reducible to $(\tilde{\psi}, \tilde{\Pi}_\sfA, \tilde{\Pi}_\sfB)$.
\end{Lem}

\begin{Remark} (On an Approximate Generalisation.) \label{rem:qapproximate} \\
	It is clear that the condition \eqref{eq:qdependent} implies that the behaviours of the two strategies $(\varrho, \Pi_\sfA, \Pi_\sfB)$ and $(\tilde{\psi}, \tilde{\Pi}_\sfA, \tilde{\Pi}_\sfB)$ are identical. Nevertheless, it has never been considered in the literature that this condition, in which the local isometries are allowed to depend on the local inputs ($x_i$), should actually be the correct notion of reducibility among quantum strategies. The result of this lemma gives a possible explanation for this. It is worth noting, however, that the relationship between the two conditions \eqref{eq:qdependent} and \eqref{eq:qindependent} is probably more subtle in the approximate case (of so-called \emph{robust} self-testing), since we use at a point in the proof a full-rank argument without regards to the fact that some Schmidt coefficients may be small.
	\end{Remark}

\begin{proof}%

	First, we argue that there exists an $x_\sfA$-independent isometry $W_\sfA : \cH_\sfA\into \tilde{\cH}_\sfA\otimes \cH^\res_\sfA$ such that the operator identity
	
	\begin{align}
	W_{\sfA} = [\bone_{\tilde{\cH}_{\sfA} \otimes \cH^\res_{\sfA}} \otimes \bra{x_{\sfA}}] V^{x_{\sfA}}_{\sfA}
	\end{align}
	
	holds  on the subspace $\up{supp}(\varrho_{\sfA})$ of $\cH_{\sfA}$, for all $x_{\sfA} \in X_{\sfA}$. (Here, $\varrho_{\sfA}$ is the $\sfA$-marginal of $\varrho$.)  
	
By summing over $y \in Y$ in condition \eqref{eq:qdependent}, we obtain the identity 
	
	\begin{align}  \label{eq:sumovera}
[	V^x  \otimes \bone_{\cP}] \ket{\psi} =  \tilde{\ket{\psi}} \otimes \ket{\psi^\res} \otimes \ket{x}
	\end{align}

for all $x \in X$. Now, let

	\begin{align}
	\ket{\psi} = \sum_{j=1}^r \sqrt{p(j)} \ket{\psi_{\sfA}(j)} \otimes \ket{\psi_{\neg \sfA}(j)} 
	\end{align}
	
	be a Schmidt decomposition of $\ket{\psi}$ relative to the factorisation $\cH_{\sfA} \otimes \cH_{\neg {\sfA}}$ where $\cH_{\neg \sfA} = \cH_\sfB \otimes \cP$, with $p(1), \ldots, p(r)> 0$. We then have by \cref{eq:sumovera}, for $x =(x_\sfA, x_\sfB)\in X$,  

	\begin{align} \label{eq:qi0collapsed}
	\begin{split}
	\tilde{\ket{\psi}} \otimes \ket{\psi^\res} \otimes \ket{x_\sfB} & = [\bone_{\tilde{\cH} \otimes \cH^\res \otimes \cP \otimes\cX_{\sfB}} \otimes \bra{x_{\sfA}}] [V^x \otimes \bone_\cP]\ket{\psi} \\ &=  \sum_{j=1}^r \sqrt{p(j)} [\bone_{\tilde{\cH}_{\sfA} \otimes \cH^\res_{\sfA}} \otimes \bra{x_{\sfA}}] V^{x_{\sfA}}_{\sfA} \ket{\psi_{\sfA}(j)} \otimes [V^{x_{\sfB}}_{\sfB}  \otimes \bone_\cP] \ket{\psi_{\neg \sfA}(j)} .
	\end{split}
	\end{align}

	The left hand side of \cref{eq:qi0collapsed} is independent of $x_{\sfA}$. The isometry $V^{x_\sfB}_\sfB$ is also independent of $x_{\sfA}$, whence the orthonormal system $\big( [V^{x_{\sfB}}_{\sfB}  \otimes \bone_\cP]\ket{\psi_{\neg \sfA}}(j)\big) _{j=1, \ldots, r}$ is independent of $x_{\sfA}$. Therefore,

	\begin{align}
	[\bone_{\tilde{\cH}_{\sfA} \otimes \cH^\res_{\sfA}} \otimes \bra{x_{\sfA}}] V^{x_{\sfA}}_{\sfA} \ket{\psi_{\sfA}(j)}
	\end{align}
	
	must be independent of $x_\sfA$, for all $j=1, \ldots, r$. (In this step we would loose something in an approximate argument, when dividing by $\sqrt{p(j)}$.) Observing that 
	
	\begin{align}
	\up{span}\{\ket{\psi_{\sfA}(j)} \mid j=1, \ldots, r  \} = \up{supp}(\varrho_{\sfA}),
	\end{align}
	
	we conclude that the operator $[\bone_{\tilde{\cH}_{\sfA} \otimes \cH^\res_{\sfA}} \otimes \bra{x_{\sfA}}] V^{x_{\sfA}}_{\sfA}$ restricted to $\up{supp}(\varrho_\sfA)$ is independent of $x_{\sfA}$. Moreover, it acts isometrically: 
	
	 By the \cref{eq:qi0collapsed}, 
	
	\begin{align}
	1= \norm*{\tilde{\ket{\psi}} \otimes \ket{\psi^\res} \otimes \ket{x_{\sfB}}}^2 = \sum_{j=1}^r p(j) \, \norm*{[\bone_{\tilde{\cH}_{\sfA} \otimes \cH^\res_{\sfA}} \otimes \bra{x_{\sfA}}] V^{x_{\sfA}}_{\sfA} \ket{\psi_{\sfA}(j)}}^2, 
	\end{align}
	
and by extremality of $1$ in the convex set $[0,1]$ this forces $\norm*{[\bone_{\tilde{\cH}_{\sfA} \otimes \cH^\res_{\sfA}} \otimes \bra{x_{\sfA}}] V^{x_{\sfA}}_{\sfA} \ket{\psi_{\sfA}(j)}}=1$ for all $j=1, \ldots, r$ (here we would lose again in the approximate version). It follows that

	\begin{align}
	\norm*{[\bone_{\tilde{\cH}_{\sfA} \otimes \cH^\res_{\sfA}} \otimes \ketbra{x_{\sfA}}] V^{x_{\sfA}}_{\sfA} \ket{\psi_{\sfA}(j)}}=1,
	\end{align}
	
	and since $\bone_{\tilde{\cH}_{\sfA} \otimes \cH^\res_{\sfA}} \otimes \ketbra{x_{\sfA}}$ is a projection and $V^{x_{\sfA}}_{\sfA} \ket{\psi_{\sfA}(j)}$ is a unit vector, we must therefore have 
	
	\begin{align}
	[\bone_{\tilde{\cH}_{\sfA} \otimes \cH^\res_{\sfA}} \otimes \ketbra{x_{\sfA}}] V^{x_{\sfA}}_{\sfA} \ket{\psi_{\sfA}(j)} =  V^{x_{\sfA}}_{\sfA} \ket{\psi_{\sfA}(j)},
	\end{align}
	
	so that 
	
	\begin{align}
	(V^{x_{\sfA}}_{\sfA})^* [\bone_{\tilde{\cH}_{\sfA} \otimes \cH^\res_{\sfA}} \otimes \ketbra{x_{\sfA}}] V^{x_{\sfA}}_{\sfA} \ket{\psi_{\sfA}(j)} =   \ket{\psi_{\sfA}(j)}
	\end{align}

for all $j=1, \ldots, r$, which is precisely the statement that 	
$[\bone_{\tilde{\cH}_{\sfA} \otimes \cH^\res_{\sfA}} \otimes \bra{x_{\sfA}}] V^{x_{\sfA}}_{\sfA}$ acts isometrically on $\up{supp}(\varrho_\sfA)$.

Altogether, we conclude as desired the existence of an isometry $W_{\sfA} : \cH_{\sfA} \into \tilde{\cH}_{\sfA} \otimes \cH^\res_{\sfA}$ such that $W_{\sfA} = [\bone_{\tilde{\cH}_{\sfA} \otimes \cH^\res_{\sfA}} \otimes \bra{x_{\sfA}}] V^{x_{\sfA}}_{\sfA}$ on $\up{supp}(\varrho_\sfA)$, for any $x_\sfA \in X_\sfA$. 
 
 By the exact same argument, we find an $x_\sfB$-independent isometry $W_\sfB : \cH_\sfB \to \tilde{\cH}_\sfB \otimes \cH^\res_\sfB$ such that  $W_{\sfB} = [\bone_{\tilde{\cH}_{\sfB} \otimes \cH^\res_{\sfB}} \otimes \bra{x_{\sfB}}] V^{x_{\sfB}}_{\sfB}$ on $\supp{\varrho_\sfB}$ for any $x_\sfB \in X_\sfB$.\\

	Next, we argue that the operator identity $W_i = [\bone_{\tilde{\cH}_i \otimes \cH^\res_i} \otimes \bra{x_i}] V^{x_i}_i$ holds not only on $\up{supp}(\varrho_i) = \up{span}\{\ket{\psi_i(j)} \mid j= 1, \ldots, r \}$, but in fact on the a priori larger space 
	
	\begin{align} \label{eq:subspace}
	\begin{split}
	&\up{span}\{\Pi^{x_i}_i(y_i)\ket{\psi_i(j)} \mid  j=1 , \ldots r, \;  x_i \in X_i, \; y_i \in Y_i\} \\ 
	&(= \up{span}\{ \Pi^{x_i}_i(y_i) \ket{\phi_i} \mid \ket{\phi_i} \in \up{supp}(\varrho_i), \; x_i \in X_i, \; y_i \in Y_i\}) .
	\end{split}
	\end{align}  
	
	This will imply, by \eqref{eq:qdependent}, that the local isometry $W := W_\sfA \otimes W_\sfB$ witnesses condition \eqref{eq:qindependent}, and thus altogether finish the proof. To demonstrate the assertion, we show that -- under the assumption that the marginal $\tilde{\varrho}_i$ of $\tilde{\psi}$ has full rank -- the subspace \eqref{eq:subspace} in fact coincides with $\up{supp}(\varrho_i)$.\footnote{It may be observed from the argument that this conclusion relies only on the weaker circumstance that $\up{supp}(\tilde{\varrho}_i)$ is \myuline{invariant} under the operators $(\tilde{\Pi}^{x_i}_i(y_i))_{y_i \in Y_i, x_i \in X_i}$, and thus we could alternatively have stated the lemma under this weaker assumption.}
	
	By rewriting \cref{eq:sumovera} in terms of density matrices (i.e. $[V^x \otimes \bone_\cP] \ketbra{\psi} [{V^x} \otimes \bone_\cP]^* =  \ketbra*{\tilde{\psi}}{\tilde{\psi}} \otimes \ketbra{\psi^\res }\otimes \ketbra{x}$) and marginalising to site $i \in \{\sfA, \sfB\}$, we obtain 
	
	\begin{align}
	V^{x_i}_i  \varrho_i (V^{x_i}_i)^* = \tilde{\varrho_i} \otimes \varrho^\res_i \otimes \ketbra{x_i},
	\end{align}
	
	which implies (since $(V^{x_i}_i)^*$ is surjective) the following equality of subspaces:
	
	\begin{align} \label{eq:rangeofstates}
	V^{x_i}_i [\up{supp}(\varrho_i)] = \up{supp}(\tilde{\varrho_i}) \otimes \up{supp}(\varrho^\res_i) \otimes \up{span}\{\ket{x_i}\}.
	\end{align}
	
	By a similar rewriting, the condition \eqref{eq:qdependent} implies that 
	
	\begin{align} 
	V^{x_i}_i  \Pi^{x_i}_i(y_i) \varrho_i   {V^{x_i}_i}^* =  \tilde{\Pi}^{x_i}_i(y_i)\tilde{\varrho_i}  \otimes \varrho^\res_i \otimes \ketbra{x_i},
	\end{align} 
	
	where this time, before marginalising to $i$, the ket side was obtained from \eqref{eq:qdependent}  by summing only over the elements of $Y_{\neg i}$, whereas the bra side was obtained by summing over all elements of $Y$. This analogously implies the following equality of subspaces:\footnote{\cref{eq:rangeofstates} actually follows from \cref{eq:rangeofprojections} by summing over $y_i \in Y_i$, but a direct argument for \eqref{eq:rangeofstates} was given first for the sake of transparency.}

	\begin{align} \label{eq:rangeofprojections}
	V^{x_i}_i  \Pi^{x_i}_i(y_i) [\up{supp}(\varrho_i)] =  \tilde{\Pi}^{x_i}_i(y_i)[ \up{supp}(\tilde{\varrho_i})]  \otimes \up{supp}(\varrho^\res_i) \otimes \up{span}\{\ket{x_i}\}.
	\end{align}
	
Now, by the full-rank assumption, $ \tilde{\Pi}^{x_i}_i(y_i)[ \up{supp}(\tilde{\varrho_i})] \subseteq  \up{supp}(\tilde{\varrho_i})$, and since $V^{x_i}_i$ has a left-inverse (indeed, it is an isometry), we thus conclude from \eqref{eq:rangeofstates} and \eqref{eq:rangeofprojections} that $\Pi^{x_i}_i(y_i) [\up{supp}(\varrho_i)] \subseteq \up{supp}(\varrho_i)$. This proves the desired assertion, and ultimately finishes the proof of the lemma. 
\end{proof}

\section{Some Implications of a Reformulation}
\label{sec:Advantages}

In this final section, I will highlight how some of the (very few) general results which are known about quantum self-testing can be easily and transparently proved using the equivalent formulation provided by \cref{thm:Bridge}. 

 The first such result is that a behaviour $P$ which self-tests some strategy is necessarily extremal in the convex set of all (tensor-product) quantum behaviours. The second and third result assert that if $P$ self-tests the strategy $(\tilde{\psi}, \tilde{\Pi}_\sfA, \tilde{\Pi}_\sfB)$, then for any strategy $(\varrho, \Pi_\sfA, \Pi_\sfB)$ with behaviour $P$, the state $\tilde{\psi}$ can be locally extracted from the state $\varrho$, and the measurement ensemble $\tilde{\Lambda}_i$ can be extracted from the  measurement ensemble $\Lambda_i$ on the local support of the state $\varrho$ (in a suitable sense). This third result clarifies the connection to other notions of self-testing of measurements, cf. Def. 3 in Ref. \cite{SB19}, as e.g. employed in Ref. \cite{Kan17}.

\subsection{Rigidity, Security and Extremality}
\label{subsec:SecExt}

It has for some time been part of the folklore in quantum self-testing that a necessary condition for $P$ to self-test a strategy is extremality of $P$ in the convex set of all behaviours realisable by quantum strategies. A formal general proof of this, however, appears to have been first given in Ref. \cite{Goh18}. As observed earlier in Ref. \cite{FFW11}, extremality in the set of quantum behaviours is equivalent to a natural notion of \emph{security}, and thus a valid way to show the implication is by demonstrating that self-testing implies security in this sense. This implication can be seen as a precise version of the remarkable insight of Ekert's in Ref. \cite{Ek91}, which was mentioned already in the general introduction to the thesis.

Below, I will give a definition of the security notion in the language of dilations, prove that it implies extremality, and then prove that self-testing implies security. This yields altogether a different proof for the necessity of extremality, and also directly the proof for security, with the interpretation that the randomness generated in a Bell-experiment is genuinely random, in the sense of being unknowable before the inputs to the channel are provided. \\

Recall from \cref{ex:Acausal} the notion of \emph{acausal side-information.}

\begin{Definition} (Security.) \\
Let $\bfD$  be a class of causal dilations of a causal channel $\scalemyQ{.8}{0.7}{0.5}{& \push{\cX_\sfA} \qw & \multigate{1}{(P, \scrC)} & \push{\cY_\sfA} \qw & \qw \\ & \push{\cX_\sfB} \qw & \ghost{(P, \scrC)} & \push{\cY_\sfB} \qw & \qw }$. We say that $(P, \scrC)$ is \emph{secure (against acausal side-information) relative to $\bfD$} if any acausal dilation $\scalemyQ{.8}{0.7}{0.5}{& \push{\cX_\sfA} \qw & \multigate{2}{(L, \scrE)} & \push{\cY_\sfA} \qw & \qw \\ & \push{\cX_\sfB} \qw & \ghost{(L, \scrE)} & \push{\cY_\sfB} \qw & \qw  \\ &\push{\bbD} \ww & \Nghost{(L, \scrE)}{\ww} & \push{\bbE} \ww & \ww}$ in $\bfD$ is trivial, i.e. factors as  $\scalemyQ{.8}{0.7}{0.5}{& \push{\cX_\sfA} \qw & \multigate{1}{(P, \scrC)} & \push{\cY_\sfA} \qw & \qw \\ & \push{\cX_\sfB} \qw & \ghost{(P, \scrC)} & \push{\cY_\sfB} \qw & \qw \\&\push{\bbD} \ww & \Ngate{(S, \scrD)}{\ww} & \push{\bbE} \ww & \ww}$ for some causal channel $(S, \scrD)$. 
\end{Definition}

The reader should have in mind the case where $\bfD$ is class of classically bound dilations; security then means that any reasonable side-information about the channel, which is present before the inputs have been supplied, must be completely independent of the channel (`reasonable' referring to classical boundedness). In particular, any randomness generated by the channel $(P, \scrC)$ is \emph{fresh}, in the sense that it must be uncorrelated with pre-existing randomness.  \\

We have the following: 

\begin{Prop} (Self-Testing implies Security.) \label{prop:SecSelfTest}\\
If $\scalemyQ{.8}{0.7}{0.5}{& \push{\cX_\sfA} \qw & \multigate{1}{(P, \scrC)} & \push{\cY_\sfA} \qw & \qw \\ & \push{\cX_\sfB} \qw & \ghost{(P, \scrC)} & \push{\cY_\sfB} \qw & \qw }$ is a Bell-channel in $\QIT$ for which $P$ self-tests some quantum strategy, then $(P, \scrC)$ is secure relative to classically bound dilations. 
	\end{Prop}

\begin{proof}
By \cref{cor:RigidSelftest}, $(P, \scrC)$  has a complete dilation for classically bound dilations, with no acausal side-information. The statement should be intuitively clear from this circumstance, but let us prove it explicitly: 

Let $\scalemyQ{.8}{0.7}{0.5}{& \push{\cX_\sfA} \qw & \multigate{2}{(K, \scrF)} & \push{\cY_\sfA} \qw & \qw \\ & \push{\cX_\sfB} \qw & \ghost{(K, \scrF)} & \push{\cY_\sfB} \qw & \qw  \\ & & \Nghost{(K, \scrF)} & \push{\bbE_0} \ww & \ww} $ be a dilation of $(P, \scrC)$ which is complete for classically bound dilations and which has no acausal side-information. By \cref{lem:DerSimpl}, any classically bound dilation is thus of the form  

\begin{align} \label{eq:AcDil}
\myQ{0.7}{0.5}{& \push{\cX_\sfA} \qw & \multigate{2}{(K, \scrF)} & \push{\cY_\sfA} \qw & \qw \\ & \push{\cX_\sfB} \qw & \ghost{(K, \scrF)} & \push{\cY_\sfB} \qw & \qw  \\ & & \Nghost{(K, \scrF)} & \push{\bbE_0} \ww & \Nmultigate{1}{(G, \scrB)}{\ww} \\ & & & \push{\bbD} \ww & \Nghost{(G, \scrB)}{\ww} & \push{\bbE} \ww & \ww} 
\end{align}

for some causal channel $(G, \scrB)$. If the dilation \eqref{eq:AcDil} is acausal, which is to say that $\scrF(\ports{\bbE_0} \cap \scrB(\ports{\bbE}))= \emptyset$, then, since $\scrF(\sfe_0) \neq \emptyset$ for all $\sfe_0 \in \ports{\bbE_0}$, we must have $\ports{\bbE_0} \cap \scrB(\ports{\bbE}) = \emptyset$, i.e. $\scrB(\ports{\bbE}) \subseteq  \ports{\bbD}$. This however implies that $(G, \scrB)$ factors as $\scalemyQ{.8}{0.7}{0.5}{& \push{\bbE_0} \ww & \Ngate{\tr}{\ww} \\ & \push{\bbD} \ww & \Ngate{(S, \scrD)}{\ww} & \push{\bbE} \ww & \ww}$, and the desired follows.	\end{proof}

\begin{Remark} (On the Significance of Security.)\\
As demonstrated by the proof, most of this statement has nothing to do with self-testing per se; it is true in \myuline{any} theory that the existence of a complete dilation with no acausal side-information implies security. What makes the statement \emph{interesting} in the context of quantum self-testing, is that a quantum behaviour can give random outputs and still have a complete dilation with no acausal side-information. 

This is not true, for example, in $\CIT$: We have seen that some Bell-channels in $\CIT$ have complete dilations (cf. \cref{ex:BipartRigidCIT} and the examples following \cref{cor:UniRigidCIT}), but any such causal channel which has a complete dilation with \myuline{no} acausal side-information must be a deterministic channel. Indeed, by a consideration as the one in \cref{ex:Acausal} (or as in \cref{prop:SecExt} below), such a channel is extremal. 
\end{Remark}

\begin{Prop} (Security and Extremality.) \label{prop:SecExt}\\
Let	$\scalemyQ{.8}{0.7}{0.5}{& \push{\cX_\sfA} \qw & \multigate{1}{(P, \scrC)} & \push{\cY_\sfA} \qw & \qw \\ & \push{\cX_\sfB} \qw & \ghost{(P, \scrC)} & \push{\cY_\sfB} \qw & \qw }$ be a Bell-channel in $\QIT$. If $(P, \scrC)$ is secure relative to classically bound dilations, then $P$ is extremal in the convex set of quantum behaviours from $\cX$ to $\cY$. 
	\end{Prop}

\begin{proof}
	First, observe that, by density of one-sided dilations (due to the constructibility assumption), $(P, \scrC)$ is secure if and only if \myuline{one-sided} acausal dilations are trivial. What we will do is to establish a correspondence  between the classically bound one-sided acausal dilations of $(P, \scrC)$ and the convex decompositions of $P$ into quantum behaviours. Conceptually, this correspondence is analogous to that in the case of $\CIT$ as expressed by \cref{thm:UniDensePrel} or \cref{thm:UniBell}. 
	
	More precisely, given a convex decomposition $P = \sum_{k=1}^n s_k P_k$ with $P_k$ a quantum behaviour and $s_1, \ldots, s_n > 0$, we may write $\scalemyQ{.8}{0.7}{0.5}{& \push{\cX_\sfA} \qw & \multigate{1}{P} & \push{\cY_\sfA} \qw & \qw \\ & \push{\cX_\sfB} \qw & \ghost{P} & \push{\cY_\sfB} \qw & \qw } = \scalemyQ{.8}{0.7}{0.5}{& \push{\cX_\sfA} \qw & \multigate{2}{\check{P}} & \push{\cY_\sfA} \qw & \qw \\ & \push{\cX_\sfB} \qw & \ghost{\check{\scrC}} & \push{\cY_\sfB} \qw & \qw \\ & \Ngate{\sigma} & \ghost{\check{P}} } $, where $\sigma$ is the state  $\sum_{k=1}^n s_k \ketbra{k}$ on $\C^n$ and $\check{P}$ the channel that reads the value $k$ in the $\C^n$-register and yields $P_k$ accordingly. Giving $\check{P}$ the obvious causal specification $\check{\scrC}$, this identity moreover holds between causal channels as well. Also, $(\check{P},\check{\scrC})$ is constructible since each $P_k$ is constructible. Letting $\scalemyQ{.8}{0.7}{0.5}{& \Nmultigate{1}{\overline{\sigma}} & \push{\C^n} \qw & \qw \\ & \Nghost{\overline{\sigma}} & \push{\C^n} \ww & \ww}$ denote the classical copy of $\sigma$, the channel 
	
	\begin{align} \label{eq:AcDilation}
\myQ{0.7}{0.5}{& \push{\cX_\sfA} \qw & \multigate{2}{(\check{P},\check{\scrC})} & \push{\cY_\sfA} \qw & \qw \\ & \push{\cX_\sfB} \qw & \ghost{(\check{P},\check{\scrC})} & \push{\cY_\sfB} \qw & \qw \\ & \Nmultigate{1}{\overline{\sigma}} & \ghost{(\check{P},\check{\scrC})}  \\ & \Nghost{\overline{\sigma}}  &  \push{\C^n} \ww & \ww & \ww }
	\end{align}
	
	is evidently a classically bound acausal dilation\footnote{Recall the tacit assumption that we only consider constructible dilations -- therefore, it is important for the argument that $(\check{P},\check{\scrC})$ is constructible.} of $(P, \scrC)$. It moreover encodes the initial convex decomposition of $P$, as it is given by $\sum_{k=1}^n s_k P_k \otimes \ketbra{k}$. Now, if $(P, \scrC)$ is secure relative to classically bound dilations, then the acausal dilation \eqref{eq:AcDilation} must factor, and this implies that $P_1=P_2 = \ldots = P_n$. Hence, $P$ is extremal. 	\end{proof}

\cref{prop:SecSelfTest} and \cref{prop:SecExt} together imply that if $P$ self-tests some strategy (i.e. is rigid a.t.t.l.), then $P$ is necessarily extremal. 

\begin{OP} (Extremality versus Rigidity) \\
Does every extremal quantum behaviour self-test some quantum strategy? 
\end{OP}

\subsection{Extraction of the Canonical State}

It is well-known that the usual definition of quantum self-testing implies that the state $\tilde{\psi}$ of a strategy $(\tilde{\psi},\tilde{\Pi}_\sfA, \tilde{\Pi}_\sfB)$ which is self-tested by $P$ can be locally extracted from the state $\varrho$ of any strategy $(\varrho,\Pi_\sfA, \Pi_\sfB)$ with behaviour $P$. Indeed, if in the reducibility condition

	\begin{align}
[	W _\sfA \Pi^{x_\sfA}_\sfA(y_\sfA) \otimes W_\sfB \Pi^{x_\sfB}_\sfB(y_\sfB)   \otimes \bone_\cP ] \ket{\psi} = [ \tilde{\Pi}^{x_\sfA}_\sfA(y_\sfA) \otimes  \tilde{\Pi}^{x_\sfB}_\sfB(y_\sfB)] \tilde{\ket{\psi}} \otimes \ket{\psi^\res} 
\end{align}

we sum over $y_\sfA \in Y_\sfA$ and $y_\sfB \in Y_\sfB$ (for any fixed $x_\sfA, x_\sfB$), then we obtain

	\begin{align}
[	W _\sfA  \otimes W_\sfB   \otimes \bone_\cP ] \ket{\psi} = \tilde{\ket{\psi}} \otimes \ket{\psi^\res} , 
\end{align}

or, more compactly,

	\begin{align}
[W  \otimes \bone_\cP ] \psi [W^* \otimes \bone_\cP ]= \tilde{\psi} \otimes \psi^\res , 
\end{align}

with $W= W_\sfA \otimes W_\sfB$. In pictures, this last condition reads 

\begin{align} \label{eq:StateExtractPure}
\myQ{0.7}{0.5}{& \Nmultigate{4}{\psi} & \push{\cH_\sfA} \qw & \multigate{1}{\Xi_\sfA} & \push{\tilde{\cH}_\sfA} \qw & \qw \\ & \Nghost{\psi} & & \Nghost{\Xi_\sfA} & \push{\cH^\res_\sfA} \qw & \qw \\ & \Nghost{\psi} & \qw & \push{\cP} \qw & \qw & \qw \\& \Nghost{\psi} & & \Nmultigate{1}{\Xi_\sfB} & \push{\cH^\res_\sfB} \qw & \qw\\ & \Nghost{\psi} & \push{\cH_\sfB} \qw & \ghost{\Xi_\sfB} & \push{\tilde{\cH}_\sfB} \qw & \qw} = \myQ{0.7}{0.5}{& \Nmultigate{4}{\tilde{\psi}} & \push{\tilde{\cH}_\sfA} \qw & \qw \\& \Nghost{\tilde{\psi}} & \Nmultigate{2}{\psi^\res} & \push{\cH^\res_\sfA} \qw & \qw  \\ & \Nghost{\tilde{\psi}}& \Nghost{\psi^\res} & \push{\cP} \qw & \qw \\ & \Nghost{\tilde{\psi}} & \Nghost{\psi^\res} & \push{\cH^\res_\sfB} \qw & \qw \\ & \Nghost{\tilde{\psi}} & \push{\tilde{\cH}_\sfB} \qw & \qw}  \quad,
\end{align}

with $\Xi_i$ the isometric channel corresponding to conjugation by $W_i$. This condition is a purified version of the condition that there exist some (not necessarily isometric) channels $M_\sfA$ and $M_\sfB$ such that

\begin{align} \label{eq:StateExtract}
\myQ{0.7}{0.5}{& \Nmultigate{1}{\varrho} & \push{\cH_\sfA} \qw & \gate{M_\sfA} & \push{\tilde{\cH}_\sfA} \qw & \qw \\& \Nghost{\varrho} & \push{\cH_\sfB} \qw & \gate{M_\sfB} & \push{\tilde{\cH}_\sfB} \qw & \qw} = \myQ{0.7}{0.5}{& \Nmultigate{1}{\tilde{\psi}} & \push{\tilde{\cH}_\sfA} \qw & \qw \\& \Nghost{\tilde{\psi}} & \push{\tilde{\cH}_\sfB} \qw & \qw} \quad,
\end{align}

asserting local extractibility of $\tilde{\psi}$ from $\varrho$. Though this argument is so simple that is hardly needs improvement, it is instructive to see how local extractibility of the canonical state $\tilde{\psi}$ is manifested in the self-testing formulation offered by \cref{thm:Bridge}. There will be no summing over $y_\sfA, y_\sfB$ in equations involving operators; another argument altogether (which would also be valid in most other theories) proves extractibility: 

\begin{Prop} (Self-Testing implies Local Extractibility of the Canonical State.) \label{prop:StateExtract}\\
Suppose that $P$ self-tests $(\tilde{\psi}, \tilde{\Pi}_\sfA, \tilde{\Pi}_\sfB)$. If $(\varrho, \Pi_\sfA, \Pi_\sfB)$ is any strategy with behaviour $P$, then there exist channels $M_\sfA$ and $M_\sfB$ such that \eqref{eq:StateExtract} holds.
	\end{Prop}

\begin{proof}
	By \cref{thm:Bridge}, there are channels $\Gamma_i$ such that
	
		\begin{align} 
	\myQ{0.7}{0.5}{
		& \push{\cX_\sfA}  \qw  & \multigate{1}{ \hat{\Lambda}_\sfA} & \push{\cY_\sfA}  \qw & \qw  \\
		& \Nmultigate{2}{\varrho}  & \ghost{\hat{\Lambda}_\sfA} &  \Ngate{\Gamma_\sfA}{\ww} & \ww\\
		& \Nghost{\varrho}   \\
		& \Nghost{\varrho} & \multigate{1}{\hat{\Lambda}_\sfB} & \Ngate{\Gamma_\sfB}{\ww} & \ww \\
		&   \push{\cX_\sfB} \qw  & \ghost{\hat{\Lambda}_\sfB} & \push{\cY_\sfB}  \qw & \qw  
	} 
	\quad =  \quad 
	\myQ{0.7}{0.5}{
		& \push{\cX_\sfA}  \qw  & \multigate{1}{ {\hat{\tilde{\Lambda}}}_\sfA} & \push{\cY_\sfA}  \qw & \qw  \\
		& \Nmultigate{2}{\tilde{\psi}}  & \ghost{{\hat{\tilde{\Lambda}}}_\sfA} &  \ww\\
		& \Nghost{\tilde{\psi}}  \\
		& \Nghost{\tilde{\psi}} & \multigate{1}{{\hat{\tilde{\Lambda}}}_\sfB} & \ww  \\
		&   \push{\cX_\sfB} \qw  & \ghost{{\hat{\tilde{\Lambda}}}_\sfB} & \push{\cY_\sfB}  \qw & \qw  
	}   \quad.
	\end{align}
	
	By applying inverses to the reversible channels $\hat{\tilde{\Lambda}}_i$ on both sides, we obtain
	
		\begin{align}
	\myQ{0.7}{0.5}{
		& \push{\cX_\sfA}  \qw  & \multigate{1}{ M'_\sfA} & \push{\cX_\sfA}  \qw & \qw  \\
		& \Nmultigate{1}{\varrho}  & \ghost{M'_\sfA} &  \push{\tilde{\cH}_\sfA} \qw & \qw \\
		& \Nghost{\varrho} & \multigate{1}{M'_\sfB} & \push{\tilde{\cH}_\sfB} \qw & \qw\\
		&   \push{\cX_\sfB} \qw  & \ghost{M'_\sfB} & \push{\cY_\sfB}  \qw & \qw  
	} 
	\quad =  \quad 
 \myQ{0.7}{0.5}{& \qw & \push{\cX_\sfA} \qw & \qw \\& \Nmultigate{1}{\tilde{\psi}} & \push{\tilde{\cH}_\sfA} \qw & \qw \\& \Nghost{\tilde{\psi}} & \push{\tilde{\cH}_\sfB} \qw & \qw\\& \qw & \push{\cX_\sfB} \qw & \qw}  \quad ,
	\end{align}
	
for some channels $M'_i$. The desired then follows by inserting arbitrary states into the systems $\cX_i$ and trashing them. 
	\end{proof}

\subsection{Extraction of Canonical Measurements}

Finally, let us prove that self-testing implies local extractibility of the canonical measurement ensembles. This connects to alternative formulations of self-testing (e.g. as in Ref. \cite{Kan17} in the case of the `tilted CHSH-behaviour'), but as far as I know the general statement below has not been presented before. The proof provided here is pictorial, in the same style as the one above for local extractibility of the state, and in principle it generalises beyond $\QIT$ to other purifiable theories.

\begin{Prop} (Self-Testing implies Local Extractibility of Canonical Measurements.)\label{prop:MeasExtract}\\
Suppose that $P$ self-tests $(\tilde{\psi}, \tilde{\Pi}_\sfA, \tilde{\Pi}_\sfB)$, and assume that $\tilde{\psi}$ has locally full rank. If $(\varrho, \Pi_\sfA, \Pi_\sfB)$ is any strategy with behaviour $P$, then there exist channels $N_\sfA$ and $N_\sfB$ such that 

\begin{align}
\myQ{0.7}{0.5}{
	& \qw	& \push{\cX_i} \qw    &  \qw &  \multigate{1}{\Lambda_i}  &   \qw & \push{\cY_i} \qw &  \qw  \\
	& \push{\tilde{\cH}_i}\qw &   \gate{N_i} & \push{\cH_i} \qw &  \ghost{\Lambda_i}  
} =
\myQ{0.7}{0.5}{
	&  \push{\cX_i}   \qw  & \multigate{1}{\tilde{\Lambda}_i}  &   \qw &  \push{\cY_i} \qw &  \qw  & \qw \\
	& \push{\tilde{\cH}_i}\qw & \ghost{\tilde{\Lambda}_i} 
} \quad
	\end{align}
	
	for $i = \sfA, \sfB$, where $\Lambda_i$ and $\tilde{\Lambda}_i$ are the measurement ensembles corresponding to $(\Pi^{x_i}_i)_{x_i \in X_i}$ and $(\tilde{\Pi}^{x_i}_i)_{x_i \in X_i}$, respectively. 
	
	\end{Prop}

\begin{Remark} (Generalisation to the Approximate Case.)\\
	As observed in the proof, approximate versions of this result (e.g. in terms of the diamond-distance) will generally lose factors, but the loss depends only on the Schmidt coefficients of the fixed and known state $\tilde{\psi}$. 
	\end{Remark}

\begin{proof}
	
It is enough to show the existence of $N_\sfA$. We know that \eqref{eq:locder} holds for some channels $\Gamma_\sfA$, $\Gamma_\sfB$. We can actually prove the desired from the weaker circumstance that occurs after purifying and moving channels to the other side, namely from the assumption that

\begin{align} \label{eq:asssim}
\myQ{0.7}{0.5}{
	& \push{\cX_\sfA} \qw  & \multigate{1}{\hat{\Lambda}_\sfA}  &   \qw & \push{\cY_\sfA} \qw &  \qw  \\
	& \Nmultigate{2}{\psi}   & \ghost{\hat{\Lambda}_\sfA} &  \ww &  \ww   \\
	& \Nghost{\psi}  &   \ww & \ww &  \ww \\
	& \Nghost{\psi}  & \multigate{1}{\hat{\Lambda}_\sfB} & \ww  & \ww   \\ 
	&   \push{\cX_\sfB} \qw  & \ghost{\hat{\Lambda}_\sfB}  &   \qw & \push{\cY_\sfB} \qw &   \qw  \\ 
}
\quad =
\quad
\myQ{0.7}{0.5}{
	&  \push{\cX_\sfA}   \qw  & \multigate{1}{\hat{\tilde{\Lambda}}_\sfA}  &   \qw &  \push{\cY_\sfA} \qw &  \qw  & \qw \\
	& \Nmultigate{4}{\tilde{\psi}}  & \ghost{\hat{\tilde{\Lambda}}_\sfA} &  \ww &  \Nmultigate{1}{\tilde{\Gamma}_\sfA}{\ww} & \ww  \\
	& \Nghost{\tilde{\psi}}   &  & \Nmultigate{2}{\psi^\res} & \Nghost{\tilde{\Gamma}_\sfA}{\dw} &  \\  
	& \Nghost{\tilde{\psi}}  &   & \Nghost{\psi^\res}    & \ww & \ww \\
	& \Nghost{\tilde{\psi}}  &  & \Nghost{\psi^\res} &   \Nmultigate{1}{\tilde{\Gamma}_\sfB}{\dw}   \\
	& \Nghost{\tilde{\psi}}   & \multigate{1}{\hat{\tilde{\Lambda}}_\sfB} &  \ww & \Nghost{\tilde{\Gamma}_\sfB}{\ww} & \ww  \\ 
	&  \push{\cX_\sfB}  \qw  & \ghost{\hat{\tilde{\Lambda}}_\sfB}   & \qw &  \push{\cY_\sfB} \qw &   \qw & \qw \\ 
}
\quad 
\end{align}

for some state $\psi^\res$ and some channels $\tilde{\Gamma}_\sfA$ and $\tilde{\Gamma}_\sfB$.

 We may assume without loss of generality that $\tilde{\Gamma}_\sfB$ is reversible. (If not, we simply replace it by a pure reversible dilation, absorb the pure state arising on the left hand side into $\psi$, noting that the modified dilated measurement channel $\hat{\Lambda}_\sfB \og \id$ is a Stinespring dilation of $\Lambda_\sfB \otimes \tr$, which is extractibility-equivalent to $\Lambda_\sfB$.)

Under this assumption, first observe that by applying a left-inverse $\hat{\Lambda}^{-}_\sfA$ to $\hat{\Lambda}_\sfA$ on both sides, and then inserting an arbitrary state on $\cX_\sfA$ and trashing $\cX_\sfA$, yields 

\begin{align} 
\myQ{0.7}{0.5}{
	& \Nmultigate{2}{\psi}   & \qw &  \qw &  \qw   \\
	& \Nghost{\psi}  &   \ww & \ww &  \ww \\
	& \Nghost{\psi}  & \multigate{1}{\hat{\Lambda}_\sfB} & \ww  & \ww   \\ 
	&   \push{\cX_\sfB} \qw  & \ghost{\hat{\Lambda}_\sfB}  &   \qw & \push{\cY_\sfB} \qw &   \qw  \\ 
}
\quad =
\quad
\myQ{0.7}{0.5}{
	& \Nmultigate{4}{\tilde{\psi}}  & \qw &  \qw &   \multigate{1}{N'_\sfA} & \qw  \\
	& \Nghost{\tilde{\psi}}   &  & \Nmultigate{2}{\psi^\res} & \Nghost{N'_\sfA}{\dw} &  \\  
	& \Nghost{\tilde{\psi}}  &   & \Nghost{\psi^\res}    & \ww & \ww \\
	& \Nghost{\tilde{\psi}}  &  & \Nghost{\psi^\res} &   \Nmultigate{1}{\tilde{\Gamma}_\sfB}{\dw}   \\
	& \Nghost{\tilde{\psi}}   & \multigate{1}{\hat{\tilde{\Lambda}}_\sfB} &  \ww & \Nghost{\tilde{\Gamma}_\sfB}{\ww} & \ww  \\ 
	&  \push{\cX_\sfB}  \qw  & \ghost{\hat{\tilde{\Lambda}}_\sfB}   & \qw &  \push{\cY_\sfB} \qw &   \qw & \qw \\ 
}
\end{align}

for some channel $N'_\sfA$. By inserting this fragment back into \cref{eq:asssim}, we find 

\begin{align}
\myQ{0.7}{0.5}{
	& \qw	& \push{\cX_\sfA} \qw    & \qw & \qw &  \multigate{1}{\hat{\Lambda}_\sfA}  &   \qw & \push{\cY_\sfA} \qw &  \qw  \\
	& \Nmultigate{4}{\tilde{\psi}}  & \qw &  \qw &   \multigate{1}{N'_\sfA} & \ghost{\hat{\Lambda}_\sfA}  & \ww & \ww\\
	& \Nghost{\tilde{\psi}}   &  & \Nmultigate{2}{\psi^\res} & \Nghost{N'_\sfA}{\dw} &  \\  
	& \Nghost{\tilde{\psi}}  &   & \Nghost{\psi^\res}    & \ww & \ww \\
	& \Nghost{\tilde{\psi}}  &  & \Nghost{\psi^\res} &   \Nmultigate{1}{\tilde{\Gamma}_\sfB}{\dw}   \\
	& \Nghost{\tilde{\psi}}   & \multigate{1}{\hat{\tilde{\Lambda}}_\sfB} &  \ww & \Nghost{\tilde{\Gamma}_\sfB}{\ww} & \ww  \\ 
	&  \push{\cX_\sfB}  \qw  & \ghost{\hat{\tilde{\Lambda}}_\sfB}   & \qw &  \push{\cY_\sfB} \qw &   \qw & \qw \\ 
}
\quad =
\quad
\myQ{0.7}{0.5}{
	&  \push{\cX_\sfA}   \qw  & \multigate{1}{\hat{\tilde{\Lambda}}_\sfA}  &   \qw &  \push{\cY_\sfA} \qw &  \qw  & \qw \\
	& \Nmultigate{4}{\tilde{\psi}}  & \ghost{\hat{\tilde{\Lambda}}_\sfA} &  \ww &  \Nmultigate{1}{\tilde{\Gamma}_\sfA}{\ww} & \ww  \\
	& \Nghost{\tilde{\psi}}   &  & \Nmultigate{2}{\psi^\res} & \Nghost{\tilde{\Gamma}_\sfA}{\dw} &  \\  
	& \Nghost{\tilde{\psi}}  &   & \Nghost{\psi^\res}    & \ww & \ww \\
	& \Nghost{\tilde{\psi}}  &  & \Nghost{\psi^\res} &   \Nmultigate{1}{\tilde{\Gamma}_\sfB}{\dw}   \\
	& \Nghost{\tilde{\psi}}   & \multigate{1}{\hat{\tilde{\Lambda}}_\sfB} &  \ww & \Nghost{\tilde{\Gamma}_\sfB}{\ww} & \ww  \\ 
	&  \push{\cX_\sfB}  \qw  & \ghost{\hat{\tilde{\Lambda}}_\sfB}   & \qw &  \push{\cY_\sfB} \qw &   \qw & \qw \\ 
}
\quad .
\end{align}

Now, cancelling $\tilde{\Gamma}_\sfB$ (which we assumed reversible) and afterwards $\hat{\tilde{\Lambda}}_\sfB$, and then inserting an arbitrary state on $\cX_\sfB$ and trashing $\cX_\sfB$, yields %

\begin{align}
\myQ{0.7}{0.5}{
	& \qw	& \push{\cX_\sfA} \qw    & \qw & \qw &  \multigate{1}{\hat{\Lambda}_\sfA}  &   \qw & \push{\cY_\sfA} \qw &  \qw  \\
	& \Nmultigate{4}{\tilde{\psi}}  & \qw &  \qw &   \multigate{1}{N'_\sfA} & \ghost{\hat{\Lambda}_\sfA}  & \ww & \ww\\
	& \Nghost{\tilde{\psi}}   &  & \Nmultigate{2}{\psi^\res} & \Nghost{N'_\sfA}{\dw} &  \\  
	& \Nghost{\tilde{\psi}}  &   & \Nghost{\psi^\res}    & \ww & \ww \\
	& \Nghost{\tilde{\psi}}  &  & \Nghost{\psi^\res}  & \dw & \dw  \\
	& \Nghost{\tilde{\psi}}   & \qw & \qw & \qw 
}
\quad =
\quad
\myQ{0.7}{0.5}{
	&  \push{\cX_\sfA}   \qw  & \multigate{1}{\hat{\tilde{\Lambda}}_\sfA}  &   \qw &  \push{\cY_\sfA} \qw &  \qw  & \qw \\
	& \Nmultigate{4}{\tilde{\psi}}  & \ghost{\hat{\tilde{\Lambda}}_\sfA} &  \ww &  \Nmultigate{1}{\tilde{\Gamma}_\sfA}{\ww} & \ww  \\
	& \Nghost{\tilde{\psi}}   &  & \Nmultigate{2}{\psi^\res} & \Nghost{\tilde{\Gamma}_\sfA}{\dw} &  \\  
	& \Nghost{\tilde{\psi}}  &   & \Nghost{\psi^\res}    & \ww & \ww \\
	& \Nghost{\tilde{\psi}}  &  & \Nghost{\psi^\res}  & \dw & \dw \\
	& \Nghost{\tilde{\psi}}   & \qw & \qw & \qw  
}
\quad .
\end{align}

Trashing furthermore the two lower systems of $\psi^\res$ finally yields 

\begin{align}
\myQ{0.7}{0.5}{
	& \qw	& \push{\cX_\sfA} \qw    & \qw &  \multigate{1}{\hat{\Lambda}_\sfA}  &   \qw & \push{\cY_\sfA} \qw &  \qw  \\
	& \Nmultigate{2}{\tilde{\psi}}  & \qw &   \gate{N_\sfA} & \ghost{\hat{\Lambda}_\sfA}  & \ww & \ww\\
	& \Nghost{\tilde{\psi}}    \\  
	& \Nghost{\tilde{\psi}}   & \qw & \qw
}
\quad =
\quad
\myQ{0.7}{0.5}{
	&  \push{\cX_\sfA}   \qw  & \multigate{1}{\hat{\tilde{\Lambda}}_\sfA}  &   \qw &  \push{\cY_\sfA} \qw &  \qw  & \qw \\
	& \Nmultigate{2}{\tilde{\psi}}  & \ghost{\hat{\tilde{\Lambda}}_\sfA} &  \ww &  \Ngate{M_\sfA}{\ww} & \ww  \\
	& \Nghost{\tilde{\psi}}     \\ 
	& \Nghost{\tilde{\psi}}   & \qw & \qw \\ 
} \quad 
\end{align}

for some channels $N_\sfA$ and $M_\sfA$. Now, since $\tilde{\psi}$ has locally full rank\footnote{This is the first argument which would not be robust to errors.} it is a universal dilation of its marginal, so we must have 

\begin{align}
\myQ{0.7}{0.5}{
	& \qw	& \push{\cX_\sfA} \qw    &  \multigate{1}{\hat{\Lambda}_\sfA}  &   \qw & \push{\cY_\sfA} \qw &  \qw  \\
	& \qw &   \gate{N_\sfA} & \ghost{\hat{\Lambda}_\sfA}  & \ww & \ww
}
\quad =
\quad
\myQ{0.7}{0.5}{
	&  \push{\cX_\sfA}   \qw  & \multigate{1}{\hat{\tilde{\Lambda}}_\sfA}  &   \qw &  \push{\cY_\sfA} \qw &  \qw  & \qw \\
	& \qw & \ghost{\hat{\tilde{\Lambda}}_\sfA} &  \ww &  \Ngate{M_\sfA}{\ww} & \ww  
} \quad ,
\end{align}

which even more strongly than desired asserts that any \myuline{dilation} of $\Lambda_\sfA$ can be extracted by means of $N_\sfA$ from some dilation of $\tilde{\Lambda}_\sfA$. \end{proof}

\section{Summary and Outlook}
\label{sec:SummarySelftest}

In this chapter, we have seen a precise connection between quantum self-testing as it is traditionally envisioned and the framework of causal dilations presented in the thesis. 

Specifically, we established that quantum strategies form a dense class of \emph{classically bound} dilations (\cref{def:ClassBound}) of the behaviour channel, and we then formulated the usual reducibility condition between strategies in terms of these dilations (\cref{thm:Bridge}). By virtue of a partial collapse in the causal-dilational ordering due to purifiability of $\QIT$, we were able to relate this condition  to the existence of a complete causal dilation with no acausal side-information, along with the existence of a simple representative in the $\cder$-equivalence class of strategies (\cref{cor:RigidSelftest}). It has moreover been conjectured that such a representative always exists and that quantum self-testing can thus be recast in purely operational terms (\cref{conj:Selftesting}).

We have also seen that the causal Stinespring dilations of a quantum behaviour can be surprisingly recharacterised by non-signalling conditions (\cref{thm:StructfromNS}) and seen how this implies a recharacterisation of the set of quantum behaviours (\cref{cor:QBRechar}).\\

Many problems are left open for future work:

\begin{enumerate}

	\item Can we give a proof of the rigidity of e.g. the CHSH-behaviour \myuline{without} reference to \cref{cor:RigidSelftest}, but instead based in the language of causal dilations rather than linear operators?  
	
	\item Do the \cref{op:SimpleRep} and the \cref{conj:Selftesting} have affirmative answers? 

	\item Does the \cref{op:CausalRigidity} have an affirmative answer?

	\item What happens to the causal-dilational ordering, and rigidity by extension, if we replace the constructible scheme $\bfE = \Cns{\QIT}$ with another? Is it feasible, and is it relevant?
	
		\item Is \cref{cor:QBRechar} useful?

\end{enumerate}

It is very interesting to observe, as mentioned in \cref{rem:qapproximate}, that the relationship between quantum strategies that seems to arise from the framework of causal dilations allows local isometries which depend on the local input. Though it happens in the exact case that this dependence can be eliminated (that is the content of \cref{lem:technical}), the situation in the \myuline{approximate} case (of \emph{robust} self-testing \cite{MYS12}) may very well be different. As such, `robustness' results could in the framework of this thesis turn out to be of an entirely different quantitative character than in the usual framework. This is probably one of the most interesting problems for future research.

\chapter*{Conclusion}
\addcontentsline{toc}{chapter}{Conclusion}

The results of this thesis were motivated by the desire to recast quantum self-testing in a new, operational language. We have seen that such a recasting is viable (\cref{chap:Selftesting}) in a framework of causally structured dilations (\cref{chap:Causal}) which applies to physical theories of a very general kind (\cref{chap:Theories}). In the course of contemplating dilations systematically, we have also seen that a rather rich theory is possible based on a handful of principles pertaining only to the structure of dilations (\cref{chap:Dilations}), and considered how a metric version of this theory can be created (\cref{chap:Metric}). \\

Each chapter has already been concluded with summaries and detailed overviews of open problems for the reader to consult (\cref{sec:SummaryTheories}, \cref{sec:SummaryDilations}, \cref{sec:SummaryMetric}, \cref{sec:SummaryCausal}, \cref{sec:SummarySelftest}). 

Let me reiterate, however, what could be the three most interesting open directions for future work. \\

First, there is the problem of understanding in more detail the causal-dilational ordering. Systematic considerations (in the style of \cref{chap:Dilations}) may be possible, but it might of course also be the case that general techniques simply lack, and that its structure depends heavily on both the theory and the causal channel in question. \\

Secondly, there is the problem of extending the metric considerations of \cref{chap:Metric} to the causal setting of \cref{chap:Causal}. Possible challenges in this regard have already been mentioned in the prelude to \cref{chap:Causal}. Such an extension is not only interesting on theoretical grounds, but also because it would seem the proper way of recasting \emph{robust} self-testing in the language of causal dilations, cf. the results of \cref{chap:Selftesting}. As mentioned in \cref{sec:SummarySelftest}, this might ultimately yield robustness results of a different kind.\\

Finally, there is the curious question of whether it might be possible to derive a self-testing result not by reference to \cref{cor:RigidSelftest}, but rather by giving a proof \myuline{within} the language of causal dilations. If such a proof were to be found, which made no reference to linear  operators, it would likely yield a new and enlightening perspective on quantum self-testing. \\

One of the most general -- but also most vague -- points that surface from the work in this thesis is that causal dilations seem to formally cover the intuitive notion of `implementations' of a physical process. We saw this already in the general introduction with the bit refreshment channel (and it was formalised in \cref{subsec:DenseCIT}), and it is also the message of the link between quantum strategies and causal dilations (\cref{thm:PVMdense}). In a sense, this correspondence between implementations and causal dilations is the moral of the thesis. It is possible that there is a precise and eloquent way of expressing this connection -- or that it is even, when properly viewed, tautological -- but at this point I do not know the answer to that question.

\bibliographystyle{amsalpha}
\clearpage
\addcontentsline{toc}{chapter}{References}
\bibliography{../../BibliographyThesis}

\newcommand{\etalchar}[1]{$^{#1}$}
\providecommand{\bysame}{\leavevmode\hbox to3em{\hrulefill}\thinspace}
\providecommand{\MR}{\relax\ifhmode\unskip\space\fi MR }
\providecommand{\MRhref}[2]{%
  \href{http://www.ams.org/mathscinet-getitem?mr=#1}{#2}
}
\providecommand{\href}[2]{#2}
\begin{thebibliography}{GKW{\etalchar{+}}18}

\bibitem[AC04]{AbCo04}
Samson Abramsky and Bob Coecke, \emph{A categorical semantics of quantum
  protocols}, Proceedings of the 19th Annual IEEE Symposium on Logic in
  Computer Science, 2004, IEEE, 2004, pp.~415--425.

\bibitem[AH16]{ScientificMethod}
Hanne Andersen and Brian Hepburn, \emph{Scientific method}, The Stanford
  Encyclopedia of Philosophy (Edward~N. Zalta, ed.), Metaphysics Research Lab,
  Stanford University, summer 2016 ed., 2016.

\bibitem[Att14]{Attal14}
Stephane Attal, \emph{Lecture 6: Quantum channels}, Lecture notes available
  online at \url{http://math.univ-lyon1.fr/~attal/chapters.html}, 2014.

\bibitem[Awo10]{Awo10}
Steve Awodey, \emph{Category theory}, Oxford university press, 2010.

\bibitem[Bae06]{Baez06}
John~C. Baez, \emph{Quantum quandaries: A category-theoretic perspective}, The
  structural foundations of quantum gravity (2006), 240--265.

\bibitem[Bar07]{Barr07}
Jonathan Barrett, \emph{Information processing in generalized probabilistic
  theories}, Physical Review A \textbf{75} (2007), no.~3, 032304.

\bibitem[BB84]{Bennett84}
Charles~H. Bennett and Gilles Brassard, \emph{Quantum cryptography: Public key
  distribution and coin tossing}, Proceedings of the International Conference
  on Computers, Systems and Signal Processing, vol.~1, 1984, pp.~175--179.

\bibitem[BCF{\etalchar{+}}96]{Barn96}
Howard Barnum, Carlton~M. Caves, Christopher~A. Fuchs, Richard Jozsa, and
  Benjamin Schumacher, \emph{Noncommuting mixed states cannot be broadcast},
  Physical Review Letters \textbf{76} (1996), no.~15, 2818--2821.

\bibitem[BCP{\etalchar{+}}14]{Brun14}
Nicolas Brunner, Daniel Cavalcanti, Stefano Pironio, Valerio Scarani, and
  Stephanie Wehner, \emph{Bell nonlocality}, Reviews of Modern Physics
  \textbf{86} (2014), no.~2, 419--478.

\bibitem[BCS12]{Buhr12}
Harry Buhrman, Matthias Christandl, and Christian Schaffner, \emph{Complete
  insecurity of quantum protocols for classical two-party computation},
  Physical review letters \textbf{109} (2012), no.~16, 160501.

\bibitem[Bel64]{Bell64}
John~S. Bell, \emph{On the {Einstein} {Podolsky} {Rosen} paradox}, Physics
  Physique Fizika \textbf{1} (1964), no.~3, 195--200.

\bibitem[BF14]{Baez14}
John~C. Baez and Tobias Fritz, \emph{A bayesian characterization of relative
  entropy}, arXiv preprint arXiv:1402.3067 (2014).

\bibitem[BGNP01]{Beck01}
David Beckman, Daniel Gottesman, Michael Nielsen, and John Preskill,
  \emph{Causal and localizable quantum operations}, Physical Review A
  \textbf{64} (2001), no.~5, 052309.

\bibitem[BMR92]{BMR92}
Samuel~L. Braunstein, Ady Mann, and Michael Revzen, \emph{Maximal violation of
  {Bell} inequalities for mixed states}, Physical Review Letters \textbf{68}
  (1992), no.~22, 3259--3261.

\bibitem[Boh52]{Bohm52}
David Bohm, \emph{A suggested interpretation of the quantum theory in terms of
  "hidden" variables. {I}}, Physical review \textbf{85} (1952), no.~2,
  166--179.

\bibitem[BP07]{Braun07}
Samuel~L Braunstein and Arun~K Pati, \emph{Quantum information cannot be
  completely hidden in correlations: implications for the black-hole
  information paradox}, Physical review letters \textbf{98} (2007), no.~8,
  080502.

\bibitem[Bru14]{Bruk14}
{\v{C}}aslav Brukner, \emph{Quantum causality}, Nature Physics \textbf{10}
  (2014), no.~4, 259--263.

\bibitem[BS10]{Rosetta}
John~C. Baez and Mike Stay, \emph{Physics, topology, logic and computation: a
  {R}osetta {S}tone}, New structures for physics, Springer, 2010, pp.~95--172.

\bibitem[BT24]{BanachTarski}
Stefan Banach and Alfred Tarski, \emph{Sur la d{\'e}composition des ensembles
  de points en parties respectivement congruentes}, Fund. math \textbf{6}
  (1924), no.~1, 244--277.

\bibitem[Bur69]{Bures69}
Donald Bures, \emph{An extension of {Kakutani's} theorem on infinite product
  measures to the tensor product of semifinite {W*}-algebras}, Transactions of
  the American Mathematical Society \textbf{135} (1969), 199--212.

\bibitem[BW11]{Barn11}
Howard Barnum and Alexander Wilce, \emph{Information processing in convex
  operational theories}, Electronic Notes in Theoretical Computer Science
  \textbf{270} (2011), no.~1, 3--15.

\bibitem[BW16]{Barn16}
\bysame, \emph{Post-classical probability theory}, Quantum Theory:
  Informational Foundations and Foils, Springer, 2016, pp.~367--420.

\bibitem[Can84]{Cantor1884}
Georg Cantor, \emph{\"{U}ber unendliche, lineare {Punktmannichfaltigkeiten}},
  Mathematische Annalen \textbf{23} (1884), no.~4, 453--488.

\bibitem[CDKW16]{Coecke16Gen}
Bob Coecke, Ross Duncan, Aleks Kissinger, and Quanlong Wang, \emph{Generalised
  compositional theories and diagrammatic reasoning}, Quantum Theory:
  Informational Foundations and Foils, Springer, 2016, pp.~309--366.

\bibitem[CDP09]{Chir09combs}
Giulio Chiribella, Giacomo~Mauro D’Ariano, and Paolo Perinotti,
  \emph{Theoretical framework for quantum networks}, Physical Review A
  \textbf{80} (2009), no.~2, 022339.

\bibitem[CDP10]{Chir10}
\bysame, \emph{Probabilistic theories with purification}, Physical Review A
  \textbf{81} (2010), no.~6, 062348.

\bibitem[CDP11]{Chir11}
\bysame, \emph{Informational derivation of quantum theory}, Physical Review A
  \textbf{84} (2011), no.~1, 012311.

\bibitem[CDPV13]{Chir13}
Giulio Chiribella, Giacomo~Mauro D’Ariano, Paolo Perinotti, and Benoit
  Valiron, \emph{Quantum computations without definite causal structure},
  Physical Review A \textbf{88} (2013), no.~2, 022318.

\bibitem[CFS16]{CFS16}
Bob Coecke, Tobias Fritz, and Robert~W. Spekkens, \emph{A mathematical theory
  of resources}, Information and Computation \textbf{250} (2016), 59--86.

\bibitem[CH17]{Cunn17}
Oscar Cunningham and Chris Heunen, \emph{Purity through factorisation}, arXiv
  preprint arXiv:1705.07652 (2017).

\bibitem[Chi14a]{Chir14dilation}
Giulio Chiribella, \emph{Dilation of states and processes in
  operational-probabilistic theories}, arXiv preprint arXiv:1412.8539 (2014).

\bibitem[Chi14b]{Chir14pure}
\bysame, \emph{Distinguishability and copiability of programs in general
  process theories}, Int J Software Informatics \textbf{1} (2014), no.~2,
  209--223.

\bibitem[Cho75]{Choi75}
Man-Duen Choi, \emph{Completely positive linear maps on complex matrices},
  Linear algebra and its applications \textbf{10} (1975), no.~3, 285--290.

\bibitem[CHSH69]{CHSH69}
John~F. Clauser, Michael~A. Horne, Abner Shimony, and Richard~A. Holt,
  \emph{Proposed experiment to test local hidden-variable theories}, Physical
  review letters \textbf{23} (1969), no.~15, 880--884.

\bibitem[Chu]{Church}
\emph{\emph{The Church of the larger Hilbert space} on quantiki.org},
  \url{https://www.quantiki.org/wiki/church-larger-hilbert-space}.

\bibitem[Cir80]{Cir80}
Boris~S. Cirel'son, \emph{Quantum generalizations of {Bell}'s inequality},
  Letters in Mathematical Physics \textbf{4} (1980), no.~2, 93--100.

\bibitem[Cir93]{Cir93}
\bysame, \emph{Some results and problems on quantum {Bell}-type inequalities},
  Hadronic Journal Supplement \textbf{8} (1993), no.~4, 329--345.

\bibitem[CL13]{CoLal13}
Bob Coecke and Raymond Lal, \emph{Causal categories: Relativistically
  interacting processes}, Foundations of Physics \textbf{43} (2013), no.~4,
  458--501.

\bibitem[Coe11]{Coecke10Guises}
Bob Coecke, \emph{A universe of processes and some of its guises}, Deep beauty:
  {Understanding} the quantum world through mathematical innovation (2011),
  129--186.

\bibitem[Coe14]{Coecke14}
\bysame, \emph{Terminality implies non-signalling}, arXiv preprint
  arXiv:1405.3681 (2014).

\bibitem[CS16]{Foils}
Giulio Chiribella and Robert~W. Spekkens, \emph{{Quantum Theory}:
  {Informational Foundations} and {Foils}}, Springer, 2016.

\bibitem[CS18]{Cola18}
Andrea Coladangelo and Jalex Stark, \emph{Unconditional separation of finite
  and infinite-dimensional quantum correlations}, arXiv preprint
  arXiv:1804.05116 (2018).

\bibitem[CS20]{Cola20}
\bysame, \emph{An inherently infinite-dimensional quantum correlation}, Nature
  communications \textbf{11} (2020), no.~1, 1--6.

\bibitem[CT09]{Christ09}
Matthias Christandl and Ben Toner, \emph{Finite de {Finetti} theorem for
  conditional probability distributions describing physical theories}, Journal
  of mathematical physics \textbf{50} (2009), no.~4, 042104.

\bibitem[Deu85]{Deutsch85}
David Deutsch, \emph{Quantum theory, the {Church}--{Turing} principle and the
  universal quantum computer}, Proceedings of the Royal Society of London.
  Series A, Mathematical and Physical Sciences \textbf{400} (1985), no.~1818,
  97--117.

\bibitem[Dir81]{Dirac81}
Paul Adrien~Maurice Dirac, \emph{The principles of quantum mechanics}, no.~27,
  Oxford university press, 1981.

\bibitem[DKSW07]{Bitcomreex}
Giacomo~Mauro D’Ariano, Dennis Kretschmann, Dirk Schlingemann, and
  Reinhard~F. Werner, \emph{Reexamination of quantum bit commitment: The
  possible and the impossible}, Physical Review A \textbf{76} (2007), no.~3,
  032328.

\bibitem[dP67]{Pill67}
John de~Pillis, \emph{Linear transformations which preserve {Hermitian} and
  positive semidefinite operators}, Pacific Journal of Mathematics \textbf{23}
  (1967), no.~1, 129--137.

\bibitem[DP16]{Dyke16}
Kenneth~J. Dykema and Vern Paulsen, \emph{Synchronous correlation matrices and
  {Connes’} embedding conjecture}, Journal of Mathematical Physics
  \textbf{57} (2016), no.~1, 015214.

\bibitem[DS05]{Devetak05}
Igor Devetak and Peter~W. Shor, \emph{The capacity of a quantum channel for
  simultaneous transmission of classical and quantum information},
  Communications in Mathematical Physics \textbf{256} (2005), no.~2, 287--303.

\bibitem[Ein05]{Einstein05}
Albert Einstein, \emph{Zur {Elektrodynamik} bewegter {K{\"o}rper}}, Annalen der
  physik \textbf{4} (1905), 891--921.

\bibitem[Ein16]{Einstein16}
A~Einstein, \emph{Die {Grundlage} der allgemeinen {Relativit{\"a}tstheorie}},
  Annalen der Physik \textbf{354} (1916), no.~7, 769--822.

\bibitem[Eke91]{Ek91}
Artur~K. Ekert, \emph{Quantum cryptography based on {Bell}’s theorem},
  Physical Review Letters \textbf{67} (1991), no.~6, 661--663.

\bibitem[EML45]{Eil45}
Samuel Eilenberg and Saunders Mac~Lane, \emph{General theory of natural
  equivalences}, Transactions of the American Mathematical Society \textbf{58}
  (1945), no.~2, 231--294.

\bibitem[End01]{Enderton01}
Herbert~B. Enderton, \emph{A mathematical introduction to logic}, Elsevier,
  2001.

\bibitem[Enr]{Enriched}
\emph{\emph{Enriched Categories} on n{Lab}},
  \url{https://ncatlab.org/nlab/show/enriched+category}.

\bibitem[EPR35]{EPR35}
Albert Einstein, Boris Podolsky, and Nathan Rosen, \emph{Can quantum-mechanical
  description of physical reality be considered complete?}, Physical review
  \textbf{47} (1935), no.~10, 777--780.

\bibitem[ESW02]{Egg02}
Tilo Eggeling, Dirk Schlingemann, and Reinhard~F Werner, \emph{Semicausal
  operations are semilocalizable}, EPL (Europhysics Letters) \textbf{57}
  (2002), no.~6, 782--788.

\bibitem[Fey82]{Feynman82}
Richard~P. Feynman, \emph{Simulating physics with computers}, Int. J. Theor.
  Phys \textbf{21} (1982), no.~6/7, 467--488.

\bibitem[Fey87]{Feyn87}
\bysame, \emph{Negative probability}, Quantum implications: {Essays} in honour
  of {David Bohm} (1987), 235--248.

\bibitem[FFW11]{FFW11}
Torsten Franz, Fabian Furrer, and Reinhard~F. Werner, \emph{Extremal quantum
  correlations and cryptographic security}, Physical review letters
  \textbf{106} (2011), no.~25, 250502.

\bibitem[Fol99]{Foll99}
Gerald~B. Folland, \emph{Real analysis: Modern techniques and their
  applications}, 2 ed., vol.~40, John Wiley \& Sons, 1999.

\bibitem[Fri20]{Fritz20Synthetic}
Tobias Fritz, \emph{A synthetic approach to {Markov} kernels, conditional
  independence and theorems on sufficient statistics}, Advances in Mathematics
  \textbf{370} (2020), 107239.

\bibitem[GKW{\etalchar{+}}18]{Goh18}
Koon~Tong Goh, J{\k{e}}drzej Kaniewski, Elie Wolfe, Tam{\'a}s V{\'e}rtesi,
  Xingyao Wu, Yu~Cai, Yeong-Cherng Liang, and Valerio Scarani, \emph{Geometry
  of the set of quantum correlations}, Physical Review A \textbf{97} (2018),
  no.~2, 022104.

\bibitem[G{\"o}d29]{GodelComplete}
Kurt G{\"o}del, \emph{{\"U}ber die {Vollst{\"a}ndigkeit} des
  {Logikkalk{\"u}ls}}, Doctoral Thesis, 1929.

\bibitem[G{\"o}d31]{GodelIncomplete}
\bysame, \emph{{\"U}ber formal unentscheidbare {S{\"a}tze} der {Principia}
  {Mathematica} und verwandter {Systeme} {I}}, Monatshefte f{\"u}r {Mathematik}
  und {Physik} \textbf{38} (1931), no.~1, 173--198.

\bibitem[Gra]{GraphProducts}
\emph{Wikipedia-page on various graph products},
  \url{https://en.wikipedia.org/wiki/Graph_product}.

\bibitem[Gri05]{Grif05}
David~J. Griffiths, \emph{Introduction to {Quantum} {Mechanics}}, 2 ed.,
  Pearson, 2005.

\bibitem[Han09]{Hansen06}
Ernst Hansen, \emph{Measure {Theory}}, 4 ed., University of Copenhagen,
  Department of Mathematical Sciences, 2009.

\bibitem[Har01]{Hard01}
Lucien Hardy, \emph{Quantum theory from five reasonable axioms}, arXiv preprint
  quant-ph/0101012 (2001).

\bibitem[Har07]{Hardy07}
Lucien Hardy, \emph{Towards quantum gravity: a framework for probabilistic
  theories with non-fixed causal structure}, Journal of Physics A: Mathematical
  and Theoretical \textbf{40} (2007), no.~12, 3081--3099.

\bibitem[Har10]{Hardy10}
\bysame, \emph{A formalism-local framework for general probabilistic theories
  including quantum theory}, arXiv preprint arXiv:1005.5164 (2010).

\bibitem[Has09]{Hase09}
Masahito Hasegawa, \emph{On traced monoidal closed categories}, Mathematical
  Structures in Computer Science \textbf{19} (2009), no.~2, 217--244.

\bibitem[HPS08]{Crypt}
Jeffrey Hoffstein, Jill Pipher, and Joseph~H. Silverman, \emph{An introduction
  to mathematical cryptography}, vol.~1, Springer, 2008.

\bibitem[Jam72]{Jam72}
Andrzej Jamio{\l}kowski, \emph{Linear transformations which preserve trace and
  positive semidefiniteness of operators}, Reports on Mathematical Physics
  \textbf{3} (1972), no.~4, 275--278.

\bibitem[JLF13]{Jiang13}
Min Jiang, Shunlong Luo, and Shuangshuang Fu, \emph{Channel-state duality},
  Physical Review A \textbf{87} (2013), no.~2, 022310.

\bibitem[JNV{\etalchar{+}}20]{MIP20}
Zhengfeng Ji, Anand Natarajan, Thomas Vidick, John Wright, and Henry Yuen,
  \emph{{MIP}*={RE}}, arXiv preprint arXiv:2001.04383 (2020).

\bibitem[JS91]{JS91}
Andr{\'e} Joyal and Ross Street, \emph{The {Geometry} of {Tensor Calculus},
  {I}}, Advances in mathematics \textbf{88} (1991), no.~1, 55--112.

\bibitem[JSV96]{JSV96}
Andr{\'e} Joyal, Ross Street, and Dominic Verity, \emph{Traced {Monoidal}
  {Categories}}, Mathematical Proceedings of the Cambridge Philosophical
  Society, Cambridge University Press, 1996, pp.~447--468.

\bibitem[Kan17]{Kan17}
J{\k{e}}drzej Kaniewski, \emph{Self-testing of binary observables based on
  commutation}, Physical Review A \textbf{95} (2017), no.~6, 062323.

\bibitem[KSW08a]{KSW08math}
Dennis Kretschmann, Dirk Schlingemann, and Reinhard~F. Werner, \emph{A
  continuity theorem for {Stinespring's} dilation}, Journal of Functional
  Analysis \textbf{255} (2008), no.~8, 1889--1904.

\bibitem[KSW08b]{KSW08phys}
\bysame, \emph{The information-disturbance tradeoff and the continuity of
  {Stinespring's} representation}, IEEE transactions on information theory
  \textbf{54} (2008), no.~4, 1708--1717.

\bibitem[KU17]{Kiss17}
Aleks Kissinger and Sander Uijlen, \emph{A categorical semantics for causal
  structure}, 2017 32nd Annual ACM/IEEE Symposium on Logic in Computer Science
  (LICS), IEEE, 2017, pp.~1--12.

\bibitem[Kuh12]{Kuhn12}
Thomas~S. Kuhn, \emph{The {Structure} of {Scientific} {Revolutions}},
  University of Chicago press, 2012.

\bibitem[Kun80]{Kunen80}
Kenneth Kunen, \emph{Set theory -- an introduction to independence proofs},
  Elsevier, 1980.

\bibitem[Lan61]{Landauer61}
Rolf Landauer, \emph{Irreversibility and heat generation in the computing
  process}, IBM journal of research and development \textbf{5} (1961), no.~3,
  183--191.

\bibitem[Law62]{Law62}
Francis~William Lawvere, \emph{The category of probabilistic mappings}, Lecture
  Notes available online at
  \url{https://ncatlab.org/nlab/files/lawvereprobability1962.pdf}, 1962.

\bibitem[Law73]{Law73}
Francis~William Lawvere, \emph{Metric spaces, generalized logic, and closed
  categories}, Rendiconti del seminario mat{\'e}matico e fisico di {Milano}
  \textbf{43} (1973), no.~1, 135--166.

\bibitem[LC97]{Lo97com}
Hoi-Kwong Lo and Hoi~Fung Chau, \emph{Is quantum bit commitment really
  possible?}, Physical Review Letters \textbf{78} (1997), no.~17, 3410--3413.

\bibitem[Lo97]{Lo97insec}
Hoi-Kwong Lo, \emph{Insecurity of quantum secure computations}, Physical Review
  A \textbf{56} (1997), no.~2, 1154--1162.

\bibitem[Mac17]{Galileo}
Peter Machamer, \emph{{Galileo} {Galilei}}, The Stanford Encyclopedia of
  Philosophy (Edward~N. Zalta, ed.), Metaphysics Research Lab, Stanford
  University, summer 2017 ed., 2017.

\bibitem[May97]{Mayers97}
Dominic Mayers, \emph{Unconditionally secure quantum bit commitment is
  impossible}, Physical review letters \textbf{78} (1997), no.~17, 3414--3417.

\bibitem[Mic]{MichelsonThomson}
\emph{Quote by {Albert} {A.} {Michelson}},
  \url{https://en.wikiquote.org/wiki/Albert_A._Michelson}.

\bibitem[ML13]{MacLane}
Saunders Mac~Lane, \emph{Categories for the {Working} {Mathematician}}, vol.~5,
  Springer Science \& Business Media, 2013.

\bibitem[MPSS18]{Modi18}
Kavan Modi, Arun~Kumar Pati, Aditi Sen(De), and Ujjwal Sen, \emph{Masking
  quantum information is impossible}, Physical review letters \textbf{120}
  (2018), no.~23, 230501.

\bibitem[MS13]{Miller13}
Carl~A. Miller and Yaoyun Shi, \emph{Optimal robust self-testing by binary
  nonlocal {XOR} games}, 8th Conference on the Theory of Quantum Computation,
  Communication and Cryptography, 2013, pp.~254--262.

\bibitem[MY98]{MY98}
Dominic Mayers and Andrew Yao, \emph{Quantum cryptography with imperfect
  apparatus}, Proceedings 39th Annual Symposium on Foundations of Computer
  Science (Cat. No. 98CB36280), IEEE, 1998, pp.~503--509.

\bibitem[MY04]{MY04}
\bysame, \emph{Self testing quantum apparatus}, Quantum Information \&
  Computation \textbf{4} (2004), no.~4, 273--286.

\bibitem[MYS12]{MYS12}
Matthew McKague, Tzyh~Haur Yang, and Valerio Scarani, \emph{Robust self-testing
  of the singlet}, Journal of Physics A: Mathematical and Theoretical
  \textbf{45} (2012), no.~45, 455304.

\bibitem[Nai40]{Neum40}
Mark Naimark, \emph{Spectral functions of a symmetric operator}, Izvestiya
  Rossiiskoi Akademii Nauk. Seriya Matematicheskaya \textbf{4} (1940), no.~3,
  277--318.

\bibitem[NC02]{NC02}
Michael Nielsen and Isaac Chuang, \emph{Quantum {Computation} and {Quantum}
  {Information}}, American Association of Physics Teachers, 2002.

\bibitem[OCB12]{Oresh12}
Ognyan Oreshkov, Fabio Costa, and {\v{C}}aslav Brukner, \emph{Quantum
  correlations with no causal order}, Nature communications \textbf{3} (2012),
  no.~1, 1--8.

\bibitem[PB00]{Pati00}
Arun~Kumar Pati and Samuel~L. Braunstein, \emph{Impossibility of deleting an
  unknown quantum state}, Nature \textbf{404} (2000), no.~6774, 164--165.

\bibitem[Pea90]{Peano1890}
Giuseppe Peano, \emph{Sur une courbe, qui remplit toute une aire plane},
  Mathematische Annalen \textbf{36} (1890), no.~1, 157--160.

\bibitem[Pen71]{Pen71}
Roger Penrose, \emph{Applications of negative dimensional tensors},
  Combinatorial mathematics and its applications \textbf{1} (1971), 221--244.

\bibitem[Per17]{Peri17}
Paolo Perinotti, \emph{Causal structures and the classification of higher order
  quantum computations}, Time in physics, Springer, 2017, pp.~103--127.

\bibitem[PHHH06]{Piani06}
Marco Piani, Michal Horodecki, Pawel Horodecki, and Ryszard Horodecki,
  \emph{Properties of quantum nonsignaling boxes}, Physical Review A
  \textbf{74} (2006), no.~1, 012305.

\bibitem[Pin18]{Pink18}
Steven Pinker, \emph{Enlightenment now: The case for reason, science, humanism,
  and progress}, Penguin, 2018.

\bibitem[PR92]{PR92}
Sandu Popescu and Daniel Rohrlich, \emph{Which states violate {Bell}'s
  inequality maximally?}, Physics Letters A \textbf{169} (1992), no.~6,
  411--414.

\bibitem[PR94]{PR94}
Sandu Popescu and Daniel Rohrlich, \emph{Quantum nonlocality as an axiom},
  Foundations of Physics \textbf{24} (1994), no.~3, 379--385.

\bibitem[R{\etalchar{+}}18]{Rosl18}
Hans Rosling et~al., \emph{Factfulness: Ten reasons we’re wrong about the
  world -- and why things are better than you think}, Flatiron Books, 2018.

\bibitem[Rus]{Russell}
\emph{Wikipedia-page on {Russell's} paradox},
  \url{https://en.wikipedia.org/wiki/Russell%27s_paradox#cite_note-2}.

\bibitem[RUV13]{RUV13}
Ben~W. Reichardt, Falk Unger, and Umesh Vazirani, \emph{Classical command of
  quantum systems}, Nature \textbf{496} (2013), no.~7446, 456--460.

\bibitem[{\v S}B19]{SB19}
Ivan {\v S}upi{\'c} and Joseph Bowles, \emph{Self-testing of quantum systems: a
  review}, arXiv preprint arXiv:1904.10042 (2019).

\bibitem[Sch]{SchroderBernstein}
\emph{Wikipedia-page on {Cantor-Schr\"{o}der-Bernstein} theorem},
  \url{https://en.wikipedia.org/wiki/Schroeder-Bernstein_theorem}.

\bibitem[Sch17]{Schill17}
Ren{\'e}~L. Schilling, \emph{Measures, {Integrals} and {Martingales}},
  Cambridge University Press, 2017.

\bibitem[Sel04]{Sel04}
Peter Selinger, \emph{Towards a semantics for higher-order quantum
  computation}, Proceedings of the 2nd International Workshop on Quantum
  Programming Languages, TUCS General Publication, vol.~33, Citeseer, 2004,
  pp.~127--143.

\bibitem[Sel10]{Sel10survey}
\bysame, \emph{A survey of graphical languages for monoidal categories}, New
  structures for physics, Springer, 2010, pp.~289--355.

\bibitem[Sem]{SemicartesianWebsite}
\emph{\emph{Semicartesian Monoidal Categories} on n{Lab}},
  \url{https://ncatlab.org/nlab/show/semicartesian+monoidal+category}.

\bibitem[{Sha}48]{Shannon48}
Claude~E. {Shannon}, \emph{A mathematical theory of communication}, The Bell
  System Technical Journal \textbf{27} (1948), no.~3, 379--423.

\bibitem[Sti55]{Stine55}
William~Forrest Stinespring, \emph{Positive functions on {C*}-algebras},
  Proceedings of the American Mathematical Society \textbf{6} (1955), no.~2,
  211--216.

\bibitem[SW87]{SW87}
Stephen~J. Summers and Reinhard~F. Werner, \emph{Maximal violation of {Bell}'s
  inequalities is generic in quantum field theory}, Communications in
  Mathematical Physics \textbf{110} (1987), no.~2, 247--259.

\bibitem[TCR10]{Toma10}
Marco Tomamichel, Roger Colbeck, and Renato Renner, \emph{Duality between
  smooth min- and max-entropies}, IEEE Transactions on information theory
  \textbf{56} (2010), no.~9, 4674--4681.

\bibitem[Thi]{ThinCat}
\emph{\emph{Thin Categories} on n{Lab}},
  \url{https://ncatlab.org/nlab/show/thin+category}.

\bibitem[Tom12]{Toma12}
Marco Tomamichel, \emph{A framework for non-asymptotic quantum information
  theory}, Ph.D. thesis, ETH Z\"{u}rich, 2012.

\bibitem[vD13]{vanD13}
Wim van Dam, \emph{Implausible consequences of superstrong nonlocality},
  Natural Computing \textbf{12} (2013), no.~1, 9--12.

\bibitem[Wat]{Wat11}
John Watrous, \emph{{CS} 766/{QIC} 820 {Theory of Quantum Information} ({Fall}
  2011)}, Lecture notes available online at
  \url{https://cs.uwaterloo.ca/~watrous/TQI-notes/TQI-notes.pdf}.

\bibitem[Wei72]{Weier1872}
Karl Weierstrass, \emph{\"{U}ber continuirliche {Functionen} eines reellen
  {Arguments}, die f\"{u}r keinen {Werth} des letzteren einen bestimmten
  {Differentailqutienten} besitzen}, Math. Werke (1872), 71--74.

\bibitem[Wie83]{Wiesner83}
Stephen Wiesner, \emph{Conjugate coding}, ACM Sigact News \textbf{15} (1983),
  no.~1, 78--88.

\bibitem[Wol19]{Wolf19}
Michael Wolf, \emph{Mathematical introduction to quantum information processing
  (growing lecture notes, {SS2019})}, Lecture notes available online at
  \url{https://www-m5.ma.tum.de/foswiki/pub/M5/Allgemeines/MA5057_2019S/QIPlecture.pdf},
  2019.

\bibitem[WZ82]{Woot82}
William~K. Wootters and Wojciech~H. Zurek, \emph{A single quantum cannot be
  cloned}, Nature \textbf{299} (1982), no.~5886, 802--803.

\end{thebibliography}

\end{document}